%% file: 3DCoupledWireCellularTopo_v5.tex
\documentclass[aps,prb,twocolumn,floatfix,superscriptaddress]{revtex4-2}
\usepackage{amsmath,amssymb}
\usepackage{graphicx}
\usepackage[bookmarks=false,colorlinks=true,linkcolor=blue,filecolor=blue,citecolor=blue,urlcolor=blue]{hyperref}
\usepackage{bm}

\allowdisplaybreaks[1]
\newcommand{\red}[1]{{\textcolor{red}{#1}}}

\newcommand{\sgn}{\, \textrm{sgn}}

\begin{document}

\title{Bridging three-dimensional coupled-wire models and cellular topological states: Solvable models for topological and fracton orders}

\author{Yohei Fuji}
\affiliation{Department of Applied Physics, University of Tokyo, Tokyo 113-8656, Japan}

\author{Akira Furusaki}
\affiliation{Condensed Matter Theory Laboratory, RIKEN CPR, Wako, Saitama 351-0198, Japan}
\affiliation{RIKEN Center for Emergent Matter Science, Wako, Saitama 351-0198, Japan}

\date{\today}

\begin{abstract}
Three-dimensional (3D) gapped topological phases with fractional excitations are divided into two subclasses: 
one has topological order with point-like and loop-like excitations fully mobile in the 3D space, and the other has fracton order with point-like excitations constrained in lower-dimensional subspaces. 
These exotic phases are often studied by exactly solvable Hamiltonians made of commuting projectors, which, however, are not capable of describing those exhibiting surface states with gapless chiral dispersion. 
Here we introduce a systematic way, based on cellular construction recently proposed for 3D topological phases, to construct another type of exactly solvable models in terms of coupled quantum wires with given inputs of cellular structure, two-dimensional Abelian topological order, and their gapped interfaces. 
We show that our models can describe both 3D topological and fracton orders (and even their hybrid) and study their universal properties such as quasiparticle statistics and topological ground-state degeneracy. 
We also apply this construction to two-dimensional coupled-wire models with ordinary topological orders and translation-symmetry-enriched topological orders. 
Our results pave the way for effective quantum field theory descriptions or microscopic model realizations of fracton orders with chiral gapless surface states.
\end{abstract}

\maketitle
\tableofcontents

\section{Introduction}

Strongly interacting quantum many-body systems can realize topological phases of matter with fractionalized quasiparticle excitations. 
The search for those topological phases is an active area of current physics research not only from fundamental physics perspectives but also from their potential applications to robust quantum memories and fault-tolerant quantum computation \cite{Kitaev03, Nayak08, Terhal15, Brown16}. 
Representative examples of topological phases are the fractional quantum Hall states \cite{Tsui82, Laughlin83}, in which electrons are fractionalized into point-like quasiparticles obeying nontrivial braiding statistics. 
As a consequence of the fractionalization, such a topological system has degenerate ground states when it is placed on a torus or a closed manifold with higher genus. 
These phenomena are known as \emph{topological order} \cite{XGWen89, XGWen90, XGWen17}. 
While quasiparticles can only be point-like objects in two-dimensional (2D) topologically ordered phases, they can be either point-like or loop-like objects with nontrivial braiding statistics in three-dimensional (3D) topologically ordered phases \cite{TLan18, TLan19, QRWang19}. 
It is believed that these topologically ordered phases are effectively described by topological quantum field theory (TQFT). 

In three dimensions, however, it has recently been recognized that there exist topological phases beyond common wisdom of the topological order \cite{Chamon05, Bravyi11, Haah11, Castelnovo12, IHKim12, Yoshida13, Haah14, Vijay15}. 
Such phases exhibit point-like quasiparticles completely immobile or mobile only within lower-dimensional subspaces of the 3D space. 
Depending on the mobility, those quasiparticles are called \emph{fractons}, \emph{lineons}, and \emph{planons}, which refer to quasiparticles mobile only within zero-, one-, and two-dimensional subspaces, respectively.
The presence of such low-dimensional quasiparticles results in ground-state degeneracy subextensively growing with the increase of the system size. 
These features are contrasted with the existence of fully mobile quasiparticles and system-size-independent ground-state degeneracy in topologically ordered phases and are coined \emph{fracton order} \cite{Nandkishore19, Pretko20}. 
The fracton phases evade an effective description in terms of TQFT, and novel quantum field theories capturing essential features of fracton order have been proposed and studied \cite{Pretko17, Slagle17b, Bulmash18a, Bulmash18b, HMa18b, Pretko18a, Pretko18b, Gromov19, Slagle19a, Slagle19b, Gorantla20, JWang20a, Seiberg20a, Seiberg20b, Shenoy20, YYou20a, YYou20b, Fontana21a, Fontana21b, Gorantla21a, Gorantla21b, Gorantla21c, JWang21a, JWang21b, MQi21, Rudelius21, Seiberg21a, Seiberg21b, Slagle21}.

In developing our understanding of basic properties of topologically ordered phases and fracton phases, exactly solvable models are of primary importance. 
They also play a pivotal role in exploring microscopic origins and experimental realizations of topological or fracton order, since strong interactions among constituent degrees of freedom substantially limit possible theoretical approaches to microscopic models. 
While exactly solvable lattice models often involve complicated multibody interactions, nontrivial 3D models exhibiting topological or fracton order with simpler interactions have been constructed from coupled 2D layers of topological phases or coupled spin chains \cite{Halasz17, HMa17, Slagle17a, THHsieh17, Vijay17, Fuji19c, Prem19, Schmitz19b, Shirley20, Williamson21, Aasen20b, XGWen20, JWang20b}. 
A natural question is to what extent such exactly solvable models are capable of describing topological and fracton orders. 
It has been shown that 2D lattice models consisting of commuting projectors cannot realize topological phases with nonzero electric or thermal Hall conductance \cite{Kapustin20a, Kapustin20b} and also that 2D frustration-free spin models cannot support gapless chiral edge states \cite{Lemm19}. 
These facts severely limit the construction of 3D lattice models from coupled layers of 2D topological phases and possibly indicate the existence of 3D topological or fracton orders that fall outside the description in terms of exactly solvable lattice models. 

However, there is yet another class of exactly solvable models defined on an array of one-dimensional (1D) quantum wires, where one direction is continuum whereas the other directions are discrete. 
This so-called coupled-wire construction \cite{Meng20} has been successfully applied to 2D topological phases, including chiral ones, such as fractional quantum Hall states \cite{Kane02, Teo14, Meng14, Neupert14, Sagi15a, Fuji17, Kane17, Kane18, Fuji19a, Sirota19, Ebisu20, PMTam20, PMTam21a} and quantum spin liquids \cite{Gorohovsky15, Meng15a, Patel16, PHHuang16, JHChen17, Lecheminant17, PHHuang17, Pereira18, JHChen19, CLi20}. 
With the helps of bosonization technique and conformal field theory (CFT) \cite{Giamarchi, GNT, Fradkin, dFMS, James18}, this construction yields exactly solvable models in the strong-coupling limit and allows us to study universal features of topological order. 
Since the continuum theory of quantum wires can in principle be derived from the low-energy limits of 1D lattice systems, the coupled-wire approach has been used to investigate microscopic realizations of 2D interacting topological phases in lattice systems \cite{Gorohovsky15, Fuji16, Fuji17, JHChen17, Lecheminant17, PHHuang17, Pereira18, JHChen19}. 
Furthermore, the coupled-wire approach allows us to derive effective gauge theory for topological phases \cite{Mross16, Mross17, Fontana19, Fuji19a, Imamura19, Leviatan20, Turker20} and thus provides a direct bridge between quantum field theory and microscopic models. 
Although it has also been used to construct 3D topological phases \cite{Meng15b, Sagi15b, Iadecola16, Volpez17, MJPark18, Raza19, Fuji19b, Iadecola19, Sullivan21a, Sullivan21b, May-Mann22, JHZhang23}, its versatility for 3D topological and fracton orders is not fully explored yet.

In this paper, we propose a systematic way to construct a family of exactly solvable coupled-wire models that describe 3D topological and fracton orders. 
Our construction is based on a recently proposed approach for building fracton topological orders, called the \emph{topological defect network} by Aasen, Bulmash, Prem, Slagle, and Williamson \cite{Aasen20b} or the \emph{cellular topological state} by Wen \cite{XGWen20} and Wang \cite{JWang20b}. 
In this approach, one first decomposes the 3D space into small 3D cells, places 2D and/or 3D topologically ordered states on each cell, and couples them via nontrivial gapped interfaces. 
The gapped interfaces dictate the mobility of quasiparticles between neighboring cells, and the resulting 3D states exhibit a variety of quasiparticle dynamics as observed in fracton models, leading to the conjecture that this approach produces all gapped fracton phases \cite{Aasen20b}. 
Aside from its conceptual significance, it will also be used to build 3D microscopic models for fracton order when exactly solvable models are known for constituent topologically ordered states and their gapped interfaces. 
The coupled-wire construction is precisely suited for this purpose and enables us to find exactly solvable models corresponding to topological defect networks/cellular topological states built upon 2D topologically ordered states. 
In particular, it can produce solvable models built from chiral topological phases, which may not be realized in lattice systems made of commuting projectors and may thus constitute a novel family of fracton order. 

In the rest of this section, we summarize the results of this paper. 
Before proceeding, a comment is in order.
Since our construction exclusively uses thin strips of 2D topologically ordered states as building blocks, we choose to use the terminology of cellular topological states throughout this paper because of the similarity to the construction in Refs.~\cite{XGWen20, JWang20b} rather than to the one in Ref.~\cite{Aasen20b}. 
However, we emphasize that the topological defect network \cite{Aasen20b} and the cellular topological states \cite{XGWen20, JWang20b} share essentially the same idea for constructing fracton phases as both use topologically ordered phases and their gapped interfaces as basic ingredients. 
The use of cellular topological states in this paper is nothing more than the choice of terminology.

\subsection*{Summary of results}

\begin{figure*}
\includegraphics[clip,width=0.8\textwidth]{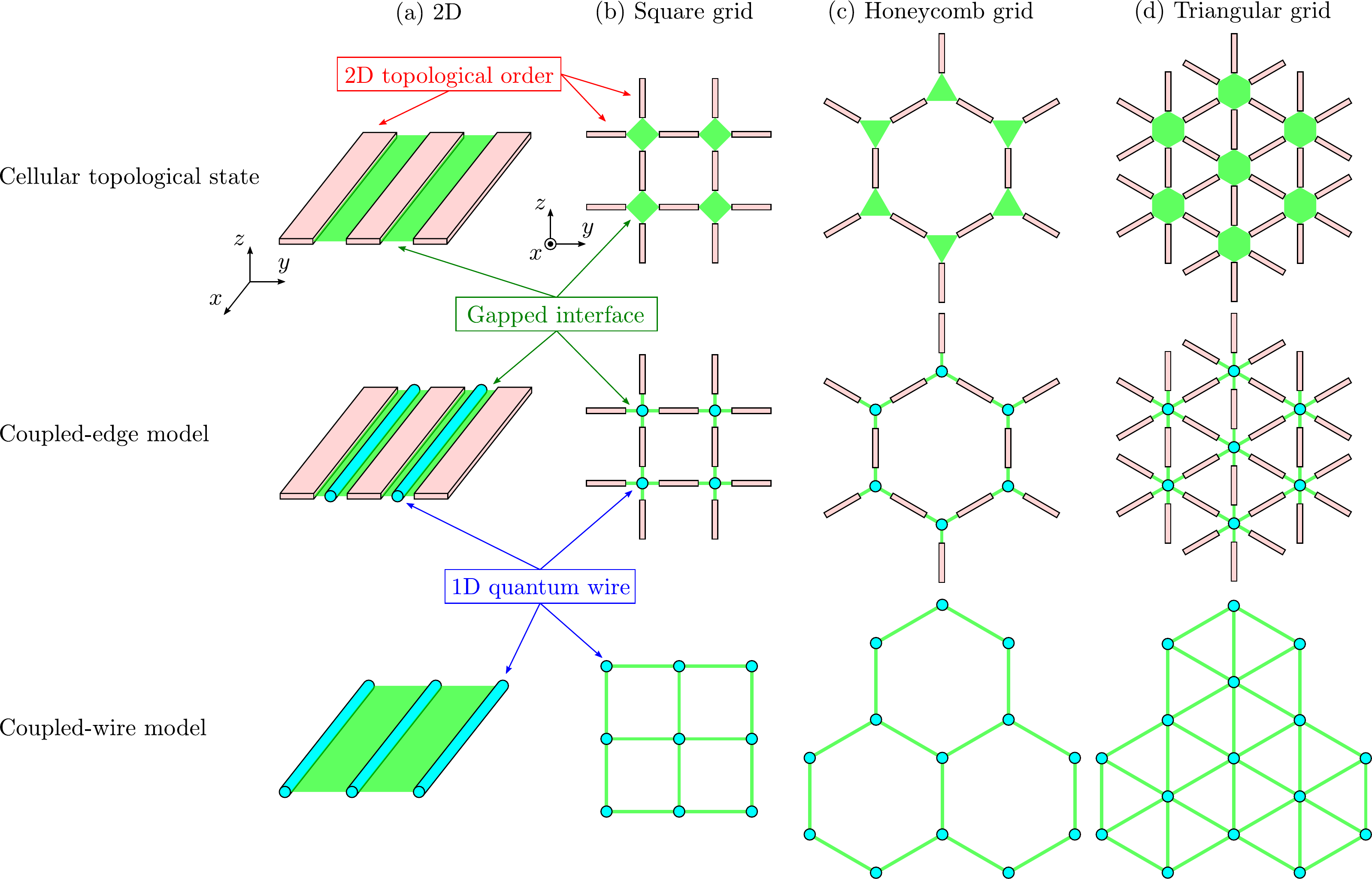}
\caption{
Generic recipe to obtain coupled-wire models from cellular topological states made of strips of 2D Abelian topological orders and their gapped interfaces. 
We consider cellular decompositions of (a) a 2D plane into an array of thin strips extended along the $x$ axis, and a 3D space into (b) the square, (c) honeycomb, and (d) triangular grid of thin strips projected onto the $yz$ plane. 
Cellular topological states may be obtained by directly coupling gapless edge states from strips meeting at each interface to open a gap and to induce the condensation of anyons. 
In general, such coupled-edge models require extra quantum wires added at each interface. 
One can choose interactions at the interface such that the edge modes from the strips are not directly coupled with each other but their interactions are only mediated by the additional quantum wires. 
At the final step, strips of 2D topological orders are shrunk and removed, yielding coupled-wire models with the same low-energy properties as those of the original cellular topological states.
}
\label{fig:CellularTopoToCW}
\end{figure*}

\emph{Coupled-wire models}---A primary goal of this paper is to provide a general framework for obtaining exactly solvable coupled-wire models for given data sets of cellular topological states. 
We consider a 3D system consisting of thin 2D strips, which extend along the $x$ axis and form a grid structure in the $yz$ plane. 
Examples of such 3D cellular topological states on the square, honeycomb, and triangular grids are illustrated in the top row of Fig.~\ref{fig:CellularTopoToCW}, in addition to a 2D cellular topological state defined on the 2D array of strips.
We place 2D topologically ordered states on each strip and couple them together by 1D gapped interfaces, at which edge modes along the $x$ axis from neighboring strips interact with each other and open an excitation gap in the energy spectrum. 
In this paper, we only consider cellular topological states made of 2D \emph{Abelian} topological orders, for which a simple algebraic framework in terms of the so-called \emph{K matrix} is available to describe quasiparticle properties \cite{XGWen95}. 
For each 1D interface labeled by its 2D coordinate in the $yz$ plane, i.e., $\bm{r}=(y,z)$, we can describe the edge modes from the neighboring 2D strips with a symmetric integer matrix $K_{\textsf{e},\bm{r}}$ and associate the 1D gapped interface with a set of integer vectors $L_{\bm{r}}$ called the \emph{Lagrangian subgroup} \cite{Kapustin11, Levin13, Barkeshli13a, Barkeshli13b, Kapustin14, JCWang15}. 
Thus, the input data for building a cellular topological state are the grid structure formed by strips and $K_{\textsf{e},\bm{r}}$ and $L_{\bm{r}}$ at each interface.

With these input data, we can construct a coupled-wire model for the corresponding cellular topological state. 
For demonstration purposes, here we consider only a cellular topological state on the square grid, but the construction can be easily extended to other grid structures. 
We place 2D topological orders described by an $N_0 \times N_0$ integer matrix $K_0$ on each strip in a translation invariant manner. 
Supposing that the strips are wide enough that the hybridization between two edges is negligible compared with the bulk gap, gapless edge modes at each interface are described by the $4N_0 \times 4N_0$ matrix $K_{\textsf{e},\mathbf{r}} \equiv K_\textsf{e}$ with
\begin{align}
K_{\textsf{e}} = \begin{pmatrix} K_0 &&& \\ & K_0 && \\ && -K_0 & \\ &&& -K_0 \end{pmatrix}, 
\end{align}
where the diagonal entries correspond to edge modes from the left, bottom, right, and top strips surrounding the interface in the ascending order. 
We then gap them out by placing a gapped interface described by the Lagrangian subgroup $L_{\bm{r}} \equiv L$, which can be generated by a set of $4N_0$-dimensional integer vectors $M = \{ \bm{m}_a \}$ as 
\begin{align}
L = \left\{ \sum_a p_a \bm{m}_a \, \middle| \, p_a \in \mathbb{Z}, \bm{m}_a \in M \right\}.
\end{align}
This gives a set of \emph{condensed quasiparticles} between neighboring strips and thus determines how quasiparticles move from one strip to another across the interface.
Given $L$, we can always find a set of $8N_0$-dimensional integer vectors $\{ \widetilde{\bm{\Lambda}}_\alpha \}$ to write down a tunneling Hamiltonian of the form
\begin{align}
\mathcal{V}_\textsf{ew} = -g \int dx \sum_{\bm{r}} \sum_{\alpha=1}^{4N_0} \cos [\widetilde{\bm{\Lambda}}_\alpha^T K_\textsf{ew} \bm{\phi}^\textsf{ew}_{\bm{r}}(x)]
\end{align}
with some coupling constant $g$. 
Here, $K_\textsf{ew}$ and $\bm{\phi}^\textsf{ew}_{\bm{r}}(x)$ represent an $8N_0 \times 8N_0$ $K$ matrix and $8N_0$-component bosonic fields, respectively, for the original $4N_0$ edge modes and extra $4N_0$ gapless modes from $2N_0$ quantum wires added at each interface.
This hybrid system of the edge modes and the additional quantum wires is regarded as a \emph{coupled-edge model} for the cellular topological state, as depicted in the middle row of Fig.~\ref{fig:CellularTopoToCW}.
This coupled-edge model provides an exactly solvable Hamiltonian in the strong-coupling limit $g \to \infty$, but it is actually redundant and lacks the microscopic description of the edge modes at the lattice level. 

In fact, we can always find a set of $4N_0$-dimensional integer vectors $\{ \bm{\Lambda}_{\textsf{w},\alpha} \}$ corresponding to $L$, which gives a tunneling Hamiltonian written only in terms of the bosonic fields in the added quantum wires, 
\begin{align}
\mathcal{V}_\textsf{w} = -g \int dx \sum_{\bm{r}} \sum_{a=1}^{N_0} \bigl( \cos \Theta^\textsf{w}_{\bm{r}+\bm{e}_y/2,a} +\cos \Theta^\textsf{w}_{\bm{r}+\bm{e}_z/2,a} \bigr),
\end{align}
where we have defined
\begin{align}
\begin{split}
\Theta^\textsf{w}_{\bm{r}+\bm{e}_y/2,a} &= \bm{\Lambda}^T_{\textsf{w},2N_0+a} K_\textsf{w} \bm{\phi}^\textsf{w}_{\bm{r}} +\bm{\Lambda}^T_{\textsf{w},a} K_\textsf{w} \bm{\phi}^\textsf{w}_{\bm{r} +\bm{e}_y}, \\
\Theta^\textsf{w}_{\bm{r}+\bm{e}_z/2,a} &= \bm{\Lambda}_{\textsf{w}, 3N_0+a}^T K_\textsf{w} \bm{\phi}^\textsf{w}_{\bm{r}} +\bm{\Lambda}_{\textsf{w}, N_0+a}^T K_\textsf{w} \bm{\phi}^\textsf{w}_{\bm{r}+\bm{e}_z},
\end{split}
\end{align}
$\bm{e}_y=(1,0)$, and $\bm{e}_z=(0,1)$. 
Here, $K_\textsf{w}$ and $\bm{\phi}^\textsf{w}_{\bm{r}}$ are a $4N_0 \times 4N_0$ $K$ matrix and $4N_0$-component bosonic fields, respectively, for $2N_0$ quantum wires at each interface. 
This gives a \emph{coupled-wire model} for the cellular topological state, as shown in the bottom row of Fig.~\ref{fig:CellularTopoToCW}, which is exactly solvable in the strong-coupling limit $g \to \infty$. 
It can be derived from the coupled-edge model with a special choice of $\{ \widetilde{\bm{\Lambda}}_\alpha \}$ and through a process of shrinking and removing the strips of 2D topological orders. 
The resulting coupled-wire model is represented either by bosonic quantum wires when $L$ contains only condensed quasiparticles with bosonic self statistics, or by fermionic quantum wires when $L$ contains at least one condensed quasiparticle with fermionic self statistics. 
Quasiparticle excitations moving across an interface are created by a vertex operator $\exp [i\bm{p} \cdot \bm{\phi}^\textsf{w}_{\bm{r}}(x)]$ with an integer vector $\bm{p}$, which has one-to-one correspondence with an element of $L$. 
On the other hand, quasiparticle excitations along a 2D strip are created in pairs by a string operator $\exp [i\bm{q} \cdot \int_{x_0}^{x_1} dx \, \partial_x \bm{\phi}^\textsf{w}_{\bm{r}}(x)]$ with some vector $\bm{q}$. 
These vertex and string operators have genuine microscopic origins in the sense that they can in principle be written as the continuum limits of some lattice operators in 1D systems.

\emph{Applications to topological and fracton orders}---We apply the above procedure to cellular topological states built from $U(1)_k$ topological orders, which are the simplest chiral topological orders with $K_0 = k$ ($1\times1$ matrix), and provide their coupled-wire models. 
To do so, we first classify all possible gapped interfaces between four $U(1)_k$ topological orders for small $k$ and those between six $U(1)_3$ topological orders, which will naturally constitute 3D cellular topological states on the square and triangular grid, respectively. 
The results are presented in the Supplemental Material (SM) \cite{suppl}.
We found several nontrivial gapped interfaces, which lead to 2D and 3D cellular topological states with intriguing quasiparticle properties as found in topological and fracton orders. 

For 2D cellular topological states, we consider the 2D array built from strips of 2D topological orders described by a symmetric integer matrix $K_0$. 
When those strips are coupled via gapped interfaces that induce the condensation of pairs of a quasiparticle and its anti-partner between neighboring strips, we find conventional 2D coupled-wire models for the 2D topological order $K_0$ as studied in Refs.~\cite{Kane02, Teo14, Patel16}. 
On the other hand, if we consider gapped interfaces that induce the condensation of nontrivial pairs of quasiparticles between neighboring strips, the 2D coupled-wire models exhibit the ground-state degeneracy depending on the linear size of a torus along the $y$ axis. 
This is a characteristic feature of \emph{translation-symmetry-enriched topological orders}, which has recently been discussed in the context of coupled-wire models \cite{PMTam21a, PMTam21b}.

For 3D cellular topological states, we adapt a classification scheme proposed by Ref.~\cite{Dua19} for 3D topological and fracton orders. 
Since our cellular topological states inevitably have quasiparticles mobile along strips, they cannot realize type-II fracton orders, which possess only completely immobile, fracton excitations. 
Nevertheless, we can still find a variety of conventional topological orders, which have constant ground-state degeneracy on a torus and are dubbed as the \emph{TQFT-type topological orders}, and type-I fracton orders with planons and lineons. 
The type-I fracton order is further divided into two subclasses: \emph{foliated} type-I fracton order characterized by subextensive ground-state degeneracy and \emph{fractal} type-I fracton order characterized by fluctuating ground-state degeneracy with subextensive envelope \cite{Dua19}. 

We present 3D coupled-wire models that exhibit the following topological and fracton orders: 
(i) foliated type-I fracton order with only planons, 
(ii) foliated type-I fracton order with both lineons and planons, for which a dipole of lineons moves along directions perpendicular to the dipole axis and thus becomes a planon, 
(iii) TQFT-type topological order with 3D point-like and 3D loop-like excitations exhibiting nontrivial mutual braiding statistics, 
(iv) hybrid of TQFT-type topological order and foliated type-I fracton order, for which not only 3D point-like excitations but also planons have nontrivial braiding statistics with 3D loop-like excitations, and
(v) fractal type-I fracton order with lineons, for which the mobility of quasiparticles is highly restricted in the $yz$ plane and leads to ground-state degeneracy with fractal-like growth in the system size. 
We remark that coupled-wire models realizing (iii) and (iv) have ground-state degeneracy extensively growing with the number of wires due to local loops of 3D point-like excitations, which can be subsequently lifted by adding appropriate local perturbations to yield genuine 3D loop-like excitations as expected in TQFT-type topological orders. 
These results are summarized in Table~\ref{tab:ListOfFractonOrder}.
\begin{table*}
\begin{ruledtabular}
\begin{tabular}{llllll}
Type & Excitations & $K_0$ & $L$ & Degeneracy on a torus & Section \\
\hline
2D TQFT-type TO & 2D point & $\begin{pmatrix} 0 & 2 \\ 2 & 0 \end{pmatrix}$ & $L^\textrm{(2TC)}_\textrm{2B;1}$ & $2^2$ & Sec.~\ref{sec:CW2dTC} \\
 & & $k$ & $\{ (l,-l) \, | \, l \in \mathbb{Z} \}$ & $k$ & Sec.~\ref{sec:CW2dU1k} \\
\hline
\begin{tabular}{l} 2D translation symmetry \\ enriched TO \end{tabular} & 2D point & $\begin{pmatrix} 0 & 2 \\ 2 & 0 \end{pmatrix}$ & $L^\textrm{(2TC)}_\textrm{2F;3}$ & $\begin{cases} 2^2 & (L_y \in 2\mathbb{Z}) \\ 2 & (L_y \in 2\mathbb{Z}+1) \end{cases}$ & Sec.~\ref{sec:CW2DTCef} \\
 & & $\begin{pmatrix} 2 & 0 \\ 0 & -2 \end{pmatrix}$ & $L^{(2,2,\overline{2},\overline{2})}_\textrm{2F;1}$  & $\begin{cases} 2^2 & (L_y \in 2\mathbb{Z}) \\ 2 & (L_y \in 2\mathbb{Z}+1) \end{cases}$ & Sec.~\ref{sec:2DFermionicDoubledSemion} \\
 & & $\begin{pmatrix} 5 & 0 \\ 0 & -5 \end{pmatrix}$ & $L^{(5,5,\overline{5},\overline{5})}_\textrm{2F;9}$ & $\begin{cases} 5^2 & (L_y \in 4\mathbb{Z}) \\ 1 & (L_y \notin 4\mathbb{Z}) \end{cases}$ & Sec.~\ref{sec:2dU5Model} \\
\hline
3D TQFT-type TO & 3D point and 3D loop & $4$ [Sq] & $L^{(4,4,\overline{4},\overline{4})}_\textrm{3B;1}$ & $4 \cdot 2^{Ly L_z} \to 2^3$ & Sec.~\ref{sec:BosonicU4Model} \\
 & & $9$ [Sq] & $L^{(9,9,\overline{9},\overline{9})}_\textrm{3F;11}$ & $9 \cdot 3^{L_yL_z} \to 3^3$ & Appendix~\ref{sec:FermionicU9Model} \\ 
\hline
3D foliated type-I FO & Planon & $2$ [Sq] & $L^{(2,2,\overline{2},\overline{2})}_\textrm{2F;1}$ & $2^{\textrm{gcd}(L_y,L_z)}$ & Sec.~\ref{sec:FermionicU2Model} \\
 & & $6$ [Sq] & $L^{(6,6,\overline{6},\overline{6})}_\textrm{2F;6}$ & $3^{L_y+L_z} \cdot 2^{\textrm{gcd}(L_y,L_z)}$ & Sec.~\ref{sec:FermionicU6Model} \\
 & & $6$ [Sq] & $L^{(6,6,\overline{6},\overline{6})}_\textrm{2B;9}$ & $3^{L_y+L_z} \cdot 2^{\textrm{gcd}(L_y,L_z)}$ & Appendix~\ref{sec:BosonicU6Model} \\ 
 & Lineon and planon & $4$ [Sq] & $L^{(4,4,\overline{4},\overline{4})}_\textrm{2F;8}$ & $2^2 \cdot 4^{L_y+L_z-2}$ & Sec.~\ref{sec:DipoleU4Model} \\
 & & $8$ [Sq] & $L^{(8,8,\overline{8},\overline{8})}_\textrm{2B;16}$ & $4^2 \cdot 8^{L_y+L_z-2}$ & Appendix~\ref{sec:DipoleU8Model} \\
\hline
\begin{tabular}{l} Hybrid of 3D TO and \\ 3D foliated type-I FO \end{tabular} & \begin{tabular}{l} 3D point, 3D loop, \\ and planon \end{tabular} & $8$ [Sq] & $L^{(8,8,\overline{8},\overline{8})}_\textrm{3B;13}$ & \begin{tabular}{l} $4 \cdot 2^{L_yL_z +L_y+L_z}$ \\ $\to 2^3 \cdot 2^{L_y+L_z}$ \end{tabular} & Sec.~\ref{sec:HybridU8Model} \\
\hline
3D fractal type-I FO & Lineon & $7$ [Sq] & $L^{(7,7,\overline{7},\overline{7})}_\textrm{2F;9}$ & Fluctuating & Sec.~\ref{sec:FermionicU7Model} \\
 & & $3$ [Tri] & $L^{(3,3,3,\overline{3},\overline{3},\overline{3})}_\textrm{3F;23}$ & Fluctuating & Sec.~\ref{sec:FermionicU3Model}
\end{tabular}
\end{ruledtabular}
\label{tab:ListOfFractonOrder}
\caption{List of topological orders (TOs) and fracton orders (FOs) constructed in this paper. 
Third column indicates the $K$ matrix for 2D topological orders placed on strips. 
For the 3D cases, the grid structures are also written in square brackets: square grid [Sq] or triangular grid [Tri]. 
Forth column indicates the Lagrangian subgroup $L$ used for gapped interfaces and symbols are adopted from the SM \cite{suppl}. 
Fifth column indicates ground-state degeneracy on an $L_x \times L_y$ torus for the 2D cases and on an $L_x \times L_y \times L_z$ torus for the 3D cases. 
Expressions after right arrows ``$\to$'' are parts of degeneracy robust under local perturbations.}
\end{table*}

\emph{Comparison with previous studies}---We here make a comparison between topological and fracton orders found in our coupled-wire models and those in some previous results. 
The dipole nature of planon excitations seen in the foliated type-I fracton order (ii) is similar to what has been found in several exactly solvable lattice models, such as the X-cube model \cite{Vijay16} and the anisotropic planon-lineon model \cite{Shirley19a}. 
Hybrids of topological and fracton orders with nontrivial planon-loop statistics as found in (iv) have recently been discussed in the context of lattice models \cite{Tantivasadakarn21a, Tantivasadakarn21b}. 
Our coupled-wire models for the fractal type-I fracton order (v) are similar to the Sierpinski fractal spin liquid, which has been proposed in Refs.~\cite{Castelnovo12, Yoshida13} and constructed from 3D cellular topological states based on the 2D or 3D toric codes in Refs.~\cite{XGWen20, Aasen20b}, as quasiparticles spread on a fractal lattice in the $yz$ plane but freely move along the $x$ axis.. 
Quasiparticle properties of our fractal type-I fracton models in the $yz$ plane can be emulated by 2D classical spin liquids and are related to 1D cellular automata, which have already been utilized in Ref.~\cite{Yoshida13} to construct commuting projector Hamiltonians for a variety of quantum fractal spin liquids.
It is also important to notice that 3D cellular topological states constructed in this paper from $U(1)_k$ topological orders on the square grid generally possess 2D gapless surface states composed of multiple 1D chiral gapless modes with the same chirality under appropriate surface terminations. 
Such surface states cannot be gapped out by any local perturbations and imply that there are no exactly solvable lattice models for these cellular topological states.

There are also several recent attempts to construct exactly solvable coupled-wire models for 3D topological and fracton orders \cite{Fuji19b, Sullivan21a, Sullivan21b, May-Mann22, JHZhang23}. 
In Ref.~\cite{Sullivan21a}, Sullivan, Iadecola, and Williamson have constructed 3D coupled-wire models for foliated type-I fracton order with fractons. 
Their models can be understood from p-string condensation mechanism \cite{HMa17, Vijay17} for two orthogonal stacks of 2D topological orders and can also be interpreted as 3D cellular topological states made of 2D topological orders placed on three orthogonal faces of cubes in the cubic lattice. 
As a result, their coupled-wire models can support immobile fracton excitations, while our coupled-wire models cannot, as any quasiparticles are mobile along the wires due to the original cellular structure. 
Another distinguishing feature is that their coupled-wire models have bulk gapless excitations on a torus, whereas our models are fully gapped. 
It has been discussed that such gapless excitations are also present in earlier coupled-wire models for higher-dimensional Abelian topological orders \cite{Iadecola16}. 
In Ref.~\cite{Sullivan21b}, Sullivan, Dua, and Cheng have constructed coupled-wire models for both gapped and gapless fracton orders. 
For gapped cases, the corresponding coupled-wire models have planon and/or lineon excitations as in foliated type-I fracton order. 
However, their models have quasiparticles with infinite-order fusion properties, which do not match with any known construction of cellular topological states. 
Thus, these coupled-wire models belong to different classes from coupled-wire models constructed in this paper. 
It has also been argued \cite{Sullivan21c} that some of the former models are described by infinite-component Chern-Simons theories \cite{XMa20}. 

Meanwhile, our models include the coupled-wire models proposed in Ref.~\cite{Fuji19b} by the present authors. 
There, we have utilized conformal embeddings for (1+1)D CFTs to construct coupled-wire models, for which the associated branching rules dictate the mobility of excitations. 
We have originally interpreted this construction as coupled-wire implementations of anyon condensation used for a coupled-layer construction of 3D topological orders \cite{CMJian14}. 
It turns out that the resulting models can be naturally understood within the framework of cellular topological states. 
Existence of a nontrivial conformal embedding implies existence of a nontrivial gapped interface between 2D topological orders associated with the embedding CFTs. 
For example, the conformal embedding $U(1)_k \times U(1)_k \supset U(1)_{2k} \times U(1)_{2k}$ implies the existence of a nontrivial gapped interface between four Abelian topological orders described by the $U(1)_{2k}$ CFT, which can be used to construct a 3D cellular topological state from strips of the $U(1)_{2k}$ topological orders on the square grid. 
We demonstrate that the corresponding coupled-wire model precisely reproduces a model constructed in Ref.~\cite{Fuji19b}. 
We further show that the same conformal embedding implies the existence of a nontrivial gapped interface between the $U(1)_k \times U(1)_k$ and $U(1)_{2k} \times U(1)_{2k}$ topological orders, which allows us to construct equivalent 3D cellular topological states from strips of the $U(1)_k$ and $U(1)_{2k}$ topological orders on the honeycomb grid. 

References~\cite{May-Mann22, JHZhang23} provided coupled-wire constructions of strongly interacting 3D topological phases protected by subsystem symmetries, i.e., symmetries acting only on lower-dimensional manifolds. 
While 2D surface states of these phases can be gapped, 1D hinge states localized along the intersection of orthogonal 2D surfaces cannot be gapped in the presence of the subsystem symmetries. 
The coupled-wire constructions yielded both unfractionalized \cite{May-Mann22, JHZhang23} and fractionalized phases \cite{May-Mann22}. 
A particular model constructed in Ref.~\cite{May-Mann22} for the latter case can actually be regarded as a cellular topological state consisting of 2D strips of the doubled $U(1)_k$ topological orders and their 1D gapped interfaces. 
It hosts lineon and planon excitations and has subextensively degenerate ground states on the 3D torus, similarly to the coupled-wire model of foliated type-I fracton order constructed in Sec.~\ref{sec:FoliatedTypeILineon}.

\subsection*{Outline of the paper}

The rest of the paper is organized as follows. 
In Sec.~\ref{sec:Preliminaries}, we review the general theory for 2D Abelian topological orders and their gapped boundaries and how they are applied to gapped interfaces. 
In Sec.~\ref{sec:2DCWModel}, we present a systematic construction of coupled-wire models from 2D cellular topological states and discuss their applications to conventional 2D topological orders and translation-symmetry-enriched topological orders. 
In Sec.~\ref{sec:CellularTopo3D}, we generalize the construction for 3D cellular topological states and provide explicit examples of coupled-wire models for 3D TQFT-type topological orders, 3D foliated and fractal type-I fracton orders, and hybrid of topological and fracton orders. 
Section~\ref{sec:Conclusion} concludes this paper with several outlooks. 

Appendix~\ref{sec:LambdaWire} provides an explicit way to construct a special form of integer vectors used for deriving tunneling interactions in coupled-wire models. 
In Appendix~\ref{sec:Some3DModels}, we discuss several 3D models that have similar quasiparticle properties with those constructed in Sec.~\ref{sec:CellularTopo3D}. 
In Appendix~\ref{sec:CellularTopoHoneycomb}, we argue that some of 3D models on the square grid can be equivalently formulated on the honeycomb grid by utilizing their conformal embedding structures. 

\section{Preliminaries}
\label{sec:Preliminaries}

We first review the general theory for 2D Abelian topological order (Sec.~\ref{sec:2dTopologicalOrder}) and its gapped boundary (Sec.~\ref{sec:GappedBoundary}). 
A gapped interface between 2D topological orders can be viewed as a gapped boundary for a stack of the 2D topological orders by the folding trick (Sec.~\ref{sec:GappedInterface}). 
These will be main building blocks for cellular topological states studied in this paper.

\subsection{2D Abelian topological order}
\label{sec:2dTopologicalOrder}

The low-energy effective theory of 2D Abelian topological order is given by an $N$-component $U(1)$ Chern-Simons theory \cite{XGWen95}, whose Lagrangian density is defined by 
\begin{align} \label{eq:ChernSimons}
\mathcal{L}_\textrm{CS} = -\sum_{I,J=1}^N \frac{K_{IJ}}{4\pi} \epsilon^{\mu \nu \lambda} a^I_\mu \partial_\nu a^J_\lambda,
\end{align}
where $K$ is a symmetric, nondegenerate, $N \times N$ integer matrix and $a^I_\mu$ are $U(1)$ gauge fields. 
For bosonic topological orders, the $K$ matrix has only even integers in the diagonal entries. 
For fermionic topological orders, the $K$ matrix has at least one odd integer in the diagonal entries. 
A quasiparticle excitation is associated with an $N$-dimensional integer vector $\bm{l}$ and carries integer charges $l_I$ of the gauge fields $a^I_\mu$. 
The statistical angle $\theta_{\bm{l}}$ of the quasiparticle $\bm{l}$ is given by 
\begin{align}
\theta_{\bm{l}} = \pi \bm{l}^T K^{-1} \bm{l}.
\end{align}
It is also related to the statistical spin $s_{\bm{l}}$ via $\theta_{\bm{l}} = 2\pi s_{\bm{l}}$. 
The mutual statistics $\theta_{\bm{l},\bm{l}'}$ between two quasiparticles $\bm{l}$ and $\bm{l}'$ is given by 
\begin{align}
\theta_{\bm{l},\bm{l}'} = 2\pi \bm{l}^T K^{-1} \bm{l}'.
\end{align}
Local bosonic or fermionic excitations must have trivial statistics and correspond to $\bm{l} \in K \mathbb{Z}^N$. 
Quasiparticle excitations with nontrivial statistics, which are called anyon excitations, are associated with $\bm{l} \notin K\mathbb{Z}^N$. 
The number of anyon excitations distinguished up to addition of local particle excitations, or equivalently the number of distinct $\bm{l}$'s up to the identification $\bm{l} \sim \bm{l} +K \mathbb{Z}^N$, is given by $|\!\det K|$, which is also equal to the number of degenerate ground states when the system is placed on a 2D torus.

When the system is placed on a manifold with a boundary, there exist gapless edge modes along the boundary, which are described by free boson CFT with the action: 
\begin{align} \label{eq:EdgeAction}
\mathcal{S}_\textrm{edge} = -\frac{1}{4\pi} \int dtdx \sum_{I,J=1}^N
 \left( K_{IJ} \partial_t \phi_I \partial_x \phi_J +v_{IJ} \partial_x \phi_I \partial_x \phi_J \right). 
\end{align}
Here, $v_{IJ}$ is a positive definite matrix associated with the velocities of the edge modes and depends on microscopic details.
The number of right-moving and left-moving bosonic modes, which we denote by $c_R$ and $c_L$, respectively, is given by the number of positive and negative eigenvalues of the matrix $K$, respectively. 
The signature of $K$, i.e., the number of positive eigenvalues minus the number of negative ones, corresponds to the chiral central charge:
\begin{align}
c=c_R-c_L.
\end{align} 
Quasiparticle excitations at the boundary are created by vertex operators: 
\begin{align}
\Psi = \exp \left( i \sum_{I=1}^N l_I \phi_I \right),
\end{align}
where $\bm{l} \in K \mathbb{Z}^N$ corresponds to a local particle excitation while $\bm{l} \notin K\mathbb{Z}^N$ to an anyon excitation. 
The conformal spin of an operator $\Psi$ associated with $\bm{l}$ is precisely given by the statistical spin $s_{\bm{l}}$. 
Therefore, those with integer conformal spins are bosonic, whereas those with half-integer conformal spins are fermionic. 

In the following discussion, we prefer to use the Hamiltonian formalism. 
The Hamiltonian corresponding to the edge action \eqref{eq:EdgeAction} is given by 
\begin{align} \label{eq:EdgeHamiltonian}
\mathcal{H}_\textrm{edge} = \frac{1}{4\pi} \int dx \sum_{I,J=1}^N v_{IJ} \partial_x \phi_I \partial_x \phi_J, 
\end{align}
where the bosonic fields $\phi_I(x)$ obey the equal-time commutation relations:
\begin{align} \label{eq:EdgeCommutationRelation}
[\partial_x \phi_I(x), \phi_J(x')] = 2\pi i K^{-1}_{IJ} \delta (x-x'),
\end{align}
where $K^{-1}_{IJ}$ is the $(I,J)$ entry of the inverse matrix of $K$.
Although the Hamiltonian \eqref{eq:EdgeHamiltonian} has gapless excitations, one can generally add interactions consisting of local particle operators. 
If the signature of $K$, or equivalently the chiral central charge $c$, vanishes, such interactions have a chance to fully gap out the edge modes; the precise condition is discussed in the next subsection. 
When $c\ne0$, there are stable chiral edge modes even in the presence of interactions. 
We call topological orders with nonvanishing $c$ chiral, whereas those with vanishing $c$ nonchiral in this paper.

We provide several examples of the 2D Abelian topological orders. 
The topological order of the 2D toric code or the 2D $Z_2$ gauge theory \cite{Kitaev03} is described by the $2 \times 2$ $K$ matrix \cite{XGWen16} of the form
\begin{align} \label{eq:KMatrixToricCode}
K = \begin{pmatrix} 0 & 2 \\ 2 & 0 \end{pmatrix}.
\end{align}
This is a nonchiral bosonic topological order.
It has three nontrivial quasiparticles $\bm{l}=(1,0)^T$, $(0,1)^T$, and $(1,1)^T$. 
The first two are bosonic but have nontrivial mutual statistics of $\pi$. 
They can thus be identified with an electric ($e$) and a magnetic ($m$) excitation. 
The last one is a fermionic ($f$) excitation obtained by fusing the $e$ and $m$ excitations. 

Another example is the Laughlin state at filling fraction $\nu = 1/k$ \cite{Laughlin83}, which is given by the $1 \times 1$ $K$ matrix \cite{XGWen95}: 
\begin{align}
K=k.
\end{align}
It is also called the $U(1)_k$ topological order in relation with the level-$k$ Chern-Simons theory as given in Eq.~\eqref{eq:ChernSimons}.
It is a chiral bosonic topological order for even $k$, whereas it is a chiral fermionic one for odd $k$. 
The nontrivial quasiparicles are simply given by $\bm{l} = 1,2,3, \cdots, k-1$, which we denote by $\bm{1}$, $\bm{2}$, $\bm{3}$, and so on. 
Its antichiral (time-reversal) partner $\overline{U(1)}_k$ is described by $K=-k$. 
We denote their quasiparticles by $\overline{\bm{1}}$, $\overline{\bm{2}}$, $\overline{\bm{3}}$, and so on. 
This simplest chiral topological order is extensively used for constructing nontrivial 2D and 3D coupled-wire models in the following sections.

\subsection{Gapped boundary}
\label{sec:GappedBoundary}

When a 2D Abelian topological order has nonchiral edge states with $c=0$ or the vanishing signature of the $K$ matrix, there is a possibility to open an excitation gap at the boundary with a suitable choice of interactions. 
In fact, the presence of nonchiral edge states is a necessary but not sufficient condition to have a gapped boundary. 
A general criterion is provided by the existence of the Lagrangian subgroup $L$ \cite{Kapustin11, Levin13, Barkeshli13a, Barkeshli13b, Kapustin14, JCWang15}, which is a set of quasiparticles satisfying the following conditions: 
\begin{enumerate}
\item All quasiparticles in $L$ have bosonic or fermionic self statistics: $\bm{l}^T K^{-1} \bm{l} \in \mathbb{Z}$ for $\bm{l} \in L$. 
\item Any two quasiparticles in $L$ have trivial mutual statistics: $\bm{l}^T K^{-1} \bm{l}' \in \mathbb{Z}$ for $\bm{l}, \bm{l}' \in L$. 
\item Quasiparticles not included in $L$ have nontrivial mutual statistics with at least one quasiparticle in $L$: for $\bm{n} \notin L$, there exits $\bm{l} \in L$ such that $\bm{n}^T K^{-1} \bm{l} \notin \mathbb{Z}$.
\end{enumerate}
A gapped boundary for a given topological order is possible if and only if there exists a Lagrangian subgroup.
Physically, the first two conditions indicate that all quasiparticles in the Lagrangian subgroup can be simultaneously condensed in the sense of anyon condensation \cite{Bais09a}. 
In contrast to the original concept of anyon condensation applied to phase transitions between topological orders, the last condition implies that there are no nontrivial quasiparticles in the resulting condensate; quasiparticles not included in the Lagrangian subgroup are all confined, that is, they become high-energy excitations above the condensate. 
Although the notion of anyon condensation is originally applied to the set of bosonic quasiparticles, it has also been extended to the set of fermionic quasiparticles, which can be condensed by supplementing local fermion excitations via fermionic anyon condensation \cite{Gaiotto16, YWan17, Aasen19, JLou21}. 
When the theory of a topological order and its gapped boundary is entirely described in a bosonic Fock space, we call such a gapped boundary bosonic. 
This is the case for gapped boundaries of bosonic topological orders obtained by bosonic anyon condensation.
On the other hand, when the theory is described in a fermionic Fock space, we call a gapped boundary fermionic. 
This is the case for gapped boundaries of bosonic topological orders obtained by fermionic anyon condensation and any gapped boundaries of fermionic topological orders. 

For gapped boundaries of Abelian topological orders, we can explicitly write down Hamiltonians for gapping potentials in terms of bosonic fields \cite{Levin13, Barkeshli13a, Barkeshli13b}. 
We here assume that $K$ is a symmetric, nondegenerate, $2N \times 2N$ integer matrix in order to have vanishing signature.
Let us denote by $M$ a subset of the Lagrangian subgroup $L$, which contains the minimal number of quasiparticle $\bm{m}$'s that generate all quasiparticles in $L$ by their linear combinations. 
Namely, 
\begin{align}
L = \left\{ \sum_a p_a \bm{m}_a \, \middle| \, p_a \in \mathbb{Z}, \bm{m}_a \in M \right\}.
\end{align}
We call the number of elements in $M$ as the rank of the Lagrangian subgroup $L$. 
If the rank of $L$ is equal to $N$ and we can choose $M= \{ \bm{m}_1, \cdots, \bm{m}_N \}$ to satisfy 
\begin{align}
\bm{m}_a^T K^{-1} \bm{m}_b =0,
\end{align}
then we can define integer vectors by 
\begin{align} \label{eq:VectorLambda}
\bm{\Lambda}_a = c_a K^{-1} \bm{m}_a,
\end{align}
where $c_a$ is the minimal integer such that $\bm{\Lambda}_a$ becomes an integer vector. 
With this set of $N$ integer vectors $\{ \bm{\Lambda}_a \}$, we can find an interaction at the boundary: 
\begin{align} \label{eq:GappingPotential}
\mathcal{V} = -g \int dx \sum_{a=1}^{N} \cos (\bm{\Lambda}^T_a K \bm{\phi}), 
\end{align}
where $g$ is some positive constant, $\bm{\phi} = (\phi_1, \cdots, \phi_{2N})^T$, and $\phi_I$'s are the bosonic fields appearing in the edge Hamiltonian \eqref{eq:EdgeHamiltonian}. 
Since the integer vectors $\bm{\Lambda}_a$ obey 
\begin{align} \label{eq:ConditionLambda}
\bm{\Lambda}_a^T K \bm{\Lambda}_b =0,
\end{align}
the bosonic fields $\bm{\Lambda}^T_a K \bm{\phi}$ commute with each other and their cosine terms too. 
Thus, the interaction $\mathcal{V}$ can open a gap for the gapless edge modes \cite{Haldane95}
by pinning the bosonic fields $\phi_I$ at the minima of the cosine potentials: 
\begin{align}
\bm{\Lambda}^T_a K \bm{\phi} = c_a \bm{m}_a^T \bm{\phi} = 2\pi n_a, 
\end{align}
where $n_a$ is an integer. 
Then, the following vertex operators have finite expectation values at the boundary: 
\begin{align}
\langle e^{i\bm{m}_a^T \bm{\phi}} \rangle = e^{2\pi i n_a/c_a} \neq 0
\end{align}
in the limit of strong coupling $g$, implying that quasiparticles in the Lagrangian subgroup $L$ are condensed at the boundary. 

We remark that the choice of integer vectors $\{ \bm{\Lambda}_a \}$ in Eq.~\eqref{eq:VectorLambda} does not necessarily give the set of primitive integer vectors \cite{Levin12, CWang13}, meaning that it may lead to a gapped boundary with spurious degeneracy due to spontaneous symmetry breaking. 
The primitivity condition is satisfied if and only if the greatest common divisor of $N \times N$ minors of the $2N \times N$ matrix $\Lambda = (\bm{\Lambda}_1,\cdots,\bm{\Lambda}_N)$ is unity. 
This condition is equivalent for the Smith normal form of $\Lambda$ to have only $\pm 1$ in its diagonal entries.
Primitive integer vectors $\{ \bm{\Lambda}_a \}$ are generally related to the Lagrangian subgroup generated by $M = \{ \bm{m}_a \}$ via the condition, 
\begin{align} \label{eq:RelationLambdaToM}
\sum_{b=1}^N B_{ab} K \bm{\Lambda}_b = c_a \bm{m}_a,
\end{align}
for some integers $B_{ab},c_a \in \mathbb{Z}$, which is weaker than Eq.~\eqref{eq:VectorLambda} but is enough to ensure the condensation of quasiparticles in $L$ at the boundary. 
Although the spurious degeneracy arising from nonprimitive vectors can be removed by adding local perturbations at the boundary, we require that the integer vectors $\{ \bm{\Lambda}_a \}$ be primitive in the following analysis for simplicity.

The above procedure of constructing gapping potentials cannot be applied to the general case of gapped boundaries. 
Such situations happen when the rank of the Lagrangian subgroup $L$ is larger than $N$ or when quasiparticles in any choice of the subset $M$ cannot have zero statistical spins. 
In particular, the latter happens for gapped boundaries obtained via fermionic anyon condensation. 
For the general case, we can still find gapping potentials by stacking $N$ trivial nonchiral bosonic topological orders on top of the original one for bosonic gapped boundaries, and by stacking $N$ trivial nonchiral fermionic topological orders for fermionic gapped boundaries. 
This is equivalent to adding $N$ purely 1D (nonanomalous) bosonic or fermionic wires at the boundary. 
Slightly extending the algorithms in Ref.~\cite{Barkeshli13b} to include gapped boundaries with fermionic anyon condensation for bosonic topological orders, we can construct gapping potentials separately for bosonic and fermionic gapped boundaries as we will show below. 

As a side remark, while the integer quantum Hall states or the Kitaev $E_8$ states \cite{Kitaev06, YMLu12} are trivial in the sense that they do not admit bulk anyon excitations, we do not use them as external resources to open a gap; they are chiral topological orders and the corresponding edge states cannot be regarded as purely 1D systems. 
They might be used as resources to construct gapped boundaries for some chiral topological orders, but we do not consider such cases in this paper.
Trivial nonchiral topological orders that we will use for constructing gapping potentials are a stack of integer quantum Hall states with the Chern numbers $+1$ and $-1$ for fermionic boundaries or a bosonic integer quantum Hall state \cite{YMLu12} for bosonic boundaries. 
Since we do not consider any global symmetry in this paper, their edge states can be realized as purely 1D fermionic or bosonic systems.

\subsubsection{Bosonic gapped boundary}
\label{sec:BosonicGappedBoundary}

Let us consider a gapped boundary of a bosonic topological order obtained via bosonic anyon condensation. 
In this case, all diagonal entries of $K$ must be even integers and $\bm{m}^T_a K^{-1} \bm{m}_a \in 2\mathbb{Z}$ for any $\bm{m}_a \in M$. 
We consider a $2N$-dimensional lattice defined by 
\begin{align} \label{eq:QuasiparticleLattice}
\Gamma = \{ \bm{l} +K \bm{q} \ | \ \bm{l} \in L, \bm{q} \in \mathbb{Z}^{2N} \}, 
\end{align}
from which we can define a $2N \times 2N$ integer matrix $U$ such that $\Gamma = U\mathbb{Z}^{2N}$. 
We then define
\begin{align} \label{eq:MatrixP}
P = U^T K^{-1} U,
\end{align}
which is a symmetric unimodular matrix with diagonal entries being all even and with vanishing signature. 
Then, there exists a unimodular matrix $W$ that brings the matrix $P$ into the following form: 
\begin{align}
W^T P W = \begin{pmatrix} 0 & I_N \\ I_N & 0 \end{pmatrix},
\end{align}
where $I_N$ is the $N \times N$ identity matrix. 
We then define the transformed matrices: 
\begin{align} 
\label{eq:MatrixUW}
U' &= W^T U W, \\
K' &= W^T K W, \\
\label{eq:MatrixPW}
P' &= W^T P W.
\end{align}
Since $W$ is unimodular and causes only the basis change, the theories described by $K$ and $K'$ are topologically equivalent and the columns of $U'$ still generate the same Lagrangian subgroup as $M$ does. 
We then introduce a $4N \times 4N$ extended $K$ matrix: 
\begin{align}
\widetilde{K}' = \begin{pmatrix} K' &&& \\ & X && \\ && \ddots & \\ &&& X \end{pmatrix}.
\end{align}
It contains $N$ diagonal blocks of the Pauli matrix $X$, 
\begin{align}
X = \begin{pmatrix} 0 & 1 \\ 1 & 0 \end{pmatrix},
\end{align}
corresponding to stacking of $N$ trivial bosonic topological orders on top of the topological order described by $K'$. 
We may write this matrix compactly as $\widetilde{K}' = K' \oplus X \oplus \cdots \oplus X$ or $\widetilde{K}' = K' \oplus X^{\oplus N}$ in the following discussion. 
In this extended theory, the Lagrangian subgroup can be generated by the following set of $4N$-dimensional integer vectors: 
\begin{align} \label{eq:VectorMW}
\begin{pmatrix} \widetilde{\bm{m}}_1^{\prime T} \\ \widetilde{\bm{m}}_2^{\prime T} \\ \widetilde{\bm{m}}_3^{\prime T} \\ \widetilde{\bm{m}}_4^{\prime T} \\ \vdots \\ \widetilde{\bm{m}}_{2N-1}^{\prime T} \\ \widetilde{\bm{m}}_{2N}^{\prime T} \end{pmatrix} = \begin{pmatrix} 
\bm{u}_1^{\prime T} & 0 & 1 & 0 & 0 & \cdots & 0 & 0 \\
\bm{u}_{N+1}^{\prime T} & -1 & 0 & 0 & 0 & \cdots & 0 & 0 \\
\bm{u}_{2}^{\prime T} & 0 & 0 & 0 & 1 & \cdots & 0 & 0 \\
\bm{u}_{N+2}^{\prime T} & 0 & 0 & -1 & 0 & \cdots & 0 & 0 \\
\vdots & \vdots &&&& \ddots && \vdots \\
\bm{u}_{N}^{\prime T} & 0 & 0 & 0 & 0 & \cdots & 0 & 1 \\
\bm{u}_{2N}^{\prime T} & 0 & 0 & 0 & 0 & \cdots & -1 & 0 \end{pmatrix}, 
\end{align}
where $\bm{u}'_I$ is the $I$th column of $U'$. 
Since the last $2N$ components of $\widetilde{\bm{m}}'_I$ correspond to additions of trivial bosonic degrees of freedom, $\{ \widetilde{\bm{m}}'_I \}$ generate the same Lagrangian subgroup as $M$ does. 
They now satisfy 
\begin{align} \label{eq:MInverseKMW}
\widetilde{\bm{m}}_I^{\prime T} \widetilde{K}^{\prime -1} \widetilde{\bm{m}}'_J =0, 
\end{align}
which can be used to construct a gapping potential. 

For our later purpose, we further consider a transformation generated by a $4N \times 4N$ unimodular matrix: 
\begin{align} \label{eq:ExtendedW}
\widetilde{W} = \begin{pmatrix} W & \\ & I_{2N} \end{pmatrix}, 
\end{align}
where $I_{2N}$ is the $2N \times 2N$ identity matrix. 
Its inverse brings the extended $K$ matrix to 
\begin{align} \label{eq:ExtendedKBoson}
\widetilde{K} = (\widetilde{W}^{-1})^T \widetilde{K}' \widetilde{W}^{-1} = \begin{pmatrix} K &&& \\ & X && \\ && \ddots & \\ &&& X \end{pmatrix}.
\end{align}
We then define 
\begin{align} \label{eq:ExtendedM}
\widetilde{\bm{m}}_I = (\widetilde{W}^{-1})^T \widetilde{\bm{m}}'_I,
\end{align}
which now satisfy
\begin{align}
\widetilde{\bm{m}}^T_I \widetilde{K}^{-1} \widetilde{\bm{m}}_J =0.
\end{align}
With this set of integer vectors $\{ \widetilde{\bm{m}}_I \}$, we can define integer vectors by 
\begin{align} \label{eq:ExtendedLambda}
\widetilde{\bm{\Lambda}}_I = \widetilde{c}_I \widetilde{K}^{-1} \widetilde{\bm{m}}_I, 
\end{align}
where $\widetilde{c}_I$ is the minimal integer such that $\widetilde{\bm{\Lambda}}_I$ becomes an integer vector. 
They satisfy 
\begin{align} \label{eq:ExtendedLambdaKLambda}
\widetilde{\bm{\Lambda}}^T_I \widetilde{K} \widetilde{\bm{\Lambda}}_J =0
\end{align}
Again, the integer vectors $\{ \widetilde{\bm{\Lambda}}_I \}$ defined in Eq.~\eqref{eq:ExtendedLambda} may not satisfy the primitivity condition, but it will be satisfied by taking their appropriate linear combinations.
We can then construct a gapping potential at the boundary:
\begin{align} \label{eq:ExtendedGappingPotential}
\widetilde{\mathcal{V}} = -g \int dx \sum_{I=1}^{2N} \cos (\widetilde{\bm{\Lambda}}_I^T \widetilde{K} \widetilde{\bm{\phi}}),
\end{align}
where $g$ is some constant and $\widetilde{\bm{\phi}} = (\phi_1, \cdots, \phi_{4N})^T$. 
Here, $\phi_I$ with $I=1,\cdots,2N$ correspond to the edge modes of the nontrivial topological order described by $K$ as defined in Eq.~\eqref{eq:EdgeHamiltonian}, whereas those with $I=2N+1,\cdots,4N$ correspond to nonchiral bosonic fields of $N$ purely 1D bosonic wires associated with the diagonal blocks of $X$. 
The advantage of this transformation becomes clear when the topological order described by $K$ itself is a decoupled stack of several topological orders. 
In such a case, each cosine term in the gapping potential is written as a coupling between a local boson operator of each constituent topological order and those of the 1D bosonic wires. 

Importantly, we find that the integer vectors $\{ \widetilde{\bm{\Lambda}}_I \}$ can always be made in the special form, 
\begin{align} \label{eq:LambdaWire}
(\widetilde{\bm{\Lambda}}_1, \cdots, \widetilde{\bm{\Lambda}}_{2N}) = \begin{pmatrix} I_{2N} \\ \Lambda_\textsf{w} \end{pmatrix},
\end{align}
where $I_{2N}$ is the $2N$-dimensional integer matrix and $\Lambda_\textsf{w}$ is a $2N \times 2N$ integer matrix. 
This form ensures that the integer vectors $\{ \widetilde{\bm{\Lambda}}_I \}$ are primitive. 
The explicit construction of $\Lambda_\textsf{w}$ is discussed in Appendix~\ref{sec:LambdaWire}. 
This choice of $\{ \widetilde{\bm{\Lambda}}_I \}$ is extremely useful for writing down minimal coupled-wire models for the associated cellular topological states with bosonic gapped interfaces as we will discuss in the next sections.

\subsubsection{Fermionic gapped boundary}
\label{sec:FermionicGappedBoundary}

Here, we consider a gapped boundary of a bosonic topological order via fermionic anyon condensation or that of a fermionic topological order. 
In the former case, all diagonal entries of $K$ are even but $\bm{m}^T_a K^{-1} \bm{m}_a \in 2\mathbb{Z}+1$ for some $\bm{m}_a \in M$. 
In the latter case, at least one of the diagonal entries of $K$ must be odd. 
In both cases, we can define a $2N$-dimensional lattice $\Gamma$ as in Eq.~\eqref{eq:QuasiparticleLattice} and also a $2N \times 2N$ integer matrix $U$ by $\Gamma = U \mathbb{Z}^{2N}$. 
We can then define the matrix $P$ as in Eq.~\eqref{eq:MatrixP}.
However, reflecting the fact that quasiparticles generating the Lagrangian subgroup now include fermionic anyons, the matrix $P$ becomes a symmetric unimodular matrix with some diagonal entry being odd and with vanishing signature. 
In this case, there exists a unimodular matrix $W$ that brings the matrix $P$ into
\begin{align} \label{eq:MatrixPWFermion}
W^T P W = \begin{pmatrix} I_N & 0 \\ 0 & -I_N \end{pmatrix},
\end{align}
where $I_N$ is again the $N \times N$ identity matrix. 
We define the transformed matrices of $U$, $K$, and $P$ as given in Eqs.~\eqref{eq:MatrixUW}--\eqref{eq:MatrixPW}. 
We then introduce a $4N \times 4N$ extended $K$ matrix by 
\begin{align}
\widetilde{K}' = \begin{pmatrix} K' &&& \\ & Z && \\ && \ddots & \\ &&& Z \end{pmatrix}. 
\end{align}
It contains $N$ diagonal blocks of the Pauli matrix $Z$, 
\begin{align}
Z = \begin{pmatrix} 1 & 0 \\ 0 & -1 \end{pmatrix}, 
\end{align}
corresponding to stacking of $N$ trivial fermionic topological orders on top of the original topological order described by $K$. 
As in the bosonic case, the extended $K$ matrix may be written as $\widetilde{K}' = K' \oplus Z \oplus \cdots \oplus Z$ or $\widetilde{K}' = K' \oplus Z^{\oplus N}$.
In this extended theory, the Lagrangian subgroup can be generated by the following set of $4N$-dimensional integer vectors: 
\begin{align} \label{eq:VectorMWFermion}
\begin{pmatrix} \widetilde{\bm{m}}_1^{\prime T} \\ \widetilde{\bm{m}}_2^{\prime T} \\ \widetilde{\bm{m}}_3^{\prime T} \\ \widetilde{\bm{m}}_4^{\prime T} \\ \vdots \\ \widetilde{\bm{m}}_{2N-1}^{\prime T} \\ \widetilde{\bm{m}}_{2N}^{\prime T} \end{pmatrix} = \begin{pmatrix} 
\bm{u}_1^{\prime T} +\bm{u}_{N+1}^{\prime T} & 1 & 1 & 0 & 0 & \cdots & 0 & 0 \\
\bm{u}_1^{\prime T} -\bm{u}_{N+1}^{\prime T} & -1 & 1 & 0 & 0 & \cdots & 0 & 0 \\
\bm{u}_2^{\prime T} +\bm{u}_{N+2}^{\prime T} & 0 & 0 & 1 & 1 & \cdots & 0 & 0 \\
\bm{u}_2^{\prime T} -\bm{u}_{N+2}^{\prime T} & 0 & 0 & -1 & 1 & \cdots & 0 & 0 \\
\vdots & \vdots &&&& \ddots && \vdots \\
\bm{u}_N^{\prime T} +\bm{u}_{2N}^{\prime T} & 0 & 0 & 0 & 0 & \cdots & 1 & 1 \\
\bm{u}_N^{\prime T} -\bm{u}_{2N}^{\prime T} & 0 & 0 & 0 & 0 & \cdots & -1 & 1 \end{pmatrix}, 
\end{align}
where $\bm{u}'_I$ is the $I$th column of $U'$. 
Since the last $2N$ components of $\widetilde{\bm{m}}'_I$ correspond to additions of trivial fermionic degrees of freedom, $\{ \widetilde{\bm{m}}'_I \}$ generate the same Lagrangian subgroup as $M$ does. 
These integer vectors now satisfy Eq.~\eqref{eq:MInverseKMW}. 
Applying the inverse of the unimodular transformation \eqref{eq:ExtendedW}, the extended $K$ matrix is brought into the form, 
\begin{align} \label{eq:ExtendedKFermion}
\widetilde{K} = (\widetilde{W}^{-1})^T \widetilde{K}' \widetilde{W}^{-1} = \begin{pmatrix} K &&& \\ & Z && \\ && \ddots & \\ &&& Z \end{pmatrix}.
\end{align}
We finally obtain the transformed integer vectors $\{ \widetilde{\bm{m}}_I \}$ and $\{ \widetilde{\bm{\Lambda}}_I \}$ and the gapping potential $\widetilde{\mathcal{V}}$ as given in Eqs.~\eqref{eq:ExtendedM}, \eqref{eq:ExtendedLambda}, and \eqref{eq:ExtendedGappingPotential}, respectively. 
Here, the last $2N$ components of $\widetilde{\bm{\phi}}$ correspond to chiral bosonic fields of $N$ purely 1D fermionic wires associated with the diagonal blocks of $Z$. 
When the topological order described by $K$ is a decoupled stack of 2D topological orders, each cosine term in the gapping potential is written as a coupling between local bosonic or fermionic operator of each constituent topological order and those of the 1D fermionic wires. 
Again, the integer vectors $\{ \widetilde{\bm{\Lambda}}_I \}$ can be made into the special form \eqref{eq:LambdaWire}, whose explicit form for the fermionic gapped boundary is presented in Appendix~\ref{sec:LambdaWire}. 
This can be used to construct coupled-wire models for the associated cellular topological states with fermionic gapped interfaces.

\subsubsection{Examples: Toric code}

As a simple example, we consider gapped boundaries of the 2D toric code, whose $K$ matrix is given in Eq.~\eqref{eq:KMatrixToricCode}. 
There are two bosonic gapped boundaries obtained by condensation of the bosonic quasiparticles $e$ or $m$, which are called the rough and smooth boundaries, respectively \cite{Kitaev12, Bravyi98}. 
The rough boundary is associated with the Lagrangian subgroup generated by 
\begin{align}
M_e = \{ (1,0)^T \},
\end{align}
and is obtained by adding a gaping potential, 
\begin{align}
\mathcal{V}_e = -g \int dx \cos (2\phi_1),
\end{align}
to the edge Hamiltonian \eqref{eq:EdgeHamiltonian}. 
The smooth boundary is associated with the Lagrangian subgroup generated by 
\begin{align}
M_m = \{ (0,1)^T \}, 
\end{align}
and its gapping potential is given by 
\begin{align}
\mathcal{V}_m = -g \int dx \cos (2\phi_2).
\end{align}
There is also a fermionic gapped boundary obtained by condensation of the fermionic quasiparticle $f$. 
The corresponding Lagrangian subgroup is generated by 
\begin{align}
M_f = \{ (1,1)^T \}.
\end{align}
In order to explicitly construct a gapping potential, we need to stack a trivial fermionic topological order on top of the toric code or to add a fermionic wire at the boundary. 
As discussed in Sec.~\ref{sec:FermionicGappedBoundary}, this leads to the extension of the $K$ matrix: 
\begin{align}
\widetilde{K} = \begin{pmatrix} 0 & 2 && \\ 2 & 0 && \\ && 1 & 0 \\ && 0 & -1 \end{pmatrix}.
\end{align}
We can then find a gapping potential, 
\begin{align}
\widetilde{\mathcal{V}}_f &= -g \int dx \bigl[ \cos (2\phi_1 -\phi_3+\phi_4) \nonumber \\
&\quad +\cos (2\phi_2 +\phi_3 +\phi_4) \bigr], 
\end{align}
where $\phi_3$ and $\phi_4$ are associated with the right-moving and left-moving fermionic modes, respectively, added at the boundary. 
This type of gapped boundary obtained via fermionic anyon condensation allows us to convert a physical fermion into an emergent fermion excitation in the bulk of topological order, as recently proposed for the non-Abelian Ising topological order \cite{Aasen20a}; there are also earlier studies in this direction \cite{Barkeshli14, Barkeshli15}.

\subsection{Gapped interface between topological orders}
\label{sec:GappedInterface}

\begin{figure}
\includegraphics[clip,width=0.45\textwidth]{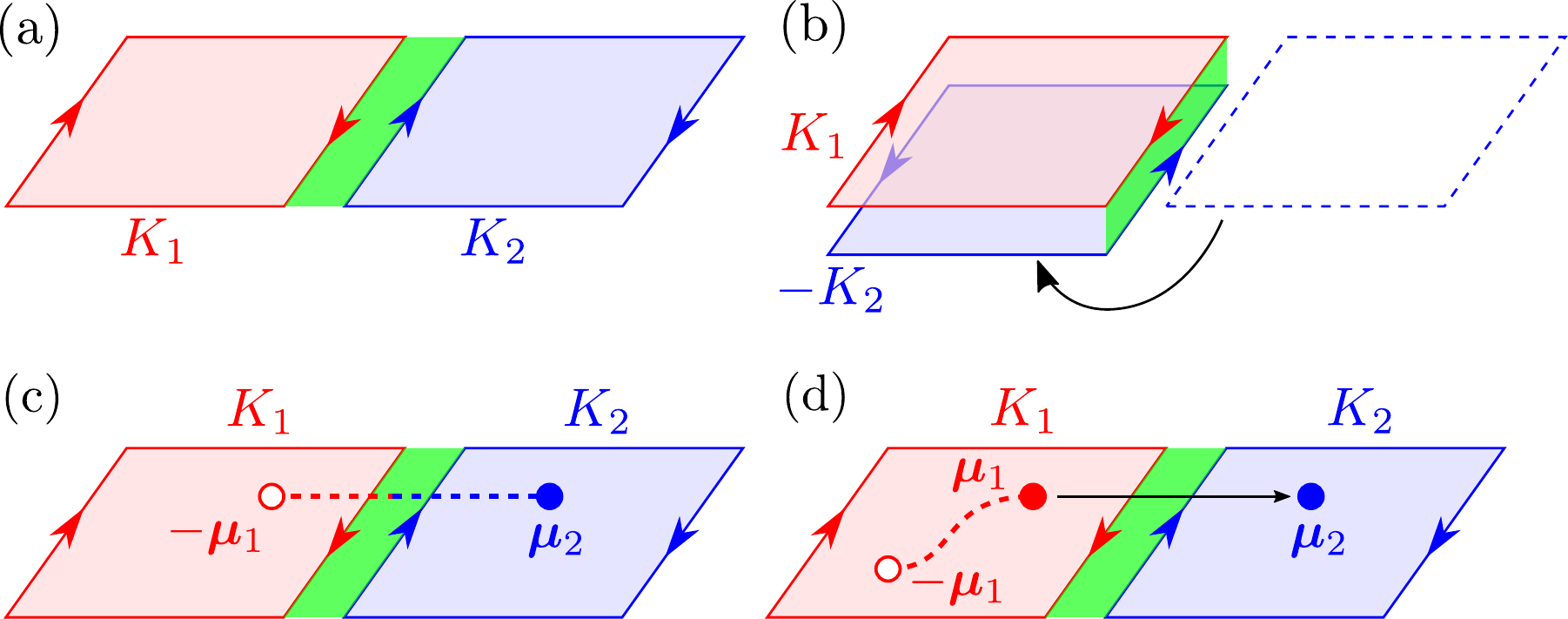}
\caption{(a) Gapped interface between two topological orders described by $K_1$ and $K_2$. 
(b) The folding trick yields a gapped boundary for a stack of topological orders $K_1$ and $-K_2$. 
When the gapped interface admits the condensation of $(-\bm{\mu}_1,\bm{\mu}_2)^T$, (c) a pair of $-\bm{\mu}_1$ and $\bm{\mu}_2$ can be created across the interface or (d) $\bm{\mu}_1$ can go across the interface and convert into $\bm{\mu}_2$.}
\label{fig:GappedInterfaceTwo}
\end{figure}

The above theory of gapped boundary for 2D Abelian topological orders is immediately applied to classification of gapped interfaces between multiple Abelian topological orders. 
Let us consider an interface between two topological orders described by matrices $K_1$ and $K_2$, which are placed on the left and right sides, respectively, of the same 2D plane as depicted in Fig.~\ref{fig:GappedInterfaceTwo}~(a). 
We employ an orientation convention for the topological orders such that their right edges host gapless modes with the commutation relations given by $K_1$ or $K_2$ [see Eq.~\eqref{eq:EdgeCommutationRelation}]. 
Then, their left edges host gapless modes described by $-K_1$ or $-K_2$. 
If $K_1$ and $K_2$ have the same signature, there are equal numbers of left-moving and right-moving edge modes at the interface. 
Hence, the interface might be gapped by adding suitable interactions between the edge modes. 
By the so-called folding trick \cite{Barkeshli13b, LKong14, YWan17}, the problem of such a gapped interface is mapped to that of a gapped boundary for a stacked topological order described by the $K$ matrix, 
\begin{align}
K_\textsf{e} = \begin{pmatrix} K_1 & \\ & -K_2 \end{pmatrix},
\end{align}
as schematically shown in Fig.~\ref{fig:GappedInterfaceTwo}~(b). 
The gapped boundary for the topological order $K_\textsf{e}$ is then classified by the Lagrangian subgroup, as reviewed in Sec.~\ref{sec:GappedBoundary}, which is nothing but a set of condensed quasiparticles at the boundary. 
Going back to the problem of gapped interface, the Lagrangian subgroup tells us which quasiparticle in the topological order $K_1$ can go across the interface and converts into a quasiparticle in the topological order $K_2$. 

Suppose that integer vectors $\bm{\mu}_1$ and $\bm{\mu}_2$ represent quasiparticles in the topological orders $K_1$ and $K_2$, respectively. 
They can be created or annihilated by local operators in pair with their antipartners $-\bm{\mu}_1$ or $-\bm{\mu}_2$. 
By the folding trick, the topological order $K_2$ becomes its mirror image $-K_2$, but we suppose that the mirror image of a quasiparticle $\bm{\mu}_2$ is still described by the same integer vector $\bm{\mu}_2$. 
We now assume that the Lagrangian subgroup for a gapped boundary of the stacked topological order $K_\textsf{e}$ contains a composite of quasiparticles $(-\bm{\mu}_1,\bm{\mu}_2)^T$. 
This means that we can create or annihilate a pair of $-\bm{\mu}_1$ and $\bm{\mu}_2$ at the boundary or at the interface by unfolding [see Fig.~\ref{fig:GappedInterfaceTwo}~(c)]. 
In other words, the quasiparticle $\bm{\mu}_1$ on the topological order $K_1$ can be brought to the interface to fuse with the condensed quasiparticle $(-\bm{\mu}_1, \bm{\mu}_2)^T$ and can be transmuted to the quasiparticle $\bm{\mu}_2$ on the topological order $K_2$ [see Fig.~\ref{fig:GappedInterfaceTwo}~(d)]. 

\begin{figure}
\includegraphics[clip,width=0.45\textwidth]{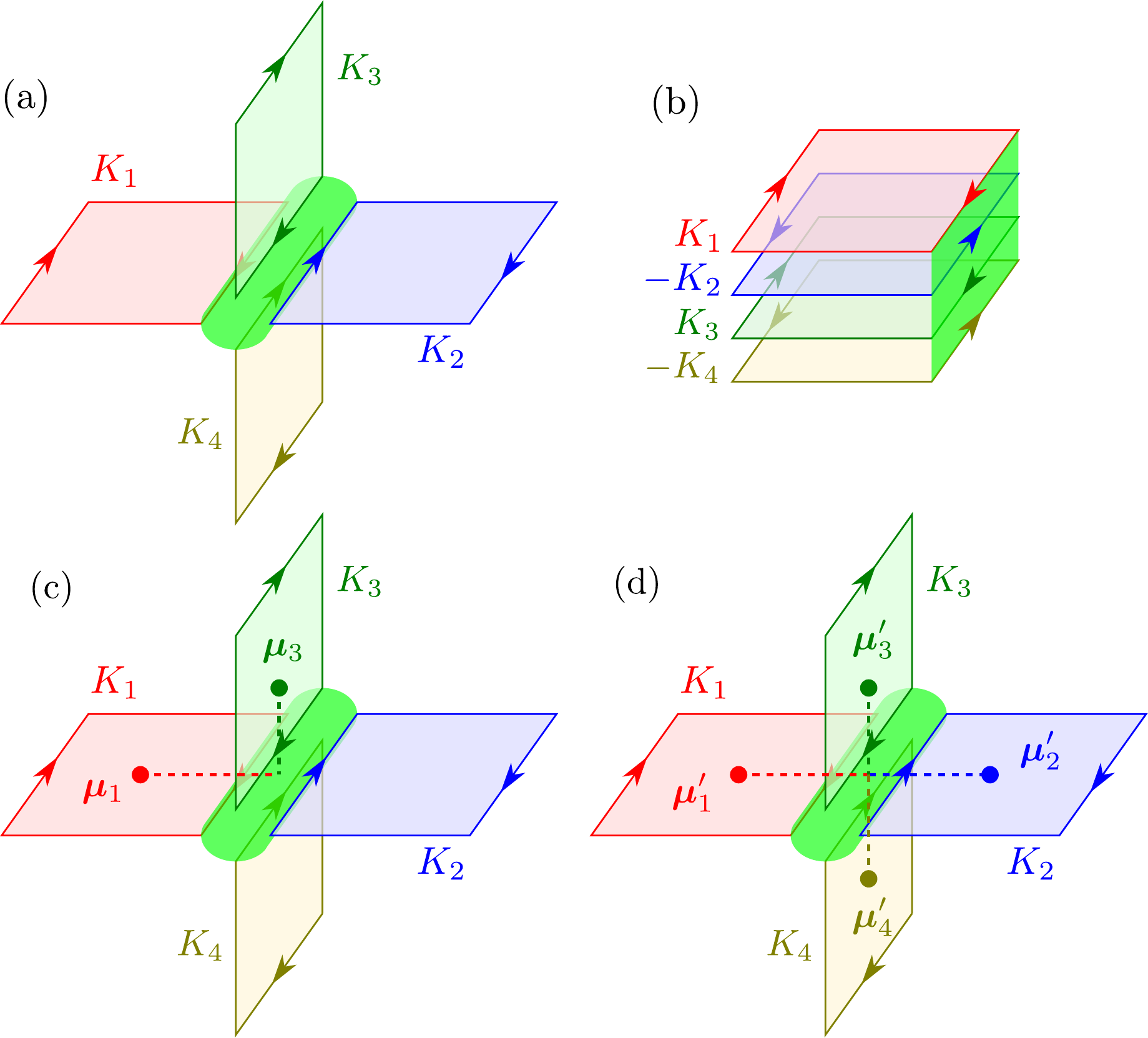}
\caption{(a) Gapped interface connecting four topological orders described by $K_1$, $K_2$, $K_3$, and $K_4$. 
(b) By the folding trick, the gapped interface is mapped to a gapped boundary of a stack of topological orders $K_1$, $-K_2$, $K_3$, and $-K_4$. 
When the gapped interface admits the condensation of $(\bm{\mu}_1, \bm{0}, \bm{\mu}_3, \bm{0})^T$ and $(\bm{\mu}'_1, \bm{\mu}'_2, \bm{\mu}'_3, \bm{\mu}'_4)^T$, (c) a pair of quasiparticles can be created between the $K_1$ and $K_3$ layers, or (d) a quadruplet of quasiparticles can be created among all four layers.}
\label{fig:GappedInterfaceFour}
\end{figure}

This consideration can be easily extended to more complicated interfaces. 
For example, we can consider an interface connecting four topological orders described by the matrices $K_1$, $K_2$, $K_3$, and $K_4$, as shown in Fig.~\ref{fig:GappedInterfaceFour}~(a). 
Here, the topological orders $K_1$ and $K_2$ are placed on the left and right sides, respectively, of the same horizontal plane, while $K_3$ and $K_4$ are placed on the top and bottom sides, respectively, of the same vertical plane. 
The four topological orders meet at the intersection of the two planes. 
We employ an orientation convention such that the right edges on the horizontal plane host gapless bosonic modes with the commutation relations given by $K_1$ or $K_2$, while the bottom edges on the vertical plane host those with the commutation relations given by $K_3$ or $K_4$. 
If all four topological orders have the same signature, the interface have equal numbers of right-moving and left-moving edge modes and thus might be gapped. 
Again by the folding trick [see Fig.~\ref{fig:GappedInterfaceFour}~(b)], the problem of finding a gapped interface is mapped to a problem of finding a gapped boundary for a stacked topological order described by the $K$ matrix, 
\begin{align}
K_\textsf{e} = \begin{pmatrix} K_1 &&& \\ & -K_2 && \\ && K_3 & \\ &&& -K_4 \end{pmatrix}.
\end{align}
Some gapped interfaces might be regarded as a stack of a gapped interface between two topological orders and that between another two, say, one between $K_1$ and $K_2$ and the other between $K_3$ and $K_4$. 
The Lagrangian subgroup corresponding to such an interface will contain bound pairs of quasiparticles, $(-\bm{\mu}_1, \bm{\mu}_2,\bm{0}, \bm{0})^T$ and $(\bm{0}, \bm{0}, -\bm{\mu}_3, \bm{\mu}_4)^T$, where we have denoted quasiparticles in each topological order by $\bm{\mu}_1$, $\bm{\mu}_2$, $\bm{\mu}_3$, and $\bm{\mu}_4$, and trivial vacuum excitations by zero vectors $\bm{0} = (0,\cdots, 0)^T$.
In this case, a pair of quasiparticles can be created between the $K_1$ and $K_2$ layers or the $K_3$ and $K_4$ layers but not between other combinations, say, between the $K_1$ and $K_3$ layers. 

However, when a gapped interface cannot be decomposed into gapped interfaces between two topological orders, pairs of quasiparticles can be created not only between two fixed pairs among four topological orders but also between other pairs. 
In such a case, quasiparticles may be created between, say, the $K_1$ and $K_3$ layers, even though they have the same nonzero signature and thus a direct gapped interface between them is impossible [see Fig.~\ref{fig:GappedInterfaceFour}~(c)]. 
Furthermore, the Lagrangian subgroup may contain bound triplets or quadruplets of quasiparticles, such as $(\bm{\mu}'_1, \bm{\mu}'_2, \bm{\mu}'_3, \bm{\mu}'_4)^T$, which cannot be written as bound objects of more elementary condensed pairs. 
In that case, quasiparticles must be created at the interface in quadruplets among all four topological orders [see Fig.~\ref{fig:GappedInterfaceFour}~(d)]. 
Such nontrivial gapped interfaces are in fact a key ingredient for cellular topological state construction of nontrivial fracton phases as we will discuss in the next sections.

In the SM \cite{suppl}, we provide a thorough classification of gapped interfaces for 2D toric codes and $U(1)_k$ topological orders with small $k$. 
Nontrivial gapped interfaces in the above sense can be found for three or four toric codes, which have been used in Refs.~\cite{Aasen20b, XGWen20, JWang20b} to construct cellular topological states for fracton phases. 
We have also found that nontrivial gapped interfaces exist between four $U(1)_k$ topological orders with $k=2,4,5,6,7,8,9$ and between six $U(1)_3$ topological orders. 
In particular, gapped interfaces between four $U(1)_7$'s or six $U(1)_3$'s admit the condensation of no pairs of quasiparticles but only triplets or quadruplets, which lead to cellular topological states for fractal type-I fracton phases with fluctuating ground-state degeneracy.

\section{2D cellular topological states as coupled-wire models}
\label{sec:2DCWModel}

Before proceeding to the construction of 3D models, we here revisit the coupled-wire construction of 2D topological orders \cite{Kane02, Teo14} in light of cellular topological states. 
There are three steps to find a coupled-wire Hamiltonian from a 2D cellular topological state: 
(i) We first construct a cellular topological state from an array of thin strips of 2D topological orders by coupling them via gapped interfaces, which may be written in terms of gapless edge modes of neighboring strips (Sec.~\ref{sec:2DCellularTopoState}). 
(ii) By adding extra bosonic or fermionic quantum wires at each interface, we construct a coupled-edge model for the cellular topological state, in which the coupling between the edge modes from neighboring strips are mediated by the additional quantum wires (Sec.~\ref{sec:2DCellularAddWire}). 
(iii) By shrinking and removing the strips of 2D topological orders, the added quantum wires between neighboring interfaces are directly coupled, yielding the coupled-wire Hamiltonian for the desired cellular topological state (Sec.~\ref{sec:2DCoupledWireHam}). 

When gapped interfaces are trivial in the sense that pairs of a quasiparticle and its antipartner are condensed, we obtain the conventional coupled-wire models for 2D topological orders, which are exemplified for the 2D toric code and the Laughlin states (Sec.~\ref{sec:2DTrivialModel}). 
On the other hand, if we choose nontrivial gapped interfaces, the resulting coupled-wire models exhibit translation-symmetry-enriched topological order with ground-state degeneracy depending on the number of wires \cite{PMTam21a, PMTam21b}, which are illustrated for the 2D toric code, doubled semion model, and doubled $U(1)_5$ model (Sec.~\ref{sec:2DNontrivialModel}). 

\subsection{2D cellular topological state}
\label{sec:2DCellularTopoState}

\begin{figure}
\includegraphics[clip,width=0.45\textwidth]{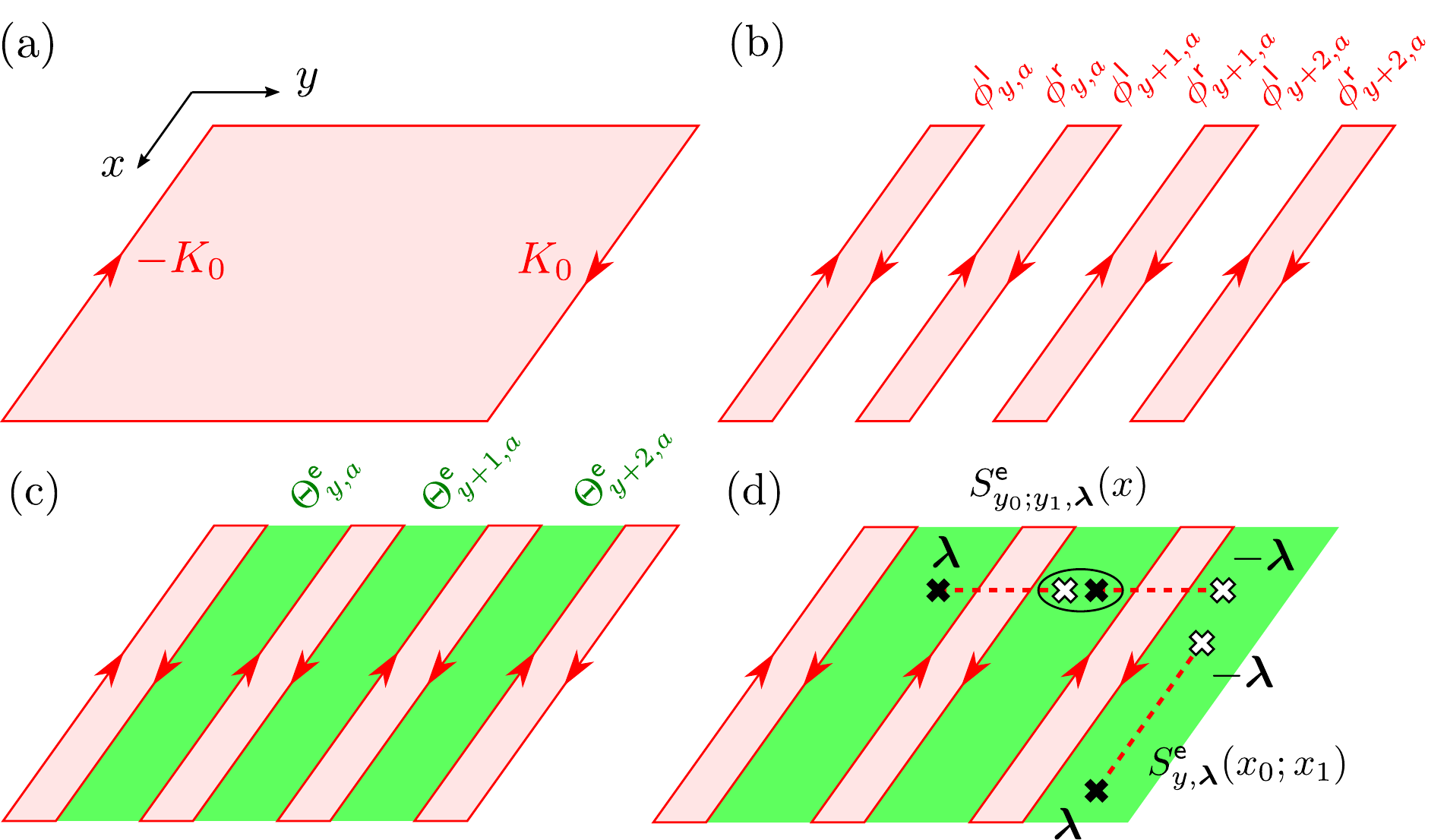}
\caption{(a) A 2D topological order placed in the $xy$ plane has gapless modes described by $K_0$ and $-K_0$ at the right and left edges, respectively. 
(b) The 2D topological order is cut apart along the $x$ axis into thin strips of the topological orders, each of which hosts gapless edge modes $\phi^\textsf{l}_{y+1,a}$ and $\phi^\textsf{r}_{y,a}$ on its right and left edges, respectively. 
(c) Tunneling terms $\cos \Theta^\textsf{e}_{y,a}$ open a gap for the edge modes and induce the condensation of pairs of quasiparticles $(\bm{\lambda}, -\bm{\lambda})^T$ between neighboring strips. 
(d) The action of string operators on the ground state creates pairs of quasiparticles associated with $\bm{\lambda}$ and $-\bm{\lambda}$, which are marked by filled and empty crosses, respectively. 
Two quasiparticles enclosed by a circle in the middle of the string are pair annihilated.}
\label{fig:CutAndGlue2D}
\end{figure}

For a 2D Abelian topological order described by an $N_0 \times N_0$ matrix $K_0$, we consider the following cut-and-glue procedure: 
We first cut the topological order placed in the $xy$ plane apart into thin strips aligned along the $x$ axis [see Figs.~\ref{fig:CutAndGlue2D}~(a) and (b)]. 
Supposing that the strips are wide enough that the hybridization between two edges is negligible compared with the bulk gap, each strip will possess gapless modes on its right and left edges, which are described by the Hamiltonian \eqref{eq:EdgeHamiltonian} corresponding to the matrices $K_0$ and $-K_0$, respectively. 
Quasiparticles created on a strip can move freely along the strip but not across strips.
In order for quasiparticles to move from one strip to another, a quasiparticle $\bm{\lambda}^\textsf{l} \in \mathbb{Z}^{N_0}$ from the left strip should be paired up with a quasiparticle $\bm{\lambda}^\textsf{r} \in \mathbb{Z}^{N_0}$ from the right strip and then condensed together at the interface. 
The corresponding gapped interface is described by the $K$ matrix, 
\begin{align} \label{eq:KmatrixEdge2D}
K_\textsf{e} = \begin{pmatrix} K_0 & \\ & -K_0 \end{pmatrix},
\end{align}
and the Lagrangian subgroup $L= \{ (\bm{\lambda}^\textsf{l}, \bm{\lambda}^\textsf{r})^T \}$. 
If all strips are connected by trivial interfaces at which quasiparticles are always paired with their antipartners, i.e., for $\bm{\lambda}^\textsf{l} = -\bm{\lambda}^\textsf{r} \equiv \bm{\lambda}$, the whole system behaves as a single topological order $K_0$ with the same topological properties as the original one. 
This may be seen as a trivial 2D cellular topological state built out of the strips of 2D topological orders and their gapped interfaces, as schematically shown in Fig.~\ref{fig:CutAndGlue2D}~(c).

In order to verify the above argument, we now explicitly construct the Hamiltonian for a trivial gapped interface $L=\{ (\bm{\lambda}, -\bm{\lambda})^T \, | \, \bm{\lambda} \in \mathbb{Z}^{N_0} \}$. 
Let $\bm{\phi}^\textsf{l}_y = (\phi^\textsf{l}_{y,1}, \cdots, \phi^\textsf{l}_{y,N_0})^T$ be an $N_0$-component bosonic fields describing the right edge modes of the left strip with respect to the $y$th interface and $\bm{\phi}^\textsf{r}_y = (\phi^\textsf{r}_{y,1}, \cdots, \phi^\textsf{r}_{y,N_0})^T$ be an $N_0$-component bosonic field describing the left edge mode of the right strip. 
We then collectively denote them as $\bm{\phi}^\textsf{e}_y = (\bm{\phi}^\textsf{l}_y, \bm{\phi}^\textsf{r}_y)^T$. 
They satisfy the commutation relations: 
\begin{align} \label{eq:CommRel2DCutGlue}
[\partial_x \phi^\textsf{e}_{y,\alpha}(x), \phi^\textsf{e}_{y',\beta}(x')] = 2\pi i \delta_{y,y'} (K_\textsf{e}^{-1})_{\alpha \beta} \delta(x-x').
\end{align}
The system is then described by the Hamiltonian for the gapless edge modes, 
\begin{align} \label{eq:EdgeHam2DCutGlue}
\mathcal{H}_\textsf{e} = \frac{v_\textsf{e}}{4\pi} \int dx \sum_{y \in \mathbb{Z}} \sum_{\alpha,\beta=1}^{2N_0} \partial_x \phi^\textsf{e}_{y,\alpha} \partial_x \phi^\textsf{e}_{y,\beta},
\end{align}
and the Hamiltonian for tunneling between the edges, 
\begin{align} \label{eq:Tunneling2DCutGlue}
\mathcal{V}_\textsf{e} &= -g_\textsf{e} \int dx \sum_{y \in \mathbb{Z}} \sum_{a=1}^{N_0} \cos \Theta^\textsf{e}_{y,a}, \\
\Theta^\textsf{e}_{y,a}(x) &= (K_0 \bm{\phi}^\textsf{l}_y(x) -K_0 \bm{\phi}^\textsf{r}_y(x))_a.
\end{align}
Here, we have assumed that the velocity $v_\textsf{e}$ and the coupling constant $g_\textsf{e}$ are uniform for simplicity, but this assumption is not necessary.
The tunneling Hamiltonian \eqref{eq:Tunneling2DCutGlue} corresponds to the gapping potential given in Eq.~\eqref{eq:GappingPotential} with $2N_0$-dimensional integer vectors $(\bm{\Lambda}_a)_\beta = \delta_{a,\beta} +\delta_{N_0+a,\beta}$.
This obviously satisfies Eq.~\eqref{eq:ConditionLambda} and Eq.~\eqref{eq:RelationLambdaToM} for $(\bm{m}_a)_\beta = \delta_{a,\beta} -\delta_{N_0+a,\beta}$ with $B_{ab} = |\det K_0| (K_0^{-1})_{ab}$ and $c_a = |\det K_0|$. 
Since $M=\{ \bm{m}_a \}$ generates the desired Lagrangian subgroup $L=\{ (\bm{\lambda}, -\bm{\lambda})^T \, | \, \bm{\lambda} \in \mathbb{Z}^{N_0} \}$, pairs of $\bm{\lambda}$ and $-\bm{\lambda}$ are condensed at each interface by the tunneling Hamiltonian \eqref{eq:Tunneling2DCutGlue}.
Such pairs of quasiparticle excitations can be created over consecutive strips between $y_0$ and $y_1$ by acting a string of vertex operators, 
\begin{align} \label{eq:YString2DCutGlue}
S^\textsf{e}_{y_0;y_1,\bm{\lambda}}(x) = \prod_{y=y_0}^{y_1-1} \exp \left[ i \bm{\lambda} \cdot ( \bm{\phi}^\textsf{r}_y(x) -\bm{\phi}^\textsf{l}_{y+1}(x) ) \right],
\end{align}
on the ground state. 
From the commutation relations in Eq.~\eqref{eq:CommRel2DCutGlue}, each factor of the string operator creates kinks at $x$ for the arguments of the cosine terms $\Theta^\textsf{e}_{y,a}$ and $\Theta^\textsf{e}_{y+1,a}$ in Eq.~\eqref{eq:Tunneling2DCutGlue}, where their ground-state expectation values jump by $2\pi \lambda_a$ and $-2\pi \lambda_a$, respectively. 
These kinks are canceled at interfaces for $y \in (y_0,y_1)$, while pairs of kinks are left at the outermost interfaces $y_0$ and $y_1$; the latter kinks are thus regarded as a pair of quasiparticles $\bm{\lambda}$ and $-\bm{\lambda}$, as schematically shown in Fig.~\ref{fig:CutAndGlue2D}~(d). 
We can also consider another string operator, 
\begin{align} \label{eq:XString2DCutGlue}
S^\textsf{e}_{y,\bm{\lambda}}(x_0;x_1) = \exp \left[ i \bm{\lambda} \cdot \int_{x_0}^{x_1} dx \, \partial_x \bm{\phi}^\textsf{l}_y(x) \right],
\end{align}
which creates a pair of $\bm{\lambda}$ at $x_1$ and $-\bm{\lambda}$ at $x_0$ along the $y$th interface [see Fig.~\ref{fig:CutAndGlue2D}~(d)]. 
We note that replacing $\phi^\textsf{l}_{y,a}$ with $\phi^\textsf{r}_{y,a}$ in Eq.~\eqref{eq:XString2DCutGlue} yields a string operator with the same role.
We can read off the statistics between two quasiparticle excitations $\bm{\lambda}$ and $\bm{\lambda}'$ by these string operators: 
When the system is placed on an $L_x \times L_y$ torus, two string operators $S^\textsf{e}_{1;L_y, \bm{\lambda}}(x)$ and $S^\textsf{e}_{y,\bm{\lambda}'}(0;L_x)$, which wind noncontractible cycles of the torus along the $x$ and $y$ axes, respectively, commute with the Hamiltonian and obey the algebra, 
\begin{align} \label{eq:CommString2DCutGlue}
S^\textsf{e}_{y,\bm{\lambda}}(0;L_x) S^\textsf{e}_{1;L_y,\bm{\lambda}'}(x) = e^{i\theta_{\bm{\lambda},\bm{\lambda}'}} S^\textsf{e}_{1;L_y,\bm{\lambda}'} (x) S^\textsf{e}_{y,\bm{\lambda}}(0;L_x),
\end{align}
where $\theta_{\bm{\lambda},\bm{\lambda}'} = 2\pi \bm{\lambda}^T K_0^{-1} \bm{\lambda}'$ and is precisely the mutual statistics between $\bm{\lambda}$ and $\bm{\lambda'}$ of the original topological order $K_0$. 
This also implies the ground-state degeneracy $|\det K_0|$ on the 2D torus, which corresponds to the minimal dimension of representations for this algebra. 

\subsection{2D coupled-edge model}
\label{sec:2DCellularAddWire}

\begin{figure}
\includegraphics[clip,width=0.45\textwidth]{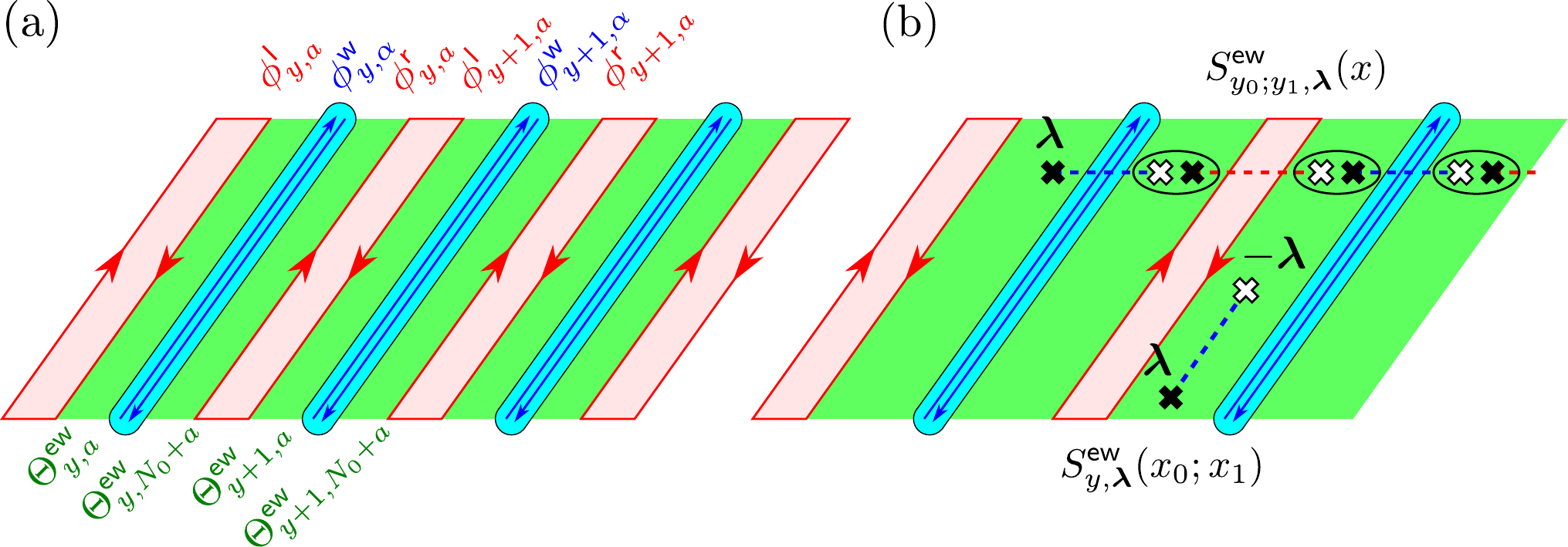}
\caption{(a) Each interface hosts the original gapless edge modes $\phi^\textsf{e}_{y,\alpha}$ from neighboring strips and extra bosonic or fermionic modes $\phi^\textsf{w}_{y,\alpha}$ from additional quantum wires, which are gapped by tunneling terms $\cos \Theta^\textsf{ew}_{y,\alpha}$. 
(b) String operator along the $y$ axis is constructed from the original one by supplementing extra vertex operators from the quantum wires. 
That along the $x$ axis is written solely in terms of bosonic fields of the quantum wires.}
\label{fig:CellularAddWire2D}
\end{figure}

Although the above model exhibits the same topological properties as those described by $K_0$, it does not immediately tell us how the model is built out of microscopic degrees of freedom, since the edge Hamiltonian cannot be regarded as a purely 1D lattice system for a nontrivial $K_0$. 
In addition, the tunneling Hamiltonian involving only edge modes, as given in Eq.~\eqref{eq:Tunneling2DCutGlue}, does not always exist for general gapped interfaces $L= \{ (\bm{\lambda}^\textsf{l}, \bm{\lambda}^\textsf{r})^T \}$ with $\bm{\lambda}^\textsf{l} \neq -\bm{\lambda}^\textsf{r}$. 
In order to connect 2D cellular topological states with ``microscopic'' models made of purely 1D bosonic or fermionic wires and to deal with general gapped interfaces, we utilize a redundant description of gapped interfaces, as discussed in Sec.~\ref{sec:GappedBoundary}, with an extended $K$ matrix of the form,
\begin{align} \label{eq:ExtendedK2D}
K_\textsf{ew} = K_\textsf{e} \oplus K_\textsf{w} = \begin{pmatrix} K_\textsf{e} & \\ & K_\textsf{w} \end{pmatrix},
\end{align}
where $K_\textsf{e}$ is the $2N_0 \times 2N_0$ matrix defined in Eq.~\eqref{eq:KmatrixEdge2D}. 
For a bosonic gapped interface, the $2N_0 \times 2N_0$ matrix $K_\textsf{w}$ is given by 
\begin{align} \label{eq:KmatrixWireBoson2D}
K_\textsf{w} = X^{\oplus N_0} = \begin{pmatrix} X && \\ & \ddots & \\ && X \end{pmatrix},
\end{align}
corresponding to the addition of $N_0$ bosonic wires at each interface. 
For a fermionic gapped interface, it is given by
\begin{align} \label{eq:KmatrixWireFermion2D}
K_\textsf{w} = Z^{\oplus N_0} = \begin{pmatrix} Z && \\ & \ddots & \\ && Z \end{pmatrix},
\end{align}
corresponding to the addition of $N_0$ fermionic wires. 
Including these additional quantum wires, the Hamiltonian for the gapless interface modes is given by
\begin{align}
\mathcal{H}_\textsf{ew} = \frac{v_\textsf{ew}}{4\pi} \int dx \sum_{y \in \mathbb{Z}} \sum_{I,J=1}^{4N_0} \partial_x \phi^\textsf{ew}_{y,I} \partial_x \phi^\textsf{ew}_{y,J},
\end{align}
where $\bm{\phi}^\textsf{ew}_y = (\bm{\phi}^\textsf{e}_y, \bm{\phi}^\textsf{w}_y)^T$ and $\bm{\phi}^\textsf{w}_y = (\phi^\textsf{w}_{y,1}, \cdots, \phi^\textsf{w}_{y,2N_0})^T$ represents a $2N_0$-component bosonic field added at the $y$th interface.
These bosonic fields obey the commutation relations, 
\begin{align}
[\partial_x \phi^\textsf{ew}_{y,I}(x), \phi^\textsf{ew}_{y',J}(x')] = 2\pi i \delta_{y,y'} (K_\textsf{ew}^{-1})_{IJ} \delta(x-x'), 
\end{align}
with the extended $K$ matrix given in Eq.~\eqref{eq:ExtendedK2D}. 
This system is schematically shown in Fig.~\ref{fig:CellularAddWire2D}~(a). 
For a general Lagrangian subgroup $L$, a tunneling Hamiltonian involving the additional quantum wires can be constructed by following the prescription given in Sec.~\ref{sec:BosonicGappedBoundary} for the bosonic case or in Sec.~\ref{sec:FermionicGappedBoundary} for the fermionic case. 
It generally takes the form of Eq.~\eqref{eq:ExtendedGappingPotential}: 
\begin{align} \label{eq:ExtendedTunneling2D}
\mathcal{V}_\textsf{ew} &= -g_\textsf{ew} \int dx \sum_{y \in \mathbb{Z}} \sum_{\alpha=1}^{2N_0} \cos \Theta^\textsf{ew}_{y,\alpha}, \\
\Theta^\textsf{ew}_{y,\alpha}(x) &= \widetilde{\bm{\Lambda}}^T_\alpha K_\textsf{ew} \bm{\phi}^\textsf{ew}_y(x), 
\end{align}
where $\widetilde{\bm{\Lambda}}_\alpha$ are $4N_0$-dimensional integer vectors. 
As discussed in Appendix~\ref{sec:LambdaWire}, these integer vectors can be chosen to take the form,
\begin{align} \label{eq:LambdaEdgeWire}
\begin{pmatrix} \widetilde{\bm{\Lambda}}_1 & \cdots & \widetilde{\bm{\Lambda}}_{2N_0} \end{pmatrix} = \begin{pmatrix} I_{2N_0} \\ \Lambda_\textsf{w} \end{pmatrix},
\end{align}
with the $2N_0 \times 2N_0$ identity matrix $I_{2N_0}$ and a $2N_0 \times 2N_0$ integer matrix $\Lambda_\textsf{w}$, each of which acts only on the bosonic fields for edge or wire modes from the block-diagonal structure of $K_\textsf{ew}$. 
This ensures that the primitivity condition is satisfied for the set of integer vectors $\{ \widetilde{\bm{\Lambda}}_I \}$, and Eq.~\eqref{eq:ExtendedLambdaKLambda} implies that
\begin{align} \label{eq:LambdaKwLambda}
\bm{\Lambda}_{\textsf{w},\alpha}^T K_\textsf{w} \bm{\Lambda}_{\textsf{w},\beta} = -(K_\textsf{e})_{\alpha \beta},
\end{align}
where $\bm{\Lambda}_{\textsf{w},\alpha}$ is the $\alpha$th column of $\Lambda_\textsf{w}$. 
We thus find 
\begin{subequations}
\begin{align}
\Theta^\textsf{ew}_{y,a} &= (K_0 \bm{\phi}^\textsf{l}_y)_a + \bm{\Lambda}^T_{\textsf{w},a} K_\textsf{w} \bm{\phi}^\textsf{w}_y, \\
\Theta^\textsf{ew}_{y,N_0+a} &= -(K_0 \bm{\phi}^\textsf{r}_y)_a + \bm{\Lambda}^T_{\textsf{w},N_0+a} K_\textsf{w} \bm{\phi}^\textsf{w}_y,
\end{align}
\end{subequations}
for $a=1,\cdots,N_0$. 
Therefore, at each interface, the edge modes of the left strip are not directly coupled with those of the right strip, but they are mediated by the additional quantum wires. 

As this tunneling Hamiltonian just gives a redundant description of the trivial gapped interface considered in Sec.~\ref{sec:2DCellularTopoState} with $L=\{ (\bm{\lambda}, -\bm{\lambda})^T \, | \, \bm{\lambda} \in \mathbb{Z}^{N_0} \}$, the model is expected to retain the same topological properties as those described by $K_0$. 
We can explicitly construct a string operator that creates a pair of quasiparticle excitations $\bm{\lambda}$ and $-\bm{\lambda}$ separated by the distance $|y_1-y_0|$ along the $y$ axis [see Fig.~\ref{fig:CellularAddWire2D}~(b)], 
\begin{align} \label{eq:YString2DExtended}
S^\textsf{ew}_{y_0;y_1,\bm{\lambda}}(x) &= \prod_{y=y_0}^{y_1-1} \exp \Bigl[ i\bm{\lambda} \cdot (\bm{\phi}^\textsf{r}_y(x) -\bm{\phi}^\textsf{l}_{y+1}(x)) \nonumber \\
&\quad +i\bm{p}_{\bm{\lambda}} \cdot \bm{\phi}^\textsf{w}_y(x) \Bigr],
\end{align}
where $\bm{p}_{\bm{\lambda}}$ is a $2N_0$-dimensional integer vector satisfying 
\begin{subequations} \label{eq:InnerProdLambdaP}
\begin{align}
\bm{\Lambda}_{\textsf{w},a} \cdot \bm{p}_{\bm{\lambda}} &= \lambda_a, \\
\bm{\Lambda}_{\textsf{w},N_0+a} \cdot \bm{p}_{\bm{\lambda}} &= -\lambda_a,
\end{align}
\end{subequations}
for $a=1,\cdots,N_0$. 
Compared with Eq.~\eqref{eq:YString2DCutGlue}, the first term in the exponent of each exponential factor of Eq.~\eqref{eq:YString2DExtended} creates kinks of $2\pi \lambda_a$ and $-2\pi \lambda_a$ in $\Theta^\textsf{ew}_{y,N_0+a}$ and $\Theta^\textsf{ew}_{y+1,a}$, respectively. 
They are now canceled by antikinks created by the additional factor $\exp (i\bm{p}_{\bm{\lambda}} \cdot \bm{\phi}^\textsf{w}_y)$ from the condition \eqref{eq:InnerProdLambdaP}, except for those of $\Theta^\textsf{ew}_{y_0,a}$ and $\Theta^\textsf{ew}_{y_1,a}$. 
The latter two kinks are regarded as a pair of quasiparticle excitations created at the $y_0$th and $y_1$th interfaces, respectively. 
The string operator defined in Eq.~\eqref{eq:XString2DCutGlue} still creates a pair of quasiparticles along the $y$th interface in the present case. 
However, we can write it solely in terms of the bosonic fields of the additional quantum wires, for example, as
\begin{align}
S^\textsf{ew}_{y,\bm{\lambda}}(x_0;x_1) = \exp \left[ -i \bm{q}_{\bm{\lambda}} \cdot \int_{x_0}^{x_1} dx \, \partial_x \bm{\phi}^\textsf{w}_y(x) \right],
\end{align}
where $\bm{q}_{\bm{\lambda}}$ is the $2N_0$-dimensional vector defined by 
\begin{align} \label{eq:DefVectorQ}
\bm{q}_{\bm{\lambda}} = \sum_{a,b=1}^{N_0} (K_0^{-1})_{ab} \lambda_b K_\textsf{w} \bm{\Lambda}_{\textsf{w},a}.
\end{align}
This string operator creates two kinks of $2\pi \lambda_a$ at $x_1$ and $-2\pi \lambda_a$ at $x_0$ in $\Theta^\textsf{ew}_{y,a}$. 
On an $L_x \times L_y$ torus, the two string operators $S^\textsf{ew}_{1;L_y,\bm{\lambda}}(x)$ and $S^\textsf{ew}_{y,\bm{\lambda}'}(0;L_x)$ commute with the Hamiltonian and obey the same algebra as Eq.~\eqref{eq:CommString2DCutGlue}: 
\begin{align}
S^\textsf{ew}_{y,\bm{\lambda}}(0;L_x) S^\textsf{ew}_{1;L_y,\bm{\lambda}'}(x) = e^{i\theta_{\bm{\lambda},\bm{\lambda}'}} S^\textsf{ew}_{1;L_y,\bm{\lambda}'} (x) S^\textsf{ew}_{y,\bm{\lambda}}(0;L_x).
\end{align}
Hence, we have confirmed that the model with additional quantum wires still exhibits the topological properties described by $K_0$.

\subsection{2D coupled-wire models}
\label{sec:2DCoupledWireHam}

\begin{figure}
\includegraphics[clip,width=0.45\textwidth]{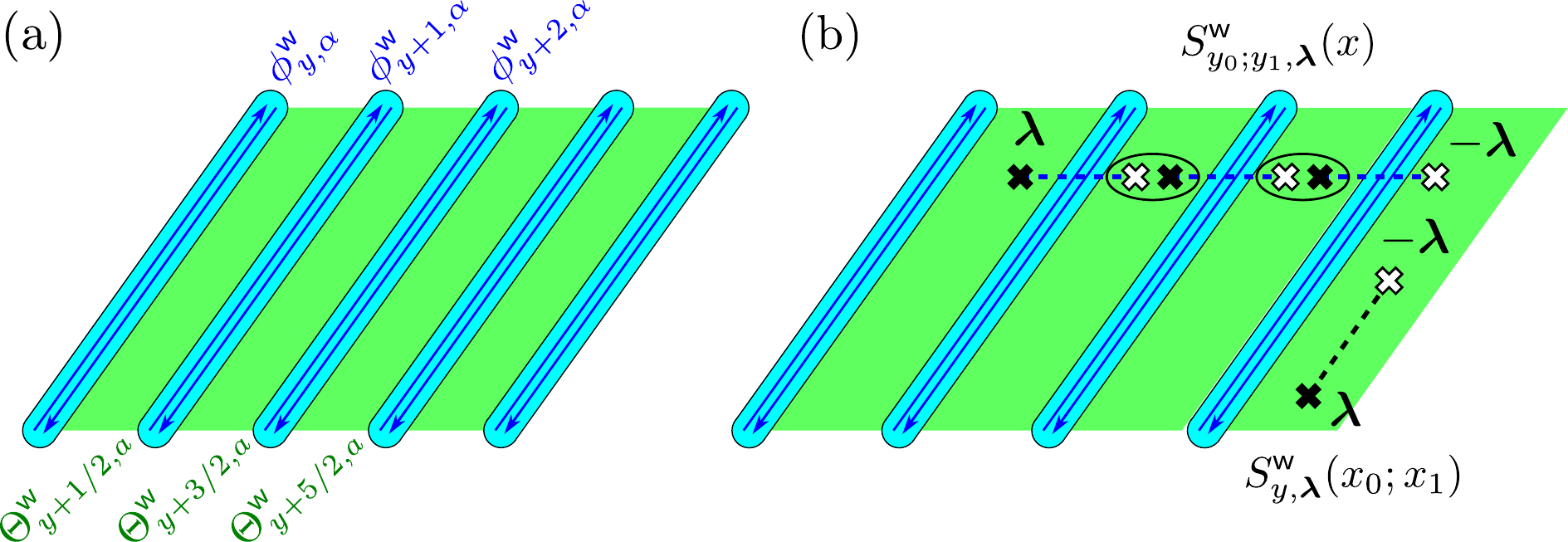}
\caption{(a) After shrinking and removing the strips of the topological orders, we end up with a model only with purely 1D gapless quantum wires described by $\phi^\textsf{w}_{y,\alpha}$, which are gapped by tunneling terms $\cos \Theta^\textsf{w}_{y+1/2,a}$. 
(b) The action of string operators in the coupled-wire model for trivial gapped interfaces with $L=\{ (\bm{\lambda}, -\bm{\lambda})^T \}$. }
\label{fig:CoupledWire2D}
\end{figure}

At the last step, we consider a process of shrinking and removing the strips of 2D topological orders with leaving only the quantum wires added to each interface, as shown in Fig.~\ref{fig:CoupledWire2D}~(a). 
With this process, the quantum wires between neighboring interfaces will directly interact with each other. 
After removing the edge modes associated with each strip, we are left with the Hamiltonian for the gapless quantum wires, 
\begin{align}
\mathcal{H}_\textsf{w} = \frac{v_\textsf{w}}{4\pi} \int dx \sum_{y \in \mathbb{Z}} \sum_{\alpha,\beta=1}^{2N_0} \partial_x \phi^\textsf{w}_{y,\alpha} \partial_x \phi^\textsf{w}_{y,\beta},
\end{align}
where $\bm{\phi}^\textsf{w}_y = (\phi^\textsf{w}_{y,1}, \cdots, \phi^\textsf{w}_{y,2N_0})^T$ is the $2N_0$-component bosonic field for the $y$th wire.
They satisfy the commutation relations, 
\begin{align} \label{eq:CommRelCW2D}
[\partial_x \phi^\textsf{w}_{y,\alpha}(x), \phi^\textsf{w}_{y',\beta}(x')] = 2\pi i \delta_{y,y'} (K_\textsf{w})_{\alpha \beta} \delta(x-x'), 
\end{align}
where $K_\textsf{w}$ is given in Eq.~\eqref{eq:KmatrixWireBoson2D} for the bosonic case and in Eq.~\eqref{eq:KmatrixWireFermion2D} for the fermionic case and we have used $K_\textsf{w}^{-1} = K_\textsf{w}$.
The bosonic fields of quantum wires that were coupled with the right strip at the $y$th interface in the tunneling Hamiltonian \eqref{eq:ExtendedTunneling2D} now interact with those coupled with the left strip at the $(y+1)$th interface, yielding the tunneling Hamiltonian with only the quantum wires, 
\begin{align} \label{eq:TunnelingCW2D}
\mathcal{V}_\textsf{w} &= -g_\textsf{w} \int dx \sum_{y \in \mathbb{Z}} \sum_{a=1}^{N_0} \cos \Theta^\textsf{w}_{y+1/2,a}, \\
\Theta^\textsf{w}_{y+1/2,a} &= \Phi^\textsf{w}_{y,N_0+a} +\Phi^\textsf{w}_{y+1,a},
\end{align}
where we have defined 
\begin{align}
\Phi^\textsf{w}_{y,\alpha} = \bm{\Lambda}^T_{\textsf{w},\alpha} K_\textsf{w} \bm{\phi}^\textsf{w}_y,
\end{align}
and $\bm{\Lambda}_{\textsf{w},\alpha}$ are the integer vectors introduced in Eq.~\eqref{eq:LambdaEdgeWire}. 
This shrinking process can also be considered as follows. 
Very thin strips will naturally have quasiparticle tunnelings, $-g_\textsf{qp} \cos (\phi^\textsf{r}_{y,a} -\phi^\textsf{l}_{y+1,a})$, connecting their edge modes. 
In the strong-coupling limit $g_\textsf{qp} \to \infty$, the bosonic fields from the edge modes are pinned at the cosine minima so that $\phi^\textsf{r}_{y,a} -\phi^\textsf{l}_{y+1,a} = 2\pi n_{y,a}$ with $n_{y,a} \in \mathbb{Z}$. 
In this limit, individual bosonic fields $\bm{\phi}^\textsf{l}_y$ and $\bm{\phi}^\textsf{r}_y$ are strongly fluctuating and thus the tunneling terms $\cos \Theta^\textsf{ew}_{y,\alpha}$ in $\mathcal{V}_\textsf{ew}$ have vanishing expectation values. 
However, second-order perturbations in $\mathcal{V}_\textsf{ew}$ generate terms of the form $\cos (\Theta^\textsf{ew}_{y+1,a} +\Theta^\textsf{ew}_{y,N_0+a})$, which contain the fluctuating fields in the pinned combination $\phi^\textsf{r}_{y,a} -\phi^\textsf{l}_{y+1,a}$; these second-order terms yield the tunneling terms in $\mathcal{V}_\textsf{w}$.

We thus end up with the Hamiltonian written solely in terms of the bosonic fields of the added quantum wires that can be realized as the low-energy limit of purely 1D bosonic or fermionic lattice systems. 
In this sense, the resulting Hamiltonian has a microscopic origin. 
From Eq.~\eqref{eq:LambdaKwLambda}, it is easy to verify that any pair of two cosine terms in $\mathcal{V}_\textsf{w}$ commute with each other. 
Thus, the tunneling Hamiltonian can open a gap for all quantum wires when the system is placed on the torus. 
On the other hand, when the system has boundaries at $y=1$ and $y=L_y$, there are unpaired gapless modes $\Phi^\textsf{w}_{1,a}$ and $\Phi^\textsf{w}_{L_y,N_0+a}$. 
Since these bosonic fields satisfy the commutation relations, 
\begin{align}
[\partial_x \Phi^\textsf{w}_{y,\alpha}(x), \Phi^\textsf{w}_{y',\beta}(x')] = -2\pi i \delta_{y,y'} (K_\textsf{e})_{\alpha \beta} \delta(x-x'), 
\end{align}
they can be identified with the edge modes of the original topological order described by $K_0$. 
We note that in terms of these bosonic fields, operators creating local particle excitations are given by vertex operators $\exp (i \Phi^\textsf{w}_{y,\alpha})$, which only contain integer multiples of $\phi^\textsf{w}_{y,\alpha}$. 
On the other hand, operators creating anyon excitations are given by $\exp [i\sum_\beta (K_\textsf{e}^{-1})_{\alpha \beta} \Phi^\textsf{w}_{y,\beta}]$, which contain fractions of $\phi^\textsf{w}_{y,\alpha}$ and cannot be regarded as local operators. 

For the trivial gapped interface $L=\{ (\bm{\lambda}, -\bm{\lambda})^T \, | \, \bm{\lambda} \in \mathbb{Z}^{N_0} \}$, pairs of bulk quasiparticles associated with $\bm{\lambda}$ and $-\bm{\lambda}$ are created by the string operators [see Fig.~\ref{fig:CoupledWire2D}~(b)], 
\begin{align}
\label{eq:YStringCW2D}
S^\textsf{w}_{y_0;y_1,\bm{\lambda}}(x) &= \prod_{y=y_0}^{y_1-1} \exp \left[ i\bm{p}_{\bm{\lambda}} \cdot \bm{\phi}^\textsf{w}_y (x) \right], \\
S^\textsf{w}_{y,\bm{\lambda}}(x_0;x_1) &= \exp \left[ -i \bm{q}_{\bm{\lambda}} \cdot \int_{x_0}^{x_1} dx \, \partial_x \bm{\phi}^\textsf{w}_y(x) \right],
\end{align}
with $\bm{p}_{\bm{\lambda}}$ and $\bm{q}_{\bm{\lambda}}$ defined in Eqs.~\eqref{eq:InnerProdLambdaP} and \eqref{eq:DefVectorQ}, respectively. 
Here, each factor in the string operator \eqref{eq:YStringCW2D} creates kinks of $2\pi \lambda_a$ and $-2\pi \lambda_a$ in the link fields $\Theta^\textsf{w}_{y-1/2,a}$ and $\Theta^\textsf{w}_{y+1/2,a}$, respectively. 
When the system is placed on the torus, these string operators obey the algebra, 
\begin{align}
S^\textsf{w}_{y,\bm{\lambda}}(0;L_x) S^\textsf{w}_{1;L_y,\bm{\lambda}'}(x) = e^{i\theta_{\bm{\lambda},\bm{\lambda}'}} S^\textsf{w}_{1;L_y,\bm{\lambda}'}(x) S^\textsf{w}_{y,\bm{\lambda}}(0;L_x).
\end{align}
This case of the trivial gapped interface is reduced to the coupled-wire construction of 2D Abelian topological orders developed in Refs.~\cite{Kane02, Teo14}. 
In particular, each factor appearing in the string operator \eqref{eq:YStringCW2D} is typically constructed from local ``backscattering'' operators in a system conserving both particle number and momentum. 

For a general gapped interface with $L = \{ (\bm{\lambda}^\textsf{l}, \bm{\lambda}^\textsf{r})^T \}$, $\bm{\lambda}^\textsf{l}$ and $\bm{\lambda}^\textsf{r}$ are not necessarily a quasiparticle and its antipartner. 
In this case, a pair of quasiparticles $\bm{l} = (\bm{\lambda}^\textsf{l}, \bm{\lambda}^\textsf{r})^T \in L$ is created from an interface by acting a local vertex operator $\exp (i\bm{p}_{\bm{l}} \cdot \bm{\phi}^\textsf{w}_y)$ with an integer vector $\bm{p}_{\bm{l}}$ satisfying 
\begin{subequations} \label{eq:InnerProdLambdaP2}
\begin{align}
\bm{\Lambda}_{w,a} \cdot \bm{p}_{\bm{l}} &= \lambda^\textsf{l}_a \\
\bm{\Lambda}_{w,N_0+a} \cdot \bm{p}_{\bm{l}} &= \lambda^\textsf{r}_a.
\end{align}
\end{subequations}
The string operator along the $y$ axis, which has been defined for $\bm{\lambda}^\textsf{l} = -\bm{\lambda}^\textsf{r}$ in Eq.~\eqref{eq:YStringCW2D}, must be modified such that kinks inside the region $(y_0,y_1)$ are properly canceled. 
On the other hand, the string operators along the $x$ axis remain unchanged so that they create a pair of quasiparticles $\bm{\lambda}$ and $-\bm{\lambda}$. 
This affects the algebra obeyed by the string operators and thereby the statistics of quasiparticles and the ground-state degeneracy on a torus.

\subsection{2D models from trivial gapped interfaces}
\label{sec:2DTrivialModel}

In this subsection, we provide several examples of the coupled-wire Hamiltonians obtained by the above procedure from 2D cellular topological states with the trivial gapped interface $L=\{ (\bm{\lambda}, -\bm{\lambda})^T \, | \, \bm{\lambda} \in \mathbb{Z}^{N_0} \}$.
We specifically consider the 2D toric code and $U(1)_k$ topological orders. 

\subsubsection{2D toric code}
\label{sec:CW2dTC}

We start with an array of strips of the 2D toric codes, whose interface hosts gapless edge modes $\bm{\phi}^\textsf{e}_y = (\phi^\textsf{e}_{y,1}, \cdots, \phi^\textsf{e}_{y,4})^T$ corresponding to the $K$ matrix: 
\begin{align} \label{eq:KmatrixTwo2DTC}
K_\textsf{e} = \begin{pmatrix} 0 & 2 && \\ 2 & 0 && \\ && 0 & -2 \\ && -2 & 0 \end{pmatrix}. 
\end{align}
We then consider a gapped interface at which pairs of quasiparticles $(e,e)$, $(m,m)$, and $(f,f)$ from neighboring strips are condensed. 
The corresponding Lagrangian subgroup is generated by $M=\{ \bm{m}_a \}$ with
\begin{align}
\begin{split}
\bm{m}_1 &= (1,0,1,0)^T, \\
\bm{m}_2 &= (0,1,0,1)^T.
\end{split}
\end{align} 
Here, we note that $e$, $m$, and $f$ are their own antiparticles in the toric code: two integer vectors $(1,0,0,0)^T$ and $(-1,0,0,0)^T$ associated with an $e$ excitation, say, are related by addition or subtraction of the local bosonic excitation $(2,0,0,0)^T$. 
Such integer vectors, however, should be distinguished when we construct string operators commuting with the Hamiltonian [this requirement has been encoded in Eq.~\eqref{eq:InnerProdLambdaP}].
While it is easy to write down the tunneling Hamiltonian only with the edge modes, we instead consider a redundant one by adding two bosonic wires at each interface. 
We thus introduce four extra bosonic fields $\bm{\phi}^\textsf{w}_y = (\phi^\textsf{w}_{y,1}, \cdots, \phi^\textsf{w}_{y,4})^T$ and the extended $K$ matrix $K_\textsf{ew} = K_\textsf{e} \oplus K_\textsf{w}$ with $K_\textsf{w} = X \oplus X$ as defined in Eq.~\eqref{eq:KmatrixWireBoson2D}. 
Following the algorithm of Sec.~\ref{sec:BosonicGappedBoundary} with the choice of matrices, 
\begin{align}
U = \begin{pmatrix} 1 & 0 & -1 & 0 \\ 0 & 1 & 0 & -1 \\ 1 & 0 & 1 & 0 \\ 0 & 1 & 0 & 1 \end{pmatrix}, \quad
W = \begin{pmatrix} -1 & 0 & 0 & 0 \\ 0 & -1 & 0 & 0 \\ 0 & 0 & 0 & 1 \\ 0 & 0 & 1 & 0 \end{pmatrix},
\end{align}
we find a set of four-dimensional integer vectors $\{ \bm{\Lambda}_{\textsf{w},\alpha} \}$,
\begin{align}
\begin{split}
\bm{\Lambda}_{\textsf{w},1} &= (0,1,-1,0)^T, \\
\bm{\Lambda}_{\textsf{w},2} &= (-1,0,0,1)^T, \\
\bm{\Lambda}_{\textsf{w},3} &= (0,1,1,0)^T, \\
\bm{\Lambda}_{\textsf{w},4} &= (1,0,0,1)^T.
\end{split}
\end{align}
Substituting this into Eq.~\eqref{eq:TunnelingCW2D} yields the desired tunneling terms for a coupled-wire model of the 2D toric code \cite{Teo14, Patel16}. 
For comparison with microscopic bosonic degrees of freedom, it might be helpful to write the bosonic fields for quantum wires as $\bm{\phi}^\textsf{w}_y = (2\theta^1_y, \varphi^1_y, 2\theta^2_y, \varphi^2_y)^T$ and relate $\theta^\sigma_y$ and $\varphi^\sigma_y$ with density and current fluctuations, respectively, of the $\sigma$th component of bosons in the $y$th wire. 
By imposing the commutation relations, 
\begin{align}
\begin{split}
[\theta^\sigma_y(x), \varphi^{\sigma'}_{y'}(x')] &= i\pi \delta_{\sigma,\sigma'} \delta_{y,y'} \Theta(x-x'), \\
[\theta^\sigma_y(x), \theta^{\sigma'}_{y'}(x')] &= [\varphi^\sigma_y(x), \varphi^{\sigma'}_{y'}(x')] =0,
\end{split}
\end{align}
with $\Theta(x)$ being the Heaviside step function, the bosonic fields $\bm{\phi}^\textsf{w}_y$ satisfy the required commutation relations in Eq.~\eqref{eq:CommRelCW2D}. 
We can then identify $\exp (-i\varphi^\sigma_y)$ as a boson creation operator and $\exp (2i\theta^\sigma_y)$ as a backscattering operator of the wave number $2k_F$ \cite{Haldane81}. 
The tunneling Hamiltonian is written as 
\begin{align}
\mathcal{V}_\textsf{w} &= -g \int dx \sum_{y \in \mathbb{Z}} \bigl[ \cos (\varphi^1_y +2\theta^2_y -\varphi^1_{y+1} +2\theta^2_{y+1}) \nonumber \\
&\quad +\cos (\varphi^2_y +2\theta^1_y -\varphi^2_{y+1} +2\theta^1_{y+1}) \bigr].
\end{align}
This will be a natural coupled-wire Hamiltonian for the 2D toric code in a system of two-component bosons with particle-number conservation for each component \cite{Patel16}. 

A pair of quasiparticles $(\bm{\lambda}, -\bm{\lambda})^T$ in neighboring links can be created by the vertex operator $\exp (i\bm{p}_{\bm{\lambda}} \cdot \bm{\phi}^\textsf{w}_y)$ with an integer vector $\bm{p}_{\bm{\lambda}}$ satisfying Eq.~\eqref{eq:InnerProdLambdaP}. 
We find $\bm{p}_{\bm{\lambda}_e} = (0,0,-1,0)^T$ for the $e$ excitation $\bm{\lambda}_e = (1,0)^T$ and $\bm{p}_{\bm{\lambda}_m} = (-1,0,0,0)^T$ for the $m$ excitation $\bm{\lambda}_m = (0,1)^T$.
We thus find string operators,
\begin{subequations}
\begin{align}
S^\textsf{w}_{y_0,y_1; \bm{\lambda}_e}(x) &= \prod_{y=y_0}^{y_1=1} \exp [ -2i\theta^2_y(x) ], \\
S^\textsf{w}_{y_0,y_1; \bm{\lambda}_m}(x) &= \prod_{y=y_0}^{y_1=1} \exp [ -2i\theta^1_y(x) ], \\
S^\textsf{w}_{y,\bm{\lambda}_e}(x_0;x_1) &= \exp \left[ \frac{i}{2} \int_{x_0}^{x_1} dx \, \partial_x (\varphi^1_y -2\theta^2_y) \right], \\
S^\textsf{w}_{y,\bm{\lambda}_m}(x_0;x_1) &= \exp \left[ \frac{i}{2} \int_{x_0}^{x_1} dx \, \partial_x (\varphi^2_y -2\theta^1_y) \right].
\end{align}
\end{subequations}
When these string operators wind nontrivial cycles of the 2D torus, they constitute logical operators acting on the $2^2$-dimensional space of degenerate ground states.

\subsubsection{$U(1)_k$ topological orders}
\label{sec:CW2dU1k}

We start with an array of strips of the $\nu=1/k$ Laughlin states or $U(1)_k$ topological orders, whose interface hosts gapless edge modes $\bm{\phi}^\textsf{e}_y = (\phi^\textsf{e}_{y,1}, \phi^\textsf{e}_{y,2})^T$ corresponding to the $K$ matrix: 
\begin{align}
K_\textsf{e} = \begin{pmatrix} k & \\ & -k \end{pmatrix}.
\end{align}
We consider a gapped interface at which pairs of quasiparticles $(\bm{\lambda}, -\bm{\lambda})^T$ with $\bm{\lambda}=1,2,\cdots,k-1$ are condensed. 
The corresponding Lagrangian subgroup is generated by 
\begin{align}
\bm{m}_1 = (1,-1)^T.
\end{align}

Let us treat the bosonic and fermionic cases separately. 
For the bosonic case, since $k$ is even, we set $k=2\kappa$ with integer $\kappa$. 
We introduce two extra bosonic fields $\bm{\phi}^\textsf{w}_y = (\phi^\textsf{w}_{y,1}, \phi^\textsf{w}_{y,2})^T$ and the extended $K$ matrix $K_\textsf{ew} = K_\textsf{e} \oplus K_\textsf{w}$ with $K_\textsf{w}=X$. 
Following the algorithm in Sec.~\ref{sec:BosonicGappedBoundary} with the choice of matrices, 
\begin{align}
U = \begin{pmatrix} 1 & \kappa \\ -1 & \kappa \end{pmatrix}, \quad
W= \begin{pmatrix} 1 & 0 \\ 0 & 1 \end{pmatrix},
\end{align}
we find a set of two-dimensional integer vectors $\{ \bm{\Lambda}_{\textsf{w},\alpha} \}$, 
\begin{align}
\begin{split}
\bm{\Lambda}_{\textsf{w},1} &= (\kappa,-1)^T, \\
\bm{\Lambda}_{\textsf{w},2} &= (\kappa, 1)^T.
\end{split}
\end{align}
Similarly to the previous case for the 2D toric code, we write $\bm{\phi}^\textsf{w}_y = (\varphi_y, 2\theta_y)^T$ and relate $\theta_y$ and $\varphi_y$ with the density and current fluctuations, respectively, of a single-component bosonic wire. 
They will satisfy the commutation relations:
\begin{align}
\begin{split}
[\theta_y(x), \varphi_{y'}(x')] &= i\pi \delta_{y,y'} \Theta(x-x'), \\
[\theta_y(x), \theta_{y'}(x')] &= [\varphi_y(x), \varphi_{y'}(x')] =0.
\end{split}
\end{align}
By substituting these expressions in Eq.~\eqref{eq:TunnelingCW2D}, we obtain the tunneling Hamiltonian, 
\begin{align}
\mathcal{V}_\textsf{w} = -g \int dx \sum_{y \in \mathbb{Z}} \cos (\varphi_y +2\kappa \theta_y -\varphi_{y+1} +2\kappa \theta_{y+1}),
\end{align}
which reproduces the coupled-wire Hamiltonian proposed for the bosonic Laughlin states \cite{Teo14}. 
String operators associated with a pair of excitations $(\bm{\lambda}_{\bm{1}}, -\bm{\lambda}_{\bm{1}})^T = (1,-1)^T$ can be constructed as 
\begin{subequations}
\begin{align}
\label{eq:YStringBosonicLaughlin}
S^\textsf{w}_{y_0;y_1, \bm{\lambda}_{\bm{1}}}(x) &= \prod_{y=y_0}^{y_1-1} \exp [-2i\theta_y(x)], \\
S^\textsf{w}_{y,\bm{\lambda}_{\bm{1}}}(x_0;x_1) &= \exp \left[ -\frac{i}{2\kappa} \int_{x_0}^{x_1} dx \, \partial_x (\varphi_y -2\kappa \theta_y) \right]. 
\end{align}
\end{subequations}
Here, each factor of Eq.~\eqref{eq:YStringBosonicLaughlin} is a backscattering operator of the wave number $-2\pi \bar{\rho}$ with $\bar{\rho}$ being the average density of bosons. 
When these operators are defined on a torus, they nontrivially act on the $2\kappa$-dimensional space of degenerate ground states.

For the fermionic case, we set $k=2\kappa+1$ with integer $\kappa$. 
We then consider the extended $K$ matrix $K_\textsf{ew} = K_\textsf{e} \oplus K_\textsf{w}$ with $K_\textsf{w}=Z$. 
Following the algorithm in Sec.~\ref{sec:FermionicGappedBoundary} with the choice of matrices, 
\begin{align}
U = \begin{pmatrix} 1 & \kappa \\ -1 & \kappa+1 \end{pmatrix}, \quad
W = \begin{pmatrix} -1 & 0 \\ -1 & -1 \end{pmatrix},
\end{align}
we arrive at a set of integer vectors $\{ \bm{\Lambda}_{\textsf{w},\alpha} \}$, 
\begin{align}
\begin{split}
\bm{\Lambda}_{\textsf{w},1} &= (\kappa, 1+\kappa)^T, \\
\bm{\Lambda}_{\textsf{w},2} &= (\kappa+1, \kappa)^T.
\end{split}
\end{align}
We then write $\bm{\phi}^\textsf{w}_y = (\phi_{y,R}, \phi_{y,L})^T$ where $\phi_{y,R}$ and $\phi_{y,L}$ correspond to right- and left-moving fermionic modes of the $y$th quantum wire, obeying the commutation relations: 
\begin{align}
\begin{split}
[\phi_{y,R}(x), \phi_{y',R}(x')] &= i\pi \delta_{y,y'} \textrm{sgn}(x-x') +i\pi \textrm{sgn}(y-y'), \\
[\phi_{y,L}(x), \phi_{y',L}(x')] &= -i\pi \delta_{y,y'} \textrm{sgn}(x-x') +i\pi \textrm{sgn}(y-y'), \\
[\phi_{y,R}(x), \phi_{y',L}(x')] &= i\pi \delta_{y,y'} +i\pi \textrm{sgn}(y-y'),
\end{split}
\end{align}
where $\textrm{sgn}(x)$ is the sign function and $\textrm{sgn}(0)=0$.
These relations ensure Eq.~\eqref{eq:CommRelCW2D} and also that any pair of fermionic operators $\exp (i\phi_{y,R})$ and $\exp (i\phi_{y,L})$ anticommute with each other. 
We then obtain the tunneling Hamiltonian, 
\begin{align}
\mathcal{V}_\textsf{w} &= -g \int dx \sum_{y \in \mathbb{Z}} \cos [ (\kappa+1) \phi_{y,R} -\kappa \phi_{y,L} \nonumber \\
&\quad +\kappa \phi_{y+1,R} -(\kappa+1) \phi_{y+1,L}],
\end{align}
which reproduces the coupled-wire Hamiltonian proposed for the fermionic Laughlin states \cite{Kane02, Teo14}. 
String operators can be constructed as 
\begin{subequations}
\begin{align}
\label{eq:YStringFermionicLaughlin}
&S^\textsf{w}_{y_0;y_1, \bm{\lambda}_{\bm{1}}}(x) = \prod_{y=y_0}^{y_1-1} \exp [-i(\phi_{y,R}(x) -\phi_{y,L}(x))], \\
&S^\textsf{w}_{y,\bm{\lambda}_{\bm{1}}}(x_0;x_1) \nonumber \\
&= \exp \left[ -\frac{i}{2\kappa+1} \int_{x_0}^{x_1} dx \, \partial_x ((\kappa+1)\phi_{y,R} -\kappa \phi_{y,L}) \right],
\end{align}
\end{subequations}
which nontrivially act on the $(2\kappa+1)$-dimensional space of degenerate ground states when they are defined on a torus. 
Each factor of Eq.~\eqref{eq:YStringFermionicLaughlin} is a backscattering operator of the wave number $-2k_F$ with $k_F$ being the Fermi momentum of the 1D quantum wire.

\subsection{2D models with nontrivial gapped interfaces}
\label{sec:2DNontrivialModel}

We here provide 2D coupled-wire models corresponding to 2D cellular topological states with nontrivial gapped interfaces. 
These models are based on the 2D toric code, doubled semion model, and doubled $U(1)_5$ model. 
The resulting coupled-wire models exhibit the ground-state degeneracy on a torus depending on the number of wires and might be regarded as translation-symmetry-enriched topological orders as discussed in Refs.~\cite{PMTam21a, PMTam21b}. 
We note that these models could also be characterized by non-Abelian zero modes localized at dislocations \cite{Bombin10}. 

\subsubsection{Toric code with $e \leftrightarrow f$ interface}
\label{sec:CW2DTCef}

We start again with an array of strips of the 2D toric codes, whose interface hosts gapless edge modes corresponding to the $K$ matrix \eqref{eq:KmatrixTwo2DTC}. 
We here consider a gapped interface at which pairs of $e$ and $f$ excitations between neighboring strips are condensed. 
In other words, an $e$ excitation from one strip is converted to an $f$ excitation in the other strip through the interface and vice versa. 
The corresponding Lagrangian subgroup is generated by $M=\{ \bm{m}_a \}$ with
\begin{align}
\begin{split}
\bm{m}_1 &= (1,0,1,1)^T, \\
\bm{m}_2 &= (1,1,1,0)^T.
\end{split}
\end{align}
As the bound objects of $e$ and $f$ are fermionic quasiparticles, this is an example of fermionic gapped interface, whose gapping potential can be constructed by adding extra fermionic wires at the interface. 
We thus consider the extended $K$ matrix $K_\textsf{ew} = K_\textsf{e} \oplus K_\textsf{w}$ with $K_\textsf{w} = Z \oplus Z$. 
Following the algorithm in Sec.~\ref{sec:FermionicGappedBoundary} with the choice of matrices, 
\begin{align}
U = \begin{pmatrix} 0 & 1 & 0 & -1 \\ 1 & 0 & -1 & 0 \\ 0 & 1 & 0 & 1 \\ 1 & 1 & 1 & -1 \end{pmatrix}, \quad
W = \begin{pmatrix} 0 & 0 & -1 & 0 \\ -1 & 0 & 0 & -1 \\ 1 & 0 & 0 & 0 \\ 0 & 1 & -1 & 0 \end{pmatrix},
\end{align}
we find a set of integer vectors $\{ \bm{\Lambda}_{\textsf{w},\alpha} \}$,
\begin{align}
\begin{split}
\bm{\Lambda}_{\textsf{w},1} &= (-1,1,1,1)^T, \\
\bm{\Lambda}_{\textsf{w},2} &= (1,1,0,0)^T, \\
\bm{\Lambda}_{\textsf{w},3} &= (1,1,1,-1)^T, \\
\bm{\Lambda}_{\textsf{w},4} &= (0,0,1,1)^T.
\end{split}
\end{align}
We then write $\bm{\phi}^\textsf{w}_y = (\phi^1_{y,R}, \phi^1_{y,L}, \phi^2_{y,R}, \phi^2_{y,L})^T$ where $\phi_{y,R}^\sigma$ and $\phi_{y,L}^\sigma$ correspond to right- and left-moving modes of the $\sigma$th component of fermions in the $y$th wire. 
The anticommutation relations between any pair of fermion operators $\exp (i\phi_{y,R}^\sigma)$ and $\exp (i\phi_{y,L}^\sigma)$ are ensured by the commutation relations for the bosonic fields: 
\begin{align} \label{eq:CommRel2DTwoCompField}
\begin{split}
[\phi^\sigma_{y,R}(x), \phi^{\sigma'}_{y',R}(x')] &= i\pi \delta_{y,y'} \delta_{\sigma,\sigma'} \textrm{sgn}(x-x') \\
&\quad +i\pi \delta_{y,y'} \textrm{sgn}(\sigma-\sigma') +i\pi \textrm{sgn}(y-y'), \\
[\phi^\sigma_{y,L}(x), \phi^{\sigma'}_{y',L}(x')] &= -i\pi \delta_{y,y'} \delta_{\sigma,\sigma'} \textrm{sgn}(x-x') \\
&\quad +i\pi \delta_{y,y'} \textrm{sgn}(\sigma-\sigma') +i\pi \textrm{sgn}(y-y'), \\
[\phi^\sigma_{y,R}(x), \phi^{\sigma'}_{y',L}(x')] &= i\pi \delta_{y,y'} \delta_{\sigma,\sigma'} +i\pi \delta_{y,y'} \textrm{sgn}(\sigma-\sigma') \\
&\quad +i\pi \textrm{sgn}(y-y').
\end{split}
\end{align}
For this two-component fermion system, we find the tunneling Hamiltonian,
\begin{align}
\mathcal{V}_\textsf{w} &= -g \int dx \sum_{y \in \mathbb{Z}} \bigl[ \cos (\phi^1_{y,R} -\phi^1_{y,L} +\phi^2_{y,R} +\phi^2_{y,L} \nonumber \\
&\quad \quad -\phi^1_{y+1,R} -\phi^1_{y+1,L} +\phi^2_{y+1,R} -\phi^2_{y+1,L}) \nonumber \\
&\quad +\cos (\phi^2_{y,R} -\phi^2_{y,L} +\phi^1_{y+1,R} -\phi^1_{y+1,L}) \bigr].
\end{align}

While this coupled-wire Hamiltonian is fully gapped and translation invariant, its ground-state degeneracy on a torus depends on the system size due to its nature of excitations. 
A pair of quasiparticles $\bm{l} = (\bm{\lambda},\bm{\lambda}')^T$ is created by the vertex operator $\exp (i\bm{p}_{\bm{l}} \cdot \bm{\phi}^\textsf{w}_y)$ with an integer vector $\bm{p}_{\bm{l}}$ satisfying Eq.~\eqref{eq:InnerProdLambdaP2}. 
For instance, we may find $\bm{p}_{(\bm{\lambda}_e,\bm{\lambda}_f)^T}=(0,0,1,0)^T$ and $\bm{p}_{(\bm{\lambda}_f,\bm{\lambda}_e)^T}=(0,1,0,0)^T$ where $\bm{\lambda}_e = (1,0)^T$ and $\bm{\lambda}_f=(1,1)^T$. 
The associated string operator is constructed by alternatively multiplying these $(\bm{\lambda}_e, \bm{\lambda}_f)$-pair and $(\bm{\lambda}_f, \bm{\lambda}_e)$-pair creation/annihilation operators along the $y$ axis; if we create a pair of kinks associated with $e$ on the $(y-1/2)$th link and $f$ on the $(y+1/2)$th link, we must create a pair of antikinks associated with $f$ on the $(y+1/2)$th link and $e$ on $(y+3/2)$th link, and so on. 

\begin{figure}
\includegraphics[clip,width=0.45\textwidth]{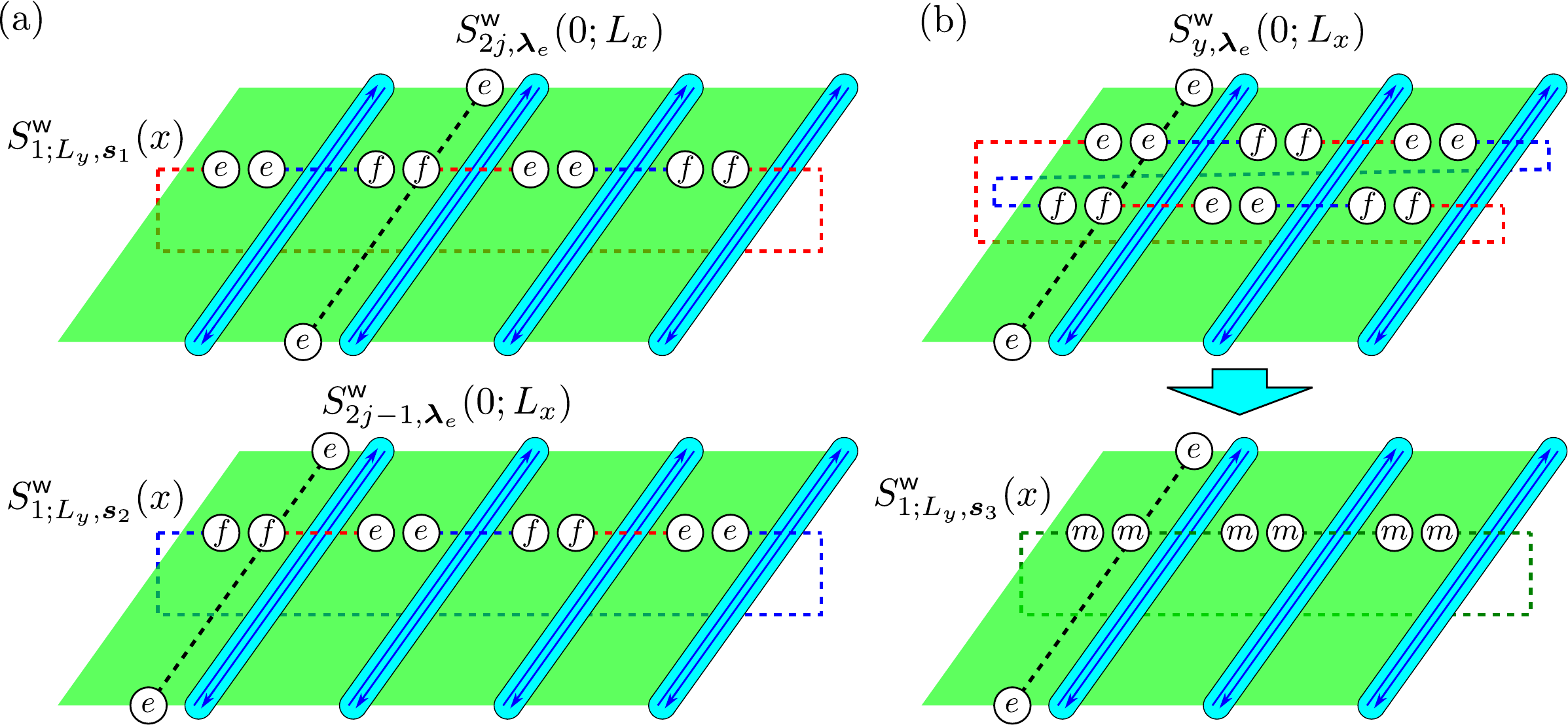}
\caption{String operators for the coupled-wire model based on the 2D toric codes with the $e \leftrightarrow f$ interfaces. 
(a) On a torus with even $L_y$, alternating excitations of $e$-$f$ and $f$-$e$ pairs lead to two string operators along the $y$ axis. 
(b) On a torus with odd $L_y$, such alternating excitations wind the torus twice and lead to one string operator creating $m$-$m$ pairs.
}
\label{fig:StringTCef}
\end{figure}

On a torus with even $L_y$, we can construct two independent string operators along the $y$ axis, 
\begin{subequations}
\begin{align}
S^\textsf{w}_{1;L_y,\bm{s}_1}(x) &= \prod_{j=1}^{L_y/2} \exp [i(\phi^2_{2j-1,R}(x) -\phi^1_{2j,L}(x))], \\
S^\textsf{w}_{1;L_y,\bm{s}_2}(x) &= \prod_{j=1}^{L_y/2} \exp [i(\phi^1_{2j-1,L}(x) -\phi^2_{2j,R}(x))],
\end{align}
\end{subequations}
which are associated with the string of excitations 
\begin{subequations}
\begin{align}
\bm{s}_1 &= \{ (\bm{\lambda}_e,\bm{\lambda}_f), -(\bm{\lambda}_f,\bm{\lambda}_e), (\bm{\lambda}_e,\bm{\lambda}_f), \cdots, -(\bm{\lambda}_f,\bm{\lambda}_e) \}, \\
\bm{s}_2 &= \{ (\bm{\lambda}_f,\bm{\lambda}_e), -(\bm{\lambda}_e,\bm{\lambda}_f), (\bm{\lambda}_f,\bm{\lambda}_e), \cdots, -(\bm{\lambda}_e,\bm{\lambda}_f) \}.
\end{align}
\end{subequations}
We can also find string operators along the $x$ axis, 
\begin{align} \label{eq:YString2DTCef}
S^\textsf{w}_{y,\bm{\lambda}_e}(0;L_x) &= \exp \left[ -\frac{i}{2} \int_0^{L_x} dx \, \partial_x (\phi^1_{y,R}(x) -\phi^1_{y,L}(x)) \right],
\end{align}
which create a pair of $e$ excitations, move one of them around a cycle of the torus along the $y$th wire, and annihilate them in pair. 
These string operators are schematically presented in Fig.~\ref{fig:StringTCef}~(a).
The string operators $S^\textsf{w}_{1;L_y,\bm{s}_1}(x)$ and $S^\textsf{w}_{2j,\bm{\lambda}_e}(0;L_x)$ form an anticommuting pair and $S^\textsf{w}_{1;L_y,\bm{s}_2}(x)$ and $S^\textsf{w}_{2j-1,\bm{\lambda}_e}(0;L_x)$ form another anticommuting pair, but any other combinations commute with each other.  
Thus, the ground-state degeneracy on a torus with even $L_y$ is four. 

On the other hand, a string of excitations of alternating $(\bm{\lambda}_e, \bm{\lambda}_f)$ and $(-\bm{\lambda}_f, -\bm{\lambda}_e)$ must wind a cycle along the $y$ axis \emph{twice} on a torus with odd $L_y$ in order to cancel kinks on all links as shown in Fig.~\ref{fig:StringTCef}~(b). 
Since $\bm{\lambda}_e-\bm{\lambda}_f = -\bm{\lambda}_m$, this process is equivalent to move $m$ excitations along the cycle and is represented by the string operator, 
\begin{align}
S^\textsf{w}_{1;L_y,\bm{s}_3}(x) = \prod_{y=1}^{L_y} \exp [-i(\phi^1_{y,L}(x) -\phi^2_{y,R}(x))],
\end{align}
which is associated with the string of excitations, 
\begin{align}
\bm{s}_3 = \{ (-\bm{\lambda}_m, \bm{\lambda}_m), (-\bm{\lambda}_m, \bm{\lambda}_m), \cdots, (-\bm{\lambda}_m, \bm{\lambda}_m) \}.
\end{align}
String operators along the $x$ axis remain the same as Eq.~\eqref{eq:YString2DTCef}. 
Since there is only one independent string operator along the $y$ axis, we have only one anticommuting pair of string operators, which leads to the ground-state degeneracy of two on a torus with odd $L_y$. 

To summarize, we find 
\begin{align}
\textrm{GSD} = \begin{cases} 2^2 & (L_y \in 2\mathbb{Z}) \\ 2 & (L_y \in 2\mathbb{Z}+1) \end{cases}.
\end{align}
As inferred from the properties of the string operators, we will have three quasiparticles in this coupled-wire model, whose properties are inherited from the 2D toric code: 
two of them fuse to the other and have the mutual statistics of $\pi$. 
However, their self statistics are obscured: as our system consists of fermionic quantum wires, the bosonic and fermionic statistics can be interchanged by adding local fermionic excitations.

\subsubsection{Doubled semion model with $\bm{1} \leftrightarrow \bar{\bm{1}}$ interface}
\label{sec:2DFermionicDoubledSemion}

We here provide another unconventional coupled-wire model starting with an array of strips of the doubled semion models, each of which is a stack of the $U(1)_2$ topological order and its antichiral counterpart $\overline{U(1)}_2$. 
Each interface hosts four gapless edge modes $\bm{\phi}^\textsf{e}_y = (\phi^\textsf{e}_{y,1}, \cdots, \phi^\textsf{e}_{y,4})^T$ corresponding to the $K$ matrix: 
\begin{align}
K_\textsf{e} = \begin{pmatrix} 2 &&& \\ & -2 && \\ && -2 & \\ &&& 2 \end{pmatrix}.
\end{align}
Since the semion $\bm{1}$ and its antichiral counterpart $\bar{\bm{1}}$, whose corresponding integer vectors are given by $\bm{\lambda}_{\bm{1}}=(1,0)^T$ and $\bm{\lambda}_{\bar{\bm{1}}}=(0,1)^T$, are their own antiparticles, a conventional coupled-wire model for the doubled semion model is obtained by condensing pairs of two $\bm{1}$'s or two $\bar{\bm{1}}$'s at each interface. 
However, we here consider a gapped interface at which pairs of $\bm{1}$ and $\bar{\bm{1}}$ from neighboring strips are condensed. 
The corresponding Lagrangian subgroup is generated by $M=\{ \bm{m}_a \}$ with 
\begin{align}
\begin{split}
\bm{m}_1 &= (1,0,0,1)^T, \\
\bm{m}_2 &= (0,1,1,0)^T.
\end{split}
\end{align}
These bound pairs of semions have the statistical spins $s=\pm 1/2$ and thus are fermionic quasiparticles.
Hence, in order to explicitly construct a gapping potential, we need to add extra fermionic wires at each interface and to consider the extended $K$ matrix $K_\textsf{ew} = K_\textsf{e} \oplus K_\textsf{w}$ with $K_\textsf{w} = Z \oplus Z$. 
Following the algorithm presented in Sec.~\ref{sec:FermionicGappedBoundary} with the choice of matrices, 
\begin{align}
U = \begin{pmatrix} 1 & 0 & 0 & -1 \\ 0 & 1 & -1 & 0 \\ 0 & 1 & 1 & 0 \\ 1 & 0 & 0 & 1 \end{pmatrix}, \quad
W = \begin{pmatrix} 0 & -1 & 0 & 0 \\ 0 & 0 & 0 & -1 \\ 0 & 0 & -1 & 0 \\ 1 & 0 & 0 & 0 \end{pmatrix},
\end{align}
we find a set of integer vectors $\{ \bm{\Lambda}_{\textsf{w},\alpha} \}$, 
\begin{align}
\begin{split}
\bm{\Lambda}_{\textsf{w},1} &= (0,1,0,1)^T, \\
\bm{\Lambda}_{\textsf{w},2} &= (-1,0,1,0)^T, \\
\bm{\Lambda}_{\textsf{w},3} &= (1,0,1,0)^T, \\
\bm{\Lambda}_{\textsf{w},4} &= (0,1,0,-1)^T.
\end{split}
\end{align}
As in Sec.~\ref{sec:CW2DTCef}, we introduce bosonic fields $\bm{\phi}^\textsf{w}_y = (\phi^1_{y,R}, \phi^1_{y,L}, \phi^2_{y,R}, \phi^2_{y,L})^T$ corresponding to two-component fermionic wires, which satisfy the commutation relations in Eq.~\eqref{eq:CommRel2DTwoCompField}. 
We then find the tunneling Hamiltonian, 
\begin{align}
\mathcal{V}_\textsf{w} &= -g \int dx \sum_{y \in \mathbb{Z}} \bigl[ \cos (\phi^1_{y,R} +\phi^2_{y,R} -\phi^1_{y+1,L} -\phi^2_{y+1,L}) \nonumber \\
&\quad +\cos (\phi^1_{y,L} -\phi^2_{y,L} -\phi^1_{y+1,R} +\phi^2_{y+1,R}) \bigr].
\end{align}

\begin{figure}
\includegraphics[clip,width=0.45\textwidth]{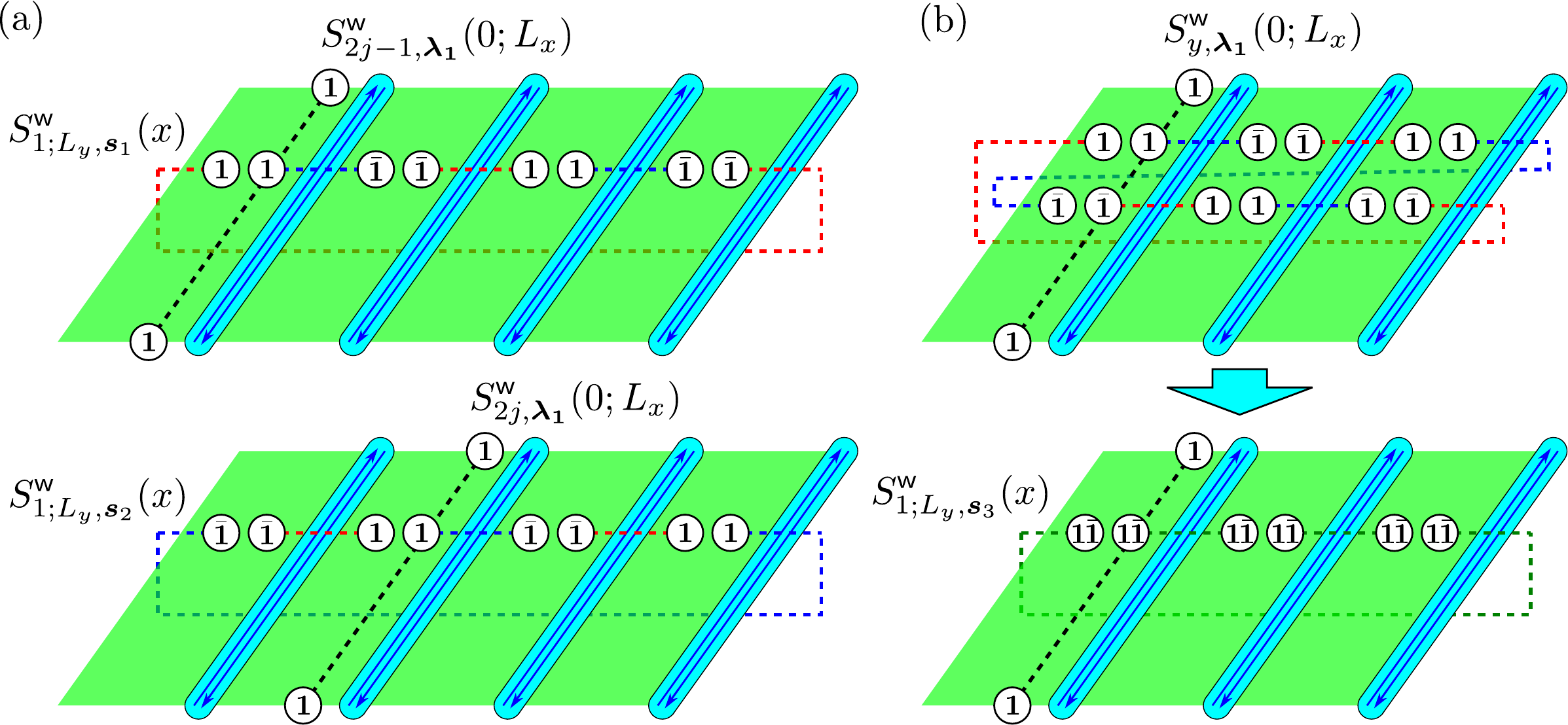}
\caption{String operators for the coupled-wire model based on the doubled semion with $\bm{1} \leftrightarrow \bar{\bm{1}}$ interfaces. 
(a) On a torus with even $L_y$, alternating excitations of $\bm{1}$-$\bar{\bm{1}}$ and $\bar{\bm{1}}$-$\bm{1}$ pairs lead to two string operators along the $y$ axis. 
(b) On a torus with odd $L_y$, such alternating excitations wind the torus twice and lead to one string operator creating pairs of bosonic quasiparticles $\bm{1} \bar{\bm{1}}$.
}
\label{fig:StringDSf}
\end{figure}

As in the previous case, this Hamiltonian is translation invariant, but its ground-state degeneracy on a torus depends on the parity of the linear size $L_y$. 
We can create kinks corresponding to a $(\bm{\lambda}_{\bm{1}}, \bm{\lambda}_{\bar{\bm{1}}})$-pair excitation on neighboring links by the vertex operator $\exp (i\bm{p} \cdot \bm{\phi}^\textsf{w}_y)$ with $\bm{p}=(0,1,0,0)^T$ and a $(\bm{\lambda}_{\bar{\bm{1}}}, \bm{\lambda}_{\bm{1}})$-pair excitation with $\bm{p}=(0,0,1,0)^T$. 
In order to construct a string operator along the $y$ axis, the two vertex operators associated with $(\bm{\lambda}_{\bm{1}}, \bm{\lambda}_{\bar{\bm{1}}})$ and $(\bm{\lambda}_{\bar{\bm{1}}}, \bm{\lambda}_{\bm{1}})$ excitations must be alternately multiplied such that semion or antichiral semion excitations are pair annihilated in the bulk of the string. 
We thus find two independent string operators on a torus with even $L_y$, 
\begin{subequations}
\begin{align}
S^\textsf{w}_{1;L_y,\bm{s}_1}(x) &= \prod_{j=1}^{L_y/2} \exp [i(\phi^1_{2j-1,L}(x) -\phi^2_{2j,R}(x))], \\
S^\textsf{w}_{1;L_y,\bm{s}_2}(x) &= \prod_{j=1}^{L_y/2} \exp [i(\phi^2_{2j-1,R}(x) -\phi^1_{2j,L}(x))],
\end{align}
\end{subequations}
which create the string of excitations, 
\begin{subequations}
\begin{align}
\bm{s}_1 &= \{ (\bm{\lambda}_{\bm{1}}, \bm{\lambda}_{\bar{\bm{1}}}), -(\bm{\lambda}_{\bar{\bm{1}}}, \bm{\lambda}_{\bm{1}}), (\bm{\lambda}_{\bm{1}}, \bm{\lambda}_{\bar{\bm{1}}}), \cdots, -(\bm{\lambda}_{\bar{\bm{1}}}, \bm{\lambda}_{\bm{1}}) \}, \\
\bm{s}_2 &= \{ (\bm{\lambda}_{\bar{\bm{1}}}, \bm{\lambda}_{\bm{1}}), -(\bm{\lambda}_{\bm{1}}, \bm{\lambda}_{\bar{\bm{1}}}), (\bm{\lambda}_{\bar{\bm{1}}}, \bm{\lambda}_{\bm{1}}), \cdots, -(\bm{\lambda}_{\bm{1}}, \bm{\lambda}_{\bar{\bm{1}}}) \},
\end{align}
\end{subequations}
as schematically shown in Fig.~\ref{fig:StringDSf}~(a).
On the other hand, such a string of excitations winds twice a cycle along the $y$ axis on a torus with odd $L_y$, resulting in creating a string of bosonic anyon excitations $\bm{\lambda}_{\bm{1}\bar{\bm{1}}}=(1,-1)^T$, which are bound states of semions and antichiral semions in the doubled semion model. 
This leads to one independent string operator [see Fig.~\ref{fig:StringDSf}~(b)], 
\begin{align}
S^\textsf{w}_{1;L_y,\bm{s}_3}(x) = \prod_{y=1}^{L_y} \exp [i(\phi^1_{y,L} -\phi^2_{y,R})],
\end{align}
with
\begin{align}
\bm{s}_3 = \{ (\bm{\lambda}_{\bm{1} \bar{\bm{1}}}, -\bm{\lambda}_{\bm{1} \bar{\bm{1}}}), (\bm{\lambda}_{\bm{1} \bar{\bm{1}}}, -\bm{\lambda}_{\bm{1} \bar{\bm{1}}}), \cdots, (\bm{\lambda}_{\bm{1} \bar{\bm{1}}}, -\bm{\lambda}_{\bm{1} \bar{\bm{1}}}) \}.
\end{align}
For both cases, we can construct string operators along the $x$ axis,
\begin{align}
S^\textsf{w}_{y,\bm{\lambda}_{\bm{1}}}(0;L_x) = \exp \left[ \frac{i}{2} \int_0^{L_x} dx \, \partial_x (\phi^1_{y,L}(x) +\phi^2_{y,L}(x)) \right],
\end{align}
which is associated with semion excitations $\bm{\lambda}_{\bm{1}}=(1,0)^T$. 
Reflecting the quasiparticle statistics of constituent doubled semion models, two pairs of the string operators $\{ S^\textsf{w}_{1;L_y,\bm{s}_1}(x), S^\textsf{w}_{2j-1,\bm{\lambda}_{\bm{1}}}(0;L_x) \}$ and $\{ S^\textsf{w}_{1;L_y,\bm{s}_2}(x), S^\textsf{w}_{2j,\bm{\lambda}_{\bm{1}}}(0;L_x) \}$ form anticommuting pairs, leading to a fourfold degenerate ground state on a torus with even $L_y$. 
On the other hand, there is only one anticommuting pair $\{ S^\textsf{w}_{1;L_y, \bm{s}_3}(x), S^\textsf{w}_{y,\bm{\lambda}_{\bm{1}}}(0;L_x) \}$ on a torus with odd $L_y$, leading to a twofold degenerate ground state. 
To summarize, we find 
\begin{align}
\textrm{GSD} = \begin{cases} 2^2 & (L_y \in 2\mathbb{Z}) \\ 2 & (L_y \in 2\mathbb{Z}+1) \end{cases}.
\end{align}

It is interesting to note that in this coupled-wire model, single local fermion excitations created by $e^{i\phi^\sigma_{y,R/L}}$ are directly fractionalized into two semion excitations, whereas in the conventional $U(1)_2$ or doubled semion topological orders, local bosonic excitations are fractionalized into two semion excitations. 
Similar excitation properties can also be found in the 3D fermionic coupled-wire model with semionic planon excitations constructed in Sec.~\ref{sec:FermionicU2Model}.

\subsubsection{Doubled $U(1)_5$ model with $\bm{1} \leftrightarrow \bar{\bm{2}}$ interface}
\label{sec:2dU5Model}

We consider a coupled-wire model starting with an array of strips made of stacks of the $U(1)_5$ and $\overline{U(1)}_5$ topological orders.
Each interface hosts gapless edge modes $\bm{\phi}^\textsf{e}_y = (\phi^\textsf{e}_{y,1}, \cdots, \phi^\textsf{e}_{y,4})^T$ corresponding to the $K$ matrix:
\begin{align}
K_\textsf{e} = \begin{pmatrix} 5 &&& \\ & -5 && \\ && -5 & \\ &&& 5 \end{pmatrix}.
\end{align}
We then consider a gapped interface at which pairs of $\bm{1}$ and $\bar{\bm{2}}$ or $\bar{\bm{1}}$ and $\bm{2}$ between neighboring strips are condensed. 
The corresponding Lagrangian subgroup is generated by $M=\{ \bm{m}_a \}$ with
\begin{align}
\begin{split}
\bm{m}_1 &= (1, 0, 0, 2)^T, \\
\bm{m}_2 &= (0, 1, 2, 0)^T,
\end{split}
\end{align}
which have $s=\pm 1/2$ and thus are fermionic quasiparticles. 
In order to explicitly construct a gapping potential, we need to add two extra fermionic wires at each interface and to consider the extended $K$ matrix $K_\textsf{ew} = K_\textsf{e} \oplus K_\textsf{w}$ with $K_\textsf{w} = Z \oplus Z$. 
Following the algorithm presented in  Sec.~\ref{sec:FermionicGappedBoundary} with the choice of matrices, 
\begin{align}
U = \begin{pmatrix} 1 & 0 & 0 & -2 \\ 0 & 1 & -2 & 0 \\ 0 & 2 & 1 & 0 \\ 2 & 0 & 0 & 1 \end{pmatrix}, \quad
W = \begin{pmatrix} 1 & 0 & 0 & 0 \\ 0 & 0 & 0 & 1 \\ 0 & 0 & 1 & 0 \\ 0 & 1 & 0 & 0 \end{pmatrix},
\end{align}
we find a set of integer vectors $\{ \bm{\Lambda}_{\textsf{w},\alpha} \}$, 
\begin{align}
\begin{split}
\bm{\Lambda}_{\textsf{w},1} &= (0, -1, 0, 2)^T, \\
\bm{\Lambda}_{\textsf{w},2} &= (2, 0, -1, 0)^T, \\
\bm{\Lambda}_{\textsf{w},3} &= (-1, 0, -2, 0)^T, \\
\bm{\Lambda}_{\textsf{w},4} &= (0, -2, 0, -1)^T.
\end{split}
\end{align}
As in Sec.~\ref{sec:CW2DTCef}, we introduce bosonic fields $\bm{\phi}^\textsf{w}_y = (\phi^1_{y,R}, \phi^1_{y,L}, \phi^2_{y,R}, \phi^2_{y,L})^T$ corresponding to two-component fermionic wires, which satisfy the commutation relations in Eq.~\eqref{eq:CommRel2DTwoCompField}. 
We then find the tunneling Hamiltonian, 
\begin{align}
\mathcal{V}_\textsf{w} &= -g \int dx \sum_{y \in \mathbb{Z}} \bigl[ \cos (\phi^1_{y,R} +2\phi^2_{y,R} +\phi^1_{y+1,L} -2\phi^2_{y+1,L}) \nonumber \\
&\quad +\cos (2\phi^1_{y,L} +\phi^2_{y,L} -2\phi^1_{y+1,R} +\phi^2_{y+1,R}) \bigr].
\end{align}

\begin{figure}
\includegraphics[clip,width=0.45\textwidth]{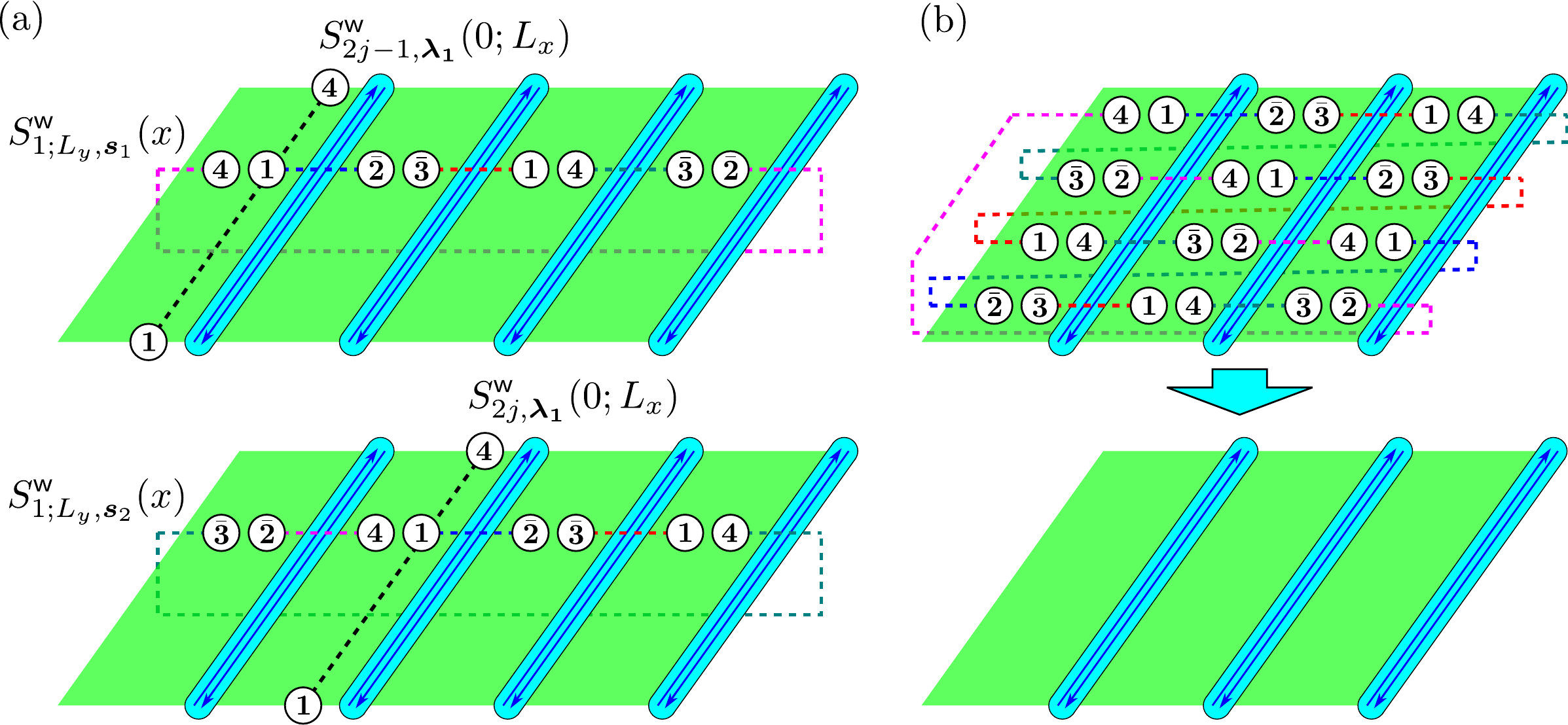}
\caption{String operators for the coupled-wire model based on the doubled $U(1)_5$ topological orders with $\bm{1} \leftrightarrow \bar{\bm{2}}$ interfaces. 
(a) On a torus with $L_y \in 4\mathbb{Z}$, a string of $\bm{1}$-$\bar{\bm{2}}$, $\bar{\bm{3}}$-$\bm{1}$, $\bm{4}$-$\bar{\bm{3}}$, and $\bar{\bm{2}}$-$\bm{4}$ pairs lead to two independent string operators along the $y$ axis. 
(b) On a torus with $L_y \in 4\mathbb{Z}+3$, the string winds the torus four times such that quasiparticles are completely canceled up to local fermionic excitations $\bm{5}$ and $\bar{\bm{5}}$.
}
\label{fig:StringU5}
\end{figure}

This Hamiltonian is translation invariant, but its ground-state degeneracy on a torus depends on the linear size $L_y$. 
To see this, let us consider a string operator along the $y$ axis, which nontrivially acts on the manifold of degenerate ground states. 
In order for excitations created by such an operator to cancel with each other, they must form a string with a pattern of the periodicity four, say, 
\begin{align}
\bm{s}_1 = \{ \bm{l}_1, \bm{l}_2, -\bm{l}_1, -\bm{l}_2, \bm{l}_1, \bm{l}_2, -\bm{l}_1, -\bm{l}_2, \cdots \}, 
\end{align}
where the excitation $\bm{l}_1 = (1,0,0,2)^T$ and $\bm{l}_2 = (0,-2,1,0)^T$ can be created by the vertex operator $\exp (i\bm{p} \cdot \bm{\phi}^\textsf{w}_y)$ with $\bm{p}=(0,-1,0,0)^T$ and $(-1,0,0,0)^T$, respectively.
Another string operator creates excitations with the same pattern but shifted by one wire: 
\begin{align}
\bm{s}_2 = \{ -\bm{l}_2, \bm{l}_1, \bm{l}_2, -\bm{l}_1, -\bm{l}_2, \bm{l}_1, \bm{l}_2, -\bm{l}_1, \cdots \}. 
\end{align}
A string of excitations obtained by further shifting $\bm{s}_2$ by one wire is equivalent to $4\bm{s}_1$ up to local fermionic excitations. 
We thus find two independent string operators along the $y$ axis for $L_y \in 4\mathbb{Z}$, 
\begin{subequations}
\begin{align}
S^\textsf{w}_{1;L_y,\bm{s}_1}(x) &= \prod_{j=1}^{L_y/4} \exp [i(-\phi^1_{4j-3,L}(x) -\phi^1_{4j-2,R}(x) \nonumber \\
&\quad +\phi^1_{4j-1, L}(x) +\phi^1_{4j, R}(x))], \\
S^\textsf{w}_{1;L_y,\bm{s}_2}(x) &= \prod_{j=1}^{L_y/4} \exp [i(\phi^1_{4j-3,R}(x) -\phi^1_{4j-2,L}(x) \nonumber \\
&\quad -\phi^1_{4j-1, R}(x) +\phi^1_{4j,L}(x))].
\end{align}
\end{subequations}
For both string operators, we can construct string operators along the $x$ axis, 
\begin{align}
S^\textsf{w}_{y, \bm{\lambda}_{\bm{1}}}(0;L_x) = \exp \left[ \frac{i}{5} \int_0^{L_x} dx \, \partial_x (\phi^1_{y,L}(x) -2\phi^2_{y,L}(x)) \right], 
\end{align}
which is associated with $s=1/10$ quasiparticle excitations $\bm{\lambda}_{\bm{1}} = (1, 0)^T$. 
These string operators are illustrated in Fig.~\ref{fig:StringU5}~(a). 
They obey the algebra,
\begin{subequations}
\begin{align}
&S^\textsf{w}_{1;L_y, \bm{s}_1}(x) S^\textsf{w}_{4j-3, \bm{\lambda}_{\bm{1}}}(0;L_x) \nonumber \\
&= e^{2\pi i/5} S^\textsf{w}_{4j-3, \bm{\lambda}_{\bm{1}}}(0;L_x) S^\textsf{w}_{1;L_y, \bm{s}_1}(x), \\
&S^\textsf{w}_{1;L_y, \bm{s}_2}(x) S^\textsf{w}_{4j-2, \bm{\lambda}_{\bm{1}}}(0;L_x) \nonumber \\
&= e^{2\pi i/5} S^\textsf{w}_{4j-2, \bm{\lambda}_{\bm{1}}}(0;L_x) S^\textsf{w}_{1;L_y, \bm{s}_2}(x), 
\end{align}
\end{subequations}
leading to the ground-state degeneracy $5^2$ on a torus with $L_y \in 4\mathbb{Z}$. 
On the other hand, on a torus with $L_y \notin 4\mathbb{Z}$, a string of excitations must wind several times a cycle along the $y$ axis to completely cancel excitations, as shown in Fig.~\ref{fig:StringU5}~(b). 
The corresponding string operator becomes trivial up to local fermionic excitations and thus leads to the unique ground state. 
To summarize, we find 
\begin{align}
\textrm{GSD} = \begin{cases} 5^2 & (L_y \in 4\mathbb{Z}) \\ 1 & (L_y \notin 4\mathbb{Z}) \end{cases}.
\end{align}

\section{3D cellular topological states as coupled-wire models}
\label{sec:CellularTopo3D}

\begin{figure}
\includegraphics[clip,width=0.45\textwidth]{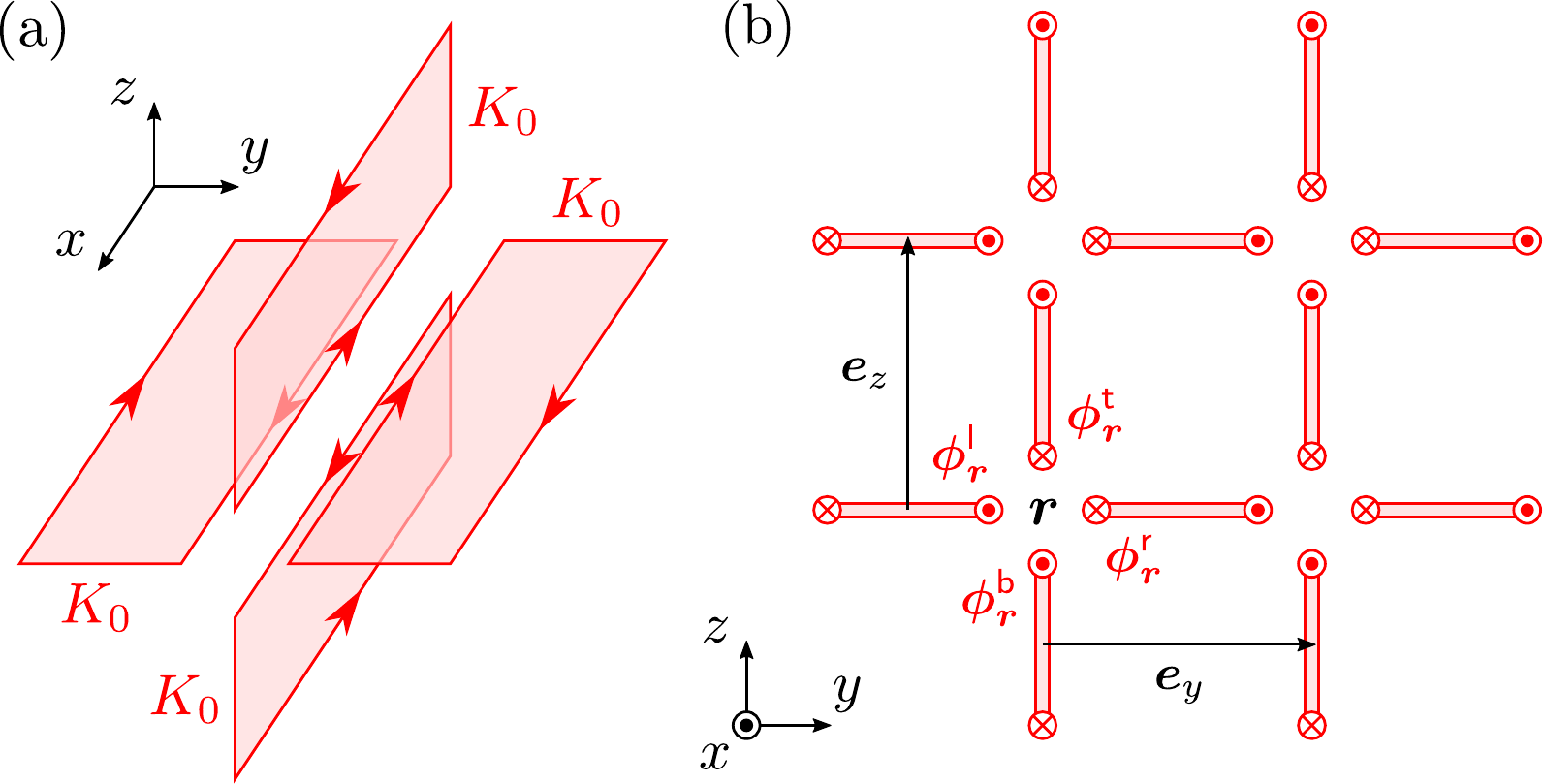}
\caption{(a) 3D cellular topological state consisting of thin strips of 2D topological orders $K_0$ along the $x$ axis. 
(b) Projection of the cellular topological state onto the $yz$ plane forms a square grid. 
}
\label{fig:CellularTopo3DSq}
\end{figure}

The construction of coupled-wire models starting from 2D cellular topological states in the previous section is readily extended to three dimensions. 
We now consider a 3D array of thin strips extended along the $x$ axis, as schematically shown in Fig.~\ref{fig:CellularTopo3DSq}~(a). 
For simplicity, we suppose that the strips form a square grid when projected onto the $yz$ plane. 
We place the same 2D Abelian topological order described by an $N_0 \times N_0$ matrix $K_0$ on each strip in both $xy$ and $xz$ planes in order to maintain translation invariance. 
Let us specify an interface between four strips by the 2D coordinates $\bm{r}=(y,z) \in \mathbb{Z}^2$ in the $yz$ plane. 
At the interface $\bm{r}$, we have $4N_0$ bosonic fields $\bm{\phi}^\textsf{e}_{\bm{r}}=(\bm{\phi}^\textsf{l}_{\bm{r}}, \bm{\phi}^\textsf{b}_{\bm{r}}, \bm{\phi}^\textsf{r}_{\bm{r}}, \bm{\phi}^\textsf{t}_{\bm{r}})^T$, where each of $\bm{\phi}^\textsf{l}_{\bm{r}}$, $\bm{\phi}^\textsf{b}_{\bm{r}}$, $\bm{\phi}^\textsf{r}_{\bm{r}}$, and $\bm{\phi}^\textsf{t}_{\bm{r}}$ is an $N_0$-component vector of bosonic fields associated with edge modes of the left, bottom, right, and top strips, respectively, surrounding the interface $\bm{r}$, as shown in Fig.~\ref{fig:CellularTopo3DSq}~(b).
We collectively denote these bosonic fields as $\bm{\phi}^\textsf{e}_{\bm{r}} = (\phi^\textsf{e}_{\bm{r}, 1}, \cdots, \phi^\textsf{e}_{\bm{r}, 4N_0})^T$. 
They obey the commutation relations, 
\begin{align}
[\partial_x \phi^\textsf{e}_{\bm{r}, \alpha}(x), \phi^\textsf{e}_{\bm{r}', \beta}(x')] = 2\pi i \delta_{\bm{r}, \bm{r}'} (K_\textsf{e}^{-1})_{\alpha \beta} \delta(x-x'), 
\end{align}
where $K_\textsf{e}$ is the $4N_0 \times 4N_0$ matrix defined by
\begin{align}
K_\textsf{e} = \begin{pmatrix} K_0 &&& \\ & K_0 && \\ && -K_0 & \\ &&& -K_0 \end{pmatrix}.
\end{align}
Here, we have assumed that each strip in the $xy$ plane has right and left edge modes described by $K_0$ and $-K_0$, respectively, and each strip in the $xz$ plane has top and bottom edge modes described by $K_0$ and $-K_0$, respectively. 
The Hamiltonian for these gapless edge modes is described by 
\begin{align}
\mathcal{H}_\textsf{e} = \frac{v_\textsf{e}}{4\pi} \int dx \sum_{\bm{r} \in \mathbb{Z}^2} \sum_{\alpha, \beta =1}^{4N_0} \partial_x \phi^\textsf{e}_{\bm{r}, \alpha} \partial_x \phi^\textsf{e}_{\bm{r}, \beta}.
\end{align}

\begin{figure}
\includegraphics[clip,width=0.45\textwidth]{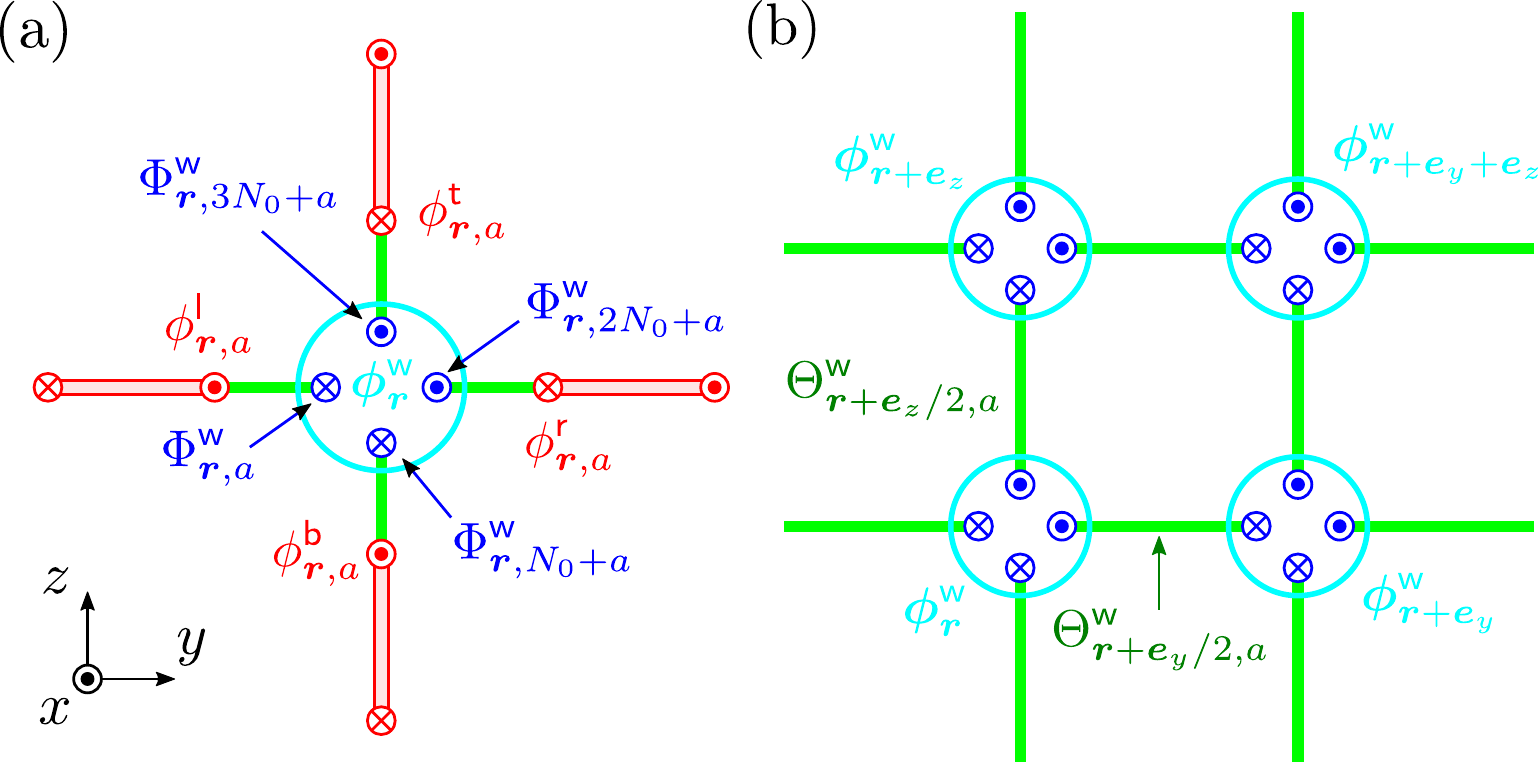}
\caption{(a) Nontrivial gapped interface can be obtained by coupling the bosonic fields $\bm{\phi}^\textsf{w}_{\bm{r}}$ from extra quantum wires with the bosonic fields $\bm{\phi}^\textsf{e}_{\bm{r}}$ from the edge modes. 
(b) After removing the strips of 2D topological orders, we obtain a coupled-wire model for the 3D cellular topological state.}
\label{fig:CellularTopo3DSqWire}
\end{figure}

We then place a gapped interface corresponding to the Lagrangian subgroup $L$ at each interface $\bm{r}$. 
As discussed in Sec.~\ref{sec:GappedBoundary}, there are some gapped interfaces whose gapping potential cannot be constructed with the associated edge modes alone. 
This is in particular the case for gapped interfaces that lead to nontrivial bulk quasiparticle properties beyond those for simple stacks of 2D topological orders. 
In such a case, we can still explicitly construct a gapping potential by adding extra $2N_0$ bosonic or fermionic wires, depending on whether the corresponding gapped interface is bosonic or fermionic, as schematically shown in Fig.~\ref{fig:CellularTopo3DSqWire}~(a).
Let us introduce $4N_0$ bosonic fields $\bm{\phi}^\textsf{w}_{\bm{r}} = (\phi^\textsf{w}_{\bm{r}, 1}, \cdots, \phi^\textsf{w}_{\bm{r}, 4N_0})^T$ associated with extra $2N_0$ quantum wires, which satisfy the commutation relations, 
\begin{align}
[\partial_x \phi^\textsf{w}_{\bm{r}, \alpha}(x), \phi^\textsf{w}_{\bm{r}', \beta}(x')] = 2\pi i \delta_{\bm{r}, \bm{r}'} (K_\textsf{w})_{\alpha \beta} \delta(x-x'), 
\end{align}
where $K_\textsf{w} = X^{\oplus 2N_0}$ for bosonic gapped interfaces and $K_\textsf{w} = Z^{\oplus 2N_0}$ for fermionic ones. 
We note that $K_\textsf{w}^{-1} = K_\textsf{w}$. 
We thus consider the Hamltonian for gapless bosonic fields, 
\begin{align} \label{eq:ExtendedDecoupledHam}
\mathcal{H}_\textsf{ew} = \frac{v_\textsf{ew}}{4\pi} \int dx \sum_{\bm{r} \in \mathbb{Z}^2} \sum_{I,J=1}^{8N_0} \partial_x \phi^\textsf{ew}_{\bm{r}, I} \partial_x \phi^\textsf{ew}_{\bm{r}, J}, 
\end{align}
where $\bm{\phi}^\textsf{ew}_{\bm{r}} = (\bm{\phi}^\textsf{e}_{\bm{r}}, \bm{\phi}^\textsf{w}_{\bm{r}})^T$. 
One can then find a gapping potential, 
\begin{align} \label{eq:GappingPotential3D}
\mathcal{V}_\textsf{ew} = -g_\textsf{ew} \int dx \sum_{\bm{r} \in \mathbb{Z}^2} \sum_{\alpha=1}^{4N_0} \cos (\widetilde{\bm{\Lambda}}_\alpha^T K_\textsf{ew} \bm{\phi}^\textsf{ew}_{\bm{r}}(x)),
\end{align}
where $\widetilde{\bm{\Lambda}}_\alpha$ are $8N_0$-dimensional integer vectors and 
\begin{align}
K_\textsf{ew} = K_\textsf{e} \oplus K_\textsf{w} = \begin{pmatrix} K_\textsf{e} & \\ & K_\textsf{w} \end{pmatrix}.
\end{align} 
Furthermore, as discussed in Sec.~\ref{sec:GappedInterface} and detailed in Appendix~\ref{sec:LambdaWire}, one can always find a set of integer vectors $\{ \widetilde{\bm{\Lambda}}_\alpha \}$ in the form, 
\begin{align}
\begin{pmatrix} \widetilde{\bm{\Lambda}}_1 & \cdots & \widetilde{\bm{\Lambda}}_{4N_0} \end{pmatrix} = \begin{pmatrix} I_{4N_0} \\ \Lambda_\textsf{w} \end{pmatrix},
\end{align}
where $\Lambda_\textsf{w}$ is a $4N_0 \times 4N_0$ integer matrix. 
We can thus rewrite Eq.~\eqref{eq:GappingPotential3D} as 
\begin{align} \label{eq:GappingPotential3D2}
\mathcal{V}_\textsf{ew} &= -g_\textsf{ew} \int dx \sum_{\bm{r} \in \mathbb{Z}^2} \sum_{a=1}^{N_0} \Bigl( \cos \Theta^\textsf{ew,l}_{\bm{r}, a} +\cos \Theta^\textsf{ew,b}_{\bm{r}, a} \nonumber \\
&\quad +\cos \Theta^\textsf{ew,r}_{\bm{r}, a} +\cos \Theta^\textsf{ew,t}_{\bm{r}, a} \Bigr),
\end{align}
where we have defined
\begin{align}
\begin{split}
\Theta^\textsf{ew,l}_{\bm{r}, a} &= (K_0 \bm{\phi}^\textsf{l}_{\bm{r}})_a +\Phi^\textsf{w}_{\bm{r},a}, \\
\Theta^\textsf{ew,b}_{\bm{r}, a} &= (K_0 \bm{\phi}^\textsf{b}_{\bm{r}})_a +\Phi^\textsf{w}_{\bm{r},N_0+a}, \\
\Theta^\textsf{ew,r}_{\bm{r}, a} &= -(K_0 \bm{\phi}^\textsf{r}_{\bm{r}})_a +\Phi^\textsf{w}_{\bm{r},2N_0+a}, \\
\Theta^\textsf{ew,t}_{\bm{r}, a} &= -(K_0 \bm{\phi}^\textsf{t}_{\bm{r}})_a +\Phi^\textsf{w}_{\bm{r},3N_0+a},
\end{split}
\end{align}
and 
\begin{align}
\Phi^\textsf{w}_{\bm{r}, \alpha} = \bm{\Lambda}_{\textsf{w},\alpha}^T K_\textsf{w} \bm{\phi}^\textsf{w}_{\bm{r}},
\end{align}
and $\bm{\Lambda}_{\textsf{w},\alpha}$ is the $\alpha$th column of $\Lambda_\textsf{w}$. 

The Hamiltonian $\mathcal{H}_\textsf{ew} +\mathcal{V}_\textsf{ew}$ defined by Eqs.~\eqref{eq:ExtendedDecoupledHam} and \eqref{eq:GappingPotential3D2} provides a coupled-edge model for 3D cellular topological states built out of thin strips of 2D topological orders $K_0$ and their gapped interface $L$ in terms of the gapless edge modes and extra quantum wires. 
However, as seen from the construction of 2D cellular topological states in Sec.~\ref{sec:2DCWModel}, this Hamiltonian is not yet minimal. 
As shown in Fig.~\ref{fig:CellularTopo3DSqWire}~(b), we can further shrink and remove the strips of 2D topological orders to obtain the Hamiltonian only with purely 1D quantum wires, 
\begin{align}
\mathcal{H}_\textsf{w} &= \frac{v_\textsf{w}}{4\pi} \int dx \sum_{\bm{r} \in \mathbb{Z}^2} \sum_{\alpha,\beta=1}^{4N_0} \partial_x \phi^\textsf{w}_{\bm{r}, \alpha} \partial_x \phi^\textsf{w}_{\bm{r}, \beta}, \\
\mathcal{V}_\textsf{w} &= -g_\textsf{w} \int dx \sum_{\bm{r} \in \mathbb{Z}^2} \sum_{a=1}^{N_0} \Bigl( \cos \Theta^\textsf{w}_{\bm{r} +\bm{e}_y/2, a} +\cos \Theta^\textsf{w}_{\bm{r} +\bm{e}_z/2, a} \Bigl),
\end{align}
where we have defined $\bm{e}_y = (1,0)$, $\bm{e}_z = (0,1)$, and
\begin{align}
\begin{split}
\Theta^\textsf{w}_{\bm{r} +\bm{e}_y/2, a} &= \Phi^\textsf{w}_{\bm{r}, 2N_0+a} +\Phi^\textsf{w}_{\bm{r} +\bm{e}_y, a}, \\
\Theta^\textsf{w}_{\bm{r} +\bm{e}_z/2, a} &= \Phi^\textsf{w}_{\bm{r}, 3N_0+a} +\Phi^\textsf{w}_{\bm{r} +\bm{e}_z, N_0+a}.
\end{split}
\end{align}
Hence, the quantum wires added at each interface are now directly coupled with each other between neighboring interfaces. 
This shrinking process can also be understood within perturbation theory as in the 2D case. 
Very thin strips have quasiparticle tunnelings, $-g_\textsf{qp} \cos (\phi^\textsf{r}_{\bm{r},a} -\phi^\textsf{l}_{\bm{r}+\bm{e}_y,a})$ and $-g_\textsf{qp} \cos (\phi^\textsf{t}_{\bm{r},a} -\phi^\textsf{b}_{\bm{r}+\bm{e}_z,a})$, connecting their edge modes. 
In the strong-coupling limit $g_\textsf{qp} \to \infty$, the bosonic fields from the edge modes are pinned at the cosine minima so that $\phi^\textsf{r}_{\bm{r},a} -\phi^\textsf{l}_{\bm{r}+\bm{e}_y,a} = 2\pi n_{\bm{r},a}$ and {$\phi^\textsf{t}_{\bm{r},a} -\phi^\textsf{b}_{\bm{r}+\bm{e}_z,a} = 2\pi m_{\bm{r},a}$ with $n_{\bm{r},a}, m_{\bm{r},a} \in \mathbb{Z}$. 
In this limit, individual bosonic fields $\bm{\phi}^\textsf{l}_{\bm{r}}$, $\bm{\phi}^\textsf{b}_{\bm{r}}$, $\bm{\phi}^\textsf{r}_{\bm{r}}$, and $\bm{\phi}^\textsf{t}_{\bm{r}}$ are strongly fluctuating and thus the tunneling terms in $\mathcal{V}_\textsf{ew}$ have vanishing expectation values. 
However, second-order perturbations in $\mathcal{V}_\textsf{ew}$ generate terms of the form $\cos (\Theta^\textsf{ew,l}_{\bm{r}+\bm{e}_y, a} +\Theta^\textsf{ew,r}_{\bm{r},a})$ or $\cos (\Theta^\textsf{ew,b}_{\bm{r}+\bm{e}_z,a} +\Theta^\textsf{ew,t}_{\bm{r},a})$, which contain the fluctuating fields in the pinned combination $\phi^\textsf{r}_{\bm{r},a} -\phi^\textsf{l}_{\bm{r}+\bm{e}_y,a}$ or $\phi^\textsf{t}_{\bm{r},a} -\phi^\textsf{b}_{\bm{r}+\bm{e}_z,a}$; these second-order terms yield the tunneling terms in $\mathcal{V}_\textsf{w}$.

When we consider a system with boundaries in the $y$ or $z$ directions, some unpaired gapless modes $\Phi^\textsf{w}_{\bm{r}, \alpha}$ are left at the boundaries. 
They satisfy the commutation relations, 
\begin{align}
[\partial_x \Phi^\textsf{w}_{\bm{r}, \alpha}(x), \Phi^\textsf{w}_{\bm{r}', \beta}(x')] = -2\pi i \delta_{\bm{r}, \bm{r}'} (K_\textsf{e})_{\alpha \beta} \delta(x-x'), 
\end{align}
and thus inherit properties of the edge states for the constituent topological order $K_0$. 
In contrast to the 2D case, even though $K_0$ is chiral, these surface modes might be gapped by short-range interactions between wires, depending on how the surface is terminated. 
In the present case for the square grid, a surface termination parallel to the $[010]$, $[001]$, or $[011]$ plane yields multiple gapless modes $\Phi^\textsf{w}_{\bm{r}.\alpha}$ with the same chirality, which can never be gapped. 
On the other hand, a surface termination parallel to the $[01\bar{1}]$ plane gives multiple gapless modes with opposite chiralities, which can potentially be gapped by local interactions.

Quasiparticle excitations created by local operators in the $yz$ plane are associated with condensed anyons $\bm{l}$ in the Lagrangian subgroup, i.e., $\bm{l} = (\bm{\lambda}^\textsf{l}, \bm{\lambda}^\textsf{b}, \bm{\lambda}^\textsf{r}, \bm{\lambda}^\textsf{t})^T \in L$, where $\bm{\lambda}^\textsf{l}$, $\bm{\lambda}^\textsf{b}$, $\bm{\lambda}^\textsf{r}$, and $\bm{\lambda}^\textsf{t}$ are $N$-dimensional integer vectors representing individual anyons in the left, bottom, right, and top strips, respectively, surrounding an interface. 
As in the case of 2D cellular topological states, kinks $2\pi \lambda^\textsf{l}_a$, $2\pi \lambda^\textsf{b}_a$, $2\pi \lambda^\textsf{r}_a$, and $2\pi \lambda^\textsf{t}_a$ in the link fields $\Theta^\textsf{w}_{\bm{r} -\bm{e}_y/2, a}$, $\Theta^\textsf{w}_{\bm{r} -\bm{e}_z/2, a}$, $\Theta^\textsf{w}_{\bm{r} +\bm{e}_y/2, a}$, and $\Theta^\textsf{w}_{\bm{r} +\bm{e}_z/2, a}$, respectively, can be regarded as quasiparticle excitations in our coupled-wire model. 
They can be created by acting a vertex operator $\exp [i\bm{p}_{\bm{l}} \cdot \bm{\phi}^\textsf{w}_{\bm{r}}(x)]$ on the ground state, where $\bm{p}_{\bm{l}}$ is a $4N_0$-dimensional integer vector satisfying 
\begin{align}
\begin{split}
\bm{\Lambda}_{\textsf{w},a} \cdot \bm{p}_{\bm{l}} &= \lambda^\textsf{l}_a, \\
\bm{\Lambda}_{\textsf{w},N+a} \cdot \bm{p}_{\bm{l}} &= \lambda^\textsf{b}_a, \\
\bm{\Lambda}_{\textsf{w},2N+a} \cdot \bm{p}_{\bm{l}} &= \lambda^\textsf{r}_a, \\
\bm{\Lambda}_{\textsf{w},3N+a} \cdot \bm{p}_{\bm{l}} &= \lambda^\textsf{t}_a.
\end{split}
\end{align}
These quasiparticle excitations are illustrated in Fig.~\ref{fig:CellularTopo3DSqQP}~(a). 
We can also prove that kinks created by a local operator $\exp [i\bm{p} \cdot \bm{\phi}^\textsf{w}_{\bm{r}}(x)]$ correspond to some condensed quasiparticle in the Lagrangian subgroup, i.e., $\bm{l} \in L$, for any integer vector $\bm{p}$. 
\begin{figure}
\includegraphics[clip,width=0.4\textwidth]{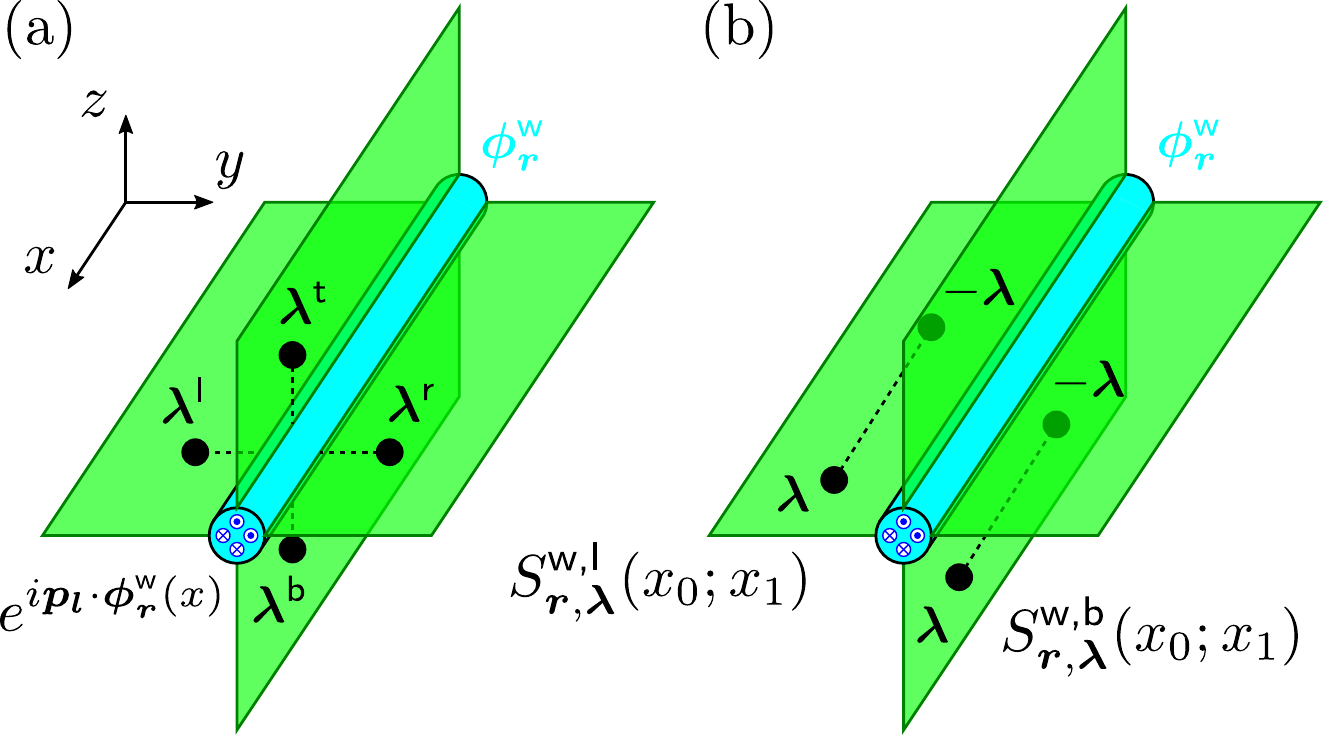}
\caption{(a) Quasiparticle excitations in the $yz$ plane associated with $\bm{l} = (\bm{\lambda}^\textsf{l}, \bm{\lambda}^\textsf{b}, \bm{\lambda}^\textsf{r}, \bm{\lambda}^\textsf{t})^T$ are created by a local operator $\exp [i\bm{p}_{\bm{l}} \cdot \bm{\phi}^\textsf{w}_{\bm{r}}(x)]$. 
(b) Pairs of lineons $\bm{\lambda}$ and $-\bm{\lambda}$ along the $x$ axis are created by the string operators $S^\textsf{w,l}_{\bm{r},\bm{\lambda}}(x_0;x_1)$ and $S^\textsf{w,b}_{\bm{r},\bm{\lambda}}(x_0;x_1)$.
}
\label{fig:CellularTopo3DSqQP}
\end{figure}
On the other hand, quasiparticles along the $x$ axis can always be created in pair on each strip. 
On a strip in the $xy$ plane, a pair of $\bm{\lambda}$ at $x_1$ and $-\bm{\lambda}$ at $x_0$, which corresponds to a pair of kinks $2\pi \lambda_a$ and $-2\pi \lambda_a$ in the link field $\Theta^\textsf{w}_{\bm{r} -\bm{e}_y/2, a}$, can be created by the string operator, 
\begin{align} \label{eq:XStringOperator3D}
S^\textsf{w,l}_{\bm{r}, \bm{\lambda}}(x_0;x_1) = \exp \left[ -i \bm{q}_{\bm{\lambda}}^\textsf{l} \cdot \int_{x_0}^{x_1} dx \, \partial_x \bm{\phi}^\textsf{w}_{\bm{r}}(x) \right],
\end{align}
where $\bm{q}_{\bm{\lambda}}^\textsf{l}$ is a $4N_0$-dimensional integer vector given by
\begin{align}
\bm{q}_{\bm{\lambda}}^\textsf{l} = \sum_{a,b=1}^{N_0} (K_0^{-1})_{ab} \lambda_b K_\textsf{w} \bm{\Lambda}_{\textsf{w},a}.
\end{align}
Similarly, on a strip in the $xz$ plane, a pair of $\bm{\lambda}$ and $-\bm{\lambda}$ corresponding to a pair of kinks in the link field $\Theta^\textsf{w}_{\bm{r} -\bm{e}_z/2, a}$ can be created by the string operator, 
\begin{align} \label{eq:XStringOperator3D2}
S^\textsf{w,b}_{\bm{r}, \bm{\lambda}}(x_0;x_1) = \exp \left[ -i \bm{q}_{\bm{\lambda}}^\textsf{b} \cdot \int_{x_0}^{x_1} dx \, \partial_x \bm{\phi}^\textsf{w}_{\bm{r}}(x) \right],
\end{align}
where $\bm{q}_{\bm{\lambda}}^\textsf{b}$ is a $4N_0$-dimensional integer vector, 
\begin{align}
\bm{q}_{\bm{\lambda}}^\textsf{b} = \sum_{a,b=1}^{N_0} (K_0^{-1})_{ab} \lambda_b K_\textsf{w} \bm{\Lambda}_{\textsf{w},N+a}.
\end{align}
These excitations along strips are illustrated in Fig.~\ref{fig:CellularTopo3DSqQP}~(b).

In the context of fracton phases, the string operators \eqref{eq:XStringOperator3D} and \eqref{eq:XStringOperator3D2} along the $x$ axis imply the existence of lineon excitations. 
As evident from the construction of 3D cellular topological states, this is a consequence of the original quasiparticles freely moving along the thin strips of 2D topological orders. 
In contrast, the mobility of quasiparticles in the $yz$ plane is more intricate and depends on the choice of gapped interfaces. 
If the corresponding Lagrangian subgroup $L$ contains pairs of quasiparticles between two strips, that is, $\bm{l} = (\bm{\lambda}^\textsf{l}, \bm{\lambda}^\textsf{b}, \bm{\lambda}^\textsf{r}, \bm{\lambda}^\textsf{t})^T \in L$ in which two of $\bm{\lambda}$'s are zero but the other two are nonzero, there will be point-like excitations transferred within the $yz$ plane by successively pair-creating and pair-annihilating excitations $\bm{l}$ along a certain path. 
Such a path will be deformable in the $yz$ plane if pairs of quasiparticles can be created between arbitrary two strips, whereas it will be undeformable if pairs are allowed only between two fixed strips. 
In combination with the lineon nature of quasiparticles along the $x$ axis, a point-like quasiparticle moving along deformable paths in the $yz$ plane implies a point-like quasiparticle fully mobile in the 3D space, while that moving along undeformable paths implies a planon mobile only within a 2D plane. 

When $\bm{l} = (\bm{\lambda}^\textsf{l}, \bm{\lambda}^\textsf{b}, \bm{\lambda}^\textsf{r}, \bm{\lambda}^\textsf{t})^T \in L$ contains three or four nonzero $\bm{\lambda}$'s and cannot be decomposed into more elementary pair excitations, the nature of quasiparticles becomes more complex. 
At a single interface, an elementary excitation associated with such $\bm{l}$ takes the form of a tiny loop. 
However, when combined with the same or other quasiparticles $\bm{l}' \in L$ created on neighboring interfaces, excitations on the whole will behave in various ways; they may be viewed as loop-like excitations fully mobile and deformable in the $yz$ plane, dipole or multipole excitations moving along undeformable paths, few point-like excitations living at the boundary of fractal-like geometric objects, and so on. 

In the case that the Lagrangian subgroup $L$ contains a pair of quasiparticles, we have possibilities to find cellular topological states for foliated type-I fracton order, TQFT-type topological order, or their hybrid. 
We construct coupled-wire models for foliated type-I fracton order with only planons in Sec.~\ref{sec:FoliatedTypeIPlanon} and for that with lineons, which can behave as planons by fusion or by forming a dipole, in Sec.~\ref{sec:FoliatedTypeILineon}. 
In both cases, there are quasiparticles whose motion in the $yz$ plane is restricted along undeformable 1D paths. 
When there are quasiparticles moving along deformable 1D paths in the $yz$ plane, we can have TQFT-type topological order, for which 3D loop-like excitations also emerge in addition to 3D point-like excitations. 
We construct the corresponding coupled-wire models in Sec.~\ref{sec:TQFTType}. 
In Sec.~\ref{sec:Hybrid}, we also provide coupled-wire models for hybrid of TQFT-type topological order and foliated type-I fracton order, in which not only 3D point-like excitations but also planons have nontrivial braiding statistics with 3D loop-like excitations.
All these models are constructed from cellular topological states on the square grid, but we find that some of them can also be constructed on the honeycomb grid with preserving quasiparticle properties by employing the conformal embedding structure of gapped interfaces, which is discussed in detail in Appendix~\ref{sec:CellularTopoHoneycomb}.

Some Lagrangian subgroup $L$ does not admit any pairs of quasiparticles but only multiplets of them. 
This is a necessary condition for the absence of point-like excitations moving along a 1D path in the $yz$ plane.
Through the classification of gapped interfaces (see the SM \cite{suppl}), we found that four $U(1)_7$ topological orders and six $U(1)_3$ topological orders admit such gapped interfaces with no condensed pairs. 
We then construct coupled-wire models from 3D cellular topological states on the square or triangular grid with those interfaces in Sec.~\ref{sec:FractalTypeI}. 
Although we do not have a rigorous proof, we argue that these coupled-wire models do not have mobile point-like excitations in the $yz$ plane. 
By examining the nature of quasiparticles and computing the ground-state degeneracy on a torus, we conclude that these models are sorts of the fractal type-I fracton models with lineon excitations. 

We remark that many of nontrivial gapped interfaces used for constructing the coupled-wire models in the present and previous sections have been found through the comprehensive classification of gapped interfaces between $U(1)_k$ topological orders, which is summarized in the SM \cite{suppl}. 
In the SM \cite{suppl}, we have also numerically computed the ground-state degeneracies of coupled-wire models, which are systematically constructed from $U(1)_k$ topological orders and their gapped interfaces, on an $L_x \times L \times L$ torus of the square and triangular grids. 
In doing so, we have employed an \textit{ab initio} approach that has been proposed in Refs.~\cite{Ganeshan16, Ganeshan17} and applied to a 3D coupled-wire model in Ref.~\cite{Fuji19b} for computing the degeneracy. 
The results coincide with those obtained in this section in a heuristic way that requires explicit construction of string and membrane operators commuting with coupled-wire Hamiltonians.

\subsection{Foliated type-I fracton models with only planons}
\label{sec:FoliatedTypeIPlanon}

We consider two models on the square grid: one consists of thin strips of the $U(1)_2$ topological orders and the other consists of those of the $U(1)_6$ topological orders. 
The corresponding coupled-wire models exhibit fractionalized quasiparticles confined in a 2D plane, namely planons. 
These models provide the simplest examples of the foliated type-I fracton model with only planons, but they are not quite trivial in the sense that they cannot be naively decomposed into decoupled stacks of 2D topological orders as we will explain.

\subsubsection{Fermionic $U(1)_2$ model}
\label{sec:FermionicU2Model}

We consider a 3D cellular topological state built out of the $U(1)_2$ topological orders, each of which is described by the $1 \times 1$ $K$ matrix $K_0=2$. 
On the square grid, each interface possesses four gapless edge modes $\bm{\phi}^\textsf{e}_{\bm{r}} = (\phi^\textsf{e}_{\bm{r},1}, \cdots, \phi^\textsf{e}_{\bm{r}, 4})^T$ corresponding to the $K$ matrix, 
\begin{align}
K_\textsf{e} = \begin{pmatrix} 2 &&& \\ & 2 && \\ && -2 & \\ &&& -2 \end{pmatrix}.
\end{align}
We then consider a fermionic gapped interface obtained by condensing pairs of semions with the same chirality, which is the same gapped interface used to construct a 2D cellular topological state from the doubled semion models in Sec.~\ref{sec:2DFermionicDoubledSemion}. 
The corresponding Lagrangian subgroup is generated by $M = \{ \bm{m}_a \}$ with 
\begin{align} \label{eq:VecMFermionicU2Model}
\begin{split}
\bm{m}_1 &= (1,1,0,0)^T, \\
\bm{m}_2 &= (0,0,1,1)^T.
\end{split}
\end{align}
In order to construct a gapping potential, we add two extra fermionic wires at each interface and consider the extended $K$ matrix $K_\textsf{ew} = K_\textsf{e} \oplus K_\textsf{w}$ with $K_\textsf{w} = Z \oplus Z$. 
Following the algorithm presented in Sec.~\ref{sec:FermionicGappedBoundary} with the choice of matrices, 
\begin{align}
U = \begin{pmatrix} 1 & -1 & 0 & 0 \\ 1 & 1 & 0 & 0 \\ 0 & 0 & 1 & -1 \\ 0 & 0 & 1 & 1 \end{pmatrix}, \quad
W = \begin{pmatrix} 1 & 0 & 0 & 0 \\ 0 & 1 & 0 & 0 \\ 0 & 0 & 1 & 0 \\ 0 & 0 & 0 & 1 \end{pmatrix},
\end{align}
we find a set of integer vectors $\{ \bm{\Lambda}_{\textsf{w},\alpha} \}$, 
\begin{align}
\begin{split}
\bm{\Lambda}_{\textsf{w},1} &= (0,-1,0,1)^T, \\
\bm{\Lambda}_{\textsf{w},2} &= (0,-1,0,-1)^T, \\
\bm{\Lambda}_{\textsf{w},3} &= (-1,0,1,0)^T, \\
\bm{\Lambda}_{\textsf{w},4} &= (-1,0,-1,0)^T.
\end{split}
\end{align}
We introduce bosonic fields $\bm{\phi}^\textsf{w}_{\bm{r}} = (\phi^1_{\bm{r},R}, \phi^1_{\bm{r},L}, \phi^2_{\bm{r},R}, \phi^2_{\bm{r},L})^T$ corresponding to two-component fermionic wires, which satisfy the commutation relations,
\begin{align} \label{eq:CommRel3DTwoCompField}
\begin{split}
[\phi^\sigma_{\bm{r},R}(x), \phi^{\sigma'}_{\bm{r}',R}(x')] &= i\pi \delta_{\bm{r},\bm{r}'} \delta_{\sigma,\sigma'} \textrm{sgn}(x-x') \\
&\quad +i\pi \delta_{\bm{r},\bm{r}'} \textrm{sgn}(\sigma-\sigma') +i\pi \textrm{sgn}(\bm{r}-\bm{r}'), \\
[\phi^\sigma_{\bm{r},L}(x), \phi^{\sigma'}_{\bm{r}',L}(x')] &= -i\pi \delta_{\bm{r},\bm{r}'} \delta_{\sigma,\sigma'} \textrm{sgn}(x-x') \\
&\quad +i\pi \delta_{\bm{r},\bm{r}'} \textrm{sgn}(\sigma-\sigma') +i\pi \textrm{sgn}(\bm{r}-\bm{r}'), \\
[\phi^\sigma_{\bm{r},R}(x), \phi^{\sigma'}_{\bm{r}',L}(x')] &= i\pi \delta_{\bm{r},\bm{r}'} \delta_{\sigma,\sigma'} +i\pi \delta_{\bm{r},\bm{r}'} \textrm{sgn}(\sigma-\sigma') \\
&\quad +i\pi \textrm{sgn}(\bm{r}-\bm{r}'),
\end{split}
\end{align}
where we have defined 
\begin{align}
\sgn(\bm{r}-\bm{r}') \equiv \delta_{z, z'} \sgn(y-y') +\sgn(z-z')
\end{align}
for $\bm{r} = (y,z)$ and $\bm{r}'=(y',z')$.
We then find the tunneling Hamiltonian, 
\begin{align} \label{eq:TunnelingHam3DSq}
\mathcal{V}_\textsf{w} &= -g \int dx \sum_{\bm{r} \in \mathbb{Z}^2} \bigl[ \cos (\Phi^-_{\bm{r}, R} +\Phi^-_{\bm{r} +\bm{e}_y, L}) \nonumber \\
&\quad +\cos (\Phi^+_{\bm{r}, R} +\Phi^+_{\bm{r} +\bm{e}_z, L}) \bigr], 
\end{align}
where
\begin{align} \label{eq:FieldFermionicU2Model}
\begin{split}
\Phi^\pm_{\bm{r}, R} &= \phi^1_{\bm{r}, R} \pm \phi^2_{\bm{r}, R}, \\
\Phi^\pm_{\bm{r}, L} &= \phi^1_{\bm{r}, L} \pm \phi^2_{\bm{r}, L}.
\end{split}
\end{align}
This is schematically shown in Fig.~\ref{fig:ModelFermionU2Sq}~(a). 
\begin{figure}
\includegraphics[clip,width=0.4\textwidth]{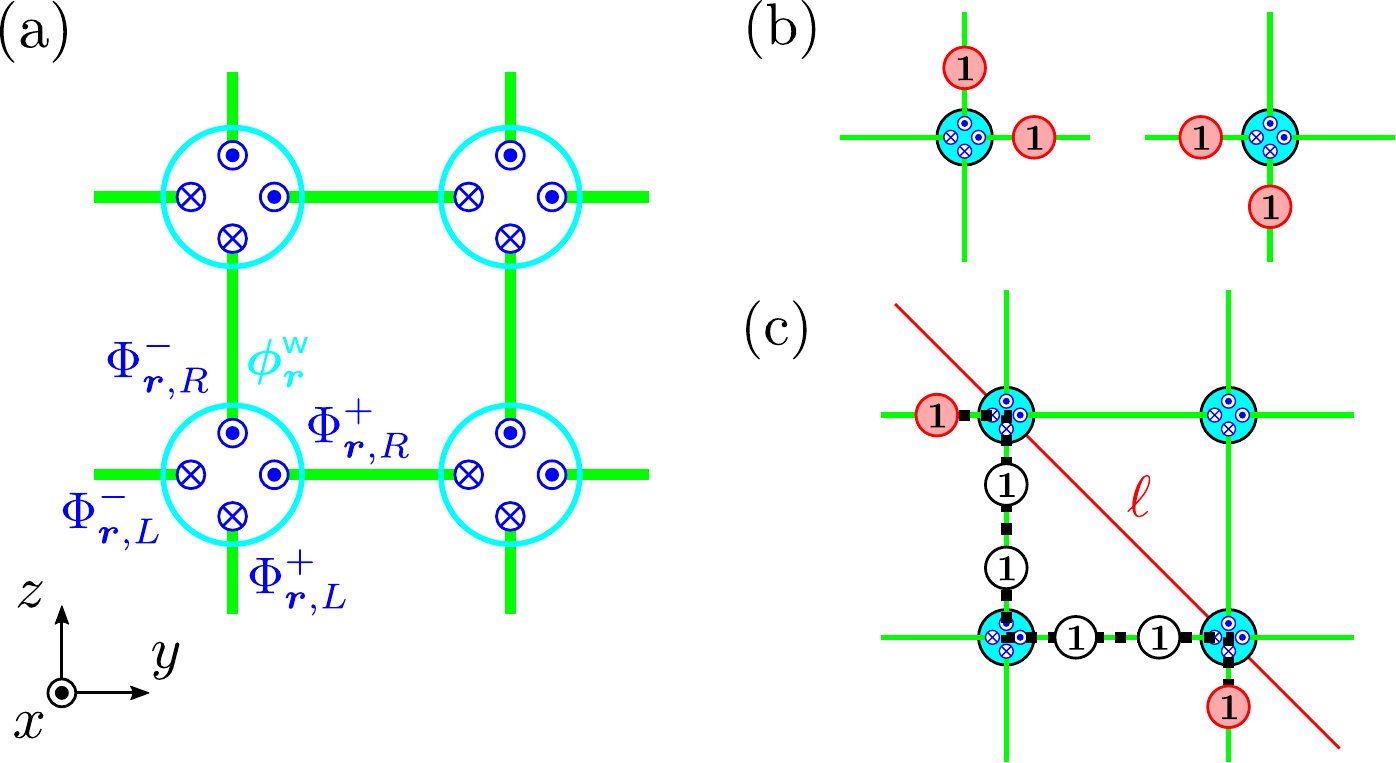}
\caption{(a) Coupled-wire model for cellular topological states built out of the $U(1)_2$ topological orders with fermionic gapped interfaces on the square grid. 
(b) Elementary excitations created by local operators in the fermionic $U(1)_2$ model. 
(c) Semionic quasiparticle $\bm{1}$ can be transferred along the zigzag path (dashed line) in the $yz$ plane, which can be specified by a diagonal line $\ell$ (red solid line). 
Red circles represent quasiparticles that cost energy, while white circles represent quasiparticles annihilated in pairs.
}
\label{fig:ModelFermionU2Sq}
\end{figure}
We note that if we regard two-component fermions are spinful electrons $\psi^\uparrow_{\bm{r}, R/L} \propto \exp (i\phi^1_{\bm{r}, R/L})$ and $\psi^\downarrow_{\bm{r}, R/L} \propto \exp (i\phi^2_{\bm{r}, R/L})$, the operators $\exp (i\Phi^+_{\bm{r}, R/L})$ and $\exp (i\Phi^-_{\bm{r}, R/L})$ correspond to charge and spin $SU(2)_1$ currents, respectively, which naturally appear in the non-Abelian bosonization of the 1D half-filled Hubbard model \cite{Fradkin}. 
Thus, the tunneling Hamiltonian \eqref{eq:TunnelingHam3DSq} is entirely written in terms of the $SU(2)_1$ current-current interactions in the array of spinful electron wires.

In this model, quasiparticle excitations are planons living in the $[011]$ plane. 
According to the Lagrangian subgroup generated by $M$, we can create a pair of semionic quasiparticles with $s= \pm 1/4$ by acting a local operator at each interface: a vertex operator $\exp (i\bm{p} \cdot \bm{\phi}^\textsf{w}_{\bm{r}})$ with $\bm{p} = (0,0,0,1)^T$ creates an excitation $\bm{l} = (1,-1,0,0)^T$, which is a pair of semionic quasiparticles on the left and bottom strips, whereas a vertex operator with $\bm{p} = (0,0,-1,0)^T$ creates an excitation $\bm{l} = (0,0,1,-1)^T$, which is a pair of semionic quasiparticles on the right and top strips. 
These elementary excitations are illustrated in Fig.~\ref{fig:ModelFermionU2Sq}~(b). 
By successively creating and annihilating these quasiparticles between orthogonal strips in a zigzag manner, we can transfer a semionic quasiparticle along a diagonal line in the $yz$ plane, as shown in Fig.~\ref{fig:ModelFermionU2Sq}~(c). 
As a single semionic quasiparticle can freely move along the $x$ axis, they behave as planons in the $[011]$ plane. 

On a torus with the linear sizes $L_x \times L_y \times L_z$, there are $\textrm{gcd}(L_y, L_z)$ such planes. 
For each plane $P$, we can define a pair of string operators: 
One is defined along the diagonal line $\ell$, which is the projection of $P$ onto the $yz$ plane,
\begin{align}
X^{[011]}_P &= \prod_{\bm{r} \in \ell} \exp \Bigl[ i\bigl( \phi^2_{\bm{r}, L}(x_0) +\phi^2_{\bm{r}-\bm{e}_z, R}(x_0) \bigr) \Bigr],
\end{align}
where $x_0$ is arbitrary. 
The other is defined along the $x$ axis on the left strip with respect to some $\bm{r}_\ell \in \ell$, 
\begin{align}
Z^{[011]}_P &= \exp \left[ -\frac{i}{2} \int_0^{L_x} dx \, \partial_x \Phi^-_{\bm{r}_\ell, L}(x) \right].
\end{align}
These string operators are illustrated in Fig.~\ref{fig:ModelFermionU2SqString} for $L_y=L_z=4$. 
\begin{figure}
\includegraphics[clip,width=0.35\textwidth]{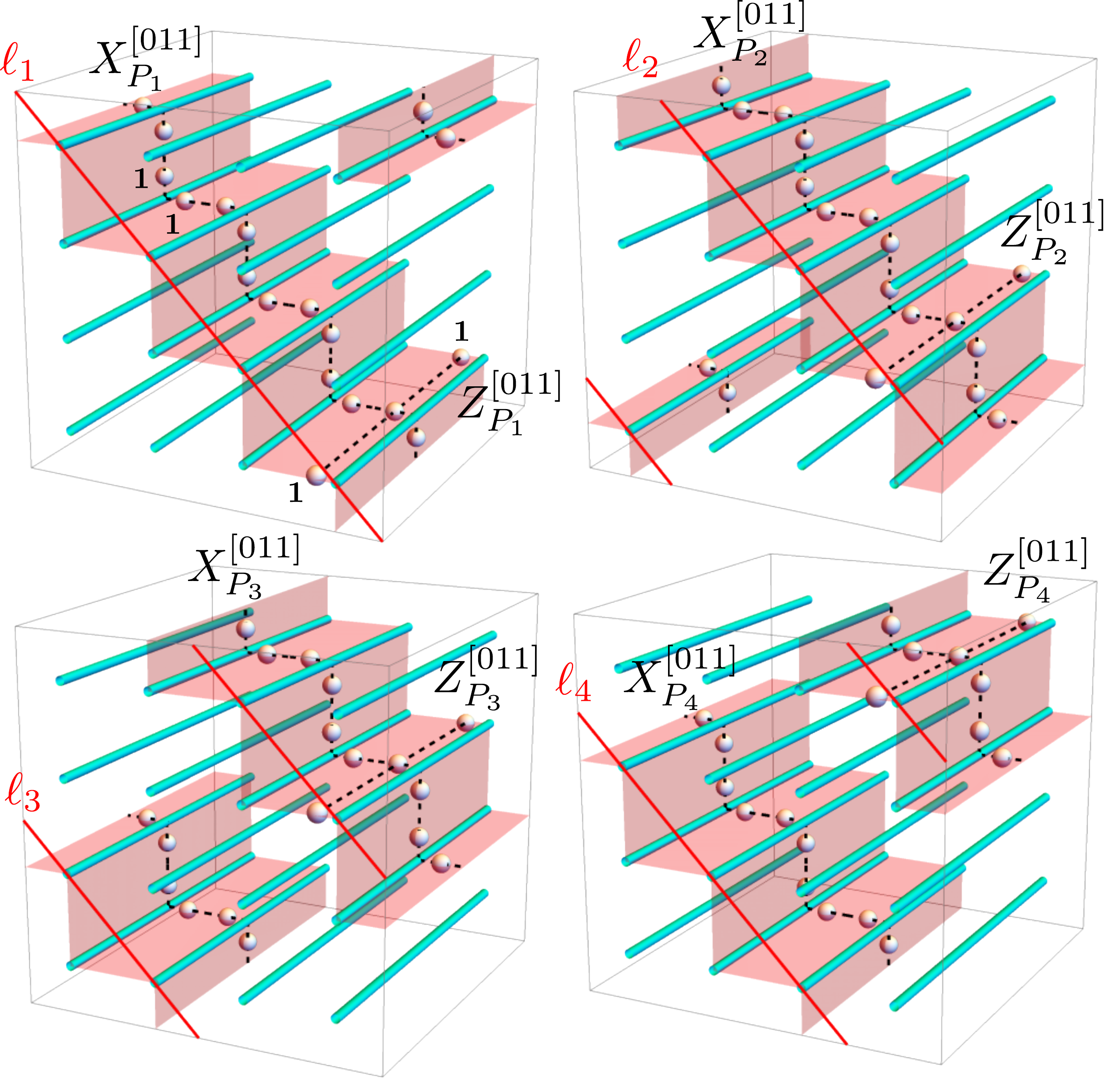}
\caption{Four sets of the string operators in the $[011]$ plane for the fermionic $U(1)_2$ model on a torus with $L_y=L_z=4$.}
\label{fig:ModelFermionU2SqString}
\end{figure}
Here, the symbols $X$ and $Z$ are used in analogy with the Pauli algebra obeyed by these string operators, 
\begin{align} \label{eq:AlgeStringU2}
Z^{[011]}_P X^{[011]}_P = -X^{[011]}_P Z^{[011]}_P.
\end{align}
While the squares of $X^{[011]}_P$ or $Z^{[011]}_P$ are not the identity operators, they trivially act on the ground state and thus become genuine Pauli operators in the subspace of degenerate ground states. 
The above string operators on different planes are independent from each other and form mutually commuting pairs of the Pauli operators. 
They thus span the ground-state manifold with degeneracy
\begin{align}
\textrm{GSD}=2^{\textrm{gcd}(L_y,L_z)}.
\end{align}

Therefore, this model might seem as a trivial fracton model, which is nothing but a decoupled stack of the 2D $U(1)_2$ topological orders in the $[011]$ direction. 
However, the model cannot be smoothly deformed into such a decoupled stack since the $U(1)_2$ topological orders are realized in a bosonic system whereas our model is microscopically described in terms of fermions. 
Indeed, local fermion excitations $\psi^\sigma_{\bm{r}, R/L} \propto \exp(i\phi^\sigma_{\bm{r}, R/L})$ are directly fractionalized into semionic quasiparticles in our model, in contrast to the standard $U(1)_2$ topological order where only local bosonic excitations are fractionalized into semions. 
In a spinful electron system, this implies that a single electron excitation is fractionalized into a (spinless) chargon and (charge neutral) spinon both with semionic statistics. 
In this sense, the present model is not considered as a simple decoupled stack of 2D topological orders.

\subsubsection{Fermionic $U(1)_6$ model}
\label{sec:FermionicU6Model}

We consider a 3D cellular topological state built out of the $U(1)_6$ topological orders, each of which is described by the $K$ matrix $K_0=6$. 
On the square grid, each interface possesses four gapless edge modes $\bm{\phi}^\textsf{e}_{\bm{r}} = (\phi^\textsf{e}_{\bm{r},1}, \cdots, \phi^\textsf{e}_{\bm{r}, 4})^T$ corresponding to the $K$ matrix, 
\begin{align} \label{eq:KmatFourU6}
K_\textsf{e} = \begin{pmatrix} 6 &&& \\ & 6 && \\ && -6 & \\ &&& -6 \end{pmatrix}.
\end{align}
We then consider a gapped interface obtained by condensing a set of quasiparticles generated by $M = \{ \bm{m}_a \}$ with 
\begin{align} \label{eq:VecMFermionicU6Model}
\begin{split}
\bm{m}_1 &= (1,3,2,0)^T, \\
\bm{m}_2 &= (0,2,3,1)^T.
\end{split}
\end{align}
Here, both $\bm{m}_1$ and $\bm{m}_2$ are fermionic quasiparticles with $s=\pm 1/2$. 
Therefore, the resulting gapped interface is fermionic. 
In order to construct a gapping potential, we add two extra fermionic wires at each interface and consider the extended $K$ matrix $K_\textsf{ew} = K_\textsf{e} \oplus K_\textsf{w}$ with $K_\textsf{w} = Z \oplus Z$. 
With the choice of matrices, 
\begin{align}
U = \begin{pmatrix} -1 & 1 & 1 & -1 \\ -1 & -1 & 1 & 1 \\ 1 & -1 & 2 & -2 \\ 1 & 1 & 2 & 2 \end{pmatrix}, \quad
W = \begin{pmatrix} -1 & 0 & 0 & 0 \\ 0 & -1 & 0 & 0 \\ 1 & 0 & -1 & 0 \\ 0 & 1 & 0 & -1 \end{pmatrix},
\end{align}
we find a set of integer vectors $\{ \bm{\Lambda}_{\textsf{w},\alpha} \}$, 
\begin{align}
\begin{split}
\bm{\Lambda}_{\textsf{w},1} &= (1,- 2, -1, 2)^T, \\
\bm{\Lambda}_{\textsf{w},2} &= (1, -2, 1, -2)^T, \\
\bm{\Lambda}_{\textsf{w},3} &= (2, -1, -2, 1)^T, \\
\bm{\Lambda}_{\textsf{w},4} &= (2, -1, 2, -1)^T.
\end{split}
\end{align}
By introducing bosonic fields $\bm{\phi}^\textsf{w}_{\bm{r}} = (\phi^1_{\bm{r},R}, \phi^1_{\bm{r},L}, \phi^2_{\bm{r},R}, \phi^2_{\bm{r},L})^T$ corresponding to two-component fermionic wires, which obey the commutation relations in Eq.~\eqref{eq:CommRel3DTwoCompField}, we find the tunneling Hamiltonian of the form \eqref{eq:TunnelingHam3DSq} with
\begin{align} \label{eq:FieldFermionicU6Model}
\begin{split}
\Phi^\pm_{\bm{r}, R} &= 2\phi^1_{\bm{r}, R} +\phi^1_{\bm{r}, L} \pm (2\phi^2_{\bm{r}, R} +\phi^2_{\bm{r}, L}), \\
\Phi^\pm_{\bm{r}, L} &= \phi^1_{\bm{r}, R} +2\phi^1_{\bm{r}, L} \pm (\phi^2_{\bm{r}, R} +2\phi^2_{\bm{r}, L}). 
\end{split}
\end{align}

Elementary excitations are dictated by the subset $M$ of the Lagrangian subgroup and are shown in Fig.~\ref{fig:ModelFermionU6Sq}~(a). 
\begin{figure}
\includegraphics[clip,width=0.4\textwidth]{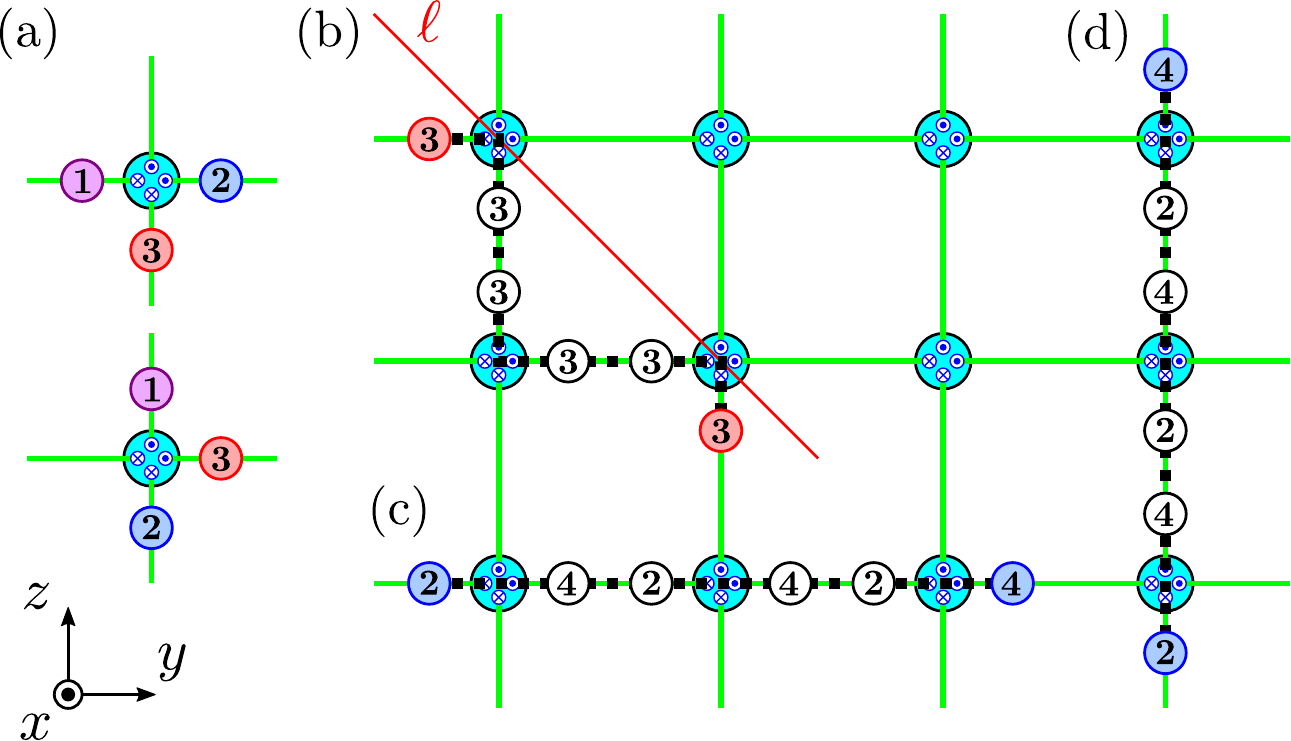}
\caption{(a) Elementary excitations created by local operators in the fermionic $U(1)_6$ model. 
(b) Semionic quasiparticle $\bm{3}$ can be transferred along a diagonal line $\ell$ in the $yz$ plane. 
Quasiparticle $\bm{2}$ or $\bm{4}$ can be transferred along (c) the $y$ axis or (d) the $z$ axis.
}
\label{fig:ModelFermionU6Sq}
\end{figure}
There are three types of planons living in the $[010]$, $[001]$, and $[011]$ planes. 
At each interface, we can create a pair of $\bm{3}$ quasiparticles, which have $s=3/4$ and thus are semionic, corresponding to $\bm{l}=(3,-3,0,0)^T$ or $(0,0,3,-3)^T$.
As in the $U(1)_2$ model, they can be transferred along a diagonal line in the $yz$ plane by successively applying local operators $\exp(i\bm{p} \cdot \bm{\phi}^\textsf{w}_{\bm{r}})$ with $\bm{p} = (0,0,1,2)^T$ or $(0,0,-2,-1)^T$ [see Fig.~\ref{fig:ModelFermionU6Sq}~(b)]. 
Combined with the mobility along the $x$ axis, semionic quasiparticles become planon excitations in the $[011]$ plane. 
On an $L_x \times L_y \times L_z$ torus, we have $\textrm{gcd}(L_y,L_z)$ such planes, for each of which we can define a pair of string operators, 
\begin{subequations}
\begin{align}
X^{[011]}_P &= \prod_{\bm{r} \in \ell} \exp \Bigl[ i\bigl( \phi^2_{\bm{r}, R}(x_0) +2\phi^2_{\bm{r}, L}(x_0) \nonumber \\
&\quad +2\phi^2_{\bm{r}-\bm{e}_z, R}(x_0) +\phi^2_{\bm{r}-\bm{e}_z, L}(x_0) \bigr) \Bigr], \\
Z^{[011]}_P &= \exp \left[ -\frac{i}{2} \int_0^{L_x} dx \, \partial_x \Phi^-_{\bm{r}_\ell, L}(x) \right], 
\end{align}
\end{subequations}
where $x_0$ is arbitrary, $\ell$ is the projection of $P$ onto the $yz$ plane, and $\bm{r}_\ell$ is some $\bm{r} \in \ell$. 
They are illustrated in Fig.~\ref{fig:ModelFermionU6SqString}~(a). 
\begin{figure}
\includegraphics[clip,width=0.4\textwidth]{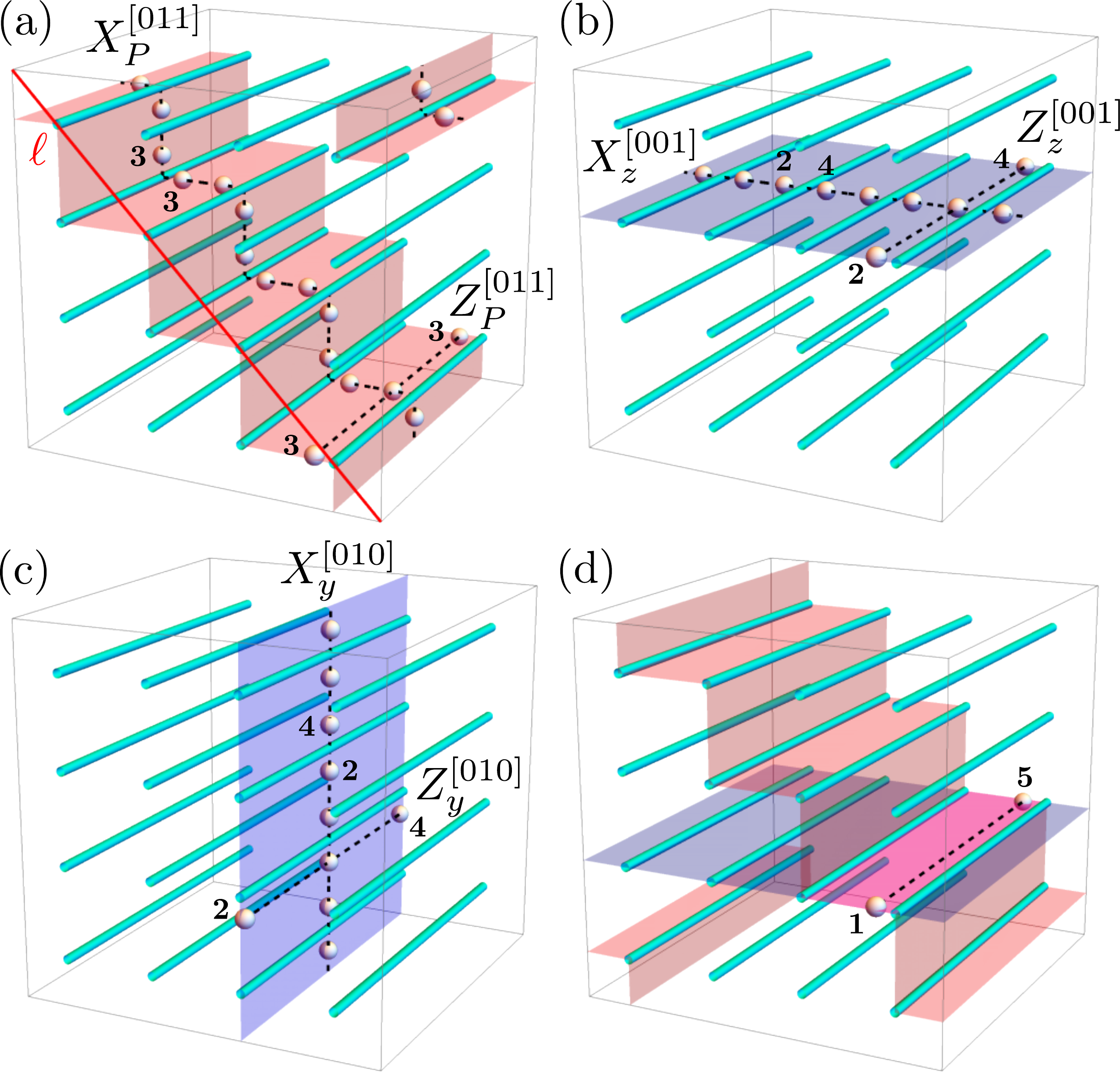}
\caption{Sets of string operators for the fermionic $U(1)_6$ model on a torus in (a) the $[011]$ plane, (b) the $[001]$ plane, and (c) the $[010]$ plane. 
(d) Lineon of $\bm{1}$ or $\bm{5}$ quasiparticle can be seen as a bound state of planons in two intersecting planes.
}
\label{fig:ModelFermionU6SqString}
\end{figure}
These string operators obey the same algebra as in Eq.~\eqref{eq:AlgeStringU2}. 

We can also create a pair of $\bm{2}$ and $\bm{4}$ quasiparticles, which have $s=1/3$ and $-1/3$, respectively, corresponding to $\bm{l} = (2,0,-2,0)^T$. 
They can be transferred along the $y$ axis by successively applying local operators $\exp(i\bm{p} \cdot \bm{\phi}^\textsf{w}_{\bm{r}})$ with $\bm{p} = (-1,-1,1,1)^T$ [see Fig.~\ref{fig:ModelFermionU6Sq}~(c)]. 
In combination with a string operator along the $x$ axis, $\bm{2}$ or $\bm{4}$ quasiparticles behave as planons in the $[001]$ plane. 
For each plane labeled by $z=1,\cdots,L_z$, we find string operators,
\begin{subequations}
\begin{align}
X^{[001]}_z &= \prod_{y=1}^{L_y} \exp \Bigl[ -i\bigl( \phi^1_{(y,z), R}(x_0) +\phi^1_{(y,z), L}(x_0) \nonumber \\
&\quad -\phi^2_{(y,z), R}(x_0) -\phi^2_{(y,z), L}(x_0) \bigr) \Bigr], \\
Z^{[001]}_z &= \exp \left[ -\frac{i}{3} \int_0^{L_x} dx \, \partial_x \Phi^-_{(y_0,z), L}(x) \right], 
\end{align}
\end{subequations}
where the choice of $x_0$ and $y_0$ is arbitrary. 
They obey the generalized Pauli algebra, 
\begin{align} \label{eq:AlgeStringU6Y}
Z^{[001]}_z X^{[001]}_z = e^{4\pi i/3} X^{[001]}_z Z^{[001]}_z.
\end{align}
This indicates that each $[001]$ plane hosts a planon with the same statistics as that of the $U(1)_3$ topological order.

Similarly, a pair of quasiparticles corresponding to $\bm{l} = (0,2,0,-2)^T$ can be created between the top and bottom strips with respect to each interface by a local operator $\exp(i\bm{p} \cdot \bm{\phi}^\textsf{w}_{\bm{r}})$ with $\bm{p} = (-1,-1,-1,-1)^T$. 
It leads to planon excitations moving within the $[010]$ plane [see Fig.~\ref{fig:ModelFermionU6Sq}~(d)]. 
Labeling such a plane by $y=1,\cdots,L_y$, we find the associated string operators, 
\begin{subequations}
\begin{align}
X^{[010]}_y &= \prod_{z=1}^{L_z} \exp \Bigl[ -i\bigl( \phi^1_{(y,z), R}(x_0) +\phi^1_{(y,z), L}(x_0) \nonumber \\
&\quad +\phi^2_{(y,z), R}(x_0) +\phi^2_{(y,z), L}(x_0) \bigr) \Bigr], \\
Z^{[010]}_y &= \exp \left[ -\frac{i}{3} \int_0^{L_x} dx \, \partial_x \Phi^+_{(y,z_0), L}(x) \right], 
\end{align}
\end{subequations}
where the choice of $x_0$ and $z_0$ is arbitrary. 
They also obey the generalized Pauli algebra, 
\begin{align} \label{eq:AlgeStringU6Z}
Z^{[010]}_y X^{[010]}_y = e^{4\pi i/3} X^{[010]}_y Z^{[010]}_y.
\end{align}

These string operators are independent from each other and form mutually commuting pairs of the (generalized) Pauli operators in the subspace of ground states.
Therefore, they span the ground-state manifold with degeneracy,
\begin{align}
\textrm{GSD} = 3^{L_y+L_z} \cdot 2^{\textrm{gcd}(L_y,L_z)}.
\end{align}
One may think that this model is made of decoupled stacks of 2D topological orders in the $[011]$, $[010]$, and $[001]$ directions. 
However, for the same reason as in the $U(1)_2$ model, we cannot decouple $[011]$ layers with semionic planons since our model is microscopically composed of fermions. 
In particular, a single fermion excitation $\psi^\sigma_{\bm{r},R/L} \propto \exp(i\phi^\sigma_{\bm{r},R/L})$ is fractionalized into a pair of semionic quasiparticles and two pairs of $\bm{2}$ and $\bm{4}$ quasiparticles. 
Lineon excitations moving on horizontal strips along the $x$ axis, which are originated from $\bm{1}$ quasiparticles with $s=1/12$ in the $U(1)_6$ topological order, are understood as bound pairs of planons between intersecting $[011]$ and $[001]$ planes as shown in Fig.~\ref{fig:ModelFermionU6SqString}~(d), whereas those moving on vertical strips are understood as bound pairs between intersecting $[011]$ and $[010]$ planes. 

We can also construct a bosonic version of the present model based on a bosonic gapped interface. 
In that model, $\bm{2}$ and $\bm{4}$ quasiparticles are planons moving in the $[001]$ and $[010]$ planes as in the present fermionic model. 
On the other hand, semionic $\bm{3}$ excitations are planons in the $[01\bar{1}]$ planes. 
The explicit construction is presented in Appendix~\ref{sec:BosonicU6Model}.

\subsection{Foliated type-I fracton model with lineons and planons}
\label{sec:FoliatedTypeILineon}

We here provide another type of foliated type-I fracton order, which is more nontrivial than the previous models with only planons. 
It exhibits both planon and lineon excitations. 
Although lineon excitations by themselves only move along the $x$ axis, they become planons by fusion of two identical lineons or by forming a dipole, as observed in the X-cube model \cite{Vijay16} or anisotropic lineon-planon model \cite{Shirley19a}.

\subsubsection{Fermionic $U(1)_4$ model}
\label{sec:DipoleU4Model}

We consider a 3D cellular topological state built out of the $U(1)_4$ topological orders, each of which is described by the $K$ matrix $K_0=4$. 
On the square grid, each interface possesses four gapless edge modes corresponding to the $K$ matrix, 
\begin{align}
K_\textsf{e} = \begin{pmatrix} 4 &&& \\ & 4 && \\ && -4 & \\ &&& -4 \end{pmatrix}.
\end{align}
We consider a gapped interface obtained by condensing a set of quasiparticles generated by $M = \{ \bm{m}_a \}$ with 
\begin{align}
\begin{split}
\bm{m}_1 &= (1,0,3,2)^T, \\
\bm{m}_2 &= (0,1,2,3)^T, 
\end{split}
\end{align}
both of which are fermionic quasiparticles with $s=-1/2$.
Hence, the corresponding gapped interface is fermionic. 
In order to construct a gapping potential, we add two extra fermionic wires at each interface and consider the extended $K$ matrix $K_\textsf{ew} = K_\textsf{e} \oplus K_\textsf{w}$ with $K_\textsf{w} = Z \oplus Z$. 
With the choice of matrices, 
\begin{align}
U = \begin{pmatrix} -1 & -1 & -1 & -2 \\ 1 & 0 & -1 & -1 \\ -1 & 1 & -1 & 0 \\ 1 & -2 & -1 & 1 \end{pmatrix}, \quad
W = \begin{pmatrix} 1 & 0 & 0 & 0 \\ 1 & 0 & 0 & 1 \\ -1 & 0 & 1 & -1 \\ 0 & 1 & -1 & 0 \end{pmatrix},
\end{align}
we find a set of integer vectors $\{ \bm{\Lambda}_{\textsf{w},\alpha} \}$, 
\begin{align}
\begin{split}
\bm{\Lambda}_{\textsf{w},1} &= (-1, 1, 0, 2)^T, \\
\bm{\Lambda}_{\textsf{w},2} &= (0, -2, -1, 1)^T, \\
\bm{\Lambda}_{\textsf{w},3} &= (1, -1, -2, 0)^T, \\
\bm{\Lambda}_{\textsf{w},4} &= (2, 0, 1, -1)^T.
\end{split}
\end{align}
By introducing bosonic fields $\bm{\phi}^\textsf{w}_{\bm{r}} = (\phi^1_{\bm{r},R}, \phi^1_{\bm{r},L}, \phi^2_{\bm{r},R}, \phi^2_{\bm{r},L})^T$ corresponding to two-component fermionic wires, which obey the commutation relations in Eq.~\eqref{eq:CommRel3DTwoCompField}, we find the tunneling Hamiltonian of the form \eqref{eq:TunnelingHam3DSq} with
\begin{align}
\begin{split}
\Phi^+_{\bm{r}, R} &= 2\phi^1_{\bm{r}, R} +\phi^2_{\bm{r}, R} +\phi^2_{\bm{r}, L}, \\
\Phi^-_{\bm{r}, R} &= \phi^1_{\bm{r}, R} +\phi^1_{\bm{r}, L} -2\phi^2_{\bm{r}, R}, \\
\Phi^+_{\bm{r}, L} &= 2\phi^1_{\bm{r}, L} -\phi^2_{\bm{r}, R} -\phi^2_{\bm{r}, L}, \\
\Phi^-_{\bm{r}, L} &= -\phi^1_{\bm{r}, R} -\phi^2_{\bm{r}, L} -2\phi^2_{\bm{r}, L}.
\end{split}
\end{align}

In this model, there are planon excitations in the $[010]$ and $[001]$ planes and also planons made of dipoles of lineon excitations moving in the same planes. 
The most elementary excitations created by local operators take the form of triplets of quasiparticles, such as $\bm{l} = (1,0,-1,2)^T$ or $(0,1,2,-1)^T$, as shown in Fig.~\ref{fig:ModelDipoleU4Sq}~(a). 
\begin{figure}
\includegraphics[clip,width=0.4\textwidth]{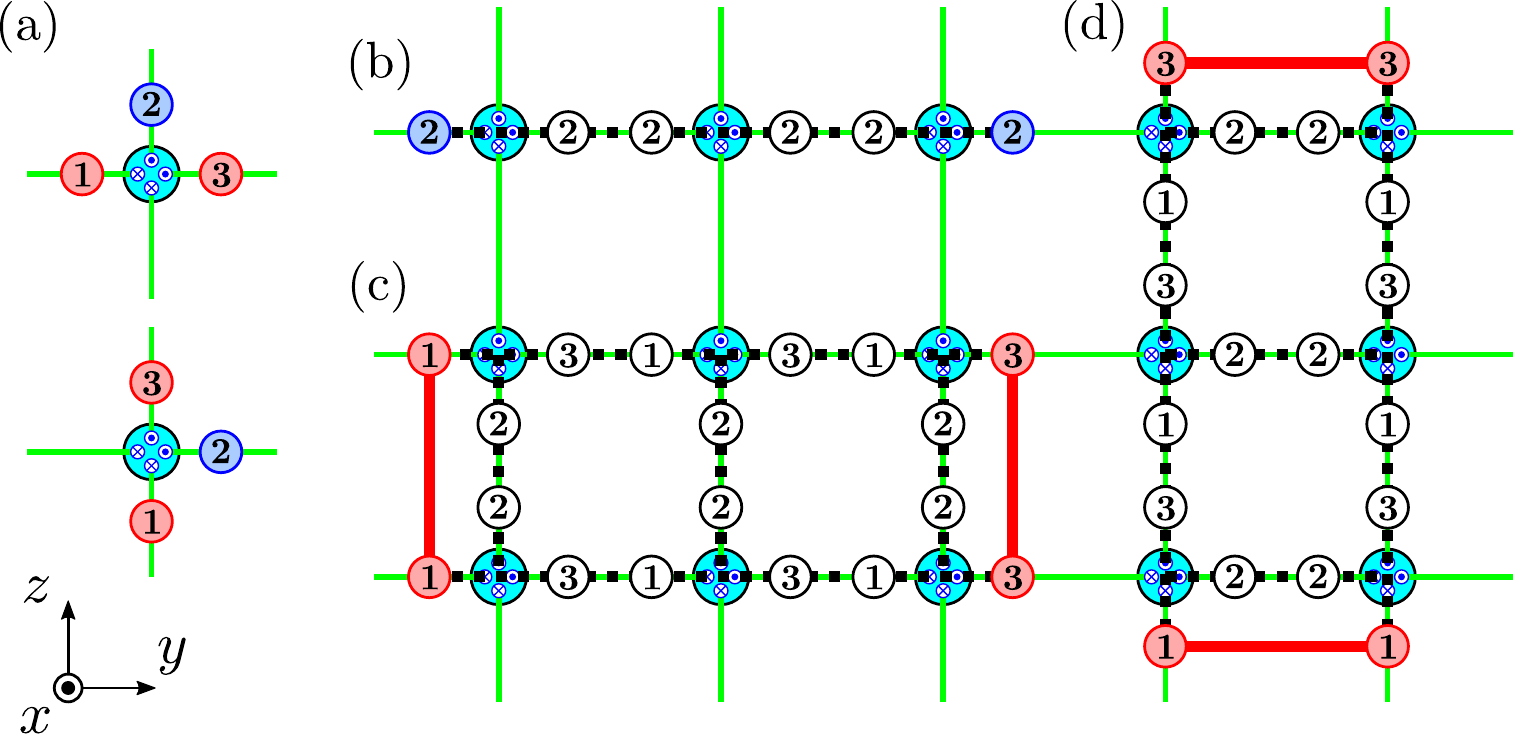}
\caption{(a) Elementary excitations created by local operators in the fermionic $U(1)_4$ model with lineons and planons. 
(b) Quasiparticle $\bm{2}$ can be transferred along the $y$ axis. 
Dipole of quasiparticles $\bm{1}$ or $\bm{3}$ can also be transferred along (c) the $y$ axis or (d) the $z$ axis.
}
\label{fig:ModelDipoleU4Sq}
\end{figure}
Fusion of two identical triplets leads to a pair of $\bm{2}$ quasiparticles with $s=\pm 1/2$, which are given by $\bm{l} = (2,0,-2,0)^T$ or $(0,2,0,-2)^T$ up to local bosonic excitations $\bm{4}$ and are created by $\exp(i\bm{p} \cdot \bm{\phi}^\textsf{w}_{\bm{r}})$ with $\bm{p} = (0,0,1,1)^T$ or $(-1,-1,0,0)^T$, respectively. 
By successively applying these operators, we can transfer $\bm{2}$ quasiparticles along the $y$ or $z$ axis [see Fig.~\ref{fig:ModelDipoleU4Sq}~(b)]. 
With the mobility along the $x$ axis, these quasiparticles become planons in the $[001]$ or $[010]$ plane. 
On each $[001]$ plane labeled by $z=1,\cdots,L_z$, we can find a string operator moving a $\bm{2}$ planon along the $y$ axis and that moving a $\bm{1}$ lineon along the $x$ axis, 
\begin{subequations}
\begin{align}
X^{[001]}_z &= \prod_{y=1}^{L_y} \exp \Bigl[ i\bigl( \phi^2_{(y,z), R}(x_0) +\phi^2_{(y,z), L}(x_0) \bigr) \Bigr], \\
Z^{[001]}_z &= \exp \left[ -\frac{i}{4} \int_0^{L_x} dx \, \partial_x \Phi^-_{(y_0,z), L}(x) \right], 
\end{align}
\end{subequations}
for arbitrary $x_0$ and $y_0$ [see Fig.~\ref{fig:ModelDipoleU4SqString}~(a)], 
\begin{figure}
\includegraphics[clip,width=0.4\textwidth]{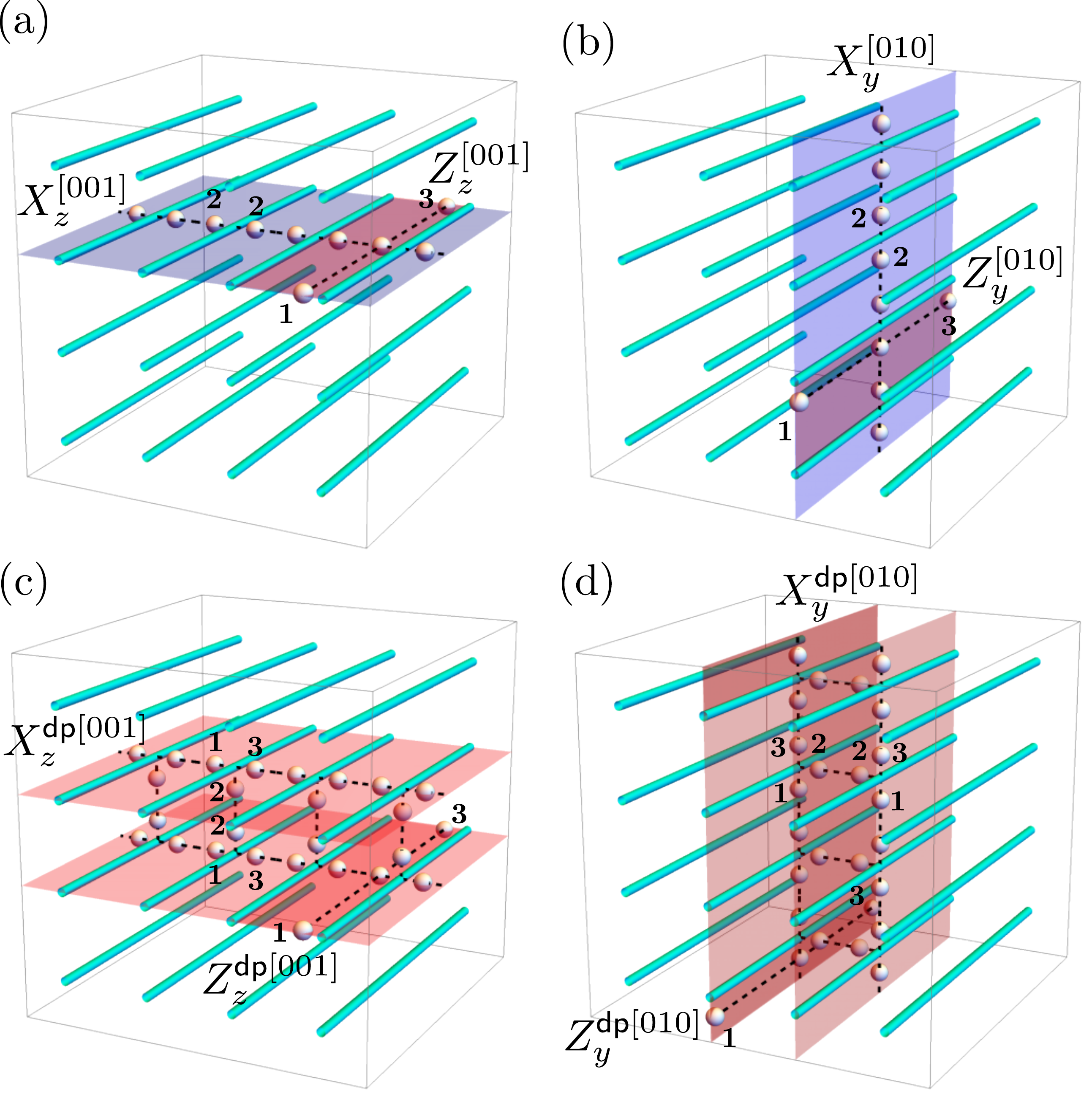}
\caption{Sets of string operators for the fermionic $U(1)_4$ model with lineons and planons on a torus. 
They are associated with planons in (a) the $[001]$ plane and (b) the $[010]$ plane and with dipole planons in (c) the $[001]$ plane and (d) the $[010]$ plane.
}
\label{fig:ModelDipoleU4SqString}
\end{figure}
while on each $[010]$ plane labeled by $y=1,\cdots,L_y$, we have a string operator moving a $\bm{2}$ planon along the $z$ axis and that moving a $\bm{1}$ lineon along the $x$ axis,
\begin{subequations}
\begin{align}
X^{[010]}_y &= \prod_{z=1}^{L_z} \exp \Bigl[ -i\bigl( \phi^1_{(y,z), R}(x_0) +\phi^1_{(y,z), L}(x_0) \bigr) \Bigr], \\
Z^{[010]}_y &= \exp \left[ -\frac{i}{4} \int_0^{L_x} dx \, \partial_x \Phi^+_{(y,z_0), L}(x) \right],
\end{align}
\end{subequations}
for arbitrary $x_0$ and $z_0$ [see Fig.~\ref{fig:ModelDipoleU4SqString}~(b)].
These string operators obey the Pauli algebra, 
\begin{align}
Z^{[001]}_z X^{[001]}_z &= -X^{[001]}_z Z^{[001]}_z, \\
Z^{[010]}_y X^{[010]}_y &= -X^{[010]}_y Z^{[010]}_y.
\end{align}

Another type of planon is formed by dipoles of $\bm{1}$ or $\bm{3}$ quasiparticles with $s=\pm 1/8$. 
Let us consider two triplets of quasiparticles $\bm{l}_1 = (1,0,-1,2)^T$ and $\bm{l}_2 = (1,-2,-1,0)^T$, which are created by local operators $\exp(i\bm{p} \cdot \bm{\phi}^\textsf{w}_{\bm{r}})$ with $\bm{p}_1 = (1,0,1,1)^T$ and $\bm{p}_2 = (0,1,0,0)^T$, respectively. 
If we create $\bm{l}_1$ at $\bm{r}$ and $\bm{l}_2$ at $\bm{r}+\bm{e}_z$, $\bm{2}$ quasiparticles on the intermediate vertical strip are pair annihilated, leaving dipoles of $\bm{1}$ or $\bm{3}$ quasiparticles at adjacent parallel strips along the $y$ axis [see Fig.~\ref{fig:ModelDipoleU4Sq}~(c)]. 
With the mobility along the $x$ axis, these dipoles coherently move in the $[001]$ plane. 
However, a single $\bm{1}$ or $\bm{3}$ quasiparticle is still a lineon moving only along the $x$ axis. 
Therefore, we can construct two string operators associated with a $\bm{1}$-$\bm{1}$ dipole along the $y$ axis and a $\bm{1}$ lineon along the $x$ axis on each $[001]$ plane: 
\begin{subequations}
\begin{align}
X^{\textsf{dp}[001]}_z &= \prod_{y=1}^{L_y} \exp \Bigr[ i\bigl( \phi^1_{(y,z), R}(x_0) +\phi^2_{(y,z), R}(x_0) \nonumber \\
&\quad +\phi^2_{(y,z), L}(x_0) +\phi^1_{(y,z+1), R}(x_0) \bigr) \Bigr], \\
Z^{\textsf{dp}[001]}_z &= \exp \left[ -\frac{i}{4} \int_0^{L_x} dx \, \partial_x \Phi^-_{(y_0,z), L}(x) \right],
\end{align}
\end{subequations}
for arbitrary $x_0$ and $y_0$ [see Fig.~\ref{fig:ModelDipoleU4SqString}~(c)].
They obey the generalized Pauli algebra, 
\begin{align}
Z^{\textsf{dp}[001]}_z X^{\textsf{dp}[001]}_z = e^{\pi i/2} X^{\textsf{dp}[001]}_z Z^{\textsf{dp}[001]}_z.
\end{align}
Similarly, we can construct a pair of $\bm{1}$-$\bm{1}$ dipoles moving along the $z$ axis by binding $\bm{l}_3 = (0,1,2,-1)^T$ at $\bm{r}$ and $\bm{l}_2 = (-2,1,0,-1)^T$ at $\bm{r}+\bm{e}_y$ [see Fig.~\ref{fig:ModelDipoleU4Sq}~(d)], which move in the $[010]$ plane when combined with the mobility along the $x$ axis. 
We can then find two string operators on each $[010]$ plane, 
\begin{subequations}
\begin{align}
X^{\textsf{dp}[010]}_y &= \prod_{y=1}^{L_y} \exp \Bigr[ -i\bigl( \phi^2_{(y,z), R}(x_0) +\phi^1_{(y+1,z), R}(x_0) \nonumber \\
&\quad +\phi^1_{(y+1,z), L}(x_0) +\phi^2_{(y+1,z), L}(x_0) \bigr) \Bigr], \\
Z^{\textsf{dp}[010]}_y &= \exp \left[ -\frac{i}{4} \int_0^{L_x} dx \, \partial_x \Phi^+_{(y,z_0), L}(x) \right], 
\end{align}
\end{subequations}
for arbitrary $x_0$ and $z_0$ [see Fig.~\ref{fig:ModelDipoleU4SqString}~(d)], which also obey the generalized Pauli algebra, 
\begin{align}
Z^{\textsf{dp}[010]}_y X^{\textsf{dp}[010]}_y = e^{\pi i/2} X^{\textsf{dp}[010]}_y Z^{\textsf{dp}[010]}_y.
\end{align}

Here, we note that these string operators do not form mutually commuting pairs of the (generalized) Pauli operators. 
It is even impossible to choose them to form commuting pairs by taking appropriate linear combinations because the string operators defined within the $yz$ plane are not independent from each other.
If we suppose that the string operators are all independent and form commuting pairs, they could span a Hilbert space of dimension $2^{L_y+L_z} \cdot 4^{L_y+L_z}$. 
However, this is not the case and they are subject to the constraints, 
\begin{subequations}
\begin{align}
(X^{\textsf{dp}[001]}_z)^2 &\sim X^{[001]}_z X^{[001]}_{z+1}, \\
(X^{\textsf{dp}[010]}_y)^2 &\sim X^{[010]}_y X^{[010]}_{y+1}, \\
\prod_{z=1}^{L_z} X^{\textsf{dp}[001]}_z &\sim \prod_{z=1}^{L_z} X^{[001]}_z \prod_{y=1}^{L_y} X^{[010]}_y, \\
\prod_{y=1}^{L_y} X^{\textsf{dp}[010]}_y &\sim \prod_{y=1}^{L_y} X^{[010]}_y \prod_{z=1}^{L_z} X^{[001]}_z.
\end{align}
\end{subequations}
We note that these identifications are valid only in the subspace of degenerate ground states and hold up to multiplications of operators creating or annihilating pairs of local bosonic excitations $\bm{4}$, which trivially act on the ground state. 
The first two constraints imply that two dipoles of $\bm{1}$ quasiparticles in the same plane fuse into two planons of $\bm{2}$ quasiparticles in the adjacent planes and reduce the dimension by factor of $2^{L_y+L_z}$. 
The last two constraints imply that only $L_y+L_z-2$ dipole string operators are independent and further reduce the dimension by factor of $2^2$. 
Overall, these constraints reduce the dimension of the Hilbert space spanned by the string operators by factor of $2^{L_y+L_z+2}$, leading to the ground-state degeneracy,
\begin{align}
\textrm{GSD} = 2^2 \cdot 4^{L_y+L_z-2}.
\end{align}
This violates a strict subextensivity of $\log \textrm{GSD}$ by a negative additive constant and proves that the present model is a nontrivial foliated fracton model, which cannot be understood as decoupled stacks of 2D topological orders.
Fully commuting pairs of the generalized Pauli operators spanning the ground-state manifold are given, for example, by $X^{\textsf{dp}[001]}_z$ and $Z^{\textsf{dp}[001]}_z$ with $z=1,\cdots,L_z-1$, $X^{\textsf{dp}[010]}_y$ and $Z^{\textsf{dp}[010]}_y$ with $y=1,\cdots,L_y-1$, $X^{[001]}_{L_z}$ and $Z^{[001]}_{L_z}$, and $X^{[010]}_{L_y}$ and $Z^{[010]}_{L_y}$. 

We can also construct a similar coupled-wire model from a bosonic gapped interface between the $U(1)_8$ topological orders. 
In that model, one can also convert lineon excitations into a planon in the $[001]$ or $[010]$ plane by fusing two of them or by taking a dipole of them. 
The explicit construction is presented in Appendix~\ref{sec:DipoleU8Model}. 

\subsection{TQFT-type model}
\label{sec:TQFTType}

We here provide 3D coupled-wire models for TQFT-type topological orders, which can be viewed as 3D cellular topological states based on the $U(1)_4$ or $U(1)_9$ topological orders. 
The former model provides a bosonic $Z_2$ gauge theory with a fermionic $Z_2$ charge and a loop-like excitation with mutual $\pi$ statistics, whereas the latter provides a fermionic $Z_3$ gauge theory with a $Z_3$ charge and a loop-like excitation with mutual $2\pi/3$ statistics. 
In both cases, the coupled-wire models possess the ground-state degeneracy growing with the number of wires due to local loop excitations. 
We can then induce the condensation of the local loop excitations by adding local perturbations, which yield the desired TQFT-type topological orders of 3D discrete gauge theories only with finite degeneracy on a torus. 
Since the phenomenology of excitations in the $U(1)_9$ model is a natural extension of the $U(1)_4$ model, we leave detailed analysis for the $U(1)_9$ model in Appendix~\ref{sec:FermionicU9Model}.

\subsubsection{Bosonic $U(1)_4$ model}
\label{sec:BosonicU4Model}

We consider a 3D cellular topological state built out of the $U(1)_4$ topological orders, each of which is described by the $K$ matrix $K_0=4$. 
On the square grid, each interface possesses four gapless edge modes corresponding to the $K$ matrix, 
\begin{align}
K_\textsf{e} = \begin{pmatrix} 4 &&& \\ & 4 && \\ && -4 & \\ &&& -4 \end{pmatrix}.
\end{align}
We consider a gapped interface obtained by condensing a set of quasiparticles generated by $M = \{ \bm{m}_a \}$ with 
\begin{align} \label{eq:VecMBosonicU4Model}
\begin{split}
\bm{m}_1 &= (1,1,3,3)^T, \\
\bm{m}_2 &= (2,2,0,0)^T, \\
\bm{m}_3 &= (2,0,2,0)^T,
\end{split}
\end{align}
which are all bosonic quasiparticles with $s=-2$, $1$, and $0$, respectively.
Hence, the corresponding gapped interface is bosonic. 
In order to construct a gapping potential, we add two extra bosonic wires at each interface and consider the extended $K$ matrix $K_\textsf{ew} = K_\textsf{e} \oplus K_\textsf{w}$ with $K_\textsf{w} = X \oplus X$. 
With the choice of matrices, 
\begin{align}
U = \begin{pmatrix} 1 & -1 & -1 & 1 \\ 1 & 1 & -1 & -1 \\ 1 & -1 & 1 & -1 \\ 1 & 1 & 1 & 1 \end{pmatrix}, \quad
W = \begin{pmatrix} -1 & 0 & 0 & 0 \\ 0 & -1 & 0 & 0 \\ 0 & 0 & 1 & 0 \\ 0 & 0 & 0 & 1 \end{pmatrix},
\end{align}
we find a set of integer vectors $\{ \bm{\Lambda}_{\textsf{w},\alpha} \}$, 
\begin{align}
\begin{split}
\bm{\Lambda}_{\textsf{w},1} &= (-1, 1, 1, -1)^T, \\
\bm{\Lambda}_{\textsf{w},2} &= (-1, 1, -1, 1)^T, \\
\bm{\Lambda}_{\textsf{w},3} &= (1, 1, -1, -1)^T, \\
\bm{\Lambda}_{\textsf{w},4} &= (1, 1, 1, 1)^T.
\end{split}
\end{align}
By introducing bosonic fields $\bm{\phi}^\textsf{w}_{\bm{r}} = (\varphi^1_{\bm{r}}, 2\theta^1_{\bm{r}}, \varphi^2_{\bm{r}}, 2\theta^2_{\bm{r}})^T$ corresponding to two-component bosonic wires, which obey the commutation relations,
\begin{align} \label{eq:CommRel3DTwoCompBoson}
[\theta^\sigma_{\bm{r}}(x), \varphi^{\sigma'}_{\bm{r}'}(x')] &= i\pi \delta_{\bm{r}, \bm{r}'} \delta_{\sigma,\sigma'} \Theta(x-x'), \\
[\theta^\sigma_{\bm{r}}(x), \theta^{\sigma'}_{\bm{r}'}(x')] &= [\varphi^\sigma_{\bm{r}}(x), \varphi^{\sigma'}_{\bm{r}'}(x')] =0.
\end{align}
we find the tunneling Hamiltonian of the form \eqref{eq:TunnelingHam3DSq} with
\begin{align} \label{eq:FieldBosonicU4Model}
\begin{split}
\Phi^\pm_{\bm{r}, R} &= \varphi^1_{\bm{r}} +2\theta^1_{\bm{r}} \pm (\varphi^2_{\bm{r}} +2\theta^2_{\bm{r}}), \\
\Phi^\pm_{\bm{r}, L} &= \varphi^1_{\bm{r}} -2\theta^1_{\bm{r}} \pm (\varphi^2_{\bm{r}} -2\theta^2_{\bm{r}}).
\end{split}
\end{align}

This model is actually a variant of the 3D coupled-wire model considered in Ref.~\cite{Fuji19b}. 
As demonstrated in Ref.~\cite{Fuji19b}, it has a 3D point-like emergent fermion excitation and a loop-like excitation with the mutual $\pi$ statistics. 
Elementary excitations created by local operators are given by the subset $M$ of the Lagrangian subgroup and are shown in Fig.~\ref{fig:ModelBosonU4Sq}~(a). 
\begin{figure}
\includegraphics[clip,width=0.4\textwidth]{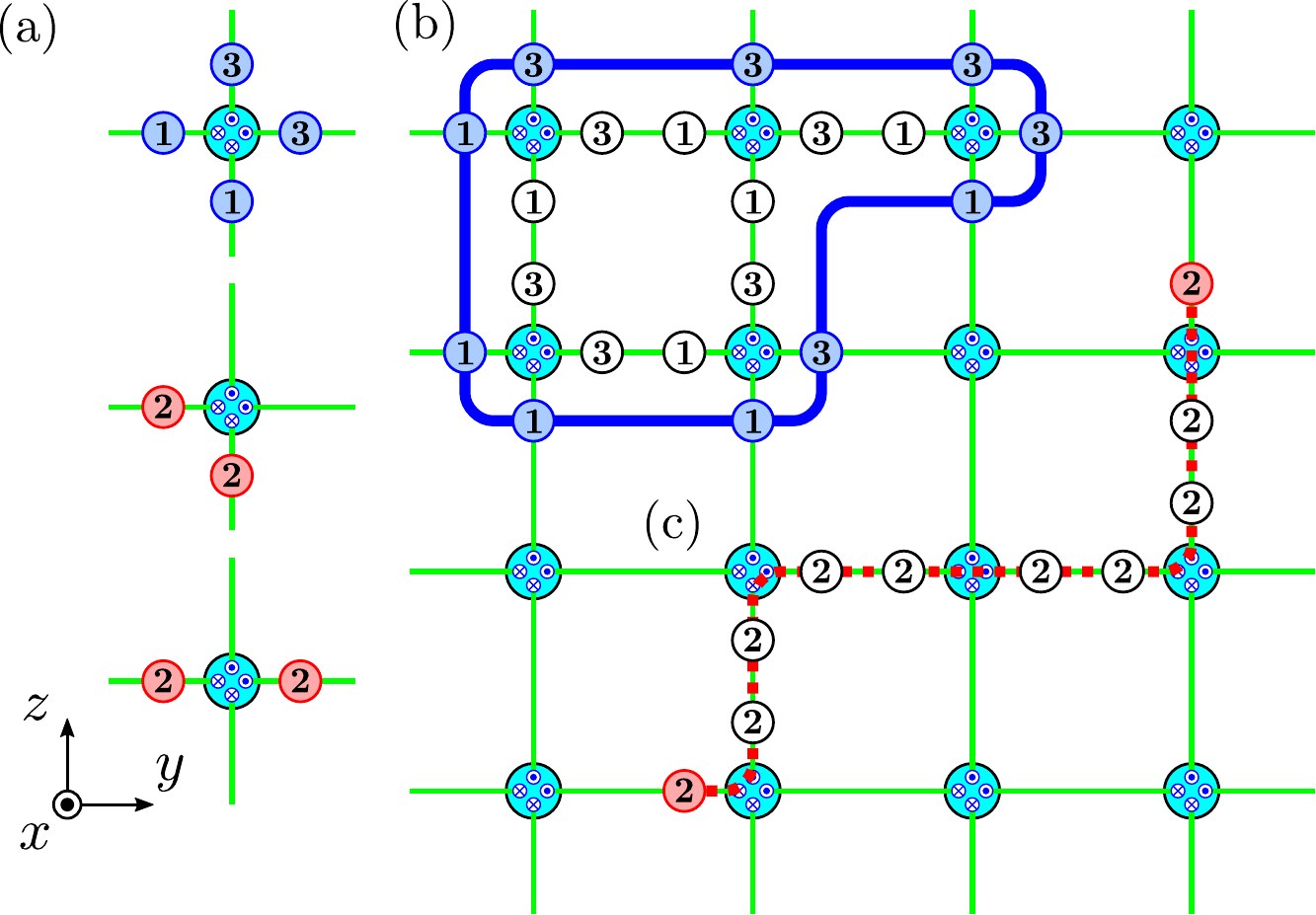}
\caption{(a) Elementary excitations created by local operators in the bosonic $U(1)_4$ model. 
(b) Quasiparticles $\bm{1}$ and $\bm{3}$ form a loop-like excitation in the $yz$ plane, while (c) fermionic quasiparticles $\bm{2}$ can freely move in the $yz$ plane.
}
\label{fig:ModelBosonU4Sq}
\end{figure}
One of them is $\bm{l} = (1,1,-1,-1)^T$, which creates $\bm{1}$ and $\bm{3}$ quasiparticles with $s=1/8$ and $-1/8$, respectively, over all strips surrounding the interface $\bm{r}$. 
The corresponding local operator is given by $\exp(i\bm{p} \cdot \bm{\phi}^\textsf{w}_{\bm{r}})$ with $\bm{p} = (-1,0,0,0)^T$. 
While single $\bm{1}$ or $\bm{3}$ quasiparticles behave as lineons along the $x$ axis, they cannot move individually in the $yz$ plane. 
However, by acting these local operators successively over a membrane in the $yz$ plane, they can form a loop-like excitation, which costs an energy proportional to the length of the boundary of the membrane [see Fig.~\ref{fig:ModelBosonU4Sq}~(b)].
When such a membrane fully covers a $yz$ plane of an $L_x \times L_y \times L_z$ torus, excitations are completely canceled and we come back to the ground state. 
Correspondingly, we can define a pair of a membrane operator in the $yz$ plane and a rigid string operator along the $x$ axis, 
\begin{subequations} \label{eq:MembraneU4}
\begin{align}
\label{eq:MembraneOpU4}
X^{\textsf{mem}} &= \prod_{\bm{r} \in \mathbb{Z}_{L_y} \times \mathbb{Z}_{L_z}} \exp \bigl[ -i\varphi^1_{\bm{r}}(x_0) \bigr], \\
Z^{\textsf{mem}} &= \exp \left[ -\frac{i}{4} \int_0^{L_x} dx \, \partial_x \Phi^-_{\bm{r}_0, L}(x) \right],
\end{align}
\end{subequations}
where the choice of $x_0$ and $\bm{r}_0$ is arbitrary. 
They are illustrated in Fig.~\ref{fig:ModelBosonU4SqString}~(a). 
\begin{figure}
\includegraphics[clip,width=0.4\textwidth]{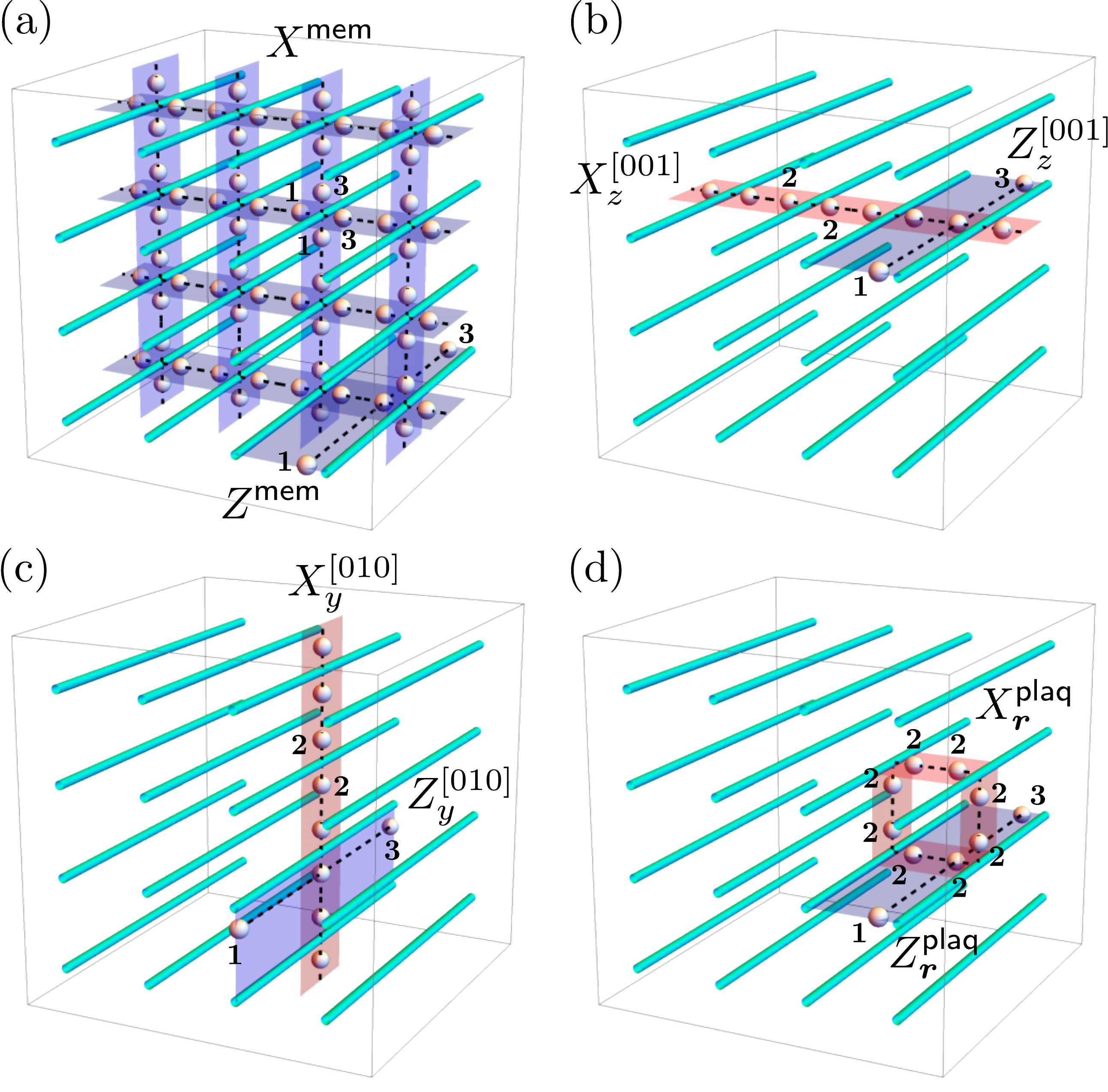}
\caption{Sets of membrane and string operators for the bosonic $U(1)_4$ model on a torus. 
They are associated with (a) loop-like excitations of quasiparticles $\bm{1}$ and $\bm{3}$ in the $yz$ plane and fermionic point-like excitations of quasiparticles $\bm{2}$ in (b) the $[001]$ plane and (c) the $[010]$ plane and (d) along plaquette loops.
}
\label{fig:ModelBosonU4SqString}
\end{figure}
They obey the generalized Pauli algebra, 
\begin{align}
Z^\textsf{mem} X^\textsf{mem} = e^{\pi i/2} X^\textsf{mem} Z^\textsf{mem}.
\end{align}

The other types of elementary excitations are pairs of $\bm{2}$ quasiparticles, which have $s=1/2$ and are thus fermions, such as $\bm{l} = (2,0,-2,0)^T$ and $(2,-2,0,0)^T$. 
These fermionic quasiparticles are created between arbitrary two strips surrounding the interface $\bm{r}$ and are thus fully mobile in the $yz$ plane [see Fig.~\ref{fig:ModelBosonU4Sq}~(c)]. 
With the mobility along the $x$ axis, they can move along an arbitrary 1D path in the 3D space. 
We may then construct three kinds of string operators defined in the $[001]$ and $[010]$ planes and on square plaquettes. 
On each $[001]$ plane labeled by $z=1,\cdots,L_z$, we find a string operator moving a fermionic quasiparticle along the $y$ axis and that moving a $\bm{1}$ lineon along the $x$ axis,
\begin{subequations} \label{eq:StringYU4}
\begin{align}
X^{[001]}_z &= \prod_{y=1}^{L_y} \exp \Bigl[ -i\bigl( \varphi^1_{(y,z)}(x_0) -\varphi^2_{(y,z)}(x_0) \bigr) \Bigr], \\
Z^{[001]}_z &= \exp \left[ -\frac{i}{4} \int_0^{L_x} dx \, \partial_x \Phi^-_{(y_0,z), L}(x) \right],
\end{align}
\end{subequations}
for arbitrary $x_0$ and $y_0$ [see Fig.~\ref{fig:ModelBosonU4SqString}~(b)].
On each $[010]$ plane labeled by $y=1,\cdots,L_y$, we find a string operator moving a fermionic quasiparticle along the $z$ axis and that moving a $\bm{1}$ lineon along the $x$ axis,
\begin{subequations} \label{eq:StringZU4}
\begin{align}
X^{[010]}_y &= \prod_{z=1}^{L_z} \exp \Bigl[ -i\bigl( \varphi^1_{(y,z)}(x_0) +\varphi^2_{(y,z)}(x_0) \bigr) \Bigr], \\
Z^{[010]}_y &= \exp \left[ -\frac{i}{4} \int_0^{L_x} dx \, \partial_x \Phi^+_{(y,z_0), L}(x) \right], 
\end{align}
\end{subequations}
for arbitrary $x_0$ and $z_0$ [see Fig.~\ref{fig:ModelBosonU4SqString}~(c)].
Finally, on each square plaquette labeled by its left bottom corner $\bm{r}$, we find a string operator moving a fermionic quasiparticle along the plaquette loop and that moving a $\bm{1}$ lineon along the $x$ axis, 
\begin{subequations}
\begin{align}
\label{eq:PlaqOpU4}
X^{\textsf{plaq}}_{\bm{r}} &= \exp \Bigl[ -i\bigl( \varphi^2_{\bm{r}}(x_0) +2\theta^2_{\bm{r}}(x_0) \nonumber \\
&\quad -\varphi^1_{\bm{r}+\bm{e}_y}(x_0) -2\theta^2_{\bm{r}+\bm{e}_y}(x_0) \nonumber \\
&\quad -\varphi^2_{\bm{r}+\bm{e}_y+\bm{e}_z}(x_0) +2\theta^2_{\bm{r}+\bm{e}_y+\bm{e}_z}(x_0) \nonumber \\
&\quad +\varphi^1_{\bm{r}+\bm{e}_z}(x_0) -2\theta^2_{\bm{r}+\bm{e}_z}(x_0) \bigr) \Bigr], \\
Z^{\textsf{plaq}}_{\bm{r}} &= \exp \left[ -\frac{i}{4} \int_0^{L_x} dx \, \partial_x \Phi^-_{\bm{r}+\bm{e}_y, L}(x) \right],
\end{align}
\end{subequations}
where the choice of $x_0$ is arbitrary [see Fig.~\ref{fig:ModelBosonU4SqString}~(d)].
They obey the Pauli algebra, 
\begin{align}
Z^{[001]}_z X^{[001]}_z &= -X^{[001]}_z Z^{[001]}_z, \\
Z^{[010]}_y X^{[010]}_y &= -X^{[010]}_y Z^{[010]}_y, \\
\label{eq:CommPlaqU4}
Z^{\textsf{plaq}}_{\bm{r}} X^{\textsf{plaq}}_{\bm{r}} &= -X^{\textsf{plaq}}_{\bm{r}} Z^{\textsf{plaq}}_{\bm{r}}.
\end{align}

If we suppose that all these membrane and string operators are independent from each other and form mutually commuting pairs of the (generalized) Pauli operators, they could span a Hilbert space of the dimension $4 \cdot 2^{L_y+L_z+L_yL_z}$. 
However, they cannot form mutually commuting pairs since the membrane and string operators defined within the $yz$ plane are not independent from each other but are subject to the constraints, 
\begin{subequations} \label{eq:ConstraintsU4}
\begin{align}
(X^\textsf{mem})^2 &\sim \prod_{y=1}^{L_y} X^{[010]}_y \prod_{z=1}^{L_z} X^{[001]}_z, \\
X^{[001]}_z X^{[001]}_{z+1} &\sim \prod_{y=1}^{L_y} X^\textsf{plaq}_{(y,z)}, \\
X^{[010]}_y X^{[010]}_{y+1} &\sim \prod_{z=1}^{L_z} X^\textsf{plaq}_{(y,z)}, \\ 
\prod_{\bm{r} \in \mathbb{Z}_{L_y} \times \mathbb{Z}_{L_z}} X^\textsf{plaq}_{\bm{r}} &\sim 1.
\end{align}
\end{subequations}
We note that these identifications hold up to multiplications of operators creating local bosonic excitations $\bm{4}$ and thus are valid only in the subspace of degenerate ground states. 
The first constraint implies that fusing two loop-like excitations yield a collection of point-like fermion excitations. 
The second and third constraints indicate that the string operators along the $y$ or $z$ axis are deformable to each other by multiplying local fermionic loops. 
These constraints reduce the dimension of the Hilbert space by factor of $2^2 \cdot 2^{L_y+L_z-2}$, leading to the ground-state degeneracy, 
\begin{align}
\textrm{GSD} = 4 \cdot 2^{L_y L_z}.
\end{align}

Here, the logarithm of GSD is proportional to the number of quantum wires or interfaces, due to degeneracy originated from local fermionic loop excitations on square plaquettes. 
As discussed in Ref.~\cite{Fuji19b}, this degeneracy can be lifted by adding the local fermionic loop operators in Eq.~\eqref{eq:PlaqOpU4} as a perturbation, 
\begin{align} \label{eq:PerturbationU4}
\mathcal{V}'_\textsf{w} = -g' \int dx \sum_{\bm{r} \in \mathbb{Z}^2} \bigl[ X^\textsf{plaq}_{\bm{r}}(x) +(X^\textsf{plaq}_{\bm{r}}(x))^\dagger ],
\end{align}
which energetically prefers states with the eigenvalues of $X^\textsf{plaq}_{\bm{r}}$ being $+1$ and thus induces the condensation of local fermionic loops. 
Hence, we consider a subspace of the degenerate ground states with $X^\textsf{plaq}_{\bm{r}} = +1$. 
This subspace is still spanned by the membrane and nonlocal string operators in Eqs.~\eqref{eq:MembraneU4}, \eqref{eq:StringYU4}, and \eqref{eq:StringZU4} and now has dimension $4 \cdot 2^{L_y+L_z}$.
The constraints in Eq.~\eqref{eq:ConstraintsU4} then become
\begin{subequations}
\begin{align}
(X^\textsf{mem})^2 &\sim \prod_{y=1}^{L_y} X^{[010]}_y \prod_{z=1}^{L_z} X^{[001]}_z, \\
X^{[001]}_z X^{[001]}_{z+1} &\sim 1, \\
X^{[010]}_y X^{[010]}_{y+1} &\sim 1.
\end{align}
\end{subequations}
Taking these constraints into account, the ground-state degeneracy after the condensation of fermionic loops becomes
\begin{align}
\textrm{GSD}' = 2^3.
\end{align}

The system-size-independent constant degeneracy implies that the resulting ground state has a TQFT-type topological order, which is described by a bosonic 3D $Z_2$ gauge theory with a $Z_2$ fermionic charge. 
This can be understood as follows. 
The condensation of fermionic loops makes different paths connecting two point-like fermion excitations indistinguishable, as desired for point-like excitations in 3D TQFT-type topological orders [see Fig.~\ref{fig:ModelBosonU4SqEx}~(a)]. 
\begin{figure}
\includegraphics[clip,width=0.4\textwidth]{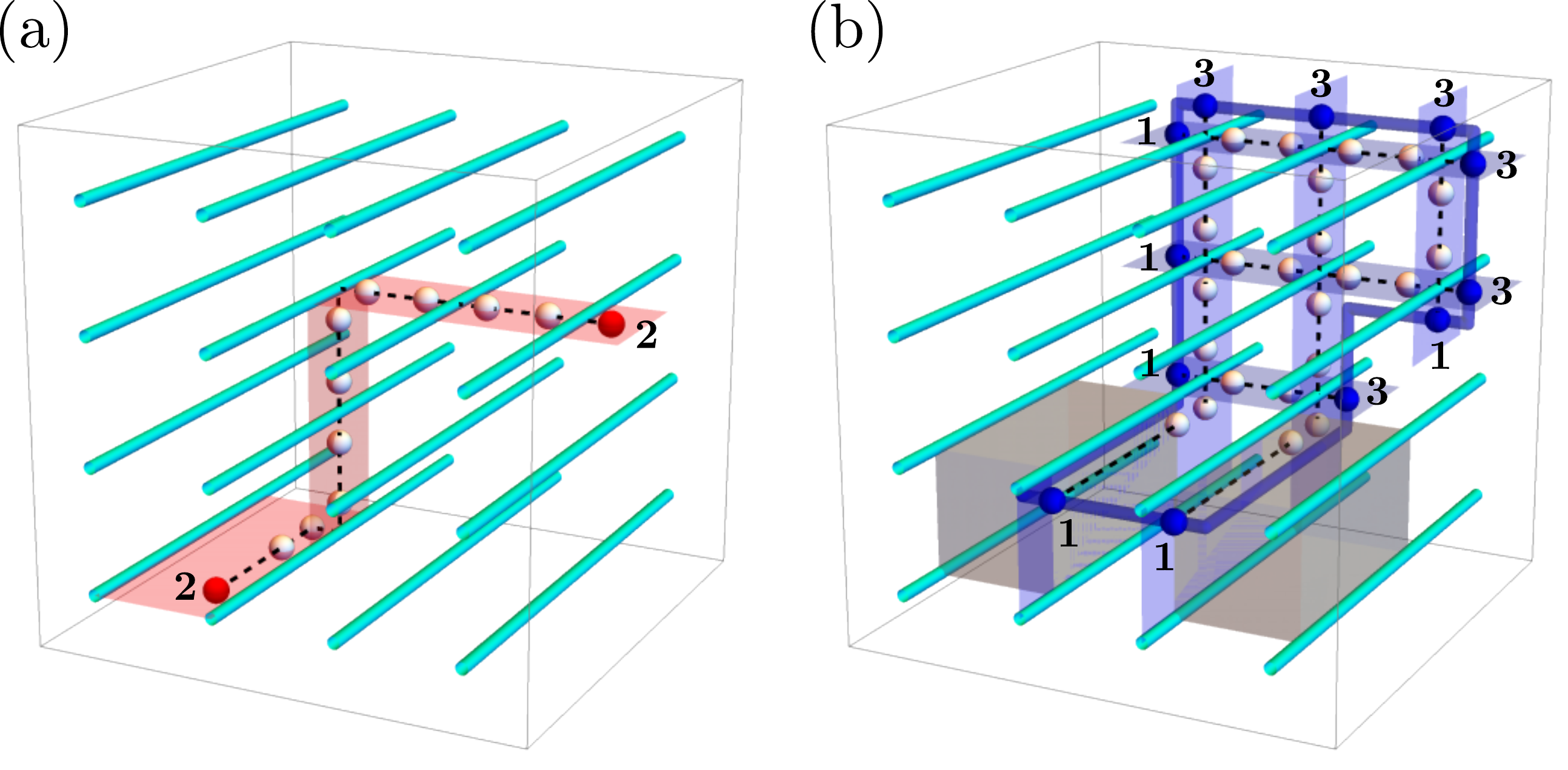}
\caption{After the condensation of local fermionic loops, the bosonic $U(1)_4$ models has (a) a point-like fermion excitation fully mobile in the 3D space and (b) a loop-like excitation fully deformable in the 3D space. 
The latter now costs energy proportional to its length along the $x$ axis since the eigenvalues of local plaquette loop operators $X^\textsf{plaq}_{\bm{r}}$ in the gray regions are flipped. 
}
\label{fig:ModelBosonU4SqEx}
\end{figure}
At the same time, it also gives energetic constraints to lineon excitations along the $x$ axis.
As indicated from Eq.~\eqref{eq:CommPlaqU4}, a string operator transferring a $\bm{1}$ lineon along the $x$ axis anticommutes with a plaquette loop operator $X^\textsf{plaq}_{\bm{r}}$ for any $x$ when the strip on which the lineon moves is shared by the square plaquette. 
Therefore, single $\bm{1}$ lineons along the $x$ axis are confined in the presence of the additional interaction $\mathcal{V}'_\textsf{w}$, as creation of a pair of them costs an energy proportional to their separation. 
However, the square of a lineon string operator, which transfers a fermionic $\bm{2}$ quasiparticle along the $x$ axis, commutes with any $X^\textsf{plaq}_{\bm{r}}$ and thus 3D fermion excitations are still deconfined. 
Furthermore, a product of two lineon string operators on different strips commute with $X^\textsf{plaq}_{\bm{r}}$ when the two strips are shared by the square plaquette. 
When a product of lineon string operators is constructed such that the corresponding strips form a closed membrane, it commutes with all plaquette loop operators $X^\textsf{plaq}_{\bm{r}}$. 
An open membrane consisting of finite-segment lineon operators costs an energy proportional to the length of its boundary and thus creates a loop-like excitation. 
Combined with a finite fraction of the membrane operator defined in Eq.~\eqref{eq:MembraneOpU4} for the $yz$ plane, we can create loop-like excitations fully deformable in the 3D space [see Fig.~\ref{fig:ModelBosonU4SqEx}~(b)], which have the mutual $\pi$ statistics with the 3D fermionic excitation [see Fig.~\ref{fig:ModelBosonU4SqBraid}~(a)]. 
\begin{figure}
\includegraphics[clip,width=0.4\textwidth]{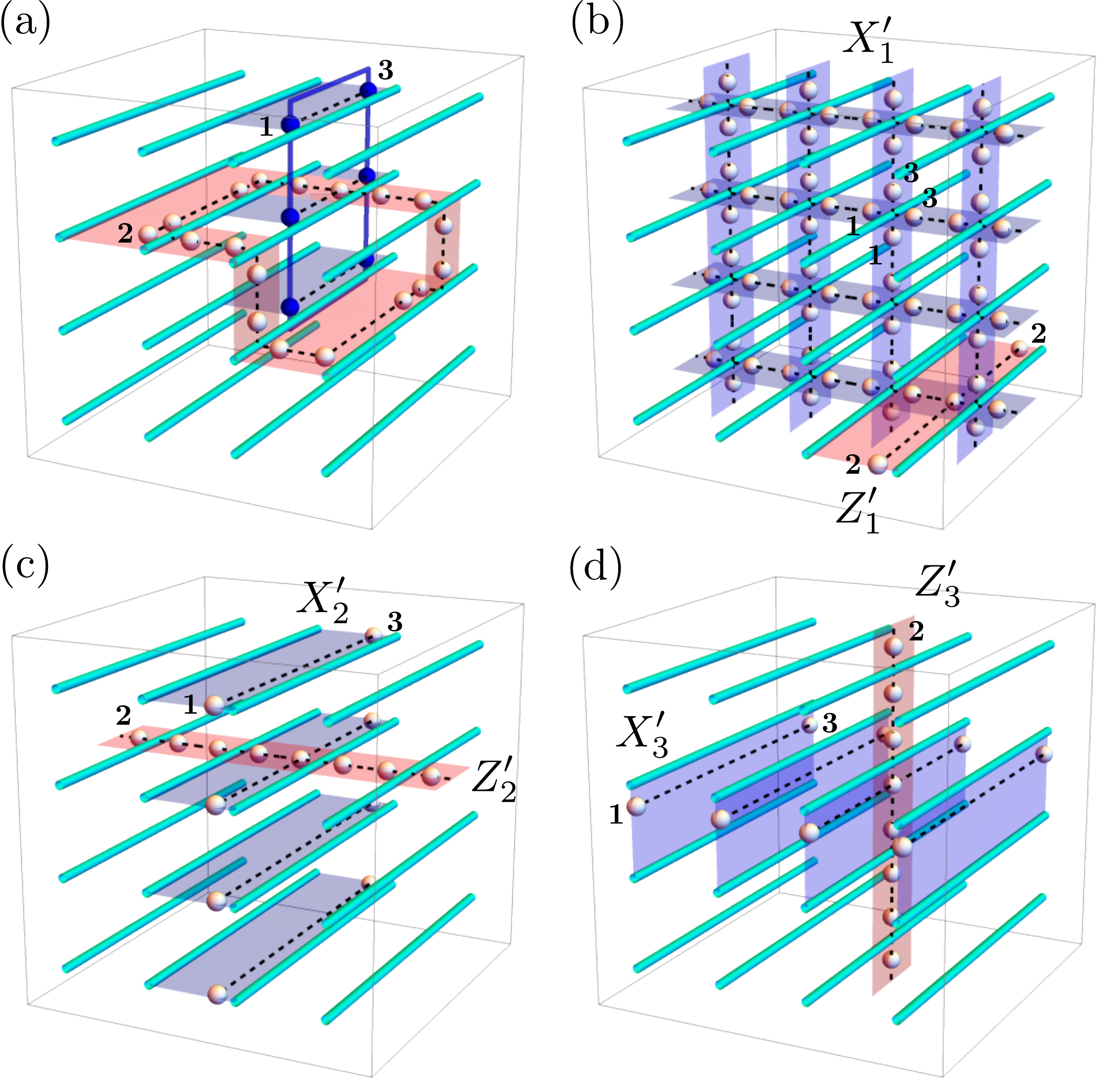}
\caption{(a) Braiding process for the bosonic $U(1)_4$ model between a point-like fermionic excitation and a loop-like excitation after the condensation of local fermionic loops, yielding the mutual $\pi$ statistics. 
There are three membrane operators on a torus in the (b) $yz$, (c) $xz$, and (d) $xy$ planes and the associated string operators.
}
\label{fig:ModelBosonU4SqBraid}
\end{figure}
A distinguished feature of our model from the ordinary 3D $Z_2$ gauge theory is that fusing two loop-like excitations yields not the ground state but a collection of point-like fermion excitations. 

Accordingly, we can find three membrane operators in the $yz$, $xz$, and $xy$ planes,
\begin{subequations}
\begin{align}
X'_1 &= \prod_{\bm{r} \in \mathbb{Z}_{L_y} \times \mathbb{Z}_{L_z}} \exp \bigl[ -i\varphi^1_{\bm{r}}(x_0) \bigr], \\
X'_2 &= \prod_{z=1}^{L_z} \exp \left[ -\frac{i}{4} \int_0^{L_x} dx \, \partial_x \Phi^-_{(y_0,z), L}(x) \right], \\
X'_3 &= \prod_{y=1}^{L_y} \exp \left[ -\frac{i}{4} \int_0^{L_x} dx \, \partial_x \Phi^+_{(y,z_0), L}(x) \right], 
\end{align}
\end{subequations}
respectively, and three fermion string operators along the $x$, $y$, and $z$ axes, 
\begin{subequations}
\begin{align}
Z'_1 &= \exp \left[ -\frac{i}{2} \int_0^{L_x} dx \, \partial_x \Phi^-_{(y_0,z_0), L}(x) \right], \\
Z'_2 &= \prod_{y=1}^{L_y} \exp \Bigl[ -i\bigl( \varphi^1_{(y,z_0)}(x_0) -\varphi^2_{(y,z_0)}(x_0) \bigr) \Bigr], \\
Z'_3 &= \prod_{z=1}^{L_z} \exp \Bigl[ -i\bigl( \varphi^1_{(y_0,z)}(x_0) +\varphi^2_{(y_0,z)}(x_0) \bigr) \Bigr],
\end{align}
\end{subequations}
respectively, where the choice of $x_0$, $y_0$, and $z_0$ is arbitrary in each expression. 
They are illustrated in Fig.~\ref{fig:ModelBosonU4SqBraid}~(b)--(d).
These operators obey the Pauli algebra, 
\begin{align}
Z'_j X'_j = -X'_j Z'_j,
\end{align}
for $j=1,2,3$. 
These membrane and string operators span the $2^3$-dimensional ground-state manifold on a torus after the condensation of local fermionic loops. 

\subsection{Hybrid of TQFT-type and foliated type-I planon models}
\label{sec:Hybrid}

We here present a coupled-wire model that exhibits a hybrid of 3D TQFT-type topological order with point-like and loop-like excitations and foliated type-I fracton order with planons. 
Interestingly, planon excitations, which on one hand characterize type-I fracton order, have nontrivial braiding statistics with 3D loop-like excitations, which on the other hand characterize TQFT-type topological order. 
This indicates that the model cannot be regarded as a decoupled stack of a TQFT-type topological order and a foliated type-I fracton order. 
Such nontrivial interplay between topological and fracton orders has also been investigated in Refs.~\cite{Tantivasadakarn21a, Tantivasadakarn21b}.

\subsubsection{Bosonic $U(1)_8$ model}
\label{sec:HybridU8Model}

We consider a 3D cellular topological state built out of the $U(1)_8$ topological orders. 
On the square grid, each interface possesses four gapless edge modes described by the $K$ matrix, 
\begin{align}
K_\textsf{e} = \begin{pmatrix} 8 &&& \\ & 8 && \\ && -8 & \\ &&& -8 \end{pmatrix}.
\end{align}
We consider a gapped interface obtained by condensing a set of quasiparticles generated by $M = \{ \bm{m}_a \}$ with 
\begin{align} \label{eq:VecMHybridU8Model}
\begin{split}
\bm{m}_1 &= (1,1,7,7)^T, \\
\bm{m}_2 &= (2,0,6,0)^T, \\
\bm{m}_3 &= (4,4,0,0)^T,
\end{split}
\end{align}
which are all bosonic quasiparticles with $s=-1$, $-2$, and $2$, respectively.
Hence, the corresponding gapped interface is bosonic. 
In order to construct a gapping potential, we add two extra bosonic wires at each interface and consider the extended $K$ matrix $K_\textsf{ew} = K_\textsf{e} \oplus K_\textsf{w}$ with $K_\textsf{w} = X \oplus X$. 
With the choice of matrices, 
\begin{align}
U = \begin{pmatrix} -1 & 1 & 2 & -2 \\ -1 & -1 & 2 & 2 \\ 1 & -1 & 2 & -2 \\ 1 & 1 & 2 & 2 \end{pmatrix}, \quad
W = \begin{pmatrix} -1 & 0 & 0 & 0 \\ 0 & -1 & 0 & 0 \\ 0 & 0 & 1 & 0 \\ 0 & 0 & 0 & 1 \end{pmatrix},
\end{align}
we find a set of integer vectors $\{ \bm{\Lambda}_{\textsf{w},\alpha} \}$, 
\begin{align}
\begin{split}
\bm{\Lambda}_{\textsf{w},1} &= (2, -1, -2, 1)^T, \\
\bm{\Lambda}_{\textsf{w},2} &= (2, -1, 2, -1)^T, \\
\bm{\Lambda}_{\textsf{w},3} &= (2, 1, -2, -1)^T, \\
\bm{\Lambda}_{\textsf{w},4} &= (2, 1, 2, 1)^T.
\end{split}
\end{align}
By introducing bosonic fields $\bm{\phi}^\textsf{w}_{\bm{r}} = (\varphi^1_{\bm{r}}, 2\theta^1_{\bm{r}}, \varphi^2_{\bm{r}}, 2\theta^2_{\bm{r}})^T$ corresponding to two-component bosonic wires, which obey the commutation relations in Eq.~\eqref{eq:CommRel3DTwoCompBoson}, we find the tunneling Hamiltonian of the form \eqref{eq:TunnelingHam3DSq} with
\begin{align} \label{eq:FieldHybridU8Model}
\begin{split}
\Phi^\pm_{\bm{r}, R} &= \varphi^1_{\bm{r}} +4\theta^1_{\bm{r}} \pm (\varphi^2_{\bm{r}} +4\theta^2_{\bm{r}}), \\
\Phi^\pm_{\bm{r}, L} &= \varphi^1_{\bm{r}} -4\theta^1_{\bm{r}} \pm (\varphi^2_{\bm{r}} -4\theta^2_{\bm{r}}).
\end{split}
\end{align}

This model hosts a hybrid of a TQFT-type topological order with 3D mobile point-like and loop-like excitations and a foliated type-I fracton order with planons in the $[001]$ and $[010]$ planes. 
There are three types of elementary excitations as shown in Fig.~\ref{fig:ModelHybridU8Sq}~(a). 
\begin{figure}
\includegraphics[clip,width=0.4\textwidth]{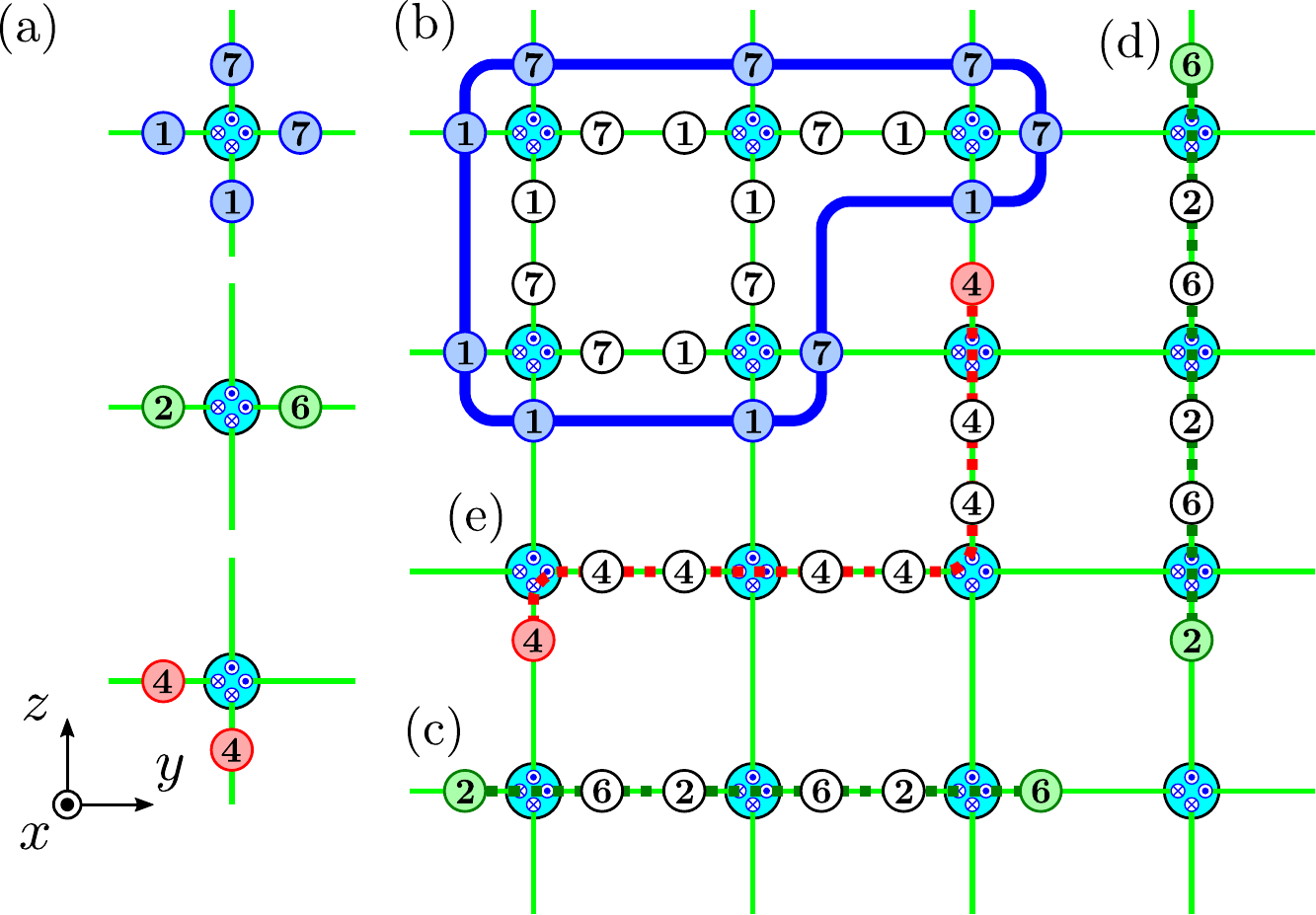}
\caption{(a) Elementary excitations created by local operators in the bosonic $U(1)_8$ model for a hybrid of TQFT-type topological order and a foliated fracton order. 
(b) Quasiparticles $\bm{1}$ and $\bm{7}$ form a loop-like excitation in the $yz$ plane. 
Semionic quasiparticles $\bm{2}$ and $\bm{6}$ can move along (c) the $y$ axis or (d) the $z$ axis. 
(e) Bosonic quasiparticles $\bm{4}$ can move freely in the $yz$ plane.
}
\label{fig:ModelHybridU8Sq}
\end{figure}
The first one is represented by $\bm{l} = (1,1,-1,-1)^T$, which creates $\bm{1}$ and $\bm{7}$ quasiparticles with $s=1/16$ and $-1/16$, respectively, on all four strips surrounding the interface $\bm{r}$. 
The corresponding local operator is given by $\exp(i\bm{p} \cdot \bm{\phi}^\textsf{w}_{\bm{r}})$ with $\bm{p} = (0,-1,0,0)^T$. 
While single $\bm{1}$ or $\bm{7}$ quasiparticles are just lineons along the $x$ axis, they can form a loop-like excitations in the $yz$ plane by successively applying these local operators over a membrane [see Fig.~\ref{fig:ModelHybridU8Sq}~(b)]. 
On an $L_x \times L_y \times L_z$ torus, we can find a pair of a membrane operator in the $yz$ plane and a lineon string operator along the $x$ axis, 
\begin{subequations} \label{eq:MembraneOpU8}
\begin{align}
X^\textsf{mem} &= \prod_{\bm{r} \in \mathbb{Z}_{L_y} \times \mathbb{Z}_{L_z}} \exp \bigl[ -i2\theta^1_{\bm{r}}(x_0) \bigr], \\
Z^\textsf{mem} &= \exp \left[ -\frac{i}{8} \int_0^{L_x} dx \, \partial_x \Phi^-_{\bm{r}_0, L}(x) \right],
\end{align}
\end{subequations}
for arbitrary $x_0$ and $\bm{r}_0$ [see Fig.~\ref{fig:ModelHybridU8SqString}~(a)]. 
\begin{figure}
\includegraphics[clip,width=0.4\textwidth]{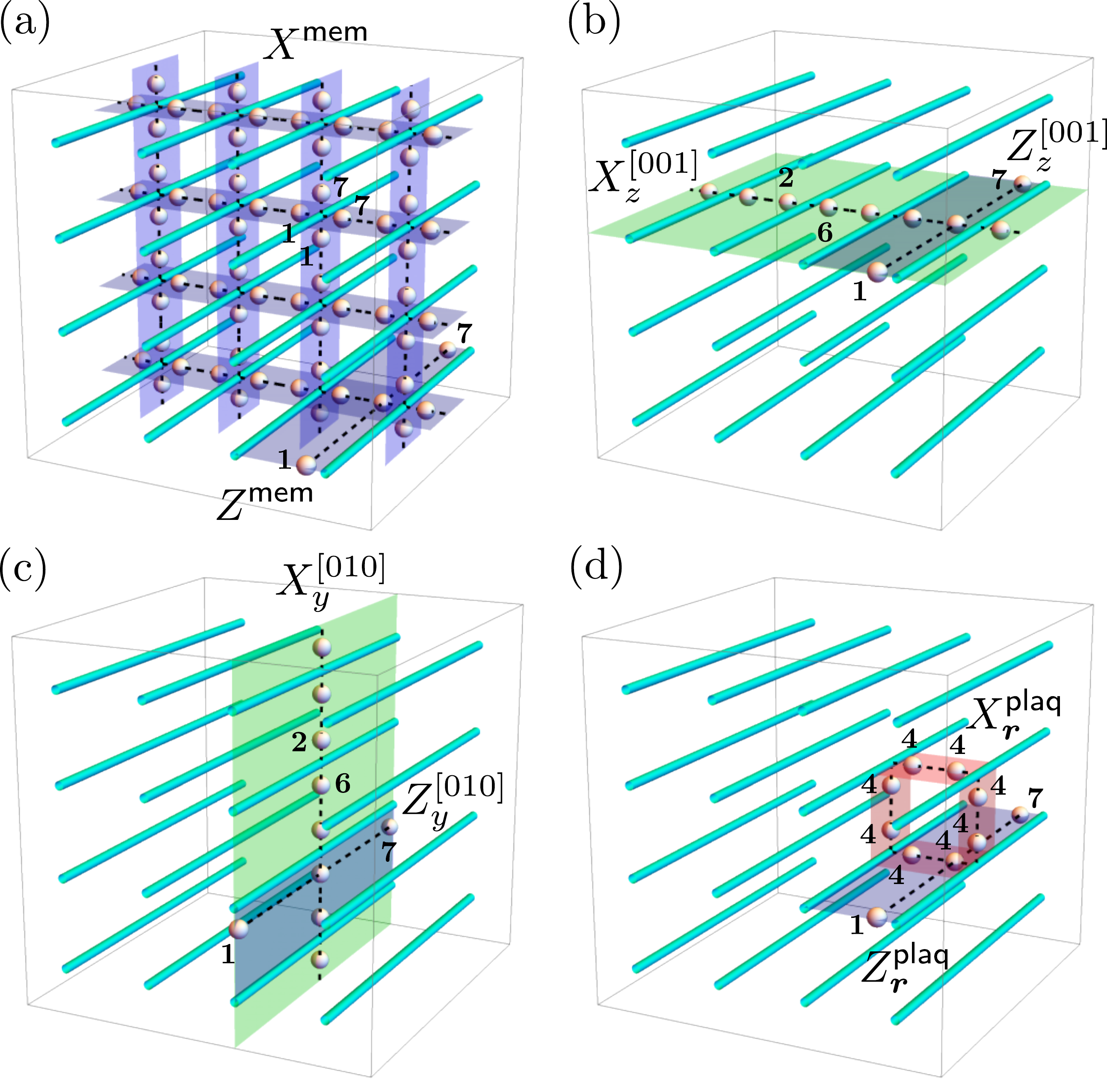}
\caption{Sets of membrane and string operators for the bosonic $U(1)_8$ model on a torus. 
They are associated with (a) loop-like excitations of quasiparticles $\bm{1}$ and $\bm{7}$ in the $yz$ plane,  planons of semionic quasiparticles $\bm{2}$ and $\bm{6}$ in (b) the $[001]$ plane and (c) the $[010]$ plane, and (d) 3D point-like excitations of bosonic quasiparticles $\bm{4}$ along plaquette loops.
}
\label{fig:ModelHybridU8SqString}
\end{figure}
They obey the generalized Pauli algebra, 
\begin{align}
Z^\textsf{mem} X^\textsf{mem} = e^{\pi i/4} X^\textsf{mem} Z^\textsf{mem}.
\end{align}

The second type of elementary excitations is represented by $\bm{l} = (2,0,-2,0)^T$ or $(0,2,0,-2)^T$, which creates a pair of $\bm{2}$ and $\bm{6}$ quasiparticles with $s=1/4$ and $-1/4$, respectively. 
These semionic quasiparticles can be transferred along the $y$ or $z$ axis by local operators $\exp (i\bm{p} \cdot \bm{\phi}^\textsf{w}_{\bm{r}})$ with $\bm{p} = (0,-1,0,1)^T$ and $(0,-1,0,-1)^T$, respectively [see Figs.~\ref{fig:ModelHybridU8Sq}~(c) and (d)]. 
With the mobility along the $x$ axis, they behave as planons in the $[001]$ or $[010]$ planes. 
On each $[001]$ plane labeled by $z=1,\cdots,L_z$, we find a pair of a string operator moving a semionic quasiparticle along the $y$ axis and that moving a $\bm{1}$ lineon along the $x$ axis, 
\begin{subequations} \label{eq:YStringOpU8}
\begin{align}
X^{[001]}_z &= \prod_{y=1}^{L_y} \exp \Bigl[ -i\bigl( 2\theta^1_{(y,z)}(x_0) -2\theta^2_{(y,z)}(x_0) \bigr) \Bigr], \\
Z^{[001]}_z &= \exp \left[ -\frac{i}{8} \int_0^{L_x} dx \, \partial_x \Phi^-_{(y_0,z), L}(x) \right],
\end{align}
\end{subequations}
for arbitrary $x_0$ and $y_0$ [see Fig.~\ref{fig:ModelHybridU8SqString}~(b)]. 
On each $[010]$ plane labeled by $y=1,\cdots,L_y$, we find a pair of a string operator moving a semionic quasiparticle along the $z$ axis and that moving a $\bm{1}$ lineon along the $x$ axis, 
\begin{subequations} \label{eq:ZStringOpU8}
\begin{align}
X^{[010]}_y &= \prod_{z=1}^{L_z} \exp \Bigl[ -i\bigl( 2\theta^1_{(y,z)}(x_0) +2\theta^2_{(y,z)}(x_0) \bigr) \Bigr], \\
Z^{[010]}_y &= \exp \left[ -\frac{i}{8} \int_0^{L_x} dx \, \partial_x \Phi^+_{(y,z_0), L}(x) \right],
\end{align}
\end{subequations}
for arbitrary $x_0$ and $z_0$ [see Fig.~\ref{fig:ModelHybridU8SqString}~(c)].
They obey the generalized Pauli algebra, 
\begin{align}
Z^{[001]}_z X^{[001]}_z &= e^{\pi i/2} X^{[001]}_z Z^{[001]}_z, \\
Z^{[010]}_y X^{[010]}_y &= e^{\pi i/2} X^{[010]}_y Z^{[010]}_y.
\end{align}

The third type of elementary excitations is represented by $\bm{l} = (4,-4,0,0)^T$ or $(0,0,4,-4)^T$, which create a pair of $\bm{4}$ quasiparticles. 
These bosonic quasiparticles can be transfered along a diagonal line in the $yz$ plane by local operators $\exp(i\bm{p} \cdot \bm{\phi}^\textsf{w}_{\bm{r}})$ with $\bm{p} = (0,0,-1,2)^T$ or $(0,0,-1,-2)^T$. 
As fusion of two planons in the $[001]$ or $[010]$ plane yields a $\bm{4}$ quasiparticle moving in the respective plane, the bosonic quasiparticles can move in any direction in the $yz$ plane [see Fig.~\ref{fig:ModelHybridU8Sq}~(e)]. 
Further combined with the mobility along the $x$ axis, they behave as point-like boson excitations fully mobile in the 3D space. 
Thus, on each square plaquette labeled by its left bottom corner $\bm{r}$, we find a pair of a string operator moving a bosonic $\bm{4}$ quasiparticle along the plaquette loop and that moving a $\bm{1}$ lineon along the $x$ axis, 
\begin{subequations}
\begin{align}
\label{eq:PlaqOpU8}
X^\textsf{plaq}_{\bm{r}} &= \exp \Bigl[ -i\bigl( \varphi^2_{\bm{r}}(x_0) +4\theta^2_{\bm{r}}(x_0) \nonumber \\
&\quad -4\theta^1_{\bm{r}+\bm{e}_y}(x_0) -\varphi^2_{\bm{r}+\bm{e}_y}(x_0) \nonumber \\
&\quad +\varphi^2_{\bm{r}+\bm{e}_y+\bm{e}_z}(x_0) -4\theta^2_{\bm{r}+\bm{e}_y+\bm{e}_z}(x_0) \nonumber \\
&\quad +4\theta^1_{\bm{r}+\bm{e}_z}(x_0) -\varphi^1_{\bm{r}+\bm{e}_z}(x_0) \bigr) \Bigr], \\
Z^\textsf{plaq}_{\bm{r}} &= \exp \left[ -\frac{i}{8} \int_0^{L_x} dx \, \partial_x \Phi^-_{\bm{r}+\bm{e}_y}(x) \right],
\end{align}
\end{subequations}
for arbitrary $x_0$ [see Fig.~\ref{fig:ModelHybridU8SqString}~(d)].
They obey the Pauli algebra, 
\begin{align} \label{eq:CommPlaqU8}
Z^\textsf{plaq}_{\bm{r}} X^\textsf{plaq}_{\bm{r}} = -X^\textsf{plaq}_{\bm{r}} Z^\textsf{plaq}_{\bm{r}}.
\end{align}

If we suppose that all membrane and string operators are independent from each other and form commuting pairs of the (generalized) Pauli operators, they could span a Hilbert space of the dimension $8 \cdot 4^{L_y+L_z} \cdot 2^{L_y L_z}$. 
However, they are not actually independent and are subject to the constraints, 
\begin{subequations}
\begin{align}
(X^\textsf{mem})^2 &\sim \prod_{y=1}^{L_y} X^{[010]}_y \prod_{z=1}^{L_z} X^{[001]}_z, \\
(X^{[001]}_z X^{[001]}_{z+1})^2 &\sim \prod_{y=1}^{L_y} X^\textsf{plaq}_{(y,z)}, \\
(X^{[010]}_y X^{[010]}_{y+1})^2 &\sim \prod_{z=1}^{L_z} X^\textsf{plaq}_{(y,z)}, \\
\prod_{\bm{r} \in \mathbb{Z}_{L_y} \times \mathbb{Z}_{L_z}} X^\textsf{plaq}_{\bm{r}} &\sim 1.
\end{align} 
\end{subequations}
These identifications hold up to multiplications of operators creating local bosonic excitations $\bm{8}$ and thus are valid only in the subspace of degenerate ground states. 
Here, the first constraint implies that fusing two loop-like excitations yields a collection of planon excitations. 
The second and third constraints imply that fusing two planon excitations yield 3D point-like boson excitations. 
These constraints reduce the dimension of the Hilbert space spanned by the membrane and string operators by factor of $2 \cdot 4 \cdot 4^{L_y+L_z-2}$, leading to the ground-state degeneracy,
\begin{align}
\textrm{GSD} = 4 \cdot 2^{L_y L_z +L_y +L_z}.
\end{align}

As in the previous models for 3D TQFT-type topological orders, the ground-state degeneracy of this model grows exponentially with the number of quantum wires or interfaces due to the presence of local bosonic loop excitations on square plaquettes. 
The degeneracy can be lifted by adding the local plaquette loop operators defined in Eq.~\eqref{eq:PlaqOpU8} as a perturbation of the form \eqref{eq:PerturbationU4}. 
The perturbation induces the condensation of the local bosonic loops and selects simultaneous eigenstates of the bosonic loop operators $X^\textsf{plaq}_{\bm{r}}$ with the eigenvalues $+1$ as a new ground state. 
The subspace of the new ground state is spanned by the membrane and string operators in Eqs.~\eqref{eq:MembraneOpU8}, \eqref{eq:YStringOpU8}, and \eqref{eq:ZStringOpU8}, which are now subject to the constraints, 
\begin{subequations}
\begin{align}
(X^\textsf{mem})^2 &\sim \prod_{y=1}^{L_y} X^{[010]}_y \prod_{z=1}^{L_z} X^{[001]}_z, \\
(X^{[001]}_z X^{[001]}_{z+1})^2 &\sim 1, \\
(X^{[010]}_y X^{[010]}_{y+1})^2 &\sim 1.
\end{align}
\end{subequations}
We thus find the ground-state degeneracy,
\begin{align}
\textrm{GSD}' = 2^3 \cdot 2^{L_y+L_z},
\end{align}
after the condensation of the local bosonic loops. 

Here, the factor $2^3$ indicates a 3D TQFT-type topological order, whereas the factor $2^{L_y+L_z}$ indicates planon excitations in each $[001]$ and $[010]$ plane. 
This can be understood as follows. 
In the condensate of the bosonic loops, $\bm{4}$ quasiparticles behave as genuine 3D point-like boson excitations as different paths connecting two point-like excitations become indistinguishable. 
Meanwhile, single $\bm{1}$ or $\bm{7}$ lineons along the $x$ axis are confined due to the anticommuting properties between the corresponding string operators and plaquette loop operators $X^\textsf{plaq}_{\bm{r}}$ in Eq.~\eqref{eq:CommPlaqU8}. 
One way to find deconfined lineon excitations is to fuse two elementary lineons to obtain $\bm{2}$ or $\bm{6}$ lineons. 
Since the corresponding string operator $(Z^\textsf{plaq}_{\bm{r}})^2$ commute with any $X^\textsf{plaq}_{\bm{r}}$, they are still deconfined along the $x$ axis. 
Combined with the mobility in the $yz$ plane, $\bm{2}$ and $\bm{6}$ quasiparticles behave as planon excitations in the $[001]$ or $[010]$ plane. 
These planons exhibit the semionic statistics on each plane. 
Another way to find deconfined excitations along the $x$ axis is to consider an appropriate product of $\bm{1}$ or $\bm{7}$ lineon string operators forming a closed membrane such that it commutes with all $X^\textsf{plaq}_{\bm{r}}$. 
Combined with the membrane operator defined in Eq.~\eqref{eq:MembraneOpU8} for the $yz$ plane, it leads to loop-like excitations consisting of $\bm{1}$ and $\bm{7}$ quasiparticles, which are fully deformable in the 3D space. 
They have the mutual $\pi$ statistics with 3D point-like boson excitations originated from $\bm{4}$ quasiparticles. 
Therefore, our model might be seen to have both TQFT-type topological order of a 3D $Z_2$ gauge theory with a $Z_2$ bosonic charge and foliated fracton order with planon excitations in the $[001]$ and $[010]$ plane.

Correspondingly, on a torus, we can construct three membrane operators in the $yz$, $xz$, and $xy$ planes, 
\begin{subequations}
\begin{align}
{X_1^\textsf{mem}}' &= \prod_{\bm{r} \in \mathbb{Z}_{L_y} \times \mathbb{Z}_{L_z}} \exp \bigl[ -i2\theta^1_{\bm{r}}(x_0) \bigr], \\
{X_2^\textsf{mem}}' &= \prod_{z=1}^{L_z} \exp \left[ -\frac{i}{8} \int_0^{L_x} dx \, \partial_x \Phi^-_{(y_0,z), L}(x) \right], \\
{X_3^\textsf{mem}}' &= \prod_{y=1}^{L_y} \exp \left[ -\frac{i}{8} \int_0^{L_x} dx \, \partial_x \Phi^+_{(y,z_0), L}(x) \right],
\end{align}
\end{subequations}
respectively, and three boson string operators along the $x$, $y$, and $z$ axes, 
\begin{subequations}
\begin{align}
{Z_1^\textsf{mem}}' &= \exp \left[ -\frac{i}{2} \int_0^{L_x} dx \, \partial_x \Phi^-_{(y_0,z_0), L}(x) \right], \\
{Z_2^\textsf{mem}}' &= \prod_{y=1}^{L_y} \exp \Bigl[ -i\bigl( 4\theta^1_{(y,z_0)}(x_0) -4\theta^2_{(y,z_0)}(x_0) \bigr) \Bigr], \\
{Z_3^\textsf{mem}}' &= \prod_{z=1}^{L_z} \exp \Bigl[ -i\bigl( 4\theta^1_{(y_0,z)}(x_0) +4\theta^2_{(y_0,z)}(x_0) \bigr) \Bigr],
\end{align}
\end{subequations}
respectively, where the choice of $x_0$, $y_0$, and $z_0$ is arbitrary in each expression [See Fig.~\ref{fig:ModelHybridU8SqString2}~(a)--(c)]. 
\begin{figure}
\includegraphics[clip,width=0.45\textwidth]{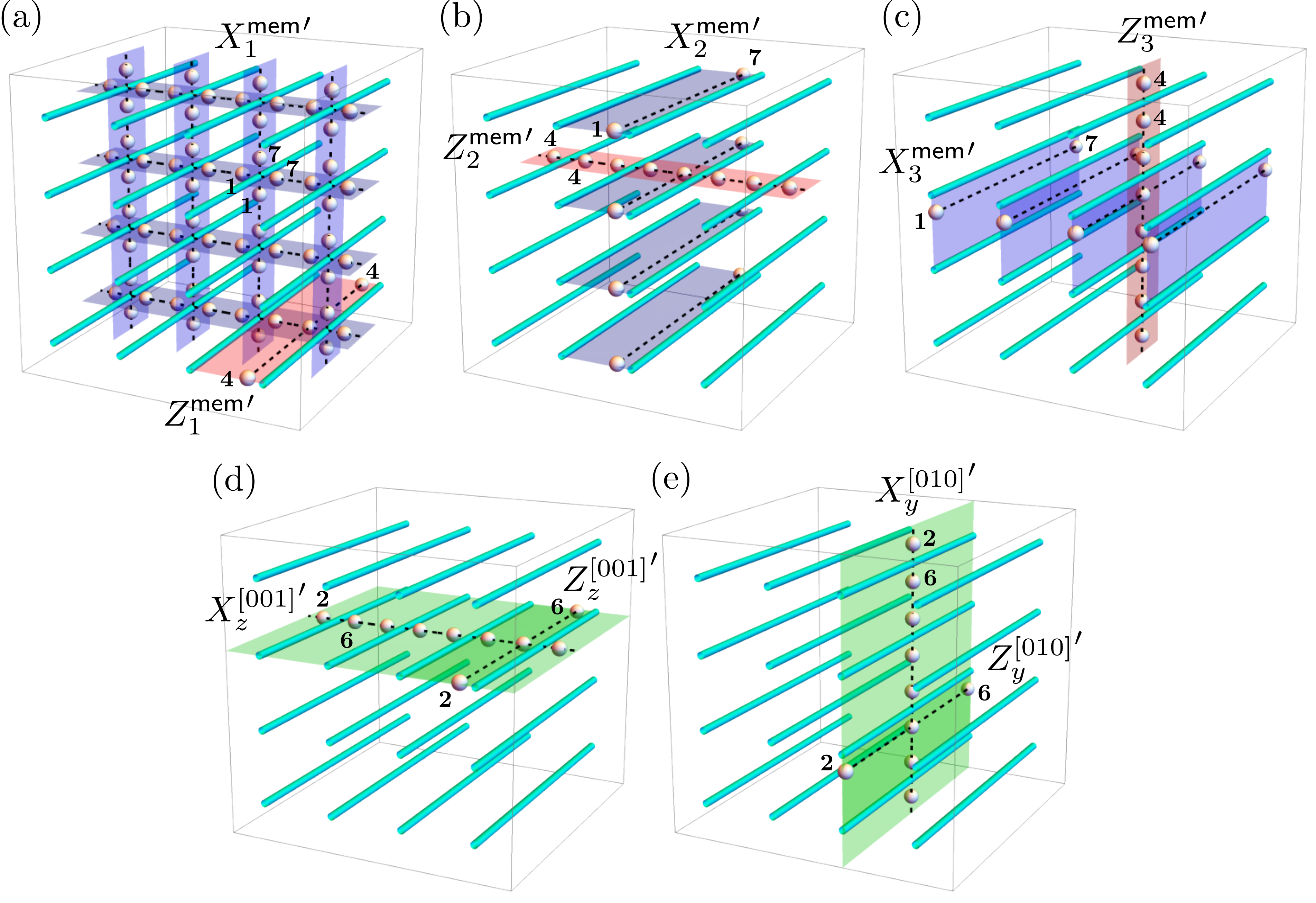}
\caption{Sets of membrane and string operators for the bosonic $U(1)_8$ model on a torus after the condensation of local bosonic loops. 
There are three membrane operators in the (a) $yz$, (b) $xz$, and (c) $xy$ planes and the associated string operators. 
There are also planon string operators in the (d) $[001]$ and (e) $[010]$ planes.
}
\label{fig:ModelHybridU8SqString2}
\end{figure}
On each $[001]$ plane, we can construct two string operators along the $y$ and $x$ axes, 
\begin{subequations}
\begin{align}
{X_z^{[001]}}' &= \prod_{y=1}^{L_y} \exp \Bigl[ -i\bigl( 2\theta^1_{(y,z)}(x_0) -2\theta^2_{(y,z)}(x_0) \bigr) \Bigr], \\
{Z_z^{[001]}}' &= \exp \left[ -\frac{i}{4} \int_0^{L_x} dx \, \partial_x \Phi^-_{(y_0,z), L}(x) \right],
\end{align}
\end{subequations}
for arbitrary $x_0$ and $y_0$ [see Fig.~\ref{fig:ModelHybridU8SqString2}~(d)], whereas on each $[010]$ plane, we can construct two string operators along the $z$ and $x$ axes, 
\begin{subequations}
\begin{align}
{X_y^{[010]}}' &= \prod_{z=1}^{L_z} \exp \Bigl[ -i\bigl( 2\theta^1_{(y,z)}(x_0) +2\theta^2_{(y,z)}(x_0) \bigr) \Bigr], \\
{Z_y^{[010]}}' &= \exp \left[ -\frac{i}{4} \int dx \, \partial_x \Phi^+_{(y,z_0), L}(x) \right],
\end{align}
\end{subequations}
for arbitrary $x_0$ and $z_0$ [see Fig.~\ref{fig:ModelHybridU8SqString2}~(e)]. 
These membrane and string operators obey the Pauli algebra, 
\begin{subequations}
\begin{align}
{Z^\textsf{mem}_j}' {X^\textsf{mem}_j}' &= -{X^\textsf{mem}_j}' {Z^\textsf{mem}_j}', \\
{Z^{[001]}_z}' {X^{[001]}_z}' &= -{X^{[001]}_z}' {Z^{[001]}_z}', \\
{Z^{[010]}_y}' {X^{[010]}_y}' &= -{X^{[010]}_y} {Z^{[010]}_y}',
\end{align}
\end{subequations}
for $j=1,2,3$, $y=1,\cdots,L_y$, and $z=1,\cdots,L_z$. 

However, it is obvious that these operators are not independent from each other and do not form mutually commuting pairs of the Pauli operators. 
This is the origin of an intriguing property of the present model: it cannot be written as a decoupled stack of a 3Dd bosonic $Z_2$ gauge theory and a foliated fracton model with planons. 
As it can be seen from the explicit forms of the membrane and string operators, fusing two loop-like excitations yields a collection of planon excitations, while fusing two planon excitations yields a 3D point-like boson excitation. 
Since a 3D point-like excitation has the mutual statistics of $\pi$ with loop-like excitations [see Fig.~\ref{fig:ModelHybridU8SqBraid}~(a)], this indicates that a planon excitation has the mutual statistics of $\pi/2$ with loop-like excitations intersecting the corresponding plane [see Fig.~\ref{fig:ModelHybridU8SqBraid}~(b)]. 
\begin{figure}
\includegraphics[clip,width=0.4\textwidth]{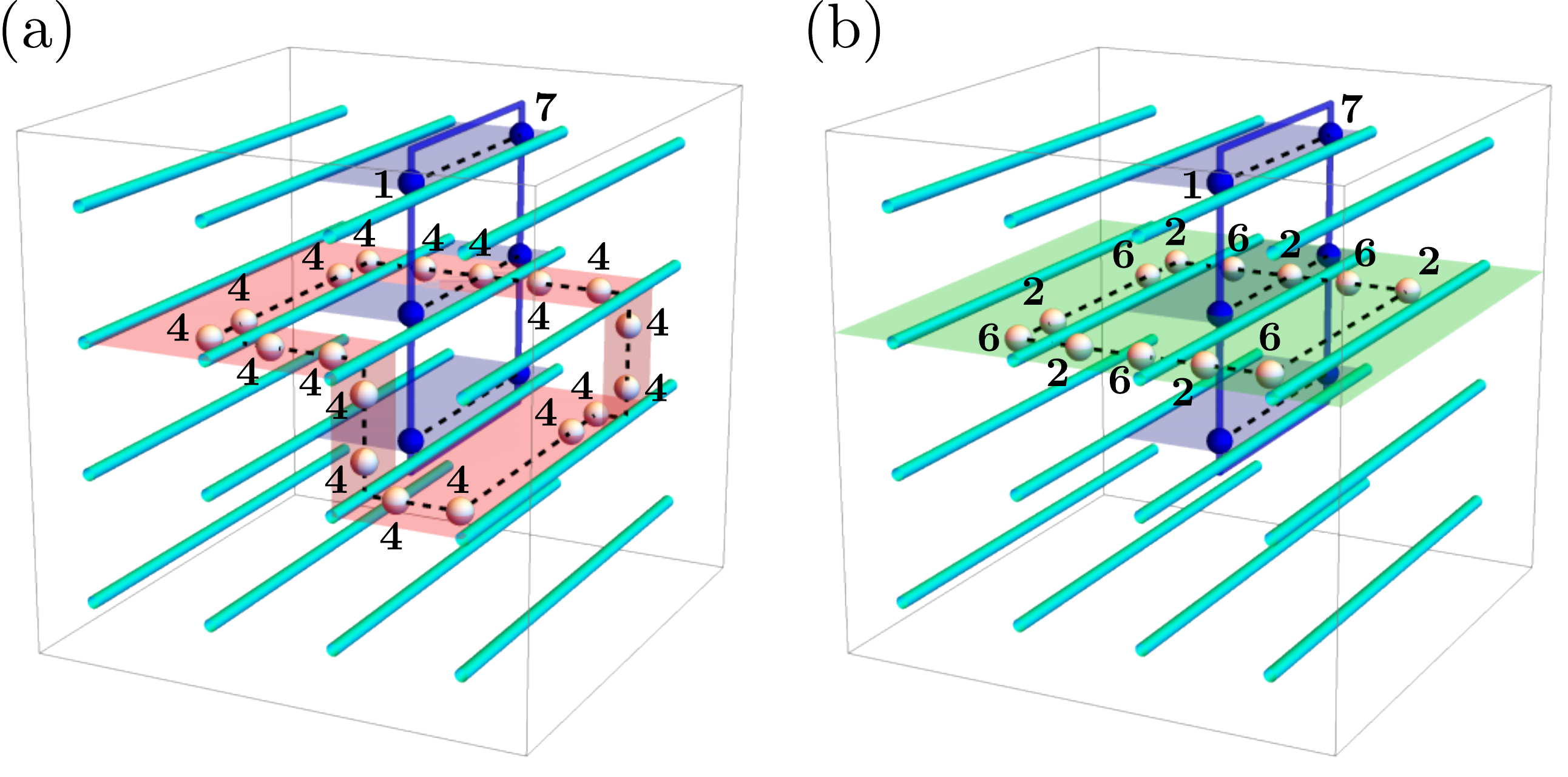}
\caption{(a) Braiding process between a loop-like excitation and a 3D point-like bosonic excitation yields the mutual $\pi$ statistics, whereas (b) that between a loop-like excitation and a planon excitation yields the mutual $\pi/2$ statistics.}
\label{fig:ModelHybridU8SqBraid}
\end{figure}
This is also evident from the generalized Pauli algebra, 
\begin{subequations}
\begin{align}
{Z^{[001]}_z}' {X^\textsf{mem}_1}' &= e^{\pi i/2} {X^\textsf{mem}_1}' {Z^{[001]}_z}', \\
{Z^{[010]}_y}' {X^\textsf{mem}_1}' &= e^{\pi i/2} {X^\textsf{mem}_1}' {Z^{[010]}_y}', \\
{X^{[001]}_z}' {X^\textsf{mem}_2}' &= e^{\pi i/2} {X^\textsf{mem}_2}' {X^{[001]}_z}', \\
{X^{[010]}_y}' {X^\textsf{mem}_3}' &= e^{\pi i/2} {X^\textsf{mem}_3}' {X^{[010]}_y}'.
\end{align}
\end{subequations}
Thus, this coupled-wire model provides a nontrivial hybrid between a 3D TQFT-type topological order and a foliated type-I fracton order.

\subsection{Fractal type-I fracton models with lineons}
\label{sec:FractalTypeI}

The cellular topological states and the corresponding coupled-wire models constructed so far admit a pair of quasiparticles created by local operators in the $yz$ plane. 
Combined with the mobility along the $x$ axis, those quasiparticles behave as planons in foliated type-I fracton orders or 3D point-like excitations in TQFT-type topological orders. 
In this subsection, we consider cellular topological states that do not exhibit any pairs of quasiparticles such that quasiparticles cannot move in the $yz$ plane in isolation. 
In order to illustrate such a restricted mobility of quasiparticles, we specifically focus on two models: one is defined on a square grid of the $U(1)_7$ topological orders, and the other is defined on a triangular grid of the $U(1)_3$ topological orders.
We expect that these cellular topological states belong to a family of fractal type-I fracton models with lineons, for which the Sierpinski fractal spin liquid is a prototypical example \cite{Castelnovo12, Yoshida13}. 

\subsubsection{Fermionic $U(1)_7$ model on square grid}
\label{sec:FermionicU7Model}

We first consider a 3D cellular topological state built out of the $U(1)_7$ topological orders. 
On the square grid, each interface possesses four gapless edge modes described by the $K$ matrix, 
\begin{align}
K_\textsf{e} = \begin{pmatrix} 7 &&& \\ & 7 && \\ && -7 & \\ &&& -7 \end{pmatrix}.
\end{align}
We consider a gapped interface obtained by condensing a set of quasiparticles generated by $M = \{ \bm{m}_a \}$ with 
\begin{align} \label{eq:CondensedMU7}
\begin{split}
\bm{m}_1 &= (1,0,2,2)^T, \\
\bm{m}_2 &= (0,1,2,5)^T, 
\end{split}
\end{align}
where $\bm{m}_1$ is a fermionic quasiparticle with $s=-1/2$ whereas $\bm{m}_2$ is a bosonic quasiparticle with $s=-2$. 
As the $U(1)_7$ topological order is fermionic, the corresponding gapped interface is also fermionic. 
In order to construct a gapping potential, we add two extra fermionic wires at each interface and consider the extended $K$ matrix $K_\textsf{ew} = K_\textsf{e} \oplus K_\textsf{w}$ with $K_\textsf{w} = Z \oplus Z$. 
With the choice of matrices, 
\begin{align}
U = \begin{pmatrix} 1 & -2 & 2 & -2 \\ 0 & 1 & -2 & -2 \\ 2 & -2 & 0 & -1 \\ 2 & 1 & 1 & 0 \end{pmatrix}, \quad
W = \begin{pmatrix} 0 & 0 & -1 & 0 \\ 0 & 0 & 0 & -1 \\ -1 & 0 & 0 & -1 \\ 0 & -1 & 0 & 0 \end{pmatrix},
\end{align}
we find a set of integer vectors $\{ \bm{\Lambda}_{\textsf{w},\alpha} \}$, 
\begin{align}
\begin{split}
\bm{\Lambda}_{\textsf{w},1} &= (1, 2, 0, -2)^T, \\
\bm{\Lambda}_{\textsf{w},2} &= (0, -2, -1, -2)^T, \\
\bm{\Lambda}_{\textsf{w},3} &= (2, 0, -2, -1)^T, \\
\bm{\Lambda}_{\textsf{w},4} &= (2, 1, 2, 0)^T.
\end{split}
\end{align}
By introducing bosonic fields $\bm{\phi}^\textsf{w}_{\bm{r}} = (\phi^1_{\bm{r},R}, \phi^1_{\bm{r},L}, \phi^2_{\bm{r},R}, \phi^2_{\bm{r},L})^T$ corresponding to two-component fermionic wires, which obey the commutation relations in Eq.~\eqref{eq:CommRel3DTwoCompField}, we find the tunneling Hamiltonian of the form \eqref{eq:TunnelingHam3DSq} with
\begin{align}
\begin{split}
\Phi^+_{\bm{r}, R} &= 2\phi^1_{\bm{r}, R} -\phi^1_{\bm{r}, L} +2\phi^2_{\bm{r}, R}, \\
\Phi^-_{\bm{r}, R} &= 2\phi^1_{\bm{r}, R} -2\phi^2_{\bm{r}, R} +\phi^2_{\bm{r}, L}, \\
\Phi^+_{\bm{r}, L} &= 2\phi^1_{\bm{r}, L} -\phi^2_{\bm{r}, R} +2\phi^2_{\bm{r}, L}, \\
\Phi^-_{\bm{r}, L} &= \phi^1_{\bm{r}, R} -2\phi^1_{\bm{r}, L} +2\phi^2_{\bm{r}, L}.
\end{split}
\end{align}

Elementary excitations created by local operators are given by the subset $M$ of the Lagrangian subgroup and are shown in Fig.~\ref{fig:ModelFermionU7Sq}~(a). 
\begin{figure}
\includegraphics[clip,width=0.45\textwidth]{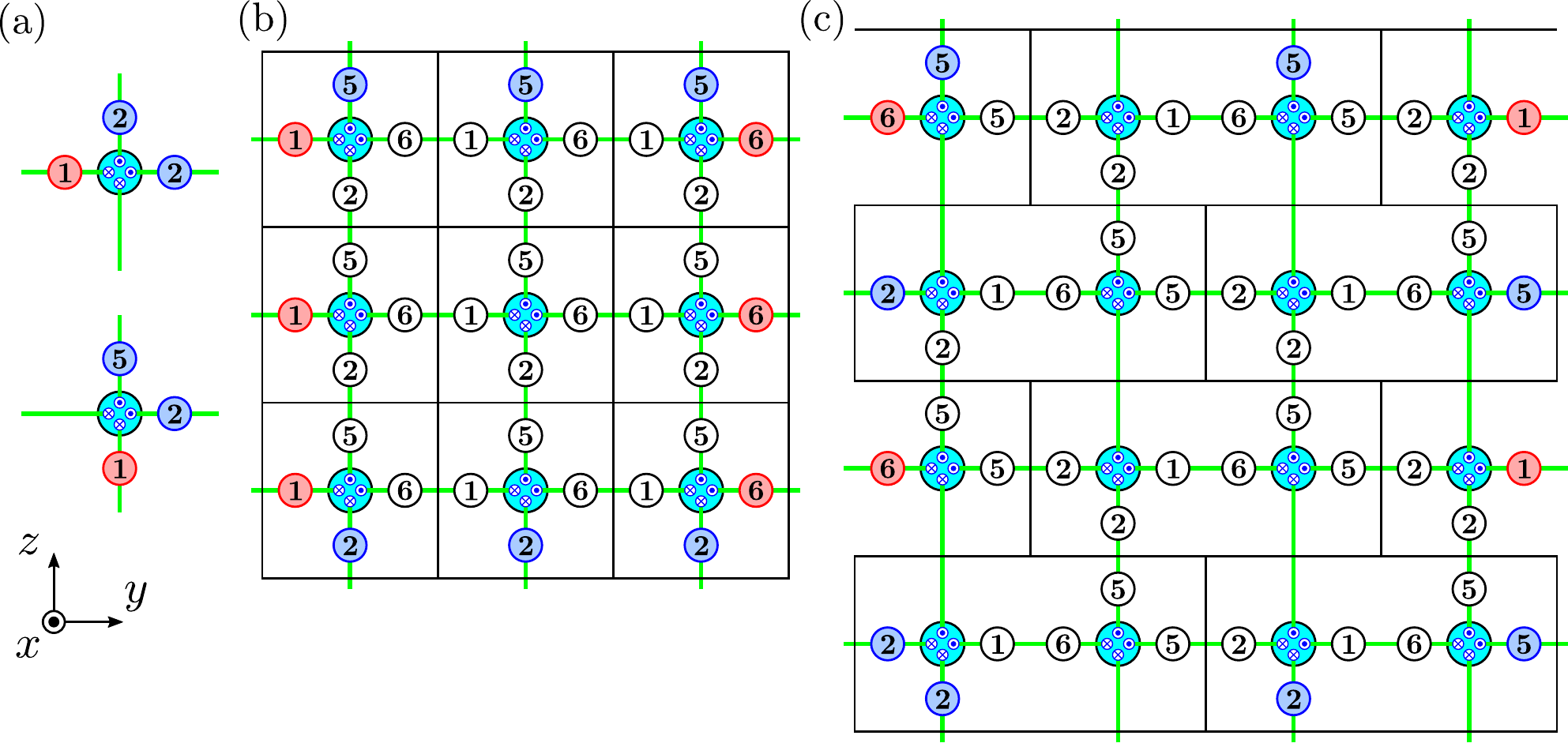}
\caption{(a) Elementary excitations created by local operators in the fermionic $U(1)_7$ model. 
Loop-like excitations can be formed (b) by $\bm{l} = (1,2,-1,-2)^T$ and (c) by combinations of $\bm{l}_1 = (2,2,1,0)^T$ and $\bm{l}_2 = (-1,0,-2,-2)$. 
The black solid lines indicate repeated patterns of quasiparticles.}
\label{fig:ModelFermionU7Sq}
\end{figure}
They are triplets of quasiparticles and any choice of their linear combinations cannot be reduced to a pair of quasiparticles. 
Hence, single quasiparticles do not behave as point-like excitations in the $yz$ plane, while they are lineons mobile along the $x$ axis. 
Furthermore, there seem no dipole or multipole excitations mobile in the $yz$ plane. 
The simplest excitations take the form of loop and can be constructed from $\bm{l} = (1,2,-1,-2)^T$, which is created by $\exp(i\bm{p} \cdot \bm{\phi}^\textsf{w}_{\bm{r}})$ with $\bm{p}=(-1,0,0,-1)^T$. 
As shown in Fig.~\ref{fig:ModelFermionU7Sq}~(b), successive application of this operator over a membrane creates a loop-like excitation in the $yz$ plane. 
It can fully cover the $yz$ plane on a torus with $L_x \times L_y \times L_z$ such that all quasiparticles are pair annihilated. 
We can thus define the set of a membrane operator and a string operator transferring a lineon $\bm{1}$ along the $x$ axis, 
\begin{subequations}
\begin{align}
X^\textsf{mem}_1 &= \prod_{\bm{r} \in \mathbb{Z}_{L_y} \times \mathbb{Z}_{L_z}} \exp \Bigl[ -i\bigl( \phi^1_{\bm{r}, R}(x_0) +\phi^2_{\bm{r}, L}(x_0) \bigr) \Bigr], \\
Z^\textsf{mem}_1 &= \exp \left[ -\frac{i}{7} \int dx \, \partial_x \Phi^-_{\bm{r}_0,L}(x) \right], 
\end{align}
\end{subequations}
for arbitrary $x_0$ and $\bm{r}_0$, which obey the generalized Pauli algebra,
\begin{align} \label{eq:U7PauliAlge}
Z^\textsf{mem}_1 X^\textsf{mem}_1 = e^{2\pi i/7} X^\textsf{mem}_1 Z^\textsf{mem}_1.
\end{align}
Hence, on a torus with any system size, the ground state is at least seven-fold degenerate. 

However, this is just a lower bound and the ground-state degeneracy actually increases nonmonotonically with the system size. 
For example, let us consider excitations $\bm{l}_1 = (2,2,1,0)^T$ and $\bm{l}_2 = (-1,0,-2,-2)^T$, which can be created by $\exp(i \bm{p} \cdot \bm{\phi}^\textsf{w}_{\bm{r}})$ with $\bm{p}_1 = (0,0,0,-1)^T$ and $\bm{p}_2 = (-1,0,0,0)^T$, respectively. 
As shown in Fig.~\ref{fig:ModelFermionU7Sq}~(c), combinations of these excitations can also form loop-like excitations, but the distribution of quasiparticles are more sparse than the previous one. 
They can fully cover the $yz$ plane on a torus only when both $L_y$ and $L_z$ are even. 
We thus find the set of a membrane and a string operator, 
\begin{subequations}
\begin{align}
X^\textsf{mem}_2 &= \prod_{\substack{\bm{r} \in \mathbb{Z}_{L_y} \times \mathbb{Z}_{L_z} \\ y+z \in 2\mathbb{Z}}} \exp \Bigl[ -i\bigl( \phi^2_{\bm{r}, L}(x_0) +\phi^1_{\bm{r}+\bm{e}_y, R}(x_0) \bigr) \Bigr], \\
Z^\textsf{mem}_2 &= \exp \left[ -\frac{4i}{7} \int dx \, \partial_x \Phi^+_{\bm{r}_0, L}(x) \right],
\end{align}
\end{subequations}
for arbitrary $x_0$ and $\bm{r}_0 = (y_0,z_0)$ such that $y_0+z_0 \in 2\mathbb{Z}$. 
They obey the same algebra as Eq.~\eqref{eq:U7PauliAlge}. 
We can construct another pair of generalized Pauli operators by translating $X^\textsf{mem}_2$ and $Z^\textsf{mem}_2$ by one site, 
\begin{subequations}
\begin{align}
X^\textsf{mem}_3 &= \prod_{\substack{\bm{r} \in \mathbb{Z}_{L_y} \times \mathbb{Z}_{L_z} \\ y+z \in 2\mathbb{Z}+1}} \exp \Bigl[ -i\bigl( \phi^2_{\bm{r}, L}(x_0) +\phi^1_{\bm{r}+\bm{e}_y, R}(x_0) \bigr) \Bigr], \\
Z^\textsf{mem}_3 &= \exp \left[ -\frac{4i}{7} \int dx \, \partial_x \Phi^+_{\bm{r}_0, L}(x) \right],
\end{align}
\end{subequations}
for arbitrary $x_0$ and $\bm{r}_0 = (y_0,z_0)$ such that $y_0+z_0 \in 2\mathbb{Z}+1$. 
Thanks to the sparse nature of the membrane operators, the lineon string operator $Z^\textsf{mem}_3$ does not intersect a strip for $X^\textsf{mem}_2$ on which quasiparticles are created and annihilated, and similarly for $Z^\textsf{mem}_2$ and $X^\textsf{mem}_3$. 
These two mutually commuting Pauli operators indicate that the ground state on a torus with even $L_y$ and $L_z$ is at least $7^2$-fold degenerate. 

As in the case of certain 3D fracton models \cite{Yoshida13}, quasiparticle properties of this coupled-wire model within the $yz$ plane are in fact described by a 2D classical spin model. 
Let us consider a square lattice each of whose vertex $\bm{r}$ is occupied by two qudits with seven states each. 
The whole Hilbert space is spanned by generalized Pauli operators $\sigma^x_{\bm{r},j}$ and $\sigma^z_{\bm{r},j}$ for $j=1,2$, which obey the algebra,
\begin{subequations}
\begin{align}
(\sigma^x_{\bm{r},j})^7 &= (\sigma^z_{\bm{r},j})^7 =1, \\
\sigma^x_{\bm{r},j} \sigma^z_{\bm{r}',j'} &= \exp ( 2\pi i \delta_{\bm{r},\bm{r}'} \delta_{j,j'}/7) \sigma^z_{\bm{r}',j'} \sigma^x_{\bm{r},j}
\end{align}
\end{subequations}
These two qudits correspond to the two components of quantum wires at each interface. 
The tunneling terms between neighboring wires are then mimicked by a classical seven-state Potts-like Hamiltonian,
\begin{align} \label{eq:ClassicalModelU7}
\mathcal{H}_\textsf{Potts}^{U(1)_7} &= -J\sum_{\bm{r} \in \mathbb{Z}^2} \bigl[ (\sigma^z_{\bm{r},1})^2 (\sigma^z_{\bm{r}, 2})^2 \sigma^z_{\bm{r}+\bm{e}_y,1} \nonumber \\
&\quad +(\sigma^z_{\bm{r},1})^2 (\sigma^z_{\bm{r},2})^5 \sigma^z_{\bm{r}+\bm{e}_z,2} +\textrm{H.c.} \bigr].
\end{align}
Since the Hamiltonian contains $\sigma^z_{\bm{r},j}$ only, each term can be simultaneously diagonalized. 
The ground state is obtained by setting the eigenvalues of all terms to be $+1$. 
Excited states are then obtained by acting $\sigma^x_{\bm{r},j}$ on the ground state, which changes the eigenvalues of interaction terms containing $(\sigma^z_{\bm{r},j})^m$ to $e^{-2\pi m i/7}$. 
As a result, the action of $\sigma^x_{\bm{r},j}$ creates excitations with energy $J \cos [2\pi (\bm{m}_j)_k/7]$ on links surrounding the vertex $\bm{r}$, where $\bm{m}_j$ are integer vectors defined in Eq.~\eqref{eq:CondensedMU7} and $k=1,2,3,4$ correspond to the left, bottom, right, and top link with respect to $\bm{r}$. 
Thus, this classical model has the same patterns of excitations as those of quasiparticles for the coupled-wire model in the $yz$ plane.

Dynamics of excitations in this classical model is most easily understood by noting that placing excitations on two links attached to a single vertex completely fixes excitations on the other two links attached to the same vertex. 
This implies that the propagation of excitations in a diagonal direction on the square lattice is given by a cellular automaton. 
Let us denote excitations on two inequivalent links by integers $a_{i,j}$ and $b_{i,j}$ and assign a square cell to each link as shown in Fig.~\ref{fig:CAU7}~(a). 
\begin{figure}
\includegraphics[clip,width=0.45\textwidth]{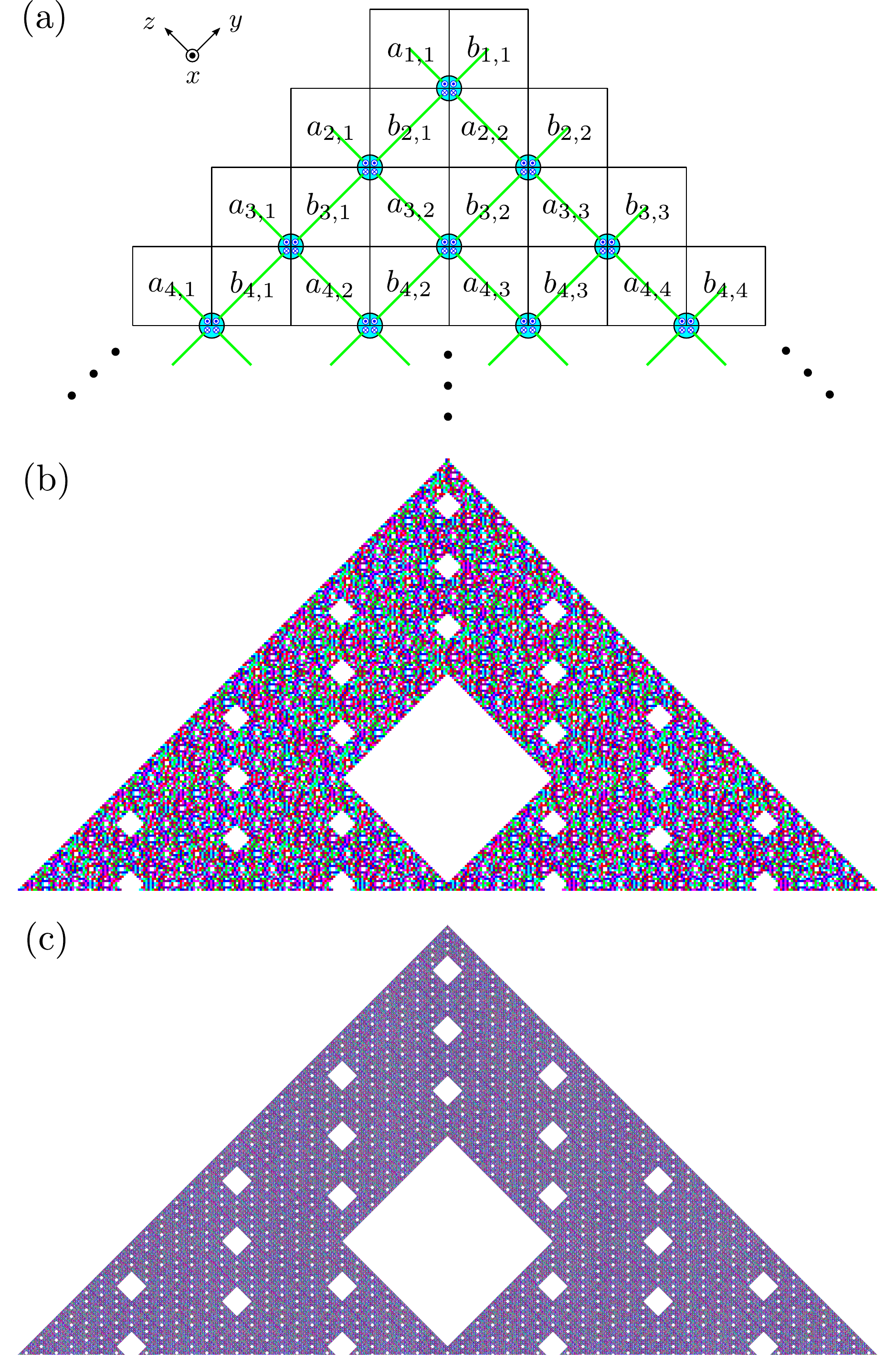}
\caption{(a) Propagation of excitations in the fermionic $U(1)_7$ model in the $[0\bar{1}\bar{1}]$ direction can be mapped to a cellular automaton. 
Here, excitations on two inequivalent links are labeled by integers $a_{i,j}$ and $b_{i,j}$ and occupy each square cell. 
Excitation patterns are generated by the automaton dynamics up to the depth (b) $i=200$ and (c) $i=1400$. 
Nontrivial excitations 1, 2, 3, 4, 5, and 6 (mod 7) are colored by red, magenta, purple, blue, cyan, and green, respectively, whereas trivial excitations 0 (mod 7) are left blank.
}
\label{fig:CAU7}
\end{figure}
The propagation of excitations in the $[0\bar{1}\bar{1}]$ direction is given by the recursion relation, 
\begin{align}
\begin{pmatrix} a_{i+1,j} \\ b_{i+1,j} \end{pmatrix} = -\begin{pmatrix} 5a_{i,j-1}+2b_{i,j-1} \\ 2a_{i,j}+2b_{i,j} \end{pmatrix} \quad \textrm{mod \ 7},
\end{align}
which describes time evolution of a linear cellular automaton on a square lattice that is $\sqrt{2}$ times smaller than and rotated by $45$ degrees from the original one.
In Figs.~\ref{fig:CAU7}~(b) and (c), we show the propagation of excitations up to the depth $i=200$ and $i=1400$, respectively, for the initial condition $(a_{1,1}, b_{1,1})=(4,1)$ and $(a_{1,j}, b_{1,j})=(0,0)$ for $j \neq 1$. 
Disregarding the species of nontrivial excitations, i.e., the colors in Figs.~\ref{fig:CAU7}~(b) and (c), excitations appear to spread over a fractal lattice, which is composed of 46 copies made by shrinking itself by a factor of 1/7, with the Hausdorff dimension $\log 46/\log 7 \sim 1.968$. 
This implies the presence of fractal operators transferring quasiparticle excitations far apart from each other as expected for fractal type-I fracton models.

We have also numerically computed the ground-state degeneracy of the classical model \eqref{eq:ClassicalModelU7} on a torus with $L_y=L_z \equiv L$. 
As shown in Fig.~\ref{fig:GSDU7}, the degeneracy increases nonmonotonically with the linear size $L$ and exhibits a self-similar structure with a scale factor of seven, as observed in the Haah's cubic code \cite{Haah13}. 
\begin{figure}
\includegraphics[clip,width=0.4\textwidth]{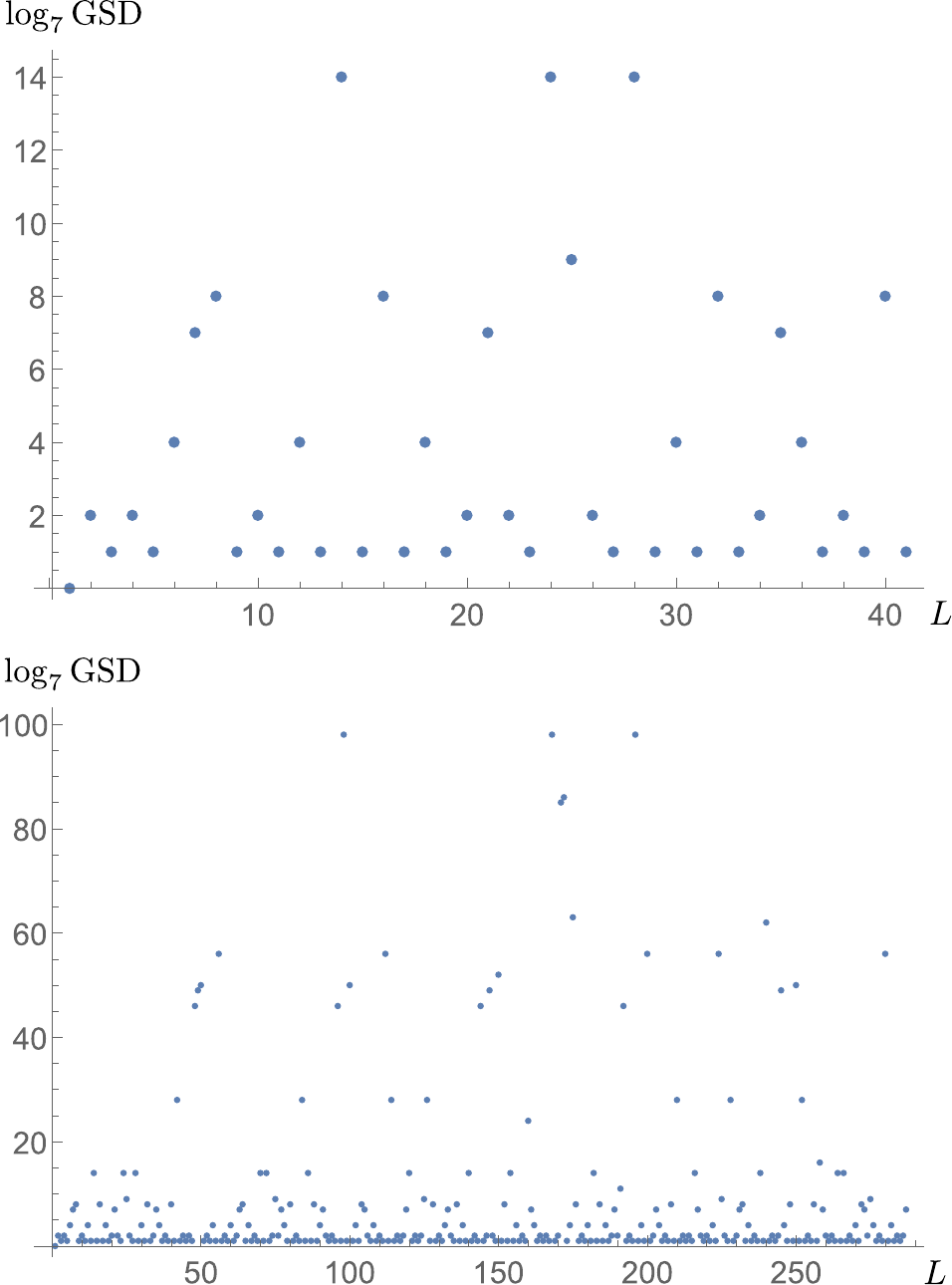}
\caption{Ground-state degeneracy of the fermionic $U(1)_7$ model on a torus with $L_y=L_z \equiv L$.}
\label{fig:GSDU7}
\end{figure}
Therefore, we conjecture that the cellular topological state built out of the $U(1)_7$ topological order on the square grid and the corresponding coupled-wire model are classified into fractal type-I fracton order with lineons.

\subsubsection{Fermionic $U(1)_3$ model on triangular grid}
\label{sec:FermionicU3Model}

We then consider a 3D cellular topological state built out of the $U(1)_3$ topological orders. 
On the triangular grid, each interface possesses six gapless edge modes $\bm{\phi}^\textsf{e}_{\bm{r}} = (\phi^\textsf{e}_{\bm{r},1}, \cdots, \phi^\textsf{e}_{\bm{r}, 6})^T$ as assigned in Fig.~\ref{fig:CellularTopo3DTr}~(a). 
\begin{figure}
\includegraphics[clip,width=0.45\textwidth]{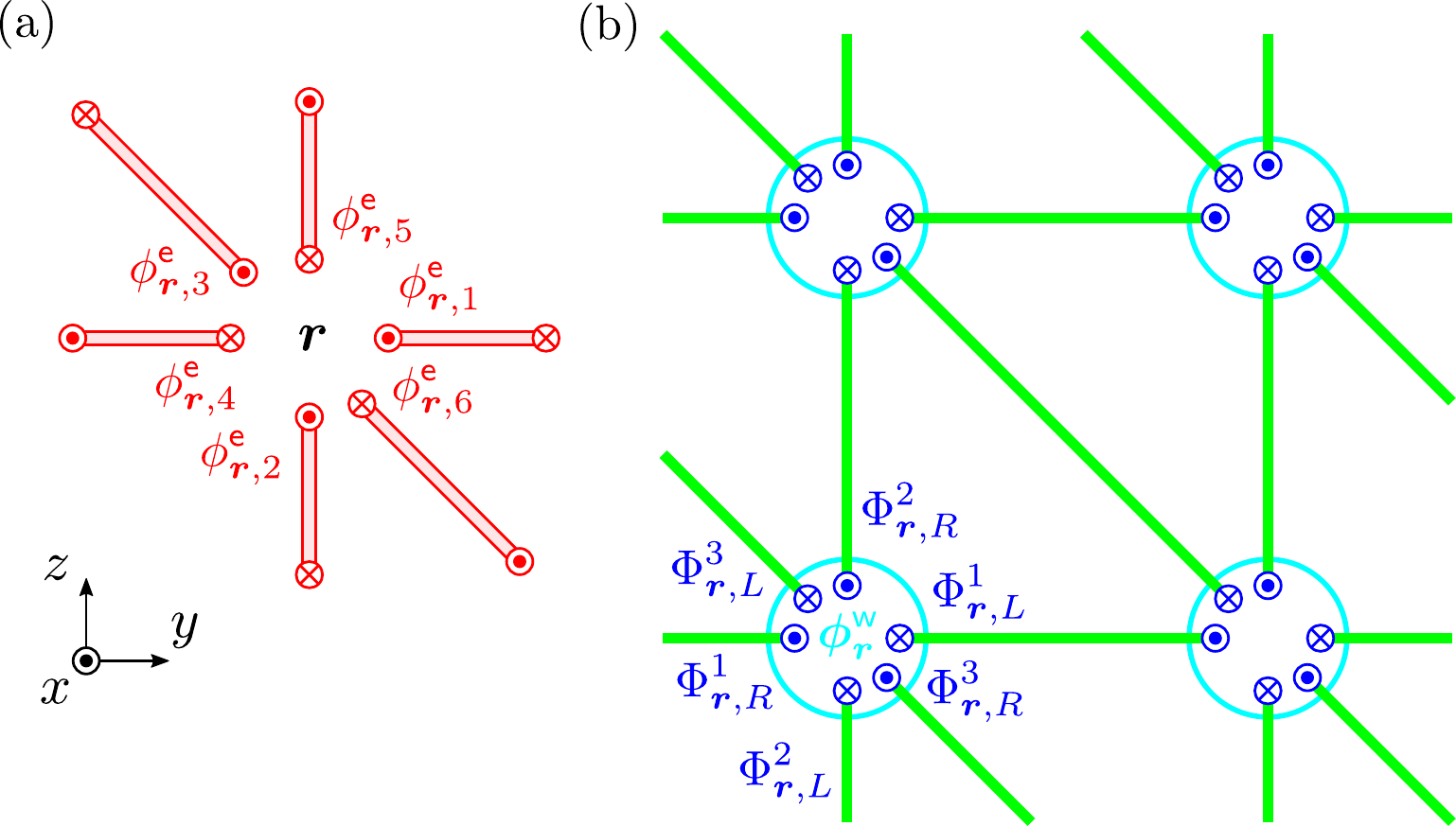}
\caption{(a) Each interface of the cellular topological state on the triangular grid has six gapless edge modes $\bm{\phi}^\textsf{e}_{\bm{r}}$. 
(b) The corresponding coupled-wire model has tunneling terms written in terms of $\Phi^\sigma_{\bm{r}, R/L}$ as given in Eq.~\eqref{eq:TunnelingU3}.}
\label{fig:CellularTopo3DTr}
\end{figure}
These bosonic fields correspond to the $K$ matrix, 
\begin{align}
K_\textsf{e} = \begin{pmatrix} 3 &&&&& \\ & 3 &&&& \\ && 3 &&& \\ &&& -3 && \\ &&&& -3 & \\ &&&&& -3 \end{pmatrix}.
\end{align}
We consider a gapped interface obtained by condensing a set of quasiparticles generated by $M = \{ \bm{m}_a \}$ with 
\begin{align}
\begin{split}
\bm{m}_1 &= (1,0,2,0,1,2)^T, \\
\bm{m}_2 &= (0,1,2,0,2,1)^T, \\
\bm{m}_3 &= (0,0,0,1,1,1)^T, 
\end{split}
\end{align}
where $\bm{m}_1$ and $\bm{m}_2$ are bosonic quasiparticles with $s=0$ whereas $\bm{m}_3$ is a fermionic quasiparticle with $s=-1/2$. 
As the $U(1)_3$ topological order is fermionic, the corresponding gapped interface is also fermionic. 
In order to construct a gapping potential, we add three extra fermionic wires at each interface and consider the extended $K$ matrix $K_\textsf{ew} = K_\textsf{e} \oplus K_\textsf{w}$ with $K_\textsf{w} = Z \oplus Z \oplus Z$. 
With the choice of matrices, 
\begin{align}
U &= \begin{pmatrix} 
0 & -1 & -1 & 1 & -1 & 1 \\ 
0 & 1 & 0 & -1 & 1 & 1 \\ 
0 & 0 & 1 & 0 & 0 & 1 \\ 
1 & 0 & 1 & -1 & -1 & 0 \\ 
1 & 1 & 0 & 1 & 0 & 0 \\ 
1 & -1 & -1 & 0 & 1 & 0 
\end{pmatrix}, \nonumber \\
W &= \begin{pmatrix} 
0 & 0 & 0 & 0 & 0 & -1 \\ 
1 & 0 & 0 & 0 & -1 & 0 \\ 
-1 & 1 & 0 & 0 & 0 & 0 \\ 
0 & 1 & 1 & 1 & -1 & 0 \\ 
0 & 1 & 0 & 1 & 0 & 0 \\ 
-1 & -1 & -1 & -1 & 1 & 0 
\end{pmatrix},
\end{align}
we find a set of integer vectors $\{ \bm{\Lambda}_{\textsf{w},\alpha} \}$, 
\begin{align}
\begin{split}
\bm{\Lambda}_{\textsf{w},1} &= (1, 1, -1, 2, 0, 0)^T, \\
\bm{\Lambda}_{\textsf{w},2} &= (1, 0, -1, 1, 0, 2)^T, \\
\bm{\Lambda}_{\textsf{w},3} &= (1, 2, -1, 0, 0, 1)^T, \\
\bm{\Lambda}_{\textsf{w},4} &= (2, 1, -1, 1, 1, 1)^T, \\
\bm{\Lambda}_{\textsf{w},5} &= (-1, -1, 2, -1, 1, -1)^T, \\
\bm{\Lambda}_{\textsf{w},6} &= (-1, 0, -1, 0, 1, 0)^T.
\end{split}
\end{align}
We then introduce bosonic fields $\bm{\phi}^\textsf{w}_{\bm{r}} = (\phi^1_{\bm{r},R}, \phi^1_{\bm{r},L}, \phi^2_{\bm{r},R}, \phi^2_{\bm{r},L}, \phi^3_{\bm{r}, R}, \phi^3_{\bm{r}, L})^T$ corresponding to three-component fermionic wires, which obey the commutation relations in Eq.~\eqref{eq:CommRel3DTwoCompField}. 
As schematically shown in Fig.~\ref{fig:CellularTopo3DTr}~(b), we find the tunneling Hamiltonian,
\begin{align} \label{eq:TunnelingU3}
\mathcal{V}_\textsf{w} &= -g \int dx \sum_{\bm{r} \in \mathbb{Z}^2} \Bigl[ \cos (\Phi^1_{\bm{r}, L} +\Phi^1_{\bm{r}+\bm{e}_y, R}) \nonumber \\
&\quad +\cos (\Phi^2_{\bm{r}, R} +\Phi^2_{\bm{r}+\bm{e}_z, L}) +\cos (\Phi^3_{\bm{r}, R} +\Phi^3_{\bm{r}+\bm{e}_y-\bm{e}_z, L}) \Bigr],
\end{align}
where
\begin{align}
\begin{split}
\Phi^\sigma_{\bm{r}, R} &= \bm{\Lambda}^T_{\textsf{w},3+\sigma} K_\textsf{w} \bm{\phi}^\textsf{w}_{\bm{r}}, \\
\Phi^\sigma_{\bm{r}, L} &= \bm{\Lambda}^T_{\textsf{w},\sigma} K_\textsf{w} \bm{\phi}^\textsf{w}_{\bm{r}}.
\end{split}
\end{align}
The explicit forms of these bosonic fields are given by
\begin{align}
\begin{split}
\Phi^1_{\bm{r}, R} &= 2\phi^1_{\bm{r},R} -\phi^1_{\bm{r}, L} -\phi^2_{\bm{r}, R} -\phi^2_{\bm{r}, L} +\phi^3_{\bm{r}, R} -\phi^3_{\bm{r}, L}, \\
\Phi^2_{\bm{r}, R} &= -\phi^1_{\bm{r}, R} +\phi^1_{\bm{r}, L} +2\phi^2_{\bm{r}, R} +\phi^2_{\bm{r}, L} +\phi^3_{\bm{r}, R} +\phi^3_{\bm{r}, L}, \\
\Phi^3_{\bm{r}, R} &= -\phi^1_{\bm{r}, R} -\phi^2_{\bm{r}, R} +\phi^3_{\bm{r}, R}, \\
\Phi^1_{\bm{r}, L} &= \phi^1_{\bm{r}, R} -\phi^1_{\bm{r}, L} -\phi^2_{\bm{r}, R} -2\phi^2_{\bm{r}, L}, \\
\Phi^2_{\bm{r}, L} &= \phi^1_{\bm{r}, R} -\phi^2_{\bm{r}, R} -\phi^2_{\bm{r}, L} -2\phi^3_{\bm{r}, L}, \\
\Phi^3_{\bm{r}, L} &= \phi^1_{\bm{r}, R} -2\phi^1_{\bm{r}, L} -\phi^2_{\bm{r}, R} -\phi^3_{\bm{r}, L}.
\end{split}
\end{align}

Similarly to the fermionic $U(1)_7$ model discussed above, elementary excitations created by local operators do not admit pairs of quasiparticles but only triplets or quadruplets as shown in Fig.~\ref{fig:ModelFermionU3Tr}~(a). 
\begin{figure}
\includegraphics[clip,width=0.45\textwidth]{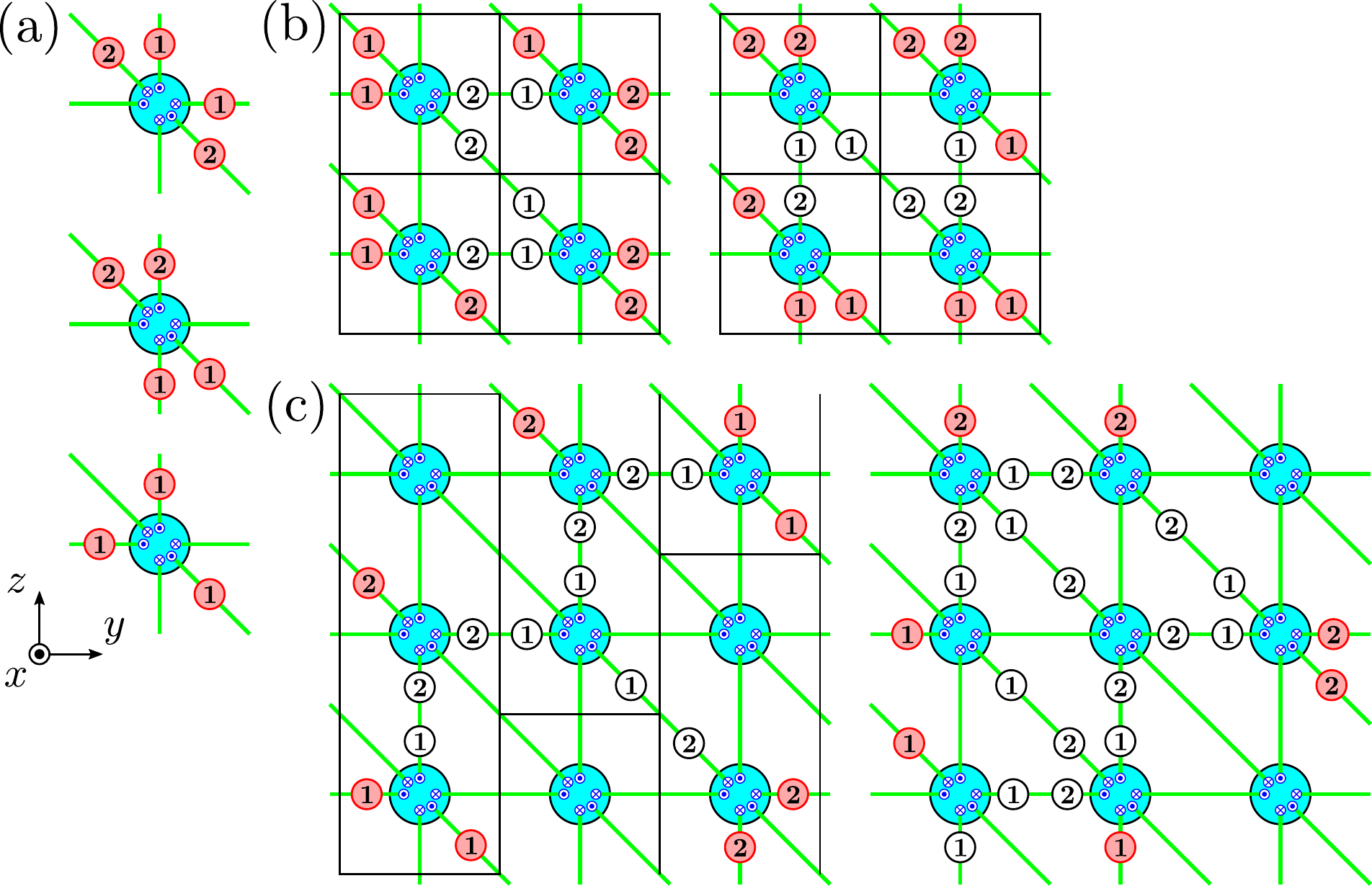}
\caption{(a) Elementary excitations in the fermionic $U(1)_3$ model on the triangular grid. 
Membrane operators in the $yz$ plane are shown for a torus with (b) $L_y=L_z=2$ and (c) $L_y=L_z=3$.}
\label{fig:ModelFermionU3Tr}
\end{figure}
Therefore, single quasiparticles cannot move in the $yz$ plane but only along the $x$ axis. 
There also seem no dipole or multipole excitations that lead to planon or 3D point-like excitations.
The simplest excitations in the $yz$ plane will be loop-like excitations created by successively applying local operators $\exp (i\bm{p} \cdot \bm{\phi}^\textsf{w}_{\bm{r}})$ with $\bm{p} = (1,0,0,-1,0,0)^T$ or $(0,-1,-1,0,0,0)^T$, which are associated with the excitations $\bm{l} = (-1,0,1,1,0,-1)^T$ or $(0,1,-1,0,-1,1)^T$, respectively, over a membrane as illustrated in Fig.~\ref{fig:ModelFermionU3Tr}~(b). 
On an $L_x \times L_y \times L_z$ torus, we can construct the corresponding membrane operators in the $yz$ plane and the string operators along the $x$ axis, 
\begin{subequations}
\begin{align}
X^\textsf{mem}_1 &= \prod_{\bm{r} \in \mathbb{Z}_{L_y} \times \mathbb{Z}_{L_z}} \exp \Bigl[ i\bigl( \phi^1_{\bm{r},R}(x_0) -\phi^2_{\bm{r}, L}(x_0) \bigr) \Bigr], \\
X^\textsf{mem}_2 &= \prod_{\bm{r} \in \mathbb{Z}_{L_y} \times \mathbb{Z}_{L_z}} \exp \Bigl[ -i\bigl( \phi^1_{\bm{r}, L}(x_0) +\phi^2_{\bm{r}, R}(x_0) \bigr) \Bigr], \\
Z^\textsf{mem}_1 &= \exp \left[ \frac{i}{3} \int_0^{L_x} dx \, \partial_x \Phi^1_{\bm{r}_0, R}(x) \right], \\
Z^\textsf{mem}_2 &= \exp \left[ -\frac{i}{3} \int_0^{L_x} dx \, \partial_x \Phi^2_{\bm{r}_0, L}(x) \right],
\end{align}
\end{subequations}
for arbitrary $x_0$ and $\bm{r}_0$, which form mutually commuting pairs of the generalized Pauli operators, 
\begin{align}
Z^\textsf{mem}_j X^\textsf{mem}_j = e^{2\pi i/3} X^\textsf{mem}_j Z^\textsf{mem}_j.
\end{align}
These operators can be defined on a torus with arbitrary linear sizes and indicate that the ground state is at least $3^2$-fold degenerate. 
However, we can further construct membrane operators with more sparse distributions of quasiparticles, which might be seen as a signature of fractal type-I fracton order. 
As an example, we show membrane operators that lead to $3^6$-fold degenerate ground states with their translations on a torus with $L_y = L_z \in 3\mathbb{Z}$ in Fig.~\ref{fig:ModelFermionU3Tr}~(c). 

As in the previous case, the energetics of quasiparticles in the $yz$ plane can be fully captured by a classical three-state Potts-like model defined on the triangular lattice. 
We here put three qutrits on each site, whose Hilbert spaces are spanned by generalized Pauli operators $\tau^x_{\bm{r},j}$ and $\tau^z_{\bm{r},j}$ for $j=1,2,3$. 
They obey the generalized Pauli algebra, 
\begin{subequations}
\begin{align}
(\tau^x_{\bm{r},j})^3 &= (\tau^z_{\bm{r},j})^3 =1, \\
\tau^x_{\bm{r},j} \tau^z_{\bm{r}',j'} &= \exp(2\pi i \delta_{\bm{r}, \bm{r}'} \delta_{j,j'}/3) \tau^z_{\bm{r}', j'} \tau^x_{\bm{r}, j}.
\end{align}
\end{subequations}
The Hamiltonian is then given by 
\begin{align} \label{eq:ClassicalModelU3}
\mathcal{H}^{U(1)_3}_\textsf{Potts} &= -J \sum_{\bm{r} \in \mathbb{Z}^2} \bigl[ \tau^z_{\bm{r},1} \tau^z_{\bm{r}+\bm{e}_y,3} +\tau^z_{\bm{r},1} (\tau^z_{\bm{r},2})^2 \tau^z_{\bm{r},3} \tau^z_{\bm{r}+\bm{e}_z,2} \nonumber \\
&\quad +(\tau^z_{\bm{r},1})^2 \tau^z_{\bm{r},2} \tau^z_{\bm{r},3} (\tau^z_{\bm{r}+\bm{e}_y-\bm{e}_z,1})^2 (\tau^z_{\bm{r}+\bm{e}_y-\bm{e}_z,2})^2 \nonumber \\
&\quad +\textrm{H.c.} \bigr].
\end{align}
Again, the ground state can be obtained by letting the eigenvalues of all terms be $+1$ as they commute with each other. 
By acting $\tau^x_{\bm{r},j}$ on the ground state, we can create the same elementary excitations as those of the above coupled-wire model in the $yz$ plane. 
Since placing excitations on three links attached to a single vertex completely fixes excitations on the other three links attached to the same vertex, the propagation of excitations in any of the six directions, $[010]$, $[001]$, $[0\bar{1}1]$, $[0\bar{1}0]$, $[00\bar{1}]$, or $[01\bar{1}]$, is described by a cellular automaton. 
Let us denote excitations on three inequivalent links by integers $a_{i,j}$, $b_{i,j}$, and $c_{i,j}$ and assign a diamond cell to each link to fully cover the $yz$ plane, as shown in Fig.~\ref{fig:CAU3}~(a). 
\begin{figure}
\includegraphics[clip,width=0.45\textwidth]{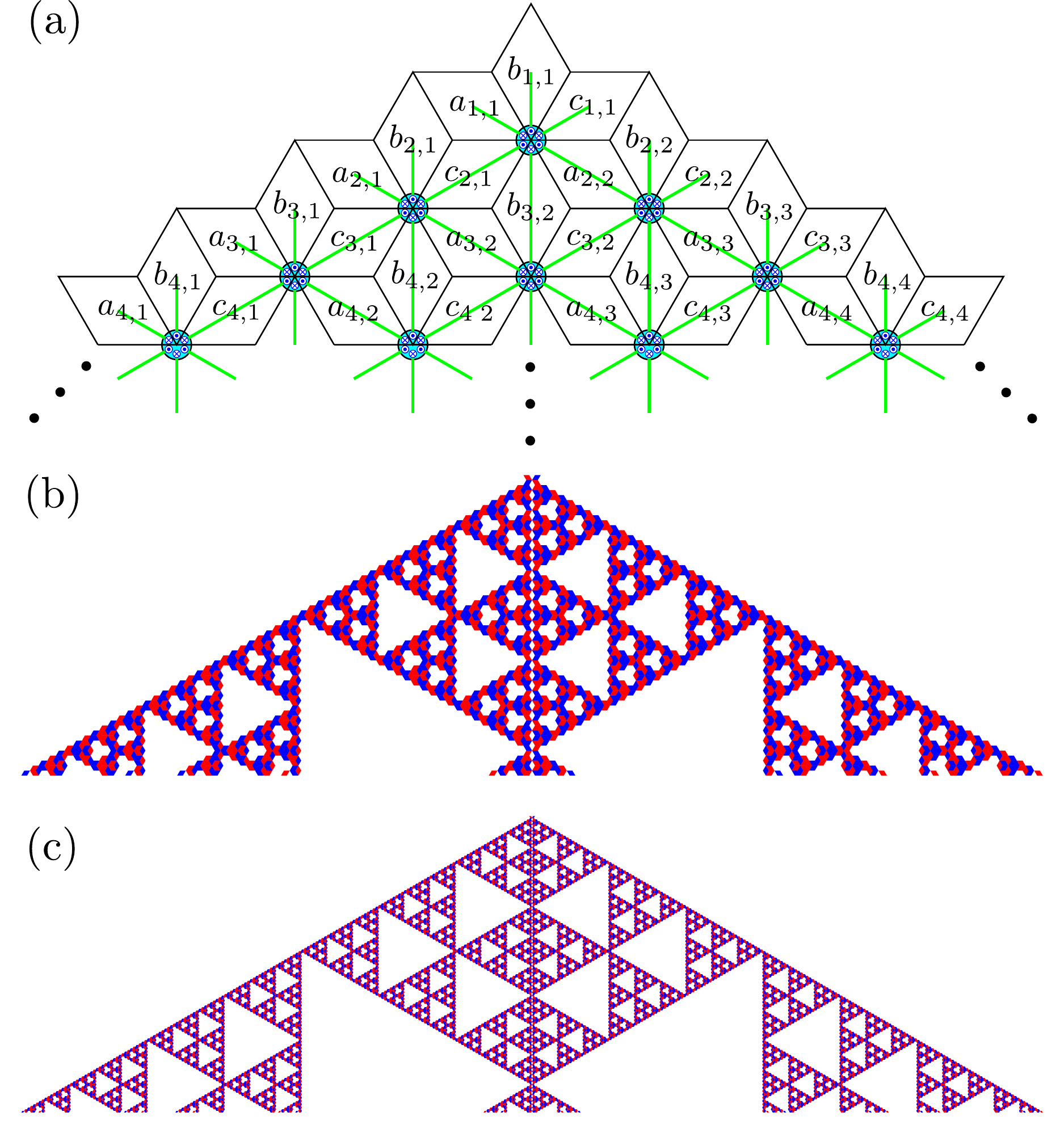}
\caption{(a) Propagation of excitations in the fermionic $U(1)_3$ model can be mapped to a cellular automaton on a dice lattice. 
Here, excitations on three inequivalent links are labeled by integers $a_{i,j}$, $b_{i,j}$, $c_{i,j}$ and occupy each diamond cell. 
Excitation patterns are generated by the automaton dynamics up to the depth (b) $i=60$ and (c) $i=180$. 
Nontrivial excitations 1 and 2 (mod 3) are colored by red and blue, respectively, whereas trivial excitations 0 (mod 3) are left blank.
}
\label{fig:CAU3}
\end{figure}
For any of the six directions mentioned above, excitations propagate via the recursion relation, 
\begin{align}
\begin{pmatrix} a_{i+1,j} \\ b_{i+1,j} \\ c_{i+1,j} \end{pmatrix}
= -\begin{pmatrix} a_{i,j-1}+b_{i,j-1}+2c_{i,j-1} \\ 2a_{i-1,j}+2b_{i-1,j} \\ 2a_{i,j}+b_{i,j}+c_{i,j} \end{pmatrix} \ \textrm{mod} \ 3,
\end{align}
which corresponds to a second-order cellular automaton on a dice lattice. 
Starting from an initial state $(a_{i,j}, b_{i,j}, c_{i,j})=(0,0,0)$ for $j \leq 1$ except with $(a_{1,1}, b_{1,1}, c_{1,1})=(1,0,2)$, we generate excitation patterns up to the depth $i=60$ and $i=180$, which are shown in Figs.~\ref{fig:CAU3}~(b) and (c), respectively. 
Disregarding the species of nontrivial excitations, excitations appear to spread over a fractal generated by the Pascal's triangle modulo three with the Hausdorff dimension $\ln 6/\ln 3 \sim 1.631$. 

We have also numerically computed the ground-state degeneracy of the classical three-state Potts model \eqref{eq:ClassicalModelU3} on a torus with $L_y = L_z \equiv L$, which is shown in Fig.~\ref{fig:GSDU3}.
\begin{figure}
\includegraphics[clip,width=0.4\textwidth]{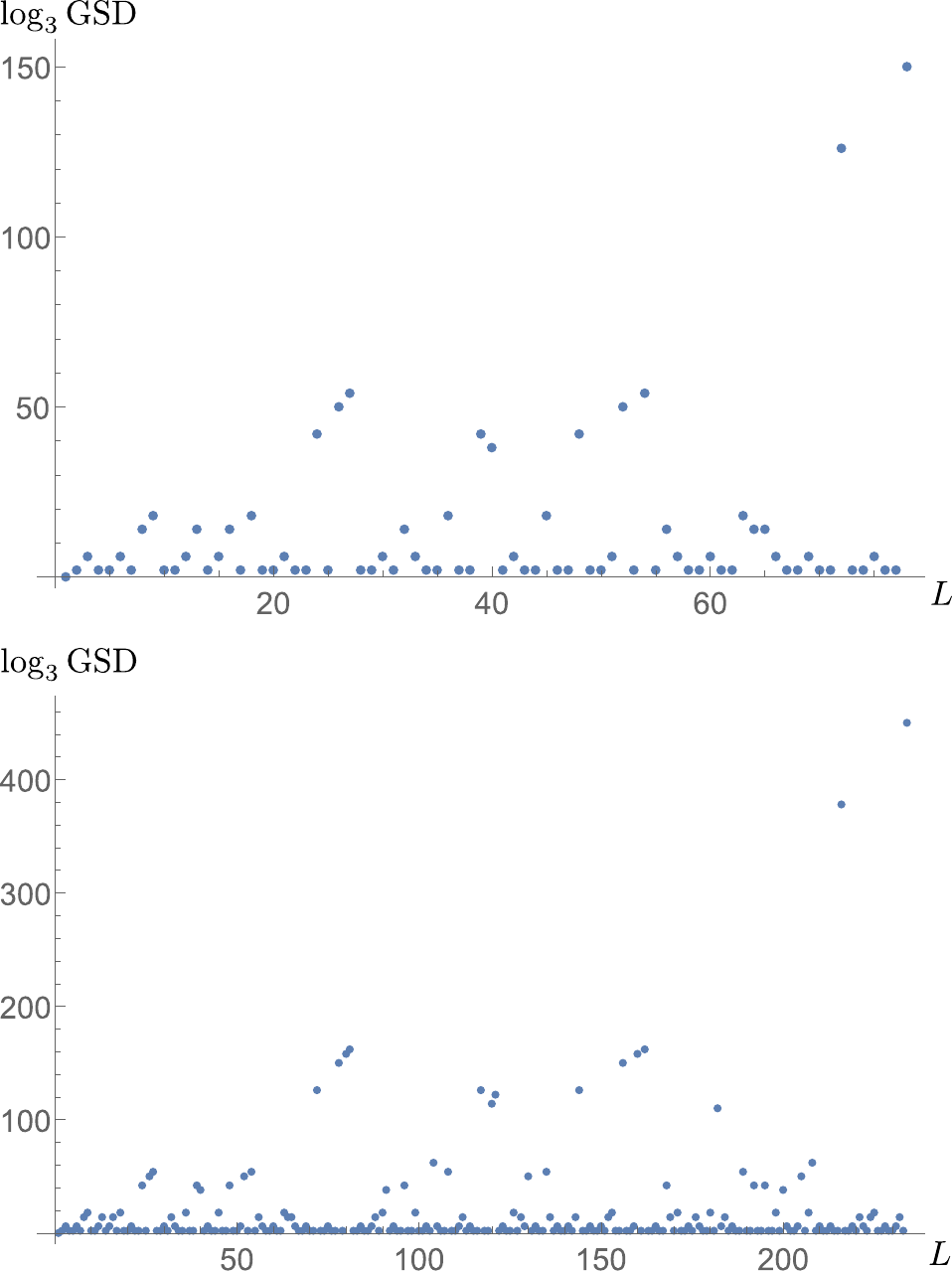}
\caption{Ground-state degeneracy of the fermionic $U(1)_3$ model on a torus with $L_y=L_z \equiv L$.}
\label{fig:GSDU3}
\end{figure}
It increases nonmonotonically with the linear size $L$ and clearly exhibits a self-similar structure with a scale factor of three. 
As in the case of the $U(1)_7$ fermionic model discussed above, these features strongly indicate that the fermionic $U(1)_3$ model on the triangular grid is also classified into fractal type-I fracton order with lineons.

\section{Conclusion and outlook}
\label{sec:Conclusion}

We have constructed 2D and 3D coupled-wire models from cellular topological states built out of strips of 2D Abelian topological orders and their nontrivial gapped interfaces. 
The resulting models are exactly solvable in the strong-coupling limit and enable us to analyze universal features of topological and fracton orders, such as quasiparticle properties and ground-state degeneracy on a torus. 
For the 2D case, we have proposed coupled-wire models for translation-symmetry-enriched topological orders in addition to the conventional coupled-wire models for ordinary 2D topological orders. 
For the 3D case, we have constructed coupled-wire models describing a variety of topological and fracton orders, including the conventional 3D (TQFT-type) topological order, foliated type-I fracton order, fractal type-I fracton order, and hybrid of topological and foliated type-I fracton orders. 
These models host fully chiral gapless surface states, which can never be gapped by local interactions, under appropriate surface terminations and thus may not have realizations in exactly solvable 3D lattice models that have been extensively used in the literature.

An interesting future direction is to extend the present construction of coupled-wire models to cellular topological states built from 2D non-Abelian topological orders and their gapped interfaces. 
For the non-Abelian case, theories for gapped interfaces have been developed in several works \cite{Bais09b, Kitaev12, LKong14, LYHung15, TLan15, Kawahigashi15, TLan20}, which might be employed to develop a systematic way to construct coupled-wire models from general inputs of the cellular structure, topological order, and gapped interface. 
Although it lacks full generality, we can also employ the conformal embedding approach developed in Ref.~\cite{Fuji19b} to construct coupled-wire models. 
The resulting models only involve current-current interactions from the Wess-Zumino-Witten CFTs and thus are still exactly solvable in the strong-coupling limit. 
While it is more complicated than the Abelian case to analyze quasiparticle statistics and ground-state degeneracy in this case, several techniques have been developed in the context of 2D coupled-wire modes \cite{Teo14, Fuji17, Iadecola19} and may allow us to study universal properties for non-Abelian generalizations of the 3D coupled-wire models.

An immediate application of the present coupled-wire approach is to investigate microscopic realizations of 3D topological and fracton orders in lattice systems beyond the paradigm of commuting-projector Hamiltonians. 
As in the 2D cases, tunneling interactions between quantum wires are often irrelevant around the fix points of decoupled Luttinger liquids and are overwhelmed by the instabilities towards conventional symmetry breaking phases. 
Thus, when this approach is applied to quasi-1D lattice systems, we generally need fine tuning of interactions for the desired tunneling terms to win. 
On the other hand, for the coupled-wire model considered in Sec.~\ref{sec:FermionicU2Model}, the tunneling interactions are marginal four-fermion interactions around the fix points of decoupled 1D fermionic chains, which might be enhanced by some chiral interactions involving three spin exchanges or pair hopping processes. 
Although the resulting fracton order is by itself somewhat boring as it only possesses planon excitations, it is still interesting to consider its lattice model realizations in order to understand the microscopic mechanism for more complicated topological and fracton orders.

Another application is to explore effective quantum field theory descriptions for fracton orders starting from 3D coupled-wire models. 
This has been achieved for 2D coupled-wire models to derive effective Chern-Simons theories for 2D Abelian topological orders \cite{Fuji19a} by utilizing nonlocal transformations for bosonic fields, which are phenomenologically understood as flux attachment and vortex duality and have been developed in Refs.~\cite{Mross16, Mross17, Leviatan20} (see also other attempts to derive the Chern-Simons theory from coupled-wire models \cite{Fontana19, Imamura19}). 
Recently, this approach has been extended to 3D coupled-wire models, which do not have the cellular structure as studied in this paper, to derive effective infinite-component Chern-Simons theories for certain fracton orders \cite{Sullivan21c}. 
Thus, this approach may potentially be applied for the present coupled-wire models to investigate effective gauge theories or response theories for some foliated and fractal type-I fracton orders.

Finally, our coupled-wire models can also be used to study entanglement properties of topological or fracton orders; some previous studies can be found, for example, in Refs.~\cite{Grover11, BShi18, HMa18a, HHe18, YZheng18, Schmitz19a, Shirley19b}. 
Since our models are sine-Gordon models with many cosine terms, we can perform quadratic expansions of the cosine terms in the strong-coupling limit to obtain free boson Hamiltonians. 
Such quadratic Hamiltonians have been used to compute topological entanglement entropy or entanglement spectrum for 2D topological phases and their interfaces \cite{Lundgren13, Cano15, Santos18, Sohal20, PKLim21}. 
These results might be generalized to the 3D coupled-wire models constructed in this paper for topological and fracton orders to study their entanglement characteristics.

\acknowledgments

Y.F. thanks Ryohei Kobayashi for useful discussions.
This work was supported in part by JSPS KAKENHI Grants No. JP20K14402 and No. JP19K03680 and by JST CREST Grant No. JPMJCR19T2.

\appendix

\section{Explicit form of $\Lambda_\textsf{w}$}
\label{sec:LambdaWire}

In Sec.~\ref{sec:BosonicGappedBoundary}, we have given a way of obtaining a set of integer vectors $\{ \widetilde{\bm{m}}'_I \}$ in Eq.~\eqref{eq:VectorMW}, which generates the same Lagrangian subgroup as $L$ for a bosonic gapped boundary. 
Let us define a $4N \times 2N$ matrix, 
\begin{align}
\widetilde{M}' \equiv (\widetilde{\bm{m}}'_1, \widetilde{\bm{m}}'_2, \cdots, \widetilde{\bm{m}}'_{2N}),
\end{align} and unimodular matrices, 
\begin{align} \label{eq:XNandZN}
X_N \equiv \begin{pmatrix} X && \\ & \ddots & \\ && X \end{pmatrix}, \quad
Z_N \equiv \begin{pmatrix} Z && \\ & \ddots & \\ && Z \end{pmatrix},
\end{align}
with $N$ diagonal blocks of the Pauli matrix $X$ or $Z$.
By permuting the columns of $\widetilde{M}'$, we can define another $4N \times 2N$ matrix $\widetilde{M}''$ that takes the form,
\begin{align}
\widetilde{M}'' \equiv \widetilde{M}' Q = \begin{pmatrix} U' \\ X_N Z_N Q \end{pmatrix},
\end{align}
where $U'$ is the $2N \times 2N$ integer matrix given in Eq.~\eqref{eq:MatrixUW} and $Q$ is a $2N \times 2N$ permutation matrix,
\begin{align} \label{eq:PermMatrixQ}
Q = \left( \begin{array}{cccc|cccc} 
1 & 0 & \cdots & 0 & 0 & 0 & \cdots & 0 \\
0 & 0 & \cdots & 0 & 1 & 0 & \cdots & 0 \\
0 & 1 & \cdots & 0 & 0 & 0 & \cdots & 0 \\
0 & 0 & \cdots & 0 & 0 & 1 & \cdots & 0 \\
\vdots & \vdots & & \vdots & \vdots & \vdots & & \vdots \\
0 & 0 & \cdots & 1 & 0 & 0 & \cdots & 0 \\
0 & 0 & \cdots & 0 & 0 & 0 & \cdots & 1 
\end{array} \right).
\end{align}
By multiplying $(\widetilde{W}^{-1})^T$, the inverse transpose of Eq.~\eqref{eq:ExtendedW}, we find 
\begin{align}
(\widetilde{W}^{-1})^T \widetilde{M}'' 
= \begin{pmatrix} (W^{-1})^T U' \\ X_N Z_N Q \end{pmatrix} 
= \begin{pmatrix} UW \\ X_N Z_N Q \end{pmatrix},
\end{align}
where we have used Eq.~\eqref{eq:MatrixUW} at the last equality. 
By further multiplying the inverse of the extended $K$ matrix \eqref{eq:ExtendedKBoson}, we find 
\begin{align}
\widetilde{K}^{-1} (\widetilde{W}^{-1})^T \widetilde{M}'' 
&= \begin{pmatrix} K^{-1} UW \\ Z_N Q \end{pmatrix} 
= \begin{pmatrix} (U^{-1})^T P W \\ Z_N Q \end{pmatrix} \nonumber \\
&= \begin{pmatrix} (U^{-1})^T (W^{-1})^T P' \\ Z_N Q \end{pmatrix},
\end{align}
where we have used Eqs.~\eqref{eq:MatrixP} and \eqref{eq:MatrixPW} at the last two equalities. 
We thus find the integer vectors $\widetilde{\Lambda} = (\widetilde{\bm{\Lambda}}_1, \cdots, \widetilde{\bm{\Lambda}}_{2N})$ in the form desired in Sec.~\ref{sec:2DCellularAddWire}, 
\begin{align} \label{eq:ExplicitLambda}
\widetilde{\Lambda} 
\equiv \widetilde{K}^{-1} (\widetilde{W}^{-1})^T \widetilde{M}'' P' W^T U^T 
= \begin{pmatrix} I_{2N} \\ \Lambda_\textsf{w} \end{pmatrix},
\end{align}
with
\begin{align} \label{eq:ExplicitLambdaWBoson}
\Lambda_\textsf{w} = Z_N Q P' W^T U^T.
\end{align}
Since $Z_N$, $Q$, $P'$, and $W$ are unimodular matrices and $U$ is an integer matrix, $\Lambda_\textsf{w}$ is an integer matrix. 
Using the integer vectors $\{ \widetilde{\bm{m}}_I \}$ in Eq.~\eqref{eq:ExtendedM}, we can also write 
\begin{align}
\widetilde{\bm{\Lambda}}_I = \sum_{J=1}^{2N} (UWP' Q^T)_{IJ} \widetilde{K}^{-1} \widetilde{\bm{m}}_J,
\end{align}
and thus $\{ \widetilde{\bm{\Lambda}}_I \}$ are linear combinations of $\{ \widetilde{K}^{-1} \widetilde{\bm{m}}_I \}$. Cosine potentials $\cos (\bm{\widetilde{\Lambda}}^T_I \widetilde{K} \widetilde{\bm{\phi}})$ then should lead to the condensation of quasipartcles in the Lagrangian subgroup generated by $\{ \widetilde{\bm{m}}_I \}$ at the interface. 

For a fermionic gapped boundary, a set of integer vectors $\{ \widetilde{\bm{m}}'_I \}$ has been given in Eq.~\eqref{eq:VectorMWFermion}. 
Again, we write them in the matrix form $\widetilde{M}' = (\widetilde{\bm{m}}'_1, \cdots, \widetilde{\bm{m}}'_{2N})$. 
By applying a linear transformation for the columns of $\widetilde{M}'$, we bring it to the following form, 
\begin{align}
\widetilde{M}'' \equiv \frac{1}{2} \widetilde{M}' (X_N+Z_N) Q = \begin{pmatrix} U' \\ X_N Q \end{pmatrix},
\end{align}
where $X_N$, $Z_N$, and $Q$ have been defined in Eqs.~\eqref{eq:XNandZN} and \eqref{eq:PermMatrixQ}. 
We can then find the integer vectors $\widetilde{\Lambda} = (\widetilde{\bm{\Lambda}}_1, \cdots, \widetilde{\bm{\Lambda}}_{2N})$ of the form in Eq.~\eqref{eq:ExplicitLambda} with 
\begin{align} \label{eq:ExplicitLambdaWFermion}
\Lambda_\textsf{w} = Z_N X_N Q P' W^T U^T,
\end{align}
where $\widetilde{K}$ and $P'$ for the fermionic case are given in Eqs.~\eqref{eq:ExtendedKFermion} and \eqref{eq:MatrixPWFermion}. 
The integer vectors $\{ \widetilde{\bm{\Lambda}}_I \}$ can also be written as 
\begin{align}
\widetilde{\bm{\Lambda}}_I = \sum_{J=1}^{2N} \left[ \frac{1}{2} UWP' Q^T (X_N+Z_N) \right]_{IJ} \widetilde{K}^{-1} \widetilde{\bm{m}}_J,
\end{align}
which makes explicit that $\{ \widetilde{\bm{\Lambda}}_I \}$ are linear combinations of $\{ \widetilde{K}^{-1} \widetilde{\bm{m}}_I \}$. 

Once written in the form \eqref{eq:ExplicitLambda}, it is easy to prove that $\{ \widetilde{\bm{\Lambda}}_I \}$ is a set of primitive integer vectors, as it has the Smith normal form, 
\begin{align}
U_\Lambda \widetilde{\Lambda} V_\Lambda = \begin{pmatrix} I_{2N} \\ O \end{pmatrix},
\end{align}
where $V_\Lambda = I_{2N}$ and 
\begin{align}
U_\Lambda = \begin{pmatrix} I_{2N} & O \\ -\Lambda_\textsf{w} & I_{2N} \end{pmatrix},
\end{align}
both of which are unimodular matrices. 

Let integer vectors $\bm{\Lambda}_{\textsf{w}, I}$ with $I=1,\cdots,2N$ be the columns of $\Lambda_\textsf{w}$.  
For a $2N$-dimensional integer vector $\bm{p}$, the vector $\bm{l}$ defined by
\begin{align} \label{eq:VecPDotLambda}
\bm{l} = \begin{pmatrix} \bm{p} \cdot \bm{\Lambda}_{\textsf{w},1} \\ \bm{p} \cdot \bm{\Lambda}_{\textsf{w},2} \\ \vdots \\ \bm{p} \cdot \bm{\Lambda}_{\textsf{w},2N} \end{pmatrix}
\end{align}
becomes an element of the Lagrangian subgroup $L$. 
This can be proven by using the explicit forms of $\Lambda_\textsf{w}$ in Eqs.~\eqref{eq:ExplicitLambdaWBoson} and \eqref{eq:ExplicitLambdaWFermion} and the fact that integer vectors $\bm{u}_I$ defined from the columns of $U$ are elements of the Lagrangian subgroup $L$. 
We then find 
\begin{align}
\bm{p} \cdot \bm{\Lambda}_{\textsf{w}, I} = \sum_{J=1}^{2N} (p'_J \bm{u}_J)_I, 
\end{align}
where $\bm{p}' = W P' Q^T Z_N \bm{p}$ for the bosonic case and $\bm{p}' = W P' Q^T X_N Z_N \bm{p}$ for the fermionic case. 
Equation~\eqref{eq:VecPDotLambda} can be written as 
\begin{align}
\bm{l} = \sum_{J=1}^{2N} p'_J \bm{u}_J.
\end{align}
Since $\bm{p}'$ is an integer vector, it is obvious that $\bm{l}$ is an element of $L$. 
As the transformation $\bm{p} \to \bm{p}'$ is unimodular, any element of $L$ will be obtained by properly choosing the integer vector $\bm{p}$ through Eq.~\eqref{eq:VecPDotLambda}. 
This proves that any local vertex operator $\exp (i\bm{p} \cdot \bm{\phi}^\textsf{w}_{\bm{r}})$ defined on a quantum wire creates some quasiparticle $\bm{l} \in L$ as discussed in the main text.

\section{Some other 3D square grid models}
\label{sec:Some3DModels}

\subsection{Bosonic $U(1)_6$ model}
\label{sec:BosonicU6Model}

We consider a 3D cellular topological state built out of the $U(1)_6$ topological orders, which can be seen as a bosonic version of the coupled-wire model constructed in Sec.~\ref{sec:FermionicU6Model}. 
It thus provides a foliated type-I fracton phase with only planons. 
For an interface described by the $K$ matrix in Eq.~\eqref{eq:KmatFourU6}, we consider a gapped interface obtained by condensing a set of quasiparticles generated by $M = \{ \bm{m}_a \}$ with 
\begin{align}
\begin{split}
\bm{m}_1 &= (1,0,2,3)^T, \\
\bm{m}_2 &= (0,1,3,2)^T.
\end{split}
\end{align}
In this case, both $\bm{m}_1$ and $\bm{m}_2$ are bosonic quasiparticles with $s=-1$ and thus the corresponding gapped interface is bosonic. 
In order to construct a gapping potential, we add two extra bosonic wires at each interface and consider the extended $K$ matrix $K_\textsf{ew} = K_\textsf{e} \oplus K_\textsf{w}$ with $K_\textsf{w} = X \oplus X$. 
With the choice of matrices, 
\begin{align}
U = \begin{pmatrix} -1 & -1 & 1 & 2 \\ 1 & -1 & -2 & 1 \\ 1 & 1 & 2 & 1 \\ -1 & 1 & -1 & 2 \end{pmatrix}, \quad
W = \begin{pmatrix} 1 & 0 & 0 & 0 \\ 0 & 1 & 0 & 0 \\ 0 & 0 & -1 & 0 \\ 0 & 0 & 0 & -1 \end{pmatrix},
\end{align}
we find a set of integer vectors $\{ \bm{\Lambda}_{\textsf{w},\alpha} \}$, 
\begin{align}
\begin{split}
\bm{\Lambda}_{\textsf{w},1} &= (-1, 1, -2, 1)^T, \\
\bm{\Lambda}_{\textsf{w},2} &= (2, -1, -1, 1)^T, \\
\bm{\Lambda}_{\textsf{w},3} &= (-2, -1, -1, -1)^T, \\
\bm{\Lambda}_{\textsf{w},4} &= (1, 1, -2, -1)^T.
\end{split}
\end{align}
We introduce bosonic fields $\bm{\phi}^\textsf{w}_{\bm{r}} = (\varphi^1_{\bm{r}}, 2\theta^1_{\bm{r}}, \varphi^2_{\bm{r}}, 2\theta^2_{\bm{r}})^T$ corresponding to two-component bosonic wires, which obey the commutation relations
\begin{align}
[\theta^\sigma_{\bm{r}}(x), \varphi^{\sigma'}_{\bm{r}'}(x')] &= i\pi \delta_{\bm{r}, \bm{r}'} \delta_{s,s'} \Theta(x-x'), \\
[\theta^\sigma_{\bm{r}}(x), \theta^{\sigma'}_{\bm{r}'}(x')] &= [\varphi^\sigma_{\bm{r}}(x), \varphi^{\sigma'}_{\bm{r}'}(x')] =0.
\end{align}
We then find the tunneling Hamiltonian of the form \eqref{eq:TunnelingHam3DSq} with
\begin{align}
\begin{split}
\Phi^+_{\bm{r}, R} &= \varphi^1_{\bm{r}} +2\theta^1_{\bm{r}} -\varphi^2_{\bm{r}} -4\theta^2_{\bm{r}}, \\
\Phi^-_{\bm{r}, R} &= -\varphi^1_{\bm{r}} -4\theta^1_{\bm{r}} -\varphi^2_{\bm{r}} -2\theta^2_{\bm{r}}, \\
\Phi^+_{\bm{r}, L} &= -\varphi^1_{\bm{r}} +4\theta^1_{\bm{r}} +\varphi^2_{\bm{r}} -2\theta^2_{\bm{r}}, \\
\Phi^-_{\bm{r}, L} &= \varphi^1_{\bm{r}} -2\theta^1_{\bm{r}} +\varphi^2_{\bm{r}} -4\theta^2_{\bm{r}}.
\end{split}
\end{align}

Similarly to the fermionic $U(1)_6$ model, this model also has three types of planons but now living in the $[010]$, $[001]$, and $[01\bar{1}]$ planes. 
Elementary excitations are given by the subset $M$ of the Lagrangian subgroup and are shown in Fig.~\ref{fig:ModelBosonU6Sq}~(a). 
\begin{figure}
\includegraphics[clip,width=0.4\textwidth]{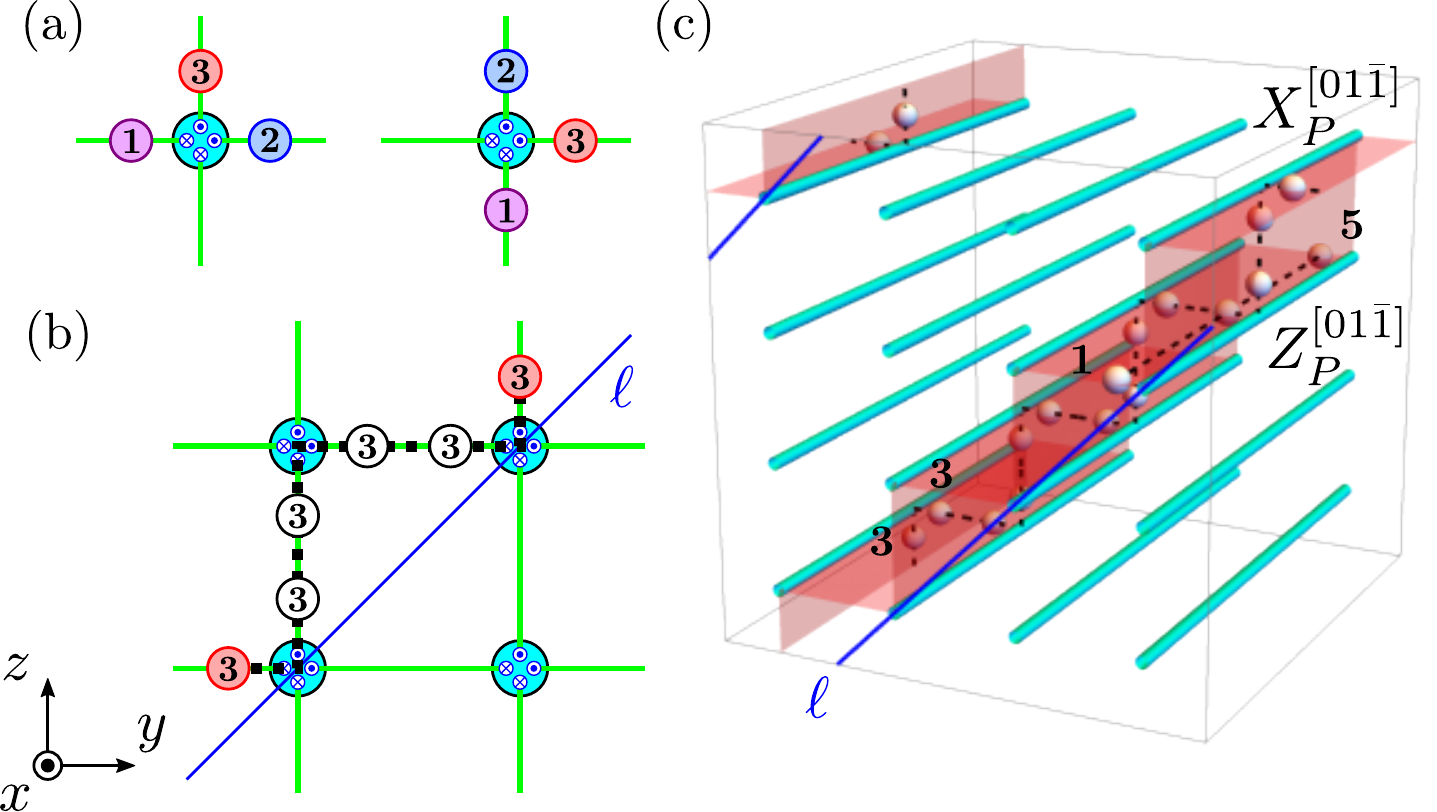}
\caption{(a) Elementary excitations created by local operators in the bosonic $U(1)_6$ model. 
(b) Semionic quasiparticle $\bm{3}$ can be transferred along a diagonal line $\ell$ in the $yz$ plane. 
(c) Set of string operators on a torus in the $[01\bar{1}]$ plane.
}
\label{fig:ModelBosonU6Sq}
\end{figure}
At each interface, we can create a pair of semionic quasiparticles corresponding to $\bm{l} = (3,0,0,-3)^T$ or $(0,3,-3,0)^T$ by applying a local operator $\exp(i\bm{p} \cdot \bm{\phi}^\textsf{w}_{\bm{r}})$ with $\bm{p} = (-1,0,0,2)^T$ or $(1,0,0,1)^T$, respectively. 
By creating and annihilating pairs of semionic quasiparticles by these operators and string operators along the $x$ axis, we can freely move semionic quasiparticles in the $[01\bar{1}]$ plane [see Fig.~\ref{fig:ModelBosonU6Sq}~(b)]. 
On the $L_x \times L_y \times L_z$ torus, we have $\textrm{gcd}(L_y,L_z)$ such planes. 
On each plane $P$, we can find string operators, 
\begin{subequations}
\begin{align}
X^{[01\bar{1}]}_P &= \prod_{\bm{r} \in \ell} \exp \Bigl[ -i\bigl( \varphi^1_{\bm{r}}(x_0) -4\theta^2_{\bm{r}'}(x_0) -\varphi^1_{\bm{r}'+\bm{e}_z}(x_0) \nonumber \\
&\quad -2\theta^2_{\bm{r}'+\bm{e}_z}(x_0) \bigr) \Bigr], \\
Z^{[01\bar{1}]}_P &= \exp \left[ -\frac{i}{2} \int_0^{L_x} dx \, \partial_x \Phi^-_{\bm{r}_\ell, L}(x) \right], 
\end{align}
\end{subequations}
where $x_0$ is arbitrary, $\ell$ is the projection of $P$ onto the $yz$ plane, and $\bm{r}_\ell$ is some $\bm{r} \in \ell$ [see Fig.~\ref{fig:ModelBosonU6Sq}~(c)]. 

We can also create a pair of $\bm{2}$ and $\bm{4}$ quasiparticles corresponding to $\bm{l} = (2,0,-2,0)^T$ and $(0,2,0,-2)^T$ by a local operator $\exp(i\bm{p} \cdot \bm{\phi}^\textsf{w}_{\bm{r}})$ with $\bm{p} = (0,1,0,1)^T$ or $(0,-1,0,1)^T$, respectively. 
These quasiparticles become planons moving in the $[010]$ or $[001]$ plane. 
On each $[001]$ plane specified by $z=1,\cdots,L_z$, we can find string operators, 
\begin{subequations}
\begin{align}
X^{[001]}_z &= \prod_{y=1}^{L_y} \exp \Bigl[ i\bigl( 2\theta^1_{(y,z)}(x_0) +2\theta^2_{(y,z)}(x_0) \bigr) \Bigr], \\
Z^{[001]}_z &= \exp \left[ -\frac{i}{3} \int_0^{L_x} dx \, \partial_x \Phi^-_{(y_0,z), L}(x) \right], 
\end{align}
\end{subequations}
for arbitrary $x_0$ and $y_0$, while on each $[010]$ plane specified by $y=1,\cdots,L_y$, we find 
\begin{subequations}
\begin{align}
X^{[010]}_y &= \prod_{z=1}^{L_z} \exp \Bigl[ -i\bigl( 2\theta^1_{(y,z)}(x_0) -2\theta^2_{(y,z)}(x_0) \bigr) \Bigr], \\
Z^{[010]}_y &= \exp \left[ -\frac{i}{3} \int_0^{L_x} dx \, \partial_x \Phi^+_{(y,z_0), L}(x) \right], 
\end{align}
\end{subequations}
for arbitrary $x_0$ and $z_0$.

These string operators obey the same algebras as those for the fermionic $U(1)_6$ model [Eqs.~\eqref{eq:AlgeStringU2}, \eqref{eq:AlgeStringU6Y}, and \eqref{eq:AlgeStringU6Z}]. 
They are independent from each other and form mutually commuting pairs of the (generalized) Pauli operators. 
They thus span the ground-state manifold with degeneracy,
\begin{align}
\textrm{GSD} = 3^{L_y+L_z} \cdot 2^{\textrm{gcd}(L_y,L_z)}.
\end{align}
In the present case, the model is microscopically built out of bosonic degrees of freedom. 
The semion topological orders in $[01\bar{1}]$ planes might be decoupled, but the $U(1)_3$ topological orders in $[010]$ and $[001]$ cannot since they are 2D fermionic topological orders. 
In this sense, this model cannot be smoothly deformed into decoupled stacks of 2D topological orders in the $[01\bar{1}]$, $[010]$, and $[001]$ directions. 

\subsection{Bosonic $U(1)_8$ model}
\label{sec:DipoleU8Model}

Here, we provide another model for foliated type-I fracton order with both lineons and planons but with different statistics of planons from that in Sec.~\ref{sec:DipoleU4Model}. 
We consider a 3D cellular topological state built out of the $U(1)_8$ topological orders, each of which is described by the $K$ matrix $K_0=8$. 
On the square grid, each interface possesses four gapless edge modes corresponding to the $K$ matrix, 
\begin{align} \label{eq:KmatFourU8}
K_\textsf{e} = \begin{pmatrix} 8 &&& \\ & 8 && \\ && -8 & \\ &&& -8 \end{pmatrix}.
\end{align}
We consider a gapped interface obtained by condensing a set of quasiparticles generated by $M = \{ \bm{m}_a \}$ with 
\begin{align}
\begin{split}
\bm{m}_1 &= (1,0,7,4)^T, \\
\bm{m}_2 &= (0,1,4,7)^T, 
\end{split}
\end{align}
both of which are bosonic quasiparticles with $s=-4$.
Hence, the corresponding gapped interface is bosonic. 
In order to construct a gapping potential, we add two extra bosonic wires at each interface and consider the extended $K$ matrix $K_\textsf{ew} = K_\textsf{e} \oplus K_\textsf{w}$ with $K_\textsf{w} = X \oplus X$. 
With the choice of matrices, 
\begin{align}
U = \begin{pmatrix} -2 & -2 & 0 & 1 \\ 1 & 0 & 2 & 2 \\ -2 & 2 & 0 & -1 \\ -1 & 0 & -2 & 2 \end{pmatrix}, \quad
W = \begin{pmatrix} 1 & 0 & 0 & 0 \\ 0 & 0 & 1 & 0 \\ 0 & 1 & 0 & 0 \\ 0 & 0 & 0 & 1 \end{pmatrix},
\end{align}
we find a set of integer vectors $\{ \bm{\Lambda}_{\textsf{w},\alpha} \}$, 
\begin{align}
\begin{split}
\bm{\Lambda}_{\textsf{w},1} &= (-2, 2, 1, 0)^T, \\
\bm{\Lambda}_{\textsf{w},2} &= (0, -1, 2, -2)^T, \\
\bm{\Lambda}_{\textsf{w},3} &= (2, 2, -1, 0)^T, \\
\bm{\Lambda}_{\textsf{w},4} &= (0, 1, 2, 2)^T.
\end{split}
\end{align}
By introducing bosonic fields $\bm{\phi}^\textsf{w}_{\bm{r}} = (\varphi^1_{\bm{r}}, 2\theta^1_{\bm{r}}, \varphi^2_{\bm{r}}, 2\theta^2_{\bm{r}})^T$ corresponding to two-component bosonic wires, which obey the commutation relations in Eq.~\eqref{eq:CommRel3DTwoCompBoson}, we find the tunneling Hamiltonian of the form \eqref{eq:TunnelingHam3DSq} with
\begin{align}
\begin{split}
\Phi^+_{\bm{r}, R} &= \varphi^1_{\bm{r}} +2\varphi^2_{\bm{r}} +4\theta^2_{\bm{r}}, \\
\Phi^-_{\bm{r}, R} &= 2\varphi^1_{\bm{r}} +4\theta^1_{\bm{r}} -2\theta^2_{\bm{r}}, \\
\Phi^+_{\bm{r}, L} &= -\varphi^1_{\bm{r}} -2\varphi^2_{\bm{r}} +4\theta^2_{\bm{r}}, \\
\Phi^-_{\bm{r}, L} &= 2\varphi^1_{\bm{r}} -4\theta^1_{\bm{r}} +2\theta^2_{\bm{r}}.
\end{split}
\end{align}

In this model, there are planon excitations in the $[010]$ and $[001]$ planes and also planons made of dipoles of lineon excitations moving in the same planes. 
The most elementary excitations created by local operators take the form of triplets of quasiparticles, such as $\bm{l} = (1,0,-1,4)^T$ or $(0,1,4,-1)^T$, as shown in Fig.~\ref{fig:ModelDipoleU8Sq}~(a). 
\begin{figure}
\includegraphics[clip,width=0.4\textwidth]{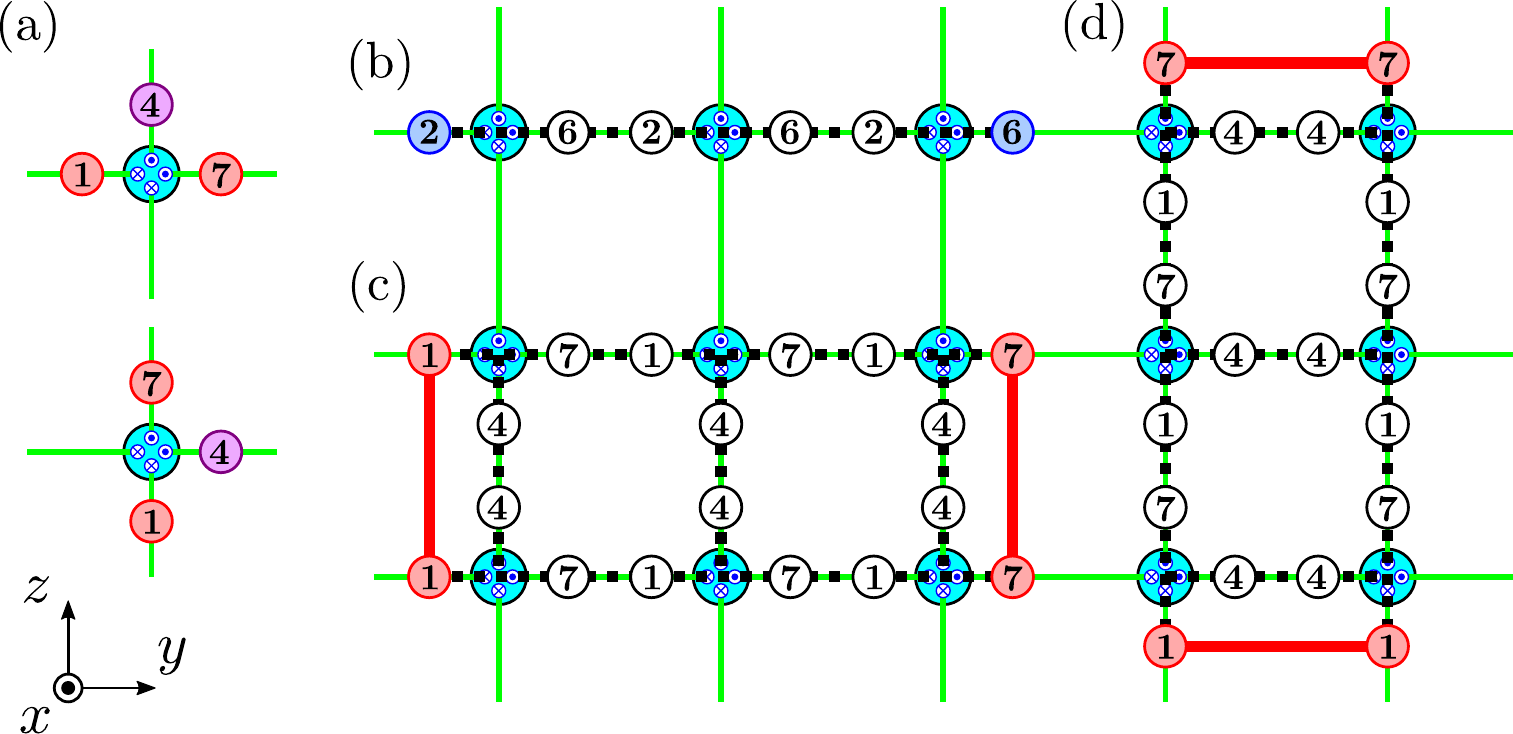}
\caption{(a) Elementary excitations created by local operators in the bosonic $U(1)_8$ model with lineons and planons. 
(b) Quasiparticle $\bm{2}$ or $\bm{6}$ can be transferred along the $y$ axis. 
Dipole of quasiparticles $\bm{1}$ or $\bm{7}$ can also be transferred along (c) the $y$ axis or (d) the $z$ axis.
}
\label{fig:ModelDipoleU8Sq}
\end{figure}
Fusion of two identical triplets leads to a pair of $\bm{2}$ and $\bm{6}$ quasiparticles with $s=\pm 1/4$, which are given by $\bm{l} = (2,0,-2,0)^T$ or $(0,2,0,-2)^T$ up to local bosonic excitations $\bm{8}$ and are created by $\exp(i\bm{p} \cdot \bm{\phi}^\textsf{w}_{\bm{r}})$ with $\bm{p} = (-1,0,0,0)^T$ or $(0,0,0,-1)^T$, respectively. 
By successively applying these operators, we can transfer $\bm{2}$ or $\bm{6}$ quasiparticles along the $y$ or $z$ axis [see Fig.~\ref{fig:ModelDipoleU8Sq}~(b)]. 
With the mobility along the $x$ axis, these quasiparticles become planons in the $[001]$ or $[010]$ plane. 
On each $[001]$ plane labeled by $z=1,\cdots,L_z$, we can find a string operator moving a $\bm{2}$ planon along the $y$ axis and that moving a $\bm{1}$ lineon along the $x$ axis, 
\begin{subequations}
\begin{align}
X^{[001]}_z &= \prod_{y=1}^{L_y} \exp \bigl[ -i\varphi^1_{(y,z)}(x_0) \bigr], \\
Z^{[001]}_z &= \exp \left[ -\frac{i}{8} \int_0^{L_x} dx \, \partial_x \Phi^-_{(y_0,z), L}(x) \right], 
\end{align}
\end{subequations}
for arbitrary $x_0$ and $y_0$ [see Fig.~\ref{fig:ModelDipoleU8SqString}~(a)], 
\begin{figure}
\includegraphics[clip,width=0.4\textwidth]{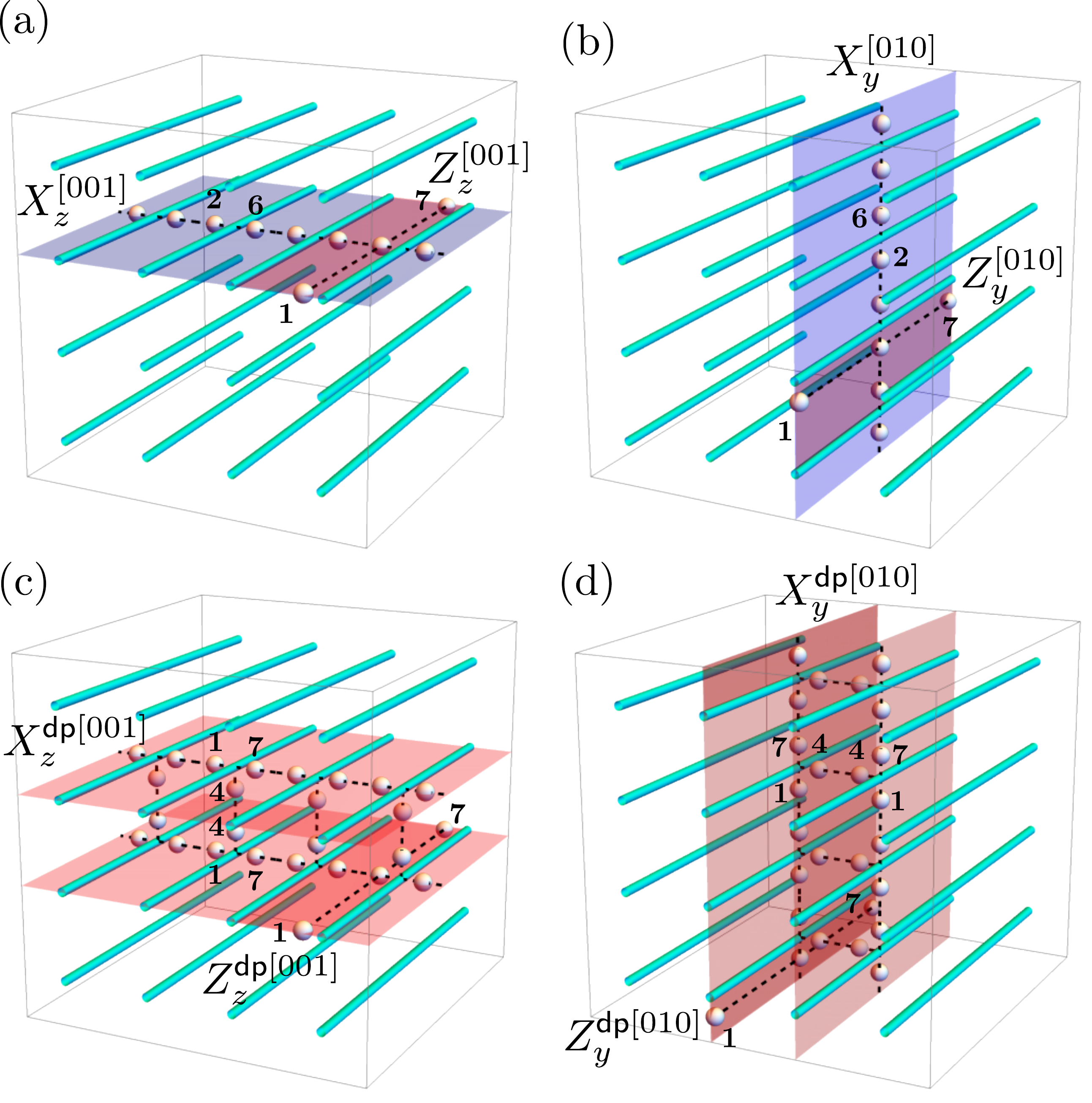}
\caption{Sets of string operators for the bosonic $U(1)_8$ model with lineons and planons on a torus. 
They are associated with planons in (a) the $[001]$ plane and (b) the $[010]$ plane and to dipole planons in (c) the $[001]$ plane and (d) the $[010]$ plane.
}
\label{fig:ModelDipoleU8SqString}
\end{figure}
while on each $[010]$ plane labeled by $y=1,\cdots,L_y$, we have a string operator moving a $\bm{2}$ planon along the $z$ axis and that moving a $\bm{1}$ lineon along the $x$ axis,
\begin{subequations}
\begin{align}
X^{[010]}_y &= \prod_{z=1}^{L_z} \exp \bigl[ -i2\theta^2_{(y,z)}(x_0) \bigr], \\
Z^{[010]}_y &= \exp \left[ -\frac{i}{8} \int_0^{L_x} dx \, \partial_x \Phi^+_{(y,z_0), L}(x) \right],
\end{align}
\end{subequations}
for arbitrary $x_0$ and $z_0$ [see Fig.~\ref{fig:ModelDipoleU8SqString}~(b)].
These string operators obey the generalized Pauli algebra, 
\begin{align}
Z^{[001]}_z X^{[001]}_z &= e^{\pi i/2} X^{[001]}_z Z^{[001]}_z, \\
Z^{[010]}_y X^{[010]}_y &= e^{\pi i/2} X^{[010]}_y Z^{[010]}_y.
\end{align}

Another type of planon is formed by dipoles of $\bm{1}$ or $\bm{7}$ quasiparticles with $s=\pm 1/16$. 
Let us consider two triplets of quasiparticles $\bm{l}_1 = (1,0,-1,4)^T$ and $\bm{l}_2 = (1,-4,-1,0)^T$, which are created by local operators $\exp(i\bm{p} \cdot \bm{\phi}^\textsf{w}_{\bm{r}})$ with $\bm{p}_1 = (0,0,1,1)^T$ and $\bm{p}_2 = (-1,0,-1,1)^T$, respectively. 
If we create $\bm{l}_1$ at $\bm{r}$ and $\bm{l}_2$ at $\bm{r}+\bm{e}_z$, $\bm{4}$ quasiparticles on the intermediate vertical strip are pair annihilated, leaving dipoles of $\bm{1}$ or $\bm{7}$ quasiparticles at adjacent parallel strips along the $y$ axis [see Fig.~\ref{fig:ModelDipoleU8Sq}~(c)]. 
With the mobility along the $x$ axis, these dipoles coherently move in the $[001]$ plane. 
However, a single $\bm{1}$ or $\bm{7}$ quasiparticle is still a lineon moving only along the $x$ axis. 
Therefore, we can construct two string operators associated with a $\bm{1}$-$\bm{1}$ dipole along the $y$ axis and a $\bm{1}$ lineon along the $x$ axis on each $[001]$ plane: 
\begin{subequations}
\begin{align}
X^{\textsf{dp}[001]}_z &= \prod_{y=1}^{L_y} \exp \Bigr[ i\bigl( \varphi^2_{(y,z)}(x_0) +2\theta^2_{(y,z)}(x_0) \nonumber \\
&\quad -\varphi^1_{(y,z+1)}(x_0) -\varphi^2_{(y,z+1)}(x_0) +2\theta^2_{(y,z+1)}(x_0) \bigr) \Bigr], \\
Z^{\textsf{dp}[001]}_z &= \exp \left[ -\frac{i}{8} \int_0^{L_x} dx \, \partial_x \Phi^-_{(y_0,z), L}(x) \right],
\end{align}
\end{subequations}
for arbitrary $x_0$ and $y_0$ [see Fig.~\ref{fig:ModelDipoleU8SqString}~(c)].
They obey the generalized Pauli algebra, 
\begin{align}
Z^{\textsf{dp}[001]}_z X^{\textsf{dp}[001]}_z = e^{\pi i/4} X^{\textsf{dp}[001]}_z Z^{\textsf{dp}[001]}_z.
\end{align}
Similarly, we can construct a pair of $\bm{1}$-$\bm{1}$ dipoles moving along the $z$ axis by binding $\bm{l}_3 = (0,1,4,-1)^T$ at $\bm{r}$ and $\bm{l}_2 = (-4,1,0,-1)^T$ at $\bm{r}+\bm{e}_y$ [see Fig.~\ref{fig:ModelDipoleU8Sq}~(d)], which then moves in the $[010]$ plane when combined with the mobility along the $x$ axis. 
We can then find two string operators on each $[010]$ plane, 
\begin{subequations}
\begin{align}
X^{\textsf{dp}[010]}_y &= \prod_{y=1}^{L_y} \exp \Bigr[ i\bigl( \varphi^1_{(y,z)}(x_0) +2\theta^1_{(y,z)}(x_0) \nonumber \\
&\quad -2\theta^2_{(y,z)}(x_0) +\varphi^1_{(y+1,z)}(x_0) -2\theta^1_{(y+1,z)}(x_0) \bigr) \Bigr], \\
Z^{\textsf{dp}[010]}_y &= \exp \left[ -\frac{i}{8} \int_0^{L_x} dx \, \partial_x \Phi^+_{(y,z_0), L}(x) \right], 
\end{align}
\end{subequations}
for arbitrary $x_0$ and $z_0$ [see Fig.~\ref{fig:ModelDipoleU8SqString}~(d)], which also obey the generalized Pauli algebra, 
\begin{align}
Z^{\textsf{dp}[010]}_y X^{\textsf{dp}[010]}_y = e^{\pi i/4} X^{\textsf{dp}[010]}_y Z^{\textsf{dp}[010]}_y.
\end{align}

Here, we note that these string operators are not linearly independent from each other and do not form mutually commuting pairs of the (generalized) Pauli operators for the same reason as in the fermionic $U(1)_4$ model in Sec.~\ref{sec:DipoleU4Model}. 
If we suppose that they are all independent and form commuting pairs, they could span a Hilbert space of dimension $4^{L_y+L_z} \cdot 8^{L_y+L_z}$. 
However, this is not the case and they are subject to the constraints, 
\begin{subequations}
\begin{align}
(X^{\textsf{dp}[001]}_z)^2 &\sim X^{[001]}_z X^{[001]}_{z+1}, \\
(X^{\textsf{dp}[010]}_y)^2 &\sim X^{[010]}_y X^{[010]}_{y+1}, \\
\prod_{z=1}^{L_z} X^{\textsf{dp}[001]}_z &\sim \prod_{z=1}^{L_z} X^{[001]}_z \prod_{y=1}^{L_y} (X^{[010]}_y)^2, \\
\prod_{y=1}^{L_y} X^{\textsf{dp}[010]}_y &\sim \prod_{y=1}^{L_y} X^{[010]}_y \prod_{z=1}^{L_z} (X^{[001]}_z)^2.
\end{align}
\end{subequations}
We note that these identifications are valid only in the subspace of degenerate ground states and hold up to multiplications of operators creating or annihilating pairs of local bosonic excitations $\bm{8}$, which trivially act on the ground state. 
The first two constraints imply that two dipoles of $\bm{1}$ quasiparticles in the same plane fuse into two planons of $\bm{2}$ quasiparticles in the adjacent planes and reduce the dimension by factor of $4^{L_y+L_z}$. 
The last two constraints imply that only $L_y+L_z-2$ dipole string operators are independent and further reduce the dimension by factor of $2^2$. 
Overall, these constraints reduce the dimension of the Hilbert space spanned by the string operators by factor of $4^{L_y+L_z+1}$, leading to the ground-state degeneracy,
\begin{align}
\textrm{GSD} = 4^2 \cdot 8^{L_y+L_z-2}.
\end{align}
This violates a strict subextensivity of $\log \textrm{GSD}$ by a negative additive constant and proves that the present model is a nontrivial foliated fracton model, which cannot be understood as decoupled stacks of 2D topological orders.
Fully commuting pairs of the generalized Pauli operators spanning the ground-state manifold are given, for example, by $X^{\textsf{dp}[001]}_z$ and $Z^{\textsf{dp}[001]}_z$ with $z=1,\cdots,L_z-1$, $X^{\textsf{dp}[010]}_y$ and $Z^{\textsf{dp}[010]}_y$ with $y=1,\cdots,L_y-1$, $X^{[001]}_{L_z}$ and $Z^{[001]}_{L_z}$, and $X^{[010]}_{L_y}$ and $Z^{[010]}_{L_y}$.

\subsection{Fermionic $U(1)_9$ model}
\label{sec:FermionicU9Model}

We here provide another model for 3D TQFT-type topological order, which can be seen as a 3D fermionic $Z_3$ gauge theory with a $Z_3$ charge and a $Z_3$ flux loop obeying nontrivial braiding statistics.
We consider a 3D cellular topological state built out of the $U(1)_9$ topological orders, each of which is described by the $K$ matrix $K_0=9$. 
On the square grid, each interface possesses four gapless edge modes corresponding to the $K$ matrix, 
\begin{align}
K_\textsf{e} = \begin{pmatrix} 9 &&& \\ & 9 && \\ && -9 & \\ &&& -9 \end{pmatrix}.
\end{align}
We consider a gapped interface obtained by condensing a set of quasiparticles generated by $M = \{ \bm{m}_a \}$ with 
\begin{align}
\begin{split}
\bm{m}_1 &= (1,1,8,8)^T, \\
\bm{m}_2 &= (3,6,0,0)^T, \\
\bm{m}_3 &= (3,0,6,0)^T,
\end{split}
\end{align}
where $\bm{m}_1$ is a bosonic quasiparticle with $s=-7$ whereas $\bm{m}_2$ and $\bm{m}_3$ are fermionic quasiparticles with $s=5/2$ and $-3/2$, respectively.
Since $U(1)_9$ is a fermionic topological order, the corresponding gapped interface is fermionic. 
In order to construct a gapping potential, we add two extra fermionic wires at each interface and consider the extended $K$ matrix $K_\textsf{ew} = K_\textsf{e} \oplus K_\textsf{w}$ with $K_\textsf{w} = Z \oplus Z$. 
With the choice of matrices, 
\begin{align}
U = \begin{pmatrix} 1 & -1 & 2 & 2 \\ -1 & 2 & -1 & 2 \\ 1 & -2 & -2 & 1 \\ 1 & 1 & 1 & 4 \end{pmatrix}, \quad
W = \begin{pmatrix} 1 & 1 & 1 & 0 \\ 1 & 0 & 1 & 0 \\ 0 & 1 & 1 & 0 \\ -1 & -1 & -1 & 1 \end{pmatrix},
\end{align}
we find a set of integer vectors $\{ \bm{\Lambda}_{\textsf{w},\alpha} \}$, 
\begin{align}
\begin{split}
\bm{\Lambda}_{\textsf{w},1} &= (2, 4, -2, 1)^T, \\
\bm{\Lambda}_{\textsf{w},2} &= (2, 1, -2, 4)^T, \\
\bm{\Lambda}_{\textsf{w},3} &= (4, 2, -1, 2)^T, \\
\bm{\Lambda}_{\textsf{w},4} &= (1, 2, -4, 2)^T.
\end{split}
\end{align}
By introducing bosonic fields $\bm{\phi}^\textsf{w}_{\bm{r}} = (\phi^1_{\bm{r},R}, \phi^1_{\bm{r},L}, \phi^2_{\bm{r},R}, \phi^2_{\bm{r},L})^T$ corresponding to two-component fermionic wires, which obey the commutation relations in Eq.~\eqref{eq:CommRel3DTwoCompField}, we find the tunneling Hamiltonian of the form \eqref{eq:TunnelingHam3DSq} with
\begin{align}
\begin{split}
\Phi^+_{\bm{r}, R} &= \phi^1_{\bm{r}, R} -2\phi^1_{\bm{r}, L} -4\phi^2_{\bm{r}, R} -2\phi^2_{\bm{r}, L}, \\
\Phi^-_{\bm{r}, R} &= 4\phi^1_{\bm{r}, R} -2\phi^1_{\bm{r}, L} -\phi^2_{\bm{r}, R} -2\phi^2_{\bm{r}, L}, \\
\Phi^+_{\bm{r}, L} &= 2\phi^1_{\bm{r}, R} -\phi^1_{\bm{r}, L} -2\phi^2_{\bm{r}, R} -4\phi^2_{\bm{r}, L}, \\
\Phi^-_{\bm{r}, L} &= 2\phi^1_{\bm{r}, R} -4\phi^1_{\bm{r}, L} -2\phi^2_{\bm{r}, R} -\phi^2_{\bm{r}, L}.
\end{split}
\end{align}

This model hosts a 3D point-like excitation and a loop-like excitation with the mutual $2\pi/3$ statistics. 
Since the present model microscopically consists of fermions, distinction between fermion or boson statistics for the point-like excitation is immaterial as we can add a physical fermion to change its statistics. 
Elementary excitations created by local operators are given by the subset $M$ of the Lagrangian subgroup and are shown in Fig.~\ref{fig:ModelFermionU9Sq}~(a). 
\begin{figure}
\includegraphics[clip,width=0.4\textwidth]{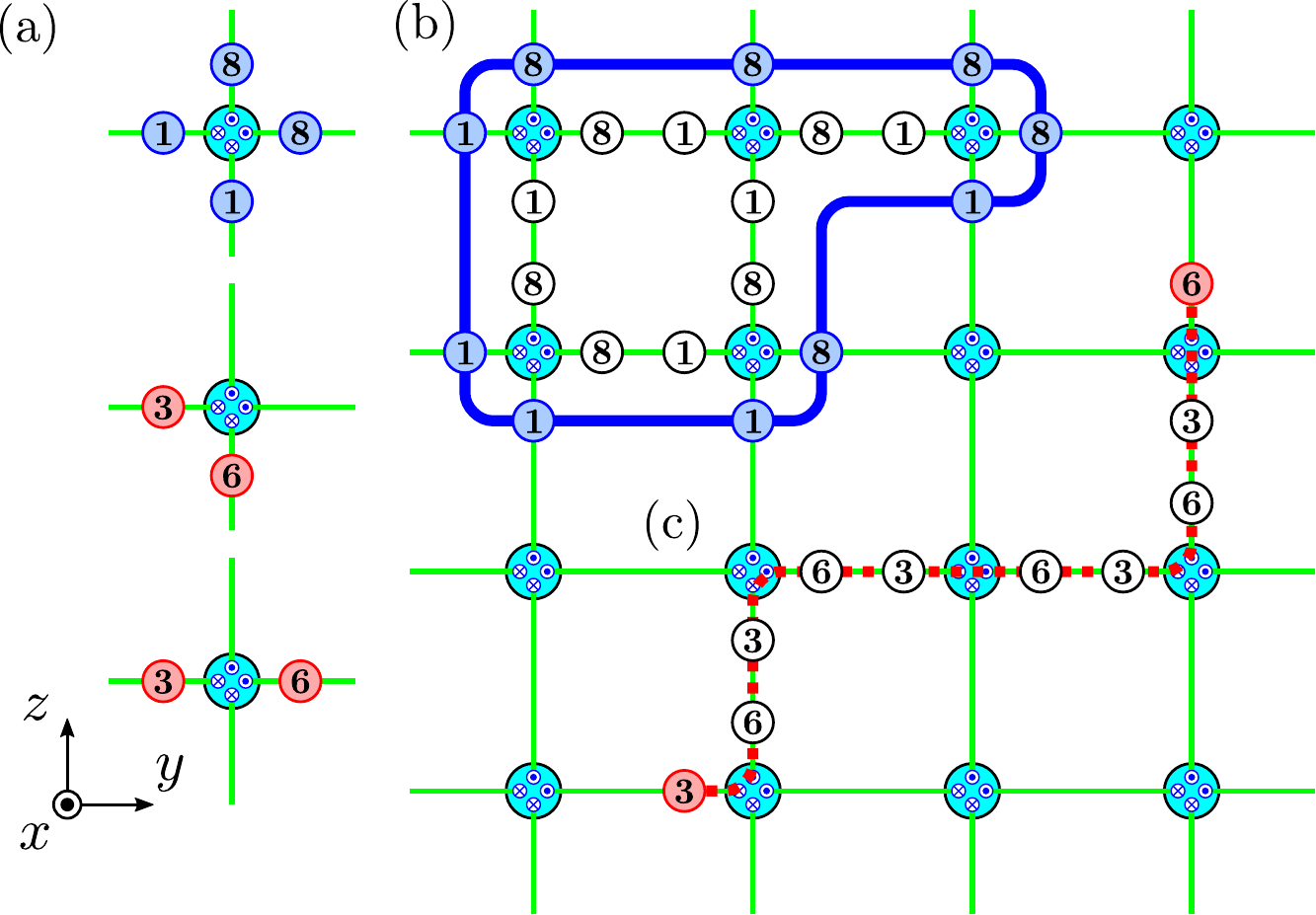}
\caption{(a) Elementary excitations created by local operators in the fermionic $U(1)_9$ model. 
(b) Quasiparticles $\bm{1}$ and $\bm{8}$ form a loop-like excitation in the $yz$ plane, while (c) quasiparticles $\bm{3}$ and $\bm{6}$ can freely move in the $yz$ plane.
}
\label{fig:ModelFermionU9Sq}
\end{figure}
One of the most elementary excitations is $\bm{l} = (1,1,-1,-1)^T$, which is composed of $\bm{1}$ and $\bm{8}$ quasiparticles with $s=1/18$ and $-1/18$, respectively, over all strips surrounding the interface $\bm{r}$ and is created by a local operator $\exp(i\bm{p} \cdot \bm{\phi}^\textsf{w}_{\bm{r}})$ with $\bm{p} = (-1,1,1,1)^T$. 
While single $\bm{1}$ or $\bm{8}$ quasiparticles only behave as a lineon along the $x$ axis, they can collectively form a loop-like excitation in the $yz$ plane by acting these local operators on the interior of the loop [see Fig.~\ref{fig:ModelFermionU9Sq}~(b)]. 
On an $L_x \times L_y \times L_z$ torus, we can find a pair of a membrane operator in the $yz$ plane and a rigid string operator along the $x$ axis, 
\begin{subequations} \label{eq:MembraneU9}
\begin{align}
\label{eq:MembraneOpU9}
X^\textsf{mem} &= \prod_{\bm{r} \in \mathbb{Z}_{L_y} \times \mathbb{Z}_{L_z}} \exp \Bigl[ -i\bigl( \phi^1_{\bm{r}, R}(x_0) -\phi^1_{\bm{r}, L}(x_0) \nonumber \\
&\quad -\phi^2_{\bm{r}, R}(x_0) -\phi^2_{\bm{r}, L}(x_0) \bigr) \Bigr], \\
Z^\textsf{mem} &= \exp \left[ -\frac{i}{9} \int_0^{L_x} dx \, \partial_x \Phi^-_{\bm{r}_0, L}(x) \right],
\end{align}
\end{subequations}
for arbitrary $x_0$ and $\bm{r}_0$ [see Fig.~\ref{fig:ModelFermionU9SqString}~(a)]. 
\begin{figure}
\includegraphics[clip,width=0.4\textwidth]{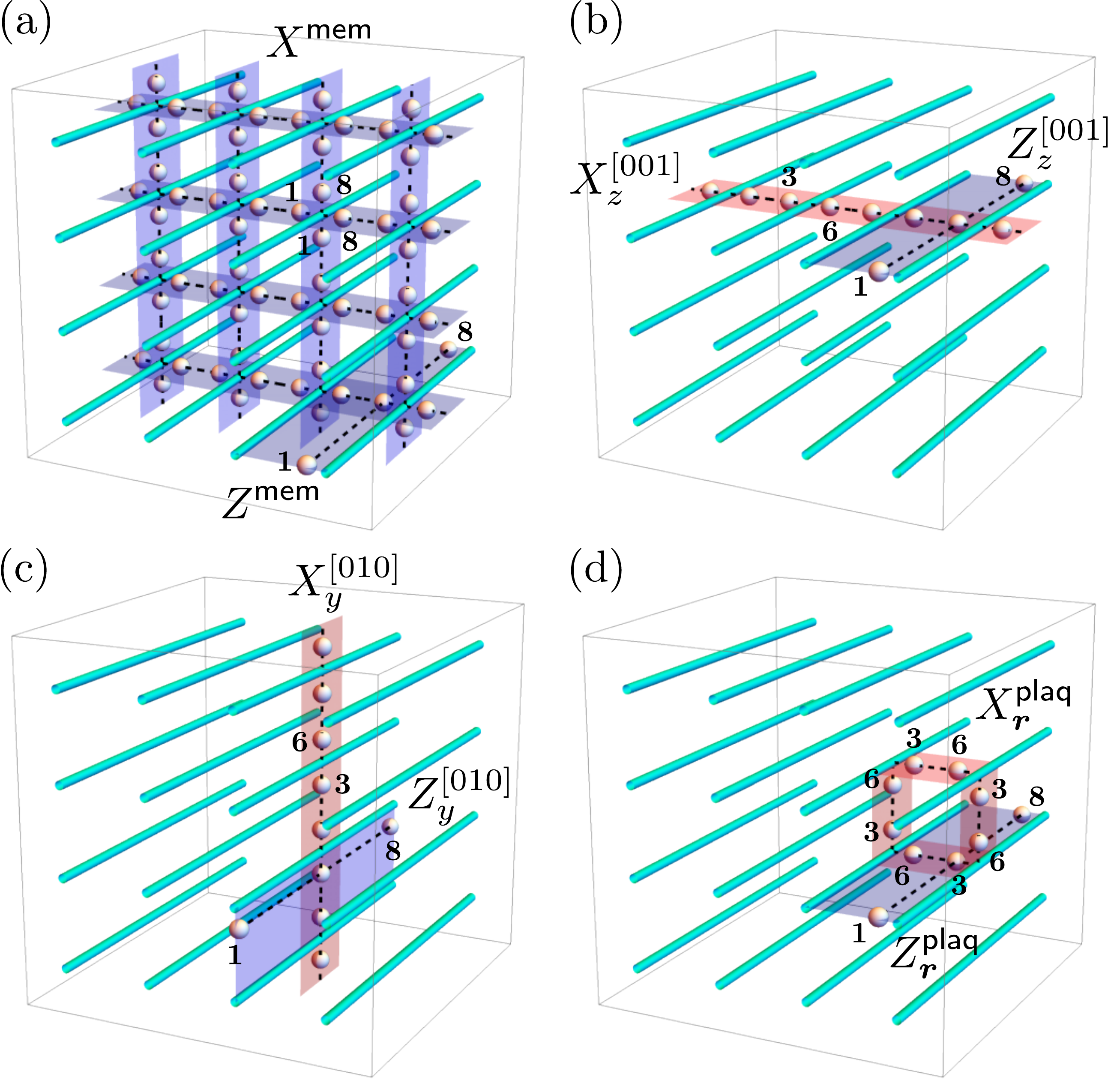}
\caption{Sets of membrane and string operators for the fermionic $U(1)_9$ model on a torus. 
They are associated with (a) loop-like excitations of quasiparticles $\bm{1}$ and $\bm{8}$ in the $yz$ plane and point-like excitations of quasiparticles $\bm{3}$ and $\bm{6}$ in (b) the $[001]$ plane and (c) the $[010]$ plane and (d) along plaquette loops.
}
\label{fig:ModelFermionU9SqString}
\end{figure}
They obey the generalized Pauli algebra, 
\begin{align}
Z^\textsf{mem} X^\textsf{mem} = e^{2\pi i/9} X^\textsf{mem} Z^\textsf{mem}.
\end{align}

The other types of excitations are pairs of $\bm{3}$ and $\bm{6}$ quasiparticles, which have $s=1/2$ and $-1/2$ and thus are fermions. 
However, we note that their fermionic statistics can be transmuted to the bosonic one by adding a local fermionic excitation. 
The corresponding excitations, for example, $\bm{l} = (3,0,-3,0)^T$ and $(3,-3,0,0)^T$ can be created by local operators $\exp(i\bm{p} \cdot \bm{\phi}^\textsf{w}_{\bm{r}})$ with $\bm{p}=(-2,2,1,1)^T$ and $(0,1,0,-1)^T$, respectively. 
These fermionic quasiparticles can be transferred in arbitrary directions within the $yz$ plane [see Fig.~\ref{fig:ModelFermionU9Sq}~(c)]. 
With the mobility along the $x$ axis, the $\bm{3}$ or $\bm{6}$ quasiparticles behave as point-like excitations fully mobile in the 3D space. 
As in the bosonic $U(1)_4$ model discussed in Sec.~\ref{sec:BosonicU4Model}, we may then construct three kinds of string operators. 
On each $[001]$ plane labeled by $z=1,\cdots,L_z$, we find a string operator moving a fermionic quasiparticle along the $y$ axis and that moving a $\bm{1}$ lineon along the $x$ axis, 
\begin{subequations} \label{eq:StringYU9}
\begin{align}
X^{[001]}_z &= \prod_{y=1}^{L_y} \exp \Bigl[ -i\bigl( 2\phi^1_{(y,z), R}(x_0) -2\phi^1_{(y,z), L}(x_0) \nonumber \\
&\quad -\phi^2_{(y,z), R}(x_0) -\phi^2_{(y,z), L}(x_0) \bigr) \Bigr], \\
Z^{[001]}_z &= \exp \left[ -\frac{i}{9} \int_0^{L_x} dx \, \partial_x \Phi^-_{(y_0,z), L}(x) \right],
\end{align}
\end{subequations}
for arbitrary $x_0$ and $y_0$. 
On each $[010]$ labeled by $y=1,\cdots,L_y$, we find a string operator moving a fermionic quasiparticle along the $z$ axis and that moving a $\bm{1}$ lineon along the $x$ axis, 
\begin{subequations} \label{eq:StringZU9}
\begin{align}
X^{[010]}_y &= \prod_{z=1}^{L_z} \exp \Bigl[ -i\bigl( \phi^1_{(y,z), R}(x_0) -\phi^1_{(y,z), L}(x_0) \nonumber \\
&\quad -2\phi^2_{(y,z), R}(x_0) -2\phi^2_{(y,z), L}(x_0) \bigr) \Bigr], \\
Z^{[010]}_y &= \exp \left[ -\frac{i}{9} \int_0^{L_x} dx \, \partial_x \Phi^+_{(y,z_0), L}(x) \right],
\end{align}
\end{subequations}
for arbitrary $x_0$ and $z_0$. 
On each square plaquette labeled by its left bottom corner $\bm{r}$, we find a string operator moving a fermionic $\bm{3}$ quasiparticle along the plaquette loop and that moving a $\bm{1}$ lineon along the $x$ axis,
\begin{subequations}
\begin{align}
\label{eq:PlaqOpU9}
X^\textsf{plaq}_{\bm{r}} &= \exp \Bigl[ -i\bigl( \phi^1_{\bm{r}, R}(x_0) +\phi^2_{\bm{r}, R}(x_0) +\phi^1_{\bm{r}+\bm{e}_y, R}(x_0) \nonumber \\
&\quad -2\phi^1_{\bm{r}+\bm{e}_y, L}(x_0) -2\phi^2_{\bm{r}+\bm{e}_y, R}(x_0) -\phi^2_{\bm{r}+\bm{e}_y, L}(x_0) \nonumber \\
&\quad +\phi^1_{\bm{r}+\bm{e}_y+\bm{e}_z, L}(x_0) -\phi^2_{\bm{r}+\bm{e}_y+\bm{e}_z, L}(x_0) \nonumber \\
&\quad -2\phi^1_{\bm{r}+\bm{e}_z, R}(x_0) +\phi^1_{\bm{r}+\bm{e}_z, L}(x_0) \nonumber \\
&\quad +\phi^2_{\bm{r}+\bm{e}_z, R}(x_0) +2\phi^2_{\bm{r}+\bm{e}_z, L}(x_0) \bigr) \Bigr], \\
Z^\textsf{plaq}_{\bm{r}} &= \exp \left[ -\frac{i}{9} \int_0^{L_x} dx \, \partial_x \Phi^-_{\bm{r}+\bm{e}_y, L}(x) \right],
\end{align}
\end{subequations}
for arbitrary $x_0$. 
These string operators obey the generalized Pauli algebra, 
\begin{align}
Z^{[001]}_z X^{[001]}_z &= e^{2\pi i/3}X^{[001]}_z Z^{[001]}_z, \\
Z^{[010]}_y X^{[010]}_y &= e^{2\pi i/3} X^{[010]}_y Z^{[010]}_y, \\
\label{eq:CommPlaqU9}
Z^{\textsf{plaq}}_{\bm{r}} X^{\textsf{plaq}}_{\bm{r}} &= e^{2\pi i/3} X^{\textsf{plaq}}_{\bm{r}} Z^{\textsf{plaq}}_{\bm{r}}.
\end{align}

If we suppose that all these membrane and string operators are independent from each other and form commuting pairs of the generalized Pauli operators, they could span a Hilbert space of the dimension $9 \cdot 3^{L_y +L_z +L_y L_z}$. 
However, it is impossible to find such mutually commuting pairs since the membrane and string operators in the $yz$ plane are not independent but are subject to the constraints, 
\begin{subequations} \label{eq:ConstraintsU9}
\begin{align}
(X^\textsf{mem})^3 &\sim \prod_{y=1}^{L_y} X^{[010]}_y \prod_{z=1}^{L_z} X^{[001]}_z, \\
X^{[001]}_z X^{[001]}_{z+1} &\sim \prod_{y=1}^{L_y} X^\textsf{plaq}_{(y,z)}, \\
X^{[010]}_y X^{[010]}_{y+1} &\sim \prod_{z=1}^{L_z} X^\textsf{plaq}_{(y,z)}, \\
\prod_{\bm{r} \in \mathbb{Z}_{L_y} \times \mathbb{Z}_{L_z}} X^\textsf{plaq}_{\bm{r}} &\sim 1.
\end{align}
\end{subequations}
These identifications hold up to multiplications of operators creating local fermionic excitations $\bm{9}$ and thus are valid only in the subspace of degenerate ground states. 
These constraints reduce the dimension of the Hilbert space by factor of $3^2 \cdot 3^{L_y+L_z-2}$, leading to the ground-state degeneracy, 
\begin{align}
\textrm{GSD} = 9 \cdot 3^{L_y L_z}.
\end{align}

As discussed for the bosonic $U(1)_4$ model in Sec.~\ref{sec:BosonicU4Model}, the degeneracy growing with the number of quantum wires or interfaces is a consequence of local fermion loop excitations on square plaquettes. 
It can be lifted by adding the local plaquette loop operators defined in Eq.~\eqref{eq:PlaqOpU9} as a perturbation of the form \eqref{eq:PerturbationU4}. 
The perturbation induces the condensation of the local fermionic loops and selects simultaneous eigenstates of the fermion loop operators $X^\textsf{plaq}_{\bm{r}}$ with the eigenvalues $+1$ as a new ground state. 
The subspace of the new ground state is spanned by the membrane and string operators in Eqs.~\eqref{eq:MembraneU9}, \eqref{eq:StringYU9}, and \eqref{eq:StringZU9}, which are now subject to the constraints,
\begin{subequations}
\begin{align}
(X^\textsf{mem})^3 &\sim \prod_{y=1}^{L_y} X^{[010]}_y \prod_{z=1}^{L_z} X^{[001]}_z, \\
X^{[001]}_z X^{[001]}_{z+1} &\sim 1, \\
X^{[010]}_y X^{[010]}_{y+1} &\sim 1.
\end{align}
\end{subequations}
We thus find that after the condensation of local fermionic loops, the ground-state degeneracy becomes a constant: 
\begin{align}
\textrm{GSD}' = 3^3.
\end{align}

This implies that the ground states is in a 3D fermionic TQFT-type topological order. 
From the noncommutativity between local plaquette loop operators and lineon string operators as given in Eq.~\eqref{eq:CommPlaqU9}, single lineon excitations of $\bm{1}$ or $\bm{8}$ along the $x$ axis are confined in the condensate. 
As the cubes of lineon operators commute with all plaquette loop operators, single $\bm{3}$ or $\bm{6}$ quasiparticles obtained by fusing three elementary lineons on the same strip behave as deconfined lineons. 
With the mobility in the $yz$ plane, they behave as 3D point-like excitations with a $Z_3$ charge. 
Another type of deconfined excitations along the $x$ axis can be constructed by taking appropriate products of lineon string operators on different strips. 
For each square plaquette, products of two lineon string operators moving two $\bm{1}$ lineons coherently between parallel strips or those moving a pair of $\bm{1}$ and $\bm{8}$ lineons between orthogonal strips commute with the plaquette loop operator $X^\textsf{plaq}_{\bm{r}}$. 
When such products of lineon operators form a closed membrane, they commute with all plaquette operators. 
Combining finite fractions of the membrane operator given in Eq.~\eqref{eq:MembraneOpU9} for the $yz$ plane and finite segments of the products of lineon operators along the $x$ axis, we can construct arbitrary shapes of closed membrane operators. 
An open membrane creates loop-like excitations whose energy is proportional to the length of its boundary. 
As these loop-like excitations consist of $\bm{1}$ and $\bm{8}$ quasiparticles, they have the mutual $2\pi/3$ statistics with 3D point-like excitations originated from $\bm{3}$ or $\bm{6}$ quasiparticles [see Fig.~\ref{fig:ModelFermionU9SqBraid}~(a)]. 
\begin{figure}
\includegraphics[clip,width=0.4\textwidth]{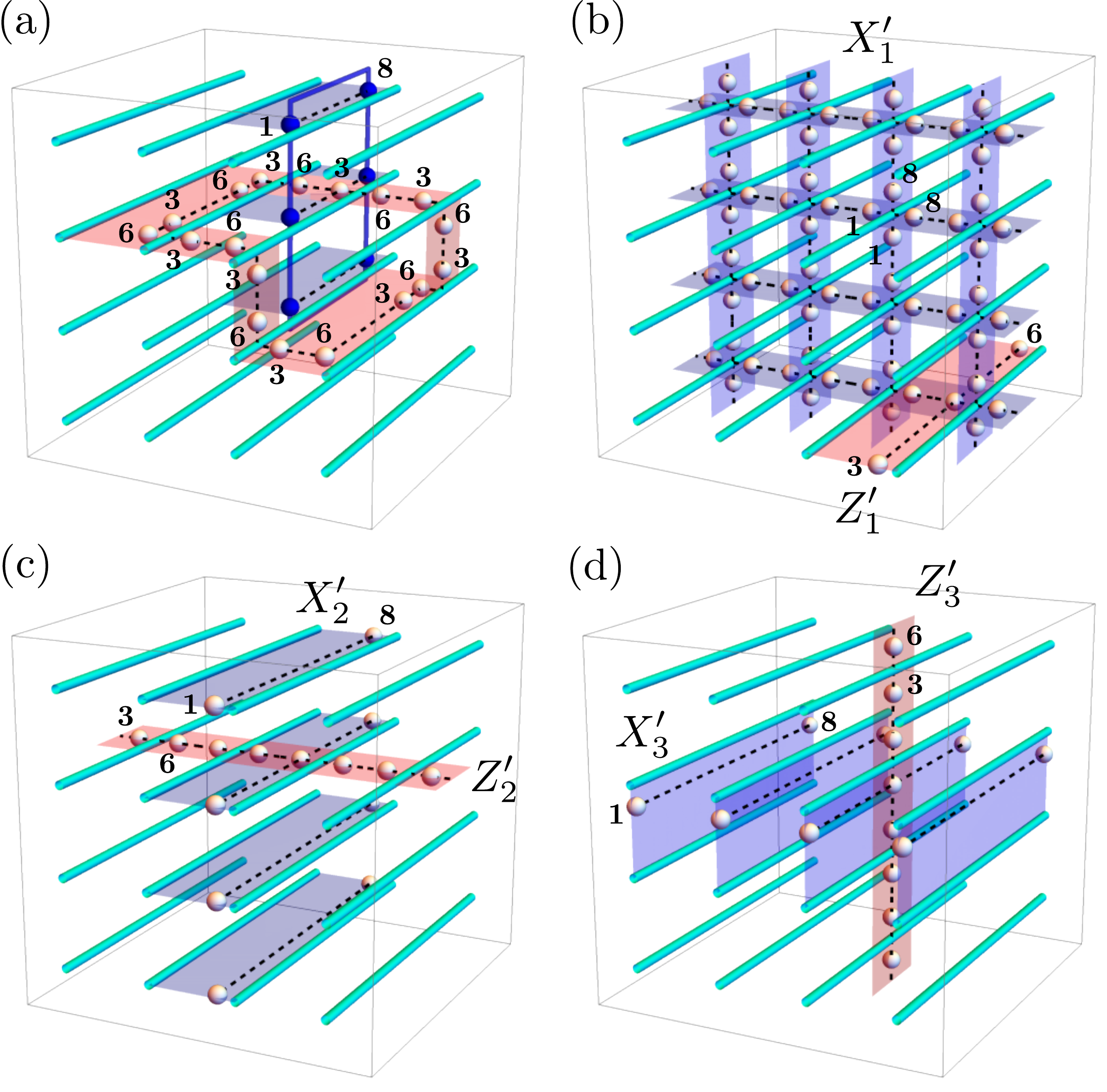}
\caption{(a) Braiding process for the fermionic $U(1)_9$ model between a point-like fermionic excitation and a loop-like excitation after the condensation of local fermionic loops, yielding the mutual $2\pi/3$ statistics. 
There are three membrane operators on a torus in the (b) $yz$, (c) $xz$, and (d) $xy$ planes and the associated string operators.
}
\label{fig:ModelFermionU9SqBraid}
\end{figure}
Thus, our model realizes a fermionic TQFT-type topological order of a 3D $Z_3$ gauge theory after the condensation of local loop excitations. 

On a torus, we can find three membrane operators in the $yz$, $xz$, and $xy$ planes, 
\begin{subequations}
\begin{align}
X'_1 &= \prod_{\bm{r} \in \mathbb{Z}_{L_y} \times \mathbb{Z}_{L_z}} \exp \Bigl[ -i\bigr( \phi^1_{\bm{r}, R}(x_0) -\phi^1_{\bm{r}, L}(x_0) \nonumber \\
&\quad -\phi^2_{\bm{r}, R}(x_0) -\phi^2_{\bm{r}, L}(x_0) \bigl) \Bigr], \\
X'_2 &= \prod_{z=1}^{L_z} \exp \left[ -\frac{i}{9} \int_0^{L_x} dx \, \partial_x \Phi^-_{(y_0,z), L}(x) \right], \\
X'_3 &= \prod_{y=1}^{L_y} \exp \left[ -\frac{i}{9} \int_0^{L_x} dx \, \partial_x \Phi^+_{(y,z_0), L}(x) \right], 
\end{align}
\end{subequations}
respectively, and three fermion string operators along in the $x$, $y$, and $z$ axes, 
\begin{subequations}
\begin{align}
Z'_1 &= \exp \left[ -\frac{i}{3} \int_0^{L_x} dx \, \partial_x \Phi^-_{(y_0,z_0), L}(x) \right], \\
Z'_2 &= \prod_{y=1}^{L_y} \exp \Bigl[ -i\bigl( 2\phi^1_{(y,z_0), R}(x_0) -2\phi^1_{(y,z_0), L}(x_0) \nonumber \\
&\quad -\phi^2_{(y,z_0), R}(x_0) -\phi^2_{(y,z_0), L}(x_0) \bigr) \Bigr], \\
Z'_3 &= \prod_{z=1}^{L_z} \exp \Bigl[ -i\bigl( \phi^1_{(y_0,z), R}(x_0) -\phi^1_{(y_0,z), L}(x_0) \nonumber \\
&\quad -2\phi^2_{(y_0,z), R}(x_0) -2\phi^2_{(y_0,z), L}(x_0) \bigr) \Bigr],
\end{align}
\end{subequations}
respectively, where the choice of $x_0$, $y_0$, and $z_0$ is arbitrary in each expression. 
They are illustrated in Fig.~\ref{fig:ModelFermionU9SqBraid}~(b)--(d). 
These operators obey the generalized Pauli algebra, 
\begin{align}
Z'_j X'_j = e^{2\pi i/3} X'_j Z'_j,
\end{align}
for $j=1,2,3$. 
These membrane and string operators form mutually commuting pairs of the generalized Pauli operators and thus span the $3^3$-dimensional ground-state manifold on a torus as expected for a 3D $Z_3$ gauge theory. 

\section{Equivalent honeycomb grid models}
\label{sec:CellularTopoHoneycomb}

Some of cellular topological states constructed in Sec.~\ref{sec:CellularTopo3D} on the square grid can also be defined on the honeycomb grid with preserving quasiparticle properties. 
Such cellular topological states are the fermionic $U(1)_2$ model in Sec.~\ref{sec:FermionicU2Model} and the fermionic $U(1)_6$ model in Sec.~\ref{sec:FermionicU6Model} for foliated type-I fracton order with only planons, the bosonic $U(1)_4$ model in Sec.~\ref{sec:BosonicU4Model} for TQFT-type topological order, and the bosonic $U(1)_8$ model in Sec.~\ref{sec:HybridU8Model} for hybrid of fracton and topological orders. 
A common feature of these models is that their constituent gapped interfaces are related to the conformal embedding \cite{dFMS} for free boson CFTs, 
\begin{align} \label{eq:EmbeddingUkU2k}
U(1)_k \times U(1)_k \supset U(1)_{2k} \times U(1)_{2k}.
\end{align}
While the conformal embedding has been fully utilized in Ref.~\cite{Fuji19b} for constructing 3D coupled-wire models, we here discuss their explicit connection with cellular topological states. 

Let us consider a strip of the $U(1)_k \times U(1)_k$ topological order sandwiched by two strips of the $U(1)_{2k} \times U(1)_{2k}$ topological orders as depicted in Fig.~\ref{fig:CellularTopo3DHoney}~(a). 
\begin{figure}
\includegraphics[clip,width=0.45\textwidth]{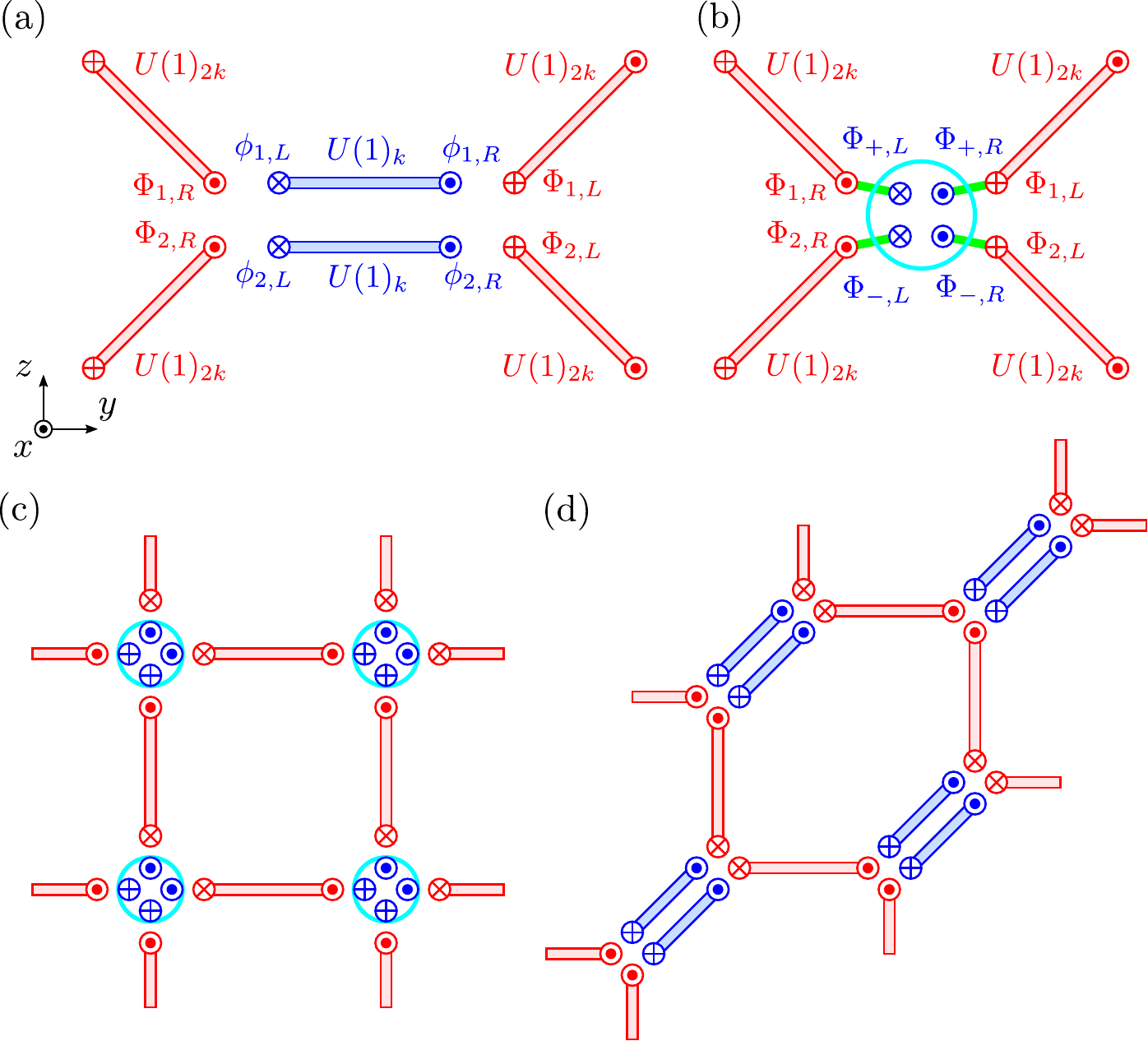}
\caption{(a) Interface between four strips of the $U(1)_{2k}$ topological order separated by a strip of the $U(1)_k \times U(1)_k$ topological order. 
(b) Shrinking the middle strip yields a single gapped interface with an additional quantum wire. 
Using such gapped interfaces, some cellular topological states on the square grid (c) can be deformed to those on the honeycomb grid (d).}
\label{fig:CellularTopo3DHoney}
\end{figure}
At the left interface, we have two left-moving edge modes $\phi_{1,L}$ and $\phi_{2,L}$ from the middle strip and two right-moving edge modes $\Phi_{1,R}$ and $\Phi_{2,R}$ from the leftmost strip. 
At the right interface, we have two right-moving edge modes $\phi_{1,R}$ and $\phi_{2,R}$ from the middle strip and two left-moving edge modes $\Phi_{1,L}$ and $\Phi_{2,L}$ from the rightmost strip. 
We assume that $\phi_{1,R}$ and $\phi_{1,L}$ belong to the same layer and $\phi_{2,R}$ and $\phi_{2,L}$ too.
Collectively denoting the bosonic fields as $\bm{\phi}^\textsf{e}_\textsf{A} = (\Phi_{1,R}, \Phi_{2,R}, \phi_{1,L}, \phi_{2,L})^T$ and $\bm{\phi}^\textsf{e}_\textsf{B} = (\phi_{1,R}, \phi_{2,R}, \Phi_{2,L}, \Phi_{1,L})^T$, 
we suppose that they obey the commutation relations, 
\begin{subequations}
\begin{align}
[\partial_x \phi^\textsf{e}_{\textsf{A},\alpha}(x), \phi^\textsf{e}_{\textsf{A},\beta}(x')] &= 2\pi i (K_\textsf{e}^\textsf{A})_{\alpha \beta} \delta(x-x'), \\
[\partial_x \phi^\textsf{e}_{\textsf{B},\alpha}(x), \phi^\textsf{e}_{\textsf{B},\beta}(x')] &= 2\pi i (K_\textsf{e}^\textsf{B})_{\alpha \beta} \delta(x-x'),
\end{align}
\end{subequations}
with the $K$ matrices, 
\begin{subequations}
\begin{align}
K_\textsf{e}^\textsf{A} &= \begin{pmatrix} 2k &&& \\ & 2k && \\ && -k & \\ &&& -k \end{pmatrix}, \\
K_\textsf{e}^\textsf{B} &= \begin{pmatrix} k &&& \\ & k && \\ && -2k & \\ &&& -2k \end{pmatrix}.
\end{align}
\end{subequations}
We can then define two $U(1)_k$ currents by $j_{\sigma,R} = e^{i\phi_{\sigma,R}}$ with conformal spin $s=k/2$, two $\overline{U(1)}_k$ currents by $j_{\sigma,L} = e^{i\phi_{\sigma,L}}$ with $s=-k/2$, two $U(1)_{2k}$ currents by $J_{\sigma,R} = e^{i\Phi_{\sigma,R}}$ with $s=k$, and two $\overline{U(1)}_{2k}$ currents by $J_{\sigma,L} = e^{i\Phi_{\sigma,L}}$ with $s=-k$. 
The conformal embedding \eqref{eq:EmbeddingUkU2k} implies that in terms of the two $U(1)_k$ currents, we can construct two $U(1)_{2k}$ currents $J_{\pm,R} = e^{i\Phi_{\pm,R}}$, where $\Phi_{\pm,R}$ are nothing but the symmetric and antisymmetric linear combinations of the bosonic fields $\Phi_{\pm,R} = \phi_{1,R} \pm \phi_{2,R}$. 
Similarly, we can construct two $\overline{U(1)}_{2k}$ currents $J_{\pm,L} = e^{i\Phi_{\pm,L}}$ by $\Phi_{\pm,L} = \phi_{1,L} \pm \phi_{2,L}$. 
Therefore, current-current interactions with $J_{\sigma,R}$ and $J_{\pm,L}$ naturally open a gap at the left interface, while those with $J_{\sigma,L}$ and $J_{\pm,R}$ open a gap at the right interface. 

We then consider a process of shrinking the middle strip, which will leave a quantum wire with two right-moving and two left-moving bosonic fields between the two strips of the $U(1)_{2k} \times U(1)_{2k}$ topological orders [Fig.~\ref{fig:CellularTopo3DHoney}~(b)]. 
Such gapped interfaces with additional quantum wires are precisely those used to construct some of the previous cellular topological states. 
Indeed, as seen form Eqs.~\eqref{eq:FieldFermionicU2Model}, \eqref{eq:FieldBosonicU4Model}, \eqref{eq:FieldFermionicU6Model}, and \eqref{eq:FieldHybridU8Model}, the corresponding tunneling Hamiltonians involve $U(1)_{2k}$ currents obtained by the symmetric and antisymmetric combinations of $U(1)_k$ currents for $k=1,2,3$, and $4$, respectively. 
We may thus consider a reverse process of stretching the quantum wires added at interfaces such that they can be viewed as strips of the $U(1)_k \times U(1)_k$ topological orders. 
Applying this process to the cellular topological states built from the $U(1)_{2k}$ topological orders on the square grid [Fig.~\ref{fig:CellularTopo3DHoney}~(c)], we can obtain equivalent cellular topological states built from both $U(1)_{2k}$ and $U(1)_k$ topological orders on the honeycomb grid without changing the quasiparticle properties of the original models [Fig.~\ref{fig:CellularTopo3DHoney}~(d)]. 

We can also interpret these results in terms of the Lagrangian subgroup. 
By collectively denoting the edge bosonic fields at an interface between four $U(1)_{2k}$ topological orders as $\bm{\phi}^\textsf{e} = (\Phi_{1,R}, \Phi_{2,R}, \Phi_{2,L}, \Phi_{1,L})^T$, gapped interfaces associated with the conformal embedding \eqref{eq:EmbeddingUkU2k} are specified by the Lagrangian subgroups generated by $\{ \bm{m}_a \}$ in Eqs.~\eqref{eq:VecMFermionicU2Model}, \eqref{eq:VecMBosonicU4Model}, \eqref{eq:VecMFermionicU6Model}, and \eqref{eq:VecMHybridU8Model}. 
Such gapped interfaces can be regarded as composite gapped interfaces obtained by inserting a strip of the $U(1)_k \times U(1)_k$ topological order, at which $\{ \bm{m}^\textsf{A}_a \}$ are condensed at the left interface and $\{ \bm{m}^\textsf{B}_a \}$ are condensed at the right interface, where
\begin{align} \label{eq:VecMHoneyA}
\begin{split}
\bm{m}^\textsf{A}_1 &= (1,1,1,0)^T, \\
\bm{m}^\textsf{A}_2 &= (1,2k-1,0,1)^T,
\end{split}
\end{align}
and
\begin{align} \label{eq:VecMHoneyB}
\begin{split}
\bm{m}^\textsf{B}_1 &= (1,0,1,1)^T, \\
\bm{m}^\textsf{B}_2 &= (0,1,1,2k-1)^T.
\end{split}
\end{align}
The former interfaces can be reproduced by pair annihilating quasiparticles at the middle strip from the composite gapped interfaces. 

It is also straightforward to write down the corresponding coupled-wire Hamiltonians on the honeycomb grid. 
We illustrate this for the bosonic $U(1)_4$ model in Sec.~\ref{sec:BosonicU4Model}.
Let us denote two sublattices on the honeycomb grid by $\textsf{A}$ and $\textsf{B}$.
We then introduce a two-component bosonic wire with the bosonic fields $\bm{\phi}^\textsf{w,A}_{\bm{r}} = (\varphi^{\textsf{A},1}_{\bm{r}}, 2\theta^{\textsf{A},1}_{\bm{r}}, \varphi^{\textsf{A},2}_{\bm{r}}, 2\theta^{\textsf{A},2}_{\bm{r}})^T$ on the $\textsf{A}$ sublattice and another two-component bosonic wires with $\bm{\phi}^\textsf{w,B}_{\bm{r}} = (\varphi^{\textsf{B},1}_{\bm{r}}, 2\theta^{\textsf{B},1}_{\bm{r}}, \varphi^{\textsf{B},2}_{\bm{r}}, 2\theta^{\textsf{B},2}_{\bm{r}})^T$ on the $\textsf{B}$ sublattice. 
Corresponding to the Lagrangian subgroups given by Eqs.~\eqref{eq:VecMHoneyA} and \eqref{eq:VecMHoneyB}, we can find two sets of integer vectors $\{ \bm{\Lambda}_{\textsf{w},\alpha}^\textsf{A} \}$ and $\{ \bm{\Lambda}_{\textsf{w},\alpha}^\textsf{B} \}$ associated with gapping potentials, which are given, for example, by
\begin{align}
\begin{split}
\bm{\Lambda}_{\textsf{w},1}^\textsf{A} &= (1,-1,-1,1)^T, \\
\bm{\Lambda}_{\textsf{w},2}^\textsf{A} &= (1,-1,1,-1)^T, \\
\bm{\Lambda}_{\textsf{w},3}^\textsf{A} &= (-1,-1,0,0)^T, \\
\bm{\Lambda}_{\textsf{w},4}^\textsf{A} &= (0,0,-1,-1)^T,
\end{split} \\
\begin{split}
\bm{\Lambda}_{\textsf{w},1}^\textsf{B} &= (1,-1,0,0)^T, \\
\bm{\Lambda}_{\textsf{w},2}^\textsf{B} &= (0,0,1,-1)^T, \\
\bm{\Lambda}_{\textsf{w},3}^\textsf{B} &= (-1,-1,1,1)^T, \\
\bm{\Lambda}_{\textsf{w},4}^\textsf{B} &= (-1,-1,-1,-1)^T.
\end{split}
\end{align}
With $K_\textsf{w} = X \oplus X$, we then find a tunneling Hamiltonian,
\begin{align}
\mathcal{V}_\textsf{w}^\textsf{honey} &= -g \int dx \sum_{\bm{r} \in \mathbb{Z}^2} \Bigl[ 
\cos \bigl( \Phi^\textsf{w,A}_{\bm{r},3} +\Phi^\textsf{w,B}_{\bm{r},1} \bigr) \nonumber \\
&\quad +\cos \bigl( \Phi^\textsf{w,A}_{\bm{r},4} +\Phi^\textsf{w,B}_{\bm{r},2} \bigr) +\cos \bigl( \Phi^\textsf{w,B}_{\bm{r},3} +\Phi^\textsf{w,A}_{\bm{r}+\bm{e}_y,1} \bigr) \nonumber \\
&\quad +\cos \bigl( \Phi^\textsf{w,B}_{\bm{r},4} +\Phi^\textsf{w,A}_{\bm{r}+\bm{e}_z,2} \bigr)
\Bigr],
\end{align}
where we have defined the bosonic fields,
\begin{subequations}
\begin{align}
\Phi^\textsf{w,A}_{\bm{r},\alpha} &= (\bm{\Lambda}^\textsf{A}_{\textsf{w},\alpha})^T K_\textsf{w} \bm{\phi}^\textsf{w,A}_{\bm{r}}, \\
\Phi^\textsf{w,B}_{\bm{r},\alpha} &= (\bm{\Lambda}^\textsf{B}_{\textsf{w},\alpha})^T K_\textsf{w} \bm{\phi}^\textsf{w,B}_{\bm{r}}.
\end{align}
\end{subequations}

\bibliography{3DCoupledWireCellularTopo}

\begin{widetext}
\newpage

\setcounter{section}{0}
\setcounter{equation}{0}
\setcounter{figure}{0}
\setcounter{table}{0}
\setcounter{page}{1}
\renewcommand{\theequation}{S\arabic{equation}}
\setcounter{figure}{0}
\renewcommand{\thefigure}{S\arabic{figure}}
\renewcommand{\thepage}{S\arabic{page}}
\renewcommand{\thesection}{S\arabic{section}}
\renewcommand{\thesubsection}{\Roman{subsection}}
\renewcommand{\thetable}{S\Roman{table}}
\makeatletter

\begin{center}
\textbf{\large Supplemental Material for ``Bridging three-dimensional coupled-wire models and cellular topological states: Solvable models for topological and fracton orders''}
\end{center}

\input{suppl.tex}
\end{widetext}

\end{document}

%% file: suppl.tex
\section{Lists of gapped interfaces}
\label{app:GappedInterface}

In this appendix, we provide lists of gapped interfaces between multiple 2D toric codes or multiple $U(1)_k$ topological orders. 
For a given $2N \times 2N$ $K$ matrix, we first look for all possible choices of the Lagrangian subgroup $L$, which can be generated by a subset $\{ \bm{m}_a \}$ of $L$ by linear combinations of its elements and by addition or subtraction of local particle excitations $K\mathbb{Z}^{2N}$. 
We denote such generating subsets $\{ \bm{m}_a \}$ in the matrix form, 
\begin{align}
M = \begin{pmatrix} \bm{m}_1 & \bm{m}_2 & \cdots & \bm{m}_{N'} \end{pmatrix},
\end{align}
where $N'$ is the rank of the Lagrangian subgroup $L$, i.e., the minimal number of $\bm{m}_a$ required to generate $L$.
Following the algorithms presented in Sec.~II B of the main text, we can always construct a gapping potential by adding $N$ extra quantum wires. 
For bosonic gapped interfaces, we add $N$ bosonic wires and extend the $K$ matrix to $\widetilde{K} = K \oplus X^{\oplus N}$. 
For fermionic gapped interfaces, we add $N$ fermionic wires and extend the $K$ matrix to $\widetilde{K} = K \oplus Z^{\oplus N}$. 
We can then find a set of $2N$ integer vectors $\{ \widetilde{\bm{\Lambda}}_\alpha \}$, which gives a gapping potential of the form, 
\begin{align}
\mathcal{V} = -g \sum_{\alpha=1}^{2N} \cos (\widetilde{\bm{\Lambda}}^T_\alpha \widetilde{K} \widetilde{\bm{\phi}}). 
\end{align}
Here, $\widetilde{\bm{\phi}} = (\phi_1, \cdots, \phi_{4N})$ represent bosonic fields defined at the interface, where the first $2N$ components correspond to the edge modes of the topological order $K$ while the last $2N$ components correspond to the extra quantum wires. 
Let us write the integer vectors $\{ \widetilde{\bm{\Lambda}}_a \}$ in the matrix form 
\begin{align}
\widetilde{\Lambda} = \begin{pmatrix} \widetilde{\bm{\Lambda}}_1 & \widetilde{\bm{\Lambda}}_2 & \cdots & \widetilde{\bm{\Lambda}}_{2N} \end{pmatrix}.
\end{align}
For our purpose, we wish to find $\widetilde{\Lambda}$ of the form,
\begin{align}
\widetilde{\Lambda} = \begin{pmatrix} I_{2N} \\ \Lambda_\textsf{w} \end{pmatrix},
\end{align}
where $\Lambda_\textsf{w}$ is a $2N \times 2N$ integer matrix whose explicit construction is presented in Appendix~A of the main text. 

In the following, we provide a list of $M$ and $\Lambda_\textsf{w}$ defined above.
In Sec.~\ref{sec:GITC}, we consider gapped interfaces of one, two, three, and four toric codes. 
We then consider gapped interfaces between two $U(1)_k$ and two $\overline{U(1)}_k$ topological orders with integer $k$ of the forms $U(1)_k \times U(1)_k \times \overline{U(1)}_k \times \overline{U(1)}_k$ in Sec.~\ref{sec:GIUkkkk}, $U(1)_k \times U(1)_l \times \overline{U(1)}_k \times \overline{U(1)}_l$ in Sec.~\ref{sec:GIUklkl}, and $U(1)_k \times U(1)_k \times \overline{U(1)}_l \times \overline{U(1)}_l$ in Sec.~\ref{sec:GIUkkll}. 
We finally consider gapped interfaces of $U(1)_3 \times U(1)_3 \times U(1)_3 \times \overline{U(1)}_3 \times \overline{U(1)}_3 \times \overline{U(1)}_3$ in Sec.~\ref{sec:GIU333333}. 

\subsection{Gapped interfaces for toric codes}
\label{sec:GITC}

\subsubsection{Single toric code}

The topological order of the 2D toric code is described by the $K$ matrix \cite{XGWen16}: 
\begin{align}
K^{(\textrm{TC})} = \begin{pmatrix} 0 & 2 \\ 2 & 0 \end{pmatrix}.
\end{align}
There are three gapped interfaces for a single toric code: two of them correspond to the condensation of the bosonic anyons $e$ or $m$ at the interface \cite{Bravyi98, Kitaev12}, known as the rough and smooth boundary conditions for the toric code, whereas the other one corresponds to condensation of the fermionic anyon $f$ \cite{Barkeshli14}. 
The corresponding integer vectors associated with the Lagrangian subgroup and a gapping potential are given by 
\begin{subequations}
\begin{align}
M^{(\textrm{TC})}_\textrm{1B;1} = \begin{pmatrix}
1 \\
0
\end{pmatrix}, \quad
\Lambda^{(\textrm{TC})}_{\textsf{w},\textrm{1B;1}} = \begin{pmatrix}
0 & 2 \\
-1 & 0
\end{pmatrix}, \\
M^{(\textrm{TC})}_\textrm{1B;2} = \begin{pmatrix}
0 \\
1
\end{pmatrix}, \quad
\Lambda^{(\textrm{TC})}_{\textsf{w},\textrm{1B;2}} = \begin{pmatrix}
2 & 0 \\
0 & -1
\end{pmatrix}, \\
M^{(\textrm{TC})}_\textrm{1F;1} = \begin{pmatrix}
1 \\
1
\end{pmatrix}, \quad
\Lambda^{(\textrm{TC})}_{\textsf{w},\textrm{1F;1}} = \begin{pmatrix}
1 & -1 \\
-1 & -1
\end{pmatrix}.
\end{align}
\end{subequations}

\subsubsection{Two toric codes}

We consider gapped interfaces for the stack of two toric codes, which is described by the $K$ matrix:
\begin{align}
K^{(2\textrm{TC})} = \begin{pmatrix} 0 & 2 && \\ 2 & 0 && \\ && 0 & 2 \\ && 2 & 0 \end{pmatrix}. 
\end{align}
There are 15 gapped interfaces in total, among which nine interfaces are obtained by just stacking gapped interfaces for two individual toric codes and are thus omitted here. 
There are two bosonic gapped interfaces obtained by condensing the following sets of bosonic anyons at the interface: 
\begin{subequations}
\begin{align}
L^{(2\textrm{TC})}_\textrm{2B;1} &= \{ \red{e_1 e_2}, \ \red{m_1 m_2}, \ f_1 f_2 \}, \\
L^{(2\textrm{TC})}_\textrm{2B;2} &= \{ \red{e_1 m_2}, \ \red{m_1 e_2}, \ f_1 f_2 \},
\end{align}
\end{subequations}
where generating subsets of anyons are highlighted by red. 
The first interface makes two layers of the toric codes into a single one, whereas the second interface interchanges $e$ and $m$ \cite{Kitaev12}. 
The corresponding integer vectors are given by
\begin{subequations}
\begin{align}
M^{(\textrm{2TC})}_\textrm{2B;1} = \begin{pmatrix}
1 & 0 \\
0 & 1 \\
1 & 0 \\
0 & 1
\end{pmatrix}, \quad
\Lambda^{(\textrm{2TC})}_{\textsf{w},\textrm{2B;1}} = \begin{pmatrix}
0 & 1 & 0 & 1 \\
-1 & 0 & -1 & 0 \\
0 & -1 & 0 & 1 \\
1 & 0 & -1 & 0
\end{pmatrix}, \\
M^{(\textrm{2TC})}_\textrm{2B;2} = \begin{pmatrix}
1 & 0 \\
0 & 1 \\
0 & 1 \\
1 & 0
\end{pmatrix}, \quad
\Lambda^{(\textrm{2TC})}_{\textsf{w},\textrm{2B;2}} = \begin{pmatrix}
0 & 1 & 1 & 0 \\
-1 & 0 & 0 & -1 \\
-1 & 0 & 0 & 1 \\
0 & 1 & -1 & 0
\end{pmatrix}.
\end{align}
\end{subequations}
The other four gapped interfaces are fermionic and correspond to condensation of the following sets of fermionic or bosonic anyons:
\begin{subequations}
\begin{align}
L^{(2\textrm{TC})}_\textrm{2F;1} &= \{ \red{e_1 e_2}, \ \red{f_1 m_2}, \ m_1 f_2 \}, \\
L^{(2\textrm{TC})}_\textrm{2F;2} &= \{ \red{e_1 m_2}, \ \red{f_1 e_2}, \ m_1 f_2 \}, \\
L^{(2\textrm{TC})}_\textrm{2F;3} &= \{ \red{f_1 e_2}, \ \red{m_1 m_2}, \ e_1 f_2 \}, \\
L^{(2\textrm{TC})}_\textrm{2F;4} &= \{ \red{f_1 m_2}, \ \red{m_1 e_2}, \ e_1 f_2 \}.
\end{align}
\end{subequations}
These interfaces interchange the fermionic anyon $f$ and either of the bosonic anyons $e$ or $m$.
The corresponding integer vectors are given by
\begin{subequations}
\begin{align}
M^{(\textrm{2TC})}_\textrm{2F;1} = \begin{pmatrix}
1 & 1 \\
0 & 1 \\
1 & 0 \\
0 & 1
\end{pmatrix}, \quad
\Lambda^{(\textrm{2TC})}_{\textsf{w},\textrm{2F;1}} = \begin{pmatrix}
-1 & 1 & 0 & 1 \\
1 & 1 & 0 & -1 \\
0 & 1 & 1 & -1 \\
0 & -1 & -1 & -1
\end{pmatrix}, \\
M^{(\textrm{2TC})}_\textrm{2F;2} = \begin{pmatrix}
1 & 1 \\
0 & 1 \\
0 & 1 \\
1 & 0
\end{pmatrix}, \quad
\Lambda^{(\textrm{2TC})}_{\textsf{w},\textrm{2F;2}} = \begin{pmatrix}
-1 & 1 & 1 & 0 \\
1 & 1 & -1 & 0 \\
0 & 1 & -1 & 1 \\
0 & -1 & -1 & -1
\end{pmatrix}, \\
M^{(\textrm{2TC})}_\textrm{2F;3} = \begin{pmatrix}
1 & 0 \\
1 & 1 \\
1 & 0 \\
0 & 1
\end{pmatrix}, \quad
\Lambda^{(\textrm{2TC})}_{\textsf{w},\textrm{2F;3}} = \begin{pmatrix}
1 & -1 & 1 & 0 \\
1 & 1 & -1 & 0 \\
1 & 0 & -1 & 1 \\
-1 & 0 & -1 & -1
\end{pmatrix}, \\
M^{(\textrm{2TC})}_\textrm{2F;4} = \begin{pmatrix}
1 & 0 \\
1 & 1 \\
0 & 1 \\
1 & 0
\end{pmatrix}, \quad
\Lambda^{(\textrm{2TC})}_{\textsf{w},\textrm{2F;4}} = \begin{pmatrix}
1 & -1 & 0 & 1 \\
1 & 1 & 0 & -1 \\
1 & 0 & 1 & -1 \\
-1 & 0 & -1 & -1
\end{pmatrix}.
\end{align}
\end{subequations}

\subsubsection{Three toric codes}

We consider gapped interfaces for the stack of three toric codes, which is described by the $K$ matrix:
\begin{align}
K^{(3\textrm{TC})} = \begin{pmatrix} 0 & 2 &&&& \\ 2 & 0 &&&& \\ && 0 & 2 && \\ && 2 & 0 && \\ &&&& 0 & 2 \\ &&&& 2 & 0 \end{pmatrix}. 
\end{align}
There are 135 gapped interfaces in total, among which 81 interfaces are obtained by just stacking gapped interfaces for single or two-layer toric codes and are thus omitted here. 
The rest of 54 interfaces have a common feature that two kinds of pair of anyons and a triplet of anyons, which are mutually commuting, are condensed to yield a gapped interface. 
We show four examples of the bosonic gapped interfaces, which correspond to the following Lagrangian subgroups:
\begin{subequations}
\begin{align}
L^{(3\textrm{TC})}_\textrm{3B;1} &=
\{ \red{e_1e_2}, \ \red{e_1e_3}, \ e_2e_3, \ \red{m_1m_2m_3}, \ f_1f_2m_3, \ f_1m_2f_3, \ m_1f_2f_3 \}, \\
L^{(3\textrm{TC})}_\textrm{3B;2} &=
\{ \red{m_1m_2}, \ \red{m_1m_3}, \ m_2m_3, \ \red{e_1e_2e_3}, \ e_1f_2f_3, \ f_1e_2f_3, \ f_1f_2e_3 \}, \\
L^{(3\textrm{TC})}_\textrm{3B;3} &= 
\{ \red{f_1f_2}, \ \red{f_1f_3}, \ f_2f_3, \ \red{e_1e_2e_3}, \ e_1m_2m_3, \ m_1e_2m_3, \ m_1m_2e_3 \}, \\
L^{(3\textrm{TC})}_\textrm{3B;4} &=
\{ \red{f_1f_2}, \ \red{f_1f_3}, \ f_2f_3, \ \red{m_1m_2m_3}, \ e_1e_2m_3, \ e_1m_2e_3, \ m_1e_2e_3 \}.
\end{align}
\end{subequations}
The corresponding integer vectors are given by
\begin{subequations}
\begin{align}
M^{(\textrm{3TC})}_\textrm{3B;1} = \begin{pmatrix}
1 & 1 & 0 \\
0 & 0 & 1 \\
1 & 0 & 0 \\
0 & 0 & 1 \\
0 & 1 & 0 \\
0 & 0 & 1
\end{pmatrix}, \quad
\Lambda^{(\textrm{3TC})}_{\textsf{w},\textrm{3B;1}} = \begin{pmatrix}
0 & 1 & 0 & 1 & 0 & 1 \\
0 & 0 & -1 & 0 & -1 & 0 \\
0 & 1 & 0 & -1 & 0 & 1 \\
-1 & 0 & 1 & 0 & 0 & 0 \\
0 & -1 & 0 & -1 & 0 & 1 \\
1 & 0 & 0 & 0 & -1 & 0
\end{pmatrix}, \\
M^{(\textrm{3TC})}_\textrm{3B;2} = \begin{pmatrix}
0 & 0 & 1 \\
1 & 1 & 0 \\
0 & 0 & 1 \\
1 & 0 & 0 \\
0 & 0 & 1 \\
0 & 1 & 0
\end{pmatrix}, \quad
\Lambda^{(\textrm{3TC})}_{\textsf{w},\textrm{3B;2}} = \begin{pmatrix}
1 & 0 & 1 & 0 & 1 & 0 \\
0 & 0 & 0 & -1 & 0 & -1 \\
1 & 0 & -1 & 0 & 1 & 0 \\
0 & -1 & 0 & 1 & 0 & 0 \\
-1 & 0 & -1 & 0 & 1 & 0 \\
0 & 1 & 0 & 0 & 0 & -1
\end{pmatrix}, \\
M^{(\textrm{3TC})}_\textrm{3B;3} = \begin{pmatrix}
1 & 1 & 1 \\
1 & 1 & 0 \\
1 & 0 & 1 \\
1 & 0 & 0 \\
0 & 1 & 1 \\
0 & 1 & 0
\end{pmatrix}, \quad
\Lambda^{(\textrm{3TC})}_{\textsf{w},\textrm{3B;3}} = \begin{pmatrix}
-1 & 0 & 1 & 0 & 1 & 0 \\
-1 & 0 & 0 & -1 & 0 & -1 \\
-1 & 0 & 0 & 1 & 0 & -1 \\
1 & 1 & -1 & 1 & 0 & 0 \\
-1 & 0 & 0 & -1 & 0 & 1 \\
0 & 1 & 0 & 1 & -1 & 0
\end{pmatrix}, \\
M^{(\textrm{3TC})}_\textrm{3B;4} = \begin{pmatrix}
1 & 1 & 0 \\
1 & 1 & 1 \\
1 & 0 & 0 \\
1 & 0 & 1 \\
0 & 1 & 0 \\
0 & 1 & 1
\end{pmatrix}, \quad
\Lambda^{(\textrm{3TC})}_{\textsf{w},\textrm{3B;4}} = \begin{pmatrix}
-1 & 0 & 1 & 0 & 0 & 1 \\
-1 & 0 & 0 & -1 & -1 & 0 \\
-1 & 0 & 0 & 1 & -1 & 0 \\
1 & 1 & -1 & 1 & 0 & 0 \\
0 & -1 & 0 & -1 & 0 & 1 \\
1 & 0 & 0 & 1 & -1 & 0
\end{pmatrix}.
\end{align}
\end{subequations}

We also show four examples of the fermionic gapped interfaces, which correspond to the following Lagrangian subgroups:
\begin{subequations}
\begin{align}
L^{(3\textrm{TC})}_\textrm{3F;1} &=
\{ \red{e_1e_2}, \ \red{e_1e_3}, \ e_2e_3, \ \red{f_1f_2f_3}, \ f_1m_2m_3, \ m_1f_2m_3, \ m_1m_2f_3 \}, \\
L^{(3\textrm{TC})}_\textrm{3F;2} &=
\{ \red{m_1m_2}, \ \red{m_1m_3}, \ m_2m_3, \ \red{f_1f_2f_3}, \ e_1e_2f_3, \ e_1f_2e_3, \ f_1e_2e_3 \}, \\
L^{(3\textrm{TC})}_\textrm{3F;3} &=
\{ \red{e_1f_2}, \ \red{e_1f_3}, \ f_2f_3, \ \red{m_1m_2m_3}, \ f_1e_2m_3, \ f_1m_2e_3, \ m_1e_2e_3 \}, \\
L^{(3\textrm{TC})}_\textrm{3F;4} &=
\{ \red{f_1e_2}, \ \red{f_1m_3}, \ e_2m_3, \ \red{m_1m_2e_3}, \ e_1f_2e_3, \ e_1m_2f_3, \ m_1f_2f_3 \}. 
\end{align}
\end{subequations}
The corresponding integer vectors are given by
\begin{subequations}
\begin{align}
M^{(\textrm{3TC})}_\textrm{3F;1} = \begin{pmatrix}
1 & 1 & 1 \\
0 & 0 & 1 \\
1 & 0 & 1 \\
0 & 0 & 1 \\
0 & 1 & 1 \\
0 & 0 & 1
\end{pmatrix}, \quad
\Lambda^{(\textrm{3TC})}_{\textsf{w},\textrm{3F;1}} = \begin{pmatrix}
1 & 1 & 1 & -1 & 1 & -1 \\
0 & 1 & 0 & -1 & 1 & 1 \\
-1 & 1 & 0 & -1 & 0 & 1 \\
1 & 1 & 0 & -1 & 0 & -1 \\
0 & 1 & 0 & 1 & 1 & -1 \\
1 & 1 & 1 & 1 & 1 & -1
\end{pmatrix}, \\
M^{(\textrm{3TC})}_\textrm{3F;2} = \begin{pmatrix}
0 & 0 & 1 \\
1 & 1 & 1 \\
0 & 0 & 1 \\
1 & 0 & 1 \\
0 & 0 & 1 \\
0 & 1 & 1
\end{pmatrix}, \quad
\Lambda^{(\textrm{3TC})}_{\textsf{w},\textrm{3F;2}} = \begin{pmatrix}
1 & 1 & -1 & 1 & -1 & 1 \\
1 & 0 & -1 & 0 & 1 & 1 \\
1 & -1 & -1 & 0 & 1 & 0 \\
1 & 1 & -1 & 0 & -1 & 0 \\
1 & 0 & 1 & 0 & -1 & 1 \\
1 & 1 & 1 & 1 & -1 & 1
\end{pmatrix}, \\
M^{(\textrm{3TC})}_\textrm{3F;3} = \begin{pmatrix}
1 & 1 & 0 \\
0 & 0 & 1 \\
1 & 0 & 0 \\
1 & 0 & 1 \\
0 & 1 & 0 \\
0 & 1 & 1
\end{pmatrix}, \quad
\Lambda^{(\textrm{3TC})}_{\textsf{w},\textrm{3F;3}} = \begin{pmatrix}
1 & 0 & 1 & -1 & 0 & 0 \\
1 & 0 & 0 & 0 & -1 & -1 \\
1 & 0 & 0 & 0 & 1 & -1 \\
1 & 1 & 0 & -1 & 1 & 0 \\
-1 & 1 & 0 & -1 & 1 & 0 \\
-1 & 0 & -1 & -1 & 0 & 0
\end{pmatrix}, \\
M^{(\textrm{3TC})}_\textrm{3F;4} = \begin{pmatrix}
1 & 1 & 0 \\
1 & 1 & 1 \\
1 & 0 & 0 \\
0 & 0 & 1 \\
0 & 0 & 1 \\
0 & 1 & 0
\end{pmatrix}, \quad
\Lambda^{(\textrm{3TC})}_{\textsf{w},\textrm{3F;4}} = \begin{pmatrix}
1 & -1 & 0 & 0 & 0 & -1 \\
1 & 1 & 1 & 0 & 0 & 0 \\
-1 & 0 & 0 & 1 & 1 & -1 \\
1 & 0 & 0 & -1 & -1 & -1 \\
1 & 1 & 1 & -2 & 0 & 0 \\
-1 & -1 & 0 & 2 & 0 & -1
\end{pmatrix}.
\end{align}
\end{subequations}
These gapped interfaces can be used to construct cellular topological states built from the 2D toric codes on the honeycomb grid \cite{XGWen20, JWang20b}, and the corresponding coupled-wire models can also be constructed with gapping potentials associated with $\Lambda_\textsf{w}$.

\subsubsection{Four toric codes}

We consider gapped interfaces for the stack of four toric codes, which is described by the $K$ matrix:
\begin{align}
K^{(4\textrm{TC})} = \begin{pmatrix} 0 & 2 &&&&&& \\ 2 & 0 &&&&&& \\ && 0 & 2 &&&& \\ && 2 & 0 &&&& \\ &&&& 0 & 2 && \\ &&&& 2 & 0 && \\ &&&&&& 0 & 2 \\ &&&&&& 2 & 0 \end{pmatrix}. 
\end{align}
There are 2295 gapped interfaces in total, among which 1161 interfaces are obtained by just stacking gapped interfaces for single, two-layer, or three-layer toric codes. 
There are two types among 1134 nontrivial gapped interfaces. 
One is obtained by condensing pairs of anyons from any two of the four layers and also quadruplets of anyons from all four layers.
There are 162 gapped interfaces in this class. 
We give three examples of gapped bosonic interfaces in this class, whose Lagrangian subgroups are given by
\begin{subequations}
\begin{align}
L^{(4\textrm{TC})}_\textrm{4B;1} &= \{ \red{e_1e_2}, \ \red{e_1e_3}, \ \red{e_1e_4}, \ e_2e_3, \ e_2e_4, \ e_3e_4, \ e_1e_2e_3e_4, \ \red{m_1m_2m_3m_4}, \ f_1f_2f_3f_4, \nonumber \\
& \quad \quad f_1f_2m_3m_4, \ f_1m_2f_3m_4, \ f_1m_2m_3f_4, \ m_1f_2f_3m_4, \ m_1f_2m_3f_4, \ m_1m_2f_3f_4 \}, \\
L^{(4\textrm{TC})}_\textrm{4B;2} &= \{ \red{m_1m_2}, \ \red{m_1m_3}, \ \red{m_1m_4}, \ m_2m_3, \ m_2m_4, \ m_3m_4, \ \red{e_1e_2e_3e_4}, \ m_1m_2m_3m_4, \ f_1f_2f_3f_4, \nonumber \\
& \quad \quad e_1e_2f_3f_4, \ e_1f_2e_3f_4, \ e_1f_2f_3e_4, \ f_1e_2e_3f_4, \ f_1e_2f_3e_4, \ f_1f_2e_3e_4 \}, \\
L^{(4\textrm{TC})}_\textrm{4B;3} &= \{ \red{f_1f_2}, \ \red{f_1f_3}, \ \red{f_1f_4}, \ f_2f_3, \ f_2f_4, \ f_3f_4, \ \red{e_1e_2e_3e_4}, \ m_1m_2m_3m_4, \ f_1f_2f_3f_4, \nonumber \\ 
& \quad \quad e_1e_2m_3m_4, \ e_1m_2e_3m_4, \ e_1m_2m_3e_4, \ m_1e_2e_3m_4, \ m_1e_2m_3e_4, \ m_1m_2e_3e_4 \}.
\end{align}
\end{subequations}
These are building blocks for cellular topological states considered in Refs.~\cite{XGWen20, JWang20b} on the square grid.
The corresponding integer vectors are given by
\begin{subequations}
\begin{align}
M^{(\textrm{4TC})}_\textrm{4B;1} = \begin{pmatrix}
1 & 1 & 1 & 0 \\
0 & 0 & 0 & 1 \\
1 & 0 & 0 & 0 \\
0 & 0 & 0 & 1 \\
0 & 1 & 0 & 0 \\
0 & 0 & 0 & 1 \\
0 & 0 & 1 & 0 \\
0 & 0 & 0 & 1
\end{pmatrix}, \quad
\Lambda^{(\textrm{4TC})}_{\textsf{w},\textrm{4B;1}} = \begin{pmatrix}
2 & 0 & 0 & 0 & 0 & 0 & 0 & 0 \\
0 & -1 & 0 & -1 & 0 & -1 & 0 & -1 \\
-1 & 0 & 1 & 0 & 0 & 0 & 0 & 0 \\
0 & 0 & 0 & -2 & 0 & 0 & 0 & 0 \\
-1 & 0 & 0 & 0 & 1 & 0 & 0 & 0 \\
0 & 0 & 0 & 0 & 0 & -2 & 0 & 0 \\
-1 & 0 & 0 & 0 & 0 & 0 & 1 & 0 \\
0 & 0 & 0 & 0 & 0 & 0 & 0 & -2
\end{pmatrix}, \\
M^{(\textrm{4TC})}_\textrm{4B;2} = \begin{pmatrix}
0 & 0 & 0 & 1 \\
1 & 1 & 1 & 0 \\
0 & 0 & 0 & 1 \\
1 & 0 & 0 & 0 \\
0 & 0 & 0 & 1 \\
0 & 1 & 0 & 0 \\
0 & 0 & 0 & 1 \\
0 & 0 & 1 & 0
\end{pmatrix}, \quad
\Lambda^{(\textrm{4TC})}_{\textsf{w},\textrm{4B;2}} = \begin{pmatrix}
0 & 2 & 0 & 0 & 0 & 0 & 0 & 0 \\
-1 & 0 & -1 & 0 & -1 & 0 & -1 & 0 \\
0 & -1 & 0 & 1 & 0 & 0 & 0 & 0 \\
0 & 0 & -2 & 0 & 0 & 0 & 0 & 0 \\
0 & -1 & 0 & 0 & 0 & 1 & 0 & 0 \\
0 & 0 & 0 & 0 & -2 & 0 & 0 & 0 \\
0 & -1 & 0 & 0 & 0 & 0 & 0 & 1 \\
0 & 0 & 0 & 0 & 0 & 0 & -2 & 0
\end{pmatrix}, \\
M^{(\textrm{4TC})}_\textrm{4B;3} = \begin{pmatrix}
1 & 1 & 1 & 1 \\
1 & 1 & 1 & 0 \\
1 & 0 & 0 & 1 \\
1 & 0 & 0 & 0 \\
0 & 1 & 0 & 1 \\
0 & 1 & 0 & 0 \\
0 & 0 & 1 & 1 \\
0 & 0 & 1 & 0
\end{pmatrix}, \quad
\Lambda^{(\textrm{4TC})}_{\textsf{w},\textrm{4B;3}} = \begin{pmatrix}
1 & 0 & 0 & -1 & 0 & 1 & 1 & 0 \\
0 & -2 & 0 & 0 & 0 & 0 & 0 & 0 \\
0 & 1 & 1 & 0 & 0 & -1 & 1 & 0 \\
0 & 0 & 0 & -2 & 0 & 0 & 0 & 0 \\
0 & -1 & 0 & 1 & 1 & 0 & 1 & 0 \\
0 & 0 & 0 & 0 & 0 & -2 & 0 & 0 \\
0 & 0 & 0 & 0 & 0 & 0 & 2 & 0 \\
0 & 1 & 0 & 1 & 0 & 1 & 0 & -1
\end{pmatrix}.
\end{align}
\end{subequations}
We also show three examples of gapped fermionic interfaces in the same class, whose Lagrangian subgroups are given by
\begin{subequations}
\begin{align}
L^{(4\textrm{TC})}_\textrm{4F;1} &= \{ \red{e_1e_2}, \ \red{e_1e_3}, \ \red{e_1e_4}, \ e_2e_3, \ e_2e_4, \ e_3e_4, \ e_1e_2e_3e_4, \ f_1f_2f_3m_4, \ f_1f_2m_3f_4, \nonumber \\
&\quad \quad f_1m_2f_3f_4, \ \red{f_1m_2m_3m_4}, \ m_1f_2f_3f_4, \ m_1f_2m_3m_4, \ m_1m_2f_3m_4, \ m_1m_2m_3f_4 \}, \\
L^{(4\textrm{TC})}_\textrm{4F;2} &= \{ \red{m_1m_2}, \ \red{m_1m_3}, \ \red{m_1m_4}, \ m_2m_3, \ m_2m_4, \ m_3m_4, \ e_1e_2e_3f_4, \ e_1e_2f_3e_4, \ e_1f_2e_3e_4, \nonumber \\
&\quad \quad e_1f_2f_3f_4, \ \red{f_1e_2e_3e_4}, \ f_1e_2f_3f_4, \ f_1f_2e_3f_4, \ f_1f_2f_3e_4, \ m_1m_2m_3m_4 \}, \\
L^{(4\textrm{TC})}_\textrm{4F;3} &= \{ \red{e_1f_2}, \ \red{e_1e_3}, \ \red{e_1e_4}, \ f_2e_3, \ f_2e_4, \ e_3e_4, \ e_1f_2e_3e_4, \ f_1e_2f_3f_4, \ f_1e_2m_3m_4, \nonumber \\ 
& f_1m_2f_3m_4, \ f_1m_2m_3f_4, \ m_1e_2f_3m_4, \ m_1e_2m_3f_4, \ m_1m_2f_3f_4, \ \red{m_1m_2m_3m_4} \}.
\end{align}
\end{subequations}
The first two examples involve fermionic anyons only in quadruplets of anyons, while the last example involves fermionic anyons both in pairs and in quadruplets.
The corresponding integer vectors are given by 
\begin{subequations}
\begin{align}
M^{(\textrm{4TC})}_\textrm{4F;1} = \begin{pmatrix}
1 & 1 & 1 & 1 \\
0 & 0 & 0 & 1 \\
1 & 0 & 0 & 0 \\
0 & 0 & 0 & 1 \\
0 & 1 & 0 & 0 \\
0 & 0 & 0 & 1 \\
0 & 0 & 1 & 0 \\
0 & 0 & 0 & 1
\end{pmatrix}, \quad
\Lambda^{(\textrm{4TC})}_{\textsf{w},\textrm{4F;1}} = \begin{pmatrix}
0 & 1 & -1 & 1 & 0 & 1 & 0 & 1 \\
0 & -1 & 1 & 1 & -1 & 1 & 1 & 1 \\
-1 & 1 & 0 & -1 & 0 & 1 & 0 & 1 \\
-1 & -1 & 0 & -1 & 0 & -1 & 0 & -1 \\
0 & 1 & -1 & -1 & 1 & -1 & -1 & 1 \\
1 & -1 & 1 & 1 & -1 & -1 & 0 & -1 \\
-1 & 1 & -1 & -1 & 1 & -1 & 0 & -1 \\
0 & 1 & -1 & -1 & 0 & -1 & 0 & 1
\end{pmatrix}, \\
M^{(\textrm{4TC})}_\textrm{4F;2} = \begin{pmatrix}
0 & 0 & 0 & 1 \\
1 & 1 & 1 & 1 \\
0 & 0 & 0 & 1 \\
1 & 0 & 0 & 0 \\
0 & 0 & 0 & 1 \\
0 & 1 & 0 & 0 \\
0 & 0 & 0 & 1 \\
0 & 0 & 1 & 0
\end{pmatrix}, \quad
\Lambda^{(\textrm{4TC})}_{\textsf{w},\textrm{4F;2}} = \begin{pmatrix}
1 & 0 & 1 & -1 & 1 & 0 & 1 & 0 \\
-1 & 0 & 1 & 1 & 1 & -1 & 1 & 1 \\
1 & -1 & -1 & 0 & 1 & 0 & 1 & 0 \\
-1 & -1 & -1 & 0 & -1 & 0 & -1 & 0 \\
1 & 0 & -1 & -1 & -1 & 1 & 1 & -1 \\
-1 & 1 & 1 & 1 & -1 & -1 & -1 & 0 \\
1 & -1 & -1 & -1 & -1 & 1 & -1 & 0 \\
1 & 0 & -1 & -1 & -1 & 0 & 1 & 0
\end{pmatrix}, \\
M^{(\textrm{4TC})}_\textrm{4F;3} = \begin{pmatrix}
1 & 1 & 1 & 0 \\
0 & 0 & 0 & 1 \\
1 & 0 & 0 & 0 \\
1 & 0 & 0 & 1 \\
0 & 1 & 0 & 0 \\
0 & 0 & 0 & 1 \\
0 & 0 & 1 & 0 \\
0 & 0 & 0 & 1
\end{pmatrix}, \quad
\Lambda^{(\textrm{4TC})}_{\textsf{w},\textrm{4F;3}} = \begin{pmatrix}
-1 & 0 & 1 & -1 & 0 & 0 & 0 & 0 \\
0 & -2 & -1 & -3 & -2 & 0 & -1 & 2 \\
0 & 1 & 1 & 2 & 2 & -1 & 1 & -1 \\
0 & 0 & -1 & -1 & -1 & 0 & -2 & 0 \\
0 & 2 & -1 & 1 & 0 & 0 & -1 & 0 \\
0 & 1 & 1 & 2 & 0 & -1 & 1 & -1 \\
0 & 2 & 1 & 3 & 1 & 0 & 2 & -2 \\
1 & 2 & -1 & 1 & 0 & 0 & 0 & 0
\end{pmatrix}.
\end{align}
\end{subequations}

The rest of 973 gapped interfaces are obtained by condensing pairs of anyons from two layers and another two layers, respectively, and also two triplets of anyons from three layers. 
We provide two examples in this class, one is bosonic and the other is fermionic, which are given by the Lagrangian subgroups:
\begin{subequations}
\begin{align}
L^{(4\textrm{TC})}_\textrm{4B;4} &= \{ \red{e_1e_2}, \ \red{e_3e_4}, \ e_1f_3f_4, \ e_1m_3m_4, \ e_2f_3f_4, \ \red{e_2m_3m_4}, \ f_1f_2e_3, \ f_1f_2e_4, \ \red{m_1m_2e_3}, \ m_1m_2e_4, \nonumber \\
&\quad \quad e_1e_2e_3e_4, \ f_1m_2f_3m_4, \ f_1m_2m_3f_4, \ m_1f_2f_3m_4, \ m_1f_2m_3f_4 \},\\
L^{(4\textrm{TC})}_\textrm{4F;4} &= \{ \red{e_1e_2}, \ \red{e_3e_4}, \ e_1f_3m_4, \ e_1m_3f_4, \ e_2f_3m_4, \ \red{e_2m_3f_4}, \ f_1m_2e_3, \ f_1m_2e_4, \ \red{m_1f_2e_3}, \ m_1f_2e_4, \nonumber \\
&\quad \quad e_1e_2e_3e_4, \ f_1f_2f_3f_4, \ f_1f_2m_3m_4, \ m_1m_2f_3f_4, \ m_1m_2m_3m_4 \}.
\end{align}
\end{subequations}
The corresponding integer vectors are given by
\begin{subequations}
\begin{align}
M^{(\textrm{4TC})}_\textrm{4B;4} = \begin{pmatrix}
1 & 0 & 0 & 0 \\
0 & 0 & 1 & 0 \\
1 & 0 & 0 & 1 \\
0 & 0 & 1 & 0 \\
0 & 1 & 1 & 0 \\
0 & 0 & 0 & 1 \\
0 & 1 & 0 & 0 \\
0 & 0 & 0 & 1
\end{pmatrix}, \quad
\Lambda^{(\textrm{4TC})}_{\textsf{w},\textrm{4B;4}} = \begin{pmatrix}
0 & -1 & 0 & 1 & 0 & 0 & -1 & 0 \\
1 & 0 & -1 & 0 & 0 & 0 & 0 & 0 \\
0 & 0 & 0 & 0 & 1 & 0 & 1 & 0 \\
1 & 0 & 0 & 0 & 0 & -1 & 0 & -1 \\
1 & 0 & 1 & 0 & 0 & 0 & 0 & 0 \\
0 & -1 & 0 & -1 & 0 & 0 & -1 & 0 \\
-1 & 0 & 0 & 0 & 0 & -1 & 0 & 1 \\
0 & 0 & 0 & 0 & 1 & 0 & -1 & 0
\end{pmatrix}, \\
M^{(\textrm{4TC})}_\textrm{4F;4} = \begin{pmatrix}
1 & 0 & 0 & 0 \\
0 & 0 & 1 & 0 \\
1 & 0 & 1 & 1 \\
0 & 0 & 1 & 0 \\
0 & 1 & 1 & 0 \\
0 & 0 & 0 & 1 \\
0 & 1 & 0 & 1 \\
0 & 0 & 0 & 1
\end{pmatrix}, \quad
\Lambda^{(\textrm{4TC})}_{\textsf{w},\textrm{4F;4}} = \begin{pmatrix}
1 & 0 & 0 & 0 & 0 & 1 & -1 & 1 \\
0 & 1 & -1 & -1 & 0 & 0 & -1 & 0 \\
0 & -1 & 1 & -1 & 0 & 0 & 1 & 0 \\
1 & 1 & 0 & 1 & 0 & 0 & -1 & 0 \\
-1 & 1 & 0 & -1 & 0 & 0 & -1 & 0 \\
1 & 0 & 0 & 0 & 0 & -1 & 1 & 1 \\
1 & 0 & 0 & 0 & 1 & -1 & 0 & 1 \\
-1 & 0 & 0 & 0 & -1 & -1 & 0 & -1
\end{pmatrix}.
\end{align}
\end{subequations}

\subsection{Gapped interface for $U(1)_k \times U(1)_k \times \overline{U(1)}_k \times \overline{U(1)}_k$}
\label{sec:GIUkkkk}

We provide the lists of gapped interfaces between $U(1)_k$ topological orders corresponding to the following $K$ matrix:
\begin{align}
K = \begin{pmatrix} k &&& \\ & k && \\ && -k & \\ &&& -k \end{pmatrix}
\end{align}
where $k$ is a positive nonzero integer such that $K$ has vanishing signature. 

\subsubsection{$U(1)_2 \times U(1)_2 \times \overline{U(1)}_2 \times \overline{U(1)}_2$}

We consider gapped interfaces of the bosonic topological order corresponding to the $K$ matrix:
\begin{align}
K^{(2,2,\overline{2},\overline{2})} = \begin{pmatrix} 2 &&& \\ & 2 && \\ && -2 & \\ &&& -2 \end{pmatrix}.
\end{align}
There are three distinct choices of the Lagrangian subgroups, which all have the rank two. 
Two of them are bosonic gapped interfaces consisting of stacks of two gapped interfaces between $U(1)_2$ and $\overline{U(1)}_2$; since there are two ways of pairing two layers with different chiralities of the edge modes, there are two interfaces in total. 
Taking the Lagrangian subgroup corresponding to $M^{(2,2,\overline{2},\overline{2})}_\textrm{2B;1}$ given below as an instance, anyons condensed at the interface are given by
\begin{align}
\red{\bm{1}_1 \overline{\bm{1}}_3}, \ \red{\bm{1}_2 \overline{\bm{1}}_4}, \ \bm{1}_1 \bm{1}_2 \overline{\bm{1}}_3 \overline{\bm{1}}_4,
\end{align}
where those generating all others are highlighted by red.
The integer vectors associated with the Lagrangian subgroups and gapping potentials are given by
\begin{subequations}
\begin{align}
M^{(2,2,\overline{2},\overline{2})}_\textrm{2B;1} = \begin{pmatrix}
1 & 0 \\
0 & 1 \\
1 & 0 \\
0 & 1
\end{pmatrix}, \quad
\Lambda^{(2,2,\overline{2},\overline{2})}_{\textsf{w},\textrm{2B;1}} = \begin{pmatrix}
1 & 0 & -1 & 0 \\
-1 & 0 & -1 & 0 \\
0 & 1 & 0 & -1 \\
0 & -1 & 0 & -1
\end{pmatrix}, \\
M^{(2,2,\overline{2},\overline{2})}_\textrm{2B;2} = \begin{pmatrix}
1 & 0 \\
0 & 1 \\
0 & 1 \\
1 & 0
\end{pmatrix}, \quad
\Lambda^{(2,2,\overline{2},\overline{2})}_{\textsf{w},\textrm{2B;2}} \
= \begin{pmatrix}
1 & 0 & 0 & -1 \\
-1 & 0 & 0 & -1 \\
0 & 1 & -1 & 0 \\
0 & -1 & -1 & 0
\end{pmatrix}.
\end{align}
\end{subequations}
The other one is a fermionic gapped interface at which a local fermionic excitation creates a pair of semions (spin-1/4 anyons) in two layers with the same chirality of the edge modes. 
Anyons condensed at the interface are given by
\begin{align}
\red{\bm{1}_1 \bm{1}_2}, \ \red{\overline{\bm{1}}_3 \overline{\bm{1}}_4}, \ \bm{1}_1 \bm{1}_2 \overline{\bm{1}}_3 \overline{\bm{1}}_4.
\end{align}
The corresponding integer vectors are given by
\begin{align}
M^{(2,2,\overline{2},\overline{2})}_\textrm{2F;1} = \begin{pmatrix}
1 & 0 \\
1 & 0 \\
0 & 1 \\
0 & 1
\end{pmatrix}, \quad
\Lambda^{(2,2,\overline{2},\overline{2})}_{\textsf{w},\textrm{2F;1}} \
= \begin{pmatrix}
0 & 0 & 1 & 1 \\
-1 & -1 & 0 & 0 \\
0 & 0 & -1 & 1 \\
1 & -1 & 0 & 0
\end{pmatrix}.
\end{align}

\subsubsection{$U(1)_3 \times U(1)_3 \times \overline{U(1)}_3 \times \overline{U(1)}_3$}

We consider gapped interfaces of the fermionic topological order corresponding to the $K$ matrix:
\begin{align}
K^{(3,3,\overline{3},\overline{3})} = \begin{pmatrix} 3 &&& \\ & 3 && \\ && -3 & \\ &&& -3 \end{pmatrix}.
\end{align}
There are eight distinct choices of the Lagrangian subgroups of rank two.
They are all obtained by stacking two fermionic gapped interfaces between $U(1)_3$ and $\overline{U(1)}_3$ with tunneling- or pairing-type interactions, which lead to the condensations of $\bm{1} \overline{\bm{2}}$ or $\bm{1} \overline{\bm{1}}$. 
Taking the Lagrangian subgroup corresponding to $M^{(3,3,\overline{3},\overline{3})}_\textrm{2F;1}$ given below as an instance, the set of condensed anyons is given by
\begin{align}
\red{\bm{1}_1\overline{\bm{1}}_3}, \ 
\bm{2}_1\overline{\bm{2}}_3, \ 
\red{\bm{1}_2\overline{\bm{1}}_4}, \ 
\bm{2}_2\overline{\bm{2}}_4, \ 
\bm{1}_1\bm{1}_2\overline{\bm{1}}_3\overline{\bm{1}}_4, \ 
\bm{1}_1\bm{2}_2\overline{\bm{1}}_3\overline{\bm{2}}_4, \ 
\bm{2}_1\bm{1}_2\overline{\bm{2}}_3\overline{\bm{1}}_4, \ 
\bm{2}_1\bm{2}_2\overline{\bm{2}}_3\overline{\bm{2}}_4.
\end{align}
The corresponding integer vectors are given by
\begin{subequations}
\begin{align}
M^{(3,3,\overline{3},\overline{3})}_\textrm{2F;1} = \begin{pmatrix}
1 & 0 \\
0 & 1 \\
1 & 0 \\
0 & 1
\end{pmatrix}, \quad
\Lambda^{(3,3,\overline{3},\overline{3})}_{\textsf{w},\textrm{2F;1}} \
= \begin{pmatrix}
1 & 0 & -2 & 0 \\
2 & 0 & -1 & 0 \\
0 & 1 & 0 & -2 \\
0 & 2 & 0 & -1
\end{pmatrix}, \\
M^{(3,3,\overline{3},\overline{3})}_\textrm{2F;2} = \begin{pmatrix}
1 & 0 \\
0 & 2 \\
1 & 0 \\
0 & 1
\end{pmatrix}, \quad
\Lambda^{(3,3,\overline{3},\overline{3})}_{\textsf{w},\textrm{2F;2}} = \begin{pmatrix}
0 & -1 & 0 & -2 \\
2 & 0 & -1 & 0 \\
1 & 0 & -2 & 0 \\
0 & 2 & 0 & 1
\end{pmatrix}, \\
M^{(3,3,\overline{3},\overline{3})}_\textrm{2F;3} = \begin{pmatrix}
2 & 0 \\
0 & 1 \\
1 & 0 \\
0 & 1
\end{pmatrix}, \quad
\Lambda^{(3,3,\overline{3},\overline{3})}_{\textsf{w},\textrm{2F;3}} = \begin{pmatrix}
-1 & 0 & -2 & 0 \\
0 & 2 & 0 & -1 \\
0 & 1 & 0 & -2 \\
2 & 0 & 1 & 0
\end{pmatrix}, \\
M^{(3,3,\overline{3},\overline{3})}_\textrm{2F;4} = \begin{pmatrix}
2 & 0 \\
0 & 2 \\
1 & 0 \\
0 & 1
\end{pmatrix}, \quad
\Lambda^{(3,3,\overline{3},\overline{3})}_{\textsf{w},\textrm{2F;4}} \
= \begin{pmatrix}
-1 & 0 & -2 & 0 \\
2 & 0 & 1 & 0 \\
0 & -1 & 0 & -2 \\
0 & 2 & 0 & 1
\end{pmatrix}, \\
M^{(3,3,\overline{3},\overline{3})}_\textrm{2F;5} = \begin{pmatrix}
0 & 1 \\
1 & 0 \\
1 & 0 \\
0 & 1
\end{pmatrix}, \quad
\Lambda^{(3,3,\overline{3},\overline{3})}_{\textsf{w},\textrm{2F;5}} = \begin{pmatrix}
1 & 0 & 0 & -2 \\
0 & 2 & -1 & 0 \\
0 & 1 & -2 & 0 \\
2 & 0 & 0 & -1
\end{pmatrix}, \\
M^{(3,3,\overline{3},\overline{3})}_\textrm{2F;6} = \begin{pmatrix}
0 & 2 \\
1 & 0 \\
1 & 0 \\
0 & 1
\end{pmatrix}, \quad
\Lambda^{(3,3,\overline{3},\overline{3})}_{\textsf{w},\textrm{2F;6}} \
= \begin{pmatrix}
-1 & 0 & 0 & -2 \\
0 & 2 & -1 & 0 \\
0 & 1 & -2 & 0 \\
2 & 0 & 0 & 1
\end{pmatrix}, \\
M^{(3,3,\overline{3},\overline{3})}_\textrm{2F;7} = \begin{pmatrix}
0 & 2 \\
1 & 0 \\
1 & 0 \\
0 & 1
\end{pmatrix}, \quad
\Lambda^{(3,3,\overline{3},\overline{3})}_{\textsf{w},\textrm{2F;7}} \
= \begin{pmatrix}
-1 & 0 & 0 & -2 \\
0 & 2 & -1 & 0 \\
0 & 1 & -2 & 0 \\
2 & 0 & 0 & 1
\end{pmatrix}, \\
M^{(3,3,\overline{3},\overline{3})}_\textrm{2F;8} = \begin{pmatrix}
0 & 2 \\
2 & 0 \\
1 & 0 \\
0 & 1
\end{pmatrix}, \quad
\Lambda^{(3,3,\overline{3},\overline{3})}_{\textsf{w},\textrm{2F;8}} \
= \begin{pmatrix}
-1 & 0 & 0 & -2 \\
0 & 2 & 1 & 0 \\
0 & -1 & -2 & 0 \\
2 & 0 & 0 & 1
\end{pmatrix}.
\end{align}
\end{subequations}

\subsubsection{$U(1)_4 \times U(1)_4 \times \overline{U(1)}_4 \times \overline{U(1)}_4$}

We consider gapped interfaces of the bosonic topological order corresponding to the $K$ matrix:
\begin{align}
K^{(4,4,\overline{4},\overline{4})} = \begin{pmatrix} 4 &&& \\ & 4 && \\ && -4 & \\ &&& -4 \end{pmatrix}.
\end{align}
There are 16 distinct choices of the Lagrangian subgroup of rank two: eight of them are bosonic, whereas another eight are fermionic. 
The bosonic gapped interfaces of the rank-2 Lagrangian subgroup are simply stacking of two independent gapped interfaces between $U(1)_4 \times \overline{U(1)}_4$. 
There are two possible choices of pairing the $U(1)_4$ and $\overline{U(1)}_4$ layers. 
Also, there are two possible ways of choosing condensed anyons independently for each of the two interfaces: $\bm{1} \overline{\bm{1}}$ or $\bm{1} \overline{\bm{3}}$, induced by pairing- or tunneling-type interactions. 
This results in the eight bosonic gapped interfaces.
Taking the Lagrangian subgroup corresponding to $M^{(4,4,\overline{4},\overline{4})}_\textrm{2B;1}$ given below as an instance, the set of condensed anyons at the interface is given by
\begin{align}
& \red{\bm{1}_1\overline{\bm{1}}_3}, \ 
\bm{2}_1\overline{\bm{2}}_3, \ 
\bm{3}_1\overline{\bm{3}}_3, \ 
\red{\bm{1}_2\overline{\bm{1}}_4}, \ 
\bm{2}_2\overline{\bm{2}}_4, \ 
\bm{3}_2\overline{\bm{3}}_4, \ 
\bm{1}_1\bm{1}_2\overline{\bm{1}}_3\overline{\bm{1}}_4, \ 
\bm{2}_1\bm{2}_2\overline{\bm{2}}_3\overline{\bm{2}}_4, \ 
\bm{3}_1\bm{3}_2\overline{\bm{3}}_3\overline{\bm{3}}_4, \nonumber \\
& \bm{1}_1\bm{2}_2\overline{\bm{1}}_3\overline{\bm{2}}_4, \ 
\bm{1}_1\bm{3}_2\overline{\bm{1}}_3\overline{\bm{3}}_4, \ 
\bm{2}_1\bm{1}_2\overline{\bm{2}}_3\overline{\bm{1}}_4, \ 
\bm{2}_1\bm{3}_2\overline{\bm{2}}_3\overline{\bm{3}}_4, \ 
\bm{3}_1\bm{1}_2\overline{\bm{3}}_3\overline{\bm{1}}_4, \ 
\bm{3}_1\bm{2}_2\overline{\bm{3}}_3\overline{\bm{2}}_4.
\end{align}
The corresponding integer vectors are given by
\begin{subequations}
\begin{align}
M^{(4,4,\overline{4},\overline{4})}_\textrm{2B;1} = \begin{pmatrix}
1 & 0 \\
0 & 1 \\
1 & 0 \\
0 & 1
\end{pmatrix}, \quad
\Lambda^{(4,4,\overline{4},\overline{4})}_{\textsf{w},\textrm{2B;1}} = \begin{pmatrix}
2 & 0 & -2 & 0 \\
-1 & 0 & -1 & 0 \\
0 & 2 & 0 & -2 \\
0 & -1 & 0 & -1
\end{pmatrix}, \\
M^{(4,4,\overline{4},\overline{4})}_\textrm{2B;2} = \begin{pmatrix}
1 & 0 \\
0 & 1 \\
1 & 0 \\
0 & 3
\end{pmatrix}, \quad
\Lambda^{(4,4,\overline{4},\overline{4})}_{\textsf{w},\textrm{2B;2}} \
= \begin{pmatrix}
2 & 0 & -2 & 0 \\
-1 & 0 & -1 & 0 \\
0 & -2 & 0 & -2 \\
0 & 1 & 0 & -1
\end{pmatrix}, \\
M^{(4,4,\overline{4},\overline{4})}_\textrm{2B;3} = \begin{pmatrix}
1 & 0 \\
0 & 1 \\
3 & 0 \\
0 & 1
\end{pmatrix}, \quad
\Lambda^{(4,4,\overline{4},\overline{4})}_{\textsf{w},\textrm{2B;3}} \
= \begin{pmatrix}
0 & 2 & 0 & -2 \\
0 & -1 & 0 & -1 \\
-2 & 0 & -2 & 0 \\
1 & 0 & -1 & 0
\end{pmatrix}, \\
M^{(4,4,\overline{4},\overline{4})}_\textrm{2B;4} = \begin{pmatrix}
1 & 0 \\
0 & 1 \\
3 & 0 \\
0 & 3
\end{pmatrix}, \quad
\Lambda^{(4,4,\overline{4},\overline{4})}_{\textsf{w},\textrm{2B;4}} = \begin{pmatrix}
-2 & 0 & -2 & 0 \\
1 & 0 & -1 & 0 \\
0 & -2 & 0 & -2 \\
0 & 1 & 0 & -1
\end{pmatrix}, \\
M^{(4,4,\overline{4},\overline{4})}_\textrm{2B;5} = \begin{pmatrix}
0 & 1 \\
1 & 0 \\
1 & 0 \\
0 & 1
\end{pmatrix}, \quad
\Lambda^{(4,4,\overline{4},\overline{4})}_{\textsf{w},\textrm{2B;5}} = \begin{pmatrix}
2 & 0 & 0 & -2 \\
-1 & 0 & 0 & -1 \\
0 & 2 & -2 & 0 \\
0 & -1 & -1 & 0
\end{pmatrix}, \\
M^{(4,4,\overline{4},\overline{4})}_\textrm{2B;6} = \begin{pmatrix}
0 & 1 \\
1 & 0 \\
3 & 0 \\
0 & 1
\end{pmatrix}, \quad
\Lambda^{(4,4,\overline{4},\overline{4})}_{\textsf{w},\textrm{2B;6}} = \begin{pmatrix}
2 & 0 & 0 & -2 \\
-1 & 0 & 0 & -1 \\
0 & -2 & -2 & 0 \\
0 & 1 & -1 & 0
\end{pmatrix}, \\
M^{(4,4,\overline{4},\overline{4})}_\textrm{2B;7} = \begin{pmatrix}
0 & 1 \\
1 & 0 \\
1 & 0 \\
0 & 3
\end{pmatrix}, \quad
\Lambda^{(4,4,\overline{4},\overline{4})}_{\textsf{w},\textrm{2B;7}} = \begin{pmatrix}
0 & 2 & -2 & 0 \\
0 & -1 & -1 & 0 \\
-2 & 0 & 0 & -2 \\
1 & 0 & 0 & -1
\end{pmatrix}, \\
M^{(4,4,\overline{4},\overline{4})}_\textrm{2B;8} = \begin{pmatrix}
0 & 1 \\
1 & 0 \\
3 & 0 \\
0 & 3
\end{pmatrix}, \quad
\Lambda^{(4,4,\overline{4},\overline{4})}_{\textsf{w},\textrm{2B;8}} \
= \begin{pmatrix}
0 & -2 & -2 & 0 \\
0 & 1 & -1 & 0 \\
-2 & 0 & 0 & -2 \\
1 & 0 & 0 & -1
\end{pmatrix}.
\end{align}
\end{subequations}

The fermionic gapped interfaces of the rank-2 Lagrangian subgroup are obtained by condensing fermionic anyons, which take the form of triplet consisting of one fermionic anyon ($\bm{2}$ or $\overline{\bm{2}}$) and two spin-1/8 anyons with opposite chiralities ($\bm{1}$, $\overline{\bm{1}}$, $\bm{3}$, or $\overline{\bm{3}}$). 
Taking the Lagrangian subgroup corresponding to $M^{(4,4,\overline{4},\overline{4})}_\textrm{2F;1}$ given below as an instance, the set of condensed anyons at the interface is given by
\begin{align}
& \bm{2}_1\overline{\bm{2}}_3, \ 
\bm{2}_2\overline{\bm{2}}_4, \ 
\bm{1}_1\bm{2}_2\overline{\bm{1}}_3, \ 
\bm{3}_1\bm{2}_2\overline{\bm{3}}_3, \ 
\bm{2}_1\bm{1}_2\overline{\bm{1}}_4, \ 
\bm{2}_1\bm{3}_2\overline{\bm{3}}_4, \ 
\red{\bm{1}_1\overline{\bm{1}}_3\overline{\bm{2}}_4}, \ 
\bm{3}_1\overline{\bm{3}}_3\overline{\bm{2}}_4, \ 
\red{\bm{1}_2\overline{\bm{2}}_3\overline{\bm{1}}_4}, \ 
\bm{3}_2\overline{\bm{2}}_3\overline{\bm{3}}_4, \nonumber \\
& \bm{1}_1\bm{1}_2\overline{\bm{3}}_3\overline{\bm{3}}_4, \ 
\bm{1}_1\bm{3}_2\overline{\bm{3}}_3\overline{\bm{1}}_4, \ 
\bm{2}_1\bm{2}_2\overline{\bm{2}}_3\overline{\bm{2}}_4, \ 
\bm{3}_1\bm{1}_2\overline{\bm{1}}_3\overline{\bm{3}}_4, \ 
\bm{3}_1\bm{3}_2\overline{\bm{1}}_3\overline{\bm{1}}_4
\end{align}
The corresponding integer vectors are given by
\begin{subequations}
\begin{align}
M^{(4,4,\overline{4},\overline{4})}_\textrm{2F;1} = \begin{pmatrix}
0 & 1 \\
1 & 0 \\
2 & 1 \\
1 & 2
\end{pmatrix}, \quad
\Lambda^{(4,4,\overline{4},\overline{4})}_{\textsf{w},\textrm{2F;1}} = \begin{pmatrix}
-1 & 0 & -1 & -2 \\
-1 & 2 & -1 & 0 \\
0 & 1 & -2 & 1 \\
2 & 1 & 0 & 1
\end{pmatrix}, \\
M^{(4,4,\overline{4},\overline{4})}_\textrm{2F;2} = \begin{pmatrix}
0 & 1 \\
1 & 0 \\
2 & 1 \\
3 & 2
\end{pmatrix}, \quad
\Lambda^{(4,4,\overline{4},\overline{4})}_{\textsf{w},\textrm{2F;2}} \
= \begin{pmatrix}
-1 & 0 & -1 & 2 \\
1 & -2 & 1 & 0 \\
0 & 1 & -2 & -1 \\
-2 & -1 & 0 & 1
\end{pmatrix}, \\
M^{(4,4,\overline{4},\overline{4})}_\textrm{2F;3} = \begin{pmatrix}
0 & 1 \\
1 & 0 \\
1 & 2 \\
2 & 1
\end{pmatrix}, \quad
\Lambda^{(4,4,\overline{4},\overline{4})}_{\textsf{w},\textrm{2F;3}} = \begin{pmatrix}
0 & -1 & -1 & -2 \\
2 & -1 & -1 & 0 \\
1 & 0 & -2 & 1 \\
1 & 2 & 0 & 1
\end{pmatrix}, \\
M^{(4,4,\overline{4},\overline{4})}_\textrm{2F;4} = \begin{pmatrix}
0 & 1 \\
1 & 0 \\
3 & 2 \\
2 & 1
\end{pmatrix}, \quad
\Lambda^{(4,4,\overline{4},\overline{4})}_{\textsf{w},\textrm{2F;4}} = \begin{pmatrix}
0 & 1 & -1 & -2 \\
1 & -2 & 0 & 1 \\
1 & 0 & -2 & 1 \\
2 & 1 & -1 & 0
\end{pmatrix}, \\
M^{(4,4,\overline{4},\overline{4})}_\textrm{2F;5} = \begin{pmatrix}
0 & 1 \\
1 & 0 \\
1 & 2 \\
2 & 3
\end{pmatrix}, \quad
\Lambda^{(4,4,\overline{4},\overline{4})}_{\textsf{w},\textrm{2F;5}} \
= \begin{pmatrix}
-1 & 0 & -2 & 1 \\
-2 & -1 & -1 & 0 \\
0 & 1 & 1 & 2 \\
1 & -2 & 0 & -1
\end{pmatrix}, \\
M^{(4,4,\overline{4},\overline{4})}_\textrm{2F;6} = \begin{pmatrix}
0 & 1 \\
1 & 0 \\
3 & 2 \\
2 & 3
\end{pmatrix}, \quad
\Lambda^{(4,4,\overline{4},\overline{4})}_{\textsf{w},\textrm{2F;6}} \
= \begin{pmatrix}
-1 & 0 & 2 & 1 \\
2 & 1 & -1 & 0 \\
0 & 1 & -1 & 2 \\
-1 & 2 & 0 & 1
\end{pmatrix}, \\
M^{(4,4,\overline{4},\overline{4})}_\textrm{2F;7} = \begin{pmatrix}
0 & 1 \\
1 & 0 \\
2 & 3 \\
1 & 2
\end{pmatrix}, \quad
\Lambda^{(4,4,\overline{4},\overline{4})}_{\textsf{w},\textrm{2F;7}} = \begin{pmatrix}
0 & 1 & 2 & 1 \\
2 & 1 & 0 & 1 \\
1 & 0 & -1 & 2 \\
1 & -2 & -1 & 0
\end{pmatrix}, \\
M^{(4,4,\overline{4},\overline{4})}_\textrm{2F;8} = \begin{pmatrix}
0 & 1 \\
1 & 0 \\
2 & 3 \\
3 & 2
\end{pmatrix}, \quad
\Lambda^{(4,4,\overline{4},\overline{4})}_{\textsf{w},\textrm{2F;8}} = \begin{pmatrix}
-1 & 0 & 1 & 2 \\
1 & -2 & -1 & 0 \\
0 & -1 & -2 & 1 \\
2 & 1 & 0 & -1
\end{pmatrix}.
\end{align}
\end{subequations}

There are 10 distinct choices of the Lagrangian subgroup of rank three: two of them are bosonic, whereas eight of them are fermionic. 
The bosonic gapped interfaces of the rank-3 Lagrangian subgroup corresponds to the condensate of a quadruplet of spin-1/8 anyons and any pairs of fermionic anyons. 
Taking the Lagrangian subgroup corresponding to $M^{(4,4,\overline{4},\overline{4})}_\textrm{3B;1}$ given below as an instance, the set of condensed anyons at the interface is given by
\begin{align}
& \bm{2}_1\bm{2}_2, \ 
\bm{2}_1\overline{\bm{2}}_3, \ 
\bm{2}_1\overline{\bm{2}}_4, \ 
\bm{2}_2\overline{\bm{2}}_3, \ 
\red{\bm{2}_2\overline{\bm{2}}_4}, \ 
\red{\overline{\bm{2}}_3\overline{\bm{2}}_4}, \ 
\red{\bm{1}_1\bm{1}_2\overline{\bm{1}}_3\overline{\bm{1}}_4},
\bm{1}_1\bm{1}_2\overline{\bm{3}}_3\overline{\bm{3}}_4, \ 
\bm{1}_1\bm{3}_2\overline{\bm{1}}_3\overline{\bm{3}}_4, \ 
\bm{1}_1\bm{3}_2\overline{\bm{3}}_3\overline{\bm{1}}_4, \nonumber \\
& \bm{2}_1\bm{2}_2\overline{\bm{2}}_3\overline{\bm{2}}_4, \ 
\bm{3}_1\bm{1}_2\overline{\bm{1}}_3\overline{\bm{3}}_4, \ 
\bm{3}_1\bm{1}_2\overline{\bm{3}}_3\overline{\bm{1}}_4, \ 
\bm{3}_1\bm{3}_2\overline{\bm{1}}_3\overline{\bm{1}}_4, \ 
\bm{3}_1\bm{3}_2\overline{\bm{3}}_3\overline{\bm{3}}_4.
\end{align}
The corresponding integer vectors are given by
\begin{subequations}
\begin{align}
M^{(4,4,\overline{4},\overline{4})}_\textrm{3B;1} = \begin{pmatrix}
0 & 0 & 1 \\
0 & 2 & 1 \\
2 & 0 & 1 \\
2 & 2 & 1
\end{pmatrix}, \quad
\Lambda^{(4,4,\overline{4},\overline{4})}_{\textsf{w},\textrm{3B;1}} = \begin{pmatrix}
1 & 1 & -1 & -1 \\
-1 & -1 & -1 & -1 \\
-1 & 1 & 1 & -1 \\
1 & -1 & 1 & -1
\end{pmatrix}, \\
M^{(4,4,\overline{4},\overline{4})}_\textrm{3B;2} = \begin{pmatrix}
0 & 0 & 1 \\
0 & 2 & 1 \\
2 & 0 & 1 \\
2 & 2 & 3
\end{pmatrix}, \quad
\Lambda^{(4,4,\overline{4},\overline{4})}_{\textsf{w},\textrm{3B;2}} = \begin{pmatrix}
-1 & 1 & 1 & 1 \\
1 & -1 & 1 & 1 \\
1 & 1 & 1 & -1 \\
-1 & -1 & 1 & -1
\end{pmatrix}.
\end{align}
\end{subequations}

The fermionic interfaces of the rank-3 Lagrangian subgroup are decoupled stacks of three gapped interfaces: a bosonic gapped interface between $U(1)_4$ and $\overline{U(1)}_4$ by condensation of $\bm{1} \overline{\bm{1}}$, a fermionic gapped interface between $U(1)_4$ and a Chern insulator $\overline{U(1)}_1$, and a fermionic gapped interface between $\overline{U(1)}_4$ and a Chern insulator $U(1)_1$. 
Taking the Lagrangian subgroup corresponding to $M^{(4,4,\overline{4},\overline{4})}_\textrm{3F;1}$ given below as an instance, the set of condensed anyons at the interface is given by
\begin{align}
& \red{\bm{2}_2}, \ 
\red{\overline{\bm{2}}_4}, \ 
\red{\bm{1}_1\overline{\bm{1}}_3}, \ 
\bm{2}_1\overline{\bm{2}}_3, \ 
\bm{3}_1\overline{\bm{3}}_3, \ 
\bm{2}_2\overline{\bm{2}}_4, \nonumber \\
& \bm{1}_1\bm{2}_2\overline{\bm{1}}_3, \ 
\bm{2}_1\bm{2}_2\overline{\bm{2}}_3, \ 
\bm{3}_1\bm{2}_2\overline{\bm{3}}_3, \ 
\bm{1}_1\overline{\bm{1}}_3\overline{\bm{2}}_4, \ 
\bm{2}_1\overline{\bm{2}}_3\overline{\bm{2}}_4, \ 
\bm{3}_1\overline{\bm{3}}_3\overline{\bm{2}}_4, \ 
\bm{1}_1\bm{2}_2\overline{\bm{1}}_3\overline{\bm{2}}_4, \ 
\bm{2}_1\bm{2}_2\overline{\bm{2}}_3\overline{\bm{2}}_4, \ 
\bm{3}_1\bm{2}_2\overline{\bm{3}}_3\overline{\bm{2}}_4
\end{align}
The corresponding integer vectors are given by
\begin{subequations}
\begin{align}
M^{(4,4,\overline{4},\overline{4})}_\textrm{3F;1} = \begin{pmatrix}
0 & 0 & 1 \\
0 & 2 & 0 \\
0 & 0 & 1 \\
2 & 0 & 0
\end{pmatrix}, \quad
\Lambda^{(4,4,\overline{4},\overline{4})}_{\textsf{w},\textrm{3F;1}} = \begin{pmatrix}
1 & 2 & -3 & 0 \\
-2 & -2 & 2 & 0 \\
0 & 0 & 0 & -2 \\
1 & -2 & 1 & 0
\end{pmatrix}, \\
M^{(4,4,\overline{4},\overline{4})}_\textrm{3F;2} = \begin{pmatrix}
0 & 0 & 1 \\
0 & 2 & 0 \\
0 & 0 & 3 \\
2 & 0 & 0
\end{pmatrix}, \quad
\Lambda^{(4,4,\overline{4},\overline{4})}_{\textsf{w},\textrm{3F;2}} \
= \begin{pmatrix}
-1 & 2 & -3 & 0 \\
2 & -2 & 2 & 0 \\
0 & 0 & 0 & -2 \\
-1 & -2 & 1 & 0
\end{pmatrix}, \\
M^{(4,4,\overline{4},\overline{4})}_\textrm{3F;3} = \begin{pmatrix}
0 & 0 & 1 \\
0 & 2 & 0 \\
2 & 0 & 0 \\
0 & 0 & 1
\end{pmatrix}, \quad
\Lambda^{(4,4,\overline{4},\overline{4})}_{\textsf{w},\textrm{3F;3}} = \begin{pmatrix}
1 & 2 & 0 & -3 \\
-2 & -2 & 0 & 2 \\
0 & 0 & -2 & 0 \\
1 & -2 & 0 & 1
\end{pmatrix}, \\
M^{(4,4,\overline{4},\overline{4})}_\textrm{3F;4} = \begin{pmatrix}
0 & 0 & 1 \\
0 & 2 & 0 \\
2 & 0 & 0 \\
0 & 0 & 3
\end{pmatrix}, \quad
\Lambda^{(4,4,\overline{4},\overline{4})}_{\textsf{w},\textrm{3F;4}} \
= \begin{pmatrix}
-1 & 2 & 0 & -3 \\
2 & -2 & 0 & 2 \\
0 & 0 & -2 & 0 \\
-1 & -2 & 0 & 1
\end{pmatrix}, \\
M^{(4,4,\overline{4},\overline{4})}_\textrm{3F;5} = \begin{pmatrix}
0 & 0 & 2 \\
0 & 1 & 0 \\
0 & 1 & 0 \\
2 & 0 & 0
\end{pmatrix}, \quad
\Lambda^{(4,4,\overline{4},\overline{4})}_{\textsf{w},\textrm{3F;5}} \
= \begin{pmatrix}
0 & 1 & 1 & -2 \\
0 & -3 & 1 & 2 \\
0 & 2 & -2 & -2 \\
2 & 0 & 0 & 0
\end{pmatrix}, \\
M^{(4,4,\overline{4},\overline{4})}_\textrm{3F;6} = \begin{pmatrix}
0 & 0 & 2 \\
0 & 1 & 0 \\
0 & 3 & 0 \\
2 & 0 & 0
\end{pmatrix}, \quad
\Lambda^{(4,4,\overline{4},\overline{4})}_{\textsf{w},\textrm{3F;6}} \
= \begin{pmatrix}
2 & -1 & -3 & 0 \\
-2 & 2 & 2 & 0 \\
0 & 0 & 0 & -2 \\
-2 & -1 & 1 & 0
\end{pmatrix}, \\
M^{(4,4,\overline{4},\overline{4})}_\textrm{3F;7} = \begin{pmatrix}
0 & 0 & 2 \\
0 & 1 & 0 \\
2 & 0 & 0 \\
0 & 1 & 0
\end{pmatrix}, \quad
\Lambda^{(4,4,\overline{4},\overline{4})}_{\textsf{w},\textrm{3F;7}} \
= \begin{pmatrix}
2 & 1 & 0 & -3 \\
-2 & -2 & 0 & 2 \\
0 & 0 & -2 & 0 \\
-2 & 1 & 0 & 1
\end{pmatrix}, \\
M^{(4,4,\overline{4},\overline{4})}_\textrm{3F;8} = \begin{pmatrix}
0 & 0 & 2 \\
0 & 1 & 0 \\
2 & 0 & 0 \\
0 & 3 & 0
\end{pmatrix}, \quad
\Lambda^{(4,4,\overline{4},\overline{4})}_{\textsf{w},\textrm{3F;8}} \
= \begin{pmatrix}
2 & -1 & 0 & -3 \\
-2 & 2 & 0 & 2 \\
0 & 0 & -2 & 0 \\
-2 & -1 & 0 & 1
\end{pmatrix}.
\end{align}
\end{subequations}

There is only one fermionic gapped interface of the rank-4 Lagrangian subgroup, which is just a stack of individual fermionic gapped interfaces between $U(1)_4$ or $\overline{U(1)}_4$ and a Chern insulator $U(1)_1$ or $\overline{U(1)}_1$. 
The set of condensed anyons at the interface is given by
\begin{align}
\red{\bm{2}_1}, \ 
\red{\bm{2}_2}, \ 
\red{\overline{\bm{2}}_3}, \ 
\red{\overline{\bm{2}}_4}, \ 
\bm{2}_1\bm{2}_2, \ 
\bm{2}_1\overline{\bm{2}}_3, \ 
\bm{2}_1\overline{\bm{2}}_4, \ 
\bm{2}_2\overline{\bm{2}}_3, \ 
\bm{2}_2\overline{\bm{2}}_4, \ 
\overline{\bm{2}}_3\overline{\bm{2}}_4, \ 
\bm{2}_1\bm{2}_2\overline{\bm{2}}_3, \ 
\bm{2}_1\bm{2}_2\overline{\bm{2}}_4, \ 
\bm{2}_1\overline{\bm{2}}_3\overline{\bm{2}}_4, \ 
\bm{2}_2\overline{\bm{2}}_3\overline{\bm{2}}_4, \ 
\bm{2}_1\bm{2}_2\overline{\bm{2}}_3\overline{\bm{2}}_4.
\end{align}
The corresponding integer vectors are given by 
\begin{align}
M^{(4,4,\overline{4},\overline{4})}_\textrm{4F;1} = \begin{pmatrix}
0 & 0 & 0 & 2 \\
0 & 0 & 2 & 0 \\
0 & 2 & 0 & 0 \\
2 & 0 & 0 & 0
\end{pmatrix}, \quad
\Lambda^{(4,4,\overline{4},\overline{4})}_{\textsf{w},\textrm{4F;1}} \
= \begin{pmatrix}
0 & 0 & -2 & 0 \\
-2 & 0 & 0 & 0 \\
0 & 0 & 0 & -2 \\
0 & -2 & 0 & 0
\end{pmatrix}.
\end{align}

\subsubsection{$U(1)_5 \times U(1)_5 \times \overline{U(1)}_5 \times \overline{U(1)}_5$}

We consider gapped interfaces of the fermionic topological order corresponding to the $K$ matrix:
\begin{align}
K^{(5,5,\overline{5},\overline{5})} = \begin{pmatrix} 5 &&& \\ & 5 && \\ && -5 & \\ &&& -5 \end{pmatrix}.
\end{align}
There are 12 distinct choices of the Lagrangian subgroups of rank two.
Eight of them are obtained by stacking two fermionic gapped interfaces between $U(1)_5$ and $\overline{U(1)}_5$ with tunneling- or pairing-type interactions, which lead to the condensations of $\bm{1} \overline{\bm{4}}$ or $\bm{1} \overline{\bm{1}}$. 
Taking the Lagrangian subgroup corresponding to $M^{(5,5,\overline{5},\overline{5})}_\textrm{2B;1}$ given below as an instance, the set of condensed anyons is given by
\begin{align}
&\red{\bm{1}_1\overline{\bm{1}}_3}, \ 
\bm{2}_1\overline{\bm{2}}_3, \ 
\bm{3}_1\overline{\bm{3}}_3, \ 
\bm{4}_1\overline{\bm{4}}_3, \ 
\red{\bm{1}_2\overline{\bm{1}}_4}, \ 
\bm{2}_2\overline{\bm{2}}_4, \ 
\bm{3}_2\overline{\bm{3}}_4, \ 
\bm{4}_2\overline{\bm{4}}_4, \nonumber \\
&\bm{1}_1\bm{1}_2\overline{\bm{1}}_3\overline{\bm{1}}_4, \ 
\bm{1}_1\bm{2}_2\overline{\bm{1}}_3\overline{\bm{2}}_4, \ 
\bm{1}_1\bm{3}_2\overline{\bm{1}}_3\overline{\bm{3}}_4, \ 
\bm{1}_1\bm{4}_2\overline{\bm{1}}_3\overline{\bm{4}}_4, \ 
\bm{2}_1\bm{1}_2\overline{\bm{2}}_3\overline{\bm{1}}_4, \ 
\bm{2}_1\bm{2}_2\overline{\bm{2}}_3\overline{\bm{2}}_4, \ 
\bm{2}_1\bm{3}_2\overline{\bm{2}}_3\overline{\bm{3}}_4, \ 
\bm{2}_1\bm{4}_2\overline{\bm{2}}_3\overline{\bm{4}}_4, \nonumber \\
&\bm{3}_1\bm{1}_2\overline{\bm{3}}_3\overline{\bm{1}}_4, \ 
\bm{3}_1\bm{2}_2\overline{\bm{3}}_3\overline{\bm{2}}_4, \ 
\bm{3}_1\bm{3}_2\overline{\bm{3}}_3\overline{\bm{3}}_4, \ 
\bm{3}_1\bm{4}_2\overline{\bm{3}}_3\overline{\bm{4}}_4, \ 
\bm{4}_1\bm{1}_2\overline{\bm{4}}_3\overline{\bm{1}}_4, \ 
\bm{4}_1\bm{2}_2\overline{\bm{4}}_3\overline{\bm{2}}_4, \ 
\bm{4}_1\bm{3}_2\overline{\bm{4}}_3\overline{\bm{3}}_4, \ 
\bm{4}_1\bm{4}_2\overline{\bm{4}}_3\overline{\bm{4}}_4.
\end{align}
The corresponding integer vectors are given by
\begin{subequations}
\begin{align}
M^{(5,5,\overline{5},\overline{5})}_\textrm{2F;1} = \begin{pmatrix}
0 & 1 \\
1 & 0 \\
0 & 1 \\
1 & 0
\end{pmatrix}, \quad
\Lambda^{(5,5,\overline{5},\overline{5})}_{\textsf{w},\textrm{2F;1}} \
= \begin{pmatrix}
2 & 0 & -3 & 0 \\
-3 & 0 & 2 & 0 \\
0 & 2 & 0 & -3 \\
0 & -3 & 0 & 2
\end{pmatrix}, \\
M^{(5,5,\overline{5},\overline{5})}_\textrm{2F;2} = \begin{pmatrix}
0 & 1 \\
1 & 0 \\
0 & 1 \\
4 & 0
\end{pmatrix}, \quad
\Lambda^{(5,5,\overline{5},\overline{5})}_{\textsf{w},\textrm{2F;2}} \
= \begin{pmatrix}
2 & 0 & -3 & 0 \\
-3 & 0 & 2 & 0 \\
0 & -2 & 0 & -3 \\
0 & -3 & 0 & -2
\end{pmatrix}, \\
M^{(5,5,\overline{5},\overline{5})}_\textrm{2F;3} = \begin{pmatrix}
0 & 1 \\
1 & 0 \\
0 & 4 \\
1 & 0
\end{pmatrix}, \quad
\Lambda^{(5,5,\overline{5},\overline{5})}_{\textsf{w},\textrm{2F;3}} \
= \begin{pmatrix}
-2 & 0 & -3 & 0 \\
0 & -3 & 0 & 2 \\
0 & 2 & 0 & -3 \\
-3 & 0 & -2 & 0
\end{pmatrix}, \\
M^{(5,5,\overline{5},\overline{5})}_\textrm{2F;4} = \begin{pmatrix}
0 & 1 \\
1 & 0 \\
0 & 4 \\
4 & 0
\end{pmatrix}, \quad
\Lambda^{(5,5,\overline{5},\overline{5})}_{\textsf{w},\textrm{2F;4}} = \begin{pmatrix}
-2 & 0 & -3 & 0 \\
-3 & 0 & -2 & 0 \\
0 & -2 & 0 & -3 \\
0 & -3 & 0 & -2
\end{pmatrix}, \\
M^{(5,5,\overline{5},\overline{5})}_\textrm{2F;5} = \begin{pmatrix}
0 & 1 \\
1 & 0 \\
1 & 0 \\
0 & 1
\end{pmatrix}, \quad
\Lambda^{(5,5,\overline{5},\overline{5})}_{\textsf{w},\textrm{2F;5}} = \begin{pmatrix}
0 & 2 & -3 & 0 \\
-3 & 0 & 0 & 2 \\
2 & 0 & 0 & -3 \\
0 & -3 & 2 & 0
\end{pmatrix}, \\
M^{(5,5,\overline{5},\overline{5})}_\textrm{2F;6} = \begin{pmatrix}
0 & 1 \\
1 & 0 \\
4 & 0 \\
0 & 1
\end{pmatrix}, \quad
\Lambda^{(5,5,\overline{5},\overline{5})}_{\textsf{w},\textrm{2F;6}} \
= \begin{pmatrix}
0 & -2 & -3 & 0 \\
-3 & 0 & 0 & 2 \\
2 & 0 & 0 & -3 \\
0 & -3 & -2 & 0
\end{pmatrix}, \\
M^{(5,5,\overline{5},\overline{5})}_\textrm{2F;7} = \begin{pmatrix}
0 & 1 \\
1 & 0 \\
1 & 0 \\
0 & 4
\end{pmatrix}, \quad
\Lambda^{(5,5,\overline{5},\overline{5})}_{\textsf{w},\textrm{2F;7}} = \begin{pmatrix}
0 & 2 & -3 & 0 \\
0 & -3 & 2 & 0 \\
-2 & 0 & 0 & -3 \\
-3 & 0 & 0 & -2
\end{pmatrix}, \\
M^{(5,5,\overline{5},\overline{5})}_\textrm{2F;8} = \begin{pmatrix}
0 & 1 \\
1 & 0 \\
4 & 0 \\
0 & 4
\end{pmatrix}, \quad
\Lambda^{(5,5,\overline{5},\overline{5})}_{\textsf{w},\textrm{2F;8}} \
= \begin{pmatrix}
0 & -2 & -3 & 0 \\
0 & -3 & -2 & 0 \\
-2 & 0 & 0 & -3 \\
-3 & 0 & 0 & -2
\end{pmatrix}.
\end{align}
\end{subequations}

The other four are obtained by stacking two gapped interfaces between $U(1)_5$'s with the same chirality of the edge modes at which $\bm{1} \bm{2}$ (fermionic anyon) or $\bm{1} \bm{3}$ (bosonic anyon) are condensed. 
Taking the Lagrangian subgroup corresponding to $M^{(5,5,\overline{5},\overline{5})}_\textrm{2F;9}$ given below as an instance, the set of condensed anyons is given by
\begin{align}
&\red{\bm{1}_1\bm{2}_2}, \ 
\bm{2}_1\bm{4}_2, \ 
\bm{3}_1\bm{1}_2, \ 
\bm{4}_1\bm{3}_2, \ 
\red{\overline{\bm{1}}_3\overline{\bm{2}}_4}, \ 
\overline{\bm{2}}_3\overline{\bm{4}}_4, \ 
\overline{\bm{3}}_3\overline{\bm{1}}_4, \ 
\overline{\bm{4}}_3\overline{\bm{3}}_4, \nonumber \\
&\bm{1}_1\bm{2}_2\overline{\bm{1}}_3\overline{\bm{2}}_4, \ 
\bm{1}_1\bm{2}_2\overline{\bm{2}}_3\overline{\bm{4}}_4, \ 
\bm{1}_1\bm{2}_2\overline{\bm{3}}_3\overline{\bm{1}}_4, \ 
\bm{1}_1\bm{2}_2\overline{\bm{4}}_3\overline{\bm{3}}_4, \ 
\bm{2}_1\bm{4}_2\overline{\bm{1}}_3\overline{\bm{2}}_4, \ 
\bm{2}_1\bm{4}_2\overline{\bm{2}}_3\overline{\bm{4}}_4, \ 
\bm{2}_1\bm{4}_2\overline{\bm{3}}_3\overline{\bm{1}}_4, \ 
\bm{2}_1\bm{4}_2\overline{\bm{4}}_3\overline{\bm{3}}_4, \nonumber \\
&\bm{3}_1\bm{1}_2\overline{\bm{1}}_3\overline{\bm{2}}_4, \ 
\bm{3}_1\bm{1}_2\overline{\bm{2}}_3\overline{\bm{4}}_4, \ 
\bm{3}_1\bm{1}_2\overline{\bm{3}}_3\overline{\bm{1}}_4, \ 
\bm{3}_1\bm{1}_2\overline{\bm{4}}_3\overline{\bm{3}}_4, \ 
\bm{4}_1\bm{3}_2\overline{\bm{1}}_3\overline{\bm{2}}_4, \ 
\bm{4}_1\bm{3}_2\overline{\bm{2}}_3\overline{\bm{4}}_4, \ 
\bm{4}_1\bm{3}_2\overline{\bm{3}}_3\overline{\bm{1}}_4, \ 
\bm{4}_1\bm{3}_2\overline{\bm{4}}_3\overline{\bm{3}}_4.
\end{align}
The corresponding integer vectors are given by
\begin{subequations}
\begin{align}
M^{(5,5,\overline{5},\overline{5})}_\textrm{2F;9} = \begin{pmatrix}
0 & 1 \\
0 & 2 \\
1 & 0 \\
2 & 0
\end{pmatrix}, \quad
\Lambda^{(5,5,\overline{5},\overline{5})}_{\textsf{w},\textrm{2F;9}} \
= \begin{pmatrix}
0 & 0 & -1 & -2 \\
-1 & -2 & 0 & 0 \\
0 & 0 & 2 & -1 \\
2 & -1 & 0 & 0
\end{pmatrix}, \\
M^{(5,5,\overline{5},\overline{5})}_\textrm{2F;10} = \begin{pmatrix}
0 & 1 \\
0 & 2 \\
1 & 0 \\
3 & 0
\end{pmatrix}, \quad
\Lambda^{(5,5,\overline{5},\overline{5})}_{\textsf{w},\textrm{2F;10}} \
= \begin{pmatrix}
0 & 0 & 2 & 1 \\
-1 & -2 & 0 & 0 \\
0 & 0 & 1 & -2 \\
2 & -1 & 0 & 0
\end{pmatrix}, \\
M^{(5,5,\overline{5},\overline{5})}_\textrm{2F;11} = \begin{pmatrix}
0 & 1 \\
0 & 3 \\
1 & 0 \\
2 & 0
\end{pmatrix}, \quad
\Lambda^{(5,5,\overline{5},\overline{5})}_{\textsf{w},\textrm{2F;11}} \
= \begin{pmatrix}
0 & 0 & -1 & -2 \\
2 & 1 & 0 & 0 \\
0 & 0 & 2 & -1 \\
1 & -2 & 0 & 0
\end{pmatrix}, \\
M^{(5,5,\overline{5},\overline{5})}_\textrm{2F;12} = \begin{pmatrix}
0 & 1 \\
0 & 3 \\
1 & 0 \\
3 & 0
\end{pmatrix}, \quad
\Lambda^{(5,5,\overline{5},\overline{5})}_{\textsf{w},\textrm{2F;12}} = \begin{pmatrix}
0 & 0 & 2 & 1 \\
2 & 1 & 0 & 0 \\
0 & 0 & 1 & -2 \\
1 & -2 & 0 & 0
\end{pmatrix}.
\end{align}
\end{subequations}

\subsubsection{$U(1)_6 \times U(1)_6 \times \overline{U(1)}_6 \times \overline{U(1)}_6$}

We consider gapped interfaces of the bosonic topological order corresponding to the $K$ matrix:
\begin{align}
K^{(6,6,\overline{6},\overline{6})} = \begin{pmatrix} 6 &&& \\ & 6 && \\ && -6 & \\ &&& -6 \end{pmatrix}.
\end{align}
There are 24 gapped interfaces corresponding to the rank-2 Lagrangian subgroups. 
Eight of them are bosonic gapped interfaces simply obtained by stacking two interfaces between $U(1)_6$ and $\overline{U(1)}_6$ with tunneling- or paring-type interactions. 
The corresponding integer vectors are given by
\begin{subequations}
\begin{align}
M^{(6,6,\overline{6},\overline{6})}_\textrm{2B;1} = \begin{pmatrix}
0 & 1 \\
1 & 0 \\
0 & 1 \\
1 & 0
\end{pmatrix}, \quad
\Lambda^{(6,6,\overline{6},\overline{6})}_{\textsf{w},\textrm{2B;1}} \
= \begin{pmatrix}
3 & 0 & -3 & 0 \\
-1 & 0 & -1 & 0 \\
0 & 3 & 0 & -3 \\
0 & -1 & 0 & -1
\end{pmatrix}, \\
M^{(6,6,\overline{6},\overline{6})}_\textrm{2B;2} = \begin{pmatrix}
0 & 1 \\
1 & 0 \\
0 & 1 \\
5 & 0
\end{pmatrix}, \quad
\Lambda^{(6,6,\overline{6},\overline{6})}_{\textsf{w},\textrm{2B;2}} = \begin{pmatrix}
3 & 0 & -3 & 0 \\
-1 & 0 & -1 & 0 \\
0 & -3 & 0 & -3 \\
0 & 1 & 0 & -1
\end{pmatrix}, \\M^{(6,6,\overline{6},\overline{6})}_\textrm{2B;3} = \begin{pmatrix}
0 & 1 \\
1 & 0 \\
0 & 5 \\
1 & 0
\end{pmatrix}, \quad
\Lambda^{(6,6,\overline{6},\overline{6})}_{\textsf{w},\textrm{2B;3}} \
= \begin{pmatrix}
0 & 3 & 0 & -3 \\
0 & -1 & 0 & -1 \\
-3 & 0 & -3 & 0 \\
1 & 0 & -1 & 0
\end{pmatrix}, \\
M^{(6,6,\overline{6},\overline{6})}_\textrm{2B;4} = \begin{pmatrix}
0 & 1 \\
1 & 0 \\
0 & 5 \\
5 & 0
\end{pmatrix}, \quad
\Lambda^{(6,6,\overline{6},\overline{6})}_{\textsf{w},\textrm{2B;4}} = \begin{pmatrix}
-3 & 0 & -3 & 0 \\
1 & 0 & -1 & 0 \\
0 & -3 & 0 & -3 \\
0 & 1 & 0 & -1
\end{pmatrix}, \\
M^{(6,6,\overline{6},\overline{6})}_\textrm{2B;5} = \begin{pmatrix}
0 & 1 \\
1 & 0 \\
1 & 0 \\
0 & 1
\end{pmatrix}, \quad
\Lambda^{(6,6,\overline{6},\overline{6})}_{\textsf{w},\textrm{2B;5}} = \begin{pmatrix}
3 & 0 & 0 & -3 \\
-1 & 0 & 0 & -1 \\
0 & 3 & -3 & 0 \\
0 & -1 & -1 & 0
\end{pmatrix}, \\
M^{(6,6,\overline{6},\overline{6})}_\textrm{2B;6} = \begin{pmatrix}
0 & 1 \\
1 & 0 \\
5 & 0 \\
0 & 1
\end{pmatrix}, \quad
\Lambda^{(6,6,\overline{6},\overline{6})}_{\textsf{w},\textrm{2B;6}} \
= \begin{pmatrix}
3 & 0 & 0 & -3 \\
-1 & 0 & 0 & -1 \\
0 & -3 & -3 & 0 \\
0 & 1 & -1 & 0
\end{pmatrix}, \\
M^{(6,6,\overline{6},\overline{6})}_\textrm{2B;7} = \begin{pmatrix}
0 & 1 \\
1 & 0 \\
1 & 0 \\
0 & 5
\end{pmatrix}, \quad
\Lambda^{(6,6,\overline{6},\overline{6})}_{\textsf{w},\textrm{2B;7}} \
= \begin{pmatrix}
0 & 3 & -3 & 0 \\
0 & -1 & -1 & 0 \\
-3 & 0 & 0 & -3 \\
1 & 0 & 0 & -1
\end{pmatrix}, \\
M^{(6,6,\overline{6},\overline{6})}_\textrm{2B;8} = \begin{pmatrix}
0 & 1 \\
1 & 0 \\
5 & 0 \\
0 & 5
\end{pmatrix}, \quad
\Lambda^{(6,6,\overline{6},\overline{6})}_{\textsf{w},\textrm{2B;8}} = \begin{pmatrix}
0 & -3 & -3 & 0 \\
0 & 1 & -1 & 0 \\
-3 & 0 & 0 & -3 \\
1 & 0 & 0 & -1
\end{pmatrix}.
\end{align}
\end{subequations}

There are also the other eight bosonic gapped interfaces, which are obtained by condensing triples of anyons such as $\bm{1} \overline{\bm{2}} \overline{\bm{3}}$. 
Taking the Lagrangian subgroup generated by $M^{(6,6,\overline{6},\overline{6})}_\textrm{2B;9}$ given below as an instance, the set of condensed anyons is given by
\begin{align}
& \bm{2}_1\overline{\bm{4}}_3, \ 
\bm{4}_1\overline{\bm{2}}_3, \ 
\bm{3}_1\overline{\bm{3}}_4, \ 
\bm{3}_2\overline{\bm{3}}_3, \ 
\bm{4}_2\overline{\bm{2}}_4, \ 
\bm{2}_2\overline{\bm{4}}_4, \nonumber \\ 
&\bm{2}_1\bm{3}_2\overline{\bm{1}}_3, \ 
\bm{4}_1\bm{3}_2\overline{\bm{5}}_3, \ 
\bm{3}_1\bm{2}_2\overline{\bm{1}}_4, \ 
\bm{3}_1\bm{4}_2\overline{\bm{5}}_4, \ 
\red{\bm{1}_1\overline{\bm{2}}_3\overline{\bm{3}}_4}, \ 
\bm{5}_1\overline{\bm{4}}_3\overline{\bm{3}}_4, \ 
\red{\bm{1}_2\overline{\bm{3}}_3\overline{\bm{2}}_4}, \ 
\bm{5}_2\overline{\bm{3}}_3\overline{\bm{4}}_4, \nonumber \\
&\bm{1}_1\bm{1}_2\overline{\bm{5}}_3\overline{\bm{5}}_4, \ 
\bm{1}_1\bm{2}_2\overline{\bm{2}}_3\overline{\bm{1}}_4, \ 
\bm{1}_1\bm{3}_2\overline{\bm{5}}_3\overline{\bm{3}}_4, \ 
\bm{1}_1\bm{4}_2\overline{\bm{2}}_3\overline{\bm{5}}_4, \ 
\bm{1}_1\bm{5}_2\overline{\bm{5}}_3\overline{\bm{1}}_4, \ 
\bm{2}_1\bm{1}_2\overline{\bm{1}}_3\overline{\bm{2}}_4, \ 
\bm{2}_1\bm{2}_2\overline{\bm{4}}_3\overline{\bm{4}}_4, \ 
\bm{2}_1\bm{4}_2\overline{\bm{4}}_3\overline{\bm{2}}_4, \ 
\bm{2}_1\bm{5}_2\overline{\bm{1}}_3\overline{\bm{4}}_4, \nonumber \\
&\bm{3}_1\bm{1}_2\overline{\bm{3}}_3\overline{\bm{5}}_4, \ 
\bm{3}_1\bm{3}_2\overline{\bm{3}}_3\overline{\bm{3}}_4, \ 
\bm{3}_1\bm{5}_2\overline{\bm{3}}_3\overline{\bm{1}}_4, \ 
\bm{4}_1\bm{1}_2\overline{\bm{5}}_3\overline{\bm{2}}_4, \ 
\bm{4}_1\bm{2}_2\overline{\bm{2}}_3\overline{\bm{4}}_4, \ 
\bm{4}_1\bm{4}_2\overline{\bm{2}}_3\overline{\bm{2}}_4, \ 
\bm{4}_1\bm{5}_2\overline{\bm{5}}_3\overline{\bm{4}}_4, \nonumber \\
&\bm{5}_1\bm{1}_2\overline{\bm{1}}_3\overline{\bm{5}}_4, \ 
\bm{5}_1\bm{2}_2\overline{\bm{4}}_3\overline{\bm{1}}_4, \ 
\bm{5}_1\bm{3}_2\overline{\bm{1}}_3\overline{\bm{3}}_4, \ 
\bm{5}_1\bm{4}_2\overline{\bm{4}}_3\overline{\bm{5}}_4, \ 
\bm{5}_1\bm{5}_2\overline{\bm{1}}_3\overline{\bm{1}}_4.
\end{align}
The corresponding integer vectors are given by
\begin{subequations}
\begin{align}
M^{(6,6,\overline{6},\overline{6})}_\textrm{2B;9} = \begin{pmatrix}
0 & 1 \\
1 & 0 \\
3 & 2 \\
2 & 3
\end{pmatrix}, \quad
\Lambda^{(6,6,\overline{6},\overline{6})}_{\textsf{w},\textrm{2B;9}} \
= \begin{pmatrix}
-1 & 2 & -2 & 1 \\
1 & -1 & -1 & 1 \\
-2 & -1 & -1 & -2 \\
1 & 1 & -1 & -1
\end{pmatrix}, \\
M^{(6,6,\overline{6},\overline{6})}_\textrm{2B;10} = \begin{pmatrix}
0 & 1 \\
1 & 0 \\
3 & 2 \\
4 & 3
\end{pmatrix}, \quad
\Lambda^{(6,6,\overline{6},\overline{6})}_{\textsf{w},\textrm{2B;10}} = \begin{pmatrix}
-1 & 2 & -2 & -1 \\
1 & -1 & -1 & -1 \\
-2 & -1 & -1 & 2 \\
1 & 1 & -1 & 1
\end{pmatrix}, \\
M^{(6,6,\overline{6},\overline{6})}_\textrm{2B;11} = \begin{pmatrix}
0 & 1 \\
1 & 0 \\
2 & 3 \\
3 & 2
\end{pmatrix}, \quad
\Lambda^{(6,6,\overline{6},\overline{6})}_{\textsf{w},\textrm{2B;11}} = \begin{pmatrix}
-1 & 2 & 1 & -2 \\
1 & -1 & 1 & -1 \\
-2 & -1 & -2 & -1 \\
1 & 1 & -1 & -1
\end{pmatrix}, \\
M^{(6,6,\overline{6},\overline{6})}_\textrm{2B;12} = \begin{pmatrix}
0 & 1 \\
1 & 0 \\
4 & 3 \\
3 & 2
\end{pmatrix}, \quad
\Lambda^{(6,6,\overline{6},\overline{6})}_{\textsf{w},\textrm{2B;12}} = \begin{pmatrix}
-2 & 1 & -2 & -1 \\
1 & -1 & -1 & -1 \\
-1 & -2 & 1 & -2 \\
1 & 1 & 1 & -1
\end{pmatrix}, \\
M^{(6,6,\overline{6},\overline{6})}_\textrm{2B;13} = \begin{pmatrix}
0 & 1 \\
1 & 0 \\
2 & 3 \\
3 & 4
\end{pmatrix}, \quad
\Lambda^{(6,6,\overline{6},\overline{6})}_{\textsf{w},\textrm{2B;13}} \
= \begin{pmatrix}
-2 & 1 & 2 & 1 \\
1 & -1 & 1 & 1 \\
-1 & -2 & -1 & 2 \\
1 & 1 & -1 & 1
\end{pmatrix}, \\
M^{(6,6,\overline{6},\overline{6})}_\textrm{2B;14} = \begin{pmatrix}
0 & 1 \\
1 & 0 \\
4 & 3 \\
3 & 4
\end{pmatrix}, \quad
\Lambda^{(6,6,\overline{6},\overline{6})}_{\textsf{w},\textrm{2B;14}} \
= \begin{pmatrix}
-2 & 1 & -2 & 1 \\
1 & -1 & -1 & 1 \\
-1 & -2 & 1 & 2 \\
1 & 1 & 1 & 1
\end{pmatrix}, \\
M^{(6,6,\overline{6},\overline{6})}_\textrm{2B;15} = \begin{pmatrix}
0 & 1 \\
1 & 0 \\
3 & 4 \\
2 & 3
\end{pmatrix}, \quad
\Lambda^{(6,6,\overline{6},\overline{6})}_{\textsf{w},\textrm{2B;15}} = \begin{pmatrix}
-1 & 2 & 2 & 1 \\
1 & -1 & 1 & 1 \\
-2 & -1 & 1 & -2 \\
1 & 1 & 1 & -1
\end{pmatrix}, \\
M^{(6,6,\overline{6},\overline{6})}_\textrm{2B;16} = \begin{pmatrix}
0 & 1 \\
1 & 0 \\
3 & 4 \\
4 & 3
\end{pmatrix}, \quad
\Lambda^{(6,6,\overline{6},\overline{6})}_{\textsf{w},\textrm{2B;16}} = \begin{pmatrix}
-2 & 1 & 1 & -2 \\
1 & -1 & 1 & -1 \\
-1 & -2 & 2 & 1 \\
1 & 1 & 1 & 1
\end{pmatrix}.
\end{align}
\end{subequations}

The last eight ones are fermionic gapped interfaces, which are obtained by condensing triples of anyons asuch as $\bm{2} \overline{\bm{1}} \overline{\bm{3}}$. 
Taking the Lagrangian subgroup generated by $M^{(6,6,\overline{6},\overline{6})}_\textrm{2F;1}$ given below as an instance, the set of condensed anyons is given by
\begin{align}
&\bm{3}_1\bm{3}_2, \ 
\bm{2}_1\overline{\bm{2}}_3, \ 
\bm{4}_1\overline{\bm{4}}_3, \ 
\bm{2}_2\overline{\bm{2}}_4, \
\bm{4}_2\overline{\bm{4}}_4, \  
\overline{\bm{3}}_3\overline{\bm{3}}_4, \nonumber \\
&\red{\bm{1}_1\bm{3}_2\overline{\bm{4}}_3}, \ 
\bm{5}_1\bm{3}_2\overline{\bm{2}}_3, \ 
\bm{3}_1\bm{1}_2\overline{\bm{4}}_4, \ 
\bm{3}_1\bm{5}_2\overline{\bm{2}}_4, \ 
\bm{2}_1\overline{\bm{5}}_3\overline{\bm{3}}_4, \ 
\bm{4}_1\overline{\bm{1}}_3\overline{\bm{3}}_4, \ 
\bm{2}_2\overline{\bm{3}}_3\overline{\bm{5}}_4, \ 
\red{\bm{4}_2\overline{\bm{3}}_3\overline{\bm{1}}_4}, \nonumber \\
&\bm{1}_1\bm{1}_2\overline{\bm{1}}_3\overline{\bm{1}}_4, \ 
\bm{1}_1\bm{1}_2\overline{\bm{4}}_3\overline{\bm{4}}_4, \ 
\bm{1}_1\bm{3}_2\overline{\bm{1}}_3\overline{\bm{3}}_4, \ 
\bm{1}_1\bm{5}_2\overline{\bm{1}}_3\overline{\bm{5}}_4, \ 
\bm{1}_1\bm{5}_2\overline{\bm{4}}_3\overline{\bm{2}}_4, \ 
\bm{2}_1\bm{2}_2\overline{\bm{2}}_3\overline{\bm{2}}_4, \ 
\bm{2}_1\bm{2}_2\overline{\bm{5}}_3\overline{\bm{5}}_4, \ 
\bm{2}_1\bm{4}_2\overline{\bm{2}}_3\overline{\bm{4}}_4, \ 
\bm{2}_1\bm{4}_2\overline{\bm{5}}_3\overline{\bm{1}}_4, \nonumber \\
&\bm{3}_1\bm{1}_2\overline{\bm{3}}_3\overline{\bm{1}}_4, \ 
\bm{3}_1\bm{3}_2\overline{\bm{3}}_3\overline{\bm{3}}_4, \ 
\bm{3}_1\bm{5}_2\overline{\bm{3}}_3\overline{\bm{5}}_4, \ 
\bm{4}_1\bm{2}_2\overline{\bm{1}}_3\overline{\bm{5}}_4, \ 
\bm{4}_1\bm{2}_2\overline{\bm{4}}_3\overline{\bm{2}}_4, \ 
\bm{4}_1\bm{4}_2\overline{\bm{1}}_3\overline{\bm{1}}_4, \ 
\bm{4}_1\bm{4}_2\overline{\bm{4}}_3\overline{\bm{4}}_4, \nonumber \\
&\bm{5}_1\bm{1}_2\overline{\bm{2}}_3\overline{\bm{4}}_4, \ 
\bm{5}_1\bm{1}_2\overline{\bm{5}}_3\overline{\bm{1}}_4, \ 
\bm{5}_1\bm{3}_2\overline{\bm{5}}_3\overline{\bm{3}}_4, \ 
\bm{5}_1\bm{5}_2\overline{\bm{2}}_3\overline{\bm{2}}_4, \ 
\bm{5}_1\bm{5}_2\overline{\bm{5}}_3\overline{\bm{5}}_4.
\end{align}
The corresponding integer vectors are given by
\begin{subequations}
\begin{align}
M^{(6,6,\overline{6},\overline{6})}_\textrm{2F;1} = \begin{pmatrix}
1 & 0 \\
3 & 4 \\
4 & 3 \\
0 & 1
\end{pmatrix}, \quad
\Lambda^{(6,6,\overline{6},\overline{6})}_{\textsf{w},\textrm{2F;1}} \
= \begin{pmatrix}
1 & 1 & -2 & -2 \\
2 & 2 & -1 & -1 \\
-1 & 1 & 2 & -2 \\
-2 & 2 & 1 & -1
\end{pmatrix}, \\
M^{(6,6,\overline{6},\overline{6})}_\textrm{2F;2} = \begin{pmatrix}
1 & 0 \\
3 & 4 \\
0 & 1 \\
4 & 3
\end{pmatrix}, \quad
\Lambda^{(6,6,\overline{6},\overline{6})}_{\textsf{w},\textrm{2F;2}} = \begin{pmatrix}
1 & 1 & -2 & -2 \\
2 & 2 & -1 & -1 \\
-1 & 1 & -2 & 2 \\
2 & -2 & 1 & -1
\end{pmatrix}, \\
M^{(6,6,\overline{6},\overline{6})}_\textrm{2F;3} = \begin{pmatrix}
1 & 0 \\
3 & 2 \\
4 & 3 \\
0 & 1
\end{pmatrix}, \quad
\Lambda^{(6,6,\overline{6},\overline{6})}_{\textsf{w},\textrm{2F;3}} = \begin{pmatrix}
-1 & 1 & 2 & 2 \\
2 & -2 & -1 & -1 \\
-1 & -1 & 2 & -2 \\
2 & 2 & -1 & 1
\end{pmatrix}, \\
M^{(6,6,\overline{6},\overline{6})}_\textrm{2F;4} = \begin{pmatrix}
1 & 0 \\
3 & 4 \\
0 & 1 \\
2 & 3
\end{pmatrix}, \quad
\Lambda^{(6,6,\overline{6},\overline{6})}_{\textsf{w},\textrm{2F;4}} = \begin{pmatrix}
1 & 1 & -2 & 2 \\
2 & -2 & 1 & 1 \\
-1 & 1 & -2 & -2 \\
-2 & -2 & 1 & -1
\end{pmatrix}, \\
M^{(6,6,\overline{6},\overline{6})}_\textrm{2F;5} = \begin{pmatrix}
1 & 0 \\
3 & 2 \\
0 & 1 \\
2 & 3
\end{pmatrix}, \quad
\Lambda^{(6,6,\overline{6},\overline{6})}_{\textsf{w},\textrm{2F;5}} \
= \begin{pmatrix}
-1 & -1 & -2 & -2 \\
-2 & 2 & 1 & -1 \\
-1 & 1 & 2 & -2 \\
2 & 2 & 1 & 1
\end{pmatrix}, \\
M^{(6,6,\overline{6},\overline{6})}_\textrm{2F;6} = \begin{pmatrix}
1 & 0 \\
3 & 2 \\
2 & 3 \\
0 & 1
\end{pmatrix}, \quad
\Lambda^{(6,6,\overline{6},\overline{6})}_{\textsf{w},\textrm{2F;6}} = \begin{pmatrix}
-1 & -1 & -2 & -2 \\
2 & 2 & 1 & 1 \\
1 & -1 & 2 & -2 \\
-2 & 2 & -1 & 1
\end{pmatrix}, \\
M^{(6,6,\overline{6},\overline{6})}_\textrm{2F;7} = \begin{pmatrix}
1 & 0 \\
3 & 2 \\
0 & 1 \\
4 & 3
\end{pmatrix}, \quad
\Lambda^{(6,6,\overline{6},\overline{6})}_{\textsf{w},\textrm{2F;7}} = \begin{pmatrix}
1 & -1 & -2 & -2 \\
2 & -2 & -1 & -1 \\
-1 & -1 & -2 & 2 \\
2 & 2 & 1 & -1
\end{pmatrix}, \\
M^{(6,6,\overline{6},\overline{6})}_\textrm{2F;8} = \begin{pmatrix}
1 & 0 \\
3 & 4 \\
2 & 3 \\
0 & 1
\end{pmatrix}, \quad
\Lambda^{(6,6,\overline{6},\overline{6})}_{\textsf{w},\textrm{2F;8}} \
= \begin{pmatrix}
-1 & 1 & -2 & -2 \\
2 & -2 & 1 & 1 \\
1 & 1 & 2 & -2 \\
2 & 2 & 1 & -1
\end{pmatrix}.
\end{align}
\end{subequations}

\subsubsection{$U(1)_7 \times U(1)_7 \times \overline{U(1)}_7 \times \overline{U(1)}_7$}
\label{sec:U7U7U7U7}

We consider gapped interfaces of the fermionic topological order corresponding to the $K$ matrix:
\begin{align}
K^{(7,7,\overline{7},\overline{7})} = \begin{pmatrix} 7 &&& \\ & 7 && \\ && -7 & \\ &&& -7 \end{pmatrix}.
\end{align}
There are 16 distinct choices of the Lagrangian subgroups of rank two.
Eight of them are obtained by stacking two fermionic gapped interfaces between $U(1)_7$ and $\overline{U(1)}_7$ with tunneling- or pairing-type interactions, which lead to the condensations of $\bm{1} \overline{\bm{6}}$ or $\bm{1} \overline{\bm{1}}$. 
The corresponding integer vectors are given by
\begin{subequations}
\begin{align}
M^{(7,7,\overline{7},\overline{7})}_\textrm{2F;1} = \begin{pmatrix}
1 & 0 \\
0 & 1 \\
1 & 0 \\
0 & 1
\end{pmatrix}, \quad
\Lambda^{(7,7,\overline{7},\overline{7})}_{\textsf{w},\textrm{2F;1}} = \begin{pmatrix}
3 & 0 & -4 & 0 \\
4 & 0 & -3 & 0 \\
0 & 3 & 0 & -4 \\
0 & 4 & 0 & -3
\end{pmatrix}, \\
M^{(7,7,\overline{7},\overline{7})}_\textrm{2F;2} = \begin{pmatrix}
1 & 0 \\
0 & 1 \\
1 & 0 \\
0 & 6
\end{pmatrix}, \quad
\Lambda^{(7,7,\overline{7},\overline{7})}_{\textsf{w},\textrm{2F;2}} \
= \begin{pmatrix}
3 & 0 & -4 & 0 \\
4 & 0 & -3 & 0 \\
0 & -3 & 0 & -4 \\
0 & 4 & 0 & 3
\end{pmatrix}, \\
M^{(7,7,\overline{7},\overline{7})}_\textrm{2F;3} = \begin{pmatrix}
1 & 0 \\
0 & 1 \\
6 & 0 \\
0 & 1
\end{pmatrix}, \quad
\Lambda^{(7,7,\overline{7},\overline{7})}_{\textsf{w},\textrm{2F;3}} = \begin{pmatrix}
-3 & 0 & -4 & 0 \\
0 & 4 & 0 & -3 \\
0 & 3 & 0 & -4 \\
4 & 0 & 3 & 0
\end{pmatrix}, \\
M^{(7,7,\overline{7},\overline{7})}_\textrm{2F;4} = \begin{pmatrix}
1 & 0 \\
0 & 1 \\
6 & 0 \\
0 & 6
\end{pmatrix}, \quad
\Lambda^{(7,7,\overline{7},\overline{7})}_{\textsf{w},\textrm{2F;4}} \
= \begin{pmatrix}
-3 & 0 & -4 & 0 \\
4 & 0 & 3 & 0 \\
0 & -3 & 0 & -4 \\
0 & 4 & 0 & 3
\end{pmatrix}, \\
M^{(7,7,\overline{7},\overline{7})}_\textrm{2F;5} = \begin{pmatrix}
1 & 0 \\
0 & 1 \\
0 & 1 \\
1 & 0
\end{pmatrix}, \quad
\Lambda^{(7,7,\overline{7},\overline{7})}_{\textsf{w},\textrm{2F;5}} = \begin{pmatrix}
3 & 0 & 0 & -4 \\
0 & 4 & -3 & 0 \\
0 & 3 & -4 & 0 \\
4 & 0 & 0 & -3
\end{pmatrix}, \\
M^{(7,7,\overline{7},\overline{7})}_\textrm{2F;6} = \begin{pmatrix}
1 & 0 \\
0 & 1 \\
0 & 6 \\
1 & 0
\end{pmatrix}, \quad
\Lambda^{(7,7,\overline{7},\overline{7})}_{\textsf{w},\textrm{2F;6}} \
= \begin{pmatrix}
0 & -3 & -4 & 0 \\
4 & 0 & 0 & -3 \\
3 & 0 & 0 & -4 \\
0 & 4 & 3 & 0
\end{pmatrix}, \\
M^{(7,7,\overline{7},\overline{7})}_\textrm{2F;7} = \begin{pmatrix}
1 & 0 \\
0 & 1 \\
0 & 1 \\
6 & 0
\end{pmatrix}, \quad
\Lambda^{(7,7,\overline{7},\overline{7})}_{\textsf{w},\textrm{2F;7}} = \begin{pmatrix}
0 & 3 & -4 & 0 \\
0 & 4 & -3 & 0 \\
-3 & 0 & 0 & -4 \\
4 & 0 & 0 & 3
\end{pmatrix}, \\
M^{(7,7,\overline{7},\overline{7})}_\textrm{2F;8} = \begin{pmatrix}
1 & 0 \\
0 & 1 \\
0 & 6 \\
6 & 0
\end{pmatrix}, \quad
\Lambda^{(7,7,\overline{7},\overline{7})}_{\textsf{w},\textrm{2F;8}} = \begin{pmatrix}
0 & -3 & -4 & 0 \\
0 & -4 & -3 & 0 \\
-3 & 0 & 0 & -4 \\
4 & 0 & 0 & 3
\end{pmatrix}.
\end{align}
\end{subequations}

The other eight are obtained by condensing triplets of anyons such as $\bm{1} \overline{\bm{2}} \overline{\bm{2}}$ (fermion) or $\bm{1} \overline{\bm{2}} \overline{\bm{5}}$ (boson) at the interface. 
Taking the Lagrangian subgroup generated by $M^{(7,7,\overline{7},\overline{7})}_\textrm{2F;9}$ given below as an instance, the set of condensed anyons is given by
\begin{align}
&\bm{1}_1\bm{1}_2\overline{\bm{4}}_3, \ 
\bm{2}_1\bm{2}_2\overline{\bm{1}}_3, \ 
\bm{3}_1\bm{3}_2\overline{\bm{5}}_3, \ 
\bm{4}_1\bm{4}_2\overline{\bm{2}}_3, \ 
\bm{5}_1\bm{5}_2\overline{\bm{6}}_3, \ 
\bm{6}_1\bm{6}_2\overline{\bm{3}}_3, \ 
\bm{1}_1\bm{6}_2\overline{\bm{4}}_4, \ 
\bm{2}_1\bm{5}_2\overline{\bm{1}}_4, \ 
\bm{3}_1\bm{4}_2\overline{\bm{5}}_4, \ 
\bm{4}_1\bm{3}_2\overline{\bm{2}}_4, \ 
\bm{5}_1\bm{2}_2\overline{\bm{6}}_4, \ 
\bm{6}_1\bm{1}_2\overline{\bm{3}}_4, \nonumber \\
&\red{\bm{1}_1\overline{\bm{2}}_3\overline{\bm{2}}_4}, \ 
\bm{2}_1\overline{\bm{4}}_3\overline{\bm{4}}_4, \ 
\bm{3}_1\overline{\bm{6}}_3\overline{\bm{6}}_4, \ 
\bm{4}_1\overline{\bm{1}}_3\overline{\bm{1}}_4, \ 
\bm{5}_1\overline{\bm{3}}_3\overline{\bm{3}}_4, \ 
\bm{6}_1\overline{\bm{5}}_3\overline{\bm{5}}_4, \ 
\red{\bm{1}_2\overline{\bm{2}}_3\overline{\bm{5}}_4}, \ 
\bm{2}_2\overline{\bm{4}}_3\overline{\bm{3}}_4, \ 
\bm{3}_2\overline{\bm{6}}_3\overline{\bm{1}}_4, \ 
\bm{4}_2\overline{\bm{1}}_3\overline{\bm{6}}_4, \ 
\bm{5}_2\overline{\bm{3}}_3\overline{\bm{4}}_4, \ 
\bm{6}_2\overline{\bm{5}}_3\overline{\bm{2}}_4, \ 
\nonumber \\
&\bm{1}_1\bm{2}_2\overline{\bm{6}}_3\overline{\bm{5}}_4, \ 
\bm{1}_1\bm{3}_2\overline{\bm{1}}_3\overline{\bm{3}}_4, \ 
\bm{1}_1\bm{4}_2\overline{\bm{3}}_3\overline{\bm{1}}_4, \ 
\bm{1}_1\bm{5}_2\overline{\bm{5}}_3\overline{\bm{6}}_4, \ 
\bm{2}_1\bm{1}_2\overline{\bm{6}}_3\overline{\bm{2}}_4, \ 
\bm{2}_1\bm{3}_2\overline{\bm{3}}_3\overline{\bm{5}}_4, \ 
\bm{2}_1\bm{4}_2\overline{\bm{5}}_3\overline{\bm{3}}_4, \ 
\bm{2}_1\bm{6}_2\overline{\bm{2}}_3\overline{\bm{6}}_4, \nonumber \\
&\bm{3}_1\bm{1}_2\overline{\bm{1}}_3\overline{\bm{4}}_4, \ 
\bm{3}_1\bm{2}_2\overline{\bm{3}}_3\overline{\bm{2}}_4, \ 
\bm{3}_1\bm{5}_2\overline{\bm{2}}_3\overline{\bm{3}}_4, \ 
\bm{3}_1\bm{6}_2\overline{\bm{4}}_3\overline{\bm{1}}_4, \ 
\bm{4}_1\bm{1}_2\overline{\bm{3}}_3\overline{\bm{6}}_4, \ 
\bm{4}_1\bm{2}_2\overline{\bm{5}}_3\overline{\bm{4}}_4, \ 
\bm{4}_1\bm{5}_2\overline{\bm{4}}_3\overline{\bm{5}}_4, \ 
\bm{4}_1\bm{6}_2\overline{\bm{6}}_3\overline{\bm{3}}_4, \nonumber \\ 
&\bm{5}_1\bm{1}_2\overline{\bm{5}}_3\overline{\bm{1}}_4, \ 
\bm{5}_1\bm{3}_2\overline{\bm{2}}_3\overline{\bm{4}}_4, \ 
\bm{5}_1\bm{4}_2\overline{\bm{4}}_3\overline{\bm{2}}_4, \ 
\bm{5}_1\bm{6}_2\overline{\bm{1}}_3\overline{\bm{5}}_4, \ 
\bm{6}_1\bm{2}_2\overline{\bm{2}}_3\overline{\bm{1}}_4, \ 
\bm{6}_1\bm{3}_2\overline{\bm{4}}_3\overline{\bm{6}}_4, \ 
\bm{6}_1\bm{4}_2\overline{\bm{6}}_3\overline{\bm{4}}_4, \ 
\bm{6}_1\bm{5}_2\overline{\bm{1}}_3\overline{\bm{2}}_4.
\end{align}
Thus, it contains only triplets or quadruplets of anyons but no pairs.
The corresponding integer vectors are given by
\begin{subequations}
\begin{align}
M^{(7,7,\overline{7},\overline{7})}_\textrm{2F;9} = \begin{pmatrix}
1 & 0 \\
0 & 1 \\
2 & 2 \\
2 & 5
\end{pmatrix}, \quad
\Lambda^{(7,7,\overline{7},\overline{7})}_{\textsf{w},\textrm{2F;9}} = \begin{pmatrix}
-1 & 0 & -2 & -2 \\
-2 & 2 & 0 & -1 \\
0 & 1 & 2 & -2 \\
2 & 2 & 1 & 0
\end{pmatrix}, \\
M^{(7,7,\overline{7},\overline{7})}_\textrm{2F;10} = \begin{pmatrix}
1 & 0 \\
0 & 1 \\
2 & 5 \\
2 & 2
\end{pmatrix}, \quad
\Lambda^{(7,7,\overline{7},\overline{7})}_{\textsf{w},\textrm{2F;10}} = \begin{pmatrix}
-1 & 0 & -2 & -2 \\
-2 & 2 & -1 & 0 \\
0 & 1 & -2 & 2 \\
2 & 2 & 0 & 1
\end{pmatrix}, \\
M^{(7,7,\overline{7},\overline{7})}_\textrm{2F;11} = \begin{pmatrix}
1 & 0 \\
0 & 1 \\
2 & 2 \\
5 & 2
\end{pmatrix}, \quad
\Lambda^{(7,7,\overline{7},\overline{7})}_{\textsf{w},\textrm{2F;11}} = \begin{pmatrix}
0 & -1 & -2 & -2 \\
2 & -2 & 0 & -1 \\
1 & 0 & 2 & -2 \\
2 & 2 & 1 & 0
\end{pmatrix}, \\
M^{(7,7,\overline{7},\overline{7})}_\textrm{2F;12} = \begin{pmatrix}
1 & 0 \\
0 & 1 \\
2 & 5 \\
5 & 5
\end{pmatrix}, \quad
\Lambda^{(7,7,\overline{7},\overline{7})}_{\textsf{w},\textrm{2F;12}} \
= \begin{pmatrix}
0 & 1 & -2 & -2 \\
2 & -2 & 1 & 0 \\
1 & 0 & 2 & -2 \\
-2 & -2 & 0 & 1
\end{pmatrix}, \\
M^{(7,7,\overline{7},\overline{7})}_\textrm{2F;13} = \begin{pmatrix}
1 & 0 \\
0 & 1 \\
5 & 2 \\
2 & 2
\end{pmatrix}, \quad
\Lambda^{(7,7,\overline{7},\overline{7})}_{\textsf{w},\textrm{2F;13}} \
= \begin{pmatrix}
0 & -1 & -2 & -2 \\
2 & -2 & -1 & 0 \\
1 & 0 & -2 & 2 \\
2 & 2 & 0 & 1
\end{pmatrix}, \\
M^{(7,7,\overline{7},\overline{7})}_\textrm{2F;14} = \begin{pmatrix}
1 & 0 \\
0 & 1 \\
5 & 5 \\
2 & 5
\end{pmatrix}, \quad
\Lambda^{(7,7,\overline{7},\overline{7})}_{\textsf{w},\textrm{2F;14}} = \begin{pmatrix}
0 & 1 & -2 & -2 \\
2 & -2 & 0 & 1 \\
1 & 0 & -2 & 2 \\
-2 & -2 & 1 & 0
\end{pmatrix}, \\
M^{(7,7,\overline{7},\overline{7})}_\textrm{2F;15} = \begin{pmatrix}
1 & 0 \\
0 & 1 \\
5 & 2 \\
5 & 5
\end{pmatrix}, \quad
\Lambda^{(7,7,\overline{7},\overline{7})}_{\textsf{w},\textrm{2F;15}} = \begin{pmatrix}
1 & 0 & -2 & -2 \\
2 & -2 & -1 & 0 \\
0 & 1 & 2 & -2 \\
-2 & -2 & 0 & 1
\end{pmatrix}, \\
M^{(7,7,\overline{7},\overline{7})}_\textrm{2F;16} = \begin{pmatrix}
1 & 0 \\
0 & 1 \\
5 & 5 \\
5 & 2
\end{pmatrix}, \quad
\Lambda^{(7,7,\overline{7},\overline{7})}_{\textsf{w},\textrm{2F;16}} = \begin{pmatrix}
1 & 0 & -2 & -2 \\
2 & -2 & 0 & -1 \\
0 & 1 & -2 & 2 \\
-2 & -2 & 1 & 0
\end{pmatrix}.
\end{align}
\end{subequations}

\subsubsection{$U(1)_8 \times U(1)_8 \times \overline{U(1)}_8 \times \overline{U(1)}_8$}

We consider gapped interfaces of the bosonic topological order corresponding to the $K$ matrix:
\begin{align}
K^{(8,8,\overline{8},\overline{8})} = \begin{pmatrix} 8 &&& \\ & 8 && \\ && -8 & \\ &&& -8 \end{pmatrix}.
\end{align}
There are 64 Lagrangian subgroups of rank two, 40 of rank three, and three of rank four. 
Among the rank-two interfaces, 16 are bosonic whereas 48 are fermionic. 
Eight bosonic gapped interfaces at rank two are obtained by just stacking two gapped interfaces between $U(1)_8$ and $\overline{U(1)}_8$ with tunneling- or pairing-type interactions, which lead to the condensations of $\bm{1} \overline{\bm{7}}$ or $\bm{1} \overline{\bm{1}}$. 
The corresponding integer vectors are given by
\begin{subequations}
\begin{align}
M^{(8,8,\overline{8},\overline{8})}_\textrm{2B;1} = \begin{pmatrix}
1 & 0 \\
0 & 1 \\
1 & 0 \\
0 & 1
\end{pmatrix}, \quad
\Lambda^{(8,8,\overline{8},\overline{8})}_{\textsf{w},\textrm{2B;1}} = \begin{pmatrix}
4 & 0 & -4 & 0 \\
-1 & 0 & -1 & 0 \\
0 & 4 & 0 & -4 \\
0 & -1 & 0 & -1
\end{pmatrix}, \\
M^{(8,8,\overline{8},\overline{8})}_\textrm{2B;2} = \begin{pmatrix}
1 & 0 \\
0 & 1 \\
1 & 0 \\
0 & 7
\end{pmatrix}, \quad
\Lambda^{(8,8,\overline{8},\overline{8})}_{\textsf{w},\textrm{2B;2}} \
= \begin{pmatrix}
4 & 0 & -4 & 0 \\
-1 & 0 & -1 & 0 \\
0 & -4 & 0 & -4 \\
0 & 1 & 0 & -1
\end{pmatrix}, \\
M^{(8,8,\overline{8},\overline{8})}_\textrm{2B;3} = \begin{pmatrix}
1 & 0 \\
0 & 1 \\
7 & 0 \\
0 & 1
\end{pmatrix}, \quad
\Lambda^{(8,8,\overline{8},\overline{8})}_{\textsf{w},\textrm{2B;3}} = \begin{pmatrix}
0 & 4 & 0 & -4 \\
0 & -1 & 0 & -1 \\
-4 & 0 & -4 & 0 \\
1 & 0 & -1 & 0
\end{pmatrix}, \\
M^{(8,8,\overline{8},\overline{8})}_\textrm{2B;4} = \begin{pmatrix}
1 & 0 \\
0 & 1 \\
7 & 0 \\
0 & 7
\end{pmatrix}, \quad
\Lambda^{(8,8,\overline{8},\overline{8})}_{\textsf{w},\textrm{2B;4}} = \begin{pmatrix}
-4 & 0 & -4 & 0 \\
1 & 0 & -1 & 0 \\
0 & -4 & 0 & -4 \\
0 & 1 & 0 & -1
\end{pmatrix}, \\
M^{(8,8,\overline{8},\overline{8})}_\textrm{2B;5} = \begin{pmatrix}
1 & 0 \\
0 & 1 \\
0 & 1 \\
1 & 0
\end{pmatrix}, \quad
\Lambda^{(8,8,\overline{8},\overline{8})}_{\textsf{w},\textrm{2B;5}} = \begin{pmatrix}
4 & 0 & 0 & -4 \\
-1 & 0 & 0 & -1 \\
0 & 4 & -4 & 0 \\
0 & -1 & -1 & 0
\end{pmatrix}, \\
M^{(8,8,\overline{8},\overline{8})}_\textrm{2B;6} = \begin{pmatrix}
1 & 0 \\
0 & 1 \\
0 & 7 \\
1 & 0
\end{pmatrix}, \quad
\Lambda^{(8,8,\overline{8},\overline{8})}_{\textsf{w},\textrm{2B;6}} \
= \begin{pmatrix}
4 & 0 & 0 & -4 \\
-1 & 0 & 0 & -1 \\
0 & -4 & -4 & 0 \\
0 & 1 & -1 & 0
\end{pmatrix}, \\
M^{(8,8,\overline{8},\overline{8})}_\textrm{2B;7} = \begin{pmatrix}
1 & 0 \\
0 & 1 \\
0 & 1 \\
7 & 0
\end{pmatrix}, \quad
\Lambda^{(8,8,\overline{8},\overline{8})}_{\textsf{w},\textrm{2B;7}} \
= \begin{pmatrix}
0 & 4 & -4 & 0 \\
0 & -1 & -1 & 0 \\
-4 & 0 & 0 & -4 \\
1 & 0 & 0 & -1
\end{pmatrix}, \\
M^{(8,8,\overline{8},\overline{8})}_\textrm{2B;8} = \begin{pmatrix}
1 & 0 \\
0 & 1 \\
0 & 7 \\
7 & 0
\end{pmatrix}, \quad
\Lambda^{(8,8,\overline{8},\overline{8})}_{\textsf{w},\textrm{2B;8}} \
= \begin{pmatrix}
0 & -4 & -4 & 0 \\
0 & 1 & -1 & 0 \\
-4 & 0 & 0 & -4 \\
1 & 0 & 0 & -1
\end{pmatrix}.
\end{align}
\end{subequations}

The other eight bosonic interfaces at rank two are obtained by condensing triplets of anyons such as $\bm{1} \bm{4} \overline{\bm{1}}$. 
Such triplets can be understood as bound objects consisting of pairs of anyons $\bm{1} \overline{\bm{1}}$, which have just appeared in the above gapped interfaces, and bosonic anyons $\bm{4}$. 
Taking the Lagrangian subgroup generated by $M^{(8,8,\overline{8},\overline{8})}_\textrm{2B;9}$ given below as an instance, the set of condensed anyons is given by
\begin{align}
&\bm{2}_1\overline{\bm{2}}_3, \ 
\bm{2}_2\overline{\bm{2}}_4, \ 
\bm{4}_1\overline{\bm{4}}_3, \ 
\bm{4}_2\overline{\bm{4}}_4, \ 
\bm{6}_1\overline{\bm{6}}_3, \ 
\bm{6}_ 2\overline{\bm{6}}_4, \ 
\bm{1}_1\bm{4}_2\overline{\bm{1}}_3, \ 
\bm{3}_1\bm{4}_2\overline{\bm{3}}_3, \ 
\bm{5}_1\bm{4}_2\overline{\bm{5}}_3, \ 
\bm{7}_1\bm{4}_2\overline{\bm{7}}_3, \nonumber \\
&\bm{4}_1\bm{1}_2\overline{\bm{1}}_4, \ 
\bm{4}_1\bm{3}_2\overline{\bm{3}}_4, \ 
\bm{4}_1\bm{5}_2\overline{\bm{5}}_4, \ 
\bm{4}_1\bm{7}_2\overline{\bm{7}}_4, \ 
\red{\bm{1}_1\overline{\bm{1}}_3\overline{\bm{4}}_4}, \ 
\bm{3}_1\overline{\bm{3}}_3\overline{\bm{4}}_4, \ 
\bm{5}_1\overline{\bm{5}}_3\overline{\bm{4}}_4, \ 
\bm{7}_1\overline{\bm{7}}_3\overline{\bm{4}}_4, \ 
\red{\bm{1}_2\overline{\bm{4}}_3\overline{\bm{1}}_4}, \
\bm{3}_2\overline{\bm{4}}_3\overline{\bm{3}}_4, \ 
\bm{5}_2\overline{\bm{4}}_3\overline{\bm{5}}_4, \ 
\bm{7}_2\overline{\bm{4}}_3\overline{\bm{7}}_4, \nonumber \\
&\bm{1}_1\bm{1}_2\overline{\bm{5}}_3\overline{\bm{5}}_4, \ 
\bm{1}_1\bm{2}_2\overline{\bm{1}}_3\overline{\bm{6}}_4, \ 
\bm{1}_1\bm{3}_2\overline{\bm{5}}_3\overline{\bm{7}}_4, \ 
\bm{1}_1\bm{5}_2\overline{\bm{5}}_3\overline{\bm{1}}_4, \ 
\bm{1}_1\bm{6}_2\overline{\bm{1}}_3\overline{\bm{2}}_4, \ 
\bm{1}_1\bm{7}_2\overline{\bm{5}}_3\overline{\bm{3}}_4, \nonumber \\
&\bm{2}_1\bm{1}_2\overline{\bm{6}}_3\overline{\bm{1}}_4, \ 
\bm{2}_1\bm{2}_2\overline{\bm{2}}_3\overline{\bm{2}}_4, \ 
\bm{2}_1\bm{3}_2\overline{\bm{6}}_3\overline{\bm{3}}_4, \ 
\bm{2}_1\bm{4}_2\overline{\bm{2}}_3\overline{\bm{4}}_4, \ 
\bm{2}_1\bm{5}_2\overline{\bm{6}}_3\overline{\bm{5}}_4, \ 
\bm{2}_1\bm{6}_2\overline{\bm{2}}_3\overline{\bm{6}}_4, \ 
\bm{2}_1\bm{7}_2\overline{\bm{6}}_3\overline{\bm{7}}_4, \nonumber \\
&\bm{3}_1\bm{1}_2\overline{\bm{7}}_3\overline{\bm{5}}_4, \ 
\bm{3}_1\bm{2}_2\overline{\bm{3}}_3\overline{\bm{6}}_4, \ 
\bm{3}_1\bm{3}_2\overline{\bm{7}}_3\overline{\bm{7}}_4, \ 
\bm{3}_1\bm{5}_2\overline{\bm{7}}_3\overline{\bm{1}}_4, \ 
\bm{3}_1\bm{6}_2\overline{\bm{3}}_3\overline{\bm{2}}_4, \ 
\bm{3}_1\bm{7}_2\overline{\bm{7}}_3\overline{\bm{3}}_4, \ 
\bm{4}_1\bm{2}_2\overline{\bm{4}}_3\overline{\bm{2}}_4, \ 
\bm{4}_1\bm{4}_2\overline{\bm{4}}_3\overline{\bm{4}}_4, \ 
\bm{4}_1\bm{6}_2\overline{\bm{4}}_3\overline{\bm{6}}_4, \nonumber \\
&\bm{5}_1\bm{1}_2\overline{\bm{1}}_3\overline{\bm{5}}_4, \ 
\bm{5}_1\bm{2}_2\overline{\bm{5}}_3\overline{\bm{6}}_4, \ 
\bm{5}_1\bm{3}_2\overline{\bm{1}}_3\overline{\bm{7}}_4, \ 
\bm{5}_1\bm{5}_2\overline{\bm{1}}_3\overline{\bm{1}}_4, \ 
\bm{5}_1\bm{6}_2\overline{\bm{5}}_3\overline{\bm{2}}_4, \ 
\bm{5}_1\bm{7}_2\overline{\bm{1}}_3\overline{\bm{3}}_4, \nonumber \\
&\bm{6}_1\bm{1}_2\overline{\bm{2}}_3\overline{\bm{1}}_4, \ 
\bm{6}_1\bm{2}_2\overline{\bm{6}}_3\overline{\bm{2}}_4, \ 
\bm{6}_1\bm{3}_2\overline{\bm{2}}_3\overline{\bm{3}}_4, \ 
\bm{6}_1\bm{4}_2\overline{\bm{6}}_3\overline{\bm{4}}_4, \ 
\bm{6}_1\bm{5}_2\overline{\bm{2}}_3\overline{\bm{5}}_4, \ 
\bm{6}_1\bm{6}_2\overline{\bm{6}}_3\overline{\bm{6}}_4, \ 
\bm{6}_1\bm{7}_2\overline{\bm{2}}_3\overline{\bm{7}}_4, \nonumber \\
&\bm{7}_1\bm{1}_2\overline{\bm{3}}_3\overline{\bm{5}}_4, \ 
\bm{7}_1\bm{2}_2\overline{\bm{7}}_3\overline{\bm{6}}_4, \ 
\bm{7}_1\bm{3}_2\overline{\bm{3}}_3\overline{\bm{7}}_4, \  
\bm{7}_1\bm{5}_2\overline{\bm{3}}_3\overline{\bm{1}}_4, \ 
\bm{7}_1\bm{6}_2\overline{\bm{7}}_3\overline{\bm{2}}_4, \ 
\bm{7}_1\bm{7}_2\overline{\bm{3}}_3\overline{\bm{3}}_4.
\end{align}
The corresponding integer vectors are given by
\begin{subequations}
\begin{align}
M^{(8,8,\overline{8},\overline{8})}_\textrm{2B;9} = \begin{pmatrix}
1 & 0 \\
0 & 1 \\
1 & 4 \\
4 & 1
\end{pmatrix}, \quad
\Lambda^{(8,8,\overline{8},\overline{8})}_{\textsf{w},\textrm{2B;9}} \
= \begin{pmatrix}
-1 & -2 & -1 & 2 \\
0 & 2 & 0 & 2 \\
-2 & 1 & 2 & 1 \\
2 & 0 & 2 & 0
\end{pmatrix}, \\
M^{(8,8,\overline{8},\overline{8})}_\textrm{2B;10} = \begin{pmatrix}
1 & 0 \\
0 & 1 \\
1 & 4 \\
4 & 7
\end{pmatrix}, \quad
\Lambda^{(8,8,\overline{8},\overline{8})}_{\textsf{w},\textrm{2B;10}} \
= \begin{pmatrix}
-2 & 0 & -2 & 0 \\
2 & -1 & -2 & 1 \\
-1 & -2 & -1 & -2 \\
0 & 2 & 0 & -2
\end{pmatrix}, \\
M^{(8,8,\overline{8},\overline{8})}_\textrm{2B;11} = \begin{pmatrix}
1 & 0 \\
0 & 1 \\
4 & 1 \\
1 & 4
\end{pmatrix}, \quad
\Lambda^{(8,8,\overline{8},\overline{8})}_{\textsf{w},\textrm{2B;11}} \
= \begin{pmatrix}
-2 & -1 & -1 & 2 \\
2 & 0 & 0 & 2 \\
1 & -2 & 2 & 1 \\
0 & 2 & 2 & 0
\end{pmatrix}, \\
M^{(8,8,\overline{8},\overline{8})}_\textrm{2B;12} = \begin{pmatrix}
1 & 0 \\
0 & 1 \\
4 & 7 \\
1 & 4
\end{pmatrix}, \quad
\Lambda^{(8,8,\overline{8},\overline{8})}_{\textsf{w},\textrm{2B;12}} = \begin{pmatrix}
-2 & 0 & 0 & -2 \\
2 & -1 & 1 & -2 \\
-1 & -2 & -2 & -1 \\
0 & 2 & -2 & 0
\end{pmatrix}, \\
M^{(8,8,\overline{8},\overline{8})}_\textrm{2B;13} = \begin{pmatrix}
1 & 0 \\
0 & 1 \\
4 & 1 \\
7 & 4
\end{pmatrix}, \quad
\Lambda^{(8,8,\overline{8},\overline{8})}_{\textsf{w},\textrm{2B;13}} = \begin{pmatrix}
0 & -2 & -2 & 0 \\
-1 & 2 & -2 & 1 \\
-2 & -1 & -1 & -2 \\
2 & 0 & 0 & -2
\end{pmatrix}, \\
M^{(8,8,\overline{8},\overline{8})}_\textrm{2B;14} = \begin{pmatrix}
1 & 0 \\
0 & 1 \\
4 & 7 \\
7 & 4
\end{pmatrix}, \quad
\Lambda^{(8,8,\overline{8},\overline{8})}_{\textsf{w},\textrm{2B;14}} = \begin{pmatrix}
0 & -2 & 2 & 0 \\
-1 & 2 & 2 & 1 \\
2 & 1 & -1 & 2 \\
-2 & 0 & 0 & 2
\end{pmatrix}, \\
M^{(8,8,\overline{8},\overline{8})}_\textrm{2B;15} = \begin{pmatrix}
1 & 0 \\
0 & 1 \\
7 & 4 \\
4 & 1
\end{pmatrix}, \quad
\Lambda^{(8,8,\overline{8},\overline{8})}_{\textsf{w},\textrm{2B;15}} = \begin{pmatrix}
0 & -2 & 0 & -2 \\
-1 & 2 & 1 & -2 \\
-2 & -1 & -2 & -1 \\
2 & 0 & -2 & 0
\end{pmatrix}, \\
M^{(8,8,\overline{8},\overline{8})}_\textrm{2B;16} = \begin{pmatrix}
1 & 0 \\
0 & 1 \\
7 & 4 \\
4 & 7
\end{pmatrix}, \quad
\Lambda^{(8,8,\overline{8},\overline{8})}_{\textsf{w},\textrm{2B;16}} \
= \begin{pmatrix}
-2 & 0 & 2 & 0 \\
2 & -1 & 2 & 1 \\
1 & 2 & -1 & 2 \\
0 & -2 & 0 & 2
\end{pmatrix}.
\end{align}
\end{subequations}

There are eight fermionic gapped interfaces at rank two obtained by stacking two fermionic gapped interfaces between $U(1)_8$ and $\overline{U(1)}_8$ with $\bm{1} \overline{\bm{3}}$ or $\bm{1} \overline{\bm{5}}$ condensation. 
Taking the Lagrangian subgroup generated by $M^{(8,8,\overline{8},\overline{8})}_\textrm{2F;1}$ given below as an instance, the set of condensed anyons is given by
\begin{align}
&\red{\bm{1}_1\overline{\bm{3}}_3}, \ 
\bm{2}_1\overline{\bm{6}}_3, \ 
\bm{3}_1\overline{\bm{1}}_3, \ 
\bm{4}_1\overline{\bm{4}}_3, \ 
\bm{5}_1\overline{\bm{7}}_3, \ 
\bm{6}_1\overline{\bm{2}}_3, \ 
\bm{7}_1\overline{\bm{5}}_3, \ 
\red{\bm{1}_2\overline{\bm{3}}_4}, \ 
\bm{2}_2\overline{\bm{6}}_4, \ 
\bm{3}_2\overline{\bm{1}}_4, \ 
\bm{4}_2\overline{\bm{4}}_4, \ 
\bm{5}_2\overline{\bm{7}}_4, \ 
\bm{6}_2\overline{\bm{2}}_4, \ 
\bm{7}_2\overline{\bm{5}}_4, \nonumber \\
&\bm{1}_1\bm{1}_2\overline{\bm{3}}_3\overline{\bm{3}}_4, \ 
\bm{1}_1\bm{2}_2\overline{\bm{3}}_3\overline{\bm{6}}_4, \ 
\bm{1}_1\bm{3}_2\overline{\bm{3}}_3\overline{\bm{1}}_4, \ 
\bm{1}_1\bm{4}_2\overline{\bm{3}}_3\overline{\bm{4}}_4, \ 
\bm{1}_1\bm{5}_2\overline{\bm{3}}_3\overline{\bm{7}}_4, \ 
\bm{1}_1\bm{6}_2\overline{\bm{3}}_3\overline{\bm{2}}_4, \ 
\bm{1}_1\bm{7}_2\overline{\bm{3}}_3\overline{\bm{5}}_4, \nonumber \\
&\bm{2}_1\bm{1}_2\overline{\bm{6}}_3\overline{\bm{3}}_4, \ 
\bm{2}_1\bm{2}_2\overline{\bm{6}}_3\overline{\bm{6}}_4, \ 
\bm{2}_1\bm{3}_2\overline{\bm{6}}_3\overline{\bm{1}}_4, \ 
\bm{2}_1\bm{4}_2\overline{\bm{6}}_3\overline{\bm{4}}_4, \ 
\bm{2}_1\bm{5}_2\overline{\bm{6}}_3\overline{\bm{7}}_4, \ 
\bm{2}_1\bm{6}_2\overline{\bm{6}}_3\overline{\bm{2}}_4, \ 
\bm{2}_1\bm{7}_2\overline{\bm{6}}_3\overline{\bm{5}}_4, \nonumber \\
&\bm{3}_1\bm{1}_2\overline{\bm{1}}_3\overline{\bm{3}}_4, \ 
\bm{3}_1\bm{2}_2\overline{\bm{1}}_3\overline{\bm{6}}_4, \ 
\bm{3}_1\bm{3}_2\overline{\bm{1}}_3\overline{\bm{1}}_4, \ 
\bm{3}_1\bm{4}_2\overline{\bm{1}}_3\overline{\bm{4}}_4, \ 
\bm{3}_1\bm{5}_2\overline{\bm{1}}_3\overline{\bm{7}}_4, \ 
\bm{3}_1\bm{6}_2\overline{\bm{1}}_3\overline{\bm{2}}_4, \ 
\bm{3}_1\bm{7}_2\overline{\bm{1}}_3\overline{\bm{5}}_4, \nonumber \\
&\bm{4}_1\bm{1}_2\overline{\bm{4}}_3\overline{\bm{3}}_4, \ 
\bm{4}_1\bm{2}_2\overline{\bm{4}}_3\overline{\bm{6}}_4, \ 
\bm{4}_1\bm{3}_2\overline{\bm{4}}_3\overline{\bm{1}}_4, \ 
\bm{4}_1\bm{4}_2\overline{\bm{4}}_3\overline{\bm{4}}_4, \ 
\bm{4}_1\bm{5}_2\overline{\bm{4}}_3\overline{\bm{7}}_4, \ 
\bm{4}_1\bm{6}_2\overline{\bm{4}}_3\overline{\bm{2}}_4, \ 
\bm{4}_1\bm{7}_2\overline{\bm{4}}_3\overline{\bm{5}}_4, \nonumber \\
&\bm{5}_1\bm{1}_2\overline{\bm{7}}_3\overline{\bm{3}}_4, \ 
\bm{5}_1\bm{2}_2\overline{\bm{7}}_3\overline{\bm{6}}_4, \ 
\bm{5}_1\bm{3}_2\overline{\bm{7}}_3\overline{\bm{1}}_4, \ 
\bm{5}_1\bm{4}_2\overline{\bm{7}}_3\overline{\bm{4}}_4, \ 
\bm{5}_1\bm{5}_2\overline{\bm{7}}_3\overline{\bm{7}}_4, \ 
\bm{5}_1\bm{6}_2\overline{\bm{7}}_3\overline{\bm{2}}_4, \ 
\bm{5}_1\bm{7}_2\overline{\bm{7}}_3\overline{\bm{5}}_4, \nonumber \\
&\bm{6}_1\bm{1}_2\overline{\bm{2}}_3\overline{\bm{3}}_4, \ 
\bm{6}_1\bm{2}_2\overline{\bm{2}}_3\overline{\bm{6}}_4, \ 
\bm{6}_1\bm{3}_2\overline{\bm{2}}_3\overline{\bm{1}}_4, \ 
\bm{6}_1\bm{4}_2\overline{\bm{2}}_3\overline{\bm{4}}_4, \ 
\bm{6}_1\bm{5}_2\overline{\bm{2}}_3\overline{\bm{7}}_4, \ 
\bm{6}_1\bm{6}_2\overline{\bm{2}}_3\overline{\bm{2}}_4, \ 
\bm{6}_1\bm{7}_2\overline{\bm{2}}_3\overline{\bm{5}}_4, \nonumber \\
&\bm{7}_1\bm{1}_2\overline{\bm{5}}_3\overline{\bm{3}}_4, \ 
\bm{7}_1\bm{2}_2\overline{\bm{5}}_3\overline{\bm{6}}_4, \ 
\bm{7}_1\bm{3}_2\overline{\bm{5}}_3\overline{\bm{1}}_4, \ 
\bm{7}_1\bm{4}_2\overline{\bm{5}}_3\overline{\bm{4}}_4, \ 
\bm{7}_1\bm{5}_2\overline{\bm{5}}_3\overline{\bm{7}}_4, \ 
\bm{7}_1\bm{6}_2\overline{\bm{5}}_3\overline{\bm{2}}_4, \ 
\bm{7}_1\bm{7}_2\overline{\bm{5}}_3\overline{\bm{5}}_4.
\end{align}
The corresponding integer vectors are given by 
\begin{subequations}
\begin{align}
M^{(8,8,\overline{8},\overline{8})}_\textrm{2F;1} = \begin{pmatrix}
1 & 0 \\
0 & 1 \\
3 & 0 \\
0 & 3
\end{pmatrix}, \quad
\Lambda^{(8,8,\overline{8},\overline{8})}_{\textsf{w},\textrm{2F;1}} \
= \begin{pmatrix}
0 & -1 & 0 & -3 \\
0 & 3 & 0 & 1 \\
-1 & 0 & -3 & 0 \\
3 & 0 & 1 & 0
\end{pmatrix}, \\
M^{(8,8,\overline{8},\overline{8})}_\textrm{2F;2} = \begin{pmatrix}
1 & 0 \\
0 & 1 \\
3 & 0 \\
0 & 5
\end{pmatrix}, \quad
\Lambda^{(8,8,\overline{8},\overline{8})}_{\textsf{w},\textrm{2F;2}} = \begin{pmatrix}
0 & -1 & 0 & 3 \\
0 & 3 & 0 & -1 \\
-1 & 0 & -3 & 0 \\
3 & 0 & 1 & 0
\end{pmatrix}, \\
M^{(8,8,\overline{8},\overline{8})}_\textrm{2F;3} = \begin{pmatrix}
1 & 0 \\
0 & 1 \\
5 & 0 \\
0 & 3
\end{pmatrix}, \quad
\Lambda^{(8,8,\overline{8},\overline{8})}_{\textsf{w},\textrm{2F;3}} = \begin{pmatrix}
0 & -1 & 0 & -3 \\
3 & 0 & -1 & 0 \\
-1 & 0 & 3 & 0 \\
0 & 3 & 0 & 1
\end{pmatrix}, \\
M^{(8,8,\overline{8},\overline{8})}_\textrm{2F;4} = \begin{pmatrix}
1 & 0 \\
0 & 1 \\
5 & 0 \\
0 & 5
\end{pmatrix}, \quad
\Lambda^{(8,8,\overline{8},\overline{8})}_{\textsf{w},\textrm{2F;4}} = \begin{pmatrix}
0 & -1 & 0 & 3 \\
0 & 3 & 0 & -1 \\
-1 & 0 & 3 & 0 \\
3 & 0 & -1 & 0
\end{pmatrix}, \\
M^{(8,8,\overline{8},\overline{8})}_\textrm{2F;5} = \begin{pmatrix}
1 & 0 \\
0 & 1 \\
0 & 3 \\
3 & 0
\end{pmatrix}, \quad
\Lambda^{(8,8,\overline{8},\overline{8})}_{\textsf{w},\textrm{2F;5}} \
= \begin{pmatrix}
-1 & 0 & 0 & -3 \\
0 & 3 & 1 & 0 \\
0 & -1 & -3 & 0 \\
3 & 0 & 0 & 1
\end{pmatrix}, \\
M^{(8,8,\overline{8},\overline{8})}_\textrm{2F;6} = \begin{pmatrix}
1 & 0 \\
0 & 1 \\
0 & 5 \\
3 & 0
\end{pmatrix}, \quad
\Lambda^{(8,8,\overline{8},\overline{8})}_{\textsf{w},\textrm{2F;6}} = \begin{pmatrix}
0 & -1 & 3 & 0 \\
3 & 0 & 0 & 1 \\
-1 & 0 & 0 & -3 \\
0 & 3 & -1 & 0
\end{pmatrix}, \\
M^{(8,8,\overline{8},\overline{8})}_\textrm{2F;7} = \begin{pmatrix}
1 & 0 \\
0 & 1 \\
0 & 3 \\
5 & 0
\end{pmatrix}, \quad
\Lambda^{(8,8,\overline{8},\overline{8})}_{\textsf{w},\textrm{2F;7}} = \begin{pmatrix}
-1 & 0 & 0 & 3 \\
3 & 0 & 0 & -1 \\
0 & -1 & -3 & 0 \\
0 & 3 & 1 & 0
\end{pmatrix}, \\
M^{(8,8,\overline{8},\overline{8})}_\textrm{2F;8} = \begin{pmatrix}
1 & 0 \\
0 & 1 \\
0 & 5 \\
5 & 0
\end{pmatrix}, \quad
\Lambda^{(8,8,\overline{8},\overline{8})}_{\textsf{w},\textrm{2F;8}} = \begin{pmatrix}
-1 & 0 & 0 & 3 \\
3 & 0 & 0 & -1 \\
0 & -1 & 3 & 0 \\
0 & 3 & -1 & 0
\end{pmatrix}.
\end{align}
\end{subequations}

There are 16 fermionic gapped interfaces at rank two obtained by stacking a bosonic gapped interface between $U(1)_8$ and $\overline{U(1)}_8$ via $\bm{1} \overline{\bm{1}}$ or $\bm{1} \overline{\bm{7}}$ condensation and a fermionic gapped interface between $U(1)_8$ and $\overline{U(1)}_8$ via $\bm{1} \overline{\bm{3}}$ or $\bm{1} \overline{\bm{5}}$ condensation. 
The corresponding integer vectors are given by
\begin{subequations}
\begin{align}
M^{(8,8,\overline{8},\overline{8})}_\textrm{2F;9} = \begin{pmatrix}
1 & 0 \\
0 & 1 \\
1 & 0 \\
0 & 3
\end{pmatrix}, \quad
\Lambda^{(8,8,\overline{8},\overline{8})}_{\textsf{w},\textrm{2F;9}} \
= \begin{pmatrix}
-1 & 1 & -1 & 3 \\
0 & 3 & 0 & 1 \\
4 & -1 & -4 & -3 \\
5 & -1 & -3 & -3
\end{pmatrix}, \\
M^{(8,8,\overline{8},\overline{8})}_\textrm{2F;10} = \begin{pmatrix}
1 & 0 \\
0 & 1 \\
1 & 0 \\
0 & 5
\end{pmatrix}, \quad
\Lambda^{(8,8,\overline{8},\overline{8})}_{\textsf{w},\textrm{2F;10}} \
= \begin{pmatrix}
-1 & 1 & -1 & -3 \\
0 & 3 & 0 & -1 \\
4 & -1 & -4 & 3 \\
5 & -1 & -3 & 3
\end{pmatrix}, \\
M^{(8,8,\overline{8},\overline{8})}_\textrm{2F;11} = \begin{pmatrix}
1 & 0 \\
0 & 1 \\
3 & 0 \\
0 & 1
\end{pmatrix}, \quad
\Lambda^{(8,8,\overline{8},\overline{8})}_{\textsf{w},\textrm{2F;11}} = \begin{pmatrix}
1 & -1 & 3 & -1 \\
3 & 0 & 1 & 0 \\
-1 & 4 & -3 & -4 \\
-1 & 5 & -3 & -3
\end{pmatrix}, \\
M^{(8,8,\overline{8},\overline{8})}_\textrm{2F;12} = \begin{pmatrix}
1 & 0 \\
0 & 1 \\
3 & 0 \\
0 & 7
\end{pmatrix}, \quad
\Lambda^{(8,8,\overline{8},\overline{8})}_{\textsf{w},\textrm{2F;12}} = \begin{pmatrix}
1 & 1 & 3 & -1 \\
3 & 0 & 1 & 0 \\
-1 & -4 & -3 & -4 \\
-1 & -5 & -3 & -3
\end{pmatrix}, \\
M^{(8,8,\overline{8},\overline{8})}_\textrm{2F;13} = \begin{pmatrix}
1 & 0 \\
0 & 1 \\
5 & 0 \\
0 & 1
\end{pmatrix}, \quad
\Lambda^{(8,8,\overline{8},\overline{8})}_{\textsf{w},\textrm{2F;13}} \
= \begin{pmatrix}
1 & -1 & -3 & -1 \\
3 & 0 & -1 & 0 \\
-1 & 4 & 3 & -4 \\
-1 & 5 & 3 & -3
\end{pmatrix}, \\
M^{(8,8,\overline{8},\overline{8})}_\textrm{2F;14} = \begin{pmatrix}
1 & 0 \\
0 & 1 \\
5 & 0 \\
0 & 7
\end{pmatrix}, \quad
\Lambda^{(8,8,\overline{8},\overline{8})}_{\textsf{w},\textrm{2F;14}} \
= \begin{pmatrix}
1 & 1 & -3 & -1 \\
3 & 0 & -1 & 0 \\
-1 & -4 & 3 & -4 \\
-1 & -5 & 3 & -3
\end{pmatrix}, \\
M^{(8,8,\overline{8},\overline{8})}_\textrm{2F;15} = \begin{pmatrix}
1 & 0 \\
0 & 1 \\
7 & 0 \\
0 & 3
\end{pmatrix}, \quad
\Lambda^{(8,8,\overline{8},\overline{8})}_{\textsf{w},\textrm{2F;15}} = \begin{pmatrix}
1 & 1 & -1 & 3 \\
0 & 3 & 0 & 1 \\
-4 & -1 & -4 & -3 \\
-5 & -1 & -3 & -3
\end{pmatrix}, \\
M^{(8,8,\overline{8},\overline{8})}_\textrm{2F;16} = \begin{pmatrix}
1 & 0 \\
0 & 1 \\
7 & 0 \\
0 & 5
\end{pmatrix}, \quad
\Lambda^{(8,8,\overline{8},\overline{8})}_{\textsf{w},\textrm{2F;16}} = \begin{pmatrix}
1 & 1 & -1 & -3 \\
0 & 3 & 0 & -1 \\
-4 & -1 & -4 & 3 \\
-5 & -1 & -3 & 3
\end{pmatrix}, \\
M^{(8,8,\overline{8},\overline{8})}_\textrm{2F;17} = \begin{pmatrix}
1 & 0 \\
0 & 1 \\
0 & 3 \\
1 & 0
\end{pmatrix}, \quad
\Lambda^{(8,8,\overline{8},\overline{8})}_{\textsf{w},\textrm{2F;17}} = \begin{pmatrix}
-1 & 1 & 3 & -1 \\
0 & 3 & 1 & 0 \\
4 & -1 & -3 & -4 \\
5 & -1 & -3 & -3
\end{pmatrix}, \\
M^{(8,8,\overline{8},\overline{8})}_\textrm{2F;18} = \begin{pmatrix}
1 & 0 \\
0 & 1 \\
0 & 5 \\
1 & 0
\end{pmatrix}, \quad
\Lambda^{(8,8,\overline{8},\overline{8})}_{\textsf{w},\textrm{2F;18}} = \begin{pmatrix}
-1 & 1 & -3 & -1 \\
0 & 3 & -1 & 0 \\
4 & -1 & 3 & -4 \\
5 & -1 & 3 & -3
\end{pmatrix}, \\
M^{(8,8,\overline{8},\overline{8})}_\textrm{2F;19} = \begin{pmatrix}
1 & 0 \\
0 & 1 \\
0 & 1 \\
3 & 0
\end{pmatrix}, \quad
\Lambda^{(8,8,\overline{8},\overline{8})}_{\textsf{w},\textrm{2F;19}} \
= \begin{pmatrix}
1 & -1 & -1 & 3 \\
3 & 0 & 0 & 1 \\
-1 & 4 & -4 & -3 \\
-1 & 5 & -3 & -3
\end{pmatrix}, \\
M^{(8,8,\overline{8},\overline{8})}_\textrm{2F;20} = \begin{pmatrix}
1 & 0 \\
0 & 1 \\
0 & 7 \\
3 & 0
\end{pmatrix}, \quad
\Lambda^{(8,8,\overline{8},\overline{8})}_{\textsf{w},\textrm{2F;20}} = \begin{pmatrix}
1 & 1 & -1 & 3 \\
3 & 0 & 0 & 1 \\
-1 & -4 & -4 & -3 \\
-1 & -5 & -3 & -3
\end{pmatrix}, \\
M^{(8,8,\overline{8},\overline{8})}_\textrm{2F;21} = \begin{pmatrix}
1 & 0 \\
0 & 1 \\
0 & 1 \\
5 & 0
\end{pmatrix}, \quad
\Lambda^{(8,8,\overline{8},\overline{8})}_{\textsf{w},\textrm{2F;21}} = \begin{pmatrix}
1 & -1 & -1 & -3 \\
3 & 0 & 0 & -1 \\
-1 & 4 & -4 & 3 \\
-1 & 5 & -3 & 3
\end{pmatrix}, \\
M^{(8,8,\overline{8},\overline{8})}_\textrm{2F;22} = \begin{pmatrix}
1 & 0 \\
0 & 1 \\
0 & 7 \\
5 & 0
\end{pmatrix}, \quad
\Lambda^{(8,8,\overline{8},\overline{8})}_{\textsf{w},\textrm{2F;22}} \
= \begin{pmatrix}
1 & 1 & -1 & -3 \\
3 & 0 & 0 & -1 \\
-1 & -4 & -4 & 3 \\
-1 & -5 & -3 & 3
\end{pmatrix}, \\
M^{(8,8,\overline{8},\overline{8})}_\textrm{2F;23} = \begin{pmatrix}
1 & 0 \\
0 & 1 \\
0 & 3 \\
7 & 0
\end{pmatrix}, \quad
\Lambda^{(8,8,\overline{8},\overline{8})}_{\textsf{w},\textrm{2F;23}} \
= \begin{pmatrix}
1 & 1 & 3 & -1 \\
0 & 3 & 1 & 0 \\
-4 & -1 & -3 & -4 \\
-5 & -1 & -3 & -3
\end{pmatrix}, \\
M^{(8,8,\overline{8},\overline{8})}_\textrm{2F;24} = \begin{pmatrix}
1 & 0 \\
0 & 1 \\
0 & 5 \\
7 & 0
\end{pmatrix}, \quad
\Lambda^{(8,8,\overline{8},\overline{8})}_{\textsf{w},\textrm{2F;24}} = \begin{pmatrix}
1 & 1 & -3 & -1 \\
0 & 3 & -1 & 0 \\
-4 & -1 & 3 & -4 \\
-5 & -1 & 3 & -3
\end{pmatrix}.
\end{align}
\end{subequations}

There are eight fermionic gapped interfaces at rank two obtained by condensing triplets of anyons such as $\bm{1} \bm{4} \overline{\bm{3}}$. 
Such triplets can be understood as bound objects consisting of pairs of fermionic anyons $\bm{1} \overline{\bm{3}}$ and bosonic anyons $\bm{4}$. 
Taking the Lagrangian subgroup generated by $M^{(8,8,\overline{8},\overline{8})}_\textrm{2F;25}$ given below as an instance, the set of condensed anyons is given by
\begin{align}
&\bm{2}_1\overline{\bm{6}}_3, \ 
\bm{4}_1\overline{\bm{4}}_3, \ 
\bm{6}_1\overline{\bm{2}}_3, \ 
\bm{2}_2\overline{\bm{6}}_4, \ 
\bm{4}_2\overline{\bm{4}}_4, \ 
\bm{6}_2\overline{\bm{2}}_4, \ 
\bm{1}_1\bm{4}_2\overline{\bm{3}}_3, \ 
\bm{3}_1\bm{4}_2\overline{\bm{1}}_3, \ 
\bm{5}_1\bm{4}_2\overline{\bm{7}}_3, \ 
\bm{7}_1\bm{4}_2\overline{\bm{5}}_3, \nonumber \\
&\bm{4}_1\bm{1}_2\overline{\bm{3}}_4, \ 
\bm{4}_1\bm{3}_2\overline{\bm{1}}_4, \ 
\bm{4}_1\bm{5}_2\overline{\bm{7}}_4, \ 
\bm{4}_1\bm{7}_2\overline{\bm{5}}_4, \ 
\red{\bm{1}_1\overline{\bm{3}}_3\overline{\bm{4}}_4}, \ 
\bm{3}_1\overline{\bm{1}}_3\overline{\bm{4}}_4, \ 
\bm{5}_1\overline{\bm{7}}_3\overline{\bm{4}}_4, \ 
\bm{7}_1\overline{\bm{5}}_3\overline{\bm{4}}_4, \ 
\red{\bm{1}_2\overline{\bm{4}}_3\overline{\bm{3}}_4}, \ 
\bm{3}_2\overline{\bm{4}}_3\overline{\bm{1}}_4, \ 
\bm{5}_2\overline{\bm{4}}_3\overline{\bm{7}}_4, \ 
\bm{7}_2\overline{\bm{4}}_3\overline{\bm{5}}_4, \nonumber \\
&\bm{1}_1\bm{1}_2\overline{\bm{7}}_3\overline{\bm{7}}_4, \ 
\bm{1}_1\bm{2}_2\overline{\bm{3}}_3\overline{\bm{2}}_4, \ 
\bm{1}_1\bm{3}_2\overline{\bm{7}}_3\overline{\bm{5}}_4, \ 
\bm{1}_1\bm{5}_2\overline{\bm{7}}_3\overline{\bm{3}}_4, \ 
\bm{1}_1\bm{6}_2\overline{\bm{3}}_3\overline{\bm{6}}_4, \ 
\bm{1}_1\bm{7}_2\overline{\bm{7}}_3\overline{\bm{1}}_4, \nonumber \\
&\bm{2}_1\bm{1}_2\overline{\bm{2}}_3\overline{\bm{3}}_4, \ 
\bm{2}_1\bm{2}_2\overline{\bm{6}}_3\overline{\bm{6}}_4, \ 
\bm{2}_1\bm{3}_2\overline{\bm{2}}_3\overline{\bm{1}}_4, \ 
\bm{2}_1\bm{4}_2\overline{\bm{6}}_3\overline{\bm{4}}_4, \ 
\bm{2}_1\bm{5}_2\overline{\bm{2}}_3\overline{\bm{7}}_4, \ 
\bm{2}_1\bm{6}_2\overline{\bm{6}}_3\overline{\bm{2}}_4, \ 
\bm{2}_1\bm{7}_2\overline{\bm{2}}_3\overline{\bm{5}}_4, \nonumber \\
&\bm{3}_1\bm{1}_2\overline{\bm{5}}_3\overline{\bm{7}}_4, \ 
\bm{3}_1\bm{2}_2\overline{\bm{1}}_3\overline{\bm{2}}_4, \ 
\bm{3}_1\bm{3}_2\overline{\bm{5}}_3\overline{\bm{5}}_4, \ 
\bm{3}_1\bm{5}_2\overline{\bm{5}}_3\overline{\bm{3}}_4, \ 
\bm{3}_1\bm{6}_2\overline{\bm{1}}_3\overline{\bm{6}}_4, \ 
\bm{3}_1\bm{7}_2\overline{\bm{5}}_3\overline{\bm{1}}_4, \ 
\bm{4}_1\bm{2}_2\overline{\bm{4}}_3\overline{\bm{6}}_4, \ 
\bm{4}_1\bm{4}_2\overline{\bm{4}}_3\overline{\bm{4}}_4, \ 
\bm{4}_1\bm{6}_2\overline{\bm{4}}_3\overline{\bm{2}}_4, \nonumber \\
&\bm{5}_1\bm{1}_2\overline{\bm{3}}_3\overline{\bm{7}}_4, \ 
\bm{5}_1\bm{2}_2\overline{\bm{7}}_3\overline{\bm{2}}_4, \ 
\bm{5}_1\bm{3}_2\overline{\bm{3}}_3\overline{\bm{5}}_4, \ 
\bm{5}_1\bm{5}_2\overline{\bm{3}}_3\overline{\bm{3}}_4, \ 
\bm{5}_1\bm{6}_2\overline{\bm{7}}_3\overline{\bm{6}}_4, \ 
\bm{5}_1\bm{7}_2\overline{\bm{3}}_3\overline{\bm{1}}_4, \nonumber \\
&\bm{6}_1\bm{1}_2\overline{\bm{6}}_3\overline{\bm{3}}_4, \ 
\bm{6}_1\bm{2}_2\overline{\bm{2}}_3\overline{\bm{6}}_4, \ 
\bm{6}_1\bm{3}_2\overline{\bm{6}}_3\overline{\bm{1}}_4, \ 
\bm{6}_1\bm{4}_2\overline{\bm{2}}_3\overline{\bm{4}}_4, \ 
\bm{6}_1\bm{5}_2\overline{\bm{6}}_3\overline{\bm{7}}_4, \ 
\bm{6}_1\bm{6}_2\overline{\bm{2}}_3\overline{\bm{2}}_4, \ 
\bm{6}_1\bm{7}_2\overline{\bm{6}}_3\overline{\bm{5}}_4, \nonumber \\
&\bm{7}_1\bm{1}_2\overline{\bm{1}}_3\overline{\bm{7}}_4, \ 
\bm{7}_1\bm{2}_2\overline{\bm{5}}_3\overline{\bm{2}}_4, \ 
\bm{7}_1\bm{3}_2\overline{\bm{1}}_3\overline{\bm{5}}_4, \ 
\bm{7}_1\bm{5}_2\overline{\bm{1}}_3\overline{\bm{3}}_4, \ 
\bm{7}_1\bm{6}_2\overline{\bm{5}}_3\overline{\bm{6}}_4, \ 
\bm{7}_1\bm{7}_2\overline{\bm{1}}_3\overline{\bm{1}}_4.
\end{align}
The corresponding integer vectors are given by 
\begin{subequations}
\begin{align}
M^{(8,8,\overline{8},\overline{8})}_\textrm{2F;25} = \begin{pmatrix}
1 & 0 \\
0 & 1 \\
3 & 4 \\
4 & 3
\end{pmatrix}, \quad
\Lambda^{(8,8,\overline{8},\overline{8})}_{\textsf{w},\textrm{2F;25}} = \begin{pmatrix}
-2 & -1 & -2 & -3 \\
-3 & -2 & -1 & -2 \\
-1 & 2 & -3 & 2 \\
-2 & 3 & -2 & 1
\end{pmatrix}, \\
M^{(8,8,\overline{8},\overline{8})}_\textrm{2F;26} = \begin{pmatrix}
1 & 0 \\
0 & 1 \\
3 & 4 \\
4 & 5
\end{pmatrix}, \quad
\Lambda^{(8,8,\overline{8},\overline{8})}_{\textsf{w},\textrm{2F;26}} = \begin{pmatrix}
-2 & 1 & -2 & -3 \\
-3 & 2 & -1 & -2 \\
-1 & -2 & -3 & 2 \\
2 & 3 & 2 & -1
\end{pmatrix}, \\
M^{(8,8,\overline{8},\overline{8})}_\textrm{2F;27} = \begin{pmatrix}
1 & 0 \\
0 & 1 \\
4 & 3 \\
3 & 4
\end{pmatrix}, \quad
\Lambda^{(8,8,\overline{8},\overline{8})}_{\textsf{w},\textrm{2F;27}} = \begin{pmatrix}
-2 & -1 & -3 & -2 \\
-3 & -2 & -2 & -1 \\
-1 & 2 & 2 & -3 \\
2 & -3 & -1 & 2
\end{pmatrix}, \\
M^{(8,8,\overline{8},\overline{8})}_\textrm{2F;28} = \begin{pmatrix}
1 & 0 \\
0 & 1 \\
4 & 5 \\
3 & 4
\end{pmatrix}, \quad
\Lambda^{(8,8,\overline{8},\overline{8})}_{\textsf{w},\textrm{2F;28}} = \begin{pmatrix}
-2 & -1 & 3 & -2 \\
3 & 2 & -2 & 1 \\
-1 & 2 & -2 & -3 \\
-2 & 3 & -1 & -2
\end{pmatrix}, \\
M^{(8,8,\overline{8},\overline{8})}_\textrm{2F;29} = \begin{pmatrix}
1 & 0 \\
0 & 1 \\
4 & 3 \\
5 & 4
\end{pmatrix}, \quad
\Lambda^{(8,8,\overline{8},\overline{8})}_{\textsf{w},\textrm{2F;29}} \
= \begin{pmatrix}
2 & -1 & -3 & -2 \\
3 & -2 & -2 & -1 \\
-1 & -2 & -2 & 3 \\
2 & 3 & 1 & -2
\end{pmatrix}, \\
M^{(8,8,\overline{8},\overline{8})}_\textrm{2F;30} = \begin{pmatrix}
1 & 0 \\
0 & 1 \\
4 & 5 \\
5 & 4
\end{pmatrix}, \quad
\Lambda^{(8,8,\overline{8},\overline{8})}_{\textsf{w},\textrm{2F;30}} = \begin{pmatrix}
-2 & 1 & -3 & 2 \\
3 & -2 & 2 & -1 \\
1 & 2 & -2 & -3 \\
2 & 3 & -1 & -2
\end{pmatrix}, \\
M^{(8,8,\overline{8},\overline{8})}_\textrm{2F;31} = \begin{pmatrix}
1 & 0 \\
0 & 1 \\
5 & 4 \\
4 & 3
\end{pmatrix}, \quad
\Lambda^{(8,8,\overline{8},\overline{8})}_{\textsf{w},\textrm{2F;31}} \
= \begin{pmatrix}
1 & -2 & -3 & -2 \\
2 & -3 & -2 & -1 \\
-2 & -1 & 2 & -3 \\
3 & 2 & -1 & 2
\end{pmatrix}, \\
M^{(8,8,\overline{8},\overline{8})}_\textrm{2F;32} = \begin{pmatrix}
1 & 0 \\
0 & 1 \\
5 & 4 \\
4 & 5
\end{pmatrix}, \quad
\Lambda^{(8,8,\overline{8},\overline{8})}_{\textsf{w},\textrm{2F;32}} = \begin{pmatrix}
-2 & 1 & 2 & -3 \\
3 & -2 & -1 & 2 \\
1 & 2 & -3 & -2 \\
2 & 3 & -2 & -1
\end{pmatrix}.
\end{align}
\end{subequations}

Finally, there are 16 fermionic gapped interfaces corresponding to the rank-two Lagrangian subgroups, which have a hybrid nature of the interfaces presented above in the sense that they are obtained by condensing, say, bosonic triplets $\bm{1} \bm{4} \overline{\bm{1}}$ and fermionic triplets $\bm{1} \bm{4} \overline{\bm{3}}$.
Taking the Lagrangian subgroup generated by $M^{(8,8,\overline{8},\overline{8})}_\textrm{2F;33}$ given below as an instance, the set of condensed anyons is given by
\begin{align}
&\bm{2}_1\overline{\bm{2}}_3, \ 
\bm{4}_1\overline{\bm{4}}_3, \ 
\bm{6}_1\overline{\bm{6}}_3, \ 
\bm{2}_2\overline{\bm{6}}_4, \ 
\bm{4}_2\overline{\bm{4}}_4, \ 
\bm{6}_2\overline{\bm{2}}_4, \ 
\bm{1}_1\bm{4}_2\overline{\bm{1}}_3, \ 
\bm{3}_1\bm{4}_2\overline{\bm{3}}_3, \ 
\bm{5}_1\bm{4}_2\overline{\bm{5}}_3, \ 
\bm{7}_1\bm{4}_2\overline{\bm{7}}_3, \nonumber \\
&\bm{4}_1\bm{1}_2\overline{\bm{3}}_4, \ 
\bm{4}_1\bm{3}_2\overline{\bm{1}}_4, \ 
\bm{4}_1\bm{5}_2\overline{\bm{7}}_4, \ 
\bm{4}_1\bm{7}_2\overline{\bm{5}}_4, \ 
\red{\bm{1}_1\overline{\bm{1}}_3\overline{\bm{4}}_4}, \ 
\bm{3}_1\overline{\bm{3}}_3\overline{\bm{4}}_4, \ 
\bm{5}_1\overline{\bm{5}}_3\overline{\bm{4}}_4, \ 
\bm{7}_1\overline{\bm{7}}_3\overline{\bm{4}}_4, \ 
\red{\bm{1}_2\overline{\bm{4}}_3\overline{\bm{3}}_4}, \ 
\bm{3}_2\overline{\bm{4}}_3\overline{\bm{1}}_4, \ 
\bm{5}_2\overline{\bm{4}}_3\overline{\bm{7}}_4, \ 
\bm{7}_2\overline{\bm{4}}_3\overline{\bm{5}}_4, \nonumber \\
&\bm{1}_1\bm{1}_2\overline{\bm{5}}_3\overline{\bm{7}}_4, \ 
\bm{1}_1\bm{2}_2\overline{\bm{1}}_3\overline{\bm{2}}_4, \ 
\bm{1}_1\bm{3}_2\overline{\bm{5}}_3\overline{\bm{5}}_4, \ 
\bm{1}_1\bm{5}_2\overline{\bm{5}}_3\overline{\bm{3}}_4, \ 
\bm{1}_1\bm{6}_2\overline{\bm{1}}_3\overline{\bm{6}}_4, \ 
\bm{1}_1\bm{7}_2\overline{\bm{5}}_3\overline{\bm{1}}_4, \nonumber \\
&\bm{2}_1\bm{1}_2\overline{\bm{6}}_3\overline{\bm{3}}_4, \ 
\bm{2}_1\bm{2}_2\overline{\bm{2}}_3\overline{\bm{6}}_4, \ 
\bm{2}_1\bm{3}_2\overline{\bm{6}}_3\overline{\bm{1}}_4, \ 
\bm{2}_1\bm{4}_2\overline{\bm{2}}_3\overline{\bm{4}}_4, \ 
\bm{2}_1\bm{5}_2\overline{\bm{6}}_3\overline{\bm{7}}_4, \ 
\bm{2}_1\bm{6}_2\overline{\bm{2}}_3\overline{\bm{2}}_4, \ 
\bm{2}_1\bm{7}_2\overline{\bm{6}}_3\overline{\bm{5}}_4, \nonumber \\
&\bm{3}_1\bm{1}_2\overline{\bm{7}}_3\overline{\bm{7}}_4, \ 
\bm{3}_1\bm{2}_2\overline{\bm{3}}_3\overline{\bm{2}}_4, \ 
\bm{3}_1\bm{3}_2\overline{\bm{7}}_3\overline{\bm{5}}_4, \ 
\bm{3}_1\bm{5}_2\overline{\bm{7}}_3\overline{\bm{3}}_4, \ 
\bm{3}_1\bm{6}_2\overline{\bm{3}}_3\overline{\bm{6}}_4, \ 
\bm{3}_1\bm{7}_2\overline{\bm{7}}_3\overline{\bm{1}}_4, \ 
\bm{4}_1\bm{2}_2\overline{\bm{4}}_3\overline{\bm{6}}_4, \ 
\bm{4}_1\bm{4}_2\overline{\bm{4}}_3\overline{\bm{4}}_4, \ 
\bm{4}_1\bm{6}_2\overline{\bm{4}}_3\overline{\bm{2}}_4, \nonumber \\
&\bm{5}_1\bm{1}_2\overline{\bm{1}}_3\overline{\bm{7}}_4, \ 
\bm{5}_1\bm{2}_2\overline{\bm{5}}_3\overline{\bm{2}}_4, \ 
\bm{5}_1\bm{3}_2\overline{\bm{1}}_3\overline{\bm{5}}_4, \ 
\bm{5}_1\bm{5}_2\overline{\bm{1}}_3\overline{\bm{3}}_4, \ 
\bm{5}_1\bm{6}_2\overline{\bm{5}}_3\overline{\bm{6}}_4, \ 
\bm{5}_1\bm{7}_2\overline{\bm{1}}_3\overline{\bm{1}}_4, \nonumber \\
&\bm{6}_1\bm{1}_2\overline{\bm{2}}_3\overline{\bm{3}}_4, \ 
\bm{6}_1\bm{2}_2\overline{\bm{6}}_3\overline{\bm{6}}_4, \ 
\bm{6}_1\bm{3}_2\overline{\bm{2}}_3\overline{\bm{1}}_4, \ 
\bm{6}_1\bm{4}_2\overline{\bm{6}}_3\overline{\bm{4}}_4, \ 
\bm{6}_1\bm{5}_2\overline{\bm{2}}_3\overline{\bm{7}}_4, \ 
\bm{6}_1\bm{6}_2\overline{\bm{6}}_3\overline{\bm{2}}_4, \ 
\bm{6}_1\bm{7}_2\overline{\bm{2}}_3\overline{\bm{5}}_4, \nonumber \\
&\bm{7}_1\bm{1}_2\overline{\bm{3}}_3\overline{\bm{7}}_4, \ 
\bm{7}_1\bm{2}_2\overline{\bm{7}}_3\overline{\bm{2}}_4, \ 
\bm{7}_1\bm{3}_2\overline{\bm{3}}_3\overline{\bm{5}}_4, \ 
\bm{7}_1\bm{5}_2\overline{\bm{3}}_3\overline{\bm{3}}_4, \ 
\bm{7}_1\bm{6}_2\overline{\bm{7}}_3\overline{\bm{6}}_4, \ 
\bm{7}_1\bm{7}_2\overline{\bm{3}}_3\overline{\bm{1}}_4,
\end{align}
The corresponding integer vectors are given by 
\begin{subequations}
\begin{align}
M^{(8,8,\overline{8},\overline{8})}_\textrm{2F;33} = \begin{pmatrix}
1 & 0 \\
0 & 1 \\
1 & 4 \\
4 & 3
\end{pmatrix}, \quad
\Lambda^{(8,8,\overline{8},\overline{8})}_{\textsf{w},\textrm{2F;33}} = \begin{pmatrix}
1 & -1 & -3 & 1 \\
3 & 1 & -1 & -1 \\
-2 & 1 & 2 & 3 \\
-2 & 3 & 2 & 1
\end{pmatrix}, \\
M^{(8,8,\overline{8},\overline{8})}_\textrm{2F;34} = \begin{pmatrix}
1 & 0 \\
0 & 1 \\
1 & 4 \\
4 & 5
\end{pmatrix}, \quad
\Lambda^{(8,8,\overline{8},\overline{8})}_{\textsf{w},\textrm{2F;34}} = \begin{pmatrix}
2 & -1 & -2 & 3 \\
3 & 1 & -1 & 1 \\
1 & -1 & -3 & -1 \\
-2 & 3 & 2 & -1
\end{pmatrix}, \\
M^{(8,8,\overline{8},\overline{8})}_\textrm{2F;35} = \begin{pmatrix}
1 & 0 \\
0 & 1 \\
3 & 4 \\
4 & 1
\end{pmatrix}, \quad
\Lambda^{(8,8,\overline{8},\overline{8})}_{\textsf{w},\textrm{2F;35}} = \begin{pmatrix}
1 & -1 & -1 & 3 \\
-3 & 2 & -1 & -2 \\
1 & -2 & 3 & 2 \\
1 & 3 & -1 & -1
\end{pmatrix}, \\
M^{(8,8,\overline{8},\overline{8})}_\textrm{2F;36} = \begin{pmatrix}
1 & 0 \\
0 & 1 \\
3 & 4 \\
4 & 7
\end{pmatrix}, \quad
\Lambda^{(8,8,\overline{8},\overline{8})}_{\textsf{w},\textrm{2F;36}} = \begin{pmatrix}
1 & -1 & -1 & -3 \\
-3 & 2 & -1 & 2 \\
1 & -2 & 3 & -2 \\
-1 & -3 & 1 & -1
\end{pmatrix}, \\
M^{(8,8,\overline{8},\overline{8})}_\textrm{2F;37} = \begin{pmatrix}
1 & 0 \\
0 & 1 \\
4 & 3 \\
1 & 4
\end{pmatrix}, \quad
\Lambda^{(8,8,\overline{8},\overline{8})}_{\textsf{w},\textrm{2F;37}} \
= \begin{pmatrix}
1 & -1 & 1 & -3 \\
2 & -3 & -1 & -2 \\
-2 & 1 & 3 & 2 \\
3 & 1 & -1 & -1
\end{pmatrix}, \\
M^{(8,8,\overline{8},\overline{8})}_\textrm{2F;38} = \begin{pmatrix}
1 & 0 \\
0 & 1 \\
4 & 5 \\
1 & 4
\end{pmatrix}, \quad
\Lambda^{(8,8,\overline{8},\overline{8})}_{\textsf{w},\textrm{2F;38}} = \begin{pmatrix}
1 & 1 & 1 & -3 \\
3 & -1 & -1 & -1 \\
2 & 1 & -3 & -2 \\
-2 & -3 & 1 & 2
\end{pmatrix}, \\
M^{(8,8,\overline{8},\overline{8})}_\textrm{2F;39} = \begin{pmatrix}
1 & 0 \\
0 & 1 \\
4 & 1 \\
3 & 4
\end{pmatrix}, \quad
\Lambda^{(8,8,\overline{8},\overline{8})}_{\textsf{w},\textrm{2F;39}} = \begin{pmatrix}
1 & -2 & 2 & 3 \\
1 & 3 & -1 & -1 \\
1 & -1 & 3 & -1 \\
-3 & 2 & -2 & -1
\end{pmatrix}, \\
M^{(8,8,\overline{8},\overline{8})}_\textrm{2F;40} = \begin{pmatrix}
1 & 0 \\
0 & 1 \\
4 & 7 \\
3 & 4
\end{pmatrix}, \quad
\Lambda^{(8,8,\overline{8},\overline{8})}_{\textsf{w},\textrm{2F;40}} = \begin{pmatrix}
-1 & -1 & -3 & 1 \\
3 & 2 & 2 & 1 \\
-1 & -2 & -2 & -3 \\
-1 & 3 & 1 & 1
\end{pmatrix}, \\
M^{(8,8,\overline{8},\overline{8})}_\textrm{2F;41} = \begin{pmatrix}
1 & 0 \\
0 & 1 \\
4 & 1 \\
5 & 4
\end{pmatrix}, \quad
\Lambda^{(8,8,\overline{8},\overline{8})}_{\textsf{w},\textrm{2F;41}} \
= \begin{pmatrix}
-1 & 2 & -2 & 3 \\
1 & 3 & -1 & 1 \\
1 & -1 & 3 & 1 \\
3 & -2 & 2 & -1
\end{pmatrix}, \\
M^{(8,8,\overline{8},\overline{8})}_\textrm{2F;42} = \begin{pmatrix}
1 & 0 \\
0 & 1 \\
4 & 7 \\
5 & 4
\end{pmatrix}, \quad
\Lambda^{(8,8,\overline{8},\overline{8})}_{\textsf{w},\textrm{2F;42}} = \begin{pmatrix}
1 & -2 & -2 & -3 \\
1 & 3 & 1 & 1 \\
1 & -1 & -3 & 1 \\
-3 & 2 & 2 & 1
\end{pmatrix}, \\
M^{(8,8,\overline{8},\overline{8})}_\textrm{2F;43} = \begin{pmatrix}
1 & 0 \\
0 & 1 \\
4 & 3 \\
7 & 4
\end{pmatrix}, \quad
\Lambda^{(8,8,\overline{8},\overline{8})}_{\textsf{w},\textrm{2F;43}} = \begin{pmatrix}
1 & 1 & -1 & 3 \\
-3 & 1 & -1 & -1 \\
-2 & -1 & -3 & -2 \\
-2 & -3 & -1 & -2
\end{pmatrix}, \\
M^{(8,8,\overline{8},\overline{8})}_\textrm{2F;44} = \begin{pmatrix}
1 & 0 \\
0 & 1 \\
4 & 5 \\
7 & 4
\end{pmatrix}, \quad
\Lambda^{(8,8,\overline{8},\overline{8})}_{\textsf{w},\textrm{2F;44}} = \begin{pmatrix}
1 & 1 & 1 & 3 \\
3 & -1 & -1 & 1 \\
2 & 1 & -3 & 2 \\
-2 & -3 & 1 & -2
\end{pmatrix}, \\
M^{(8,8,\overline{8},\overline{8})}_\textrm{2F;45} = \begin{pmatrix}
1 & 0 \\
0 & 1 \\
5 & 4 \\
4 & 1
\end{pmatrix}, \quad
\Lambda^{(8,8,\overline{8},\overline{8})}_{\textsf{w},\textrm{2F;45}} \
= \begin{pmatrix}
1 & -1 & 1 & 3 \\
1 & 3 & 1 & -1 \\
1 & -2 & -3 & 2 \\
-3 & 2 & 1 & -2
\end{pmatrix}, \\
M^{(8,8,\overline{8},\overline{8})}_\textrm{2F;46} = \begin{pmatrix}
1 & 0 \\
0 & 1 \\
5 & 4 \\
4 & 7
\end{pmatrix}, \quad
\Lambda^{(8,8,\overline{8},\overline{8})}_{\textsf{w},\textrm{2F;46}} \
= \begin{pmatrix}
1 & -1 & 1 & -3 \\
1 & 3 & 1 & 1 \\
1 & -2 & -3 & -2 \\
-3 & 2 & 1 & 2
\end{pmatrix}, \\
M^{(8,8,\overline{8},\overline{8})}_\textrm{2F;47} = \begin{pmatrix}
1 & 0 \\
0 & 1 \\
7 & 4 \\
4 & 3
\end{pmatrix}, \quad
\Lambda^{(8,8,\overline{8},\overline{8})}_{\textsf{w},\textrm{2F;47}} = \begin{pmatrix}
2 & -1 & 2 & -3 \\
3 & 1 & 1 & -1 \\
1 & -1 & 3 & 1 \\
-2 & 3 & -2 & 1
\end{pmatrix}, \\
M^{(8,8,\overline{8},\overline{8})}_\textrm{2F;48} = \begin{pmatrix}
1 & 0 \\
0 & 1 \\
7 & 4 \\
4 & 5
\end{pmatrix}, \quad
\Lambda^{(8,8,\overline{8},\overline{8})}_{\textsf{w},\textrm{2F;48}} = \begin{pmatrix}
-1 & 1 & -3 & 1 \\
3 & 1 & 1 & 1 \\
2 & -1 & 2 & 3 \\
-2 & 3 & -2 & -1
\end{pmatrix}.
\end{align}
\end{subequations}

Among 40 gapped interfaces of rank three, 16 are bosonic whereas 24 are fermionic. 
There are eight bosonic gapped interfaces obtained by stacking a gapped interface between $U(1)_8$ and $\overline{U(1)}_8$ via the condensation of $\bm{1} \overline{\bm{1}}$ or $\bm{1} \overline{\bm{7}}$ between two layers and a gapped interface via the condensation of $\bm{2} \overline{\bm{2}}$ and $\bm{4}$ or $\overline{\bm{4}}$ between another two layers.
Taking the Lagrangian subgroup generated by $M^{(8,8,\overline{8},\overline{8})}_\textrm{3B;1}$ given below as an instance, the set of condensed anyons is given by
\begin{align}
&\red{\bm{4}_2}, \ 
\overline{\bm{4}}_4, \ 
\red{\bm{1}_1\overline{\bm{1}}_3}, \ 
\bm{3}_1\overline{\bm{3}}_3, \ 
\bm{4}_1\overline{\bm{4}}_3, \ 
\bm{5}_1\overline{\bm{5}}_3, \ 
\bm{6}_1\overline{\bm{6}}_3, \ 
\bm{7}_1\overline{\bm{7}}_3, \ 
\bm{2}_1\overline{\bm{2}}_3, \ 
\red{\bm{2}_2\overline{\bm{2}}_4}, \ 
\bm{2}_2\overline{\bm{6}}_4, \ 
\bm{4}_2\overline{\bm{4}}_4, \ 
\bm{6}_2\overline{\bm{2}}_4, \ 
\bm{6}_2\overline{\bm{6}}_4, \nonumber \\
&\bm{1}_1\bm{4}_2\overline{\bm{1}}_3, \ 
\bm{2}_1\bm{4}_2\overline{\bm{2}}_3, \ 
\bm{3}_1\bm{4}_2\overline{\bm{3}}_3, \ 
\bm{4}_1\bm{4}_2\overline{\bm{4}}_3, \ 
\bm{5}_1\bm{4}_2\overline{\bm{5}}_3, \ 
\bm{6}_1\bm{4}_2\overline{\bm{6}}_3, \ 
\bm{7}_1\bm{4}_2\overline{\bm{7}}_3, \nonumber \\
&\bm{1}_1\overline{\bm{1}}_3\overline{\bm{4}}_4, \ 
\bm{2}_1\overline{\bm{2}}_3\overline{\bm{4}}_4, \ 
\bm{3}_1\overline{\bm{3}}_3\overline{\bm{4}}_4, \ 
\bm{4}_1\overline{\bm{4}}_3\overline{\bm{4}}_4, \ 
\bm{5}_1\overline{\bm{5}}_3\overline{\bm{4}}_4, \ 
\bm{6}_1\overline{\bm{6}}_3\overline{\bm{4}}_4, \ 
\bm{7}_1\overline{\bm{7}}_3\overline{\bm{4}}_4, \nonumber \\
&\bm{1}_1\bm{2}_2\overline{\bm{1}}_3\overline{\bm{2}}_4, \ 
\bm{1}_1\bm{2}_2\overline{\bm{1}}_3\overline{\bm{6}}_4, \ 
\bm{1}_1\bm{4}_2\overline{\bm{1}}_3\overline{\bm{4}}_4, \ 
\bm{1}_1\bm{6}_2\overline{\bm{1}}_3\overline{\bm{2}}_4, \ 
\bm{1}_1\bm{6}_2\overline{\bm{1}}_3\overline{\bm{6}}_4, \ 
\bm{2}_1\bm{2}_2\overline{\bm{2}}_3\overline{\bm{2}}_4, \ 
\bm{2}_1\bm{2}_2\overline{\bm{2}}_3\overline{\bm{6}}_4, \ 
\bm{2}_1\bm{4}_2\overline{\bm{2}}_3\overline{\bm{4}}_4, \ 
\bm{2}_1\bm{6}_2\overline{\bm{2}}_3\overline{\bm{2}}_4, \ 
\bm{2}_1\bm{6}_2\overline{\bm{2}}_3\overline{\bm{6}}_4, \nonumber \\ 
&\bm{3}_1\bm{2}_2\overline{\bm{3}}_3\overline{\bm{2}}_4, \ 
\bm{3}_1\bm{2}_2\overline{\bm{3}}_3\overline{\bm{6}}_4, \ 
\bm{3}_1\bm{4}_2\overline{\bm{3}}_3\overline{\bm{4}}_4, \ 
\bm{3}_1\bm{6}_2\overline{\bm{3}}_3\overline{\bm{2}}_4, \ 
\bm{3}_1\bm{6}_2\overline{\bm{3}}_3\overline{\bm{6}}_4, \ 
\bm{4}_1\bm{2}_2\overline{\bm{4}}_3\overline{\bm{2}}_4, \ 
\bm{4}_1\bm{2}_2\overline{\bm{4}}_3\overline{\bm{6}}_4, \ 
\bm{4}_1\bm{4}_2\overline{\bm{4}}_3\overline{\bm{4}}_4, \ 
\bm{4}_1\bm{6}_2\overline{\bm{4}}_3\overline{\bm{2}}_4, \ 
\bm{4}_1\bm{6}_2\overline{\bm{4}}_3\overline{\bm{6}}_4, \nonumber \\
&\bm{5}_1\bm{2}_2\overline{\bm{5}}_3\overline{\bm{2}}_4, \ 
\bm{5}_1\bm{2}_2\overline{\bm{5}}_3\overline{\bm{6}}_4, \ 
\bm{5}_1\bm{4}_2\overline{\bm{5}}_3\overline{\bm{4}}_4, \ 
\bm{5}_1\bm{6}_2\overline{\bm{5}}_3\overline{\bm{2}}_4, \ 
\bm{5}_1\bm{6}_2\overline{\bm{5}}_3\overline{\bm{6}}_4, \ 
\bm{6}_1\bm{2}_2\overline{\bm{6}}_3\overline{\bm{2}}_4, \ 
\bm{6}_1\bm{2}_2\overline{\bm{6}}_3\overline{\bm{6}}_4, \ 
\bm{6}_1\bm{4}_2\overline{\bm{6}}_3\overline{\bm{4}}_4, \ 
\bm{6}_1\bm{6}_2\overline{\bm{6}}_3\overline{\bm{2}}_4, \ 
\bm{6}_1\bm{6}_2\overline{\bm{6}}_3\overline{\bm{6}}_4, \nonumber \\
&\bm{7}_1\bm{2}_2\overline{\bm{7}}_3\overline{\bm{2}}_4, \ 
\bm{7}_1\bm{2}_2\overline{\bm{7}}_3\overline{\bm{6}}_4, \ 
\bm{7}_1\bm{4}_2\overline{\bm{7}}_3\overline{\bm{4}}_4, \ 
\bm{7}_1\bm{6}_2\overline{\bm{7}}_3\overline{\bm{2}}_4, \ 
\bm{7}_1\bm{6}_2\overline{\bm{7}}_3\overline{\bm{6}}_4.
\end{align}
The corresponding integer vectors are given by 
\begin{subequations}
\begin{align}
M^{(8,8,\overline{8},\overline{8})}_\textrm{3B;1} = \begin{pmatrix}
1 & 0 & 0 \\
0 & 2 & 4 \\
1 & 0 & 0 \\
0 & 2 & 0
\end{pmatrix}, \quad
\Lambda^{(8,8,\overline{8},\overline{8})}_{\textsf{w},\textrm{3B;1}} \
= \begin{pmatrix}
4 & 0 & -4 & 0 \\
-1 & 0 & -1 & 0 \\
0 & 2 & 0 & -2 \\
0 & -2 & 0 & -2
\end{pmatrix}, \\
M^{(8,8,\overline{8},\overline{8})}_\textrm{3B;2} = \begin{pmatrix}
1 & 0 & 0 \\
0 & 2 & 4 \\
7 & 0 & 0 \\
0 & 2 & 0
\end{pmatrix}, \quad
\Lambda^{(8,8,\overline{8},\overline{8})}_{\textsf{w},\textrm{3B;2}} = \begin{pmatrix}
-4 & 0 & -4 & 0 \\
1 & 0 & -1 & 0 \\
0 & 2 & 0 & -2 \\
0 & -2 & 0 & -2
\end{pmatrix}, \\
M^{(8,8,\overline{8},\overline{8})}_\textrm{3B;3} = \begin{pmatrix}
1 & 0 & 0 \\
0 & 2 & 4 \\
0 & 2 & 0 \\
1 & 0 & 0
\end{pmatrix}, \quad
\Lambda^{(8,8,\overline{8},\overline{8})}_{\textsf{w},\textrm{3B;3}} \
= \begin{pmatrix}
4 & 0 & 0 & -4 \\
-1 & 0 & 0 & -1 \\
0 & 2 & -2 & 0 \\
0 & -2 & -2 & 0
\end{pmatrix}, \\
M^{(8,8,\overline{8},\overline{8})}_\textrm{3B;4} = \begin{pmatrix}
1 & 0 & 0 \\
0 & 2 & 4 \\
0 & 2 & 0 \\
7 & 0 & 0
\end{pmatrix}, \quad
\Lambda^{(8,8,\overline{8},\overline{8})}_{\textsf{w},\textrm{3B;4}} \
= \begin{pmatrix}
-4 & 0 & 0 & -4 \\
1 & 0 & 0 & -1 \\
0 & 2 & -2 & 0 \\
0 & -2 & -2 & 0
\end{pmatrix}, \\
M^{(8,8,\overline{8},\overline{8})}_\textrm{3B;5} = \begin{pmatrix}
2 & 4 & 0 \\
0 & 0 & 1 \\
2 & 0 & 0 \\
0 & 0 & 1
\end{pmatrix}, \quad
\Lambda^{(8,8,\overline{8},\overline{8})}_{\textsf{w},\textrm{3B;5}} = \begin{pmatrix}
0 & 4 & 0 & -4 \\
0 & -1 & 0 & -1 \\
2 & 0 & -2 & 0 \\
-2 & 0 & -2 & 0
\end{pmatrix}, \\
M^{(8,8,\overline{8},\overline{8})}_\textrm{3B;6} = \begin{pmatrix}
2 & 4 & 0 \\
0 & 0 & 1 \\
2 & 0 & 0 \\
0 & 0 & 7
\end{pmatrix}, \quad
\Lambda^{(8,8,\overline{8},\overline{8})}_{\textsf{w},\textrm{3B;6}} \
= \begin{pmatrix}
0 & -4 & 0 & -4 \\
0 & 1 & 0 & -1 \\
2 & 0 & -2 & 0 \\
-2 & 0 & -2 & 0
\end{pmatrix}, \\
M^{(8,8,\overline{8},\overline{8})}_\textrm{3B;7} = \begin{pmatrix}
2 & 4 & 0 \\
0 & 0 & 1 \\
0 & 0 & 1 \\
2 & 0 & 0
\end{pmatrix}, \quad
\Lambda^{(8,8,\overline{8},\overline{8})}_{\textsf{w},\textrm{3B;7}} = \begin{pmatrix}
0 & 4 & -4 & 0 \\
0 & -1 & -1 & 0 \\
2 & 0 & 0 & -2 \\
-2 & 0 & 0 & -2
\end{pmatrix}, \\
M^{(8,8,\overline{8},\overline{8})}_\textrm{3B;8} = \begin{pmatrix}
2 & 4 & 0 \\
0 & 0 & 1 \\
0 & 0 & 7 \\
2 & 0 & 0
\end{pmatrix}, \quad
\Lambda^{(8,8,\overline{8},\overline{8})}_{\textsf{w},\textrm{3B;8}} = \begin{pmatrix}
0 & -4 & -4 & 0 \\
0 & 1 & -1 & 0 \\
2 & 0 & 0 & -2 \\
-2 & 0 & 0 & -2
\end{pmatrix}.
\end{align}
\end{subequations}

There are also eight bosonic gapped interfaces at rank three obtained by condensing pairs of $\bm{4}$ or $\overline{\bm{4}}$ from any two layers, pairs of $\bm{2}$ and $\overline{\bm{2}}$ between the layers 1 and 3 or 1 and 4 and between the other two layers, and quadruplets of anyons such as $\bm{1} \bm{1} \overline{\bm{1}} \overline{\bm{1}}$.
Taking the Lagrangian subgroup generated by $M^{(8,8,\overline{8},\overline{8})}_\textrm{3B;9}$ given below as an instance, the set of condensed anyons is given by
\begin{align}
&\red{\bm{4}_1\bm{4}_2}, \ 
\red{\bm{2}_1\overline{\bm{2}}_3}, \ 
\bm{4}_1\overline{\bm{4}}_3, \ 
\bm{6}_1\overline{\bm{6}}_3, \ 
\bm{4}_1\overline{\bm{4}}_4, \ 
\bm{4}_2\overline{\bm{4}}_3, \ 
\bm{2}_2\overline{\bm{2}}_4, \ 
\bm{4}_2\overline{\bm{4}}_4, \ 
\bm{6}_2\overline{\bm{6}}_4, \ 
\overline{\bm{4}}_3\overline{\bm{4}}_4, \nonumber \\
&\bm{2}_1\bm{4}_2\overline{\bm{6}}_3, \ 
\bm{6}_1\bm{4}_2\overline{\bm{2}}_3, \ 
\bm{4}_1\bm{2}_2\overline{\bm{6}}_4, \ 
\bm{4}_1\bm{6}_2\overline{\bm{2}}_4, \ 
\bm{2}_1\overline{\bm{6}}_3\overline{\bm{4}}_4, \ 
\bm{6}_1\overline{\bm{2}}_3\overline{\bm{4}}_4, \ 
\bm{6}_2\overline{\bm{4}}_3\overline{\bm{2}}_4, \ 
\bm{2}_2\overline{\bm{4}}_3\overline{\bm{6}}_4, \nonumber \\
&\red{\bm{1}_1\bm{1}_2\overline{\bm{1}}_3\overline{\bm{1}}_4}, \ 
\bm{1}_1\bm{1}_2\overline{\bm{5}}_3\overline{\bm{5}}_4, \ 
\bm{1}_1\bm{3}_2\overline{\bm{1}}_3\overline{\bm{3}}_4, \ 
\bm{1}_1\bm{3}_2\overline{\bm{5}}_3\overline{\bm{7}}_4, \ 
\bm{1}_1\bm{5}_2\overline{\bm{1}}_3\overline{\bm{5}}_4, \ 
\bm{1}_1\bm{5}_2\overline{\bm{5}}_3\overline{\bm{1}}_4, \ 
\bm{1}_1\bm{7}_2\overline{\bm{1}}_3\overline{\bm{7}}_4, \ 
\bm{1}_1\bm{7}_2\overline{\bm{5}}_3\overline{\bm{3}}_4, \nonumber \\
&\bm{2}_1\bm{2}_2\overline{\bm{2}}_3\overline{\bm{2}}_4, \ 
\bm{2}_1\bm{2}_2\overline{\bm{6}}_3\overline{\bm{6}}_4, \ 
\bm{2}_1\bm{4}_2\overline{\bm{2}}_3\overline{\bm{4}}_4, \ 
\bm{2}_1\bm{6}_2\overline{\bm{2}}_3\overline{\bm{6}}_4, \ 
\bm{2}_1\bm{6}_2\overline{\bm{6}}_3\overline{\bm{2}}_4, \nonumber \\
&\bm{3}_1\bm{1}_2\overline{\bm{3}}_3\overline{\bm{1}}_4, \ 
\bm{3}_1\bm{1}_2\overline{\bm{7}}_3\overline{\bm{5}}_4, \ 
\bm{3}_1\bm{3}_2\overline{\bm{3}}_3\overline{\bm{3}}_4, \ 
\bm{3}_1\bm{3}_2\overline{\bm{7}}_3\overline{\bm{7}}_4, \ 
\bm{3}_1\bm{5}_2\overline{\bm{3}}_3\overline{\bm{5}}_4, \ 
\bm{3}_1\bm{5}_2\overline{\bm{7}}_3\overline{\bm{1}}_4, \ 
\bm{3}_1\bm{7}_2\overline{\bm{3}}_3\overline{\bm{7}}_4, \ 
\bm{3}_1\bm{7}_2\overline{\bm{7}}_3\overline{\bm{3}}_4, \nonumber \\
&\bm{4}_1\bm{2}_2\overline{\bm{4}}_3\overline{\bm{2}}_4, \ 
\bm{4}_1\bm{4}_2\overline{\bm{4}}_3\overline{\bm{4}}_4, \ 
\bm{4}_1\bm{6}_2\overline{\bm{4}}_3\overline{\bm{6}}_4, \nonumber \\
&\bm{5}_1\bm{1}_2\overline{\bm{1}}_3\overline{\bm{5}}_4, \ 
\bm{5}_1\bm{1}_2\overline{\bm{5}}_3\overline{\bm{1}}_4, \ 
\bm{5}_1\bm{3}_2\overline{\bm{1}}_3\overline{\bm{7}}_4, \ 
\bm{5}_1\bm{3}_2\overline{\bm{5}}_3\overline{\bm{3}}_4, \ 
\bm{5}_1\bm{5}_2\overline{\bm{1}}_3\overline{\bm{1}}_4, \ 
\bm{5}_1\bm{5}_2\overline{\bm{5}}_3\overline{\bm{5}}_4, \ 
\bm{5}_1\bm{7}_2\overline{\bm{1}}_3\overline{\bm{3}}_4, \ 
\bm{5}_1\bm{7}_2\overline{\bm{5}}_3\overline{\bm{7}}_4, \nonumber \\
&\bm{6}_1\bm{2}_2\overline{\bm{2}}_3\overline{\bm{6}}_4, \ 
\bm{6}_1\bm{2}_2\overline{\bm{6}}_3\overline{\bm{2}}_4, \ 
\bm{6}_1\bm{4}_2\overline{\bm{6}}_3\overline{\bm{4}}_4, \ 
\bm{6}_1\bm{6}_2\overline{\bm{2}}_3\overline{\bm{2}}_4, \ 
\bm{6}_1\bm{6}_2\overline{\bm{6}}_3\overline{\bm{6}}_4, \nonumber \\
&\bm{7}_1\bm{1}_2\overline{\bm{3}}_3\overline{\bm{5}}_4, \ 
\bm{7}_1\bm{1}_2\overline{\bm{7}}_3\overline{\bm{1}}_4, \ 
\bm{7}_1\bm{3}_2\overline{\bm{3}}_3\overline{\bm{7}}_4, \ 
\bm{7}_1\bm{3}_2\overline{\bm{7}}_3\overline{\bm{3}}_4, \ 
\bm{7}_1\bm{5}_2\overline{\bm{3}}_3\overline{\bm{1}}_4, \ 
\bm{7}_1\bm{5}_2\overline{\bm{7}}_3\overline{\bm{5}}_4, \ 
\bm{7}_1\bm{7}_2\overline{\bm{3}}_3\overline{\bm{3}}_4, \ 
\bm{7}_1\bm{7}_2\overline{\bm{7}}_3\overline{\bm{7}}_4. 
\end{align}
The corresponding integer vectors are given by 
\begin{subequations}
\begin{align}
M^{(8,8,\overline{8},\overline{8})}_\textrm{3B;9} = \begin{pmatrix}
1 & 2 & 4 \\
1 & 0 & 4 \\
1 & 2 & 0 \\
1 & 0 & 0
\end{pmatrix}, \quad
\Lambda^{(8,8,\overline{8},\overline{8})}_{\textsf{w},\textrm{3B;9}} \
= \begin{pmatrix}
2 & 2 & -2 & -2 \\
-1 & -1 & -1 & -1 \\
-2 & 2 & 2 & -2 \\
1 & -1 & 1 & -1
\end{pmatrix}, \\
M^{(8,8,\overline{8},\overline{8})}_\textrm{3B;10} = \begin{pmatrix}
1 & 2 & 4 \\
1 & 0 & 4 \\
1 & 0 & 0 \\
1 & 2 & 0
\end{pmatrix}, \quad
\Lambda^{(8,8,\overline{8},\overline{8})}_{\textsf{w},\textrm{3B;10}} \
= \begin{pmatrix}
2 & 2 & -2 & -2 \\
-1 & -1 & -1 & -1 \\
-2 & 2 & -2 & 2 \\
1 & -1 & -1 & 1
\end{pmatrix}, \\
M^{(8,8,\overline{8},\overline{8})}_\textrm{3B;11} = \begin{pmatrix}
1 & 2 & 4 \\
1 & 0 & 4 \\
1 & 2 & 0 \\
7 & 0 & 0
\end{pmatrix}, \quad
\Lambda^{(8,8,\overline{8},\overline{8})}_{\textsf{w},\textrm{3B;11}} \
= \begin{pmatrix}
-2 & 2 & 2 & 2 \\
1 & -1 & 1 & 1 \\
2 & 2 & -2 & 2 \\
-1 & -1 & -1 & 1
\end{pmatrix}, \\
M^{(8,8,\overline{8},\overline{8})}_\textrm{3B;12} = \begin{pmatrix}
1 & 2 & 4 \\
1 & 0 & 4 \\
1 & 0 & 0 \\
7 & 6 & 0
\end{pmatrix}, \quad
\Lambda^{(8,8,\overline{8},\overline{8})}_{\textsf{w},\textrm{3B;12}} = \begin{pmatrix}
-2 & 2 & -2 & -2 \\
1 & -1 & -1 & -1 \\
-2 & -2 & 2 & -2 \\
1 & 1 & 1 & -1
\end{pmatrix}, \\
M^{(8,8,\overline{8},\overline{8})}_\textrm{3B;13} = \begin{pmatrix}
1 & 2 & 4 \\
1 & 0 & 4 \\
3 & 6 & 0 \\
3 & 0 & 0
\end{pmatrix}, \quad
\Lambda^{(8,8,\overline{8},\overline{8})}_{\textsf{w},\textrm{3B;13}} \
= \begin{pmatrix}
-2 & -2 & -2 & -2 \\
1 & 1 & -1 & -1 \\
-2 & 2 & -2 & 2 \\
1 & -1 & -1 & 1
\end{pmatrix}, \\
M^{(8,8,\overline{8},\overline{8})}_\textrm{3B;14} = \begin{pmatrix}
1 & 2 & 4 \\
1 & 0 & 4 \\
3 & 0 & 0 \\
3 & 6 & 0
\end{pmatrix}, \quad
\Lambda^{(8,8,\overline{8},\overline{8})}_{\textsf{w},\textrm{3B;14}} = \begin{pmatrix}
-2 & 2 & 2 & -2 \\
1 & -1 & 1 & -1 \\
2 & 2 & 2 & 2 \\
-1 & -1 & 1 & 1
\end{pmatrix}, \\
M^{(8,8,\overline{8},\overline{8})}_\textrm{3B;15} = \begin{pmatrix}
1 & 2 & 4 \\
1 & 0 & 4 \\
3 & 6 & 0 \\
5 & 0 & 0
\end{pmatrix}, \quad
\Lambda^{(8,8,\overline{8},\overline{8})}_{\textsf{w},\textrm{3B;15}} = \begin{pmatrix}
-2 & -2 & -2 & 2 \\
1 & 1 & -1 & 1 \\
2 & -2 & 2 & 2 \\
-1 & 1 & 1 & 1
\end{pmatrix}, \\
M^{(8,8,\overline{8},\overline{8})}_\textrm{3B;16} = \begin{pmatrix}
1 & 2 & 4 \\
1 & 0 & 4 \\
3 & 0 & 0 \\
5 & 2 & 0
\end{pmatrix}, \quad
\Lambda^{(8,8,\overline{8},\overline{8})}_{\textsf{w},\textrm{3B;16}} = \begin{pmatrix}
-2 & 2 & 2 & 2 \\
1 & -1 & 1 & 1 \\
-2 & -2 & -2 & 2 \\
1 & 1 & -1 & 1
\end{pmatrix}.
\end{align}
\end{subequations}

There are eight fermionic gapped interfaces at rank three obtained by stacking a fermionic gapped interface between $U(1)_8$ and $\overline{U(1)}_8$ with $\bm{1} \overline{\bm{3}}$ or $\bm{1} \overline{\bm{5}}$ condensation and a bosonic gapped interface with $\bm{2} \overline{\bm{2}}$ and $\bm{4}$ condensation. 
Taking the Lagrangian subgroup generated by $M^{(8,8,\overline{8},\overline{8})}_\textrm{3F;1}$ given below as an instance, the set of condensed anyons is given by 
\begin{align}
&\red{\bm{4}_2}, \ 
\overline{\bm{4}}_4, \ 
\red{\bm{1}_1\overline{\bm{3}}_3}, \ 
\bm{2}_1\overline{\bm{6}}_3, \ 
\bm{3}_1\overline{\bm{1}}_3, \ 
\bm{4}_1\overline{\bm{4}}_3, \ 
\bm{5}_1\overline{\bm{7}}_3, \ 
\bm{6}_1\overline{\bm{2}}_3, \ 
\bm{7}_1\overline{\bm{5}}_3, \ 
\red{\bm{2}_2\overline{\bm{2}}_4}, \ 
\bm{2}_2\overline{\bm{6}}_4, \ 
\bm{4}_2\overline{\bm{4}}_4, \ 
\bm{6}_2\overline{\bm{2}}_4, \ 
\bm{6}_2\overline{\bm{6}}_4, \nonumber \\
&\bm{1}_1\bm{4}_2\overline{\bm{3}}_3, \ 
\bm{2}_1\bm{4}_2\overline{\bm{6}}_3, \ 
\bm{3}_1\bm{4}_2\overline{\bm{1}}_3, \ 
\bm{4}_1\bm{4}_2\overline{\bm{4}}_3, \ 
\bm{5}_1\bm{4}_2\overline{\bm{7}}_3, \ 
\bm{6}_1\bm{4}_2\overline{\bm{2}}_3, \ 
\bm{7}_1\bm{4}_2\overline{\bm{5}}_3, \nonumber \\
&\bm{1}_1\overline{\bm{3}}_3\overline{\bm{4}}_4, \ 
\bm{2}_1\overline{\bm{6}}_3\overline{\bm{4}}_4, \ 
\bm{3}_1\overline{\bm{1}}_3\overline{\bm{4}}_4, \ 
\bm{4}_1\overline{\bm{4}}_3\overline{\bm{4}}_4, \ 
\bm{5}_1\overline{\bm{7}}_3\overline{\bm{4}}_4, \ 
\bm{6}_1\overline{\bm{2}}_3\overline{\bm{4}}_4, \ 
\bm{7}_1\overline{\bm{5}}_3\overline{\bm{4}}_4, \nonumber \\
&\bm{1}_1\bm{2}_2\overline{\bm{3}}_3\overline{\bm{2}}_4, \ 
\bm{1}_1\bm{2}_2\overline{\bm{3}}_3\overline{\bm{6}}_4, \ 
\bm{1}_1\bm{4}_2\overline{\bm{3}}_3\overline{\bm{4}}_4, \ 
\bm{1}_1\bm{6}_2\overline{\bm{3}}_3\overline{\bm{2}}_4, \ 
\bm{1}_1\bm{6}_2\overline{\bm{3}}_3\overline{\bm{6}}_4, \ 
\bm{2}_1\bm{2}_2\overline{\bm{6}}_3\overline{\bm{2}}_4, \ 
\bm{2}_1\bm{2}_2\overline{\bm{6}}_3\overline{\bm{6}}_4, \ 
\bm{2}_1\bm{4}_2\overline{\bm{6}}_3\overline{\bm{4}}_4, \ 
\bm{2}_1\bm{6}_2\overline{\bm{6}}_3\overline{\bm{2}}_4, \ 
\bm{2}_1\bm{6}_2\overline{\bm{6}}_3\overline{\bm{6}}_4, \nonumber \\
&\bm{3}_1\bm{2}_2\overline{\bm{1}}_3\overline{\bm{2}}_4, \ 
\bm{3}_1\bm{2}_2\overline{\bm{1}}_3\overline{\bm{6}}_4, \ 
\bm{3}_1\bm{4}_2\overline{\bm{1}}_3\overline{\bm{4}}_4, \ 
\bm{3}_1\bm{6}_2\overline{\bm{1}}_3\overline{\bm{2}}_4, \ 
\bm{3}_1\bm{6}_2\overline{\bm{1}}_3\overline{\bm{6}}_4, \ 
\bm{4}_1\bm{2}_2\overline{\bm{4}}_3\overline{\bm{2}}_4, \ 
\bm{4}_1\bm{2}_2\overline{\bm{4}}_3\overline{\bm{6}}_4, \ 
\bm{4}_1\bm{4}_2\overline{\bm{4}}_3\overline{\bm{4}}_4, \ 
\bm{4}_1\bm{6}_2\overline{\bm{4}}_3\overline{\bm{2}}_4, \ 
\bm{4}_1\bm{6}_2\overline{\bm{4}}_3\overline{\bm{6}}_4, \nonumber \\
&\bm{5}_1\bm{2}_2\overline{\bm{7}}_3\overline{\bm{2}}_4, \ 
\bm{5}_1\bm{2}_2\overline{\bm{7}}_3\overline{\bm{6}}_4, \ 
\bm{5}_1\bm{4}_2\overline{\bm{7}}_3\overline{\bm{4}}_4, \ 
\bm{5}_1\bm{6}_2\overline{\bm{7}}_3\overline{\bm{2}}_4, \ 
\bm{5}_1\bm{6}_2\overline{\bm{7}}_3\overline{\bm{6}}_4, \ 
\bm{6}_1\bm{2}_2\overline{\bm{2}}_3\overline{\bm{2}}_4, \ 
\bm{6}_1\bm{2}_2\overline{\bm{2}}_3\overline{\bm{6}}_4, \ 
\bm{6}_1\bm{4}_2\overline{\bm{2}}_3\overline{\bm{4}}_4, \ 
\bm{6}_1\bm{6}_2\overline{\bm{2}}_3\overline{\bm{2}}_4, \ 
\bm{6}_1\bm{6}_2\overline{\bm{2}}_3\overline{\bm{6}}_4, \nonumber \\
&\bm{7}_1\bm{2}_2\overline{\bm{5}}_3\overline{\bm{2}}_4, \ 
\bm{7}_1\bm{2}_2\overline{\bm{5}}_3\overline{\bm{6}}_4, \ 
\bm{7}_1\bm{4}_2\overline{\bm{5}}_3\overline{\bm{4}}_4, \ 
\bm{7}_1\bm{6}_2\overline{\bm{5}}_3\overline{\bm{2}}_4, \ 
\bm{7}_1\bm{6}_2\overline{\bm{5}}_3\overline{\bm{6}}_4.
\end{align}
The corresponding integer vectors are given by 
\begin{subequations}
\begin{align}
M^{(8,8,\overline{8},\overline{8})}_\textrm{3F;1} = \begin{pmatrix}
1 & 0 & 0 \\
0 & 2 & 4 \\
3 & 0 & 0 \\
0 & 2 & 0
\end{pmatrix}, \quad
\Lambda^{(8,8,\overline{8},\overline{8})}_{\textsf{w},\textrm{3F;1}} = \begin{pmatrix}
-1 & -2 & -3 & -2 \\
1 & 4 & 3 & 0 \\
1 & 2 & 3 & -2 \\
3 & 0 & 1 & 0
\end{pmatrix}, \\
M^{(8,8,\overline{8},\overline{8})}_\textrm{3F;2} = \begin{pmatrix}
1 & 0 & 0 \\
0 & 2 & 4 \\
5 & 0 & 0 \\
0 & 2 & 0
\end{pmatrix}, \quad
\Lambda^{(8,8,\overline{8},\overline{8})}_{\textsf{w},\textrm{3F;2}} = \begin{pmatrix}
-3 & 0 & 1 & 4 \\
3 & -2 & -1 & -2 \\
-1 & 0 & 3 & 0 \\
3 & 2 & -1 & -2
\end{pmatrix}, \\
M^{(8,8,\overline{8},\overline{8})}_\textrm{3F;3} = \begin{pmatrix}
1 & 0 & 0 \\
0 & 2 & 4 \\
0 & 2 & 0 \\
3 & 0 & 0
\end{pmatrix}, \quad
\Lambda^{(8,8,\overline{8},\overline{8})}_{\textsf{w},\textrm{3F;3}} \
= \begin{pmatrix}
-1 & 0 & 0 & -3 \\
3 & 2 & -2 & 1 \\
-3 & 0 & 4 & -1 \\
3 & -2 & -2 & 1
\end{pmatrix}, \\
M^{(8,8,\overline{8},\overline{8})}_\textrm{3F;4} = \begin{pmatrix}
1 & 0 & 0 \\
0 & 2 & 4 \\
0 & 2 & 0 \\
5 & 0 & 0
\end{pmatrix}, \quad
\Lambda^{(8,8,\overline{8},\overline{8})}_{\textsf{w},\textrm{3F;4}} \
= \begin{pmatrix}
-1 & 2 & -2 & 3 \\
3 & 0 & 0 & -1 \\
1 & -2 & -2 & -3 \\
1 & -4 & 0 & -3
\end{pmatrix}, \\
M^{(8,8,\overline{8},\overline{8})}_\textrm{3F;5} = \begin{pmatrix}
2 & 4 & 0 \\
0 & 0 & 1 \\
2 & 0 & 0 \\
0 & 0 & 3
\end{pmatrix}, \quad
\Lambda^{(8,8,\overline{8},\overline{8})}_{\textsf{w},\textrm{3F;5}} = \begin{pmatrix}
2 & -1 & -2 & -3 \\
0 & 3 & 0 & 1 \\
-2 & 1 & -2 & 3 \\
-4 & 1 & 0 & 3
\end{pmatrix}, \\
M^{(8,8,\overline{8},\overline{8})}_\textrm{3F;6} = \begin{pmatrix}
2 & 4 & 0 \\
0 & 0 & 1 \\
2 & 0 & 0 \\
0 & 0 & 5
\end{pmatrix}, \quad
\Lambda^{(8,8,\overline{8},\overline{8})}_{\textsf{w},\textrm{3F;6}} \
= \begin{pmatrix}
2 & -1 & -2 & 3 \\
0 & 3 & 0 & -1 \\
-2 & 1 & -2 & -3 \\
-4 & 1 & 0 & -3
\end{pmatrix}, \\
M^{(8,8,\overline{8},\overline{8})}_\textrm{3F;7} = \begin{pmatrix}
2 & 4 & 0 \\
0 & 0 & 1 \\
0 & 0 & 3 \\
2 & 0 & 0
\end{pmatrix}, \quad
\Lambda^{(8,8,\overline{8},\overline{8})}_{\textsf{w},\textrm{3F;7}} \
= \begin{pmatrix}
0 & -1 & -3 & 0 \\
2 & 3 & 1 & -2 \\
0 & -3 & -1 & 4 \\
-2 & 3 & 1 & -2
\end{pmatrix}, \\
M^{(8,8,\overline{8},\overline{8})}_\textrm{3F;8} = \begin{pmatrix}
2 & 4 & 0 \\
0 & 0 & 1 \\
0 & 0 & 5 \\
2 & 0 & 0
\end{pmatrix}, \quad
\Lambda^{(8,8,\overline{8},\overline{8})}_{\textsf{w},\textrm{3F;8}} = \begin{pmatrix}
2 & -1 & 3 & -2 \\
0 & 3 & -1 & 0 \\
-2 & 1 & -3 & -2 \\
-4 & 1 & -3 & 0
\end{pmatrix}.
\end{align}
\end{subequations}

The rest of 16 fermionic gapped interfaces at rank three are similar to the bosonic gapped interfaces $M^{(8,8,\overline{8},\overline{8})}_\textrm{3B;9--16}$, but the Lagrangian subgroups of the former involve fermionic quadruplets (e.g., $\bm{1} \bm{1} \overline{\bm{1}} \overline{\bm{3}}$) or pairs (e.g., $\bm{2} \bm{2}$), apart from pairs of $\bm{4}$ or $\overline{\bm{4}}$ from any two layers.
Taking the Lagrangian subgroup generated by $M^{(8,8,\overline{8},\overline{8})}_\textrm{3F;9}$ given below as an instance, the set of condensed anyons is given by 
\begin{align}
&\red{\bm{2}_1\bm{6}_2}, \ 
\bm{4}_1\bm{4}_2, \ 
\bm{6}_1\bm{2}_2, \ 
\red{\bm{4}_1\overline{\bm{4}}_3}, \ 
\bm{4}_1\overline{\bm{4}}_4, \ 
\bm{4}_2\overline{\bm{4}}_3, \ 
\bm{4}_2\overline{\bm{4}}_4, \ 
\overline{\bm{2}}_3\overline{\bm{6}}_4, \ 
\overline{\bm{4}}_3\overline{\bm{4}}_4, \ 
\overline{\bm{6}}_3\overline{\bm{2}}_4, \nonumber \\
&\bm{2}_1\bm{2}_2\overline{\bm{4}}_3, \ 
\bm{6}_1\bm{6}_2\overline{\bm{4}}_3, \ 
\bm{2}_1\bm{2}_2\overline{\bm{4}}_4, \ 
\bm{6}_1\bm{6}_2\overline{\bm{4}}_4, \ 
\bm{4}_1\overline{\bm{2}}_3\overline{\bm{2}}_4, \ 
\bm{4}_1\overline{\bm{6}}_3\overline{\bm{6}}_4, \ 
\bm{4}_2\overline{\bm{2}}_3\overline{\bm{2}}_4, \ 
\bm{4}_2\overline{\bm{6}}_3\overline{\bm{6}}_4, \nonumber \\
&\red{\bm{1}_1\bm{1}_2\overline{\bm{1}}_3\overline{\bm{1}}_4}, \ 
\bm{1}_1\bm{1}_2\overline{\bm{3}}_3\overline{\bm{7}}_4, \ 
\bm{1}_1\bm{1}_2\overline{\bm{5}}_3\overline{\bm{5}}_4, \ 
\bm{1}_1\bm{1}_2\overline{\bm{7}}_3\overline{\bm{3}}_4, \ 
\bm{1}_1\bm{5}_2\overline{\bm{1}}_3\overline{\bm{5}}_4, \ 
\bm{1}_1\bm{5}_2\overline{\bm{3}}_3\overline{\bm{3}}_4, \ 
\bm{1}_1\bm{5}_2\overline{\bm{5}}_3\overline{\bm{1}}_4, \ 
\bm{1}_1\bm{5}_2\overline{\bm{7}}_3\overline{\bm{7}}_4, \nonumber \\
&\bm{2}_1\bm{2}_2\overline{\bm{2}}_3\overline{\bm{2}}_4, \ 
\bm{2}_1\bm{2}_2\overline{\bm{6}}_3\overline{\bm{6}}_4, \ 
\bm{2}_1\bm{6}_2\overline{\bm{2}}_3\overline{\bm{6}}_4, \ 
\bm{2}_1\bm{6}_2\overline{\bm{4}}_3\overline{\bm{4}}_4, \ 
\bm{2}_1\bm{6}_2\overline{\bm{6}}_3\overline{\bm{2}}_4, \nonumber \\
&\bm{3}_1\bm{3}_2\overline{\bm{1}}_3\overline{\bm{5}}_4, \ 
\bm{3}_1\bm{3}_2\overline{\bm{3}}_3\overline{\bm{3}}_4, \ 
\bm{3}_1\bm{3}_2\overline{\bm{5}}_3\overline{\bm{1}}_4, \ 
\bm{3}_1\bm{3}_2\overline{\bm{7}}_3\overline{\bm{7}}_4, \ 
\bm{3}_1\bm{7}_2\overline{\bm{1}}_3\overline{\bm{1}}_4, \ 
\bm{3}_1\bm{7}_2\overline{\bm{3}}_3\overline{\bm{7}}_4, \ 
\bm{3}_1\bm{7}_2\overline{\bm{5}}_3\overline{\bm{5}}_4, \ 
\bm{3}_1\bm{7}_2\overline{\bm{7}}_3\overline{\bm{3}}_4, \nonumber \\
&\bm{4}_1\bm{4}_2\overline{\bm{2}}_3\overline{\bm{6}}_4, \ 
\bm{4}_1\bm{4}_2\overline{\bm{4}}_3\overline{\bm{4}}_4, \ 
\bm{4}_1\bm{4}_2\overline{\bm{6}}_3\overline{\bm{2}}_4, \nonumber \\
&\bm{5}_1\bm{1}_2\overline{\bm{1}}_3\overline{\bm{5}}_4, \ 
\bm{5}_1\bm{1}_2\overline{\bm{3}}_3\overline{\bm{3}}_4, \ 
\bm{5}_1\bm{1}_2\overline{\bm{5}}_3\overline{\bm{1}}_4, \ 
\bm{5}_1\bm{1}_2\overline{\bm{7}}_3\overline{\bm{7}}_4, \ 
\bm{5}_1\bm{5}_2\overline{\bm{1}}_3\overline{\bm{1}}_4, \ 
\bm{5}_1\bm{5}_2\overline{\bm{3}}_3\overline{\bm{7}}_4, \ 
\bm{5}_1\bm{5}_2\overline{\bm{5}}_3\overline{\bm{5}}_4, \ 
\bm{5}_1\bm{5}_2\overline{\bm{7}}_3\overline{\bm{3}}_4, \nonumber \\
&\bm{6}_1\bm{2}_2\overline{\bm{2}}_3\overline{\bm{6}}_4, \ 
\bm{6}_1\bm{2}_2\overline{\bm{4}}_3\overline{\bm{4}}_4, \ 
\bm{6}_1\bm{2}_2\overline{\bm{6}}_3\overline{\bm{2}}_4, \ 
\bm{6}_1\bm{6}_2\overline{\bm{2}}_3\overline{\bm{2}}_4, \ 
\bm{6}_1\bm{6}_2\overline{\bm{6}}_3\overline{\bm{6}}_4, \nonumber \\
&\bm{7}_1\bm{3}_2\overline{\bm{1}}_3\overline{\bm{1}}_4, \ 
\bm{7}_1\bm{3}_2\overline{\bm{3}}_3\overline{\bm{7}}_4, \ 
\bm{7}_1\bm{3}_2\overline{\bm{5}}_3\overline{\bm{5}}_4, \ 
\bm{7}_1\bm{3}_2\overline{\bm{7}}_3\overline{\bm{3}}_4, \ 
\bm{7}_1\bm{7}_2\overline{\bm{1}}_3\overline{\bm{5}}_4, \ 
\bm{7}_1\bm{7}_2\overline{\bm{3}}_3\overline{\bm{3}}_4, \ 
\bm{7}_1\bm{7}_2\overline{\bm{5}}_3\overline{\bm{1}}_4, \ 
\bm{7}_1\bm{7}_2\overline{\bm{7}}_3\overline{\bm{7}}_4.
\end{align}
The corresponding integer vectors are given by 
\begin{subequations}
\begin{align}
M^{(8,8,\overline{8},\overline{8})}_\textrm{3F;9} = \begin{pmatrix}
1 & 2 & 4 \\
1 & 6 & 0 \\
1 & 0 & 4 \\
1 & 0 & 0
\end{pmatrix}, \quad
\Lambda^{(8,8,\overline{8},\overline{8})}_{\textsf{w},\textrm{3F;9}} \
= \begin{pmatrix}
-1 & 3 & -3 & -3 \\
-3 & 1 & -1 & -1 \\
0 & 0 & 2 & -2 \\
0 & 4 & -2 & -2
\end{pmatrix}, \\
M^{(8,8,\overline{8},\overline{8})}_\textrm{3F;10} = \begin{pmatrix}
1 & 2 & 4 \\
1 & 6 & 0 \\
1 & 0 & 4 \\
3 & 0 & 0
\end{pmatrix}, \quad
\Lambda^{(8,8,\overline{8},\overline{8})}_{\textsf{w},\textrm{3F;10}} = \begin{pmatrix}
-3 & 1 & -3 & 3 \\
-1 & 3 & -1 & 1 \\
0 & 0 & -2 & -2 \\
-4 & 0 & -2 & 2
\end{pmatrix}, \\
M^{(8,8,\overline{8},\overline{8})}_\textrm{3F;11} = \begin{pmatrix}
1 & 2 & 4 \\
1 & 0 & 4 \\
1 & 2 & 0 \\
3 & 0 & 0
\end{pmatrix}, \quad
\Lambda^{(8,8,\overline{8},\overline{8})}_{\textsf{w},\textrm{3F;11}} \
= \begin{pmatrix}
1 & -1 & 1 & -3 \\
3 & 1 & -1 & -1 \\
1 & 1 & -3 & -1 \\
-1 & 3 & -1 & 1
\end{pmatrix}, \\
M^{(8,8,\overline{8},\overline{8})}_\textrm{3F;12} = \begin{pmatrix}
1 & 2 & 4 \\
1 & 0 & 4 \\
1 & 0 & 0 \\
3 & 6 & 0
\end{pmatrix}, \quad
\Lambda^{(8,8,\overline{8},\overline{8})}_{\textsf{w},\textrm{3F;12}} \
= \begin{pmatrix}
-1 & 1 & 1 & -3 \\
-1 & -3 & 1 & 1 \\
1 & 1 & -3 & -1 \\
3 & -1 & -1 & 1
\end{pmatrix}, \\
M^{(8,8,\overline{8},\overline{8})}_\textrm{3F;13} = \begin{pmatrix}
1 & 2 & 4 \\
1 & 6 & 0 \\
1 & 0 & 4 \\
5 & 0 & 0
\end{pmatrix}, \quad
\Lambda^{(8,8,\overline{8},\overline{8})}_{\textsf{w},\textrm{3F;13}} = \begin{pmatrix}
-3 & 1 & -3 & -3 \\
-1 & 3 & -1 & -1 \\
0 & 0 & 2 & -2 \\
-4 & 0 & -2 & -2
\end{pmatrix}, \\
M^{(8,8,\overline{8},\overline{8})}_\textrm{3F;14} = \begin{pmatrix}
1 & 2 & 4 \\
1 & 0 & 4 \\
1 & 2 & 0 \\
5 & 0 & 0
\end{pmatrix}, \quad
\Lambda^{(8,8,\overline{8},\overline{8})}_{\textsf{w},\textrm{3F;14}} = \begin{pmatrix}
-1 & 1 & -1 & -3 \\
3 & 1 & -1 & 1 \\
1 & 1 & -3 & 1 \\
-1 & 3 & -1 & -1
\end{pmatrix}, \\
M^{(8,8,\overline{8},\overline{8})}_\textrm{3F;15} = \begin{pmatrix}
1 & 2 & 4 \\
1 & 0 & 4 \\
1 & 0 & 0 \\
5 & 2 & 0
\end{pmatrix}, \quad
\Lambda^{(8,8,\overline{8},\overline{8})}_{\textsf{w},\textrm{3F;15}} \
= \begin{pmatrix}
-1 & -1 & -1 & 3 \\
1 & -3 & 1 & 1 \\
-1 & 1 & -3 & -1 \\
3 & 1 & 1 & -1
\end{pmatrix}, \\
M^{(8,8,\overline{8},\overline{8})}_\textrm{3F;16} = \begin{pmatrix}
1 & 2 & 4 \\
1 & 6 & 0 \\
1 & 0 & 4 \\
7 & 0 & 0
\end{pmatrix}, \quad
\Lambda^{(8,8,\overline{8},\overline{8})}_{\textsf{w},\textrm{3F;16}} = \begin{pmatrix}
-3 & 1 & 3 & -3 \\
-1 & 3 & 1 & -1 \\
0 & 0 & -2 & -2 \\
-4 & 0 & 2 & -2
\end{pmatrix}, \\
M^{(8,8,\overline{8},\overline{8})}_\textrm{3F;17} = \begin{pmatrix}
1 & 2 & 4 \\
1 & 0 & 4 \\
3 & 6 & 0 \\
1 & 0 & 0
\end{pmatrix}, \quad
\Lambda^{(8,8,\overline{8},\overline{8})}_{\textsf{w},\textrm{3F;17}} \
= \begin{pmatrix}
-1 & 1 & -3 & 1 \\
-1 & -3 & 1 & 1 \\
1 & 1 & -1 & -3 \\
-3 & 1 & -1 & 1
\end{pmatrix}, \\
M^{(8,8,\overline{8},\overline{8})}_\textrm{3F;18} = \begin{pmatrix}
1 & 2 & 4 \\
1 & 0 & 4 \\
3 & 0 & 0 \\
1 & 2 & 0
\end{pmatrix}, \quad
\Lambda^{(8,8,\overline{8},\overline{8})}_{\textsf{w},\textrm{3F;18}} = \begin{pmatrix}
-1 & -1 & -3 & -1 \\
1 & 3 & 1 & 1 \\
1 & -1 & 1 & -3 \\
-3 & 1 & -1 & 1
\end{pmatrix}, \\
M^{(8,8,\overline{8},\overline{8})}_\textrm{3F;19} = \begin{pmatrix}
1 & 2 & 4 \\
1 & 0 & 4 \\
3 & 6 & 0 \\
7 & 0 & 0
\end{pmatrix}, \quad
\Lambda^{(8,8,\overline{8},\overline{8})}_{\textsf{w},\textrm{3F;19}} = \begin{pmatrix}
1 & 1 & 3 & -1 \\
-3 & -1 & -1 & 1 \\
1 & -1 & -1 & -3 \\
-1 & 3 & 1 & 1
\end{pmatrix}, \\
M^{(8,8,\overline{8},\overline{8})}_\textrm{3F;20} = \begin{pmatrix}
1 & 2 & 4 \\
1 & 0 & 4 \\
3 & 0 & 0 \\
7 & 6 & 0
\end{pmatrix}, \quad
\Lambda^{(8,8,\overline{8},\overline{8})}_{\textsf{w},\textrm{3F;20}} = \begin{pmatrix}
1 & -1 & -3 & -1 \\
3 & 1 & -1 & 1 \\
1 & 1 & -1 & 3 \\
1 & -3 & -1 & -1
\end{pmatrix}, \\
M^{(8,8,\overline{8},\overline{8})}_\textrm{3F;21} = \begin{pmatrix}
1 & 2 & 4 \\
3 & 2 & 0 \\
1 & 0 & 4 \\
1 & 0 & 0
\end{pmatrix}, \quad
\Lambda^{(8,8,\overline{8},\overline{8})}_{\textsf{w},\textrm{3F;21}} = \begin{pmatrix}
3 & 1 & 3 & 3 \\
1 & 3 & 1 & 1 \\
0 & 0 & 2 & -2 \\
4 & 0 & 2 & 2
\end{pmatrix}, \\
M^{(8,8,\overline{8},\overline{8})}_\textrm{3F;22} = \begin{pmatrix}
1 & 2 & 4 \\
3 & 2 & 0 \\
1 & 0 & 4 \\
3 & 0 & 0
\end{pmatrix}, \quad
\Lambda^{(8,8,\overline{8},\overline{8})}_{\textsf{w},\textrm{3F;22}} = \begin{pmatrix}
1 & 3 & 3 & -3 \\
3 & 1 & 1 & -1 \\
0 & 0 & -2 & -2 \\
0 & 4 & 2 & -2
\end{pmatrix}, \\
M^{(8,8,\overline{8},\overline{8})}_\textrm{3F;23} = \begin{pmatrix}
1 & 2 & 4 \\
3 & 2 & 0 \\
1 & 0 & 4 \\
5 & 0 & 0
\end{pmatrix}, \quad
\Lambda^{(8,8,\overline{8},\overline{8})}_{\textsf{w},\textrm{3F;23}} = \begin{pmatrix}
-3 & -1 & 3 & 3 \\
-1 & -3 & 1 & 1 \\
0 & 0 & 2 & -2 \\
-4 & 0 & 2 & 2
\end{pmatrix}, \\
M^{(8,8,\overline{8},\overline{8})}_\textrm{3F;24} = \begin{pmatrix}
1 & 2 & 4 \\
3 & 2 & 0 \\
1 & 0 & 4 \\
7 & 0 & 0
\end{pmatrix}, \quad
\Lambda^{(8,8,\overline{8},\overline{8})}_{\textsf{w},\textrm{3F;24}} = \begin{pmatrix}
-3 & -1 & -3 & 3 \\
-1 & -3 & -1 & 1 \\
0 & 0 & 2 & 2 \\
-4 & 0 & -2 & 2
\end{pmatrix}.
\end{align}
\end{subequations}

There are three gapped interfaces at rank four, among which two are bosonic whereas one is fermionic.
The two bosonic gapped interfaces at rank four are obtained by stacking two bosonic interfaces between $U(1)_8$ and $\overline{U(1)}_8$ via the condensation of $\bm{2} \overline{\bm{2}}$ and $\bm{4}$ or $\overline{\bm{4}}$.
Taking the Lagrangian subgroup generated by $M^{(8,8,\overline{8},\overline{8})}_\textrm{4B;1}$ given below as an instance, the set of condensed anyons is given by 
\begin{align}
&\red{\bm{4}_1}, \ 
\red{\bm{4}_2}, \ 
\overline{\bm{4}}_3, \ 
\overline{\bm{4}}_4, \ 
\bm{4}_1\bm{4}_2, \ 
\red{\bm{2}_1\overline{\bm{2}}_3}, \ 
\bm{2}_1\overline{\bm{6}}_3, \ 
\bm{4}_1\overline{\bm{4}}_3, \ 
\bm{6}_1\overline{\bm{2}}_3, \ 
\bm{6}_1\overline{\bm{6}}_3, \ 
\bm{4}_1\overline{\bm{4}}_4, \ 
\bm{4}_2\overline{\bm{4}}_3, \ 
\red{\bm{2}_2\overline{\bm{2}}_4}, \ 
\bm{2}_2\overline{\bm{6}}_4, \ 
\bm{4}_2\overline{\bm{4}}_4, \ 
\bm{6}_2\overline{\bm{2}}_4, \ 
\bm{6}_2\overline{\bm{6}}_4, \ 
\overline{\bm{4}}_3\overline{\bm{4}}_4, \nonumber \\
&\bm{2}_1\bm{4}_2\overline{\bm{2}}_3, \ 
\bm{2}_1\bm{4}_2\overline{\bm{6}}_3, \ 
\bm{4}_1\bm{4}_2\overline{\bm{4}}_3, \ 
\bm{6}_1\bm{4}_2\overline{\bm{2}}_3, \ 
\bm{6}_1\bm{4}_2\overline{\bm{6}}_3, \ 
\bm{4}_1\bm{2}_2\overline{\bm{2}}_4, \ 
\bm{4}_1\bm{2}_2\overline{\bm{6}}_4, \ 
\bm{4}_1\bm{4}_2\overline{\bm{4}}_4, \ 
\bm{4}_1\bm{6}_2\overline{\bm{2}}_4, \ 
\bm{4}_1\bm{6}_2\overline{\bm{6}}_4, \nonumber \\
&\bm{2}_1\overline{\bm{2}}_3\overline{\bm{4}}_4, \ 
\bm{2}_1\overline{\bm{6}}_3\overline{\bm{4}}_4, \ 
\bm{4}_1\overline{\bm{4}}_3\overline{\bm{4}}_4, \ 
\bm{6}_1\overline{\bm{2}}_3\overline{\bm{4}}_4, \ 
\bm{6}_1\overline{\bm{6}}_3\overline{\bm{4}}_4, \ 
\bm{2}_2\overline{\bm{4}}_3\overline{\bm{2}}_4, \ 
\bm{2}_2\overline{\bm{4}}_3\overline{\bm{6}}_4, \ 
\bm{4}_2\overline{\bm{4}}_3\overline{\bm{4}}_4, \ 
\bm{6}_2\overline{\bm{4}}_3\overline{\bm{2}}_4, \ 
\bm{6}_2\overline{\bm{4}}_3\overline{\bm{6}}_4, \nonumber \\
&\bm{2}_1\bm{2}_2\overline{\bm{2}}_3\overline{\bm{2}}_4, \ 
\bm{2}_1\bm{2}_2\overline{\bm{2}}_3\overline{\bm{6}}_4, \ 
\bm{2}_1\bm{2}_2\overline{\bm{6}}_3\overline{\bm{2}}_4, \ 
\bm{2}_1\bm{2}_2\overline{\bm{6}}_3\overline{\bm{6}}_4, \ 
\bm{2}_1\bm{4}_2\overline{\bm{2}}_3\overline{\bm{4}}_4, \ 
\bm{2}_1\bm{4}_2\overline{\bm{6}}_3\overline{\bm{4}}_4, \ 
\bm{2}_1\bm{6}_2\overline{\bm{2}}_3\overline{\bm{2}}_4, \ 
\bm{2}_1\bm{6}_2\overline{\bm{2}}_3\overline{\bm{6}}_4, \ 
\bm{2}_1\bm{6}_2\overline{\bm{6}}_3\overline{\bm{2}}_4, \ 
\bm{2}_1\bm{6}_2\overline{\bm{6}}_3\overline{\bm{6}}_4, \nonumber \\
&\bm{4}_1\bm{2}_2\overline{\bm{4}}_3\overline{\bm{2}}_4, \ 
\bm{4}_1\bm{2}_2\overline{\bm{4}}_3\overline{\bm{6}}_4, \ 
\bm{4}_1\bm{4}_2\overline{\bm{4}}_3\overline{\bm{4}}_4, \ 
\bm{4}_1\bm{6}_2\overline{\bm{4}}_3\overline{\bm{2}}_4, \ 
\bm{4}_1\bm{6}_2\overline{\bm{4}}_3\overline{\bm{6}}_4, \nonumber \\
&\bm{6}_1\bm{2}_2\overline{\bm{2}}_3\overline{\bm{2}}_4, \ 
\bm{6}_1\bm{2}_2\overline{\bm{2}}_3\overline{\bm{6}}_4, \ 
\bm{6}_1\bm{2}_2\overline{\bm{6}}_3\overline{\bm{2}}_4, \ 
\bm{6}_1\bm{2}_2\overline{\bm{6}}_3\overline{\bm{6}}_4, \ 
\bm{6}_1\bm{4}_2\overline{\bm{2}}_3\overline{\bm{4}}_4, \ 
\bm{6}_1\bm{4}_2\overline{\bm{6}}_3\overline{\bm{4}}_4, \ 
\bm{6}_1\bm{6}_2\overline{\bm{2}}_3\overline{\bm{2}}_4, \ 
\bm{6}_1\bm{6}_2\overline{\bm{2}}_3\overline{\bm{6}}_4, \ 
\bm{6}_1\bm{6}_2\overline{\bm{6}}_3\overline{\bm{2}}_4, \ 
\bm{6}_1\bm{6}_2\overline{\bm{6}}_3\overline{\bm{6}}_4.
\end{align}
The corresponding integer vectors are given by 
\begin{subequations}
\begin{align}
M^{(8,8,\overline{8},\overline{8})}_\textrm{4B;1} = \begin{pmatrix}
2 & 4 & 0 & 0 \\
0 & 0 & 2 & 4 \\
2 & 0 & 0 & 0 \\
0 & 0 & 2 & 0
\end{pmatrix}, \quad
\Lambda^{(8,8,\overline{8},\overline{8})}_{\textsf{w},\textrm{4B;1}} = \begin{pmatrix}
2 & 0 & -2 & 0 \\
-2 & 0 & -2 & 0 \\
0 & 2 & 0 & -2 \\
0 & -2 & 0 & -2
\end{pmatrix}, \\
M^{(8,8,\overline{8},\overline{8})}_\textrm{4B;2} = \begin{pmatrix}
2 & 4 & 0 & 0 \\
0 & 0 & 2 & 4 \\
0 & 0 & 2 & 0 \\
2 & 0 & 0 & 0
\end{pmatrix}, \quad
\Lambda^{(8,8,\overline{8},\overline{8})}_{\textsf{w},\textrm{4B;2}} = \begin{pmatrix}
2 & 0 & 0 & -2 \\
-2 & 0 & 0 & -2 \\
0 & -2 & 2 & 0 \\
0 & 2 & 2 & 0
\end{pmatrix}.
\end{align}
\end{subequations}
The fermionic gapped interface at rank four is obtained by stacking a fermionic interface between two $U(1)_8$ layers via the condensation of $\bm{2} \bm{2}$ and $\bm{4}$ from each layer and its antichiral counterpart. 
The corresponding integer vectors are given by 
\begin{align}
M^{(8,8,\overline{8},\overline{8})}_\textrm{4F;1} = \begin{pmatrix}
2 & 4 & 0 & 0 \\
2 & 0 & 0 & 0 \\
0 & 0 & 2 & 4 \\
0 & 0 & 2 & 0
\end{pmatrix}, \quad
\Lambda^{(8,8,\overline{8},\overline{8})}_{\textsf{w},\textrm{4F;1}} = \begin{pmatrix}
0 & 0 & -2 & -2 \\
-2 & -2 & 0 & 0 \\
0 & 0 & 2 & -2 \\
2 & -2 & 0 & 0
\end{pmatrix}.
\end{align}

\subsubsection{$U(1)_9 \times U(1)_9 \times \overline{U(1)}_9 \times \overline{U(1)}_9$}

We consider gapped interfaces of the fermionic topological order corresponding to the $K$ matrix: 
\begin{align}
K^{(9,9,\overline{9},\overline{9})} = \begin{pmatrix} 9 &&& \\ & 9 && \\ && -9 & \\ &&& -9 \end{pmatrix}.
\end{align}
There are 24 Lagrangian subgroups of rank two, 16 of rank three, and one of rank four. 
Eight gapped interfaces at rank two are obtained by stacking two fermionic gapped interfaces between $U(1)_9$ and $\overline{U(1)}_9$ with tunneling- or pairing-type interactions, which lead to the condensations of $\bm{1} \overline{\bm{8}}$ or $\bm{1} \overline{\bm{1}}$. 
The corresponding integer vectors are given by 
\begin{subequations}
\begin{align}
M^{(9,9,\overline{9},\overline{9})}_\textrm{2F;1} = \begin{pmatrix}
1 & 0 \\
0 & 1 \\
1 & 0 \\
0 & 1
\end{pmatrix}, \quad
\Lambda^{(9,9,\overline{9},\overline{9})}_{\textsf{w},\textrm{2F;1}} = \begin{pmatrix}
4 & 0 & -5 & 0 \\
-5 & 0 & 4 & 0 \\
0 & 4 & 0 & -5 \\
0 & -5 & 0 & 4
\end{pmatrix}, \\
M^{(9,9,\overline{9},\overline{9})}_\textrm{2F;2} = \begin{pmatrix}
1 & 0 \\
0 & 1 \\
1 & 0 \\
0 & 8
\end{pmatrix}, \quad
\Lambda^{(9,9,\overline{9},\overline{9})}_{\textsf{w},\textrm{2F;2}} \
= \begin{pmatrix}
4 & 0 & -5 & 0 \\
-5 & 0 & 4 & 0 \\
0 & -4 & 0 & -5 \\
0 & -5 & 0 & -4
\end{pmatrix}, \\
M^{(9,9,\overline{9},\overline{9})}_\textrm{2F;3} = \begin{pmatrix}
1 & 0 \\
0 & 1 \\
8 & 0 \\
0 & 1
\end{pmatrix}, \quad
\Lambda^{(9,9,\overline{9},\overline{9})}_{\textsf{w},\textrm{2F;3}} = \begin{pmatrix}
-4 & 0 & -5 & 0 \\
0 & -5 & 0 & 4 \\
0 & 4 & 0 & -5 \\
-5 & 0 & -4 & 0
\end{pmatrix}, \\
M^{(9,9,\overline{9},\overline{9})}_\textrm{2F;4} = \begin{pmatrix}
1 & 0 \\
0 & 1 \\
8 & 0 \\
0 & 8
\end{pmatrix}, \quad
\Lambda^{(9,9,\overline{9},\overline{9})}_{\textsf{w},\textrm{2F;4}} = \begin{pmatrix}
-4 & 0 & -5 & 0 \\
-5 & 0 & -4 & 0 \\
0 & -4 & 0 & -5 \\
0 & -5 & 0 & -4
\end{pmatrix}, \\
M^{(9,9,\overline{9},\overline{9})}_\textrm{2F;5} = \begin{pmatrix}
1 & 0 \\
0 & 1 \\
0 & 1 \\
1 & 0
\end{pmatrix}, \quad
\Lambda^{(9,9,\overline{9},\overline{9})}_{\textsf{w},\textrm{2F;5}} \
= \begin{pmatrix}
0 & 4 & -5 & 0 \\
-5 & 0 & 0 & 4 \\
4 & 0 & 0 & -5 \\
0 & -5 & 4 & 0
\end{pmatrix}, \\
M^{(9,9,\overline{9},\overline{9})}_\textrm{2F;6} = \begin{pmatrix}
1 & 0 \\
0 & 1 \\
0 & 8 \\
1 & 0
\end{pmatrix}, \quad
\Lambda^{(9,9,\overline{9},\overline{9})}_{\textsf{w},\textrm{2F;6}} = \begin{pmatrix}
0 & -4 & -5 & 0 \\
-5 & 0 & 0 & 4 \\
4 & 0 & 0 & -5 \\
0 & -5 & -4 & 0
\end{pmatrix}, \\
M^{(9,9,\overline{9},\overline{9})}_\textrm{2F;7} = \begin{pmatrix}
1 & 0 \\
0 & 1 \\
0 & 1 \\
8 & 0
\end{pmatrix}, \quad
\Lambda^{(9,9,\overline{9},\overline{9})}_{\textsf{w},\textrm{2F;7}} \
= \begin{pmatrix}
0 & 4 & -5 & 0 \\
0 & -5 & 4 & 0 \\
-4 & 0 & 0 & -5 \\
-5 & 0 & 0 & -4
\end{pmatrix}, \\
M^{(9,9,\overline{9},\overline{9})}_\textrm{2F;8} = \begin{pmatrix}
1 & 0 \\
0 & 1 \\
0 & 8 \\
8 & 0
\end{pmatrix}, \quad
\Lambda^{(9,9,\overline{9},\overline{9})}_{\textsf{w},\textrm{2F;8}} \
= \begin{pmatrix}
0 & -4 & -5 & 0 \\
0 & -5 & -4 & 0 \\
-4 & 0 & 0 & -5 \\
-5 & 0 & 0 & -4
\end{pmatrix}.
\end{align}
\end{subequations}

There are 16 gapped interfaces at rank two, which are obtained by condensing triplets of anyons $\bm{1} \bm{3} \overline{\bm{1}}$. 
Such triplets can be understood as bound objects consisting of pairs of spin-1/18 anyons with different chiralities ($\bm{1} \overline{\bm{1}}$, $\bm{1} \overline{\bm{8}}$, $\bm{8} \overline{\bm{1}}$, or $\bm{8} \overline{\bm{8}}$) and a fermionic or bosonic anyon ($\bm{3}$, $\bm{6}$, $\overline{\bm{3}}$, or $\overline{\bm{8}}$). 
Taking the Lagrangian subgroup generated by $M^{(9,9,\overline{9},\overline{9})}_\textrm{2F;9}$ given below as an instance, the set of condensed anyons is given by 
\begin{align}
&\bm{3}_1\overline{\bm{6}}_4, \ 
\bm{6}_1\overline{\bm{3}}_4, \ 
\bm{3}_2\overline{\bm{3}}_3, \ 
\bm{6}_2\overline{\bm{6}}_3, \nonumber \\
&\bm{3}_1\bm{1}_2\overline{\bm{1}}_3, \ 
\bm{3}_1\bm{4}_2\overline{\bm{4}}_3, \ 
\bm{3}_1\bm{7}_2\overline{\bm{7}}_3, \ 
\bm{6}_1\bm{2}_2\overline{\bm{2}}_3, \ 
\bm{6}_1\bm{5}_2\overline{\bm{5}}_3, \ 
\bm{6}_1\bm{8}_2\overline{\bm{8}}_3, \ 
\bm{1}_1\bm{6}_2\overline{\bm{8}}_4, \ 
\bm{2}_1\bm{3}_2\overline{\bm{7}}_4, \ 
\bm{4}_1\bm{6}_2\overline{\bm{5}}_4, \ 
\bm{5}_1\bm{3}_2\overline{\bm{4}}_4, \ 
\bm{7}_1\bm{6}_2\overline{\bm{2}}_4, \ 
\bm{8}_1\bm{3}_2\overline{\bm{1}}_4, \nonumber \\
&\red{\bm{1}_1\overline{\bm{3}}_3\overline{\bm{8}}_4}, \ 
\bm{2}_1\overline{\bm{6}}_3\overline{\bm{7}}_4, \ 
\bm{4}_1\overline{\bm{3}}_3\overline{\bm{5}}_4, \ 
\bm{5}_1\overline{\bm{6}}_3\overline{\bm{4}}_4, \ 
\bm{7}_1\overline{\bm{3}}_3\overline{\bm{2}}_4, \ 
\bm{8}_1\overline{\bm{6}}_3\overline{\bm{1}}_4, \ 
\red{\bm{1}_2\overline{\bm{1}}_3\overline{\bm{3}}_4}, \ 
\bm{2}_2\overline{\bm{2}}_3\overline{\bm{6}}_4, \ 
\bm{4}_2\overline{\bm{4}}_3\overline{\bm{3}}_4, \ 
\bm{5}_2\overline{\bm{5}}_3\overline{\bm{6}}_4, \ 
\bm{7}_2\overline{\bm{7}}_3\overline{\bm{3}}_4, \ 
\bm{8}_2\overline{\bm{8}}_3\overline{\bm{6}}_4, \nonumber \\
&\bm{1}_1\bm{1}_2\overline{\bm{4}}_3\overline{\bm{2}}_4, \ 
\bm{1}_1\bm{2}_2\overline{\bm{5}}_3\overline{\bm{5}}_4, \ 
\bm{1}_1\bm{3}_2\overline{\bm{6}}_3\overline{\bm{8}}_4, \ 
\bm{1}_1\bm{4}_2\overline{\bm{7}}_3\overline{\bm{2}}_4, \ 
\bm{1}_1\bm{5}_2\overline{\bm{8}}_3\overline{\bm{5}}_4, \ 
\bm{1}_1\bm{7}_2\overline{\bm{1}}_3\overline{\bm{2}}_4, \ 
\bm{1}_1\bm{8}_2\overline{\bm{2}}_3\overline{\bm{5}}_4, \nonumber \\
&\bm{2}_1\bm{1}_2\overline{\bm{7}}_3\overline{\bm{1}}_4, \ 
\bm{2}_1\bm{2}_2\overline{\bm{8}}_3\overline{\bm{4}}_4, \ 
\bm{2}_1\bm{4}_2\overline{\bm{1}}_3\overline{\bm{1}}_4, \ 
\bm{2}_1\bm{5}_2\overline{\bm{2}}_3\overline{\bm{4}}_4, \ 
\bm{2}_1\bm{6}_2\overline{\bm{3}}_3\overline{\bm{7}}_4, \ 
\bm{2}_1\bm{7}_2\overline{\bm{4}}_3\overline{\bm{1}}_4, \ 
\bm{2}_1\bm{8}_2\overline{\bm{5}}_3\overline{\bm{4}}_4, \nonumber \\
&\bm{3}_1\bm{2}_2\overline{\bm{2}}_3\overline{\bm{3}}_4, \ 
\bm{3}_1\bm{3}_2\overline{\bm{3}}_3\overline{\bm{6}}_4, \ 
\bm{3}_1\bm{5}_2\overline{\bm{5}}_3\overline{\bm{3}}_4, \ 
\bm{3}_1\bm{6}_2\overline{\bm{6}}_3\overline{\bm{6}}_4, \ 
\bm{3}_1\bm{8}_2\overline{\bm{8}}_3\overline{\bm{3}}_4, \nonumber \\
&\bm{4}_1\bm{1}_2\overline{\bm{4}}_3\overline{\bm{8}}_4, \ 
\bm{4}_1\bm{2}_2\overline{\bm{5}}_3\overline{\bm{2}}_4, \ 
\bm{4}_1\bm{3}_2\overline{\bm{6}}_3\overline{\bm{5}}_4, \ 
\bm{4}_1\bm{4}_2\overline{\bm{7}}_3\overline{\bm{8}}_4, \ 
\bm{4}_1\bm{5}_2\overline{\bm{8}}_3\overline{\bm{2}}_4, \ 
\bm{4}_1\bm{7}_2\overline{\bm{1}}_3\overline{\bm{8}}_4, \ 
\bm{4}_1\bm{8}_2\overline{\bm{2}}_3\overline{\bm{2}}_4, \nonumber \\
&\bm{5}_1\bm{1}_2\overline{\bm{7}}_3\overline{\bm{7}}_4, \ 
\bm{5}_1\bm{2}_2\overline{\bm{8}}_3\overline{\bm{1}}_4, \ 
\bm{5}_1\bm{4}_2\overline{\bm{1}}_3\overline{\bm{7}}_4, \ 
\bm{5}_1\bm{5}_2\overline{\bm{2}}_3\overline{\bm{1}}_4, \ 
\bm{5}_1\bm{6}_2\overline{\bm{3}}_3\overline{\bm{4}}_4, \ 
\bm{5}_1\bm{7}_2\overline{\bm{4}}_3\overline{\bm{7}}_4, \ 
\bm{5}_1\bm{8}_2\overline{\bm{5}}_3\overline{\bm{1}}_4, \nonumber \\
&\bm{6}_1\bm{1}_2\overline{\bm{1}}_3\overline{\bm{6}}_4, \ 
\bm{6}_1\bm{3}_2\overline{\bm{3}}_3\overline{\bm{3}}_4, \ 
\bm{6}_1\bm{4}_2\overline{\bm{4}}_3\overline{\bm{6}}_4, \ 
\bm{6}_1\bm{6}_2\overline{\bm{6}}_3\overline{\bm{3}}_4, \ 
\bm{6}_1\bm{7}_2\overline{\bm{7}}_3\overline{\bm{6}}_4, \nonumber \\
&\bm{7}_1\bm{1}_2\overline{\bm{4}}_3\overline{\bm{5}}_4, \ 
\bm{7}_1\bm{2}_2\overline{\bm{5}}_3\overline{\bm{8}}_4, \ 
\bm{7}_1\bm{3}_2\overline{\bm{6}}_3\overline{\bm{2}}_4, \ 
\bm{7}_1\bm{4}_2\overline{\bm{7}}_3\overline{\bm{5}}_4, \ 
\bm{7}_1\bm{5}_2\overline{\bm{8}}_3\overline{\bm{8}}_4, \ 
\bm{7}_1\bm{7}_2\overline{\bm{1}}_3\overline{\bm{5}}_4, \ 
\bm{7}_1\bm{8}_2\overline{\bm{2}}_3\overline{\bm{8}}_4, \nonumber \\
&\bm{8}_1\bm{1}_2\overline{\bm{7}}_3\overline{\bm{4}}_4, \ 
\bm{8}_1\bm{2}_2\overline{\bm{8}}_3\overline{\bm{7}}_4, \ 
\bm{8}_1\bm{4}_2\overline{\bm{1}}_3\overline{\bm{4}}_4, \ 
\bm{8}_1\bm{5}_2\overline{\bm{2}}_3\overline{\bm{7}}_4, \ 
\bm{8}_1\bm{6}_2\overline{\bm{3}}_3\overline{\bm{1}}_4, \ 
\bm{8}_1\bm{7}_2\overline{\bm{4}}_3\overline{\bm{4}}_4, \ 
\bm{8}_1\bm{8}_2\overline{\bm{5}}_3\overline{\bm{7}}_4.
\end{align}
The corresponding integer vectors are given by
\begin{subequations}
\begin{align}
M^{(9,9,\overline{9},\overline{9})}_\textrm{2F;9} = \begin{pmatrix}
0 & 1 \\
1 & 0 \\
1 & 3 \\
3 & 8
\end{pmatrix}, \quad
\Lambda^{(9,9,\overline{9},\overline{9})}_{\textsf{w},\textrm{2F;9}} \
= \begin{pmatrix}
1 & 0 & 3 & -1 \\
3 & 1 & 1 & 0 \\
0 & -1 & -1 & -3 \\
-1 & 3 & 0 & 1
\end{pmatrix}, \\
M^{(9,9,\overline{9},\overline{9})}_\textrm{2F;10} = \begin{pmatrix}
0 & 1 \\
1 & 0 \\
1 & 6 \\
3 & 1
\end{pmatrix}, \quad
\Lambda^{(9,9,\overline{9},\overline{9})}_{\textsf{w},\textrm{2F;10}} = \begin{pmatrix}
-1 & 0 & 3 & -1 \\
1 & 3 & 0 & 1 \\
0 & -1 & -1 & -3 \\
3 & -1 & -1 & 0
\end{pmatrix}, \\
M^{(9,9,\overline{9},\overline{9})}_\textrm{2F;11} = \begin{pmatrix}
0 & 1 \\
1 & 0 \\
1 & 6 \\
6 & 8
\end{pmatrix}, \quad
\Lambda^{(9,9,\overline{9},\overline{9})}_{\textsf{w},\textrm{2F;11}} \
= \begin{pmatrix}
1 & 0 & -3 & -1 \\
-1 & -3 & 0 & 1 \\
0 & 1 & 1 & -3 \\
3 & -1 & -1 & 0
\end{pmatrix}, \\
M^{(9,9,\overline{9},\overline{9})}_\textrm{2F;12} = \begin{pmatrix}
0 & 1 \\
1 & 0 \\
1 & 3 \\
6 & 1
\end{pmatrix}, \quad
\Lambda^{(9,9,\overline{9},\overline{9})}_{\textsf{w},\textrm{2F;12}} = \begin{pmatrix}
0 & -1 & -1 & 3 \\
1 & -3 & 0 & 1 \\
-1 & 0 & -3 & -1 \\
3 & 1 & 1 & 0
\end{pmatrix}, \\
M^{(9,9,\overline{9},\overline{9})}_\textrm{2F;13} = \begin{pmatrix}
0 & 1 \\
1 & 0 \\
3 & 1 \\
1 & 6
\end{pmatrix}, \quad
\Lambda^{(9,9,\overline{9},\overline{9})}_{\textsf{w},\textrm{2F;13}} = \begin{pmatrix}
-1 & 0 & -1 & 3 \\
-3 & 1 & 0 & 1 \\
0 & -1 & -3 & -1 \\
1 & 3 & 1 & 0
\end{pmatrix}, \\
M^{(9,9,\overline{9},\overline{9})}_\textrm{2F;14} = \begin{pmatrix}
0 & 1 \\
1 & 0 \\
3 & 8 \\
1 & 3
\end{pmatrix}, \quad
\Lambda^{(9,9,\overline{9},\overline{9})}_{\textsf{w},\textrm{2F;14}} \
= \begin{pmatrix}
1 & 0 & -1 & 3 \\
3 & 1 & 0 & 1 \\
0 & -1 & -3 & -1 \\
-1 & 3 & 1 & 0
\end{pmatrix}, \\
M^{(9,9,\overline{9},\overline{9})}_\textrm{2F;15} = \begin{pmatrix}
0 & 1 \\
1 & 3 \\
3 & 1 \\
8 & 0
\end{pmatrix}, \quad
\Lambda^{(9,9,\overline{9},\overline{9})}_{\textsf{w},\textrm{2F;15}} = \begin{pmatrix}
0 & 1 & 3 & -1 \\
1 & 3 & 1 & 0 \\
-1 & 0 & -1 & -3 \\
3 & -1 & 0 & 1
\end{pmatrix}, \\
M^{(9,9,\overline{9},\overline{9})}_\textrm{2F;16} = \begin{pmatrix}
0 & 1 \\
1 & 0 \\
3 & 8 \\
8 & 6
\end{pmatrix}, \quad
\Lambda^{(9,9,\overline{9},\overline{9})}_{\textsf{w},\textrm{2F;16}} = \begin{pmatrix}
0 & -1 & -3 & 1 \\
-3 & -1 & 0 & 1 \\
1 & 0 & -1 & -3 \\
1 & -3 & -1 & 0
\end{pmatrix}, \\
M^{(9,9,\overline{9},\overline{9})}_\textrm{2F;17} = \begin{pmatrix}
0 & 1 \\
1 & 0 \\
6 & 1 \\
1 & 3
\end{pmatrix}, \quad
\Lambda^{(9,9,\overline{9},\overline{9})}_{\textsf{w},\textrm{2F;17}} \
= \begin{pmatrix}
0 & -1 & 3 & -1 \\
3 & 1 & 0 & 1 \\
-1 & 0 & -1 & -3 \\
-1 & 3 & -1 & 0
\end{pmatrix}, \\
M^{(9,9,\overline{9},\overline{9})}_\textrm{2F;18} = \begin{pmatrix}
0 & 1 \\
1 & 0 \\
6 & 8 \\
1 & 6
\end{pmatrix}, \quad
\Lambda^{(9,9,\overline{9},\overline{9})}_{\textsf{w},\textrm{2F;18}} = \begin{pmatrix}
-1 & 0 & 1 & 3 \\
3 & -1 & 0 & -1 \\
0 & 1 & -3 & 1 \\
1 & 3 & -1 & 0
\end{pmatrix}, \\
M^{(9,9,\overline{9},\overline{9})}_\textrm{2F;19} = \begin{pmatrix}
0 & 1 \\
1 & 0 \\
6 & 1 \\
8 & 6
\end{pmatrix}, \quad
\Lambda^{(9,9,\overline{9},\overline{9})}_{\textsf{w},\textrm{2F;19}} = \begin{pmatrix}
1 & 0 & 1 & -3 \\
-1 & 3 & -1 & 0 \\
0 & -1 & 3 & 1 \\
3 & 1 & 0 & -1
\end{pmatrix}, \\
M^{(9,9,\overline{9},\overline{9})}_\textrm{2F;20} = \begin{pmatrix}
0 & 1 \\
1 & 0 \\
6 & 8 \\
8 & 3
\end{pmatrix}, \quad
\Lambda^{(9,9,\overline{9},\overline{9})}_{\textsf{w},\textrm{2F;20}} \
= \begin{pmatrix}
0 & 1 & -3 & -1 \\
-1 & -3 & 1 & 0 \\
-1 & 0 & 1 & -3 \\
-3 & 1 & 0 & -1
\end{pmatrix}, \\
M^{(9,9,\overline{9},\overline{9})}_\textrm{2F;21} = \begin{pmatrix}
0 & 1 \\
1 & 0 \\
8 & 6 \\
3 & 8
\end{pmatrix}, \quad
\Lambda^{(9,9,\overline{9},\overline{9})}_{\textsf{w},\textrm{2F;21}} = \begin{pmatrix}
1 & 0 & -3 & -1 \\
-3 & -1 & 1 & 0 \\
0 & -1 & 1 & -3 \\
-1 & 3 & 0 & 1
\end{pmatrix}, \\
M^{(9,9,\overline{9},\overline{9})}_\textrm{2F;22} = \begin{pmatrix}
0 & 1 \\
1 & 0 \\
8 & 3 \\
3 & 1
\end{pmatrix}, \quad
\Lambda^{(9,9,\overline{9},\overline{9})}_{\textsf{w},\textrm{2F;22}} = \begin{pmatrix}
0 & 1 & -1 & 3 \\
1 & 3 & 0 & 1 \\
-1 & 0 & -3 & -1 \\
3 & -1 & 1 & 0
\end{pmatrix}, \\
M^{(9,9,\overline{9},\overline{9})}_\textrm{2F;23} = \begin{pmatrix}
0 & 1 \\
1 & 0 \\
8 & 3 \\
6 & 8
\end{pmatrix}, \quad
\Lambda^{(9,9,\overline{9},\overline{9})}_{\textsf{w},\textrm{2F;23}} = \begin{pmatrix}
-1 & 0 & -3 & 1 \\
-1 & -3 & 0 & 1 \\
0 & 1 & -1 & -3 \\
3 & -1 & 1 & 0
\end{pmatrix}, \\M^{(9,9,\overline{9},\overline{9})}_\textrm{2F;24} = \begin{pmatrix}
0 & 1 \\
1 & 0 \\
8 & 6 \\
6 & 1
\end{pmatrix}, \quad
\Lambda^{(9,9,\overline{9},\overline{9})}_{\textsf{w},\textrm{2F;24}} \
= \begin{pmatrix}
0 & 1 & -1 & -3 \\
1 & -3 & 0 & 1 \\
1 & 0 & -3 & 1 \\
3 & 1 & -1 & 0
\end{pmatrix}.
\end{align}
\end{subequations}

Eight gapped interfaces at rank three are obtained by stacking a gapped interface between $U(1)_9$ and $\overline{U(1)}_9$ via the condensation of $\bm{1} \overline{\bm{1}}$ or $\bm{1} \overline{\bm{8}}$ between two layers, a gapped interface between $U(1)_9$ and a Chern insulator $\overline{U(1)}_1$ via the condensation of $\bm{3}$, and a gapped interface between $\overline{U(1)}_9$ and a Chern insulator $\overline{U(1)}_1$ via the condensation of $\overline{\bm{3}}$. 
Taking the Lagrangian subgroup generated by $M^{(9,9,\overline{9},\overline{9})}_\textrm{3F;1}$ given below as an instance, the set of condensed anyons is given by
\begin{align}
&\red{\bm{3}_1}, \ 
\red{\bm{6}_1}, \ 
\overline{\bm{3}}_3, \ 
\overline{\bm{6}}_3, \ 
\bm{3}_1\overline{\bm{3}}_3, \ 
\bm{3}_1\overline{\bm{6}}_3, \ 
\bm{6}_1\overline{\bm{3}}_3, \ 
\bm{6}_1\overline{\bm{6}}_3, \ 
\red{\bm{1}_2\overline{\bm{1}}_4}, \ 
\bm{2}_2\overline{\bm{2}}_4, \ 
\bm{3}_2\overline{\bm{3}}_4, \ 
\bm{4}_2\overline{\bm{4}}_4, \ 
\bm{5}_2\overline{\bm{5}}_4, \ 
\bm{6}_2\overline{\bm{6}}_4, \ 
\bm{7}_2\overline{\bm{7}}_4, \ 
\bm{8}_2\overline{\bm{8}}_4, \nonumber \\
&\bm{3}_1\bm{1}_2\overline{\bm{1}}_4, \ 
\bm{3}_1\bm{2}_2\overline{\bm{2}}_4, \ 
\bm{3}_1\bm{3}_2\overline{\bm{3}}_4, \ 
\bm{3}_1\bm{4}_2\overline{\bm{4}}_4, \ 
\bm{3}_1\bm{5}_2\overline{\bm{5}}_4, \ 
\bm{3}_1\bm{6}_2\overline{\bm{6}}_4, \ 
\bm{3}_1\bm{7}_2\overline{\bm{7}}_4, \ 
\bm{3}_1\bm{8}_2\overline{\bm{8}}_4, \nonumber \\
&\bm{6}_1\bm{1}_2\overline{\bm{1}}_4, \ 
\bm{6}_1\bm{2}_2\overline{\bm{2}}_4, \ 
\bm{6}_1\bm{3}_2\overline{\bm{3}}_4, \ 
\bm{6}_1\bm{4}_2\overline{\bm{4}}_4, \ 
\bm{6}_1\bm{5}_2\overline{\bm{5}}_4, \ 
\bm{6}_1\bm{6}_2\overline{\bm{6}}_4, \ 
\bm{6}_1\bm{7}_2\overline{\bm{7}}_4, \ 
\bm{6}_1\bm{8}_2\overline{\bm{8}}_4, \nonumber \\
&\bm{1}_2\overline{\bm{3}}_3\overline{\bm{1}}_4, \ 
\bm{1}_2\overline{\bm{6}}_3\overline{\bm{1}}_4, \ 
\bm{2}_2\overline{\bm{3}}_3\overline{\bm{2}}_4, \ 
\bm{2}_2\overline{\bm{6}}_3\overline{\bm{2}}_4, \ 
\bm{3}_2\overline{\bm{3}}_3\overline{\bm{3}}_4, \ 
\bm{3}_2\overline{\bm{6}}_3\overline{\bm{3}}_4, \ 
\bm{4}_2\overline{\bm{3}}_3\overline{\bm{4}}_4, \ 
\bm{4}_2\overline{\bm{6}}_3\overline{\bm{4}}_4, \nonumber \\
&\bm{5}_2\overline{\bm{3}}_3\overline{\bm{5}}_4, \ 
\bm{5}_2\overline{\bm{6}}_3\overline{\bm{5}}_4, \ 
\bm{6}_2\overline{\bm{3}}_3\overline{\bm{6}}_4, \ 
\bm{6}_2\overline{\bm{6}}_3\overline{\bm{6}}_4, \ 
\bm{7}_2\overline{\bm{3}}_3\overline{\bm{7}}_4, \ 
\bm{7}_2\overline{\bm{6}}_3\overline{\bm{7}}_4, \ 
\bm{8}_2\overline{\bm{3}}_3\overline{\bm{8}}_4, \ 
\bm{8}_2\overline{\bm{6}}_3\overline{\bm{8}}_4, \nonumber \\
&\bm{3}_1\bm{1}_2\overline{\bm{3}}_3\overline{\bm{1}}_4, \ 
\bm{3}_1\bm{1}_2\overline{\bm{6}}_3\overline{\bm{1}}_4, \ 
\bm{3}_1\bm{2}_2\overline{\bm{3}}_3\overline{\bm{2}}_4, \ 
\bm{3}_1\bm{2}_2\overline{\bm{6}}_3\overline{\bm{2}}_4, \ 
\bm{3}_1\bm{3}_2\overline{\bm{3}}_3\overline{\bm{3}}_4, \ 
\bm{3}_1\bm{3}_2\overline{\bm{6}}_3\overline{\bm{3}}_4, \ 
\bm{3}_1\bm{4}_2\overline{\bm{3}}_3\overline{\bm{4}}_4, \ 
\bm{3}_1\bm{4}_2\overline{\bm{6}}_3\overline{\bm{4}}_4, \nonumber \\
&\bm{3}_1\bm{5}_2\overline{\bm{3}}_3\overline{\bm{5}}_4, \ 
\bm{3}_1\bm{5}_2\overline{\bm{6}}_3\overline{\bm{5}}_4, \ 
\bm{3}_1\bm{6}_2\overline{\bm{3}}_3\overline{\bm{6}}_4, \ 
\bm{3}_1\bm{6}_2\overline{\bm{6}}_3\overline{\bm{6}}_4, \ 
\bm{3}_1\bm{7}_2\overline{\bm{3}}_3\overline{\bm{7}}_4, \ 
\bm{3}_1\bm{7}_2\overline{\bm{6}}_3\overline{\bm{7}}_4, \ 
\bm{3}_1\bm{8}_2\overline{\bm{3}}_3\overline{\bm{8}}_4, \ 
\bm{3}_1\bm{8}_2\overline{\bm{6}}_3\overline{\bm{8}}_4, \nonumber \\
&\bm{6}_1\bm{1}_2\overline{\bm{3}}_3\overline{\bm{1}}_4, \ 
\bm{6}_1\bm{1}_2\overline{\bm{6}}_3\overline{\bm{1}}_4, \ 
\bm{6}_1\bm{2}_2\overline{\bm{3}}_3\overline{\bm{2}}_4, \ 
\bm{6}_1\bm{2}_2\overline{\bm{6}}_3\overline{\bm{2}}_4, \ 
\bm{6}_1\bm{3}_2\overline{\bm{3}}_3\overline{\bm{3}}_4, \ 
\bm{6}_1\bm{3}_2\overline{\bm{6}}_3\overline{\bm{3}}_4, \ 
\bm{6}_1\bm{4}_2\overline{\bm{3}}_3\overline{\bm{4}}_4, \ 
\bm{6}_1\bm{4}_2\overline{\bm{6}}_3\overline{\bm{4}}_4, \nonumber \\
&\bm{6}_1\bm{5}_2\overline{\bm{3}}_3\overline{\bm{5}}_4, \ 
\bm{6}_1\bm{5}_2\overline{\bm{6}}_3\overline{\bm{5}}_4, \ 
\bm{6}_1\bm{6}_2\overline{\bm{3}}_3\overline{\bm{6}}_4, \ 
\bm{6}_1\bm{6}_2\overline{\bm{6}}_3\overline{\bm{6}}_4, \ 
\bm{6}_1\bm{7}_2\overline{\bm{3}}_3\overline{\bm{7}}_4, \ 
\bm{6}_1\bm{7}_2\overline{\bm{6}}_3\overline{\bm{7}}_4, \ 
\bm{6}_1\bm{8}_2\overline{\bm{3}}_3\overline{\bm{8}}_4, \ 
\bm{6}_1\bm{8}_2\overline{\bm{6}}_3\overline{\bm{8}}_4.
\end{align}
The corresponding integer vectors are given by
\begin{subequations}
\begin{align}
M^{(9,9,\overline{9},\overline{9})}_\textrm{3F;1} = \begin{pmatrix}
3 & 0 & 0 \\
0 & 0 & 1 \\
0 & 3 & 0 \\
0 & 0 & 1
\end{pmatrix}, \quad
\Lambda^{(9,9,\overline{9},\overline{9})}_{\textsf{w},\textrm{3F;1}} = \begin{pmatrix}
0 & 0 & -3 & 0 \\
0 & -5 & 0 & 4 \\
0 & 4 & 0 & -5 \\
-3 & 0 & 0 & 0
\end{pmatrix}, \\
M^{(9,9,\overline{9},\overline{9})}_\textrm{3F;2} = \begin{pmatrix}
3 & 0 & 0 \\
0 & 0 & 1 \\
0 & 3 & 0 \\
0 & 0 & 8
\end{pmatrix}, \quad
\Lambda^{(9,9,\overline{9},\overline{9})}_{\textsf{w},\textrm{3F;2}} = \begin{pmatrix}
0 & 0 & -3 & 0 \\
0 & -5 & 0 & -4 \\
0 & -4 & 0 & -5 \\
-3 & 0 & 0 & 0
\end{pmatrix}, \\
M^{(9,9,\overline{9},\overline{9})}_\textrm{3F;3} = \begin{pmatrix}
0 & 0 & 1 \\
3 & 0 & 0 \\
0 & 0 & 1 \\
0 & 3 & 0
\end{pmatrix}, \quad
\Lambda^{(9,9,\overline{9},\overline{9})}_{\textsf{w},\textrm{3F;3}} \
= \begin{pmatrix}
0 & 0 & 0 & -3 \\
-5 & 0 & 4 & 0 \\
4 & 0 & -5 & 0 \\
0 & -3 & 0 & 0
\end{pmatrix}, \\
M^{(9,9,\overline{9},\overline{9})}_\textrm{3F;4} = \begin{pmatrix}
0 & 0 & 1 \\
3 & 0 & 0 \\
0 & 0 & 8 \\
0 & 3 & 0
\end{pmatrix}, \quad
\Lambda^{(9,9,\overline{9},\overline{9})}_{\textsf{w},\textrm{3F;4}} = \begin{pmatrix}
0 & 0 & 0 & -3 \\
-5 & 0 & -4 & 0 \\
-4 & 0 & -5 & 0 \\
0 & -3 & 0 & 0
\end{pmatrix}, \\
M^{(9,9,\overline{9},\overline{9})}_\textrm{3F;5} = \begin{pmatrix}
3 & 0 & 0 \\
0 & 0 & 1 \\
0 & 0 & 1 \\
0 & 3 & 0
\end{pmatrix}, \quad
\Lambda^{(9,9,\overline{9},\overline{9})}_{\textsf{w},\textrm{3F;5}} = \begin{pmatrix}
0 & 0 & 0 & -3 \\
0 & -5 & 4 & 0 \\
0 & 4 & -5 & 0 \\
-3 & 0 & 0 & 0
\end{pmatrix}, \\
M^{(9,9,\overline{9},\overline{9})}_\textrm{3F;6} = \begin{pmatrix}
3 & 0 & 0 \\
0 & 0 & 1 \\
0 & 0 & 8 \\
0 & 3 & 0
\end{pmatrix}, \quad
\Lambda^{(9,9,\overline{9},\overline{9})}_{\textsf{w},\textrm{3F;6}} \
= \begin{pmatrix}
0 & 0 & 0 & -3 \\
0 & -5 & -4 & 0 \\
0 & -4 & -5 & 0 \\
-3 & 0 & 0 & 0
\end{pmatrix}, \\
M^{(9,9,\overline{9},\overline{9})}_\textrm{3F;7} = \begin{pmatrix}
0 & 0 & 1 \\
3 & 0 & 0 \\
0 & 3 & 0 \\
0 & 0 & 1
\end{pmatrix}, \quad
\Lambda^{(9,9,\overline{9},\overline{9})}_{\textsf{w},\textrm{3F;7}} = \begin{pmatrix}
0 & 0 & -3 & 0 \\
-5 & 0 & 0 & 4 \\
4 & 0 & 0 & -5 \\
0 & -3 & 0 & 0
\end{pmatrix}, \\
M^{(9,9,\overline{9},\overline{9})}_\textrm{3F;8} = \begin{pmatrix}
0 & 0 & 1 \\
3 & 0 & 0 \\
0 & 3 & 0 \\
0 & 0 & 8
\end{pmatrix}, \quad
\Lambda^{(9,9,\overline{9},\overline{9})}_{\textsf{w},\textrm{3F;8}} = \begin{pmatrix}
0 & 0 & -3 & 0 \\
-5 & 0 & 0 & -4 \\
-4 & 0 & 0 & -5 \\
0 & -3 & 0 & 0
\end{pmatrix}.
\end{align}
\end{subequations}

Another eight gapped interfaces at rank three are obtained by condensing pairs of $\bm{3}$, $\bm{6}$, $\overline{\bm{3}}$, and/or $\overline{\bm{6}}$ between arbitrary two layers and quadruplets of elementary anyons such as $\bm{1} \bm{1} \overline{\bm{1}} \overline{\bm{1}}$. 
Taking the Lagrangian subgroup generated by $M^{(9,9,\overline{9},\overline{9})}_\textrm{3F;9}$ given below as an instance, the set of condensed anyons is given by
\begin{align}
&\bm{3}_1\bm{6}_2, \ 
\bm{6}_1\bm{3}_2, \ 
\bm{3}_1\overline{\bm{3}}_3, \ 
\bm{6}_1\overline{\bm{6}}_3, \ 
\bm{3}_1\overline{\bm{3}}_4, \ 
\bm{6}_1\overline{\bm{6}}_4, \ 
\red{\bm{3}_2\overline{\bm{3}}_3}, \ 
\bm{6}_2\overline{\bm{6}}_3, \ 
\bm{3}_2\overline{\bm{3}}_4, \ 
\bm{6}_2\overline{\bm{6}}_4, \ 
\red{\overline{\bm{3}}_3\overline{\bm{6}}_4}, \ 
\overline{\bm{6}}_3\overline{\bm{3}}_4, \nonumber \\
&\bm{3}_1\bm{3}_2\overline{\bm{6}}_3, \ 
\bm{6}_1\bm{6}_2\overline{\bm{3}}_3, \ 
\bm{3}_1\bm{3}_2\overline{\bm{6}}_4, \ 
\bm{6}_1\bm{6}_2\overline{\bm{3}}_4, \ 
\bm{3}_1\overline{\bm{6}}_3\overline{\bm{6}}_4, \ 
\bm{6}_1\overline{\bm{3}}_3\overline{\bm{3}}_4, \ 
\bm{3}_2\overline{\bm{6}}_3\overline{\bm{6}}_4, \ 
\bm{6}_2\overline{\bm{3}}_3\overline{\bm{3}}_4, \nonumber \\
&\red{\bm{1}_1\bm{1}_2\overline{\bm{1}}_3\overline{\bm{1}}_4}, \ 
\bm{1}_1\bm{1}_2\overline{\bm{4}}_3\overline{\bm{7}}_4, \ 
\bm{1}_1\bm{1}_2\overline{\bm{7}}_3\overline{\bm{4}}_4, \ 
\bm{1}_1\bm{4}_2\overline{\bm{1}}_3\overline{\bm{4}}_4, \ 
\bm{1}_1\bm{4}_2\overline{\bm{4}}_3\overline{\bm{1}}_4, \ 
\bm{1}_1\bm{4}_2\overline{\bm{7}}_3\overline{\bm{7}}_4, \ 
\bm{1}_1\bm{7}_2\overline{\bm{1}}_3\overline{\bm{7}}_4, \ 
\bm{1}_1\bm{7}_2\overline{\bm{4}}_3\overline{\bm{4}}_4, \ 
\bm{1}_1\bm{7}_2\overline{\bm{7}}_3\overline{\bm{1}}_4, \nonumber \\
&\bm{2}_1\bm{2}_2\overline{\bm{2}}_3\overline{\bm{2}}_4, \ 
\bm{2}_1\bm{2}_2\overline{\bm{5}}_3\overline{\bm{8}}_4, \ 
\bm{2}_1\bm{2}_2\overline{\bm{8}}_3\overline{\bm{5}}_4, \ 
\bm{2}_1\bm{5}_2\overline{\bm{2}}_3\overline{\bm{5}}_4, \ 
\bm{2}_1\bm{5}_2\overline{\bm{5}}_3\overline{\bm{2}}_4, \ 
\bm{2}_1\bm{5}_2\overline{\bm{8}}_3\overline{\bm{8}}_4, \ 
\bm{2}_1\bm{8}_2\overline{\bm{2}}_3\overline{\bm{8}}_4, \ 
\bm{2}_1\bm{8}_2\overline{\bm{5}}_3\overline{\bm{5}}_4, \ 
\bm{2}_1\bm{8}_2\overline{\bm{8}}_3\overline{\bm{2}}_4, \nonumber \\
&\bm{3}_1\bm{3}_2\overline{\bm{3}}_3\overline{\bm{3}}_4, \ 
\bm{3}_1\bm{6}_2\overline{\bm{3}}_3\overline{\bm{6}}_4, \ 
\bm{3}_1\bm{6}_2\overline{\bm{6}}_3\overline{\bm{3}}_4, \nonumber \\
&\bm{4}_1\bm{1}_2\overline{\bm{1}}_3\overline{\bm{4}}_4, \ 
\bm{4}_1\bm{1}_2\overline{\bm{4}}_3\overline{\bm{1}}_4, \ 
\bm{4}_1\bm{1}_2\overline{\bm{7}}_3\overline{\bm{7}}_4, \ 
\bm{4}_1\bm{4}_2\overline{\bm{1}}_3\overline{\bm{7}}_4, \ 
\bm{4}_1\bm{4}_2\overline{\bm{4}}_3\overline{\bm{4}}_4, \ 
\bm{4}_1\bm{4}_2\overline{\bm{7}}_3\overline{\bm{1}}_4, \ 
\bm{4}_1\bm{7}_2\overline{\bm{1}}_3\overline{\bm{1}}_4, \ 
\bm{4}_1\bm{7}_2\overline{\bm{4}}_3\overline{\bm{7}}_4, \ 
\bm{4}_1\bm{7}_2\overline{\bm{7}}_3\overline{\bm{4}}_4, \nonumber \\
&\bm{5}_1\bm{2}_2\overline{\bm{2}}_3\overline{\bm{5}}_4, \ 
\bm{5}_1\bm{2}_2\overline{\bm{5}}_3\overline{\bm{2}}_4, \ 
\bm{5}_1\bm{2}_2\overline{\bm{8}}_3\overline{\bm{8}}_4, \ 
\bm{5}_1\bm{5}_2\overline{\bm{2}}_3\overline{\bm{8}}_4, \ 
\bm{5}_1\bm{5}_2\overline{\bm{5}}_3\overline{\bm{5}}_4, \ 
\bm{5}_1\bm{5}_2\overline{\bm{8}}_3\overline{\bm{2}}_4, \ 
\bm{5}_1\bm{8}_2\overline{\bm{2}}_3\overline{\bm{2}}_4, \ 
\bm{5}_1\bm{8}_2\overline{\bm{5}}_3\overline{\bm{8}}_4, \ 
\bm{5}_1\bm{8}_2\overline{\bm{8}}_3\overline{\bm{5}}_4, \nonumber \\
&\bm{6}_1\bm{3}_2\overline{\bm{3}}_3\overline{\bm{6}}_4, \ 
\bm{6}_1\bm{3}_2\overline{\bm{6}}_3\overline{\bm{3}}_4, \ 
\bm{6}_1\bm{6}_2\overline{\bm{6}}_3\overline{\bm{6}}_4, \nonumber \\
&\bm{7}_1\bm{1}_2\overline{\bm{1}}_3\overline{\bm{7}}_4, \ 
\bm{7}_1\bm{1}_2\overline{\bm{4}}_3\overline{\bm{4}}_4, \ 
\bm{7}_1\bm{1}_2\overline{\bm{7}}_3\overline{\bm{1}}_4, \ 
\bm{7}_1\bm{4}_2\overline{\bm{1}}_3\overline{\bm{1}}_4, \ 
\bm{7}_1\bm{4}_2\overline{\bm{4}}_3\overline{\bm{7}}_4, \ 
\bm{7}_1\bm{4}_2\overline{\bm{7}}_3\overline{\bm{4}}_4, \ 
\bm{7}_1\bm{7}_2\overline{\bm{1}}_3\overline{\bm{4}}_4, \ 
\bm{7}_1\bm{7}_2\overline{\bm{4}}_3\overline{\bm{1}}_4, \ 
\bm{7}_1\bm{7}_2\overline{\bm{7}}_3\overline{\bm{7}}_4, \nonumber \\
&\bm{8}_1\bm{2}_2\overline{\bm{2}}_3\overline{\bm{8}}_4, \ 
\bm{8}_1\bm{2}_2\overline{\bm{5}}_3\overline{\bm{5}}_4, \ 
\bm{8}_1\bm{2}_2\overline{\bm{8}}_3\overline{\bm{2}}_4, \ 
\bm{8}_1\bm{5}_2\overline{\bm{2}}_3\overline{\bm{2}}_4, \ 
\bm{8}_1\bm{5}_2\overline{\bm{5}}_3\overline{\bm{8}}_4, \ 
\bm{8}_1\bm{5}_2\overline{\bm{8}}_3\overline{\bm{5}}_4, \ 
\bm{8}_1\bm{8}_2\overline{\bm{2}}_3\overline{\bm{5}}_4, \ 
\bm{8}_1\bm{8}_2\overline{\bm{5}}_3\overline{\bm{2}}_4, \ 
\bm{8}_1\bm{8}_2\overline{\bm{8}}_3\overline{\bm{8}}_4.
\end{align}
The corresponding integer vectors are given by
\begin{subequations}
\begin{align}
M^{(9,9,\overline{9},\overline{9})}_\textrm{3F;9} = \begin{pmatrix}
0 & 0 & 1 \\
0 & 3 & 1 \\
3 & 3 & 1 \\
6 & 0 & 1
\end{pmatrix}, \quad
\Lambda^{(9,9,\overline{9},\overline{9})}_{\textsf{w},\textrm{3F;9}} \
= \begin{pmatrix}
4 & 1 & 1 & -5 \\
-3 & -3 & 0 & 3 \\
3 & 0 & -3 & -3 \\
-5 & 1 & 1 & 4
\end{pmatrix}, \\
M^{(9,9,\overline{9},\overline{9})}_\textrm{3F;10} = \begin{pmatrix}
0 & 0 & 1 \\
0 & 3 & 8 \\
3 & 3 & 8 \\
6 & 0 & 8
\end{pmatrix}, \quad
\Lambda^{(9,9,\overline{9},\overline{9})}_{\textsf{w},\textrm{3F;10}} \
= \begin{pmatrix}
-1 & 4 & 1 & -5 \\
1 & 5 & -1 & -4 \\
0 & 3 & -3 & -3 \\
-3 & 3 & 0 & -3
\end{pmatrix}, \\
M^{(9,9,\overline{9},\overline{9})}_\textrm{3F;11} = \begin{pmatrix}
0 & 0 & 1 \\
0 & 3 & 1 \\
3 & 6 & 8 \\
6 & 0 & 8
\end{pmatrix}, \quad
\Lambda^{(9,9,\overline{9},\overline{9})}_{\textsf{w},\textrm{3F;11}} = \begin{pmatrix}
3 & 0 & 3 & 3 \\
3 & 3 & 0 & 3 \\
-4 & -1 & 1 & -5 \\
-5 & 1 & -1 & -4
\end{pmatrix}, \\
M^{(9,9,\overline{9},\overline{9})}_\textrm{3F;12} = \begin{pmatrix}
0 & 0 & 1 \\
0 & 3 & 8 \\
3 & 6 & 1 \\
6 & 0 & 1
\end{pmatrix}, \quad
\Lambda^{(9,9,\overline{9},\overline{9})}_{\textsf{w},\textrm{3F;12}} \
= \begin{pmatrix}
3 & 0 & -3 & -3 \\
3 & -3 & 0 & -3 \\
4 & -1 & 1 & -5 \\
5 & 1 & -1 & -4
\end{pmatrix}, \\
M^{(9,9,\overline{9},\overline{9})}_\textrm{3F;13} = \begin{pmatrix}
0 & 0 & 1 \\
0 & 3 & 1 \\
3 & 3 & 1 \\
3 & 0 & 8
\end{pmatrix}, \quad
\Lambda^{(9,9,\overline{9},\overline{9})}_{\textsf{w},\textrm{3F;13}} = \begin{pmatrix}
1 & 4 & -5 & -1 \\
-3 & -3 & 3 & 0 \\
0 & -3 & 3 & -3 \\
1 & -5 & 4 & -1
\end{pmatrix}, \\
M^{(9,9,\overline{9},\overline{9})}_\textrm{3F;14} = \begin{pmatrix}
0 & 0 & 1 \\
0 & 3 & 8 \\
3 & 3 & 8 \\
3 & 0 & 1
\end{pmatrix}, \quad
\Lambda^{(9,9,\overline{9},\overline{9})}_{\textsf{w},\textrm{3F;14}} = \begin{pmatrix}
1 & -4 & -1 & -5 \\
3 & -3 & 0 & -3 \\
0 & -3 & 3 & -3 \\
-1 & -5 & 1 & -4
\end{pmatrix}, \\
M^{(9,9,\overline{9},\overline{9})}_\textrm{3F;15} = \begin{pmatrix}
0 & 0 & 1 \\
0 & 3 & 1 \\
3 & 6 & 8 \\
3 & 0 & 1
\end{pmatrix}, \quad
\Lambda^{(9,9,\overline{9},\overline{9})}_{\textsf{w},\textrm{3F;15}} = \begin{pmatrix}
3 & 0 & 3 & -3 \\
3 & 3 & 0 & -3 \\
4 & 1 & -1 & -5 \\
5 & -1 & 1 & -4
\end{pmatrix}, \\
M^{(9,9,\overline{9},\overline{9})}_\textrm{3F;16} = \begin{pmatrix}
0 & 0 & 1 \\
0 & 3 & 8 \\
3 & 6 & 1 \\
3 & 0 & 8
\end{pmatrix}, \quad
\Lambda^{(9,9,\overline{9},\overline{9})}_{\textsf{w},\textrm{3F;16}} = \begin{pmatrix}
0 & -3 & -3 & 3 \\
3 & -3 & 0 & 3 \\
-1 & 4 & -1 & -5 \\
1 & 5 & 1 & -4
\end{pmatrix}.
\end{align}
\end{subequations}

There is one gapped interface at rank four, which is obtained by stacking four individual interfaces between $U(1)_9$ or $\overline{U(1)}_9$ and a Chern insulator $U(1)_1$ or $\overline{U(1)}_1$. 
The set of condensed anyons is given by
\begin{align}
&\red{\bm{3}_1}, \ 
\bm{6}_1, \ 
\red{\bm{3}_2}, \ 
\bm{6}_2, \ 
\red{\overline{\bm{3}}_3}, \ 
\overline{\bm{6}}_3, \ 
\red{\overline{\bm{3}}_4}, \ 
\overline{\bm{6}}_4, \nonumber \\
&\bm{3}_1\bm{3}_2, \ 
\bm{3}_1\bm{6}_2, \ 
\bm{6}_1\bm{3}_2, \ 
\bm{6}_1\bm{6}_2, \ 
\bm{3}_1\overline{\bm{3}}_3, \ 
\bm{3}_1\overline{\bm{6}}_3, \ 
\bm{6}_1\overline{\bm{3}}_3, \ 
\bm{6}_1\overline{\bm{6}}_3, \ 
\bm{3}_1\overline{\bm{3}}_4, \ 
\bm{3}_1\overline{\bm{6}}_4, \ 
\bm{6}_1\overline{\bm{3}}_4, \ 
\bm{6}_1\overline{\bm{6}}_4, \nonumber \\
&\bm{3}_2\overline{\bm{3}}_3, \ 
\bm{3}_2\overline{\bm{6}}_3, \ 
\bm{6}_2\overline{\bm{3}}_3, \ 
\bm{6}_2\overline{\bm{6}}_3, \ 
\bm{3}_2\overline{\bm{3}}_4, \ 
\bm{3}_2\overline{\bm{6}}_4, \ 
\bm{6}_2\overline{\bm{3}}_4, \ 
\bm{6}_2\overline{\bm{6}}_4, \ 
\overline{\bm{3}}_3\overline{\bm{3}}_4, \ 
\overline{\bm{3}}_3\overline{\bm{6}}_4, \ 
\overline{\bm{6}}_3\overline{\bm{3}}_4, \ 
\overline{\bm{6}}_3\overline{\bm{6}}_4, \nonumber \\
&\bm{3}_1\bm{3}_2\overline{\bm{3}}_3, \ 
\bm{3}_1\bm{3}_2\overline{\bm{6}}_3, \ 
\bm{3}_1\bm{6}_2\overline{\bm{3}}_3, \ 
\bm{3}_1\bm{6}_2\overline{\bm{6}}_3, \ 
\bm{6}_1\bm{3}_2\overline{\bm{3}}_3, \ 
\bm{6}_1\bm{3}_2\overline{\bm{6}}_3, \ 
\bm{6}_1\bm{6}_2\overline{\bm{3}}_3, \ 
\bm{6}_1\bm{6}_2\overline{\bm{6}}_3, \nonumber \\
&\bm{3}_1\bm{3}_2\overline{\bm{3}}_4, \ 
\bm{3}_1\bm{3}_2\overline{\bm{6}}_4, \ 
\bm{3}_1\bm{6}_2\overline{\bm{3}}_4, \ 
\bm{3}_1\bm{6}_2\overline{\bm{6}}_4, \ 
\bm{6}_1\bm{3}_2\overline{\bm{3}}_4, \ 
\bm{6}_1\bm{3}_2\overline{\bm{6}}_4, \ 
\bm{6}_1\bm{6}_2\overline{\bm{3}}_4, \ 
\bm{6}_1\bm{6}_2\overline{\bm{6}}_4, \nonumber \\
&\bm{3}_1\overline{\bm{3}}_3\overline{\bm{3}}_4, \ 
\bm{3}_1\overline{\bm{3}}_3\overline{\bm{6}}_4, \ 
\bm{3}_1\overline{\bm{6}}_3\overline{\bm{3}}_4, \ 
\bm{3}_1\overline{\bm{6}}_3\overline{\bm{6}}_4, \ 
\bm{6}_1\overline{\bm{3}}_3\overline{\bm{3}}_4, \ 
\bm{6}_1\overline{\bm{3}}_3\overline{\bm{6}}_4, \ 
\bm{6}_1\overline{\bm{6}}_3\overline{\bm{3}}_4, \ 
\bm{6}_1\overline{\bm{6}}_3\overline{\bm{6}}_4, \nonumber \\
&\bm{3}_2\overline{\bm{3}}_3\overline{\bm{3}}_4, \ 
\bm{3}_2\overline{\bm{3}}_3\overline{\bm{6}}_4, \ 
\bm{3}_2\overline{\bm{6}}_3\overline{\bm{3}}_4, \ 
\bm{3}_2\overline{\bm{6}}_3\overline{\bm{6}}_4, \ 
\bm{6}_2\overline{\bm{3}}_3\overline{\bm{3}}_4, \ 
\bm{6}_2\overline{\bm{3}}_3\overline{\bm{6}}_4, \ 
\bm{6}_2\overline{\bm{6}}_3\overline{\bm{3}}_4, \ 
\bm{6}_2\overline{\bm{6}}_3\overline{\bm{6}}_4, \nonumber \\
&\bm{3}_1\bm{3}_2\overline{\bm{3}}_3\overline{\bm{3}}_4, \ 
\bm{3}_1\bm{3}_2\overline{\bm{3}}_3\overline{\bm{6}}_4, \ 
\bm{3}_1\bm{3}_2\overline{\bm{6}}_3\overline{\bm{3}}_4, \ 
\bm{3}_1\bm{3}_2\overline{\bm{6}}_3\overline{\bm{6}}_4, \ 
\bm{3}_1\bm{6}_2\overline{\bm{3}}_3\overline{\bm{3}}_4, \ 
\bm{3}_1\bm{6}_2\overline{\bm{3}}_3\overline{\bm{6}}_4, \ 
\bm{3}_1\bm{6}_2\overline{\bm{6}}_3\overline{\bm{3}}_4, \ 
\bm{3}_1\bm{6}_2\overline{\bm{6}}_3\overline{\bm{6}}_4, \nonumber \\
&\bm{6}_1\bm{3}_2\overline{\bm{3}}_3\overline{\bm{3}}_4, \ 
\bm{6}_1\bm{3}_2\overline{\bm{3}}_3\overline{\bm{6}}_4, \ 
\bm{6}_1\bm{3}_2\overline{\bm{6}}_3\overline{\bm{3}}_4, \ 
\bm{6}_1\bm{3}_2\overline{\bm{6}}_3\overline{\bm{6}}_4, \ 
\bm{6}_1\bm{6}_2\overline{\bm{3}}_3\overline{\bm{3}}_4, \ 
\bm{6}_1\bm{6}_2\overline{\bm{3}}_3\overline{\bm{6}}_4, \ 
\bm{6}_1\bm{6}_2\overline{\bm{6}}_3\overline{\bm{3}}_4, \ 
\bm{6}_1\bm{6}_2\overline{\bm{6}}_3\overline{\bm{6}}_4.
\end{align}
The corresponding integer vectors are given by
\begin{align}
M^{(9,9,\overline{9},\overline{9})}_\textrm{4F;1} = \begin{pmatrix}
3 & 0 & 0 & 0 \\
0 & 3 & 0 & 0 \\
0 & 0 & 3 & 0 \\
0 & 0 & 0 & 3
\end{pmatrix}, \quad
\Lambda^{(9,9,\overline{9},\overline{9})}_{\textsf{w},\textrm{4F;1}} \
= \begin{pmatrix}
0 & 0 & -3 & 0 \\
-3 & 0 & 0 & 0 \\
0 & 0 & 0 & -3 \\
0 & -3 & 0 & 0
\end{pmatrix}.
\end{align}

\subsubsection{Other notable examples}

Gapped interfaces that do not admit the condensation of pairs anyons from two layers, as found in Sec.~\ref{sec:U7U7U7U7}, may also appear in the $U(1)_k \times U(1)_k \times \overline{U(1)}_k \times \overline{U(1)}_k$ topological order with $k$ being prime. 
We find 16 such interfaces for $k=11$, 16 interfaces for $k=13$, 24 interfaces for $k=17$, 32 interfaces for $k=19$, and 40 interfaces for $k=23$. 
Integer vectors associated with the Lagrangian subgroups and gapping potentials are given, for example, by 
\begin{subequations}
\begin{align}
M^{(11,11,\overline{11},\overline{11})}_\textrm{2F} = \begin{pmatrix}
0 & 1 \\
1 & 0 \\
3 & 5 \\
6 & 3
\end{pmatrix}, \quad
\Lambda^{(11,11,\overline{11},\overline{11})}_{\textsf{w},\textrm{2F}} = \begin{pmatrix}
1 & -1 & 2 & -3 \\
-2 & 3 & -1 & 1 \\
1 & 1 & -3 & -2 \\
-3 & -2 & 1 & 1
\end{pmatrix}, \\
M^{(13,13,\overline{13},\overline{13})}_\textrm{2F} = \begin{pmatrix}
0 & 1 \\
1 & 0 \\
2 & 7 \\
6 & 2
\end{pmatrix}, \quad
\Lambda^{(13,13,\overline{13},\overline{13})}_{\textsf{w},\textrm{2F}} = \begin{pmatrix}
-2 & 0 & -1 & -4 \\
-4 & 1 & 0 & -2 \\
0 & -2 & -4 & 1 \\
-1 & -4 & -2 & 0
\end{pmatrix}, \\
M^{(17,17,\overline{17},\overline{17})}_\textrm{2F} = \begin{pmatrix}
0 & 1 \\
1 & 0 \\
3 & 3 \\
3 & 14
\end{pmatrix}, \quad
\Lambda^{(17,17,\overline{17},\overline{17})}_{\textsf{w},\textrm{2F}} = \begin{pmatrix}
0 & -1 & -3 & -3 \\
-3 & 3 & 0 & 1 \\
1 & 0 & 3 & -3 \\
3 & 3 & 1 & 0
\end{pmatrix}, \\
M^{(19,19,\overline{19},\overline{19})}_\textrm{2F} = \begin{pmatrix}
0 & 1 \\
1 & 0 \\
2 & 4 \\
4 & 17
\end{pmatrix}, \quad
\Lambda^{(19,19,\overline{19},\overline{19})}_{\textsf{w},\textrm{2F}} = \begin{pmatrix}
1 & 0 & 4 & -2 \\
4 & 2 & 1 & 0 \\
0 & -1 & -2 & -4 \\
2 & -4 & 0 & -1
\end{pmatrix}, \\
M^{(23,23,\overline{23},\overline{23})}_\textrm{2F} = \begin{pmatrix}
0 & 1 \\
1 & 0 \\
4 & 10 \\
10 & 19
\end{pmatrix}, \quad
\Lambda^{(23,23,\overline{23},\overline{23})}_{\textsf{w},\textrm{2F}} = \begin{pmatrix}
-1 & 4 & 6 & -2 \\
-2 & 6 & 4 & -1 \\
4 & 1 & -2 & -6 \\
6 & 2 & -1 & -4
\end{pmatrix}.
\end{align}
\end{subequations}

For $k=10$, there is a bosonic gapped interface at which pairs of $\bm{2}$ and $\overline{\bm{8}}$ are condensed between the layer 1 and 3 or  2 and 4 whereas pairs of $\bm{5}$ and $\overline{\bm{5}}$ are condensed between the layer 1 and 4 or 2 and 3, apart from some triplets of anyons. 
The corresponding integer vectors are given by
\begin{align}
M^{(10,10,\overline{10},\overline{10})}_\textrm{2B} = \begin{pmatrix}
0 & 1 \\
1 & 0 \\
5 & 4 \\
4 & 5
\end{pmatrix}, \quad
\Lambda^{(10,10,\overline{10},\overline{10})}_{\textsf{w},\textrm{2B}} = \begin{pmatrix}
-2 & 3 & -3 & 2 \\
1 & -1 & -1 & 1 \\
-3 & -2 & -2 & -3 \\
1 & 1 & -1 & -1
\end{pmatrix}.
\end{align}
There is also a fermionic gapped interface for $k=10$ at which pairs of $\bm{2}$ and $\overline{\bm{8}}$ are condensed between the layer 1 and 3 or  2 and 4 whereas pairs of $\bm{5}$ and $\bm{5}$ or its antichiral counterparts are condensed between the layer 1 and 2 or 3 and 3, apart from some triplets of anyons. 
Such a gapped interface is related to the conformal embedding $U(1)_5 \times U(1)_5 \supset U(1)_{10} \times U(1)_{10}$. 
The corresponding integer vectors are given, for example, by
\begin{align}
M^{(10,10,\overline{10},\overline{10})}_\textrm{2F} = \begin{pmatrix}
0 & 1 \\
2 & 5 \\
5 & 4 \\
3 & 0
\end{pmatrix}, \quad
\Lambda^{(10,10,\overline{10},\overline{10})}_{\textsf{w},\textrm{2F}} = \begin{pmatrix}
-2 & -2 & -3 & -3 \\
3 & 3 & 2 & 2 \\
2 & -2 & 3 & -3 \\
3 & -3 & 2 & -2
\end{pmatrix}.
\end{align}

For $k=12$, there is a bosonic gapped interface at which pairs of $\bm{6}$'s or $\overline{\bm{6}}$'s from any two layers are condensed, apart from $\bm{2} \overline{\bm{2}}$ pairs between the layer 1 and 3 or 2 and 4 and $\bm{1} \bm{1} \overline{\bm{1}} \overline{\bm{1}}$ quadruplets. 
Such a gapped interface is related to the conformal embedding $U(1)_6 \times U(1)_6 \supset U(1)_{12} \times U(1)_{12}$. 
The corresponding integer vectors are given, for example, by
\begin{align}
M^{(12,12,\overline{12},\overline{12})}_\textrm{3B} = \begin{pmatrix}
0 & 0 & 1 \\
0 & 2 & 1 \\
6 & 0 & 1 \\
6 & 2 & 1
\end{pmatrix}, \quad
\Lambda^{(12,12,\overline{12},\overline{12})}_{\textsf{w},\textrm{3B}} = \begin{pmatrix}
3 & 3 & -3 & -3 \\
-1 & -1 & -1 & -1 \\
-3 & 3 & 3 & -3 \\
1 & -1 & 1 & -1
\end{pmatrix}.
\end{align}

\subsection{Gapped interface for $U(1)_k \times U(1)_l \times \overline{U(1)}_k \times \overline{U(1)}_l$}
\label{sec:GIUklkl}

We provide the lists of gapped interfaces between $U(1)_k$ topological orders corresponding to the following $K$ matrix:
\begin{align}
K = \begin{pmatrix} k &&& \\ & l && \\ && -k & \\ &&& -l \end{pmatrix}
\end{align}
where $k$ and $l$ are positive nonzero integers from 2 to 8. 
In the following discussion, we do not consider gapped interfaces that are simply obtained by stacking individual interfaces for $U(1)_k \times \overline{U(1)}_k$ and $U(1)_l \times \overline{U(1)}_l$. 
We list only nontrivial gapped interfaces that cannot be obtained by stacking if exist. 

\subsubsection{$U(1)_2 \times U(1)_6 \times \overline{U(1)}_2 \times \overline{U(1)}_6$}

We consider gapped interfaces of the bosonic topological order corresponding to the $K$ matrix:
\begin{align}
K^{(2,6,\overline{2},\overline{6})} = \begin{pmatrix} 2 &&& \\ & 6 && \\ && -2 & \\ &&& -6 \end{pmatrix}.
\end{align}
Apart from two trivial interfaces, there are four nontrivial gapped interfaces, among which two are bosonic whereas another two are fermionic. 
The two bosonic interfaces admit the condensation of pairs $\bm{1} \bm{3}$ between $U(1)_2$ and $U(1)_6$ and its antichiral counterpart between $\overline{U(1)}_2$ and $\overline{U(1)}_6$, pairs $\bm{2} \overline{\bm{4}}$ between $U(1)_6$ and $\overline{U(1)}_6$, and some triplets of anyons.
Taking the Lagrangian subgroup generated by $M^{(2,6,\overline{2},\overline{6})}_\textrm{2B;1}$ given below as an instance, the set of condensed anyons is given by 
\begin{align}
\bm{1}_1\bm{3}_2, \ 
\bm{2}_2\overline{\bm{4}}_4, \ 
\bm{4}_2\overline{\bm{2}}_4, \ 
\red{\overline{\bm{1}}_3\overline{\bm{3}}_4}, \ 
\red{\bm{1}_1\bm{1}_2\overline{\bm{2}}_4}, \ 
\bm{1}_1\bm{5}_2\overline{\bm{4}}_4, \ 
\bm{2}_2\overline{\bm{1}}_3\overline{\bm{1}}_4, \ 
\bm{4}_2\overline{\bm{1}}_3\overline{\bm{5}}_4, \ 
\bm{1}_1\bm{1}_2\overline{\bm{1}}_3\overline{\bm{5}}_4, \ 
\bm{1}_1\bm{3}_2\overline{\bm{1}}_3\overline{\bm{3}}_4, \ 
\bm{1}_1\bm{5}_2\overline{\bm{1}}_3\overline{\bm{1}}_4.
\end{align}
Integer vectors associated with the Lagrangian subgroups and gapping potentials are given by
\begin{subequations}
\begin{align}
M^{(2,6,\overline{2},\overline{6})}_\textrm{2B;1} = \begin{pmatrix}
1 & 0 \\
1 & 0 \\
0 & 1 \\
2 & 3
\end{pmatrix}, \quad
\Lambda^{(2,6,\overline{2},\overline{6})}_{\textsf{w},\textrm{2B;1}} = \begin{pmatrix}
-1 & 1 & -1 & -1 \\
0 & -2 & -1 & -1 \\
-1 & -1 & 0 & -2 \\
1 & 1 & 1 & -1
\end{pmatrix}, \\
M^{(2,6,\overline{2},\overline{6})}_\textrm{2B;2} = \begin{pmatrix}
1 & 0 \\
1 & 0 \\
0 & 1 \\
4 & 3
\end{pmatrix}, \quad
\Lambda^{(2,6,\overline{2},\overline{6})}_{\textsf{w},\textrm{2B;2}} = \begin{pmatrix}
-1 & -1 & -1 & -1 \\
0 & 2 & -1 & -1 \\
-1 & 1 & 0 & -2 \\
1 & -1 & 1 & -1
\end{pmatrix}.
\end{align}
\end{subequations}

The two fermionic gapped interfaces admit the condensation of pairs $\bm{1} \overline{\bm{3}}$ between $U(1)_2$ and $\overline{U(1)}_6$, $\bm{3} \overline{\bm{1}}$ between $U(1)_6$ and $\overline{U(1)}_2$, $\bm{2} \overline{\bm{4}}$ between $U(1)_6$ and $\overline{U(1)}_6$, and some triplets of anyons. 
Taking the Lagrangian subgroup generated by $M^{(2,6,\overline{2},\overline{6})}_\textrm{2F;1}$ given below as an instance, the set of condensed anyons is given by 
\begin{align}
\red{\bm{1}_1\overline{\bm{3}}_4}, \ 
\bm{2}_2\overline{\bm{4}}_4, \ 
\bm{3}_2\overline{\bm{1}}_3, \ 
\bm{4}_2\overline{\bm{2}}_4, \ 
\bm{1}_1\bm{2}_2\overline{\bm{1}}_4, \ 
\bm{1}_1\bm{4}_2\overline{\bm{5}}_4, \ 
\red{\bm{1}_2\overline{\bm{1}}_3\overline{\bm{2}}_4}, \ 
\bm{5}_2\overline{\bm{1}}_3\overline{\bm{4}}_4, \ 
\bm{1}_1\bm{1}_2\overline{\bm{1}}_3\overline{\bm{5}}_4, \ 
\bm{1}_1\bm{3}_2\overline{\bm{1}}_3\overline{\bm{3}}_4, \ 
\bm{1}_1\bm{5}_2\overline{\bm{1}}_3\overline{\bm{1}}_4.
\end{align}
The corresponding integer vectors are given by 
\begin{subequations}
\begin{align}
M^{(2,6,\overline{2},\overline{6})}_\textrm{2F;1} = \begin{pmatrix}
1 & 0 \\
0 & 1 \\
0 & 1 \\
3 & 2
\end{pmatrix}, \quad
\Lambda^{(2,6,\overline{2},\overline{6})}_{\textsf{w},\textrm{2F;1}} = \begin{pmatrix}
0 & -1 & -1 & -2 \\
1 & -2 & 0 & -1 \\
0 & -1 & 1 & -2 \\
1 & 2 & 0 & 1
\end{pmatrix}, \\
M^{(2,6,\overline{2},\overline{6})}_\textrm{2F;2} = \begin{pmatrix}
1 & 0 \\
0 & 1 \\
0 & 1 \\
3 & 4
\end{pmatrix}, \quad
\Lambda^{(2,6,\overline{2},\overline{6})}_{\textsf{w},\textrm{2F;2}} = \begin{pmatrix}
0 & 1 & -1 & -2 \\
1 & 2 & 0 & -1 \\
0 & -1 & -1 & 2 \\
-1 & 2 & 0 & -1
\end{pmatrix}.
\end{align}
\end{subequations}

\subsubsection{$U(1)_2 \times U(1)_8 \times \overline{U(1)}_2 \times \overline{U(1)}_8$}

We consider gapped interfaces of the bosonic topological order corresponding to the $K$ matrix:
\begin{align}
K^{(2,8,\overline{2},\overline{8})} = \begin{pmatrix} 2 &&& \\ & 8 && \\ && -2 & \\ &&& -8 \end{pmatrix}.
\end{align}
Apart from five trivial interfaces, there are two nontrivial gapped interfaces, among which one is bosonic whereas another is fermionic. 
The nontrivial bosonic interface is obtained by stacking a bosonic gapped interface with the condensation of $\bm{1} \overline{\bm{2}}$ between $U(1)_2$ and $\overline{U(1)}_8$ and its antichiral counterpart.
The corresponding integer vectors are given by 
\begin{align}
M^{(2,8,\overline{2},\overline{8})}_\textrm{2B;1} = \begin{pmatrix}
1 & 0 \\
0 & 2 \\
0 & 1 \\
2 & 0
\end{pmatrix}, \quad
\Lambda^{(2,8,\overline{2},\overline{8})}_{\textsf{w},\textrm{2B;1}} = \begin{pmatrix}
1 & 0 & 0 & -2 \\
-1 & 0 & 0 & -2 \\
0 & 2 & -1 & 0 \\
0 & -2 & -1 & 0
\end{pmatrix}
\end{align}
The nontrivial fermionic interface is obtained by stacking a fermionic gapped interface with the condensation of $\bm{1} \bm{2}$ between $U(1)_2$ and $U(1)_8$ and its antichiral counterpart.
The corresponding integer vectors are given by
\begin{align}
M^{(2,8,\overline{2},\overline{8})}_\textrm{2F;1} = \begin{pmatrix}
1 & 0 \\
2 & 0 \\
0 & 1 \\
0 & 2
\end{pmatrix}, \quad
\Lambda^{(2,8,\overline{2},\overline{8})}_{\textsf{w},\textrm{2F;1}} = \begin{pmatrix}
0 & 0 & -1 & -2 \\
-1 & -2 & 0 & 0 \\
0 & 0 & 1 & -2 \\
1 & -2 & 0 & 0
\end{pmatrix}.
\end{align}

\subsubsection{$U(1)_3 \times U(1)_6 \times \overline{U(1)}_3 \times \overline{U(1)}_6$}

We consider gapped interfaces of the fermionic topological order corresponding to the $K$ matrix:
\begin{align}
K^{(3,6,\overline{3},\overline{6})} = \begin{pmatrix} 3 &&& \\ & 6 && \\ && -3 & \\ &&& -6 \end{pmatrix}.
\end{align}
Apart from four trivial interfaces, there are four nontrivial gapped interfaces obtained by condensing a pair $\bm{1} \bm{2}$ or $\bm{1} \bm{4}$ between $U(1)_3$ and $U(1)_6$ and its antichiral counterpart, a pair $\bm{3} \overline{\bm{3}}$ between $U(1)_3$ and $\overline{U(1)}_4$, and some triplets of anyons.
Taking the Lagrangian subgroup generated by $M^{(3,6,\overline{3},\overline{6})}_\textrm{2F;1}$ given below as an instance, the set of condensed anyons is given by 
\begin{align}
&\bm{1}_1\bm{4}_2, \ 
\bm{2}_1\bm{2}_2, \ 
\bm{3}_2\overline{\bm{3}}_4, \ 
\red{\overline{\bm{1}}_3\overline{\bm{2}}_4}, \ 
\overline{\bm{2}}_3\overline{\bm{4}}_4, \ 
\red{\bm{1}_1\bm{1}_2\overline{\bm{3}}_4}, \ 
\bm{2}_1\bm{5}_2\overline{\bm{3}}_4, \ 
\bm{3}_2\overline{\bm{1}}_3\overline{\bm{5}}_4, \ 
\bm{3}_2\overline{\bm{2}}_3\overline{\bm{1}}_4, \nonumber \\
&\bm{1}_1\bm{1}_2\overline{\bm{1}}_3\overline{\bm{5}}_4, \ 
\bm{1}_1\bm{1}_2\overline{\bm{2}}_3\overline{\bm{1}}_4, \ 
\bm{1}_1\bm{4}_2\overline{\bm{1}}_3\overline{\bm{2}}_4, \ 
\bm{1}_1\bm{4}_2\overline{\bm{2}}_3\overline{\bm{4}}_4, \ 
\bm{2}_1\bm{2}_2\overline{\bm{1}}_3\overline{\bm{2}}_4, \ 
\bm{2}_1\bm{2}_2\overline{\bm{2}}_3\overline{\bm{4}}_4, \ 
\bm{2}_1\bm{5}_2\overline{\bm{1}}_3\overline{\bm{5}}_4, \ 
\bm{2}_1\bm{5}_2\overline{\bm{2}}_3\overline{\bm{1}}_4.
\end{align}
The corresponding integer vectors are given by
\begin{subequations}
\begin{align}
M^{(3,6,\overline{3},\overline{6})}_\textrm{2F;1} = \begin{pmatrix}
1 & 0 \\
1 & 0 \\
0 & 1 \\
3 & 2
\end{pmatrix}, \quad
\Lambda^{(3,6,\overline{3},\overline{6})}_{\textsf{w},\textrm{2F;1}} = \begin{pmatrix}
1 & 1 & -2 & -1 \\
1 & -2 & 0 & 0 \\
1 & 1 & 0 & -3 \\
-2 & -2 & 1 & 2
\end{pmatrix}, \\
M^{(3,6,\overline{3},\overline{6})}_\textrm{2F;2} = \begin{pmatrix}
1 & 0 \\
1 & 0 \\
0 & 1 \\
3 & 4
\end{pmatrix}, \quad
\Lambda^{(3,6,\overline{3},\overline{6})}_{\textsf{w},\textrm{2F;2}} = \begin{pmatrix}
-1 & 2 & 2 & 2 \\
0 & 3 & 1 & 1 \\
0 & 0 & 1 & -2 \\
2 & -1 & -1 & -1
\end{pmatrix}, \\
M^{(3,6,\overline{3},\overline{6})}_\textrm{2F;3} = \begin{pmatrix}
1 & 0 \\
5 & 0 \\
0 & 1 \\
3 & 2
\end{pmatrix}, \quad
\Lambda^{(3,6,\overline{3},\overline{6})}_{\textsf{w},\textrm{2F;3}} = \begin{pmatrix}
-1 & 1 & -2 & -1 \\
-1 & -2 & 0 & 0 \\
-1 & 1 & 0 & -3 \\
2 & -2 & 1 & 2
\end{pmatrix}, \\
M^{(3,6,\overline{3},\overline{6})}_\textrm{2F;4} = \begin{pmatrix}
1 & 0 \\
5 & 0 \\
0 & 1 \\
3 & 4
\end{pmatrix}, \quad
\Lambda^{(3,6,\overline{3},\overline{6})}_{\textsf{w},\textrm{2F;4}} = \begin{pmatrix}
1 & 2 & 2 & 2 \\
0 & 3 & 1 & 1 \\
0 & 0 & 1 & -2 \\
-2 & -1 & -1 & -1
\end{pmatrix}.
\end{align}
\end{subequations}

\subsubsection{$U(1)_4 \times U(1)_8 \times \overline{U(1)}_4 \times \overline{U(1)}_8$}

We consider gapped interfaces of the bosonic topological order corresponding to the $K$ matrix:
\begin{align}
K^{(4,8,\overline{4},\overline{8})} = \begin{pmatrix} 4 &&& \\ & 8 && \\ && -4 & \\ &&& -8 \end{pmatrix}.
\end{align}
Apart from 15 trivial interfaces, there are four nontrivial bosonic gapped interfaces of rank two, four fermionic ones of rank two, and four fermionic ones of rank three. 
The four bosonic gapped interfaces of rank two admit the condensation of pairs $\bm{2} \overline{\bm{2}}$ between $U(1)_4$ and $\overline{U(1)}_4$ and $\bm{2} \overline{\bm{2}}$ or $\bm{2} \overline{\bm{6}}$ between $U(1)_8$ and $\overline{U(1)}_8$, as a result of the condensation of some triplets of bosonic anyons. 
Taking the Lagrangian subgroup generated by $M^{(4,8,\overline{4},\overline{8})}_\textrm{2B;1}$ given below as an instance, the set of condensed anyons is given by 
\begin{align}
&\bm{2}_1\overline{\bm{2}}_3, \ 
\bm{2}_2\overline{\bm{6}}_4, \ 
\bm{4}_2\overline{\bm{4}}_4, \ 
\bm{6}_2\overline{\bm{2}}_4, \ 
\bm{1}_1\bm{4}_2\overline{\bm{1}}_3, \ 
\bm{3}_1\bm{4}_2\overline{\bm{3}}_3, \ 
\bm{2}_1\bm{1}_2\overline{\bm{3}}_4, \ 
\bm{2}_1\bm{3}_2\overline{\bm{1}}_4, \ 
\bm{2}_1\bm{5}_2\overline{\bm{7}}_4, \ 
\bm{2}_1\bm{7}_2\overline{\bm{5}}_4, \nonumber \\
&\red{\bm{1}_1\overline{\bm{1}}_3\overline{\bm{4}}_4}, \ 
\bm{3}_1\overline{\bm{3}}_3\overline{\bm{4}}_4, \ 
\red{\bm{1}_2\overline{\bm{2}}_3\overline{\bm{3}}_4}, \ 
\bm{3}_2\overline{\bm{2}}_3\overline{\bm{1}}_4, \ 
\bm{5}_2\overline{\bm{2}}_3\overline{\bm{7}}_4, \ 
\bm{7}_2\overline{\bm{2}}_3\overline{\bm{5}}_4, \nonumber \\
&\bm{1}_1\bm{1}_2\overline{\bm{3}}_3\overline{\bm{7}}_4, \ 
\bm{1}_1\bm{2}_2\overline{\bm{1}}_3\overline{\bm{2}}_4, \ 
\bm{1}_1\bm{3}_2\overline{\bm{3}}_3\overline{\bm{5}}_4, \ 
\bm{1}_1\bm{5}_2\overline{\bm{3}}_3\overline{\bm{3}}_4, \ 
\bm{1}_1\bm{6}_2\overline{\bm{1}}_3\overline{\bm{6}}_4, \ 
\bm{1}_1\bm{7}_2\overline{\bm{3}}_3\overline{\bm{1}}_4, \ 
\bm{2}_1\bm{2}_2\overline{\bm{2}}_3\overline{\bm{6}}_4, \ 
\bm{2}_1\bm{4}_2\overline{\bm{2}}_3\overline{\bm{4}}_4, \ 
\bm{2}_1\bm{6}_2\overline{\bm{2}}_3\overline{\bm{2}}_4, \nonumber \\
&\bm{3}_1\bm{1}_2\overline{\bm{1}}_3\overline{\bm{7}}_4, \ 
\bm{3}_1\bm{2}_2\overline{\bm{3}}_3\overline{\bm{2}}_4, \ 
\bm{3}_1\bm{3}_2\overline{\bm{1}}_3\overline{\bm{5}}_4, \ 
\bm{3}_1\bm{5}_2\overline{\bm{1}}_3\overline{\bm{3}}_4, \ 
\bm{3}_1\bm{6}_2\overline{\bm{3}}_3\overline{\bm{6}}_4, \ 
\bm{3}_1\bm{7}_2\overline{\bm{1}}_3\overline{\bm{1}}_4.
\end{align}
The corresponding integer vectors are given by
\begin{subequations}
\begin{align}
M^{(4,8,\overline{4},\overline{8})}_\textrm{2B;1} = \begin{pmatrix}
1 & 0 \\
0 & 1 \\
1 & 2 \\
4 & 3
\end{pmatrix}, \quad
\Lambda^{(4,8,\overline{4},\overline{8})}_{\textsf{w},\textrm{2B;1}} = \begin{pmatrix}
-1 & 2 & -1 & 2 \\
0 & -2 & 0 & 2 \\
-2 & 0 & -2 & 0 \\
1 & 1 & -1 & -1
\end{pmatrix}, \\
M^{(4,8,\overline{4},\overline{8})}_\textrm{2B;2} = \begin{pmatrix}
1 & 0 \\
0 & 1 \\
1 & 2 \\
4 & 5
\end{pmatrix}, \quad
\Lambda^{(4,8,\overline{4},\overline{8})}_{\textsf{w},\textrm{2B;2}} = \begin{pmatrix}
-1 & 2 & -1 & -2 \\
0 & -2 & 0 & -2 \\
-2 & 0 & -2 & 0 \\
1 & 1 & -1 & 1
\end{pmatrix}, \\
M^{(4,8,\overline{4},\overline{8})}_\textrm{2B;3} = \begin{pmatrix}
1 & 0 \\
0 & 1 \\
3 & 2 \\
4 & 3
\end{pmatrix}, \quad
\Lambda^{(4,8,\overline{4},\overline{8})}_{\textsf{w},\textrm{2B;3}} = \begin{pmatrix}
-1 & 2 & 1 & 2 \\
0 & -2 & 0 & 2 \\
-2 & 0 & 2 & 0 \\
1 & 1 & 1 & -1
\end{pmatrix}, \\
M^{(4,8,\overline{4},\overline{8})}_\textrm{2B;4} = \begin{pmatrix}
1 & 0 \\
0 & 1 \\
3 & 2 \\
4 & 5
\end{pmatrix}, \quad
\Lambda^{(4,8,\overline{4},\overline{8})}_{\textsf{w},\textrm{2B;4}} = \begin{pmatrix}
-1 & 2 & 1 & -2 \\
0 & -2 & 0 & -2 \\
-2 & 0 & 2 & 0 \\
1 & 1 & 1 & 1
\end{pmatrix}.
\end{align}
\end{subequations}

The four fermionic gapped interfaces of rank two also admit the condensation of pairs $\bm{2} \overline{\bm{2}}$ between $U(1)_4$ and $\overline{U(1)}_4$ and $\bm{2} \overline{\bm{2}}$ or $\bm{2} \overline{\bm{6}}$ between $U(1)_8$ and $\overline{U(1)}_8$, but here as a result of the condensation of some triplets of fermionic anyons. 
Taking the Lagrangian subgroup generated by $M^{(4,8,\overline{4},\overline{8})}_\textrm{2F;1}$ given below as an instance, the set of condensed anyons is given by 
\begin{align}
&\bm{2}_1\overline{\bm{2}}_3, \ 
\bm{2}_2\overline{\bm{2}}_4, \ 
\bm{4}_2\overline{\bm{4}}_4, \ 
\bm{6}_2\overline{\bm{6}}_4, \ 
\bm{1}_1\bm{4}_2\overline{\bm{1}}_3, \ 
\bm{3}_1\bm{4}_2\overline{\bm{3}}_3, \ 
\bm{2}_1\bm{1}_2\overline{\bm{1}}_4, \ 
\bm{2}_1\bm{3}_2\overline{\bm{3}}_4, \ 
\bm{2}_1\bm{5}_2\overline{\bm{5}}_4, \ 
\bm{2}_1\bm{7}_2\overline{\bm{7}}_4, \nonumber \\
&\red{\bm{1}_1\overline{\bm{1}}_3\overline{\bm{4}}_4}, \ 
\bm{3}_1\overline{\bm{3}}_3\overline{\bm{4}}_4, \ 
\red{\bm{1}_2\overline{\bm{2}}_3\overline{\bm{1}}_4}, \ 
\bm{3}_2\overline{\bm{2}}_3\overline{\bm{3}}_4, \ 
\bm{5}_2\overline{\bm{2}}_3\overline{\bm{5}}_4, \ 
\bm{7}_2\overline{\bm{2}}_3\overline{\bm{7}}_4, \nonumber \\
&\bm{1}_1\bm{1}_2\overline{\bm{3}}_3\overline{\bm{5}}_4, \ 
\bm{1}_1\bm{2}_2\overline{\bm{1}}_3\overline{\bm{6}}_4, \ 
\bm{1}_1\bm{3}_2\overline{\bm{3}}_3\overline{\bm{7}}_4, \ 
\bm{1}_1\bm{5}_2\overline{\bm{3}}_3\overline{\bm{1}}_4, \ 
\bm{1}_1\bm{6}_2\overline{\bm{1}}_3\overline{\bm{2}}_4, \ 
\bm{1}_1\bm{7}_2\overline{\bm{3}}_3\overline{\bm{3}}_4, \ 
\bm{2}_1\bm{2}_2\overline{\bm{2}}_3\overline{\bm{2}}_4, \ 
\bm{2}_1\bm{4}_2\overline{\bm{2}}_3\overline{\bm{4}}_4, \ 
\bm{2}_1\bm{6}_2\overline{\bm{2}}_3\overline{\bm{6}}_4, \nonumber \\
&\bm{3}_1\bm{1}_2\overline{\bm{1}}_3\overline{\bm{5}}_4, \ 
\bm{3}_1\bm{2}_2\overline{\bm{3}}_3\overline{\bm{6}}_4, \ 
\bm{3}_1\bm{3}_2\overline{\bm{1}}_3\overline{\bm{7}}_4, \ 
\bm{3}_1\bm{5}_2\overline{\bm{1}}_3\overline{\bm{1}}_4, \ 
\bm{3}_1\bm{6}_2\overline{\bm{3}}_3\overline{\bm{2}}_4, \ 
\bm{3}_1\bm{7}_2\overline{\bm{1}}_3\overline{\bm{3}}_4.
\end{align}
The corresponding integer vectors are given by
\begin{subequations}
\begin{align}
M^{(4,8,\overline{4},\overline{8})}_\textrm{2F;1} = \begin{pmatrix}
1 & 0 \\
0 & 1 \\
1 & 2 \\
4 & 1
\end{pmatrix}, \quad
\Lambda^{(4,8,\overline{4},\overline{8})}_{\textsf{w},\textrm{2F;1}} = \begin{pmatrix}
0 & -1 & -2 & -1 \\
2 & 1 & 0 & 1 \\
1 & -1 & -1 & 3 \\
-1 & 3 & 1 & -1
\end{pmatrix}, \\
M^{(4,8,\overline{4},\overline{8})}_\textrm{2F;2} = \begin{pmatrix}
1 & 0 \\
0 & 1 \\
1 & 2 \\
4 & 7
\end{pmatrix}, \quad
\Lambda^{(4,8,\overline{4},\overline{8})}_{\textsf{w},\textrm{2F;2}} = \begin{pmatrix}
0 & 1 & -2 & -1 \\
2 & -1 & 0 & 1 \\
-1 & -1 & 1 & -3 \\
-1 & -3 & 1 & -1
\end{pmatrix}, \\
M^{(4,8,\overline{4},\overline{8})}_\textrm{2F;3} = \begin{pmatrix}
1 & 0 \\
0 & 1 \\
3 & 2 \\
4 & 1
\end{pmatrix}, \quad
\Lambda^{(4,8,\overline{4},\overline{8})}_{\textsf{w},\textrm{2F;3}} = \begin{pmatrix}
0 & 1 & -2 & 1 \\
1 & -3 & 1 & 1 \\
-1 & 1 & -1 & -3 \\
2 & 1 & 0 & 1
\end{pmatrix}, \\
M^{(4,8,\overline{4},\overline{8})}_\textrm{2F;4} = \begin{pmatrix}
1 & 0 \\
0 & 1 \\
3 & 2 \\
4 & 7
\end{pmatrix}, \quad
\Lambda^{(4,8,\overline{4},\overline{8})}_{\textsf{w},\textrm{2F;4}} = \begin{pmatrix}
0 & 1 & 2 & -1 \\
2 & -1 & 0 & 1 \\
-1 & -1 & -1 & -3 \\
1 & 3 & 1 & 1
\end{pmatrix}.
\end{align}
\end{subequations}

The four fermionic gapped interfaces of rank three admit the condensation of any pairs of fermionic anyons $\bm{2}$ from $U(1)_4$ and $\overline{\bm{2}}$ from $\overline{U(1)}_4$ and bosonic anyons $\bm{4}$ from $U(1)_8$ and $\overline{\bm{4}}$ from $\overline{U(1)}_8$, apart from a pair such as $\bm{2} \overline{\bm{2}}$ between $U(1)_8$ and $\overline{U(1)}_8$ and a quadruplet $\bm{1} \bm{2} \overline{\bm{1}} \overline{\bm{1}}$. 
Taking the Lagrangian subgroup generated by $M^{(4,8,\overline{4},\overline{8})}_\textrm{3F;1}$ given below as an instance, the set of condensed anyons is given by 
\begin{align}
&\bm{2}_1\bm{4}_2, \ 
\bm{2}_1\overline{\bm{2}}_3, \ 
\bm{2}_1\overline{\bm{4}}_4, \ 
\bm{4}_2\overline{\bm{2}}_3, \ 
\red{\bm{2}_2\overline{\bm{2}}_4}, \ 
\bm{4}_2\overline{\bm{4}}_4, \ 
\bm{6}_2\overline{\bm{6}}_4, \ 
\red{\overline{\bm{2}}_3\overline{\bm{4}}_4}, \ 
\bm{2}_1\bm{2}_2\overline{\bm{6}}_4, \ 
\bm{2}_1\bm{6}_2\overline{\bm{2}}_4, \ 
\bm{2}_2\overline{\bm{2}}_3\overline{\bm{6}}_4, \ 
\bm{6}_2\overline{\bm{2}}_3\overline{\bm{2}}_4, \nonumber \\
&\red{\bm{1}_1\bm{1}_2\overline{\bm{1}}_3\overline{\bm{1}}_4}, \ 
\bm{1}_1\bm{1}_2\overline{\bm{3}}_3\overline{\bm{5}}_4, \ 
\bm{1}_1\bm{3}_2\overline{\bm{1}}_3\overline{\bm{3}}_4, \ 
\bm{1}_1\bm{3}_2\overline{\bm{3}}_3\overline{\bm{7}}_4, \ 
\bm{1}_1\bm{5}_2\overline{\bm{1}}_3\overline{\bm{5}}_4, \ 
\bm{1}_1\bm{5}_2\overline{\bm{3}}_3\overline{\bm{1}}_4, \ 
\bm{1}_1\bm{7}_2\overline{\bm{1}}_3\overline{\bm{7}}_4, \ 
\bm{1}_1\bm{7}_2\overline{\bm{3}}_3\overline{\bm{3}}_4, \nonumber \\
&\bm{2}_1\bm{2}_2\overline{\bm{2}}_3\overline{\bm{2}}_4, \ 
\bm{2}_1\bm{4}_2\overline{\bm{2}}_3\overline{\bm{4}}_4, \ 
\bm{2}_1\bm{6}_2\overline{\bm{2}}_3\overline{\bm{6}}_4, \nonumber \\
&\bm{3}_1\bm{1}_2\overline{\bm{1}}_3\overline{\bm{5}}_4, \ 
\bm{3}_1\bm{1}_2\overline{\bm{3}}_3\overline{\bm{1}}_4, \ 
\bm{3}_1\bm{3}_2\overline{\bm{1}}_3\overline{\bm{7}}_4, \ 
\bm{3}_1\bm{3}_2\overline{\bm{3}}_3\overline{\bm{3}}_4, \ 
\bm{3}_1\bm{5}_2\overline{\bm{1}}_3\overline{\bm{1}}_4, \ 
\bm{3}_1\bm{5}_2\overline{\bm{3}}_3\overline{\bm{5}}_4, \ 
\bm{3}_1\bm{7}_2\overline{\bm{1}}_3\overline{\bm{3}}_4, \ 
\bm{3}_1\bm{7}_2\overline{\bm{3}}_3\overline{\bm{7}}_4.
\end{align}
The corresponding integer vectors are given by
\begin{subequations}
\begin{align}
M^{(4,8,\overline{4},\overline{8})}_\textrm{3F;1} = \begin{pmatrix}
0 & 0 & 1 \\
0 & 2 & 1 \\
2 & 0 & 1 \\
4 & 2 & 1
\end{pmatrix}, \quad
\Lambda^{(4,8,\overline{4},\overline{8})}_{\textsf{w},\textrm{3F;1}} = \begin{pmatrix}
0 & -2 & -2 & 2 \\
-1 & 3 & 1 & -1 \\
1 & 1 & -1 & -3 \\
2 & 2 & 0 & -2
\end{pmatrix}, \\
M^{(4,8,\overline{4},\overline{8})}_\textrm{3F;2} = \begin{pmatrix}
0 & 0 & 1 \\
0 & 2 & 1 \\
2 & 0 & 1 \\
4 & 6 & 3
\end{pmatrix}, \quad
\Lambda^{(4,8,\overline{4},\overline{8})}_{\textsf{w},\textrm{3F;2}} = \begin{pmatrix}
1 & -1 & 1 & -3 \\
1 & 3 & 1 & 1 \\
0 & 2 & 2 & 2 \\
-2 & 2 & 0 & 2
\end{pmatrix}, \\
M^{(4,8,\overline{4},\overline{8})}_\textrm{3F;3} = \begin{pmatrix}
0 & 0 & 1 \\
0 & 2 & 1 \\
2 & 0 & 1 \\
4 & 2 & 5
\end{pmatrix}, \quad
\Lambda^{(4,8,\overline{4},\overline{8})}_{\textsf{w},\textrm{3F;3}} = \begin{pmatrix}
0 & 2 & -2 & -2 \\
-1 & 3 & -1 & -1 \\
1 & 1 & 1 & -3 \\
-2 & -2 & 0 & 2
\end{pmatrix}, \\
M^{(4,8,\overline{4},\overline{8})}_\textrm{3F;3} = \begin{pmatrix}
0 & 0 & 1 \\
0 & 2 & 1 \\
2 & 0 & 1 \\
4 & 2 & 5
\end{pmatrix}, \quad
\Lambda^{(4,8,\overline{4},\overline{8})}_{\textsf{w},\textrm{3F;3}} = \begin{pmatrix}
0 & 2 & -2 & -2 \\
-1 & 3 & -1 & -1 \\
1 & 1 & 1 & -3 \\
-2 & -2 & 0 & 2
\end{pmatrix}.
\end{align}
\end{subequations}

\subsubsection{$U(1)_6 \times U(1)_8 \times \overline{U(1)}_6 \times \overline{U(1)}_8$}

We consider gapped interfaces of the bosonic topological order corresponding to the $K$ matrix:
\begin{align}
K^{(6,8,\overline{6},\overline{8})} = \begin{pmatrix} 6 &&& \\ & 8 && \\ && -6 & \\ &&& -8 \end{pmatrix}.
\end{align}
Apart from 10 trivial interfaces, there are two nontrivial bosonic gapped interfaces and two fermionic ones. 
The two bosonic gapped interfaces admit the condensation of a pair $\bm{2} \overline{\bm{2}}$ or $\bm{2} \overline{\bm{4}}$ between $U(1)_6$ and $\overline{U(1)}_6$, a pair $\bm{3} \bm{2}$ between $U(1)_6$ and $U(1)_8$ and its antichiral counterpart, and single anyons $\bm{4}$ and $\overline{\bm{4}}$, apart from some triplets of anyons.
Taking the Lagrangian subgroup generated by $M^{(6,8,\overline{6},\overline{8})}_\textrm{2B;1}$ given below as an instance, the set of condensed anyons is given by 
\begin{align}
&\bm{4}_2, \ 
\overline{\bm{4}}_4, \ 
\bm{3}_1\bm{2}_2, \ 
\bm{3}_1\bm{6}_2, \ 
\bm{2}_1\overline{\bm{4}}_3, \ 
\bm{4}_1\overline{\bm{2}}_3, \ 
\bm{4}_2\overline{\bm{4}}_4, \ 
\red{\overline{\bm{3}}_3\overline{\bm{2}}_4}, \ 
\overline{\bm{3}}_3\overline{\bm{6}}_4, \nonumber \\
&\red{\bm{1}_1\bm{2}_2\overline{\bm{2}}_3}, \ 
\bm{1}_1\bm{6}_2\overline{\bm{2}}_3, \ 
\bm{2}_1\bm{4}_2\overline{\bm{4}}_3, \ 
\bm{4}_1\bm{4}_2\overline{\bm{2}}_3, \ 
\bm{5}_1\bm{2}_2\overline{\bm{4}}_3, \ 
\bm{5}_1\bm{6}_2\overline{\bm{4}}_3, \ 
\bm{3}_1\bm{2}_2\overline{\bm{4}}_4, \ 
\bm{3}_1\bm{6}_2\overline{\bm{4}}_4, \nonumber \\
&\bm{2}_1\overline{\bm{1}}_3\overline{\bm{2}}_4, \ 
\bm{2}_1\overline{\bm{1}}_3\overline{\bm{6}}_4, \ 
\bm{2}_1\overline{\bm{4}}_3\overline{\bm{4}}_4, \ 
\bm{4}_1\overline{\bm{2}}_3\overline{\bm{4}}_4, \ 
\bm{4}_1\overline{\bm{5}}_3\overline{\bm{2}}_4, \ 
\bm{4}_1\overline{\bm{5}}_3\overline{\bm{6}}_4, \ 
\bm{4}_2\overline{\bm{3}}_3\overline{\bm{2}}_4, \ 
\bm{4}_2\overline{\bm{3}}_3\overline{\bm{6}}_4, \nonumber \\
&\bm{1}_1\bm{2}_2\overline{\bm{2}}_3\overline{\bm{4}}_4, \ 
\bm{1}_1\bm{2}_2\overline{\bm{5}}_3\overline{\bm{2}}_4, \ 
\bm{1}_1\bm{2}_2\overline{\bm{5}}_3\overline{\bm{6}}_4, \ 
\bm{1}_1\bm{6}_2\overline{\bm{2}}_3\overline{\bm{4}}_4, \ 
\bm{1}_1\bm{6}_2\overline{\bm{5}}_3\overline{\bm{2}}_4, \ 
\bm{1}_1\bm{6}_2\overline{\bm{5}}_3\overline{\bm{6}}_4, \ 
\bm{2}_1\bm{4}_2\overline{\bm{1}}_3\overline{\bm{2}}_4, \ 
\bm{2}_1\bm{4}_2\overline{\bm{1}}_3\overline{\bm{6}}_4, \ 
\bm{2}_1\bm{4}_2\overline{\bm{4}}_3\overline{\bm{4}}_4, \nonumber \\
&\bm{3}_1\bm{2}_2\overline{\bm{3}}_3\overline{\bm{2}}_4, \ 
\bm{3}_1\bm{2}_2\overline{\bm{3}}_3\overline{\bm{6}}_4, \ 
\bm{3}_1\bm{6}_2\overline{\bm{3}}_3\overline{\bm{2}}_4, \ 
\bm{3}_1\bm{6}_2\overline{\bm{3}}_3\overline{\bm{6}}_4, \ 
\bm{4}_1\bm{4}_2\overline{\bm{2}}_3\overline{\bm{4}}_4, \ 
\bm{4}_1\bm{4}_2\overline{\bm{5}}_3\overline{\bm{2}}_4, \ 
\bm{4}_1\bm{4}_2\overline{\bm{5}}_3\overline{\bm{6}}_4, \nonumber \\
&\bm{5}_1\bm{2}_2\overline{\bm{1}}_3\overline{\bm{2}}_4, \ 
\bm{5}_1\bm{2}_2\overline{\bm{1}}_3\overline{\bm{6}}_4, \ 
\bm{5}_1\bm{2}_2\overline{\bm{4}}_3\overline{\bm{4}}_4, \ 
\bm{5}_1\bm{6}_2\overline{\bm{1}}_3\overline{\bm{2}}_4, \ 
\bm{5}_1\bm{6}_2\overline{\bm{1}}_3\overline{\bm{6}}_4, \ 
\bm{5}_1\bm{6}_2\overline{\bm{4}}_3\overline{\bm{4}}_4.
\end{align}
The corresponding integer vectors are given by
\begin{subequations}
\begin{align}
M^{(6,8,\overline{6},\overline{8})}_\textrm{2B;1} = \begin{pmatrix}
1 & 0 \\
2 & 0 \\
2 & 3 \\
0 & 2
\end{pmatrix}, \quad
\Lambda^{(6,8,\overline{6},\overline{8})}_{\textsf{w},\textrm{2B;1}} = \begin{pmatrix}
-1 & 2 & -2 & 0 \\
-1 & -2 & -2 & 0 \\
-2 & 0 & -1 & -2 \\
2 & 0 & 1 & -2
\end{pmatrix}, \\
M^{(6,8,\overline{6},\overline{8})}_\textrm{2B;2} = \begin{pmatrix}
1 & 0 \\
2 & 0 \\
4 & 3 \\
0 & 2
\end{pmatrix}, \quad
\Lambda^{(6,8,\overline{6},\overline{8})}_{\textsf{w},\textrm{2B;2}} = \begin{pmatrix}
-1 & 2 & -1 & -2 \\
-1 & -2 & -1 & -2 \\
-2 & 0 & -2 & 0 \\
2 & 0 & -1 & 2
\end{pmatrix}.
\end{align}
\end{subequations}

The two fermionic gapped interfaces admit the condensation of pairs $\bm{2} \overline{\bm{2}}$ or $\bm{2} \overline{\bm{4}}$ between $U(1)_6$ and $\overline{U(1)}_6$, $\bm{3} \overline{\bm{2}}$ between $U(1)_6$ and $\overline{U(1)}_8$, and $\bm{3} \overline{\bm{2}}$ between $U(1)_8$ and $\overline{U(1)}_6$, and single anyons $\bm{4}$ and $\overline{\bm{4}}$, apart from some triplets of anyons.
Taking the Lagrangian subgroup generated by $M^{(6,8,\overline{6},\overline{8})}_\textrm{2F;1}$ given below as an instance, the set of condensed anyons is given by 
\begin{align}
&\bm{4}_2, \ 
\overline{\bm{4}}_4, \ 
\bm{4}_1\overline{\bm{2}}_3, \ 
\bm{2}_1\overline{\bm{4}}_3, \ 
\bm{3}_1\overline{\bm{2}}_4, \ 
\bm{3}_1\overline{\bm{6}}_4, \ 
\red{\bm{2}_2\overline{\bm{3}}_3}, \ 
\bm{6}_2\overline{\bm{3}}_3, \ 
\bm{4}_2\overline{\bm{4}}_4, \nonumber \\
&\bm{2}_1\bm{2}_2\overline{\bm{1}}_3, \ 
\bm{2}_1\bm{4}_2\overline{\bm{4}}_3, \ 
\bm{2}_1\bm{6}_2\overline{\bm{1}}_3, \ 
\bm{4}_1\bm{2}_2\overline{\bm{5}}_3, \ 
\bm{4}_1\bm{4}_2\overline{\bm{2}}_3, \ 
\bm{4}_1\bm{6}_2\overline{\bm{5}}_3, \ 
\bm{3}_1\bm{4}_2\overline{\bm{2}}_4, \ 
\bm{3}_1\bm{4}_2\overline{\bm{6}}_4, \nonumber \\
&\red{\bm{1}_1\overline{\bm{2}}_3\overline{\bm{2}}_4}, \ 
\bm{1}_1\overline{\bm{2}}_3\overline{\bm{6}}_4, \ 
\bm{2}_1\overline{\bm{4}}_3\overline{\bm{4}}_4, \ 
\bm{4}_1\overline{\bm{2}}_3\overline{\bm{4}}_4, \ 
\bm{5}_1\overline{\bm{4}}_3\overline{\bm{2}}_4, \ 
\bm{5}_1\overline{\bm{4}}_3\overline{\bm{6}}_4, \ 
\bm{2}_2\overline{\bm{3}}_3\overline{\bm{4}}_4, \ 
\bm{6}_2\overline{\bm{3}}_3\overline{\bm{4}}_4, \nonumber \\
&\bm{1}_1\bm{2}_2\overline{\bm{5}}_3\overline{\bm{2}}_4, \ 
\bm{1}_1\bm{2}_2\overline{\bm{5}}_3\overline{\bm{6}}_4, \ 
\bm{1}_1\bm{4}_2\overline{\bm{2}}_3\overline{\bm{2}}_4, \ 
\bm{1}_1\bm{4}_2\overline{\bm{2}}_3\overline{\bm{6}}_4, \ 
\bm{1}_1\bm{6}_2\overline{\bm{5}}_3\overline{\bm{2}}_4, \ 
\bm{1}_1\bm{6}_2\overline{\bm{5}}_3\overline{\bm{6}}_4, \ 
\bm{2}_1\bm{2}_2\overline{\bm{1}}_3\overline{\bm{4}}_4, \ 
\bm{2}_1\bm{4}_2\overline{\bm{4}}_3\overline{\bm{4}}_4, \ 
\bm{2}_1\bm{6}_2\overline{\bm{1}}_3\overline{\bm{4}}_4, \nonumber \\
&\bm{3}_1\bm{2}_2\overline{\bm{3}}_3\overline{\bm{2}}_4, \ 
\bm{3}_1\bm{2}_2\overline{\bm{3}}_3\overline{\bm{6}}_4, \ 
\bm{3}_1\bm{6}_2\overline{\bm{3}}_3\overline{\bm{2}}_4, \ 
\bm{3}_1\bm{6}_2\overline{\bm{3}}_3\overline{\bm{6}}_4, \ 
\bm{4}_1\bm{2}_2\overline{\bm{5}}_3\overline{\bm{4}}_4, \ 
\bm{4}_1\bm{4}_2\overline{\bm{2}}_3\overline{\bm{4}}_4, \ 
\bm{4}_1\bm{6}_2\overline{\bm{5}}_3\overline{\bm{4}}_4, \nonumber \\
&\bm{5}_1\bm{2}_2\overline{\bm{1}}_3\overline{\bm{2}}_4, \ 
\bm{5}_1\bm{2}_2\overline{\bm{1}}_3\overline{\bm{6}}_4, \ 
\bm{5}_1\bm{4}_2\overline{\bm{4}}_3\overline{\bm{2}}_4, \ 
\bm{5}_1\bm{4}_2\overline{\bm{4}}_3\overline{\bm{6}}_4, \ 
\bm{5}_1\bm{6}_2\overline{\bm{1}}_3\overline{\bm{2}}_4, \ 
\bm{5}_1\bm{6}_2\overline{\bm{1}}_3\overline{\bm{6}}_4.
\end{align}
The corresponding integer vectors are given by
\begin{subequations}
\begin{align}
M^{(6,8,\overline{6},\overline{8})}_\textrm{2F;1} = \begin{pmatrix}
1 & 0 \\
0 & 2 \\
2 & 3 \\
2 & 0
\end{pmatrix}, \quad
\Lambda^{(6,8,\overline{6},\overline{8})}_{\textsf{w},\textrm{2F;1}} = \begin{pmatrix}
-1 & 0 & -2 & -2 \\
2 & 2 & 1 & 0 \\
-1 & 0 & -2 & 2 \\
-2 & 2 & -1 & 0
\end{pmatrix}, \\
M^{(6,8,\overline{6},\overline{8})}_\textrm{2F;2} = \begin{pmatrix}
1 & 0 \\
0 & 2 \\
4 & 3 \\
2 & 0
\end{pmatrix}, \quad
\Lambda^{(6,8,\overline{6},\overline{8})}_{\textsf{w},\textrm{2F;2}} = \begin{pmatrix}
-1 & 0 & 2 & 2 \\
2 & 2 & -1 & 0 \\
1 & 0 & -2 & 2 \\
2 & -2 & -1 & 0
\end{pmatrix}.
\end{align}
\end{subequations}

\subsection{Gapped interface for $U(1)_k \times U(1)_k \times \overline{U(1)}_l \times \overline{U(1)}_l$}
\label{sec:GIUkkll}

We provide the lists of gapped interfaces between $U(1)_k$ topological orders corresponding to the following $K$ matrix:
\begin{align}
K = \begin{pmatrix} k &&& \\ & k && \\ && -l & \\ &&& -l \end{pmatrix}
\end{align}
where $k$ and $l$ are positive nonzero integers from 2 to 8. 
In the following discussion, we do not consider gapped interfaces that are simply obtained by stacking individual interfaces for $U(1)_k \times U(1)_k$ and $\overline{U(1)}_l \times \overline{U(1)}_l$ or those for $U(1)_k \times \overline{U(1)}_l$. 
We list only nontrivial gapped interfaces that cannot be obtained by stacking if exist. 

\subsubsection{$U(1)_2 \times U(1)_2 \times \overline{U(1)}_4 \times \overline{U(1)}_4$}

We consider gapped interfaces of the bosonic topological order corresponding to the $K$ matrix:
\begin{align}
K^{(2,2,\overline{4},\overline{4})} = \begin{pmatrix} 2 &&& \\ & 2 && \\ && -4 & \\ &&& -4 \end{pmatrix}.
\end{align}
Apart from one trivial gapped interface, there are two bosonic gapped interfaces at which triplets consisting of an $s=1/4$ anyon $\bm{1}$ from $U(1)_2$ and two $s=-1/8 \ (\textrm{mod} \ 1)$ anyons $\overline{\bm{1}}$ or $\overline{\bm{3}}$ from $\overline{U(1)}_4$ are condensed. 
These interfaces are a consequence of the conformal embedding $U(1)_2 \times U(1)_2 \supset U(1)_4 \times U(1)_4$.
Taking the Lagrangian subgroup generated by $M^{(2,2,\overline{4},\overline{4})}_\textrm{2B;1}$ given below as an instance, the set of condensed anyons is given by 
\begin{align}
\overline{\bm{2}}_3\overline{\bm{2}}_4, \ 
\bm{1}_1\bm{1}_2\overline{\bm{2}}_3, \ 
\bm{1}_1\bm{1}_2\overline{\bm{2}}_4, \ 
\red{\bm{1}_1\overline{\bm{1}}_3\overline{\bm{3}}_4}, \ 
\bm{1}_1\overline{\bm{3}}_3\overline{\bm{1}}_4, \ 
\red{\bm{1}_2\overline{\bm{1}}_3\overline{\bm{1}}_4}, \ 
\bm{1}_2\overline{\bm{3}}_3\overline{\bm{3}}_4.
\end{align}
Integer vectors associated with the Lagrangian subgroups and gapping potentials are given by
\begin{subequations}
\begin{align}
M^{(2,2,\overline{4},\overline{4})}_\textrm{2B;1} = \begin{pmatrix}
1 & 0 \\
0 & 1 \\
1 & 1 \\
3 & 1
\end{pmatrix}, \quad
\Lambda^{(2,2,\overline{4},\overline{4})}_{\textsf{w},\textrm{2B;1}} = \begin{pmatrix}
0 & 1 & -1 & -1 \\
0 & -1 & -1 & -1 \\
1 & 0 & 1 & -1 \\
-1 & 0 & 1 & -1
\end{pmatrix}, \\
M^{(2,2,\overline{4},\overline{4})}_\textrm{2B;2} = \begin{pmatrix}
1 & 0 \\
0 & 1 \\
1 & 1 \\
1 & 3
\end{pmatrix}, \quad
\Lambda^{(2,2,\overline{4},\overline{4})}_{\textsf{w},\textrm{2B;2}} = \begin{pmatrix}
1 & 0 & -1 & -1 \\
-1 & 0 & -1 & -1 \\
0 & 1 & 1 & -1 \\
0 & -1 & 1 & -1
\end{pmatrix}.
\end{align}
\end{subequations}

\subsubsection{$U(1)_3 \times U(1)_3 \times \overline{U(1)}_6 \times \overline{U(1)}_6$}

We consider gapped interfaces of the fermionic topological order corresponding to the $K$ matrix:
\begin{align}
K^{(3,3,\overline{6},\overline{6})} = \begin{pmatrix} 3 &&& \\ & 3 && \\ && -6 & \\ &&& -6 \end{pmatrix}.
\end{align}
There are eight fermionic gapped interfaces at which triplets consisting of an $s=1/6 \ (\textrm{mod} \ 1/2)$ anyon $\bm{1}$ or $\bm{2}$ from $U(1)_3$ and two $s=-1/12 \ (\textrm{mod} \ 1/2)$ anyons $\overline{\bm{1}}$ or $\overline{\bm{5}}$ from $\overline{U(1)}_6$ are condensed. 
These interfaces are a consequence of the conformal embedding $U(1)_3 \times U(1)_3 \supset U(1)_6 \times U(1)_6$.
Taking the Lagrangian subgroup generated by $M^{(3,3,\overline{6},\overline{6})}_\textrm{2F;1}$ given below as an instance, the set of condensed anyons is given by 
\begin{align}
&\overline{\bm{3}}_3\overline{\bm{3}}_4, \ 
\bm{1}_1\bm{1}_2\overline{\bm{2}}_3, \ 
\bm{2}_1\bm{2}_2\overline{\bm{4}}_3, \ 
\bm{1}_1\bm{2}_2\overline{\bm{2}}_4, \ 
\bm{2}_1\bm{1}_2\overline{\bm{4}}_4, \ 
\red{\bm{1}_1\overline{\bm{1}}_3\overline{\bm{1}}_4}, \ 
\bm{1}_1\overline{\bm{4}}_3\overline{\bm{4}}_4, \ 
\bm{2}_1\overline{\bm{2}}_3\overline{\bm{2}}_4, \ 
\bm{2}_1\overline{\bm{5}}_3\overline{\bm{5}}_4, \nonumber \\
&\red{\bm{1}_2\overline{\bm{1}}_3\overline{\bm{5}}_4}, \ 
\bm{1}_2\overline{\bm{4}}_3\overline{\bm{2}}_4, \ 
\bm{2}_2\overline{\bm{2}}_3\overline{\bm{4}}_4, \ 
\bm{2}_2\overline{\bm{5}}_3\overline{\bm{1}}_4, \ 
\bm{1}_1\bm{1}_2\overline{\bm{5}}_3\overline{\bm{3}}_4, \ 
\bm{1}_1\bm{2}_2\overline{\bm{3}}_3\overline{\bm{5}}_4, \ 
\bm{2}_1\bm{1}_2\overline{\bm{3}}_3\overline{\bm{1}}_4, \ 
\bm{2}_1\bm{2}_2\overline{\bm{1}}_3\overline{\bm{3}}_4.
\end{align}
The corresponding integer vectors are given by
\begin{subequations}
\begin{align}
M^{(3,3,\overline{6},\overline{6})}_\textrm{2F;1} = \begin{pmatrix}
1 & 0 \\
0 & 1 \\
1 & 1 \\
1 & 5
\end{pmatrix}, \quad
\Lambda^{(3,3,\overline{6},\overline{6})}_{\textsf{w},\textrm{2F;1}} = \begin{pmatrix}
1 & 0 & -2 & -2 \\
2 & 0 & -1 & -1 \\
0 & -1 & 2 & -2 \\
0 & -2 & 1 & -1
\end{pmatrix}, \\
M^{(3,3,\overline{6},\overline{6})}_\textrm{2F;2} = \begin{pmatrix}
1 & 0 \\
0 & 1 \\
1 & 2 \\
1 & 4
\end{pmatrix}, \quad
\Lambda^{(3,3,\overline{6},\overline{6})}_{\textsf{w},\textrm{2F;2}} = \begin{pmatrix}
1 & 0 & -2 & -2 \\
2 & 0 & -1 & -1 \\
0 & -1 & -2 & 2 \\
0 & 2 & 1 & -1
\end{pmatrix}, \\
M^{(3,3,\overline{6},\overline{6})}_\textrm{2F;3} = \begin{pmatrix}
1 & 0 \\
0 & 1 \\
1 & 1 \\
5 & 1
\end{pmatrix}, \quad
\Lambda^{(3,3,\overline{6},\overline{6})}_{\textsf{w},\textrm{2F;3}} = \begin{pmatrix}
0 & 1 & -2 & -2 \\
0 & 2 & -1 & -1 \\
-1 & 0 & 2 & -2 \\
-2 & 0 & 1 & -1
\end{pmatrix}, \\
M^{(3,3,\overline{6},\overline{6})}_\textrm{2F;4} = \begin{pmatrix}
1 & 0 \\
0 & 1 \\
1 & 2 \\
5 & 2
\end{pmatrix}, \quad
\Lambda^{(3,3,\overline{6},\overline{6})}_{\textsf{w},\textrm{2F;4}} = \begin{pmatrix}
0 & -1 & -2 & -2 \\
0 & -2 & -1 & -1 \\
-1 & 0 & 2 & -2 \\
-2 & 0 & 1 & -1
\end{pmatrix}, \\
M^{(3,3,\overline{6},\overline{6})}_\textrm{2F;5} = \begin{pmatrix}
1 & 0 \\
0 & 1 \\
2 & 1 \\
2 & 5
\end{pmatrix}, \quad
\Lambda^{(3,3,\overline{6},\overline{6})}_{\textsf{w},\textrm{2F;5}} = \begin{pmatrix}
-1 & 0 & -2 & -2 \\
-2 & 0 & -1 & -1 \\
0 & -1 & 2 & -2 \\
0 & -2 & 1 & -1
\end{pmatrix}, \\
M^{(3,3,\overline{6},\overline{6})}_\textrm{2F;6} = \begin{pmatrix}
1 & 0 \\
0 & 1 \\
2 & 1 \\
4 & 1
\end{pmatrix}, \quad
\Lambda^{(3,3,\overline{6},\overline{6})}_{\textsf{w},\textrm{2F;6}} = \begin{pmatrix}
0 & 1 & -2 & -2 \\
0 & 2 & -1 & -1 \\
-1 & 0 & -2 & 2 \\
2 & 0 & 1 & -1
\end{pmatrix}, \\
M^{(3,3,\overline{6},\overline{6})}_\textrm{2F;7} = \begin{pmatrix}
1 & 0 \\
0 & 1 \\
5 & 2 \\
1 & 2
\end{pmatrix}, \quad
\Lambda^{(3,3,\overline{6},\overline{6})}_{\textsf{w},\textrm{2F;7}} = \begin{pmatrix}
-1 & 0 & -2 & 2 \\
0 & -2 & -1 & -1 \\
0 & 1 & 2 & 2 \\
2 & 0 & 1 & -1
\end{pmatrix}, \\
M^{(3,3,\overline{6},\overline{6})}_\textrm{2F;8} = \begin{pmatrix}
1 & 0 \\
0 & 1 \\
5 & 2 \\
5 & 4
\end{pmatrix}, \quad
\Lambda^{(3,3,\overline{6},\overline{6})}_{\textsf{w},\textrm{2F;8}} = \begin{pmatrix}
-1 & 0 & -2 & -2 \\
-2 & 0 & -1 & -1 \\
0 & 1 & 2 & -2 \\
0 & 2 & 1 & -1
\end{pmatrix}.
\end{align}
\end{subequations}

\subsubsection{$U(1)_4 \times U(1)_4 \times \overline{U(1)}_8 \times \overline{U(1)}_8$}

We consider gapped interfaces of the bosonic topological order corresponding to the $K$ matrix:
\begin{align}
K^{(4,4,\overline{8},\overline{8})} = \begin{pmatrix} 4 &&& \\ & 4 && \\ && -8 & \\ &&& -8 \end{pmatrix}.
\end{align}
Apart from one trivial gapped interfaces, there are eight bosonic gapped interfaces at rank two, eight fermionic ones at rank two, two bosonic ones at rank three, and eight fermionic ones at rank three.
The eight bosonic gapped interfaces at rank two are obtained by condensing triplets consisting of an $s=1/8 \ (\textrm{mod} \ 1)$ anyon $\bm{1}$ or $\bm{3}$ from $U(1)_4$ and two $s=-1/16 \ (\textrm{mod} \ 1)$ anyons $\overline{\bm{1}}$ or $\overline{\bm{7}}$ from $\overline{U(1)}_8$. 
These interfaces are a consequence of the conformal embedding $U(1)_4 \times U(1)_4 \supset U(1)_8 \times U(1)_8$.
Taking the Lagrangian subgroup generated by $M^{(4,4,\overline{8},\overline{8})}_\textrm{2B;1}$ given below as an instance, the set of condensed anyons is given by 
\begin{align}
&\overline{\bm{4}}_3\overline{\bm{4}}_4, \ 
\bm{1}_1\bm{1}_2\overline{\bm{2}}_3, \ 
\bm{2}_1\bm{2}_2\overline{\bm{4}}_3, \ 
\bm{3}_1\bm{3}_2\overline{\bm{6}}_3, \ 
\bm{1}_1\bm{3}_2\overline{\bm{2}}_4, \ 
\bm{2}_1\bm{2}_2\overline{\bm{4}}_4, \ 
\bm{3}_1\bm{1}_2\overline{\bm{6}}_4, \nonumber \\
&\red{\bm{1}_1\overline{\bm{1}}_3\overline{\bm{1}}_4}, \ 
\bm{1}_1\overline{\bm{5}}_3\overline{\bm{5}}_4, \ 
\bm{2}_1\overline{\bm{2}}_3\overline{\bm{2}}_4, \ 
\bm{2}_1\overline{\bm{6}}_3\overline{\bm{6}}_4, \ 
\bm{3}_1\overline{\bm{3}}_3\overline{\bm{3}}_4, \ 
\bm{3}_1\overline{\bm{7}}_3\overline{\bm{7}}_4, \ 
\red{\bm{1}_2\overline{\bm{1}}_3\overline{\bm{7}}_4}, \ 
\bm{1}_2\overline{\bm{5}}_3\overline{\bm{3}}_4, \ 
\bm{2}_2\overline{\bm{2}}_3\overline{\bm{6}}_4, \ 
\bm{2}_2\overline{\bm{6}}_3\overline{\bm{2}}_4, \ 
\bm{3}_2\overline{\bm{3}}_3\overline{\bm{5}}_4, \ 
\bm{3}_2\overline{\bm{7}}_3\overline{\bm{1}}_4, \nonumber \\
&\bm{1}_1\bm{1}_2\overline{\bm{6}}_3\overline{\bm{4}}_4, \ 
\bm{1}_1\bm{2}_2\overline{\bm{3}}_3\overline{\bm{7}}_4, \ 
\bm{1}_1\bm{2}_2\overline{\bm{7}}_3\overline{\bm{3}}_4, \ 
\bm{1}_1\bm{3}_2\overline{\bm{4}}_3\overline{\bm{6}}_4, \ 
\bm{2}_1\bm{1}_2\overline{\bm{3}}_3\overline{\bm{1}}_4, \ 
\bm{2}_1\bm{1}_2\overline{\bm{7}}_3\overline{\bm{5}}_4, \ 
\bm{2}_1\bm{3}_2\overline{\bm{1}}_3\overline{\bm{3}}_4, \ 
\bm{2}_1\bm{3}_2\overline{\bm{5}}_3\overline{\bm{7}}_4, \nonumber \\
&\bm{3}_1\bm{1}_2\overline{\bm{4}}_3\overline{\bm{2}}_4, \ 
\bm{3}_1\bm{2}_2\overline{\bm{1}}_3\overline{\bm{5}}_4, \ 
\bm{3}_1\bm{2}_2\overline{\bm{5}}_3\overline{\bm{1}}_4, \ 
\bm{3}_1\bm{3}_2\overline{\bm{2}}_3\overline{\bm{4}}_4.
\end{align}
The corresponding integer vectors are given by
\begin{subequations}
\begin{align}
M^{(4,4,\overline{8},\overline{8})}_\textrm{2B;1} = \begin{pmatrix}
1 & 0 \\
0 & 1 \\
1 & 1 \\
1 & 7
\end{pmatrix}, \quad
\Lambda^{(4,4,\overline{8},\overline{8})}_{\textsf{w},\textrm{2B;1}} = \begin{pmatrix}
2 & 0 & -2 & -2 \\
-1 & 0 & -1 & -1 \\
0 & -2 & 2 & -2 \\
0 & 1 & 1 & -1
\end{pmatrix}, \\
M^{(4,4,\overline{8},\overline{8})}_\textrm{2B;2} = \begin{pmatrix}
1 & 0 \\
0 & 1 \\
1 & 3 \\
1 & 5
\end{pmatrix}, \quad
\Lambda^{(4,4,\overline{8},\overline{8})}_{\textsf{w},\textrm{2B;2}} = \begin{pmatrix}
2 & 0 & -2 & -2 \\
-1 & 0 & -1 & -1 \\
0 & -2 & -2 & 2 \\
0 & 1 & -1 & 1
\end{pmatrix}, \\
M^{(4,4,\overline{8},\overline{8})}_\textrm{2B;3} = \begin{pmatrix}
1 & 0 \\
0 & 1 \\
1 & 1 \\
7 & 1
\end{pmatrix}, \quad
\Lambda^{(4,4,\overline{8},\overline{8})}_{\textsf{w},\textrm{2B;3}} = \begin{pmatrix}
0 & 2 & -2 & -2 \\
0 & -1 & -1 & -1 \\
-2 & 0 & 2 & -2 \\
1 & 0 & 1 & -1
\end{pmatrix}, \\
M^{(4,4,\overline{8},\overline{8})}_\textrm{2B;4} = \begin{pmatrix}
1 & 0 \\
0 & 1 \\
1 & 3 \\
7 & 3
\end{pmatrix}, \quad
\Lambda^{(4,4,\overline{8},\overline{8})}_{\textsf{w},\textrm{2B;4}} = \begin{pmatrix}
0 & -2 & -2 & -2 \\
0 & 1 & -1 & -1 \\
-2 & 0 & 2 & -2 \\
1 & 0 & 1 & -1
\end{pmatrix}, \\
M^{(4,4,\overline{8},\overline{8})}_\textrm{2B;5} = \begin{pmatrix}
1 & 0 \\
0 & 1 \\
3 & 1 \\
3 & 7
\end{pmatrix}, \quad
\Lambda^{(4,4,\overline{8},\overline{8})}_{\textsf{w},\textrm{2B;5}} = \begin{pmatrix}
-2 & 0 & -2 & -2 \\
1 & 0 & -1 & -1 \\
0 & -2 & 2 & -2 \\
0 & 1 & 1 & -1
\end{pmatrix}, \\
M^{(4,4,\overline{8},\overline{8})}_\textrm{2B;6} = \begin{pmatrix}
1 & 0 \\
0 & 1 \\
3 & 3 \\
3 & 5
\end{pmatrix}, \quad
\Lambda^{(4,4,\overline{8},\overline{8})}_{\textsf{w},\textrm{2B;6}} = \begin{pmatrix}
0 & -2 & -2 & 2 \\
0 & 1 & -1 & 1 \\
-2 & 0 & -2 & -2 \\
1 & 0 & -1 & -1
\end{pmatrix}, \\
M^{(4,4,\overline{8},\overline{8})}_\textrm{2B;7} = \begin{pmatrix}
1 & 0 \\
0 & 1 \\
3 & 1 \\
5 & 1
\end{pmatrix}, \quad
\Lambda^{(4,4,\overline{8},\overline{8})}_{\textsf{w},\textrm{2B;7}} = \begin{pmatrix}
0 & 2 & -2 & -2 \\
0 & -1 & -1 & -1 \\
-2 & 0 & -2 & 2 \\
1 & 0 & -1 & 1
\end{pmatrix}, \\
M^{(4,4,\overline{8},\overline{8})}_\textrm{2B;8} = \begin{pmatrix}
1 & 0 \\
0 & 1 \\
3 & 3 \\
5 & 3
\end{pmatrix}, \quad
\Lambda^{(4,4,\overline{8},\overline{8})}_{\textsf{w},\textrm{2B;8}} = \begin{pmatrix}
-2 & 0 & -2 & 2 \\
1 & 0 & -1 & 1 \\
0 & 2 & 2 & 2 \\
0 & -1 & 1 & 1
\end{pmatrix}.
\end{align}
\end{subequations}

The eight fermionic gapped interfaces at rank two are obtained condensing fermionic triplets consisting of an $s=1/8 \ (\textrm{mod} \ 1)$ anyon $\bm{1}$ or $\bm{3}$ from $U(1)_4$, an $s=-1/16 \ (\textrm{mod} \ 1)$ anyon $\overline{\bm{1}}$ or $\overline{\bm{7}}$ from $\overline{U(1)}_8$, and an $s=-9/16$ anyon $\overline{\bm{3}}$ or $\overline{\bm{5}}$ from another $\overline{U(1)}_8$. 
Taking the Lagrangian subgroup generated by $M^{(4,4,\overline{8},\overline{8})}_\textrm{2F;1}$ given below as an instance, the set of condensed anyons is given by 
\begin{align}
&\overline{\bm{4}}_3\overline{\bm{4}}_4, \ 
\bm{1}_1\bm{1}_2\overline{\bm{2}}_3, \ 
\bm{2}_1\bm{2}_2\overline{\bm{4}}_3, \ 
\bm{3}_1\bm{3}_2\overline{\bm{6}}_3, \ 
\bm{1}_1\bm{3}_2\overline{\bm{6}}_4, \ 
\bm{2}_1\bm{2}_2\overline{\bm{4}}_4, \ 
\bm{3}_1\bm{1}_2\overline{\bm{2}}_4, \nonumber \\
&\red{\bm{1}_1\overline{\bm{1}}_3\overline{\bm{3}}_4}, \ 
\bm{1}_1\overline{\bm{5}}_3\overline{\bm{7}}_4, \ 
\bm{2}_1\overline{\bm{2}}_3\overline{\bm{6}}_4, \ 
\bm{2}_1\overline{\bm{6}}_3\overline{\bm{2}}_4, \ 
\bm{3}_1\overline{\bm{3}}_3\overline{\bm{1}}_4, \ 
\bm{3}_1\overline{\bm{7}}_3\overline{\bm{5}}_4, \ 
\red{\bm{1}_2\overline{\bm{1}}_3\overline{\bm{5}}_4}, \ 
\bm{1}_2\overline{\bm{5}}_3\overline{\bm{1}}_4, \ 
\bm{2}_2\overline{\bm{2}}_3\overline{\bm{2}}_4, \ 
\bm{2}_2\overline{\bm{6}}_3\overline{\bm{6}}_4, \ 
\bm{3}_2\overline{\bm{3}}_3\overline{\bm{7}}_4, \ 
\bm{3}_2\overline{\bm{7}}_3\overline{\bm{3}}_4, \nonumber \\
&\bm{1}_1\bm{1}_2\overline{\bm{6}}_3\overline{\bm{4}}_4, \ 
\bm{1}_1\bm{2}_2\overline{\bm{3}}_3\overline{\bm{5}}_4, \ 
\bm{1}_1\bm{2}_2\overline{\bm{7}}_3\overline{\bm{1}}_4, \ 
\bm{1}_1\bm{3}_2\overline{\bm{4}}_3\overline{\bm{2}}_4, \ 
\bm{2}_1\bm{1}_2\overline{\bm{3}}_3\overline{\bm{3}}_4, \ 
\bm{2}_1\bm{1}_2\overline{\bm{7}}_3\overline{\bm{7}}_4, \ 
\bm{2}_1\bm{3}_2\overline{\bm{1}}_3\overline{\bm{1}}_4, \ 
\bm{2}_1\bm{3}_2\overline{\bm{5}}_3\overline{\bm{5}}_4, \nonumber \\
&\bm{3}_1\bm{1}_2\overline{\bm{4}}_3\overline{\bm{6}}_4, \ 
\bm{3}_1\bm{2}_2\overline{\bm{1}}_3\overline{\bm{7}}_4, \ 
\bm{3}_1\bm{2}_2\overline{\bm{5}}_3\overline{\bm{3}}_4, \ 
\bm{3}_1\bm{3}_2\overline{\bm{2}}_3\overline{\bm{4}}_4.
\end{align}
The corresponding integer vectors are given by
\begin{subequations}
\begin{align}
M^{(4,4,\overline{8},\overline{8})}_\textrm{2F;1} = \begin{pmatrix}
1 & 0 \\
0 & 1 \\
1 & 1 \\
3 & 5
\end{pmatrix}, \quad
\Lambda^{(4,4,\overline{8},\overline{8})}_{\textsf{w},\textrm{2F;1}} = \begin{pmatrix}
0 & -1 & 3 & -1 \\
1 & 2 & -1 & 1 \\
-1 & 0 & -1 & -3 \\
-2 & 1 & -1 & -1
\end{pmatrix}, \\
M^{(4,4,\overline{8},\overline{8})}_\textrm{2F;2} = \begin{pmatrix}
1 & 0 \\
0 & 1 \\
1 & 3 \\
3 & 7
\end{pmatrix}, \quad
\Lambda^{(4,4,\overline{8},\overline{8})}_{\textsf{w},\textrm{2F;2}} = \begin{pmatrix}
-1 & 0 & -1 & -3 \\
1 & -2 & -1 & 1 \\
0 & -1 & -3 & 1 \\
-2 & -1 & -1 & -1
\end{pmatrix}, \\
M^{(4,4,\overline{8},\overline{8})}_\textrm{2F;3} = \begin{pmatrix}
1 & 0 \\
0 & 1 \\
1 & 1 \\
5 & 3
\end{pmatrix}, \quad
\Lambda^{(4,4,\overline{8},\overline{8})}_{\textsf{w},\textrm{2F;3}} = \begin{pmatrix}
-1 & 0 & 3 & -1 \\
2 & 1 & -1 & 1 \\
0 & 1 & 1 & 3 \\
1 & -2 & -1 & -1
\end{pmatrix}, \\
M^{(4,4,\overline{8},\overline{8})}_\textrm{2F;4} = \begin{pmatrix}
1 & 0 \\
0 & 1 \\
1 & 3 \\
5 & 1
\end{pmatrix}, \quad
\Lambda^{(4,4,\overline{8},\overline{8})}_{\textsf{w},\textrm{2F;4}} = \begin{pmatrix}
0 & 1 & 3 & 1 \\
1 & -2 & -1 & -1 \\
-1 & 0 & -1 & 3 \\
2 & 1 & 1 & -1
\end{pmatrix}, \\
M^{(4,4,\overline{8},\overline{8})}_\textrm{2F;5} = \begin{pmatrix}
1 & 0 \\
0 & 1 \\
3 & 1 \\
1 & 5
\end{pmatrix}, \quad
\Lambda^{(4,4,\overline{8},\overline{8})}_{\textsf{w},\textrm{2F;5}} = \begin{pmatrix}
1 & 0 & 3 & 1 \\
-2 & 1 & -1 & -1 \\
0 & -1 & -1 & 3 \\
1 & 2 & 1 & -1
\end{pmatrix}, \\
M^{(4,4,\overline{8},\overline{8})}_\textrm{2F;6} = \begin{pmatrix}
1 & 0 \\
0 & 1 \\
3 & 3 \\
1 & 7
\end{pmatrix}, \quad
\Lambda^{(4,4,\overline{8},\overline{8})}_{\textsf{w},\textrm{2F;6}} = \begin{pmatrix}
1 & 0 & -1 & -3 \\
2 & -1 & -1 & -1 \\
0 & 1 & 3 & -1 \\
-1 & -2 & -1 & 1
\end{pmatrix}, \\
M^{(4,4,\overline{8},\overline{8})}_\textrm{2F;7} = \begin{pmatrix}
1 & 0 \\
0 & 1 \\
3 & 1 \\
7 & 3
\end{pmatrix}, \quad
\Lambda^{(4,4,\overline{8},\overline{8})}_{\textsf{w},\textrm{2F;7}} = \begin{pmatrix}
0 & -1 & -1 & -3 \\
-2 & 1 & -1 & 1 \\
-1 & 0 & -3 & 1 \\
-1 & -2 & -1 & -1
\end{pmatrix}, \\
M^{(4,4,\overline{8},\overline{8})}_\textrm{2F;8} = \begin{pmatrix}
1 & 0 \\
0 & 1 \\
3 & 3 \\
7 & 1
\end{pmatrix}, \quad
\Lambda^{(4,4,\overline{8},\overline{8})}_{\textsf{w},\textrm{2F;8}} = \begin{pmatrix}
0 & 1 & -1 & -3 \\
-1 & 2 & -1 & -1 \\
1 & 0 & 3 & -1 \\
-2 & -1 & -1 & 1
\end{pmatrix}.
\end{align}
\end{subequations}

The two bosonic gapped interfaces at rank three are obtained by condensing triplets consisting of two $s=1/8 \ (\textrm{mod} \ 1)$ anyons $\bm{1}$ or $\bm{3}$ from $U(1)_4$ and an $s=-1/4 \ (\textrm{mod} \ 1)$ anyon $\overline{\bm{2}}$ or $\overline{\bm{6}}$ from $\overline{U(1)}_8$ along with a pair $\bm{2} \bm{2}$ and single $\overline{\bm{4}}$'s. 
Taking the Lagrangian subgroup generated by $M^{(4,4,\overline{8},\overline{8})}_\textrm{3B;1}$ given below as an instance, the set of condensed anyons is given by 
\begin{align}
&\overline{\bm{4}}_3, \ 
\red{\overline{\bm{4}}_4}, \ 
\bm{2}_1\bm{2}_2, \ 
\overline{\bm{4}}_3\overline{\bm{4}}_4, \ 
\bm{1}_1\bm{3}_2\overline{\bm{2}}_3, \ 
\bm{1}_1\bm{3}_2\overline{\bm{6}}_3, \ 
\bm{2}_1\bm{2}_2\overline{\bm{4}}_3, \ 
\bm{3}_1\bm{1}_2\overline{\bm{2}}_3, \ 
\bm{3}_1\bm{1}_2\overline{\bm{6}}_3, \ 
\red{\bm{1}_1\bm{1}_2\overline{\bm{2}}_4}, \ 
\bm{1}_1\bm{1}_2\overline{\bm{6}}_4, \ 
\bm{2}_1\bm{2}_2\overline{\bm{4}}_4, \ 
\bm{3}_1\bm{3}_2\overline{\bm{2}}_4, \ 
\bm{3}_1\bm{3}_2\overline{\bm{6}}_4, \nonumber \\
&\red{\bm{2}_1\overline{\bm{2}}_3\overline{\bm{2}}_4}, \ 
\bm{2}_1\overline{\bm{2}}_3\overline{\bm{6}}_4, \ 
\bm{2}_1\overline{\bm{6}}_3\overline{\bm{2}}_4, \ 
\bm{2}_1\overline{\bm{6}}_3\overline{\bm{6}}_4, \ 
\bm{2}_2\overline{\bm{2}}_3\overline{\bm{2}}_4, \ 
\bm{2}_2\overline{\bm{2}}_3\overline{\bm{6}}_4, \ 
\bm{2}_2\overline{\bm{6}}_3\overline{\bm{2}}_4, \ 
\bm{2}_2\overline{\bm{6}}_3\overline{\bm{6}}_4, \nonumber \\
&\bm{1}_1\bm{1}_2\overline{\bm{4}}_3\overline{\bm{2}}_4, \ 
\bm{1}_1\bm{1}_2\overline{\bm{4}}_3\overline{\bm{6}}_4, \ 
\bm{1}_1\bm{3}_2\overline{\bm{2}}_3\overline{\bm{4}}_4, \ 
\bm{1}_1\bm{3}_2\overline{\bm{6}}_3\overline{\bm{4}}_4, \ 
\bm{2}_1\bm{2}_2\overline{\bm{4}}_3\overline{\bm{4}}_4, \ 
\bm{3}_1\bm{1}_2\overline{\bm{2}}_3\overline{\bm{4}}_4, \ 
\bm{3}_1\bm{1}_2\overline{\bm{6}}_3\overline{\bm{4}}_4, \ 
\bm{3}_1\bm{3}_2\overline{\bm{4}}_3\overline{\bm{2}}_4, \ 
\bm{3}_1\bm{3}_2\overline{\bm{4}}_3\overline{\bm{6}}_4.
\end{align}
The corresponding integer vectors are given by
\begin{subequations}
\begin{align}
M^{(4,4,\overline{8},\overline{8})}_\textrm{3B;1} = \begin{pmatrix}
0 & 0 & 1 \\
0 & 2 & 1 \\
0 & 2 & 0 \\
4 & 2 & 2
\end{pmatrix}, \quad
\Lambda^{(4,4,\overline{8},\overline{8})}_{\textsf{w},\textrm{3B;1}} = \begin{pmatrix}
1 & 1 & 0 & -2 \\
-1 & -1 & 0 & -2 \\
-1 & 1 & -2 & 0 \\
1 & -1 & -2 & 0
\end{pmatrix}, \\
M^{(4,4,\overline{8},\overline{8})}_\textrm{3B;2} = \begin{pmatrix}
0 & 0 & 1 \\
0 & 2 & 1 \\
0 & 2 & 2 \\
4 & 2 & 0
\end{pmatrix}, \quad
\Lambda^{(4,4,\overline{8},\overline{8})}_{\textsf{w},\textrm{3B;2}} = \begin{pmatrix}
1 & 1 & -2 & 0 \\
-1 & -1 & -2 & 0 \\
-1 & 1 & 0 & -2 \\
1 & -1 & 0 & -2
\end{pmatrix}.
\end{align}
\end{subequations}

The eight fermionic gapped interfaces at rank three are obtained by condensing fermionic pairs of $\bm{2}$ from another $U(1)_4$ and $\overline{\bm{4}}$ from $\overline{U(1)}_8$, a fermionic pair $\overline{\bm{2}} \overline{\bm{2}}$ or $\overline{\bm{2}} \overline{\bm{6}}$ from two $\overline{U(1)}_8$'s, and some triplets of anyons. 
Taking the Lagrangian subgroup generated by $M^{(4,4,\overline{8},\overline{8})}_\textrm{3F;1}$ given below as an instance, the set of condensed anyons is given by 
\begin{align}
&\red{\bm{2}_2}, \ 
\bm{2}_1\overline{\bm{4}}_3, \ 
\bm{2}_1\overline{\bm{4}}_4, \ 
\red{\overline{\bm{2}}_3\overline{\bm{6}}_4}, \ 
\overline{\bm{4}}_3\overline{\bm{4}}_4, \ 
\overline{\bm{6}}_3\overline{\bm{2}}_4, \ 
\bm{2}_1\bm{2}_2\overline{\bm{4}}_3, \ 
\bm{2}_1\bm{2}_2\overline{\bm{4}}_4, \nonumber \\
&\red{\bm{1}_1\overline{\bm{1}}_3\overline{\bm{1}}_4}, \ 
\bm{1}_1\overline{\bm{3}}_3\overline{\bm{7}}_4, \ 
\bm{1}_1\overline{\bm{5}}_3\overline{\bm{5}}_4, \ 
\bm{1}_1\overline{\bm{7}}_3\overline{\bm{3}}_4, \ 
\bm{2}_1\overline{\bm{2}}_3\overline{\bm{2}}_4, \ 
\bm{2}_1\overline{\bm{6}}_3\overline{\bm{6}}_4, \ 
\bm{3}_1\overline{\bm{1}}_3\overline{\bm{5}}_4, \ 
\bm{3}_1\overline{\bm{3}}_3\overline{\bm{3}}_4, \ 
\bm{3}_1\overline{\bm{5}}_3\overline{\bm{1}}_4, \ 
\bm{3}_1\overline{\bm{7}}_3\overline{\bm{7}}_4, \ 
\bm{2}_2\overline{\bm{2}}_3\overline{\bm{6}}_4, \ 
\bm{2}_2\overline{\bm{4}}_3\overline{\bm{4}}_4, \ 
\bm{2}_2\overline{\bm{6}}_3\overline{\bm{2}}_4, \nonumber \\
&\bm{1}_1\bm{2}_2\overline{\bm{1}}_3\overline{\bm{1}}_4, \ 
\bm{1}_1\bm{2}_2\overline{\bm{3}}_3\overline{\bm{7}}_4, \ 
\bm{1}_1\bm{2}_2\overline{\bm{5}}_3\overline{\bm{5}}_4, \ 
\bm{1}_1\bm{2}_2\overline{\bm{7}}_3\overline{\bm{3}}_4, \ 
\bm{2}_1\bm{2}_2\overline{\bm{2}}_3\overline{\bm{2}}_4, \ 
\bm{2}_1\bm{2}_2\overline{\bm{6}}_3\overline{\bm{6}}_4, \ 
\bm{3}_1\bm{2}_2\overline{\bm{1}}_3\overline{\bm{5}}_4, \ 
\bm{3}_1\bm{2}_2\overline{\bm{3}}_3\overline{\bm{3}}_4, \ 
\bm{3}_1\bm{2}_2\overline{\bm{5}}_3\overline{\bm{1}}_4, \ 
\bm{3}_1\bm{2}_2\overline{\bm{7}}_3\overline{\bm{7}}_4.
\end{align}
The corresponding integer vectors are given by
\begin{subequations}
\begin{align}
M^{(4,4,\overline{8},\overline{8})}_\textrm{3F;1} = \begin{pmatrix}
0 & 0 & 1 \\
0 & 2 & 0 \\
2 & 0 & 1 \\
6 & 0 & 1
\end{pmatrix}, \quad
\Lambda^{(4,4,\overline{8},\overline{8})}_{\textsf{w},\textrm{3F;1}} = \begin{pmatrix}
-1 & 0 & -3 & 1 \\
-3 & 0 & -1 & 3 \\
2 & 0 & 0 & -4 \\
0 & -2 & 0 & 0
\end{pmatrix}, \\
M^{(4,4,\overline{8},\overline{8})}_\textrm{3F;2} = \begin{pmatrix}
0 & 0 & 1 \\
0 & 2 & 0 \\
2 & 0 & 1 \\
2 & 0 & 3
\end{pmatrix}, \quad
\Lambda^{(4,4,\overline{8},\overline{8})}_{\textsf{w},\textrm{3F;2}} = \begin{pmatrix}
1 & 0 & -3 & -1 \\
-3 & 0 & 1 & 3 \\
2 & 0 & 0 & -4 \\
0 & -2 & 0 & 0
\end{pmatrix}, \\
M^{(4,4,\overline{8},\overline{8})}_\textrm{3F;3} = \begin{pmatrix}
0 & 0 & 1 \\
0 & 2 & 0 \\
2 & 0 & 1 \\
6 & 0 & 5
\end{pmatrix}, \quad
\Lambda^{(4,4,\overline{8},\overline{8})}_{\textsf{w},\textrm{3F;3}} = \begin{pmatrix}
1 & 0 & -3 & 1 \\
3 & 0 & -1 & 3 \\
-2 & 0 & 0 & -4 \\
0 & -2 & 0 & 0
\end{pmatrix}, \\
M^{(4,4,\overline{8},\overline{8})}_\textrm{3F;4} = \begin{pmatrix}
0 & 0 & 1 \\
0 & 2 & 0 \\
2 & 0 & 1 \\
2 & 0 & 7
\end{pmatrix}, \quad
\Lambda^{(4,4,\overline{8},\overline{8})}_{\textsf{w},\textrm{3F;4}} = \begin{pmatrix}
1 & 0 & 3 & 1 \\
3 & 0 & 1 & 3 \\
-2 & 0 & 0 & -4 \\
0 & -2 & 0 & 0
\end{pmatrix}, \\
M^{(4,4,\overline{8},\overline{8})}_\textrm{3F;5} = \begin{pmatrix}
0 & 2 & 0 \\
0 & 0 & 1 \\
2 & 0 & 1 \\
6 & 0 & 1
\end{pmatrix}, \quad
\Lambda^{(4,4,\overline{8},\overline{8})}_{\textsf{w},\textrm{3F;5}} = \begin{pmatrix}
0 & -1 & -3 & 1 \\
0 & -3 & -1 & 3 \\
0 & 2 & 0 & -4 \\
-2 & 0 & 0 & 0
\end{pmatrix}, \\
M^{(4,4,\overline{8},\overline{8})}_\textrm{3F;6} = \begin{pmatrix}
0 & 2 & 0 \\
0 & 0 & 1 \\
2 & 0 & 1 \\
2 & 0 & 3
\end{pmatrix}, \quad
\Lambda^{(4,4,\overline{8},\overline{8})}_{\textsf{w},\textrm{3F;6}} = \begin{pmatrix}
0 & 1 & -3 & -1 \\
0 & -3 & 1 & 3 \\
0 & 2 & 0 & -4 \\
-2 & 0 & 0 & 0
\end{pmatrix}, \\
M^{(4,4,\overline{8},\overline{8})}_\textrm{3F;7} = \begin{pmatrix}
0 & 2 & 0 \\
0 & 0 & 1 \\
2 & 0 & 1 \\
6 & 0 & 5
\end{pmatrix}, \quad
\Lambda^{(4,4,\overline{8},\overline{8})}_{\textsf{w},\textrm{3F;7}} = \begin{pmatrix}
0 & 1 & -3 & 1 \\
0 & 3 & -1 & 3 \\
0 & -2 & 0 & -4 \\
-2 & 0 & 0 & 0
\end{pmatrix}, \\
M^{(4,4,\overline{8},\overline{8})}_\textrm{3F;8} = \begin{pmatrix}
0 & 2 & 0 \\
0 & 0 & 1 \\
2 & 0 & 1 \\
2 & 0 & 7
\end{pmatrix}, \quad
\Lambda^{(4,4,\overline{8},\overline{8})}_{\textsf{w},\textrm{3F;8}} = \begin{pmatrix}
0 & 1 & 3 & 1 \\
0 & 3 & 1 & 3 \\
0 & -2 & 0 & -4 \\
-2 & 0 & 0 & 0
\end{pmatrix}.
\end{align}
\end{subequations}

\subsection{Gapped interface for $U(1)_3 \times U(1)_3 \times U(1)_3 \times \overline{U(1)}_3 \times \overline{U(1)}_3 \times \overline{U(1)}_3$}
\label{sec:GIU333333}

We consider gapped interfaces for fermionic topological orders corresponding to the following $K$ matrix:
\begin{align}
K^{(3,3,3,\overline{3},\overline{3},\overline{3})} = \begin{pmatrix} 3 &&&&& \\ & 3 &&&& \\ && 3 &&& \\ &&& -3 && \\ &&&& -3 & \\ &&&&& -3 \end{pmatrix}.
\end{align}
There are 80 gapped interfaces, among which 48 interfaces are obtained by just stacking three individual gapped interfaces between $U(1)_3$ and $\overline{U(1)}_3$. 
The other 32 interfaces do not admit the condensation of pairs of anyons but only triplets or quadruplets.
Taking the Lagrangian subgroup generated by $M^{(3,3,3,\overline{3},\overline{3},\overline{3})}_\textrm{3F;23}$ given below as an instance, the set of condensed anyons is given by 
\begin{align}
&\bm{1}_1\bm{1}_2\bm{1}_3, \ 
\bm{2}_1\bm{2}_2\bm{2}_3, \ 
\red{\overline{\bm{1}}_4\overline{\bm{1}}_5\overline{\bm{1}}_6}, \ 
\overline{\bm{2}}_4\overline{\bm{2}}_5\overline{\bm{2}}_6, \nonumber \\
&\bm{1}_1\bm{2}_2\overline{\bm{2}}_4\overline{\bm{1}}_5, \ 
\bm{2}_1\bm{1}_2\overline{\bm{1}}_4\overline{\bm{2}}_5, \ 
\bm{1}_1\bm{2}_2\overline{\bm{1}}_4\overline{\bm{2}}_6, \ 
\bm{2}_1\bm{1}_2\overline{\bm{2}}_4\overline{\bm{1}}_6, \ 
\bm{1}_1\bm{2}_2\overline{\bm{2}}_5\overline{\bm{1}}_6, \ 
\bm{2}_1\bm{1}_2\overline{\bm{1}}_5\overline{\bm{2}}_6, \ 
\bm{1}_1\bm{2}_3\overline{\bm{1}}_4\overline{\bm{2}}_5, \ 
\bm{2}_1\bm{1}_3\overline{\bm{2}}_4\overline{\bm{1}}_5, \ 
\bm{1}_1\bm{2}_3\overline{\bm{2}}_4\overline{\bm{1}}_6, \ 
\bm{2}_1\bm{1}_3\overline{\bm{1}}_4\overline{\bm{2}}_6, \nonumber \\
&\red{\bm{1}_1\bm{2}_3\overline{\bm{1}}_5\overline{\bm{2}}_6}, \ 
\bm{2}_1\bm{1}_3\overline{\bm{2}}_5\overline{\bm{1}}_6, \ 
\bm{1}_2\bm{2}_3\overline{\bm{2}}_4\overline{\bm{1}}_5, \ 
\bm{2}_2\bm{1}_3\overline{\bm{1}}_4\overline{\bm{2}}_5, \ 
\bm{1}_2\bm{2}_3\overline{\bm{1}}_4\overline{\bm{2}}_6, \ 
\bm{2}_2\bm{1}_3\overline{\bm{2}}_4\overline{\bm{1}}_6, \ 
\bm{1}_2\bm{2}_3\overline{\bm{2}}_5\overline{\bm{1}}_6, \ 
\red{\bm{2}_2\bm{1}_3\overline{\bm{1}}_5\overline{\bm{2}}_6}, \nonumber \\
&\bm{1}_1\bm{1}_2\bm{1}_3\overline{\bm{1}}_4\overline{\bm{1}}_5\overline{\bm{1}}_6, \ 
\bm{1}_1\bm{1}_2\bm{1}_3\overline{\bm{2}}_4\overline{\bm{2}}_5\overline{\bm{2}}_6, \ 
\bm{2}_1\bm{2}_2\bm{2}_3\overline{\bm{1}}_4\overline{\bm{1}}_5\overline{\bm{1}}_6, \ 
\bm{2}_1\bm{2}_2\bm{2}_3\overline{\bm{2}}_4\overline{\bm{2}}_5\overline{\bm{2}}_6.
\end{align}
Integer vectors associated with the Lagrangian subgroups and gapping potentials are given by
\begin{subequations}
\begin{align}
M^{(3,3,3,\overline{3},\overline{3},\overline{3})}_\textrm{3F;1} = \begin{pmatrix}
1 & 0 & 0 \\
0 & 1 & 0 \\
1 & 1 & 0 \\
0 & 0 & 1 \\
1 & 2 & 1 \\
1 & 2 & 2
\end{pmatrix}, \quad
\Lambda^{(3,3,3,\overline{3},\overline{3},\overline{3})}_{\textsf{w},\textrm{3F;1}} = \begin{pmatrix}
1 & 1 & -1 & 1 & -2 & -1 \\
1 & 0 & -2 & 1 & -1 & 0 \\
-1 & -1 & 1 & -2 & 1 & -1 \\
2 & 1 & 0 & 1 & -1 & 0 \\
0 & 0 & 0 & -1 & -1 & 1 \\
0 & 2 & -1 & 1 & -1 & 0
\end{pmatrix}, \\
M^{(3,3,3,\overline{3},\overline{3},\overline{3})}_\textrm{3F;2} = \begin{pmatrix}
1 & 0 & 0 \\
0 & 1 & 0 \\
1 & 1 & 0 \\
0 & 0 & 1 \\
1 & 2 & 2 \\
1 & 2 & 1
\end{pmatrix}, \quad
\Lambda^{(3,3,3,\overline{3},\overline{3},\overline{3})}_{\textsf{w},\textrm{3F;2}} = \begin{pmatrix}
1 & 1 & -1 & 1 & -1 & -2 \\
1 & 0 & -2 & 1 & 0 & -1 \\
-1 & -1 & 1 & -2 & -1 & 1 \\
2 & 1 & 0 & 1 & 0 & -1 \\
0 & 0 & 0 & 1 & -1 & 1 \\
0 & 2 & -1 & 1 & 0 & -1
\end{pmatrix}, \\
M^{(3,3,3,\overline{3},\overline{3},\overline{3})}_\textrm{3F;3} = \begin{pmatrix}
1 & 0 & 0 \\
0 & 1 & 0 \\
1 & 2 & 0 \\
0 & 0 & 1 \\
1 & 1 & 1 \\
1 & 1 & 2
\end{pmatrix}, \quad
\Lambda^{(3,3,3,\overline{3},\overline{3},\overline{3})}_{\textsf{w},\textrm{3F;3}} = \begin{pmatrix}
1 & -1 & -1 & 1 & -2 & -1 \\
0 & -2 & -1 & 1 & -1 & 0 \\
-1 & 1 & 1 & -2 & 1 & -1 \\
2 & -1 & 0 & 1 & -1 & 0 \\
0 & 0 & 0 & -1 & -1 & 1 \\
1 & 0 & -2 & 1 & -1 & 0
\end{pmatrix}, \\
M^{(3,3,3,\overline{3},\overline{3},\overline{3})}_\textrm{3F;4} = \begin{pmatrix}
1 & 0 & 0 \\
0 & 1 & 0 \\
1 & 2 & 0 \\
0 & 0 & 1 \\
1 & 1 & 2 \\
1 & 1 & 1
\end{pmatrix}, \quad
\Lambda^{(3,3,3,\overline{3},\overline{3},\overline{3})}_{\textsf{w},\textrm{3F;4}} = \begin{pmatrix}
1 & -1 & -1 & 1 & -1 & -2 \\
0 & -2 & -1 & 1 & 0 & -1 \\
-1 & 1 & 1 & -2 & -1 & 1 \\
2 & -1 & 0 & 1 & 0 & -1 \\
0 & 0 & 0 & 1 & -1 & 1 \\
1 & 0 & -2 & 1 & 0 & -1
\end{pmatrix}, \\
M^{(3,3,3,\overline{3},\overline{3},\overline{3})}_\textrm{3F;5} = \begin{pmatrix}
1 & 0 & 0 \\
0 & 1 & 0 \\
1 & 1 & 0 \\
0 & 0 & 1 \\
1 & 2 & 1 \\
2 & 1 & 1
\end{pmatrix}, \quad
\Lambda^{(3,3,3,\overline{3},\overline{3},\overline{3})}_{\textsf{w},\textrm{3F;5}} = \begin{pmatrix}
1 & 1 & -1 & 1 & 1 & -2 \\
1 & 0 & -2 & 0 & 1 & -1 \\
-1 & -1 & 1 & 1 & -2 & 1 \\
2 & 1 & 0 & 0 & 1 & -1 \\
0 & 0 & 0 & -1 & -1 & -1 \\
0 & 2 & -1 & 0 & 1 & -1
\end{pmatrix}, \\
M^{(3,3,3,\overline{3},\overline{3},\overline{3})}_\textrm{3F;6} = \begin{pmatrix}
1 & 0 & 0 \\
0 & 1 & 0 \\
1 & 1 & 0 \\
0 & 0 & 1 \\
1 & 2 & 2 \\
2 & 1 & 2
\end{pmatrix}, \quad
\Lambda^{(3,3,3,\overline{3},\overline{3},\overline{3})}_{\textsf{w},\textrm{3F;6}} = \begin{pmatrix}
-1 & -1 & 1 & -2 & -1 & -1 \\
2 & 0 & -1 & 1 & 1 & 0 \\
0 & 0 & 0 & 1 & -1 & -1 \\
0 & -1 & 2 & -1 & -1 & 0 \\
-1 & -1 & 1 & -1 & -2 & 1 \\
1 & 2 & 0 & 1 & 1 & 0
\end{pmatrix}, \\
M^{(3,3,3,\overline{3},\overline{3},\overline{3})}_\textrm{3F;7} = \begin{pmatrix}
1 & 0 & 0 \\
0 & 1 & 0 \\
1 & 2 & 0 \\
0 & 0 & 1 \\
1 & 1 & 1 \\
2 & 2 & 1
\end{pmatrix}, \quad
\Lambda^{(3,3,3,\overline{3},\overline{3},\overline{3})}_{\textsf{w},\textrm{3F;7}} = \begin{pmatrix}
1 & -1 & -1 & 2 & -1 & -1 \\
0 & -2 & -1 & 1 & -1 & 0 \\
-1 & 1 & 1 & -1 & 2 & -1 \\
2 & -1 & 0 & 1 & -1 & 0 \\
0 & 0 & 0 & -1 & -1 & -1 \\
1 & 0 & -2 & 1 & -1 & 0
\end{pmatrix}, \\
M^{(3,3,3,\overline{3},\overline{3},\overline{3})}_\textrm{3F;8} = \begin{pmatrix}
1 & 0 & 0 \\
0 & 1 & 0 \\
1 & 2 & 0 \\
0 & 0 & 1 \\
1 & 1 & 2 \\
2 & 2 & 2
\end{pmatrix}, \quad
\Lambda^{(3,3,3,\overline{3},\overline{3},\overline{3})}_{\textsf{w},\textrm{3F;8}} = \begin{pmatrix}
1 & -1 & -1 & -1 & -2 & 1 \\
2 & 0 & -1 & 0 & -1 & 1 \\
-1 & 1 & 1 & -1 & 1 & -2 \\
1 & -2 & 0 & 0 & -1 & 1 \\
0 & 0 & 0 & 1 & -1 & -1 \\
0 & -1 & -2 & 0 & -1 & 1
\end{pmatrix}, \\
M^{(3,3,3,\overline{3},\overline{3},\overline{3})}_\textrm{3F;9} = \begin{pmatrix}
1 & 0 & 0 \\
0 & 1 & 0 \\
1 & 1 & 0 \\
0 & 0 & 1 \\
2 & 1 & 1 \\
1 & 2 & 1
\end{pmatrix}, \quad
\Lambda^{(3,3,3,\overline{3},\overline{3},\overline{3})}_{\textsf{w},\textrm{3F;9}} = \begin{pmatrix}
0 & 0 & 0 & -1 & -1 & -1 \\
1 & 0 & -2 & 1 & 0 & -1 \\
-1 & -1 & 1 & -2 & 1 & 1 \\
0 & 2 & -1 & 1 & 0 & -1 \\
-1 & -1 & 1 & -1 & -1 & 2 \\
2 & 1 & 0 & 1 & 0 & -1
\end{pmatrix}, \\
M^{(3,3,3,\overline{3},\overline{3},\overline{3})}_\textrm{3F;10} = \begin{pmatrix}
1 & 0 & 0 \\
0 & 1 & 0 \\
1 & 1 & 0 \\
0 & 0 & 1 \\
2 & 1 & 2 \\
1 & 2 & 2
\end{pmatrix}, \quad
\Lambda^{(3,3,3,\overline{3},\overline{3},\overline{3})}_{\textsf{w},\textrm{3F;10}} = \begin{pmatrix}
1 & 0 & 1 & 0 & -1 & -2 \\
1 & 0 & -2 & 1 & 1 & 0 \\
1 & -1 & 0 & 2 & -1 & 0 \\
1 & 2 & 0 & 0 & 1 & -1 \\
-1 & -1 & 1 & -1 & -2 & 1 \\
-2 & 1 & -1 & -1 & 1 & 1
\end{pmatrix}, \\
M^{(3,3,3,\overline{3},\overline{3},\overline{3})}_\textrm{3F;11} = \begin{pmatrix}
1 & 0 & 0 \\
0 & 1 & 0 \\
1 & 2 & 0 \\
0 & 0 & 1 \\
2 & 2 & 1 \\
1 & 1 & 1
\end{pmatrix}, \quad
\Lambda^{(3,3,3,\overline{3},\overline{3},\overline{3})}_{\textsf{w},\textrm{3F;11}} = \begin{pmatrix}
1 & -1 & -1 & 2 & -1 & -1 \\
0 & -2 & -1 & 1 & 0 & -1 \\
-1 & 1 & 1 & -1 & -1 & 2 \\
2 & -1 & 0 & 1 & 0 & -1 \\
0 & 0 & 0 & -1 & -1 & -1 \\
1 & 0 & -2 & 1 & 0 & -1
\end{pmatrix}, \\
M^{(3,3,3,\overline{3},\overline{3},\overline{3})}_\textrm{3F;12} = \begin{pmatrix}
1 & 0 & 0 \\
0 & 1 & 0 \\
1 & 2 & 0 \\
0 & 0 & 1 \\
2 & 2 & 2 \\
1 & 1 & 2
\end{pmatrix}, \quad
\Lambda^{(3,3,3,\overline{3},\overline{3},\overline{3})}_{\textsf{w},\textrm{3F;12}} = \begin{pmatrix}
1 & -1 & -1 & 2 & 1 & 1 \\
0 & -2 & -1 & 1 & 1 & 0 \\
-1 & 1 & 1 & -1 & -2 & 1 \\
1 & 0 & -2 & 1 & 1 & 0 \\
0 & 0 & 0 & 1 & -1 & -1 \\
2 & -1 & 0 & 1 & 1 & 0
\end{pmatrix}, \\
M^{(3,3,3,\overline{3},\overline{3},\overline{3})}_\textrm{3F;13} = \begin{pmatrix}
1 & 0 & 0 \\
0 & 1 & 0 \\
1 & 1 & 0 \\
0 & 0 & 1 \\
2 & 1 & 1 \\
2 & 1 & 2
\end{pmatrix}, \quad
\Lambda^{(3,3,3,\overline{3},\overline{3},\overline{3})}_{\textsf{w},\textrm{3F;13}} = \begin{pmatrix}
1 & 1 & -1 & 1 & -2 & -1 \\
0 & 1 & -2 & 1 & -1 & 0 \\
-1 & -1 & 1 & -2 & 1 & -1 \\
1 & 2 & 0 & 1 & -1 & 0 \\
0 & 0 & 0 & -1 & -1 & 1 \\
2 & 0 & -1 & 1 & -1 & 0
\end{pmatrix}, \\
M^{(3,3,3,\overline{3},\overline{3},\overline{3})}_\textrm{3F;14} = \begin{pmatrix}
1 & 0 & 0 \\
0 & 1 & 0 \\
1 & 1 & 0 \\
0 & 0 & 1 \\
2 & 1 & 2 \\
2 & 1 & 1
\end{pmatrix}, \quad
\Lambda^{(3,3,3,\overline{3},\overline{3},\overline{3})}_{\textsf{w},\textrm{3F;14}} = \begin{pmatrix}
1 & 1 & -1 & 1 & -1 & -2 \\
0 & 1 & -2 & 1 & 0 & -1 \\
-1 & -1 & 1 & -2 & -1 & 1 \\
1 & 2 & 0 & 1 & 0 & -1 \\
0 & 0 & 0 & 1 & -1 & 1 \\
2 & 0 & -1 & 1 & 0 & -1
\end{pmatrix}, \\
M^{(3,3,3,\overline{3},\overline{3},\overline{3})}_\textrm{3F;15} = \begin{pmatrix}
1 & 0 & 0 \\
0 & 1 & 0 \\
1 & 2 & 0 \\
0 & 0 & 1 \\
2 & 2 & 1 \\
2 & 2 & 2
\end{pmatrix}, \quad
\Lambda^{(3,3,3,\overline{3},\overline{3},\overline{3})}_{\textsf{w},\textrm{3F;15}} = \begin{pmatrix}
1 & -1 & -1 & 1 & -2 & -1 \\
0 & -2 & -1 & 0 & -1 & -1 \\
-1 & 1 & 1 & 1 & 1 & 2 \\
2 & -1 & 0 & 0 & -1 & -1 \\
0 & 0 & 0 & -1 & -1 & 1 \\
1 & 0 & -2 & 0 & -1 & -1
\end{pmatrix}, \\
M^{(3,3,3,\overline{3},\overline{3},\overline{3})}_\textrm{3F;16} = \begin{pmatrix}
1 & 0 & 0 \\
0 & 1 & 0 \\
1 & 2 & 0 \\
0 & 0 & 1 \\
2 & 2 & 2 \\
2 & 2 & 1
\end{pmatrix}, \quad
\Lambda^{(3,3,3,\overline{3},\overline{3},\overline{3})}_{\textsf{w},\textrm{3F;16}} = \begin{pmatrix}
1 & -1 & -1 & 2 & 1 & -1 \\
0 & -2 & -1 & 1 & 1 & 0 \\
-1 & 1 & 1 & -1 & -2 & -1 \\
2 & -1 & 0 & 1 & 1 & 0 \\
0 & 0 & 0 & 1 & -1 & 1 \\
1 & 0 & -2 & 1 & 1 & 0
\end{pmatrix}, \\
M^{(3,3,3,\overline{3},\overline{3},\overline{3})}_\textrm{3F;17} = \begin{pmatrix}
1 & 0 & 0 \\
0 & 1 & 0 \\
2 & 1 & 0 \\
0 & 0 & 1 \\
1 & 1 & 1 \\
1 & 1 & 2
\end{pmatrix}, \quad
\Lambda^{(3,3,3,\overline{3},\overline{3},\overline{3})}_{\textsf{w},\textrm{3F;17}} = \begin{pmatrix}
1 & -1 & 1 & 1 & -2 & -1 \\
0 & -2 & 1 & 1 & -1 & 0 \\
-1 & 1 & -1 & -2 & 1 & -1 \\
1 & 0 & 2 & 1 & -1 & 0 \\
0 & 0 & 0 & -1 & -1 & 1 \\
2 & -1 & 0 & 1 & -1 & 0
\end{pmatrix}, \\
M^{(3,3,3,\overline{3},\overline{3},\overline{3})}_\textrm{3F;18} = \begin{pmatrix}
1 & 0 & 0 \\
0 & 1 & 0 \\
2 & 1 & 0 \\
0 & 0 & 1 \\
1 & 1 & 2 \\
1 & 1 & 1
\end{pmatrix}, \quad
\Lambda^{(3,3,3,\overline{3},\overline{3},\overline{3})}_{\textsf{w},\textrm{3F;18}} = \begin{pmatrix}
1 & -1 & 1 & 1 & -1 & -2 \\
0 & -2 & 1 & 1 & 0 & -1 \\
-1 & 1 & -1 & -2 & -1 & 1 \\
1 & 0 & 2 & 1 & 0 & -1 \\
0 & 0 & 0 & 1 & -1 & 1 \\
2 & -1 & 0 & 1 & 0 & -1
\end{pmatrix}, \\
M^{(3,3,3,\overline{3},\overline{3},\overline{3})}_\textrm{3F;19} = \begin{pmatrix}
1 & 0 & 0 \\
0 & 1 & 0 \\
2 & 2 & 0 \\
0 & 0 & 1 \\
1 & 2 & 1 \\
1 & 2 & 2
\end{pmatrix}, \quad
\Lambda^{(3,3,3,\overline{3},\overline{3},\overline{3})}_{\textsf{w},\textrm{3F;19}} = \begin{pmatrix}
-1 & -1 & -1 & 1 & -2 & -1 \\
-1 & -2 & 0 & 1 & -1 & 0 \\
1 & 1 & 1 & -2 & 1 & -1 \\
-2 & 0 & -1 & 1 & -1 & 0 \\
0 & 0 & 0 & -1 & -1 & 1 \\
0 & -1 & -2 & 1 & -1 & 0
\end{pmatrix}, \\
M^{(3,3,3,\overline{3},\overline{3},\overline{3})}_\textrm{3F;20} = \begin{pmatrix}
1 & 0 & 0 \\
0 & 1 & 0 \\
2 & 2 & 0 \\
0 & 0 & 1 \\
1 & 2 & 2 \\
1 & 2 & 1
\end{pmatrix}, \quad
\Lambda^{(3,3,3,\overline{3},\overline{3},\overline{3})}_{\textsf{w},\textrm{3F;20}} = \begin{pmatrix}
-1 & -1 & -1 & 1 & -1 & -2 \\
-1 & -2 & 0 & 1 & 0 & -1 \\
1 & 1 & 1 & -2 & -1 & 1 \\
-2 & 0 & -1 & 1 & 0 & -1 \\
0 & 0 & 0 & 1 & -1 & 1 \\
0 & -1 & -2 & 1 & 0 & -1
\end{pmatrix}, \\
M^{(3,3,3,\overline{3},\overline{3},\overline{3})}_\textrm{3F;21} = \begin{pmatrix}
1 & 0 & 0 \\
0 & 1 & 0 \\
2 & 1 & 0 \\
0 & 0 & 1 \\
1 & 1 & 1 \\
2 & 2 & 1
\end{pmatrix}, \quad
\Lambda^{(3,3,3,\overline{3},\overline{3},\overline{3})}_{\textsf{w},\textrm{3F;21}} = \begin{pmatrix}
1 & -1 & 1 & 2 & -1 & -1 \\
0 & -2 & 1 & 1 & -1 & 0 \\
-1 & 1 & -1 & -1 & 2 & -1 \\
1 & 0 & 2 & 1 & -1 & 0 \\
0 & 0 & 0 & -1 & -1 & -1 \\
2 & -1 & 0 & 1 & -1 & 0
\end{pmatrix}, \\
M^{(3,3,3,\overline{3},\overline{3},\overline{3})}_\textrm{3F;22} = \begin{pmatrix}
1 & 0 & 0 \\
0 & 1 & 0 \\
2 & 1 & 0 \\
0 & 0 & 1 \\
1 & 1 & 2 \\
2 & 2 & 2
\end{pmatrix}, \quad
\Lambda^{(3,3,3,\overline{3},\overline{3},\overline{3})}_{\textsf{w},\textrm{3F;22}} = \begin{pmatrix}
1 & -1 & 1 & -1 & -2 & 1 \\
2 & 0 & 1 & 0 & -1 & 1 \\
-1 & 1 & -1 & -1 & 1 & -2 \\
0 & -1 & 2 & 0 & -1 & 1 \\
0 & 0 & 0 & 1 & -1 & -1 \\
1 & -2 & 0 & 0 & -1 & 1
\end{pmatrix}, \\
M^{(3,3,3,\overline{3},\overline{3},\overline{3})}_\textrm{3F;23} = \begin{pmatrix}
1 & 0 & 0 \\
0 & 1 & 0 \\
2 & 2 & 0 \\
0 & 0 & 1 \\
1 & 2 & 1 \\
2 & 1 & 1
\end{pmatrix}, \quad
\Lambda^{(3,3,3,\overline{3},\overline{3},\overline{3})}_{\textsf{w},\textrm{3F;23}} = \begin{pmatrix}
-1 & -1 & -1 & -2 & 1 & 1 \\
-1 & 0 & -2 & -1 & 1 & 0 \\
1 & 1 & 1 & 1 & -2 & 1 \\
-2 & -1 & 0 & -1 & 1 & 0 \\
0 & 0 & 0 & -1 & -1 & -1 \\
0 & -2 & -1 & -1 & 1 & 0
\end{pmatrix}, \\
M^{(3,3,3,\overline{3},\overline{3},\overline{3})}_\textrm{3F;24} = \begin{pmatrix}
1 & 0 & 0 \\
0 & 1 & 0 \\
2 & 2 & 0 \\
0 & 0 & 1 \\
1 & 2 & 2 \\
2 & 1 & 2
\end{pmatrix}, \quad
\Lambda^{(3,3,3,\overline{3},\overline{3},\overline{3})}_{\textsf{w},\textrm{3F;24}} = \begin{pmatrix}
0 & 1 & -1 & -2 & 1 & 0 \\
1 & 1 & -2 & -1 & 1 & 1 \\
0 & 1 & -1 & 0 & 2 & 1 \\
1 & -2 & 1 & 1 & -1 & -1 \\
0 & 1 & -1 & -1 & 0 & 2 \\
-1 & -1 & -1 & 0 & 0 & 0
\end{pmatrix}, \\
M^{(3,3,3,\overline{3},\overline{3},\overline{3})}_\textrm{3F;25} = \begin{pmatrix}
1 & 0 & 0 \\
0 & 1 & 0 \\
2 & 1 & 0 \\
0 & 0 & 1 \\
2 & 2 & 1 \\
1 & 1 & 1
\end{pmatrix}, \quad
\Lambda^{(3,3,3,\overline{3},\overline{3},\overline{3})}_{\textsf{w},\textrm{3F;25}} = \begin{pmatrix}
1 & -1 & 1 & 2 & -1 & -1 \\
0 & -2 & 1 & 1 & 0 & -1 \\
-1 & 1 & -1 & -1 & -1 & 2 \\
1 & 0 & 2 & 1 & 0 & -1 \\
0 & 0 & 0 & -1 & -1 & -1 \\
2 & -1 & 0 & 1 & 0 & -1
\end{pmatrix}, \\
M^{(3,3,3,\overline{3},\overline{3},\overline{3})}_\textrm{3F;26} = \begin{pmatrix}
1 & 0 & 0 \\
0 & 1 & 0 \\
2 & 1 & 0 \\
0 & 0 & 1 \\
2 & 2 & 2 \\
1 & 1 & 2
\end{pmatrix}, \quad
\Lambda^{(3,3,3,\overline{3},\overline{3},\overline{3})}_{\textsf{w},\textrm{3F;26}} = \begin{pmatrix}
1 & -1 & 1 & 2 & 1 & 1 \\
0 & -2 & 1 & 1 & 1 & 0 \\
-1 & 1 & -1 & -1 & -2 & 1 \\
2 & -1 & 0 & 1 & 1 & 0 \\
0 & 0 & 0 & 1 & -1 & -1 \\
1 & 0 & 2 & 1 & 1 & 0
\end{pmatrix}, \\
M^{(3,3,3,\overline{3},\overline{3},\overline{3})}_\textrm{3F;27} = \begin{pmatrix}
1 & 0 & 0 \\
0 & 1 & 0 \\
2 & 2 & 0 \\
0 & 0 & 1 \\
2 & 1 & 1 \\
1 & 2 & 1
\end{pmatrix}, \quad
\Lambda^{(3,3,3,\overline{3},\overline{3},\overline{3})}_{\textsf{w},\textrm{3F;27}} = \begin{pmatrix}
-1 & -1 & -1 & -1 & 2 & -1 \\
-1 & 0 & -2 & 0 & 1 & -1 \\
1 & 1 & 1 & -1 & -1 & 2 \\
-2 & -1 & 0 & 0 & 1 & -1 \\
0 & 0 & 0 & -1 & -1 & -1 \\
0 & -2 & -1 & 0 & 1 & -1
\end{pmatrix}, \\
M^{(3,3,3,\overline{3},\overline{3},\overline{3})}_\textrm{3F;28} = \begin{pmatrix}
1 & 0 & 0 \\
0 & 1 & 0 \\
2 & 2 & 0 \\
0 & 0 & 1 \\
2 & 1 & 2 \\
1 & 2 & 2
\end{pmatrix}, \quad
\Lambda^{(3,3,3,\overline{3},\overline{3},\overline{3})}_{\textsf{w},\textrm{3F;28}} = \begin{pmatrix}
1 & -1 & 0 & -2 & 0 & 1 \\
2 & -1 & -1 & -1 & 1 & 1 \\
1 & -1 & 0 & 0 & 1 & 2 \\
-1 & 2 & -1 & 1 & -1 & -1 \\
1 & -1 & 0 & -1 & 2 & 0 \\
-1 & -1 & -1 & 0 & 0 & 0
\end{pmatrix}, \\
M^{(3,3,3,\overline{3},\overline{3},\overline{3})}_\textrm{3F;29} = \begin{pmatrix}
1 & 0 & 0 \\
0 & 1 & 0 \\
2 & 1 & 0 \\
0 & 0 & 1 \\
2 & 2 & 1 \\
2 & 2 & 2
\end{pmatrix}, \quad
\Lambda^{(3,3,3,\overline{3},\overline{3},\overline{3})}_{\textsf{w},\textrm{3F;29}} = \begin{pmatrix}
1 & -1 & 1 & 1 & -2 & -1 \\
0 & -2 & 1 & 0 & -1 & -1 \\
-1 & 1 & -1 & 1 & 1 & 2 \\
1 & 0 & 2 & 0 & -1 & -1 \\
0 & 0 & 0 & -1 & -1 & 1 \\
2 & -1 & 0 & 0 & -1 & -1
\end{pmatrix}, \\
M^{(3,3,3,\overline{3},\overline{3},\overline{3})}_\textrm{3F;30} = \begin{pmatrix}
1 & 0 & 0 \\
0 & 1 & 0 \\
2 & 1 & 0 \\
0 & 0 & 1 \\
2 & 2 & 2 \\
2 & 2 & 1
\end{pmatrix}, \quad
\Lambda^{(3,3,3,\overline{3},\overline{3},\overline{3})}_{\textsf{w},\textrm{3F;30}} = \begin{pmatrix}
1 & -1 & 1 & 2 & 1 & -1 \\
0 & -2 & 1 & 1 & 1 & 0 \\
-1 & 1 & -1 & -1 & -2 & -1 \\
1 & 0 & 2 & 1 & 1 & 0 \\
0 & 0 & 0 & 1 & -1 & 1 \\
2 & -1 & 0 & 1 & 1 & 0
\end{pmatrix}, \\
M^{(3,3,3,\overline{3},\overline{3},\overline{3})}_\textrm{3F;31} = \begin{pmatrix}
1 & 0 & 0 \\
0 & 1 & 0 \\
2 & 2 & 0 \\
0 & 0 & 1 \\
2 & 1 & 1 \\
2 & 1 & 2
\end{pmatrix}, \quad
\Lambda^{(3,3,3,\overline{3},\overline{3},\overline{3})}_{\textsf{w},\textrm{3F;31}} = \begin{pmatrix}
-1 & -1 & -1 & 1 & -2 & -1 \\
-2 & -1 & 0 & 1 & -1 & 0 \\
1 & 1 & 1 & -2 & 1 & -1 \\
0 & -2 & -1 & 1 & -1 & 0 \\
0 & 0 & 0 & -1 & -1 & 1 \\
-1 & 0 & -2 & 1 & -1 & 0
\end{pmatrix}, \\
M^{(3,3,3,\overline{3},\overline{3},\overline{3})}_\textrm{3F;32} = \begin{pmatrix}
1 & 0 & 0 \\
0 & 1 & 0 \\
2 & 2 & 0 \\
0 & 0 & 1 \\
2 & 1 & 2 \\
2 & 1 & 1
\end{pmatrix}, \quad
\Lambda^{(3,3,3,\overline{3},\overline{3},\overline{3})}_{\textsf{w},\textrm{3F;32}} = \begin{pmatrix}
-1 & -1 & -1 & 1 & -1 & -2 \\
-2 & -1 & 0 & 1 & 0 & -1 \\
1 & 1 & 1 & -2 & -1 & 1 \\
0 & -2 & -1 & 1 & 0 & -1 \\
0 & 0 & 0 & 1 & -1 & 1 \\
-1 & 0 & -2 & 1 & 0 & -1
\end{pmatrix}.
\end{align}
\end{subequations}

\section{Ground-state degeneracy of cellular topological states}
\label{sec:GSD}

\begin{figure}
\includegraphics[clip,width=0.45\textwidth]{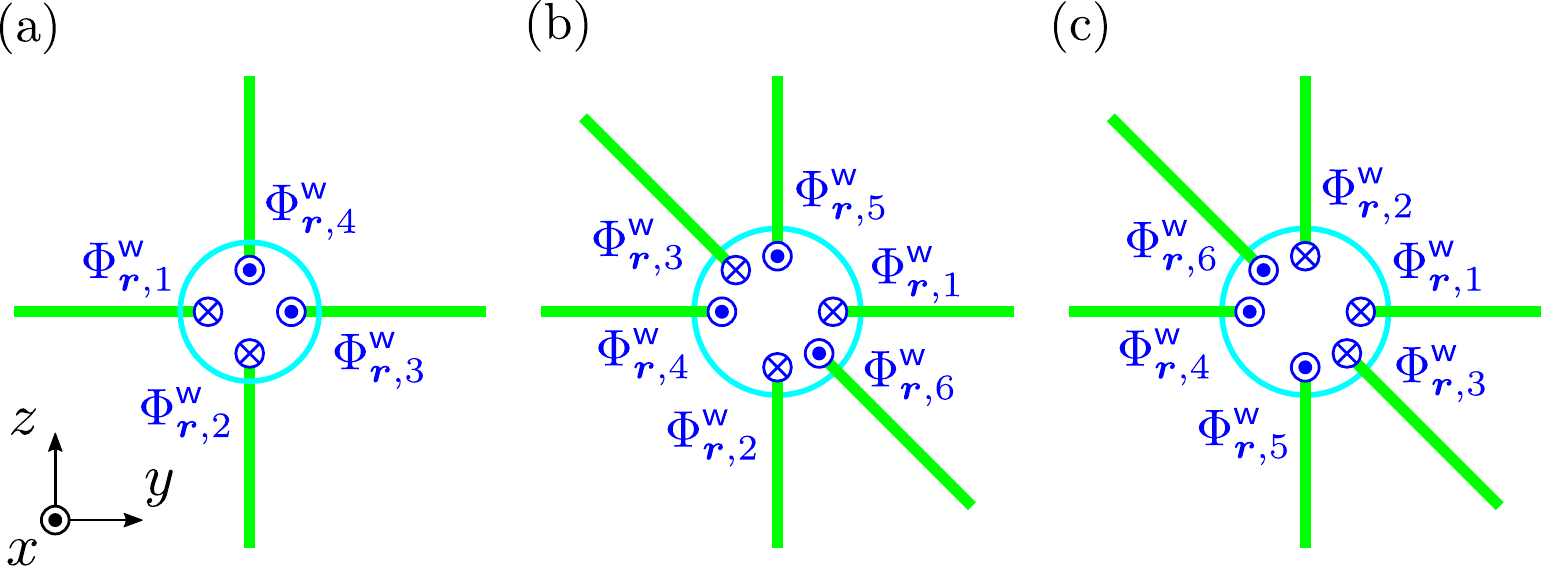}
\caption{Arrangement of the bosonic fields $\Phi^\textsf{w}_{\bm{r},\alpha}$ in the tunneling Hamiltonian for (a) the square grid and the triangular grid with (b) the pattern $\textsf{A}$ and (c) the pattern $\textsf{B}$.}
\label{fig:CellularTopo3DGSD}
\end{figure}

In this appendix, we provide ground-state degeneracies of the cellular topological states built from the $U(1)_k$ topological orders. 
For ceullar topological states on the square grid, we use gapped interfaces presented in Sec.~\ref{sec:GIUkkkk} between four $U(1)_k$ topological orders for $k=2,\cdots,9$. 
For each gapped interface corresponding to the Lagrangian subgroup $L$, we have a set of four integer vectors $\{ \bm{ \Lambda}_{\textsf{w},\alpha} \}$. 
Let us define a linear combination of the bosonic fields $\phi^\textsf{w}_{\bm{r},\alpha}$ by
\begin{align}
\Phi^\textsf{w}_{\bm{r},\alpha} = \bm{\Lambda}^T_{\textsf{w},\alpha} K_\textsf{w} \bm{\phi}^\textsf{w}_{\bm{r}}.
\end{align}
As discussed in Sec.~IV of the main text, we can then construct a tunneling Hamiltonian for the coupled-wire model [see Fig.~\ref{fig:CellularTopo3DGSD}~(a)],
\begin{align}
\mathcal{V}^\textsf{sq}_\textsf{w} = -g \int dx \sum_{\bm{r}} \left[ \cos (\Phi^\textsf{w}_{\bm{r},3} +\Phi^\textsf{w}_{\bm{r}+\bm{e}_y,1}) +\cos (\Phi^\textsf{w}_{\bm{r},4} +\Phi^\textsf{w}_{\bm{r}+\bm{e}_z,2}) \right].
\end{align}
For ceullar topological states on the triangular grid, we use gapped interfaces presented in Sec.~\ref{sec:GIU333333} between six $U(1)_3$ topological orders. 
For each gapped interface corresponding to the Lagrangian subgroup $L$, we have a set of six integer vectors $\{ \bm{ \Lambda}_{\textsf{w},\alpha} \}$. 
In this case, we consider two patterns of the arrangement of strips, which lead to different coupled-wire models. 
The first pattern denoted by $\textsf{A}$ gives the coupled-wire model [see Fig.~\ref{fig:CellularTopo3DGSD}~(b)],
\begin{align}
\mathcal{V}^\textsf{tri(A)}_\textsf{w} = -g \int dx \sum_{\bm{r}} \left[ \cos (\Phi^\textsf{w}_{\bm{r},1} +\Phi^\textsf{w}_{\bm{r}+\bm{e}_y,4}) +\cos (\Phi^\textsf{w}_{\bm{r},5} +\Phi^\textsf{w}_{\bm{r}+\bm{e}_z,2}) +\cos (\Phi^\textsf{w}_{\bm{r},6} +\Phi^\textsf{w}_{\bm{r}+\bm{e}_y-\bm{e}_z,3}) \right],
\end{align}
whereas the second pattern denoted by $\textsf{B}$ gives the coupled-wire model [see Fig.~\ref{fig:CellularTopo3DGSD}~(c)],
\begin{align}
\mathcal{V}^\textsf{tri(B)}_\textsf{w} = -g \int dx \sum_{\bm{r}} \left[ \cos (\Phi^\textsf{w}_{\bm{r},1} +\Phi^\textsf{w}_{\bm{r}+\bm{e}_y,4}) +\cos (\Phi^\textsf{w}_{\bm{r},2} +\Phi^\textsf{w}_{\bm{r}+\bm{e}_z,5}) +\cos (\Phi^\textsf{w}_{\bm{r},3} +\Phi^\textsf{w}_{\bm{r}+\bm{e}_y-\bm{e}_z,6}) \right].
\end{align}

We place these coupled-wire models on a $L_x \times L \times L$ torus and compute the ground-state degeneracy. 
To this end, we employ an \textit{ab initio} method proposed in Refs.~\cite{Ganeshan16, Ganeshan17}, which has been applied for a 3D coupled-wire model in Ref.~\cite{Fuji19b}. 
Specifically, we introduce the operator,
\begin{align}
N^\textsf{w}_{\bm{r},\alpha} = \frac{1}{2\pi} \int_0^{L_x} dx \, \partial_x \phi^\textsf{w}_{\bm{r},\alpha}(x),
\end{align}
which corresponds to a zero-mode part in the mode expansion of the bosonic field $\phi^\textsf{w}_{\bm{r},\alpha}(x)$ for quantum wires. 
These operators obey the commutation relations, 
\begin{align}
\begin{split}
[N^\textsf{w}_{\bm{r},\alpha}, \phi^\textsf{w}_{\bm{r}',\beta}(x')] &= i\delta_{\bm{r},\bm{r}'} (K_\textsf{w})_{\alpha \beta}, \\
[N^\textsf{w}_{\bm{r},\alpha}, N^\textsf{w}_{\bm{r}',\beta}] &=0.
\end{split}
\end{align}
We then consider the additional term, 
\begin{align}
\mathcal{V}_U = -U \sum_{\bm{r}} \sum_\alpha \cos (2\pi N^\textsf{w}_{\bm{r},\alpha}), 
\end{align}
to the tunneling Hamiltonian $\mathcal{H}_\textsf{w} +\mathcal{V}_\textsf{w}$, which dynamically enforces the compactification conditions, 
\begin{align}
\phi^\textsf{w}_{\bm{r},\alpha} \sim \phi^\textsf{w}_{\bm{r},\alpha} +2\pi \mathbb{Z},
\end{align}
in the limit $U \to \infty$. 
Writing the interaction Hamiltonian in the form, 
\begin{subequations}
\begin{align}
\mathcal{V}_\textsf{w}^\textsf{sq} + \mathcal{V}_U &= -g \sum_{I=1}^{2L_y L_z} \cos (C_I(x)) -U \sum_{I=2L_yL_z+1}^{6L_yL_z} \cos (C_I), \\
\mathcal{V}_\textsf{w}^\textsf{tri(X)} + \mathcal{V}_U &= -g \sum_{I=1}^{3L_y L_z} \cos (C_I(x)) -U \sum_{I=3L_yL_z+1}^{9L_yL_z} \cos (C_I), 
\end{align}
\end{subequations}
with $\textsf{X} = \textsf{A},\textsf{B}$, we define the skew-symmetric integer matrix, 
\begin{align}
Z_{IJ} = \frac{1}{2\pi i} [C_I, C_J].
\end{align}
Once we find the smith normal form of $Z$, which may be written as $\textrm{diag} (d_1, d_1, d_2, d_2, \cdots, d_r, d_r, 0, \cdots,0)$ with $r$ being the rank of $Z$, we obtain the ground-state degeneracy, 
\begin{align}
\textrm{GSD} = \prod_{j=1}^r d_j.
\end{align}
The results are presented in Tables~\ref{tab:GSDU2}--\ref{tab:GSDU9} for cellular topological states on the square grid and in Tables~\ref{tab:GSDU3triA} and \ref{tab:GSDU3triB} for those on the triangular grid.

\begin{table}
\begin{ruledtabular}
\begin{tabular}{llllllllllllllllllll}
& $L$ \\
Interface & 2 & 3 & 4 & 5 & 6 & 7 & 8 & 9 & 10 & 11 & 12 & 13 & 14 & 15 & 16 & 17 & 18 & 19 & 20 \\
\hline
$L^{(2,2,\overline{2},\overline{2})}_\textrm{2B;1}$ & $2^{4}$ & $2^{6}$ & $2^{8}$ & $2^{10}$ & $2^{12}$ & $2^{14}$ & $2^{16}$ & $2^{18}$ & $2^{20}$ & $2^{22}$ & $2^{24}$ & $2^{26}$ & $2^{28}$ & $2^{30}$ & $2^{32}$ & $2^{34}$ & $2^{36}$ & $2^{38}$ & $2^{40}$ \\
$L^{(2,2,\overline{2},\overline{2})}_\textrm{2B;2}$ & $2^{2}$ & $2^{3}$ & $2^{4}$ & $2^{5}$ & $2^{6}$ & $2^{7}$ & $2^{8}$ & $2^{9}$ & $2^{10}$ & $2^{11}$ & $2^{12}$ & $2^{13}$ & $2^{14}$ & $2^{15}$ & $2^{16}$ & $2^{17}$ & $2^{18}$ & $2^{19}$ & $2^{20}$ \\
\hline
$L^{(2,2,\overline{2},\overline{2})}_\textrm{2F;1}$ & $2^{2}$ & $2^{3}$ & $2^{4}$ & $2^{5}$ & $2^{6}$ & $2^{7}$ & $2^{8}$ & $2^{9}$ & $2^{10}$ & $2^{11}$ & $2^{12}$ & $2^{13}$ & $2^{14}$ & $2^{15}$ & $2^{16}$ & $2^{17}$ & $2^{18}$ & $2^{19}$ & $2^{20}$ \\
\end{tabular}
\end{ruledtabular}
\caption{Ground-state degeneracy of cellular topological states built from the $U(1)_2$ topological orders on a $L_x \times L \times L$ torus of the square grid.}
\end{table}

\begin{table}
\begin{ruledtabular}
\begin{tabular}{llllllllllllllllllll}
& $L$ \\
Interface & 2 & 3 & 4 & 5 & 6 & 7 & 8 & 9 & 10 & 11 & 12 & 13 & 14 & 15 & 16 & 17 & 18 & 19 & 20 \\
\hline
$L^{(3,3,\overline{3},\overline{3})}_\textrm{2F;1}$ & $3^{4}$ & $1$ & $3^{8}$ & $1$ & $3^{12}$ & $1$ & $3^{16}$ & $1$ & $3^{20}$ & $1$ & $3^{24}$ & $1$ & $3^{28}$ & $1$ & $3^{32}$ & $1$ & $3^{36}$ & $1$ & $3^{40}$ \\
$L^{(3,3,\overline{3},\overline{3})}_\textrm{2F;2}$ & $3^{4}$ & $3^{3}$ & $3^{8}$ & $3^{5}$ & $3^{12}$ & $3^{7}$ & $3^{16}$ & $3^{9}$ & $3^{20}$ & $3^{11}$ & $3^{24}$ & $3^{13}$ & $3^{28}$ & $3^{15}$ & $3^{32}$ & $3^{17}$ & $3^{36}$ & $3^{19}$ & $3^{40}$ \\
$L^{(3,3,\overline{3},\overline{3})}_\textrm{2F;3}$ & $3^{4}$ & $3^{3}$ & $3^{8}$ & $3^{5}$ & $3^{12}$ & $3^{7}$ & $3^{16}$ & $3^{9}$ & $3^{20}$ & $3^{11}$ & $3^{24}$ & $3^{13}$ & $3^{28}$ & $3^{15}$ & $3^{32}$ & $3^{17}$ & $3^{36}$ & $3^{19}$ & $3^{40}$ \\
$L^{(3,3,\overline{3},\overline{3})}_\textrm{2F;4}$ & $3^{4}$ & $3^{6}$ & $3^{8}$ & $3^{10}$ & $3^{12}$ & $3^{14}$ & $3^{16}$ & $3^{18}$ & $3^{20}$ & $3^{22}$ & $3^{24}$ & $3^{26}$ & $3^{28}$ & $3^{30}$ & $3^{32}$ & $3^{34}$ & $3^{36}$ & $3^{38}$ & $3^{40}$ \\
$L^{(3,3,\overline{3},\overline{3})}_\textrm{2F;5}$ & $3^{2}$ & $3^{3}$ & $3^{4}$ & $3^{5}$ & $3^{6}$ & $3^{7}$ & $3^{8}$ & $3^{9}$ & $3^{10}$ & $3^{11}$ & $3^{12}$ & $3^{13}$ & $3^{14}$ & $3^{15}$ & $3^{16}$ & $3^{17}$ & $3^{18}$ & $3^{19}$ & $3^{20}$ \\
$L^{(3,3,\overline{3},\overline{3})}_\textrm{2F;6}$ & $3^{2}$ & $1$ & $3^{4}$ & $1$ & $3^{6}$ & $1$ & $3^{8}$ & $1$ & $3^{10}$ & $1$ & $3^{12}$ & $1$ & $3^{14}$ & $1$ & $3^{16}$ & $1$ & $3^{18}$ & $1$ & $3^{20}$ \\
\end{tabular}
\end{ruledtabular}
\caption{Ground-state degeneracy of cellular topological states built from the $U(1)_3$ topological orders on a $L_x \times L \times L$ torus of the square grid.}
\label{tab:GSDU2}
\end{table}

\begin{table}
\begin{ruledtabular}
\begin{tabular}{llllllllllllllllllll}
& $L$ \\
Interface & 2 & 3 & 4 & 5 & 6 & 7 & 8 & 9 & 10 & 11 & 12 & 13 & 14 & 15 & 16 & 17 & 18 & 19 & 20 \\
\hline
$L^{(4,4,\overline{4},\overline{4})}_\textrm{2B;1}$ & $2^{8}$ & $2^{6}$ & $2^{16}$ & $2^{10}$ & $2^{24}$ & $2^{14}$ & $2^{32}$ & $2^{18}$ & $2^{40}$ & $2^{22}$ & $2^{48}$ & $2^{26}$ & $2^{56}$ & $2^{30}$ & $2^{64}$ & $2^{34}$ & $2^{72}$ & $2^{38}$ & $2^{80}$ \\
$L^{(4,4,\overline{4},\overline{4})}_\textrm{2B;2}$ & $2^{8}$ & $2^{9}$ & $2^{16}$ & $2^{15}$ & $2^{24}$ & $2^{21}$ & $2^{32}$ & $2^{27}$ & $2^{40}$ & $2^{33}$ & $2^{48}$ & $2^{39}$ & $2^{56}$ & $2^{45}$ & $2^{64}$ & $2^{51}$ & $2^{72}$ & $2^{57}$ & $2^{80}$ \\
$L^{(4,4,\overline{4},\overline{4})}_\textrm{2B;3}$ & $2^{8}$ & $2^{9}$ & $2^{16}$ & $2^{15}$ & $2^{24}$ & $2^{21}$ & $2^{32}$ & $2^{27}$ & $2^{40}$ & $2^{33}$ & $2^{48}$ & $2^{39}$ & $2^{56}$ & $2^{45}$ & $2^{64}$ & $2^{51}$ & $2^{72}$ & $2^{57}$ & $2^{80}$ \\
$L^{(4,4,\overline{4},\overline{4})}_\textrm{2B;4}$ & $2^{8}$ & $2^{12}$ & $2^{16}$ & $2^{20}$ & $2^{24}$ & $2^{28}$ & $2^{32}$ & $2^{36}$ & $2^{40}$ & $2^{44}$ & $2^{48}$ & $2^{52}$ & $2^{56}$ & $2^{60}$ & $2^{64}$ & $2^{68}$ & $2^{72}$ & $2^{76}$ & $2^{80}$ \\
$L^{(4,4,\overline{4},\overline{4})}_\textrm{2B;5}$ & $2^{4}$ & $2^{6}$ & $2^{8}$ & $2^{10}$ & $2^{12}$ & $2^{14}$ & $2^{16}$ & $2^{18}$ & $2^{20}$ & $2^{22}$ & $2^{24}$ & $2^{26}$ & $2^{28}$ & $2^{30}$ & $2^{32}$ & $2^{34}$ & $2^{36}$ & $2^{38}$ & $2^{40}$ \\
$L^{(4,4,\overline{4},\overline{4})}_\textrm{2B;6}$ & $2^{4}$ & $2^{3}$ & $2^{8}$ & $2^{5}$ & $2^{12}$ & $2^{7}$ & $2^{16}$ & $2^{9}$ & $2^{20}$ & $2^{11}$ & $2^{24}$ & $2^{13}$ & $2^{28}$ & $2^{15}$ & $2^{32}$ & $2^{17}$ & $2^{36}$ & $2^{19}$ & $2^{40}$ \\
\end{tabular}
\end{ruledtabular}
\caption{Ground-state degeneracy of cellular topological states built from the $U(1)_4$ topological orders on a $L_x \times L \times L$ torus of the square grid.}
\end{table}

\begin{table}
\begin{ruledtabular}
\begin{tabular}{llllllllllllllllllll}
& $L$ \\
Interface & 2 & 3 & 4 & 5 & 6 & 7 & 8 & 9 & 10 & 11 & 12 & 13 & 14 & 15 & 16 & 17 & 18 & 19 & 20 \\
\hline
$L^{(5,5,\overline{5},\overline{5})}_\textrm{2F;1}$ & $5^{4}$ & $1$ & $5^{8}$ & $1$ & $5^{12}$ & $1$ & $5^{16}$ & $1$ & $5^{20}$ & $1$ & $5^{24}$ & $1$ & $5^{28}$ & $1$ & $5^{32}$ & $1$ & $5^{36}$ & $1$ & $5^{40}$ \\
$L^{(5,5,\overline{5},\overline{5})}_\textrm{2F;2}$ & $5^{4}$ & $5^{3}$ & $5^{8}$ & $5^{5}$ & $5^{12}$ & $5^{7}$ & $5^{16}$ & $5^{9}$ & $5^{20}$ & $5^{11}$ & $5^{24}$ & $5^{13}$ & $5^{28}$ & $5^{15}$ & $5^{32}$ & $5^{17}$ & $5^{36}$ & $5^{19}$ & $5^{40}$ \\
$L^{(5,5,\overline{5},\overline{5})}_\textrm{2F;3}$ & $5^{4}$ & $5^{3}$ & $5^{8}$ & $5^{5}$ & $5^{12}$ & $5^{7}$ & $5^{16}$ & $5^{9}$ & $5^{20}$ & $5^{11}$ & $5^{24}$ & $5^{13}$ & $5^{28}$ & $5^{15}$ & $5^{32}$ & $5^{17}$ & $5^{36}$ & $5^{19}$ & $5^{40}$ \\
$L^{(5,5,\overline{5},\overline{5})}_\textrm{2F;4}$ & $5^{4}$ & $5^{6}$ & $5^{8}$ & $5^{10}$ & $5^{12}$ & $5^{14}$ & $5^{16}$ & $5^{18}$ & $5^{20}$ & $5^{22}$ & $5^{24}$ & $5^{26}$ & $5^{28}$ & $5^{30}$ & $5^{32}$ & $5^{34}$ & $5^{36}$ & $5^{38}$ & $5^{40}$ \\
$L^{(5,5,\overline{5},\overline{5})}_\textrm{2F;5}$ & $5^{2}$ & $5^{3}$ & $5^{4}$ & $5^{5}$ & $5^{6}$ & $5^{7}$ & $5^{8}$ & $5^{9}$ & $5^{10}$ & $5^{11}$ & $5^{12}$ & $5^{13}$ & $5^{14}$ & $5^{15}$ & $5^{16}$ & $5^{17}$ & $5^{18}$ & $5^{19}$ & $5^{20}$ \\
$L^{(5,5,\overline{5},\overline{5})}_\textrm{2F;6}$ & $5^{2}$ & $1$ & $5^{4}$ & $1$ & $5^{6}$ & $1$ & $5^{8}$ & $1$ & $5^{10}$ & $1$ & $5^{12}$ & $1$ & $5^{14}$ & $1$ & $5^{16}$ & $1$ & $5^{18}$ & $1$ & $5^{20}$ \\
\end{tabular}
\end{ruledtabular}
\caption{Ground-state degeneracy of cellular topological states built from the $U(1)_5$ topological orders on a $L_x \times L \times L$ torus of the square grid.}
\end{table}

\begin{table}
\begin{ruledtabular}
\begin{tabular}{llllllllllllllll}
& $L$ \\
Interface & 2 & 3 & 4 & 5 & 6 & 7 & 8 & 9 & 10 & 11 & 12 & 13 & 14 & 15 & 16 \\
\hline
$L^{(6,6,\overline{6},\overline{6})}_\textrm{2B;1}$ & $2^{4}3^{4}$ & $2^{6}$ & $2^{8}3^{8}$ & $2^{10}$ & $2^{12}3^{12}$ & $2^{14}$ & $2^{16}3^{16}$ & $2^{18}$ & $2^{20}3^{20}$ & $2^{22}$ & $2^{24}3^{24}$ & $2^{26}$ & $2^{28}3^{28}$ & $2^{30}$ & $2^{32}3^{32}$ \\
$L^{(6,6,\overline{6},\overline{6})}_\textrm{2B;2}$ & $2^{4}3^{4}$ & $2^{6}3^{3}$ & $2^{8}3^{8}$ & $2^{10}3^{5}$ & $2^{12}3^{12}$ & $2^{14}3^{7}$ & $2^{16}3^{16}$ & $2^{18}3^{9}$ & $2^{20}3^{20}$ & $2^{22}3^{11}$ & $2^{24}3^{24}$ & $2^{26}3^{13}$ & $2^{28}3^{28}$ & $2^{30}3^{15}$ & $2^{32}3^{32}$ \\
$L^{(6,6,\overline{6},\overline{6})}_\textrm{2B;3}$ & $2^{4}3^{4}$ & $2^{6}3^{3}$ & $2^{8}3^{8}$ & $2^{10}3^{5}$ & $2^{12}3^{12}$ & $2^{14}3^{7}$ & $2^{16}3^{16}$ & $2^{18}3^{9}$ & $2^{20}3^{20}$ & $2^{22}3^{11}$ & $2^{24}3^{24}$ & $2^{26}3^{13}$ & $2^{28}3^{28}$ & $2^{30}3^{15}$ & $2^{32}3^{32}$ \\
$L^{(6,6,\overline{6},\overline{6})}_\textrm{2B;4}$ & $2^{4}3^{4}$ & $2^{6}3^{6}$ & $2^{8}3^{8}$ & $2^{10}3^{10}$ & $2^{12}3^{12}$ & $2^{14}3^{14}$ & $2^{16}3^{16}$ & $2^{18}3^{18}$ & $2^{20}3^{20}$ & $2^{22}3^{22}$ & $2^{24}3^{24}$ & $2^{26}3^{26}$ & $2^{28}3^{28}$ & $2^{30}3^{30}$ & $2^{32}3^{32}$ \\
$L^{(6,6,\overline{6},\overline{6})}_\textrm{2B;5}$ & $2^{2}3^{2}$ & $2^{3}3^{3}$ & $2^{4}3^{4}$ & $2^{5}3^{5}$ & $2^{6}3^{6}$ & $2^{7}3^{7}$ & $2^{8}3^{8}$ & $2^{9}3^{9}$ & $2^{10}3^{10}$ & $2^{11}3^{11}$ & $2^{12}3^{12}$ & $2^{13}3^{13}$ & $2^{14}3^{14}$ & $2^{15}3^{15}$ & $2^{16}3^{16}$ \\
$L^{(6,6,\overline{6},\overline{6})}_\textrm{2B;6}$ & $2^{2}3^{2}$ & $2^{3}$ & $2^{4}3^{4}$ & $2^{5}$ & $2^{6}3^{6}$ & $2^{7}$ & $2^{8}3^{8}$ & $2^{9}$ & $2^{10}3^{10}$ & $2^{11}$ & $2^{12}3^{12}$ & $2^{13}$ & $2^{14}3^{14}$ & $2^{15}$ & $2^{16}3^{16}$ \\
\end{tabular}
\end{ruledtabular}
\caption{Ground-state degeneracy of cellular topological states built from the $U(1)_6$ topological orders on a $L_x \times L \times L$ torus of the square grid.}
\end{table}

\begin{table}
\begin{ruledtabular}
\begin{tabular}{llllllllllllllllllll}
& $L$ \\
Interface & 2 & 3 & 4 & 5 & 6 & 7 & 8 & 9 & 10 & 11 & 12 & 13 & 14 & 15 & 16 & 17 & 18 & 19 & 20 \\
\hline
$L^{(7,7,\overline{7},\overline{7})}_\textrm{2F;1}$ & $7^{4}$ & $1$ & $7^{8}$ & $1$ & $7^{12}$ & $1$ & $7^{16}$ & $1$ & $7^{20}$ & $1$ & $7^{24}$ & $1$ & $7^{28}$ & $1$ & $7^{32}$ & $1$ & $7^{36}$ & $1$ & $7^{40}$ \\
$L^{(7,7,\overline{7},\overline{7})}_\textrm{2F;2}$ & $7^{4}$ & $7^{3}$ & $7^{8}$ & $7^{5}$ & $7^{12}$ & $7^{7}$ & $7^{16}$ & $7^{9}$ & $7^{20}$ & $7^{11}$ & $7^{24}$ & $7^{13}$ & $7^{28}$ & $7^{15}$ & $7^{32}$ & $7^{17}$ & $7^{36}$ & $7^{19}$ & $7^{40}$ \\
$L^{(7,7,\overline{7},\overline{7})}_\textrm{2F;3}$ & $7^{4}$ & $7^{3}$ & $7^{8}$ & $7^{5}$ & $7^{12}$ & $7^{7}$ & $7^{16}$ & $7^{9}$ & $7^{20}$ & $7^{11}$ & $7^{24}$ & $7^{13}$ & $7^{28}$ & $7^{15}$ & $7^{32}$ & $7^{17}$ & $7^{36}$ & $7^{19}$ & $7^{40}$ \\
$L^{(7,7,\overline{7},\overline{7})}_\textrm{2F;4}$ & $7^{4}$ & $7^{6}$ & $7^{8}$ & $7^{10}$ & $7^{12}$ & $7^{14}$ & $7^{16}$ & $7^{18}$ & $7^{20}$ & $7^{22}$ & $7^{24}$ & $7^{26}$ & $7^{28}$ & $7^{30}$ & $7^{32}$ & $7^{34}$ & $7^{36}$ & $7^{38}$ & $7^{40}$ \\
$L^{(7,7,\overline{7},\overline{7})}_\textrm{2F;5}$ & $7^{2}$ & $7^{3}$ & $7^{4}$ & $7^{5}$ & $7^{6}$ & $7^{7}$ & $7^{8}$ & $7^{9}$ & $7^{10}$ & $7^{11}$ & $7^{12}$ & $7^{13}$ & $7^{14}$ & $7^{15}$ & $7^{16}$ & $7^{17}$ & $7^{18}$ & $7^{19}$ & $7^{20}$ \\
$L^{(7,7,\overline{7},\overline{7})}_\textrm{2F;6}$ & $7^{2}$ & $1$ & $7^{4}$ & $1$ & $7^{6}$ & $1$ & $7^{8}$ & $1$ & $7^{10}$ & $1$ & $7^{12}$ & $1$ & $7^{14}$ & $1$ & $7^{16}$ & $1$ & $7^{18}$ & $1$ & $7^{20}$ \\
\end{tabular}
\end{ruledtabular}
\caption{Ground-state degeneracy of cellular topological states built from the $U(1)_7$ topological orders on a $L_x \times L \times L$ torus of the square grid.}
\end{table}

\begin{table}
\begin{ruledtabular}
\begin{tabular}{llllllllllllllllllll}
& $L$ \\
Interface & 2 & 3 & 4 & 5 & 6 & 7 & 8 & 9 & 10 & 11 & 12 & 13 & 14 & 15 & 16 & 17 & 18 & 19 & 20 \\
\hline
$L^{(8,8,\overline{8},\overline{8})}_\textrm{2B;1}$ & $2^{12}$ & $2^{6}$ & $2^{24}$ & $2^{10}$ & $2^{36}$ & $2^{14}$ & $2^{48}$ & $2^{18}$ & $2^{60}$ & $2^{22}$ & $2^{72}$ & $2^{26}$ & $2^{84}$ & $2^{30}$ & $2^{96}$ & $2^{34}$ & $2^{108}$ & $2^{38}$ & $2^{120}$ \\
$L^{(8,8,\overline{8},\overline{8})}_\textrm{2B;2}$ & $2^{12}$ & $2^{12}$ & $2^{24}$ & $2^{20}$ & $2^{36}$ & $2^{28}$ & $2^{48}$ & $2^{36}$ & $2^{60}$ & $2^{44}$ & $2^{72}$ & $2^{52}$ & $2^{84}$ & $2^{60}$ & $2^{96}$ & $2^{68}$ & $2^{108}$ & $2^{76}$ & $2^{120}$ \\
$L^{(8,8,\overline{8},\overline{8})}_\textrm{2B;3}$ & $2^{12}$ & $2^{12}$ & $2^{24}$ & $2^{20}$ & $2^{36}$ & $2^{28}$ & $2^{48}$ & $2^{36}$ & $2^{60}$ & $2^{44}$ & $2^{72}$ & $2^{52}$ & $2^{84}$ & $2^{60}$ & $2^{96}$ & $2^{68}$ & $2^{108}$ & $2^{76}$ & $2^{120}$ \\
$L^{(8,8,\overline{8},\overline{8})}_\textrm{2B;4}$ & $2^{12}$ & $2^{18}$ & $2^{24}$ & $2^{30}$ & $2^{36}$ & $2^{42}$ & $2^{48}$ & $2^{54}$ & $2^{60}$ & $2^{66}$ & $2^{72}$ & $2^{78}$ & $2^{84}$ & $2^{90}$ & $2^{96}$ & $2^{102}$ & $2^{108}$ & $2^{114}$ & $2^{120}$ \\
$L^{(8,8,\overline{8},\overline{8})}_\textrm{2B;5}$ & $2^{6}$ & $2^{9}$ & $2^{12}$ & $2^{15}$ & $2^{18}$ & $2^{21}$ & $2^{24}$ & $2^{27}$ & $2^{30}$ & $2^{33}$ & $2^{36}$ & $2^{39}$ & $2^{42}$ & $2^{45}$ & $2^{48}$ & $2^{51}$ & $2^{54}$ & $2^{57}$ & $2^{60}$ \\
$L^{(8,8,\overline{8},\overline{8})}_\textrm{2B;6}$ & $2^{6}$ & $2^{3}$ & $2^{12}$ & $2^{5}$ & $2^{18}$ & $2^{7}$ & $2^{24}$ & $2^{9}$ & $2^{30}$ & $2^{11}$ & $2^{36}$ & $2^{13}$ & $2^{42}$ & $2^{15}$ & $2^{48}$ & $2^{17}$ & $2^{54}$ & $2^{19}$ & $2^{60}$ \\
\end{tabular}
\end{ruledtabular}
\caption{Ground-state degeneracy of cellular topological states built from the $U(1)_8$ topological orders on a $L_x \times L \times L$ torus of the square grid.}
\end{table}

\begin{table}
\begin{ruledtabular}
\begin{tabular}{llllllllllllllllllll}
& $L$ \\
Interface & 2 & 3 & 4 & 5 & 6 & 7 & 8 & 9 & 10 & 11 & 12 & 13 & 14 & 15 & 16 & 17 & 18 & 19 & 20 \\
\hline
$L^{(8,8,\overline{8},\overline{8})}_\textrm{2F;25}$ & $2^{10}$ & $2^{13}$ & $2^{22}$ & $2^{21}$ & $2^{34}$ & $2^{29}$ & $2^{46}$ & $2^{37}$ & $2^{58}$ & $2^{45}$ & $2^{70}$ & $2^{53}$ & $2^{82}$ & $2^{61}$ & $2^{94}$ & $2^{69}$ & $2^{106}$ & $2^{77}$ & $2^{118}$ \\
$L^{(8,8,\overline{8},\overline{8})}_\textrm{2F;26}$ & $2^{10}$ & $2^{9}$ & $2^{22}$ & $2^{15}$ & $2^{34}$ & $2^{21}$ & $2^{46}$ & $2^{27}$ & $2^{58}$ & $2^{33}$ & $2^{70}$ & $2^{39}$ & $2^{82}$ & $2^{45}$ & $2^{94}$ & $2^{51}$ & $2^{106}$ & $2^{57}$ & $2^{118}$ \\
$L^{(8,8,\overline{8},\overline{8})}_\textrm{2F;27}$ & $2^{6}$ & $2^{7}$ & $2^{12}$ & $2^{11}$ & $2^{18}$ & $2^{15}$ & $2^{24}$ & $2^{19}$ & $2^{30}$ & $2^{23}$ & $2^{36}$ & $2^{27}$ & $2^{42}$ & $2^{31}$ & $2^{48}$ & $2^{35}$ & $2^{54}$ & $2^{39}$ & $2^{60}$ \\
$L^{(8,8,\overline{8},\overline{8})}_\textrm{2F;28}$ & $2^{6}$ & $2^{3}$ & $2^{12}$ & $2^{5}$ & $2^{18}$ & $2^{7}$ & $2^{24}$ & $2^{9}$ & $2^{30}$ & $2^{11}$ & $2^{36}$ & $2^{13}$ & $2^{42}$ & $2^{15}$ & $2^{48}$ & $2^{17}$ & $2^{54}$ & $2^{19}$ & $2^{60}$ \\
$L^{(8,8,\overline{8},\overline{8})}_\textrm{2F;29}$ & $2^{6}$ & $2^{3}$ & $2^{12}$ & $2^{5}$ & $2^{18}$ & $2^{7}$ & $2^{24}$ & $2^{9}$ & $2^{30}$ & $2^{11}$ & $2^{36}$ & $2^{13}$ & $2^{42}$ & $2^{15}$ & $2^{48}$ & $2^{17}$ & $2^{54}$ & $2^{19}$ & $2^{60}$ \\
$L^{(8,8,\overline{8},\overline{8})}_\textrm{2F;30}$ & $2^{6}$ & $2^{7}$ & $2^{12}$ & $2^{11}$ & $2^{18}$ & $2^{15}$ & $2^{24}$ & $2^{19}$ & $2^{30}$ & $2^{23}$ & $2^{36}$ & $2^{27}$ & $2^{42}$ & $2^{31}$ & $2^{48}$ & $2^{35}$ & $2^{54}$ & $2^{39}$ & $2^{60}$ \\
\end{tabular}
\end{ruledtabular}
\caption{(Continued) Ground-state degeneracy of cellular topological states built from the $U(1)_8$ topological orders on a $L_x \times L \times L$ torus of the square grid.}
\end{table}

\begin{table}
\begin{ruledtabular}
\begin{tabular}{llllllllllllllllllll}
& $L$ \\
Interface & 2 & 3 & 4 & 5 & 6 & 7 & 8 & 9 & 10 & 11 & 12 & 13 & 14 & 15 & 16 & 17 & 18 & 19 & 20 \\
\hline
$L^{(8,8,\overline{8},\overline{8})}_\textrm{3F;1}$ & $2^{12}$ & $2^{18}$ & $2^{32}$ & $2^{40}$ & $2^{60}$ & $2^{70}$ & $2^{96}$ & $2^{108}$ & $2^{140}$ & $2^{154}$ & $2^{192}$ & $2^{208}$ & $2^{252}$ & $2^{270}$ & $2^{320}$ & $2^{340}$ & $2^{396}$ & $2^{418}$ & $2^{480}$ \\
$L^{(8,8,\overline{8},\overline{8})}_\textrm{3F;2}$ & $2^{12}$ & $2^{15}$ & $2^{32}$ & $2^{35}$ & $2^{60}$ & $2^{63}$ & $2^{96}$ & $2^{99}$ & $2^{140}$ & $2^{143}$ & $2^{192}$ & $2^{195}$ & $2^{252}$ & $2^{255}$ & $2^{320}$ & $2^{323}$ & $2^{396}$ & $2^{399}$ & $2^{480}$ \\
$L^{(8,8,\overline{8},\overline{8})}_\textrm{3F;3}$ & $2^{6}$ & $2^{12}$ & $2^{20}$ & $2^{30}$ & $2^{42}$ & $2^{56}$ & $2^{72}$ & $2^{90}$ & $2^{110}$ & $2^{132}$ & $2^{156}$ & $2^{182}$ & $2^{210}$ & $2^{240}$ & $2^{272}$ & $2^{306}$ & $2^{342}$ & $2^{380}$ & $2^{420}$ \\
$L^{(8,8,\overline{8},\overline{8})}_\textrm{3F;4}$ & $2^{6}$ & $2^{12}$ & $2^{20}$ & $2^{30}$ & $2^{42}$ & $2^{56}$ & $2^{72}$ & $2^{90}$ & $2^{110}$ & $2^{132}$ & $2^{156}$ & $2^{182}$ & $2^{210}$ & $2^{240}$ & $2^{272}$ & $2^{306}$ & $2^{342}$ & $2^{380}$ & $2^{420}$ \\
$L^{(8,8,\overline{8},\overline{8})}_\textrm{3F;5}$ & $2^{12}$ & $2^{18}$ & $2^{32}$ & $2^{40}$ & $2^{60}$ & $2^{70}$ & $2^{96}$ & $2^{108}$ & $2^{140}$ & $2^{154}$ & $2^{192}$ & $2^{208}$ & $2^{252}$ & $2^{270}$ & $2^{320}$ & $2^{340}$ & $2^{396}$ & $2^{418}$ & $2^{480}$ \\
$L^{(8,8,\overline{8},\overline{8})}_\textrm{3F;6}$ & $2^{12}$ & $2^{15}$ & $2^{32}$ & $2^{35}$ & $2^{60}$ & $2^{63}$ & $2^{96}$ & $2^{99}$ & $2^{140}$ & $2^{143}$ & $2^{192}$ & $2^{195}$ & $2^{252}$ & $2^{255}$ & $2^{320}$ & $2^{323}$ & $2^{396}$ & $2^{399}$ & $2^{480}$ \\
\end{tabular}
\end{ruledtabular}
\caption{(Continued) Ground-state degeneracy of cellular topological states built from the $U(1)_8$ topological orders on a $L_x \times L \times L$ torus of the square grid.}
\end{table}

\begin{table}
\begin{ruledtabular}
\begin{tabular}{llllllllllllllllllll}
& $L$ \\
Interface & 2 & 3 & 4 & 5 & 6 & 7 & 8 & 9 & 10 & 11 & 12 & 13 & 14 & 15 & 16 & 17 & 18 & 19 & 20 \\
\hline
$L^{(9,9,\overline{9},\overline{9})}_\textrm{2F;1}$ & $3^{8}$ & $1$ & $3^{16}$ & $1$ & $3^{24}$ & $1$ & $3^{32}$ & $1$ & $3^{40}$ & $1$ & $3^{48}$ & $1$ & $3^{56}$ & $1$ & $3^{64}$ & $1$ & $3^{72}$ & $1$ & $3^{80}$ \\
$L^{(9,9,\overline{9},\overline{9})}_\textrm{2F;2}$ & $3^{8}$ & $3^{6}$ & $3^{16}$ & $3^{10}$ & $3^{24}$ & $3^{14}$ & $3^{32}$ & $3^{18}$ & $3^{40}$ & $3^{22}$ & $3^{48}$ & $3^{26}$ & $3^{56}$ & $3^{30}$ & $3^{64}$ & $3^{34}$ & $3^{72}$ & $3^{38}$ & $3^{80}$ \\
$L^{(9,9,\overline{9},\overline{9})}_\textrm{2F;3}$ & $3^{8}$ & $3^{6}$ & $3^{16}$ & $3^{10}$ & $3^{24}$ & $3^{14}$ & $3^{32}$ & $3^{18}$ & $3^{40}$ & $3^{22}$ & $3^{48}$ & $3^{26}$ & $3^{56}$ & $3^{30}$ & $3^{64}$ & $3^{34}$ & $3^{72}$ & $3^{38}$ & $3^{80}$ \\
$L^{(9,9,\overline{9},\overline{9})}_\textrm{2F;4}$ & $3^{8}$ & $3^{12}$ & $3^{16}$ & $3^{20}$ & $3^{24}$ & $3^{28}$ & $3^{32}$ & $3^{36}$ & $3^{40}$ & $3^{44}$ & $3^{48}$ & $3^{52}$ & $3^{56}$ & $3^{60}$ & $3^{64}$ & $3^{68}$ & $3^{72}$ & $3^{76}$ & $3^{80}$ \\
$L^{(9,9,\overline{9},\overline{9})}_\textrm{2F;5}$ & $3^{4}$ & $3^{6}$ & $3^{8}$ & $3^{10}$ & $3^{12}$ & $3^{14}$ & $3^{16}$ & $3^{18}$ & $3^{20}$ & $3^{22}$ & $3^{24}$ & $3^{26}$ & $3^{28}$ & $3^{30}$ & $3^{32}$ & $3^{34}$ & $3^{36}$ & $3^{38}$ & $3^{40}$ \\
$L^{(9,9,\overline{9},\overline{9})}_\textrm{2F;6}$ & $3^{4}$ & $1$ & $3^{8}$ & $1$ & $3^{12}$ & $1$ & $3^{16}$ & $1$ & $3^{20}$ & $1$ & $3^{24}$ & $1$ & $3^{28}$ & $1$ & $3^{32}$ & $1$ & $3^{36}$ & $1$ & $3^{40}$ \\
\end{tabular}
\end{ruledtabular}
\caption{Ground-state degeneracy of cellular topological states built from the $U(1)_9$ topological orders on a $L_x \times L \times L$ torus of the square grid.}
\label{tab:GSDU9}
\end{table}

\begin{table}
\begin{ruledtabular}
\begin{tabular}{llllllllllllllllllll}
& $L$ \\
Interface & 2 & 3 & 4 & 5 & 6 & 7 & 8 & 9 & 10 & 11 & 12 & 13 & 14 & 15 & 16 & 17 & 18 & 19 & 20 \\
\hline
$L^{(3,3,3,\overline{3},\overline{3},\overline{3})}_\textrm{3F;1}$ & $3^{2}$ & $3^{6}$ & $3^{2}$ & $3^{2}$ & $3^{6}$ & $3^{2}$ & $3^{14}$ & $3^{18}$ & $3^{2}$ & $3^{2}$ & $3^{6}$ & $3^{14}$ & $3^{2}$ & $3^{6}$ & $3^{14}$ & $3^{2}$ & $3^{18}$ & $3^{2}$ & $3^{2}$ \\
$L^{(3,3,3,\overline{3},\overline{3},\overline{3})}_\textrm{3F;2}$ & $3^{4}$ & $3^{3}$ & $3^{10}$ & $3^{5}$ & $3^{16}$ & $3^{7}$ & $3^{22}$ & $3^{9}$ & $3^{28}$ & $3^{11}$ & $3^{34}$ & $3^{13}$ & $3^{40}$ & $3^{15}$ & $3^{46}$ & $3^{17}$ & $3^{52}$ & $3^{19}$ & $3^{58}$ \\
$L^{(3,3,3,\overline{3},\overline{3},\overline{3})}_\textrm{3F;3}$ & $3^{4}$ & $3^{3}$ & $3^{10}$ & $3^{5}$ & $3^{16}$ & $3^{7}$ & $3^{22}$ & $3^{9}$ & $3^{28}$ & $3^{11}$ & $3^{34}$ & $3^{13}$ & $3^{40}$ & $3^{15}$ & $3^{46}$ & $3^{17}$ & $3^{52}$ & $3^{19}$ & $3^{58}$ \\
$L^{(3,3,3,\overline{3},\overline{3},\overline{3})}_\textrm{3F;4}$ & $3^{2}$ & $1$ & $3^{2}$ & $1$ & $3^{6}$ & $1$ & $3^{14}$ & $1$ & $3^{2}$ & $1$ & $3^{6}$ & $3^{12}$ & $3^{2}$ & $1$ & $3^{14}$ & $1$ & $3^{18}$ & $1$ & $3^{2}$ \\
$L^{(3,3,3,\overline{3},\overline{3},\overline{3})}_\textrm{3F;5}$ & $3^{4}$ & $3^{3}$ & $3^{10}$ & $3^{5}$ & $3^{16}$ & $3^{7}$ & $3^{22}$ & $3^{9}$ & $3^{28}$ & $3^{11}$ & $3^{34}$ & $3^{13}$ & $3^{40}$ & $3^{15}$ & $3^{46}$ & $3^{17}$ & $3^{52}$ & $3^{19}$ & $3^{58}$ \\
$L^{(3,3,3,\overline{3},\overline{3},\overline{3})}_\textrm{3F;6}$ & $3^{2}$ & $1$ & $3^{2}$ & $1$ & $3^{6}$ & $1$ & $3^{14}$ & $1$ & $3^{2}$ & $1$ & $3^{6}$ & $3^{12}$ & $3^{2}$ & $1$ & $3^{14}$ & $1$ & $3^{18}$ & $1$ & $3^{2}$ \\
\end{tabular}
\end{ruledtabular}
\caption{Ground-state degeneracy of cellular topological states built from the $U(1)_3$ topological orders on a $L_x \times L \times L$ torus of the triangular grid with the pattern A.}
\label{tab:GSDU3triA}
\end{table}

\begin{table}
\begin{ruledtabular}
\begin{tabular}{llllllllllllllllllll}
& $L$ \\
Interface & 2 & 3 & 4 & 5 & 6 & 7 & 8 & 9 & 10 & 11 & 12 & 13 & 14 & 15 & 16 & 17 & 18 & 19 & 20 \\
\hline
$L^{(3,3,3,\overline{3},\overline{3},\overline{3})}_\textrm{3F;1}$ & $3^{2}$ & $3^{6}$ & $3^{6}$ & $3^{2}$ & $3^{6}$ & $3^{2}$ & $3^{6}$ & $3^{18}$ & $3^{10}$ & $3^{2}$ & $3^{18}$ & $3^{2}$ & $3^{2}$ & $3^{6}$ & $3^{6}$ & $3^{2}$ & $3^{18}$ & $3^{2}$ & $3^{14}$ \\
$L^{(3,3,3,\overline{3},\overline{3},\overline{3})}_\textrm{3F;2}$ & $3^{4}$ & $3^{3}$ & $3^{6}$ & $3^{5}$ & $3^{12}$ & $3^{7}$ & $3^{14}$ & $3^{9}$ & $3^{20}$ & $3^{11}$ & $3^{22}$ & $3^{13}$ & $3^{28}$ & $3^{15}$ & $3^{30}$ & $3^{17}$ & $3^{36}$ & $3^{19}$ & $3^{38}$ \\
$L^{(3,3,3,\overline{3},\overline{3},\overline{3})}_\textrm{3F;3}$ & $3^{4}$ & $3^{3}$ & $3^{6}$ & $3^{5}$ & $3^{12}$ & $3^{7}$ & $3^{14}$ & $3^{9}$ & $3^{20}$ & $3^{11}$ & $3^{22}$ & $3^{13}$ & $3^{28}$ & $3^{15}$ & $3^{30}$ & $3^{17}$ & $3^{36}$ & $3^{19}$ & $3^{38}$ \\
$L^{(3,3,3,\overline{3},\overline{3},\overline{3})}_\textrm{3F;4}$ & $3^{2}$ & $1$ & $3^{6}$ & $1$ & $3^{6}$ & $1$ & $3^{6}$ & $1$ & $3^{10}$ & $1$ & $3^{18}$ & $3^{12}$ & $3^{2}$ & $1$ & $3^{6}$ & $1$ & $3^{18}$ & $1$ & $3^{14}$ \\
$L^{(3,3,3,\overline{3},\overline{3},\overline{3})}_\textrm{3F;5}$ & $3^{4}$ & $3^{3}$ & $3^{6}$ & $3^{5}$ & $3^{12}$ & $3^{7}$ & $3^{14}$ & $3^{9}$ & $3^{20}$ & $3^{11}$ & $3^{22}$ & $3^{13}$ & $3^{28}$ & $3^{15}$ & $3^{30}$ & $3^{17}$ & $3^{36}$ & $3^{19}$ & $3^{38}$ \\
$L^{(3,3,3,\overline{3},\overline{3},\overline{3})}_\textrm{3F;6}$ & $3^{2}$ & $1$ & $3^{6}$ & $1$ & $3^{6}$ & $1$ & $3^{6}$ & $1$ & $3^{10}$ & $1$ & $3^{18}$ & $3^{12}$ & $3^{2}$ & $1$ & $3^{6}$ & $1$ & $3^{18}$ & $1$ & $3^{14}$ \\
\end{tabular}
\end{ruledtabular}
\caption{Ground-state degeneracy of cellular topological states built from the $U(1)_3$ topological orders on a $L_x \times L \times L$ torus of the triangular grid with the pattern B.}
\label{tab:GSDU3triB}
\end{table}